\documentclass[a4paper,11pt,twoside]{ThesisStyle}

\usepackage{lscape}

\usepackage{cancel}

\usepackage{amsmath,amssymb}             
\usepackage[latin1]{inputenc}
\usepackage[T1]{fontenc}
\usepackage[left=1.5in,right=1.3in,top=1.1in,bottom=1.1in,includefoot,includehead,headheight=13.6pt]{geometry}

\usepackage[nottoc, notlof, notlot]{tocbibind}
\usepackage{minitoc}
\usepackage{aecompl}

\usepackage[intoc]{nomencl}

\makenomenclature

\usepackage{ifpdf}

\ifpdf
  \usepackage[pdftex]{graphicx}
  \DeclareGraphicsExtensions{.jpg}
  \usepackage[a4paper,pagebackref,hyperindex=true]{hyperref}
\else
  \usepackage{graphicx}
  \DeclareGraphicsExtensions{.ps,.eps}
  \usepackage[a4paper,dvipdfm,pagebackref,hyperindex=true]{hyperref}
\fi

\graphicspath{{.}{images/}}

\usepackage{color}
\definecolor{linkcol}{rgb}{0,0,0.4} 
\definecolor{citecol}{rgb}{0.5,0,0} 

\hypersetup
{
bookmarksopen=true,
pdftitle="Probing eV-scale Axions with a MICROMEGAS Detector in the CAST Experiment.",
pdfauthor="Javier GALAN", 
pdfsubject="CAST MICROMEGAS Detectors", 
pdfmenubar=true, 
pdfhighlight=/O, 
colorlinks=true, 
pdfpagemode=None, 
pdfpagelayout=SinglePage, 
pdffitwindow=true, 
linkcolor=linkcol, 
citecolor=citecol, 
urlcolor=linkcol 
}

\usepackage{rotating}                    
\usepackage{fancyhdr}                    

\let\headruleORIG\headrule
\renewcommand{\headrule}{\color{black} \headruleORIG}

\usepackage{colortbl}
\arrayrulecolor{black}

\fancypagestyle{plain}{
  \fancyhead{}
  \fancyfoot{}
  
}

\usepackage{algorithm}
\usepackage[noend]{algorithmic}

\makeatletter

\def\cleardoublepage{\clearpage\if@twoside \ifodd\c@page\else%
  \hbox{}%
  \thispagestyle{empty}
  \newpage%
  \if@twocolumn\hbox{}\newpage\fi\fi\fi}

\makeatother
 
{%

\hrulefill
\vspace*{0.5cm}%
\end{minipage}
}

\let\minitocORIG\minitoc
\renewcommand{\minitoc}{\minitocORIG \vspace{1.5em}}

\usepackage{multirow}
\usepackage{slashbox}

{ \begin{list}%
	{$\bullet$}%
	{\setlength{\labelwidth}{25pt}%
	 \setlength{\leftmargin}{30pt}%
	 \setlength{\itemsep}{\parsep}}}%
{ \end{list} }

\renewcommand{\epsilon}{\varepsilon}

\begin{document}

\begin{titlepage}
\begin{center}
\noindent {\LARGE \textbf{Probing eV-mass scale Axions with a Micromegas\\
\vspace*{0.2cm}
\noindent Detector in the CAST Experiment.}} \\
\vspace{2.2cm}
\noindent \large memoria presentada por\\
\noindent \large {\bf Javier A. Gal\'an Lacarra}\\
\noindent \large para optar al grado de doctor\\
\noindent \large en F\'isica\\
\vspace*{1.5cm}
{\centering \resizebox{0.3\textwidth}{!} {\includegraphics{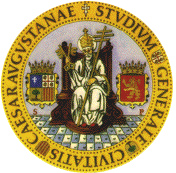}} \par} 
\vspace*{1.5cm}
\noindent \large {\bf Laboratorio de F\'isica Nuclear y Astropart\'iculas}\\
\noindent \large \'Area de F\'isica At\'omica, Molecular y Nuclear\\
\noindent \large Departamento de F\'isica Te\'orica\\
\vspace*{0.5cm}
\noindent \large {\bf Universidad de Zaragoza}\\
\vspace*{0.2cm}
\noindent \large { 28 de Enero de 2011}\\

\end{center}
\end{titlepage}
\sloppy

\titlepage

\dominitoc

\pagenumbering{roman}

 \cleardoublepage



\tableofcontents

\mainmatter

\chapter{The Axion search}
\label{chap:intro}
\minitoc

\section{Introduction}

The axion is a hypothetical neutral pseudoscalar particle which was already predicted around 1977. This new weakly interacting particle came out as a simple solution to the CP problem of strong interactions~\cite{PhysRevLett.38.1440}.

\vspace{0.2cm}

The particular properties of the axion can be restricted by the actual observational consequences that its existence would imply in astrophysics and cosmology. Moreover, the fact that the axion would have a non vanishing mass places it as good dark matter candidate under some conditions. The idea of the existence of the axion is particularly attractive because of the way it brings cosmology, astrophysics and particle physics together so elegantly.

\vspace{0.2cm}

During last years several experiments have been competing for the search of the axion, based on laboratory searches, galactic halo axion searches, and solar axion searches, which will be briefly reviewed in this chapter. Furthermore, the basis for solar axion search with an axion helioscope which is the main topic of this thesis will be discussed.

\section{The strong CP problem.}

The theory of Quantum Chromodynamics (QCD) describes the mechanism of the strong interactions. The QCD theory is non-abelian since gluons carry color charge causing them to interact with each other in a more complicated structure, SU(3), than in other field theories like Quantum Electrodynamics (QED), U(1). However, a problem arises in QCD and is called the strong CP problem. As in weak interactions, the CP-symmetry is expected to be violated in strong interactions. The CP-violation in QCD is not observed experimentally, and the fact that QCD theory does not forbid a CP-violating term leads to a fine tuning problem in its mathematical description. 

\subsection{The U(1)$_A$ problem.}
The theory of Quantum Chromodynamics describes the interactions of quarks and gluons which carry color charge, the Lagrangian of QCD can then be written as

\begin{equation}
\mathcal{L}_{QCD} = \sum_n \bar{q}(\gamma^\mu i D_{\mu} - m) q_n - \frac{1}{4} G^a_{\mu\nu} G_a^{\mu\nu}
\end{equation}

\noindent where $q_n$ are the quark fields with the quark flavor $n$, $m$ the current quark masses and $G_a$ is the gluons field tensor with the gluons color index $a = 1, ... , 8$. $D_\mu$ is the covariant derivative and it is defined as $D_\mu = \partial_\mu - igT^aG_\mu^a$ with the coupling constant $g$ and the generators of the SU(3) group $T^a$.

\vspace{0.2cm}

In the limit of vanishing quark masses $\mathcal{L}_{QCD}$ is invariant under global axial and vector transformations. Experimentally one finds that the vector symmetry $U(1)_V$ is a good symmetry of nature and leads to baryon number conservation. The axial symmetry $U(1)_A$ should lead to a symmetry between left and right handed quarks, which has not been observed in nature.

\vspace{0.2cm}

The expected spontaneous symmetry breaking of $U(1)_A$ leads to the existence of eight massless Goldstone bosons. In the case of non vanishing quark masses the expected Goldstone bosons acquire mass, and the the pseudoscalar octet can be related with the experimentally observed bosons $\pi$, $K$ and $\eta$. However, the theory also predicts the existence of another lightest isoscalar pseudoscalar particle which mass must be less than $m_L \leq \sqrt{3}m_\pi$.
\vspace{0.2cm}

The best candidate for the missing boson would be the $\eta'$ which has the right quantum numbers, however is too heavy since its mass $m_{\eta'} = 957.78$\,MeV disagrees with the mass prediction since $m_\pi \sim 135$\,MeV. This is known as the $U(1)_A$ problem~\cite{PhysRevD.11.3583}. The non observed boson could produce the required corrections in the form of a change in the bare quark mass matrix leading to conserve parity, strangeness, charm, etc. Moreover, the theory does not discard the existence of any extra unitary pseudoscalar particles, which would constrain even more the mass limit imposed in $m_L$.

\vspace{0.2cm}

\subsection{The strong CP problem and its solutions}

The $U(1)_A$ problem can be solved by adding a more complex structure to the QCD Lagrangian description. A solution was proposed by 't Hooft~\cite{PhysRevD.14.3432,PhysRevLett.37.8} that bypassed the problem by introducing an anomalous breaking of $U(1)_A$, which resulted in an additional term $\mathcal{L}_\theta$ to the Lagrangian

\begin{equation}
\mathcal{L}_\theta = \theta \frac{g^2}{32\pi^2} G_{\mu\nu}^a \tilde{G}^{\mu\nu}_a
\end{equation}

\vspace{0.2cm}
\noindent where $g$ is the coupling constant and $G_{\mu\nu}^a$ is the gluons field strength tensor. Its dual $\tilde{G}_a^{\mu\nu}$ is given by

\begin{equation}
\tilde{G}_a^{\mu\nu} = \frac{1}{2} \epsilon^{\mu\nu\rho\sigma} G^a_{\rho\sigma}
\end{equation}

\noindent and the arbitrary parameter $\theta$ is an angle between $0$ and $2\pi$, which arises from the fact that the ground state is a superposition of an infinite number of degenerate states $|n \rangle$~\cite{Turner199067}, denominated $\theta$-vacuum, and that can be expressed as

\begin{equation}
|\theta\rangle = \sum_{n=-\infty}^{\infty} e^{-in\theta}|n\rangle
\end{equation}

\noindent with the winding number $n$ which characterizes each vacuum. In order to take into account electroweak interactions the parameter $\theta$ must be transformed as follows,

\begin{equation}\label{eq:theta}
\bar{\theta} = \theta + \theta_{weak} = \theta + \mbox{arg} \left(  \mbox{det} M \right)
\end{equation}

\noindent where $M$ denotes the quark mass matrix, which transforms the new additional term of the Lagrangian

\begin{equation}
\mathcal{L}_{\bar{\theta}} = \bar{\theta} \frac{g^2}{32\pi^2} G_{\mu\nu}^a \tilde{G}^{\mu\nu}_a
\end{equation}

\vspace{0.2cm}

\noindent that is non invariant under CP-transformations, thus the effects of this CP violating term would be large unless the $\bar{\theta}$ parameter would be really small. Experimentally, this is the case, CP violation is not observed in strong interactions. The electric dipole moment of the neutron (EDMN), denoted by $d_n$, is closely related with the $\bar{\theta}$ parameter, an estimate is given in~\cite{Pospelov:2005pr},

\begin{equation}
d_{n} \sim \frac{e\bar{\theta} m_q}{m^2_n}
\end{equation}

\vspace{0.2cm}
\noindent where $m_q = m_um_d/(m_u+m_d)$ and $m_n$ is the neutron mass. The EDMN has an experimental bound of $\left|d_n\right| < 2.9\cdot10^{-26}e$\,cm (90\% C.L.)~\cite{PhysRevLett.97.131801}, which constrains the value of the parameter to be below $\bar{\theta} \leq 10^{-10}$. Such low value of $\bar{\theta}$ is allowed, however it would imply either both contributions to $\bar{\theta}$ in expression~\ref{eq:theta} are really small or that both contributions cancel them selves, resulting in a fine-tuning of both parameters. The strong CP problem arises by placing the question why the $\bar{\theta}$ parameter acquires such small value. Or in other words, why strong interactions seem to do \emph{not} violate CP when CP violation is not forbidden in the theory. 

\vspace{0.2cm}

There are mainly three possible solutions to the strong CP problem~\cite{Kim:2009xp}. The first and most unlikely idea entails zero quark masses, the second possibility is to impose CP symmetry on the QCD Lagrangian by setting $\theta = 0$. The third and most accepted option introduces an additional chiral symmetry introduced by Peccei and Quinn in 1977~\cite{PhysRevLett.38.1440}. The idea proposed to explain the strong CP problem consists in treating $\bar{\theta}$ not as a parameter but as a dynamical variable, which allows different states with a vacuum state at $\bar{\theta} = 0$. This is achieved by introducing a new global, chiral symmetry $U(1)_{PQ}$, which is spontaneously broken at the energy scale, $f_a$, producing a Goldstone boson with a non vanishing mass, the axion~\cite{PhysRevLett.40.223,PhysRevLett.40.279}. The resulting Lagrangian, which introduces the axion field $a$ and its coupling to gluons is given by 

\begin{equation}
\mathcal{L}_a = \frac{a}{f_a}\xi \frac{g^2}{32\pi^2} G_a^{\mu\nu} \tilde{G}^a_{\mu\nu} \end{equation}

\vspace{0.2cm}

\noindent where $\xi$ is a theory model-dependent parameter, and $f_a$ is a free parameter in the theory, the so-called Peccei-Quinn scale. $L_a$ is the additional contribution of the axion field to the effective potential $V_{eff}$ of the QCD Lagrangian that reaches its minimum at the axion field expectation value $\langle a \rangle$ given by

\begin{equation}
\langle a \rangle = -\frac{f_a}{\xi}\bar{\theta}
\end{equation}

\noindent which cancels the $\bar{\theta}$ term and provides a dynamical solution to the strong CP problem. Moreover, the expansion of $V_{eff}$ around its minimum does not vanish and the potential curvature leads to the following relation for the axion mass

\begin{equation}
m_a^2 = \left< \frac{\partial^2 V_{eff}}{\partial a^2} \right> = -\frac{\xi}{f_a} \frac{g^2}{32\pi^2} \frac{\partial}{\partial a} \left< G_b^{\mu\nu} \tilde{G}^b_{\mu\nu} \right>\Big{|}_{\left< a \right> = -\bar{\theta} f_a/\xi}
\end{equation}

\vspace{0.2cm}
\noindent which places the axion as the most elegant solution to the strong CP problem since the discovery of the axion would be the definitive proof that the theory is correct.

\section{The axion properties}

The axion is a neutral pseudoscalar particle, with a light mass and weakly interactions with matter, since it has not been detected yet. The properties of the new hypothetical particle only depend on the Peccei-Quinn scale $f_a$ which was introduced to solve the strong CP-Problem. The axion properties are described by the axion mass $m_a$ and the coupling to other particles $g_{ai}$, which are inversely proportional to~$f_a$

\begin{equation}
m_a \propto \frac{1}{f_a} \quad\quad g_{ai} \propto \frac{1}{f_a}.
\end{equation}

\subsection{Axion interactions}\label{sc:axionInteractions}

The different axion models include axion interactions with several fundamental particles. The axion interacts mainly with gluons and photons. However, depending on the model axions would also couple with fermions, like electrons or nucleons~\cite{Raffelt:2006cw}. The axion theory allows to describe these interactions and to express their coupling strengths as a function of $f_a$.

\subsubsection{Coupling to gluons}

The coupling of axions to gluons is described by the interaction term in the Lagrangian, by the relation

\begin{equation}
\mathcal{L}_{aG} = -\frac{\alpha_s}{8\pi f_a} a G_a^{\mu\nu} \tilde{G}^a_{\mu\nu}
\end{equation}

\vspace{0.2cm}

\noindent expressed as a function of the fine structure constant of strong interactions $\alpha_s$. The coupling to gluons makes possible the mixing with pions, which allows to obtain the axion mass by using the following expression

\begin{equation}
m_a = \frac{m_{\pi^0} f_\pi}{f_a} \left( \frac{z}{(1+z+w)(1+z)} \right)^{1/2}
\end{equation}

\vspace{0.2cm}
\noindent where $m_{\pi^0} = 135$\,MeV is the pion mass and $f_{\pi^0} = 93$\,MeV its decay constant, the parameters $z$ and $w$ are the quarks mass ratios which experimental values are~\cite{1996PhLB..378..313L}

\begin{eqnarray}
\label{eq:m_d_quarkmass}
z \equiv m_u/m_d = 0.568 \pm 0.042 \\
\label{eq:m_s_quarkmass}
w \equiv m_u/m_s = 0.029 \pm 0.003
\end{eqnarray}

\noindent values which vary depending on the specific model, and roughly lead to the axion mass relation with $f_a$

\begin{equation}
m_a \simeq 0.60 eV \frac{10^7\,\mbox{GeV}}{f_a}
\end{equation}

\subsubsection{Coupling to photons}

The mixing of axions with pions, due to the coupling to gluons, permits to describe the coupling of axions to photons by the Primakoff effect. The Lagrangian of the interaction

\begin{equation}
\mathcal{L}_{int} = -\frac{1}{4} g_{a\gamma} F_{\mu\nu} \tilde{F}^{\mu\nu}_a = g_{a\gamma} \Vec{E} \Vec{B} a
\end{equation}

\vspace{0.2cm}
\noindent where $g_{a\gamma}$ is the axion-photon coupling constant and the axion field is denoted by $a$. Further contributions appear in models in which standard fermions carry additional PQ-charges. The axion-photon coupling strength is given by the relation

\begin{equation}
g_{a\gamma} = \frac{\alpha}{2\pi f_a} \left| \frac{E}{N} - \frac{2\left( 4 + z + w \right)}{3\left( 1 + z + w \right)} \right|
\end{equation}

\vspace{0.2cm}
\noindent where $\alpha$ denotes the fine structure constant and $z$ and $w$ are the quark mass ratios mentioned before. By entering these values as given in equation~\ref{eq:m_d_quarkmass} and~\ref{eq:m_s_quarkmass} one obtains

\begin{equation}
g_{a\gamma} = \frac{\alpha}{2\pi f_a} \left| \frac{E}{N} - 1.92 \pm 0.08 \right| = \frac{\alpha}{2\pi f_a} C_\gamma
\end{equation}

\vspace{0.2cm}

The actual strength of the coupling depends on a model dependent factor $\frac{E}{N}$ which consists of the color anomaly $N \equiv \sum_f X_f$ and the electromagnetic anomaly $E \equiv 2\sum_f X_f Q_f^2 D_f$, where $X_f$ and $Q_f$ stand for the PQ and electric charges respectively. For quarks (color triplets) $D_f = 3$ and for charged leptons (color singlets) $D_f = 1$. Thus, the axion-photon coupling can be enhanced for large values of $E/N$ or suppressed if $E/N \sim 2$.

\subsubsection{Coupling to fermions}

Depending on the model, axions couple as well to fermions, like electrons and nucleons. The interaction with a fermion $f$ is described by

\begin{equation}
\mathcal{L}_{af} = \frac{C_{f}}{2f_a}\bar{\Psi}_f \gamma^\mu \gamma_5 \Psi_f \partial_\mu a
\end{equation}

\noindent where $C_f$ is an effective PQ charge of order unity, and $g_{af} = C_f m_f/f_a$ plays the role of a Yukawa coupling with the fermion mass $m_f$.

\vspace{0.2cm}

A direct coupling (at tree level) to electrons is only possible if electrons carry PQ charge and thus $C_e \neq 0$, which is the case of some axion models while in others this coupling is absent (see section~\ref{sc:axionModels}). For models which assume coupling to electrons at the tree level the contribution to the QCD Lagrangian leads to the coupling~\cite{Raffelt:2006cw}

\begin{equation}
g_{ae}^{tree} = \frac{C_e m_e}{f_a} = C_e \cdot 0.85\cdot10^{-10} m_a
\end{equation}

\vspace{0.2cm}
\noindent where $m_a$ is expressed in eV, and $C_e$ is the model dependent parameter.
\vspace{0.2cm}

Free quarks do not exist below the QCD scale $\Lambda_{QCD} \simeq 200\,$MeV, however the mixing with $\pi^0$ and $\eta$ leads to an effective coupling with nucleons with an equivalent PQ charge for protons $X'_p$ and neutrons $X'_n$ which result from $C_f \equiv X'_f/N$ where $N=\sum_{quarks} X_f$ is introduced in order to absorb the color anomaly in their definition.

\vspace{0.2cm}

The coupling strength to nucleons is related with the axion mass $m_a$ by the following expression

\begin{equation}
g_{aN} = \frac{C_N m_N}{f_a} = C_N \cdot 1.56\cdot 10^{-7} m_a
\end{equation}

\vspace{0.2cm}
\noindent where $m_a$ is expressed in eV. In the DSFZ model where $C_e \neq 0$,  the couplings to protons and neutrons are given by~\cite{Raffelt:book}

\begin{equation}
C_p = -0.10 - 0.45\,\mbox{cos}^2\beta \quad\quad C_n = -0.18 + 0.39\,\mbox{cos}^2\beta
\end{equation}

\noindent where $\beta$ is a parameterization of $C_e = \mbox{cos}^2\beta/N_f$ assuming the number of families $N_f = 3$. In the KSVZ model which only assumes axion interactions with hadrons are given by

\begin{equation}
C_p = -0.39\quad\quad C_n = -0.04.
\end{equation}

\vspace{0.2cm}
It must be noticed that in any model $C_p$ and $C_n$ never seem to vanish simultaneously.

\subsection{Axion models.}

Axion properties related to the coupling to matter vary slightly depending on the axion model used. Axions generally mix with pions, thus the axion mass and its coupling to photons must be roughly $f_\pi/f_a$ times those of $\pi^0$. However, these properties depend on the way the PQ mechanism is implemented in the model which might slightly vary the axion mass and coupling strengths. One can distinguish between two main axion model branches, which differ mainly in the size of the PQ scale $f_a$. If $f_a$ is small and thus $m_a$ large, we speak of visible axion models, if $f_a$ is large and $m_a$ small of invisible axion models.

\subsubsection{The visible axion model}

The first axion model proposed was the so-called visible or PQWW-axion\footnote{Peccei, Quinn, Weinberg, Wilczek}~\cite{PhysRevLett.40.223,PhysRevLett.40.279,PhysRevD.16.1791}. The model assumes that the axion has a decay constant which is related to the electroweak constant $f_{weak} \sim 250$\,GeV. In order to describe the PQ mechanism in this model it is required to introduce \emph{two} independent Higgs fields $\Phi_1$ and $\Phi_2$ with vacuum expectation values $f_1/\sqrt{2}$ and $f_2/\sqrt{2}$ which obey $\sqrt{f_1^2 + f_2^2} = f_{weak}$. In the PQWW model, $\Phi_1$ gives mass to the up-quarks and $\Phi_2$ to the down-quarks. The PQ scale factor is then calculated to be $f_a \lesssim 42$\,GeV, which implies an axion mass of $m_a \approx 200$\,keV. This axion model has been excluded by experiments and by astrophysical considerations, as the estimation of the theoretical branching ratio~(BR) of the $K^+$ decay~\cite{Antoniadis198267}, that in the context of the PQWW model is given by,

\begin{equation}
BR\left(K^+ \rightarrow \pi^+ + a\right) \gtrsim 3.5\cdot10^{-5}
\end{equation}

\vspace{0.2cm}
\noindent which is much higher than the experimental limit

\begin{equation}
BR\left(K^+ \rightarrow \pi^+ + a\right) \lesssim 1.4\cdot10^{-6}
\end{equation}

\vspace{0.2cm}
\noindent excluding the possible existence of the PQWW axion.

\subsubsection{Invisible axion models}\label{sc:axionModels}

Since the PQWW has not been observed invisible axion models introduce a higher scale factor $f_a \gg f_{weak}$, which can be achieved by introducing an electroweak Higgs field $\Phi$ with a vacuum expectation value $f_a/\sqrt{2}$. Due to the higher $f_a$ which acquires in these models, the axion interacts even weaker, its mass would be very small and its lifetime quite long, which makes its experimental detection more difficult. There are \emph{two} main models of invisible axions, which mainly differ in the description of the coupling to matter.

\vspace{0.4cm}
\noindent{\bf The KSVZ model}
\vspace{0.2cm}

The first invisible axion model was introduced by Kim, Shifman, Vainshtein and Zakharov~\cite{Shifman1980493,PhysRevLett.43.103}. In this model axions decouple completely from ordinary particles, implying that interactions with ordinary matter take place through the generic axion-gluon coupling introduced by PQ. This model requires an exotic heavy quark $\mathcal{Q}$ to be introduced in the theory which couples directly to the axion, and in fact, all the loop interactions involve an interaction with this new quark. These axions are also called hadronic axions because they do not interact with electrons at tree level. In general, the coupling strength (see section~\ref{sc:axionInteractions}) depends on the model assumptions. Different implementations of the KSVZ axion leads to different values of the parameter $E/N$, which is related with the electric charge of the new heavy quark introduced

\begin{equation}
\frac{E}{N} = 6 Q_{em}^2
\end{equation}

\vspace{0.2cm}
\noindent where $Q_{em}$ can take values $Q_{em} = 2/3,-1/3,1,0$~\cite{Cheng:1995fd} which allows for the ratio of the anomalies $E/N$ to be between $0$ and $6$. The ratios commonly used are $E/N = 2$ and $E/N=6$, and the standard KSVZ model with $E/N = 0$.

\vspace{0.2cm}
The main disadvantage of this model resides in that there is no physical motivation for the introduction of a new heavy quark.

\vspace{0.4cm}
\noindent{\bf The DFSZ model}
\vspace{0.2cm}

\begin{figure}[!ht]
{\centering \resizebox{0.9\textwidth}{!} {\includegraphics{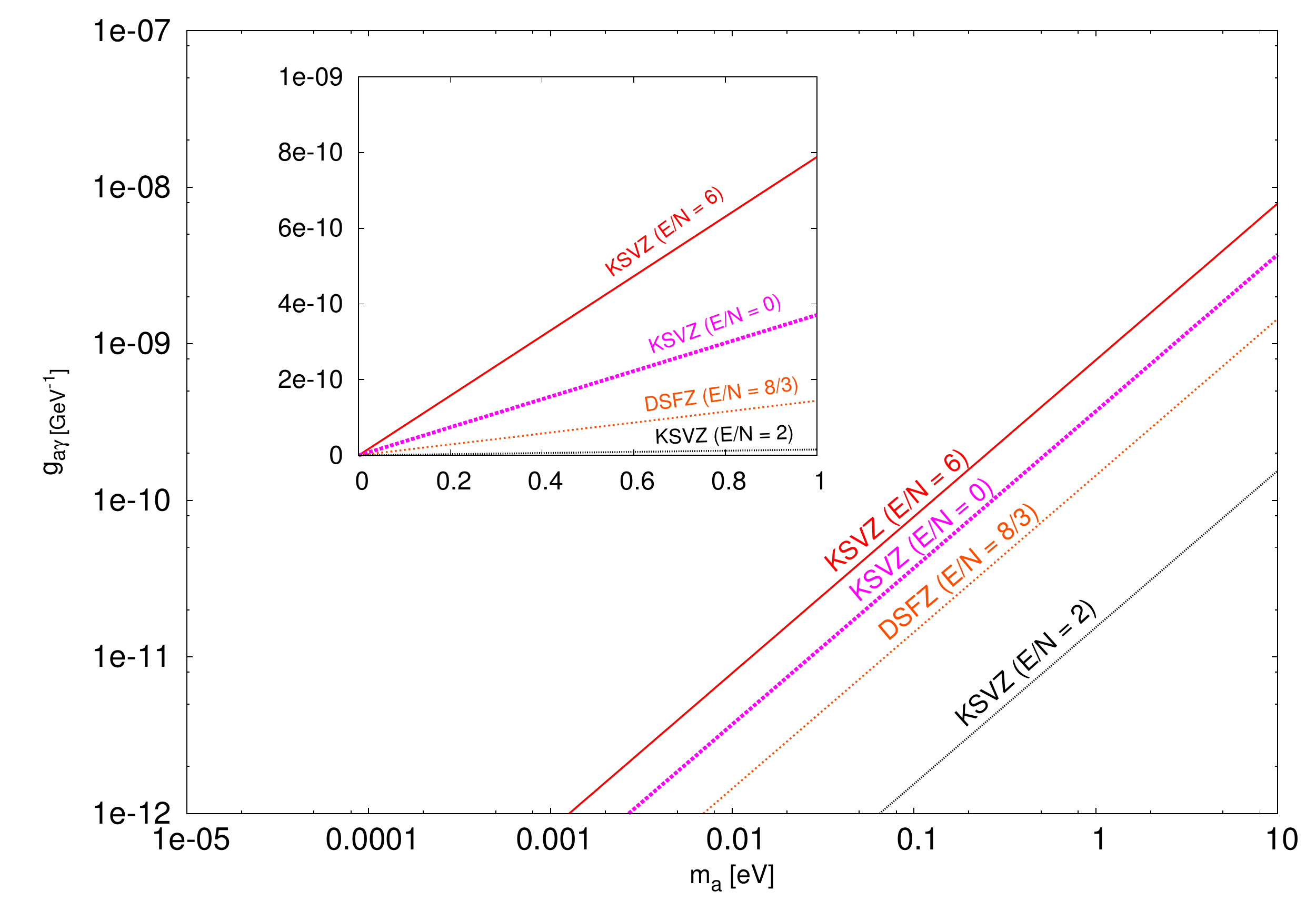}} \par}
\caption{\fontfamily{ptm}\selectfont{\normalsize{ Axion to photon coupling predicted by different axion models as a function of the axion mass. The KSVZ model is presented for the commonly used $E/N$ ratios, including the standard KSVZ model with $E/N = 1$, while the DSFZ model is obtained for the expected value $E/N = 8/3$ in GUT theories.  }}}
\label{fi:axionModels}
\end{figure}

The DFSZ model introduced by Dine, Fischler, Srednicki \cite{Dine1981199} combines properties of the standard axion model and the KSVZ model. In this model axions couple to charged leptons in addition to nucleons and photons. Because here the fundamental fermions carry PQ charge, no exotic quarks are needed and the axion couples directly to the Standard Model matter fields. One advantage of the DSFZ model is that can be easily introduced in Grand Unified Theories (GUT). The coupling of axions to photons is given by $E/N = 8/3$ for any GUT theory. Figure~\ref{fi:axionModels} shows the different axion models coupling strength to axions as a function of the axion mass.

\subsection{The axion as a candidate for dark matter.}

The Universe curvature is believed to be flat, supported by numerous cosmological and astrophysical studies. The total energy density $\rho_{tot}= 3H_o^2/8\pi G_N$ is approximately equal to the critical density to achieve this flatness $\rho_c \approx \rho_{total}$~\cite{Steffen:1141157}. Here $H_o$ is the present Hubble expansion rate and $G_N$ is Newton's constant. The contribution to the total density $\rho_c$ is coming from different sources; dark energy ($\Lambda$), non-baryonic dark matter ($dm$), baryons ($b$), photons ($\gamma$), and neutrinos ($\nu$) which fractional abundance is known~\cite{0067-0049-180-2-330}

\begin{eqnarray*}
\Omega_\Lambda \simeq 72\%,\quad\Omega_{dm} \simeq 23\%,\quad\Omega_b \simeq 4.6\%	\\
\Omega_\gamma \simeq 0.005\%,\quad0.1\% \lesssim\Omega_{\nu} \lesssim 1.5\%
\end{eqnarray*}

\noindent where the dominant dark energy fraction $\Omega_\Lambda$ is the most exotic component which is related with the expansion rate of the Universe, while the most standard component of matter, composed by baryons $\Omega_b$, still entails some unexplained phenomena related with matter-antimatter asymmetry that cannot be understood within the Standard Model. The major contribution to the neutrino component $\Omega_\nu$ is coming from cosmic neutrino background which detection is a challenging search, its uncertainty to the total contribution resides in the non well known neutrino masses which are bounded by $0.05$\,eV$\lesssim\sum_i m_{\nu_i} \lesssim \mathcal{O}(1$\,eV$)$. The last contribution, the non-baryonic matter $\Omega_{dm}$, must be composed by one or more species, denominated \emph{dark matter}, which have not been observed yet. The existence of dark matter is well fundamented in astrophysical and cosmological considerations~\cite{Bergstrom:2000pn}.

\vspace{0.2cm}

Therefore most part of the Universe is composed by an essence of unknown nature. There are many candidates for dark matter, like for example baryonic dark matter, such as MAssive Compact Halo Objects (MACHOs) and exotic particles like Weakly Interacting Massive Particles (WIMPs). The axion properties place this new particle as a promising candidate for dark matter which could coexist with those other candidates.

\vspace{0.2cm}

A dark matter candidate should be electrically neutral, color neutral, and to be stable over long periods of time, with lifetimes comparable to the age of the Universe today, $t_o \simeq 14$\,Gyr, which is about $4\cdot10^{17}$\,s. The weak axion coupling to matter allows for a low decay rate value, i.e. the lifetime of the axion governed by de decay $a\rightarrow \gamma\gamma$ is~\cite{Chen:2003gz}

\begin{equation}\label{eq:axionLifetime}
\tau_a \simeq 4.6\cdot10^{40} s\left( \frac{E}{N} - 1.92) \right)^{-2} \left( \frac{f_a/N}{10^{10}\,\mbox{GeV}} \right)^5
\end{equation}

\vspace{0.2cm}
\noindent which for $E/N = 0$ leads to an axion lifetime higher than the Universe age $\tau_a \gtrsim t_o$ for $f_a/N \gtrsim 3\cdot 10^{5}$\,GeV, and thus $m_a \lesssim 20$\,eV, region which is favored by cosmological and astrophysical constrains. Furthermore, for shorter axion lifetimes the axion population would decrease and the decay into photons would cause observable consequences, which allows to constrain the axion mass.

\vspace{0.2cm}
The influence that the axion properties would have in the large-structure formation of the Universe allows to constrain even more the axion mass, by comparing the structure formation with that induced from the Cosmic Microwave Background (CMB) provided by WMAP~\cite{1475-7516-2007-08-015} data, a bound in the axion mass is found at $m_a < 1.02$\,eV (95\% CL)~\cite{1475-7516-2010-08-001}. 

\vspace{0.2cm}

In principle, since the axion mass is an arbitrary parameter, axions could lead to contributions to the hot dark matter (HDM) component if its coupling strength is large enough, or equivalently large masses, or to the cold dark matter (CDM) component, for a larger PQ scale factor, $f_a$.

\vspace{0.2cm}

The HDM contribution from axions keeps a close analogy with thermal neutrinos production. These could have been produced in the early universe and thermalized before the QCD phase transition, which occurs at $T \sim200$\,MeV, if $f_a \lesssim 10^8$\,GeV. They could have been also produced, afterwards the QCD epoch, by thermal pions by the process $\pi + \pi \leftrightarrow \pi + a$ till start disappearing~\cite{springerlink2} at about $T \lesssim 30$\,MeV, thus these thermal axions would provide a component of the HDM background.

\vspace{0.2cm}

A deeper cosmological interest resides in the role of axions as CDM since it could account alone for a dominant contribution. Axions could have been also produced in a posterior epoch by different mechanisms, in particular they could have been abundantly produced by the \emph{misalignment mechanism}~\cite{Preskill1983127,Dine1983137,Abbott1983133}, by coherent oscillations of the axion field. The production of axions by this mechanism leads to a non negligible contribution to the cosmic dark matter density for some values of $f_a$, which is given by~\cite{PhysRevD.33.889}

\begin{equation}\label{eq:cdmDensity}
\Omega_a h^2 \approx 0.3\left( \frac{f_a}{10^{12}} \right)^{7/6}
\end{equation}

\noindent which can be compared to the total cold dark matter density $\Omega_{\mbox{\tiny CDM}}h^2 \approx 0.13$ implying that axions with mass $m_a \approx 10$\,$\mu$eV would not be a dark matter candidate but \emph{the candidate}. This value might slightly vary depending on the contribution of different production mechanisms (axionic strings, domain walls, and non-zero momentum modes of the axion field), anyhow, a conservative limit places the axion mass to be $m_a \gtrsim 10^{-5}$\,eV.

\subsection{Astrophysical bounds}

The existence of the axion would slightly affect physics related with stars evolution and well established physical processes. The axion properties can be bounded by the discrepancies that some axion parameters would entail in deeply studied astrophysical processes. The main argument resides in the fact that axions can be produced in hot and dense environments like in stars and other galactic objects. The axion production, which depends on $f_a$, provides an additional energy loss channel for the source~\cite{Raffelt:2006cw,Raffelt:book,Raffelt:998034,RevModPhys.82.557,Raffelt:1999tx}. 

\subsubsection{Solar model constrains}

The Sun, like other stars, fulfills the requirements to produce axions via the Primakoff effect. The axions scape from the Sun being an additional energy loss channel which can lead to an increased consumption of nuclear fuel and thus shorten the lifetime of the star. Since the Sun is in its halfway through its hydrogen burning phase a first constrain in the coupling $g_{a\gamma}$ is imposed by the expected axion luminosity\footnote{$g_{10} = g_{a\gamma}\, 10^{-10}$}, $L_a = g_{10}^2\,1.85\cdot10^{-3}\,L_{Sun}$ (detailed on section~\ref{sc:axionFlux}), which cannot exceed the photon luminosity, constraining $g_{a\gamma} < 3\cdot10^{-9}$\,GeV$^{-1}$. For example, an axion with coupling of about $g_{a\gamma} \approx 2-5\cdot 10^{-6}$\,GeV$^{-1}$ would imply that the Sun could live only for about 1000 years, that highlights the constraining arguments.

\vspace{0.2cm}
A more restrictive limit in the axion-photon coupling is obtained by a deeper study of the Sun physics model. The effect of the existence of axions with strong couplings would change substantially the sound speed-profile of the star, and would imply modifications in the actual abundance of helium in the core of the Sun. These arguments allow to set a lower limit at $g_{a\gamma} \lesssim 1\cdot 10^{-9}$\,GeV$^{-1}$~\cite{Schlattl:360710}. Moreover, the coupling has implications in the enhanced nuclear burning, and thus the Sun temperature distribution, which would imply an increase of the $^8$B neutrino flux. The all-flavour measured value for $^8$B neutrino flux is $4.94\cdot10^{6}$\,cm$^{-2}$s$^{-1}$ which allows to further constrain the coupling by $g_{a\gamma} \lesssim 5\cdot10^{-10}$\,GeV$^{-1}$~\cite{Bahcall:2004pz}.

\subsubsection{ Globular clusters}

\begin{figure}[!ht]
{\centering \resizebox{0.8\textwidth}{!} {\includegraphics{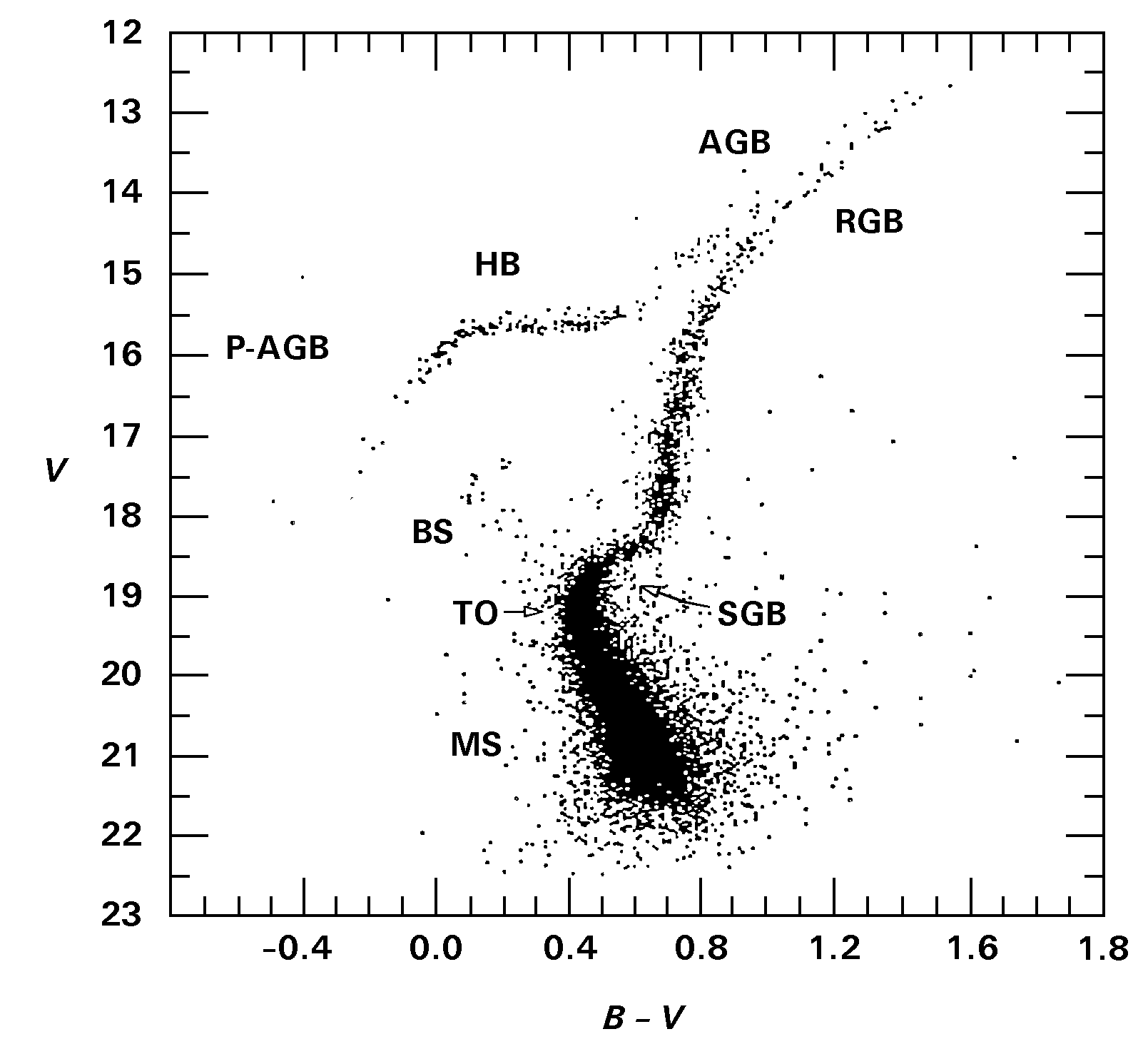}} \par}
\caption{\fontfamily{ptm}\selectfont{\normalsize{A globular cluster color-magnitude diagram where the different evolutionary stages of stars have a representative pattern distribution. The y-axis represents the brightness of the star, while the x-axis is related with the surface temperature of the star, hot stars laying at the left of this plot. In this map can be distinguished different types of stars, main sequence (MS): core hydrogen is burning, main-sequence turnoff (TO): central hydrogen is exhausted, red-giant branch (RGB): growing radius till helium ignites, and horizontal branch (HB) stars: helium burning in the core, between others. Extracted from Ref.~\cite{Raffelt:2006cw}. }}}
\label{fi:globularClusters}
\end{figure}

A globular cluster is a gravitationally linked system of stars which were formed at about the same time. This particular system is extremely useful for the study of stellar evolution since it allows to relate the evolutionary state of the star with its properties, like its mass and surface temperature. Stars can be classified by using a color-magnitude diagram (see Fig.~\ref{fi:globularClusters}) where the surface temperature, which is related with the star color luminosity, and represented by $B-V$, is compared with the total luminosity or brightness of the star, represented by $V$.

\vspace{0.2cm}
In helium burning stars, the accelerated consumption of helium due to axion production is related with the axion coupling. This consumption can be easily calculated in horizontal branch (HB) stars since their core is mainly composed of helium, and would reduce the HB lifetime by a factor $1/(1+3g_{10}^2/8)$~\cite{Raffelt:2006cw}. The lifetime of the HB stars can be measured relative to the red giant branch (RGB) time evolution, which is given by the ratio between HB stars and RGB stars. In any globular cluster this number is in agreement within a $20-40\%$, an uncertainty that is mainly coming from statistical errors, since only about 100 HB stars for each globular cluster were taken. The helium burning lifetime agrees with expectations within a 10\%~\cite{Raffelt:book} which leads to the coupling limit

\begin{equation}
g_{a\gamma\gamma} < 10^{-10}\,\mbox{GeV}^{-1}
\end{equation}

\vspace{0.2cm}
\noindent which is the most restrictive limit given by astrophysical arguments to the axion-photon coupling.

\subsubsection{ White dwarf cooling}

When a HB star reaches is helium burning end, it ascends through the red giant branch evolving to the asymptotic giant branch (AGB). AGB stars have a degenerate carbon-oxygen core and helium burning in a shell, and they might become in a white dwarf star, which first cools by neutrino emission and later by surface photon emission.

\vspace{0.2cm}
White dwarfs could also increase its cooling rate by an additional channel, in this particular case, the dominant contribution would be coming from bremsstrahlung processes where an axion is involved $e + Ze \rightarrow Ze + e + a$. The theoretical luminosity function for these stars agrees with the observed cooling rate, which is related with the measured decrease of its rotational period $\dot{P}/P$. This allows to constrain the axion coupling to electrons which describe bremsstrahlung processes due to axion losses

\begin{equation}
g_{aee} < 1.3\cdot10^{-13}
\end{equation}

\vspace{0.2cm}
\noindent which is the most restrictive limit on the axion-electron interaction.

\subsubsection{Supernova 1987A}

The core collapse of a massive star produces a proto-neutron star, a solar-mass object of such high density that even neutrinos are trapped. The collapse produces in few seconds as much radiation as the Sun will produce during all its lifetime. This process is known as supernova (SN), and produces so much light that could eclipse a galaxy luminosity.

\vspace{0.2cm}

\begin{figure}[!ht]
{\centering \resizebox{0.6\textwidth}{!} {\includegraphics{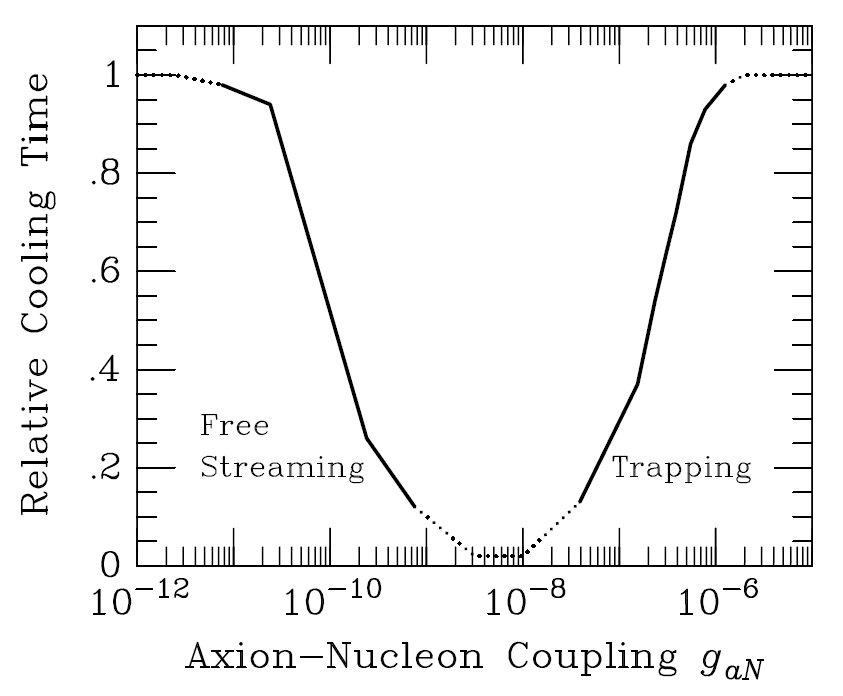}} \par}
\caption{\fontfamily{ptm}\selectfont{\normalsize{Burst duration as a function of the axion-nucleon coupling. To the left, the region mean free path of axions is much larger than the proto-neutron core, \emph{Free streaming}, while to the right, the generated axion population does not scape from the core, \emph{Trapping}. Extracted from Ref.~\cite{Raffelt:2006cw}.   }}}
\label{fi:burstDuration}
\end{figure}

Axion production in proto-neutron stars, emerging from supernovas of type II, can happen via axion-nucleon bremsstrahlung emission $N + N \rightarrow N + N + a$ in the dense nuclear medium, depending on its effective nucleon coupling $g_{aN}$. The cooling process time could be affected by this additional channel, thus shortening the burst emission duration. For very small axion-nucleon couplings the axion emission is to weak to affect significatively to the burst duration. As the coupling becomes larger the burst duration becomes shorter due to the increased axion emission, it reaches the minimum burst duration when the mean free path of axions is about the size of the SN, when the axions start to be trapped and axion emission decreases as the coupling increases, restoring the maximum burst duration (see Fig.~\ref{fi:burstDuration}).

\vspace{0.2cm}
In 1987, a supernova emission was observed at Earth, the neutrino flux going through was so high that in a short period of time a total of $20$ neutrinos $\bar{\nu}_e$ were detected, equally shared between IMB and Kamiokande II detectors. The burst duration of this neutrino flux, taking into account the statistical likelihood of arrival times and energies, was found to be within the theoretical expectation. Even though there are some uncertainties due to the low statistics, the compatibility of observations and theory allowed to discard a range of axion-nucleon couplings~\cite{Raffelt:book},

\begin{equation}
3\cdot10^{-10} \lesssim g_{aN} \lesssim 3\cdot10^{-7}
\end{equation}

\noindent since, as mentioned before, low couplings would not lead to a weak axion emission, and larger couplings would produce recombination and axion would be trapped inside the stars core.

\vspace{0.2cm}

\begin{figure}[!ht]
{\centering \resizebox{0.9\textwidth}{!} {\includegraphics[angle=270]{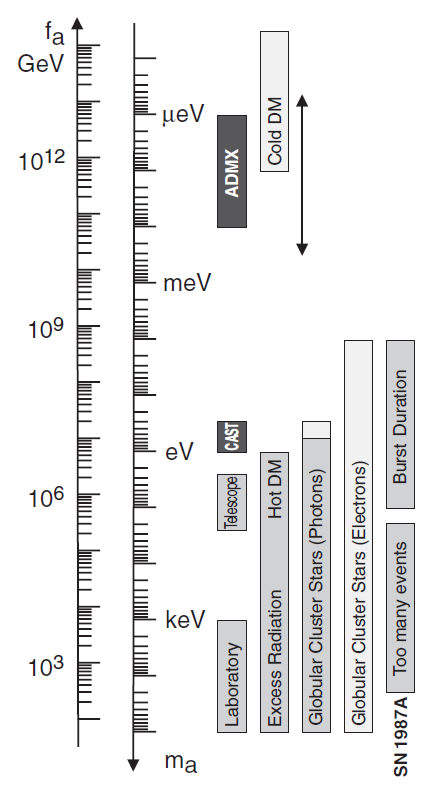}} \par}
\caption{\fontfamily{ptm}\selectfont{\normalsize{Excluded axion mass regions by different astrophysical considerations and axion searches. Dark grey bars highlight CAST and ADMX searches. Light gray exclusion bars are very model dependent. Extracted from Ref.~\cite{Raffelt:2006cw}.  }}}
\label{fi:axionRestrictions}
\end{figure}

The combination of astrophysical and cosmological bounds on axion properties, leaves only a small window for the axion mass. Except for the constrains given by supernovae, CAST is searching in that allowed axion mass window. Anyhow, it must be noticed that the low statistics and the incompatibilities in the detection of the supernova neutrinos measured by the two different detectors might lead to some uncertainties in the axion mass range excluded by supernovae. In figure~\ref{fi:axionRestrictions}  the astrophysical and cosmological bounds are summarized together with the sensitivity regions of some experimental searches.

\section{Axion searches}

The Primakoff effect describes the conversion of photons into axions in the presence of virtual photons provided by external electric or magnetic fields. These fields can be the Coulomb field of a nucleus, the electric field of charged particles in the hot plasma of stars or a magnetic field in the laboratory~\cite{PhysRevD.39.2089}. The inverse process, in which an axion is converted into a photon, provides the basic principle for most axion search experiments. First experiments to search for the invisible axion were suggested by Sikivie~\cite{PhysRevLett.51.1415} in 1983.

\vspace{0.2cm}

We can distinguish between three type of experiments that search for axions; those from galactic origin (axion haloscopes), those produced in the Sun (axion helioscopes), and also pure laboratory experiments, in which the axions are produced by lasers in strong magnetic fields.

\subsection{Haloscope searches}

Haloscope searches are mainly performed with microwave cavities. Searches based on this technique are by far the most sensitive for low mass dark matter axions. These experiments rely on converting relic axions which would be gravitationally linked to our galaxy. The axion halo mean energy is given by $E \approx m_ac^2 +\frac{1}{2}m_ac^2\beta^2$ with an expected velocity dispersion of $\Delta\beta \sim 10^{-3}$. The experimental idea is to set-up a resonant cavity where an intense magnetic field is applied at a given frequency associated to a cavity size for which axion conversion is enhanced for a narrow axion mass range. The cavity is re-sizable allowing for slight shifts on axion mass. The converted photons are measured by sensitive microwave receivers which measure the increase of power related with the increased number of photons due to axion conversion in the cavity. Moreover, the resonant cavity quality factor $Q_L$ would allow to accurately derive the velocity dispersion $\Delta\beta$ of the axion galactic flow from the signal resonance width.

\vspace{0.2cm}

A motivating feature of these experiments is that the axion mass range coverage is in the range of $\mu$eV, which is the range where the axion would be a dominant candidate for dark matter, given by relation~(\ref{eq:cdmDensity}). Therefore, the limit on the axion-photon coupling given by these experiments assumes that the dark matter halo density $\rho_{halo} \sim 5\cdot10^{-25}$g/cm$^3$ is only given by axions $\rho_a \simeq \rho_{halo}$.

\vspace{0.2cm}

\begin{figure}[!ht]
\begin{center}
{\centering \resizebox{0.9\textwidth}{!} {\includegraphics{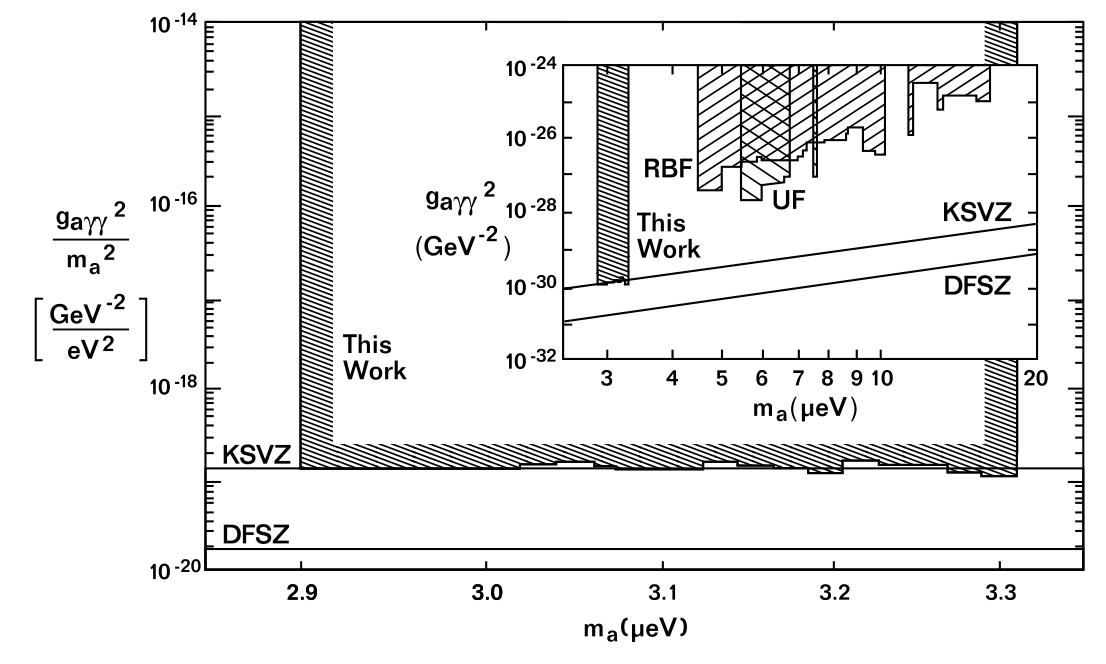}} \par} 
\end{center}
\caption{\fontfamily{ptm}\selectfont{\normalsize{ First ADMX results on the axion-photon coupling exclusion reaching the theoretical model lines, together with RBF and UF searches. Given in Ref.~\cite{PhysRevLett.80.2043}. }}}
\label{fi:admxSearch}
\end{figure}

The first experiments were carried out by Rochester-Brookhaven-Fermilab (RFB) and the University of Florida (UF) in the 1980s. RBF covered axion masses between $5.4-5.9$\,$\mu$eV~\cite{PhysRevD.40.3153}, and UF covered the range $4.5-16.3$\,$\mu$eV~\cite{PhysRevD.42.1297}, however, they were not sensitive enough to reach the theoretical axion models (see Fig.~\ref{fi:admxSearch}).

\begin{figure}[!ht]
\begin{center}
{\centering \resizebox{0.9\textwidth}{!} {\includegraphics{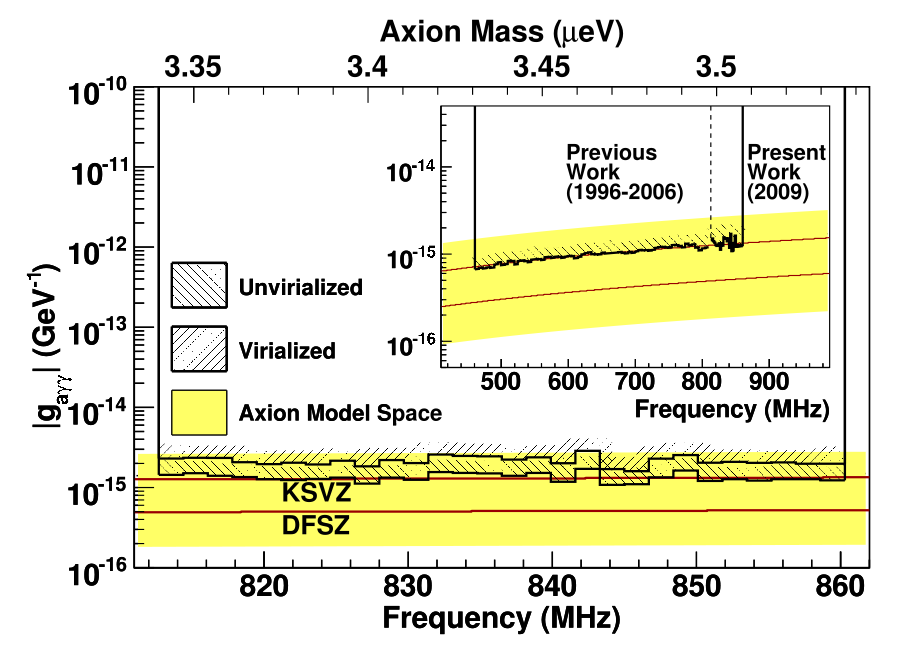}} \par} 
\end{center}
\caption{\fontfamily{ptm}\selectfont{\normalsize{ADMX axion-photon coupling exclusion for a 90\% confidence level. A local dark matter density of $0.45$\,GeV/cm$^3$ is assumed for two dark matter distribution models. Extracted from Ref.~\cite{Asztalos:2009yp}. }}}
\label{fi:admxSearch2009}
\end{figure}

Later on, the Axion Dark Matter eXperiment (ADMX) at Laurence Livermore National Laboratory (LLNL) was the first experiment providing enough sensitivity to test theoretical axion models~\cite{RevModPhys.75.777,PhysRevLett.95.091304,PhysRevD.74.012006}. A first axion mass scanning excluded the mass range~\cite{PhysRevLett.80.2043}

\begin{equation}
2.9\,\mu\mbox{eV} < m_a < 3.3\,\mu\mbox{eV}
\end{equation}

\noindent reaching the KSVZ models border line (see Fig.~\ref{fi:admxSearch}). The experiment was upgraded in order to extend the axion mass range coverage to higher masses. During this upgrade, the microwave receivers were substituted by a SQUID (Superconducting QUantum Interference Device) which offered an improvement in the scan rate by 2 orders of magnitude. The results of this new scanning phase exclude a new mass region from the favored theoretical models~\cite{Asztalos:2009yp}

\begin{equation}
3.3\,\mu\mbox{eV} < m_a < 3.53\,\mu\mbox{eV}
\end{equation}

\noindent where the axion-photon coupling limit reached for each mass is presented in figure~\ref{fi:admxSearch2009} for two different models for the axion halo density distribution.

\subsection{Laser experiments}

The advantage of laboratory experiments is their model-independence because they do not rely on physical processes and conditions in the Sun or in stars~\cite{Battesti:2007um}. Laboratory experiments make use of the fact that a small fraction of photons of an intense laser beam can convert to axions in the presence of a strong magnetic field via the Primakoff effect.

\vspace{0.2cm}

In addition to axion searches, laser-based experiments can also perform precise tests of QED and search for other exotic particles which couple to photons.  

\subsubsection{Polarization experiments}

\begin{figure}[!h]
{\centering \resizebox{0.75\textwidth}{!} {\includegraphics{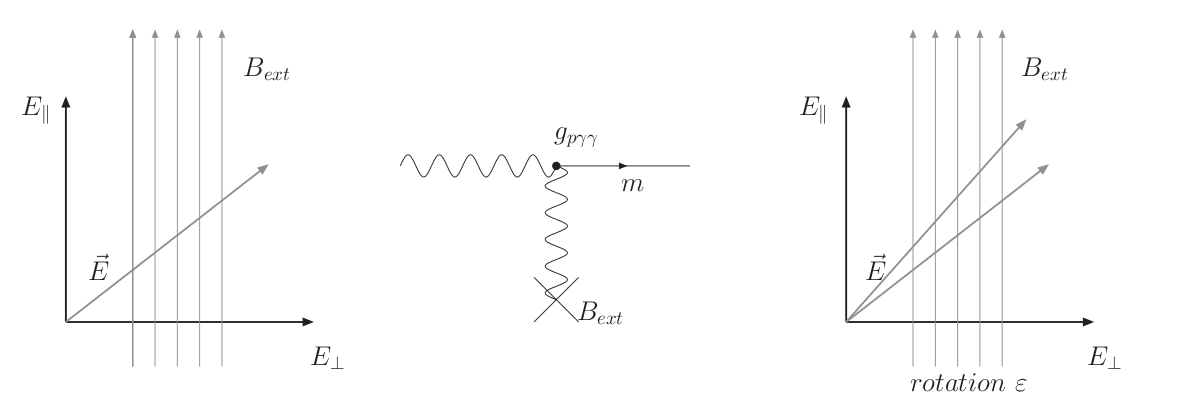}} \par}
{\centering \resizebox{0.75\textwidth}{!} {\includegraphics{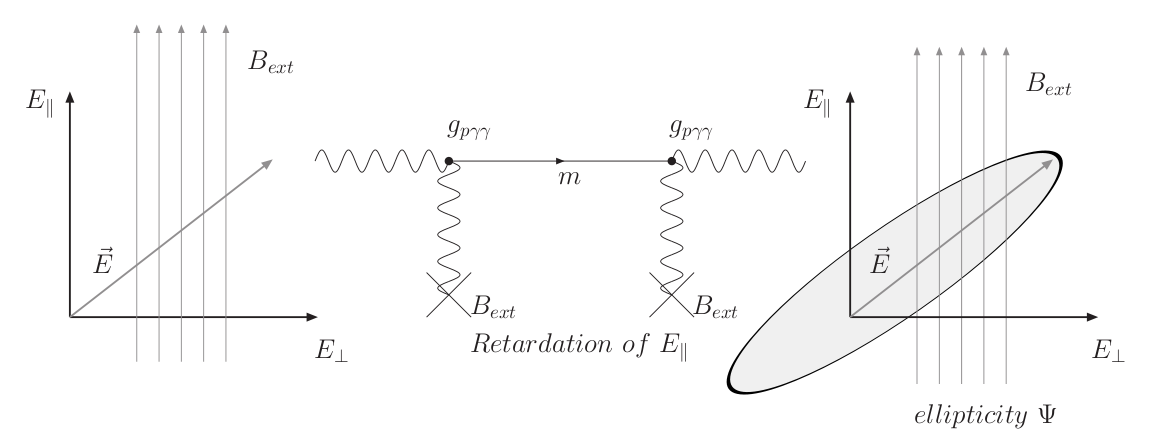}} \par}
\caption{\fontfamily{ptm}\selectfont{\normalsize{ Detection principle of laser polarization experiments. On top the expected rotation of the laser beam polarization by loss beam related with the magnetic field orientation. On bottom, the double reconversion from photon-axion and axion-photon would produce a retarding on one of the photon beam components.  }}}
\label{fi:opticalEffects}
\end{figure}

The polarization of a laser beam propagating in vacuum through a transverse magnetic field can be affected by axions and other scalar or pseudoscalar particles because of their coupling to photons. The production of axions inside the magnetic field would introduce of a small ellipticity and rotation~(see Fig.~\ref{fi:opticalEffects}), due to the higher conversion probability for the perpendicular polarization component of the photon beam. Some of the experiments are able to search for these effects and the bound obtained by the BFRT collaboration in the mass range $m_a < 5 \cdot 10^{-4}$\,eV is~\cite{PhysRevD.47.3707}

\begin{equation}
g_{a\gamma} < 3.6\cdot10^{-7} GeV^{-1} .
\end{equation}

The Polarizzazione del Vuoto LASer (PVLAS) collaboration~\cite{2008LNP...741..157C} claimed to have found a signal in 2006 but it was identified as an instrumental artifact later.

\subsubsection{Regeneration experiments}

\begin{figure}[!ht]
{\centering \resizebox{0.9\textwidth}{!} {\includegraphics{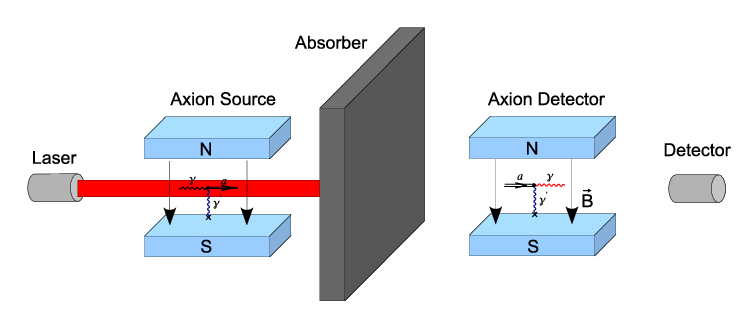}} \par}
\caption{\fontfamily{ptm}\selectfont{\normalsize{A conceptual drawing of "shinning through wall" experiments, where photons would be converted to axions in a first magnetic region and then re-converted back to photons in a second magnetic region. }}}
\label{fi:shinningWall}
\end{figure}

The so-called "shining light through a wall experiments" use the principle of photon regeneration.  If a polarized laser beam that propagates in a transverse magnetic field is blocked by a wall, there is the chance to detect photons of the same wavelength as the laser light behind the wall. This is the case if weakly-interacting particles are created before and can pass the wall, where they get into a second magnetic field responsible for the reconversion into photons (see Fig.~\ref{fi:shinningWall}).

With the first experiment of that kind the Brookhaven-Fermilab-Rutherford-Trieste (BFRT) collaboration could set an upper limit of~\cite{PhysRevD.47.3707}

\begin{equation}
g_{a\gamma} < 6.7 \cdot 10^{-7}\,\mbox{GeV}\quad\mbox{for}\quad m_a < 10^{-3}\,\mbox{eV}
\end{equation}

\vspace{0.2cm}

A new experiment like this has been started at CERN. The Optical Search for QED vacuum birefringence, Axions and photon Regeneration (OSQAR) experiment uses two $15$\,m long LHC superconducting magnets and is expected to improve the limits from BFRT by two orders of magnitude. First results of this experiment rule out the PVLAS signal~\cite{Pugnat:1076861}.

\vspace{0.2cm}

Besides the axion, additional extensions beyond the Standard Model (SM) predict the existence of a possible large number of axion-like particles (ALPs). The ALPS collaboration searches for "Weakly Interacting Sub-eV Particles" (WISPs) by using a HERA dipole magnet placed at DESY~\cite{Ehret:2010mh}. The experiment is able to detect photon-WISP-photon conversions of a few $\times10^{-25}$ resulting in the most restringent laboratory constrains in the existence of low mass axion-like particles. The motivation for the search of ALPS experiment resides in the direct experimental test of extra-dimensions properties predicted by string theory. Furthermore, in future these kind of experiments could establish the road-map for the search of QCD axions at $\sim$meV masses.

\subsection{Helioscope searches}
Helioscope searches are based on the well established Standard Solar Model. The Sun fulfills the requirements to produce an abundant amount of axions in its core. The expected solar axion flux due to the axion production at the Sun core will be detailed on section~\ref{sc:axionFlux}.

\vspace{0.2cm}
Moreover, the axion could be related with additional physic processes taking place in the Sun, as the Sun corona heating problem~\cite{DiLella2002175}, and solar axion flux could be related with the Earth magnetic field modulation~\cite{Rusov:2010ze}.

\subsubsection{Solar axion searches via Brag scattering}

One possible detection principle is to use the intense Coulomb field of nuclei in a crystal lattice to convert axions into photons.  The expected mean energy of solar axions is of the order of $\approx 4$\,keV, which results in an axion wavelength comparable to the inverse lattice spacing of the crystal.

\vspace{0.2cm}

Thus a Bragg pattern, with maxima where the Bragg condition is fulfilled, should be visible. The constructive interference results in an enhancement of the signal of the order of $10^4$ compared to the scattering of a single atom in the crystal~\cite{Paschos1994367,1998PhLB..427..235C}. As the Sun is a moving source the signal would change with time because of a varying entrance angle.  

\vspace{0.2cm}

There are several experiments located underground using different detector material such as Germanium (SOLAX) or Natriumiodide (DAMA). The achieved bounds on the axion to photon coupling~\cite{PhysRevLett.103.141802,Bernabei20016,Avignone:1997th,Morales:484536}

\vspace{0.2cm}

\begin{eqnarray*}
& g_{a\gamma} < & 2.7 \cdot 10^{-9} \mbox{GeV}^{-1}\,\mbox{(SOLAX)}\,\,\mbox{for}\quad m_a \leq 1\,\mbox{keV}\\
& g_{a\gamma} < & 1.7 \cdot 10^{-9} \mbox{GeV}^{-1}\,\mbox{(DAMA)}\\
& g_{a\gamma} < & 2.4 \cdot 10^{-9} \mbox{GeV}^{-1}\,\mbox{(CDMS)} \\
& g_{a\gamma} < & 2.8 \cdot 10^{-9} \mbox{GeV}^{-1}\,\mbox{(COSME)} 
\end{eqnarray*}

\vspace{0.2cm}

\noindent are all in the same range but cannot compete with those of e.g. helioscopes. The main advantage is that they practically do not depend on the axion mass.

\vspace{0.2cm}

CDMS uses Germanium and Silicon detectors and in addition to solar can also be used to limit galactic axions to \cite{PhysRevLett.103.141802}

\begin{equation*}
g_{ae} < 1.4 \cdot 10^{-12}\,\mbox{GeV}^{-1}
\end{equation*}

\subsubsection{Magnet searches}

\begin{figure}[!ht]
{\centering \resizebox{0.9\textwidth}{!} {\includegraphics{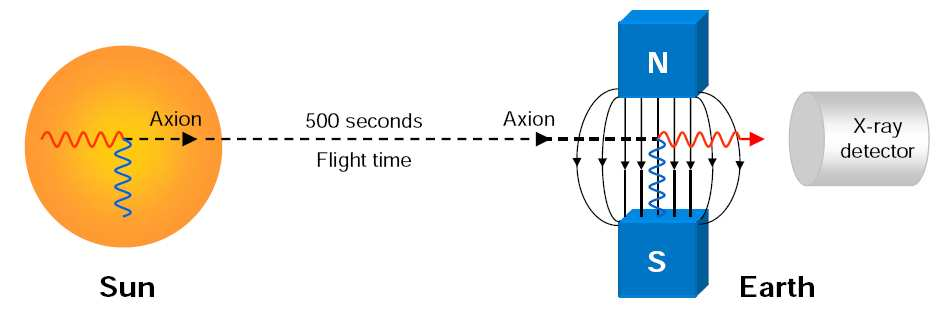}} \par}
\caption{\fontfamily{ptm}\selectfont{\normalsize{ Detection concept for the magnet helioscope idea. Where the Sun is used an intense source of axions. }}}
\label{fi:helioscopeConcept}
\end{figure}

Magnet helioscopes consist of a powerful magnet, that can be pointed to the Sun, and low-energy X-ray detectors for detecting the re-converted photons (which additionally can implement X-ray optics for signal magnification, therefore this kind of experiments are usually denominated \emph{axion telescopes}). Figure~\ref{fi:helioscopeConcept} shows the concept of magnet search detection. They are the most sensitive experiments in a wide axion mass range, this technique allows to be sensitive for axion masses up to $1$\,eV.

The probability for axions to convert into detectable X-rays depends on the length $L$ and strength $B$ of the magnetic field but is limited to certain masses due to a coherence condition (detailed on section~\ref{sc:probConv}). To be sensitive to other axion masses a buffer gas, which gives the virtual photons an effective mass, is inserted in the conversion volume. Then coherence is restored in a narrow mass range for a certain gas pressure.

\vspace{0.2cm}

The first experiment to use this technique was performed by Lazarus et al.~\cite{PhysRevLett.69.2333} reaching an axion-photon coupling sensitivity that allowed to exclude $g_{a\gamma} \lesssim 3.6\cdot 10^{-9}$\,GeV$^{-1}$ for axion masses bellow $m_a \lesssim 0.03$\,eV and $g_{a\gamma} \lesssim 7.7\cdot10^{-9}$\,GeV$^{-1}$ for the axion mass range $0.03$\,eV$\lesssim m_a \lesssim 0.11$\,eV.

\vspace{0.2cm}

The Tokyo Axion Helioscope continued the search of solar axions using this technique, this experiment provides an improved sensitivity with respect to the previous searches, setting the more restrictive limit $g_{a\gamma} \lesssim 6.0\cdot10^{-10}$\,GeV$^{-1}$ for $m_a \lesssim 0.03$\,eV~\cite{Moriyama1998147}. Furthermore, the experiment continuous the search of axions in other axion mass ranges~\cite{Inoue200218,Inoue200893}.

\vspace{0.2cm}

The CERN Axion Solar Telescope (CAST) experiment is the most sensitive axion helioscope up to date. CAST has provided the most restrictive experimental limit, which by first time was below the most restrictive astrophysical limit of HB stars. The CAST experiment is the main topic of this thesis and it will be fully described on chapter~\ref{chap:cast}, where the research physics program is described and coupling limit values reached are given.

\vspace{0.2cm}

Furthermore, recent studies look for the possibility of using the geomagnetic field as a converter of solar axions to photons~\cite{Davoudiasl:2008fy}, which could be detected by orbiting satellites implementing X-ray telescopes. Somehow, this kind of experiment pretends to "observe the Sun core through the Earth".

\section{Details on helioscope axion detection.}\label{sc:detectionDetails}

Because of its vicinity, the Sun is the most dominant axion-source in the sky. Axions would be constantly produced by the Primakoff effect in the nuclear core of the Sun. They can stream out freely so that they reach the earth after $\approx 500$\,s as an almost parallel axion beam. In a strong magnetic field the inverse Primakoff effect $a + \gamma_{virtual} \rightarrow \gamma$ reconverts them into real photons which have the same energy and momentum as the axion. The detection of these photons would be a proof for the existence of axions.

The sensitivity in the detection of axions coming from the Sun is related with the expected axion flux from the source together with the probability of detecting an axion to photon conversion.

\vspace{0.2cm}

The theoretical axion flux at Earth is very well known due to the great knowledge of our closest star, which is well described by the Standard Solar Model. The Primakoff conversion mechanism in the core of the stars is combined with the Sun properties to obtain the axion flux.

\vspace{0.2cm}

Thus, the probability of converting axions to photons, by using an intense magnetic field, will be described on detail.

\subsection{The Solar Axion Model}\label{sc:axionFlux}

Solar axions searches are based on the well established Standard Solar Model. The hot and dense plasma found in the core of stars have the right conditions for producing a non negligible amount of axions by the Primakoff effect, allowing for the conversion $a \leftrightarrow \gamma$. In stars the Primakoff process takes place thanks to the electric field of nuclei and electrons. The Primakoff process turns out to be important for non-relativistic scenarios where the energy of the generated axion is lower than the electron $T \ll m_e$, and both, electrons and nuclei can be treated as "heavy". Ignoring recoil effects the differential cross section is given by

\begin{equation}
\frac{d\sigma_{\gamma\rightarrow a}}{d\Omega} = \frac{g_{a\gamma}^2 Z^2 \alpha}{8\pi} \frac{\left| \Vec{k}_\gamma \times \Vec{k}_a \right|}{\left|\Vec{q}\right|^4}
\end{equation}

\noindent where $\Vec{q} = \Vec{k}_\gamma - \Vec{k}_a$ is the momentum transferred; while the axion and photon energies are the same. Thus, the maximum differential cross section is reached for a transversal axion-photon interaction.

\vspace{0.2cm}

In the plasma, the long range Coulomb interactions are cut off by screening effects~\cite{PhysRevD.33.897} modifying the differential cross section by a factor $\left|\Vec{q}\right|^2/(k_s^2 + \left|\Vec{q}\right|^2)$. In a non degenerate medium the screening factor is given by the Debye-H$\ddot{\mbox{u}}$ckel formula

\vspace{0.2cm}

\begin{equation}
k_s^2 = \frac{4\pi\alpha}{T} n_B \left( Y_e + \sum_j Z^2_j Y_j \right)
\end{equation}

\vspace{0.2cm}

\noindent where $n_B$ is the baryon density while $Y_e$ and $Y_j$ are the fraction of electrons and different nuclear species $j$ per baryon. The introduction of screening effects allowed to void some mathematical divergences in the calculation of the total cross section, and derive the decay rate of a photon of energy $E$ into an axion of the same energy~\cite{PhysRevD.37.1356},

\begin{equation}
\Gamma_{\gamma\rightarrow a} = \frac{g_{a\gamma}^2 T k_s^2}{32\pi}\left[ \left( 1 + \frac{k_s^2}{4E^2} \right) \mbox{log} \left( 1 + \frac{4E^2}{k_s^2} \right) - 1 \right]
\end{equation}

\vspace{0.2cm}

\noindent where the effective photon mass and the axion mass are considered negligible relative to the energy $E$. The energy loss per unit volume can be obtained by integrating the decay rate by the momentum $\Vec{k}_\gamma$ density distribution at a temperature $T$,

\begin{equation}\label{eq:energyLoss}
Q = \int \frac{2d^3\Vec{k}_\gamma}{\left(2\pi\right)^3} \frac{\Gamma_{\gamma\rightarrow a} E}{e^{E/T} - 1} = \frac{g_{a\gamma}^2 T^7}{4\pi} F(k^2)
\end{equation}

\vspace{0.2cm}

\noindent where $k \equiv k_s/2T$, and $F(k^2)$ is a value of order unity. The parameter $k$ depends mainly in the solar model, for a standard solar model as the Sun $k^2 \approx 12$ ($F = 0.98$) with variations less than $15\%$, while for the core of an HB star it is $k^2 \approx 2.5$ ($F = 1.84$).

\vspace{0.2cm}

The axion energy loss per unit volume given by equation~\ref{eq:energyLoss} can be integrated to the solar temperature distribution given by the Standard Solar Model to calculate the expected flux of axions at Earth coming from the Primakoff conversion. This calculation leads to a total axion flux of

\begin{equation}
L_a = g_{10}^2 1.7\cdot10^{-3} L_{Sun}
\end{equation} 

\begin{figure}[!ht]
{\centering \resizebox{0.95\textwidth}{!} {\includegraphics{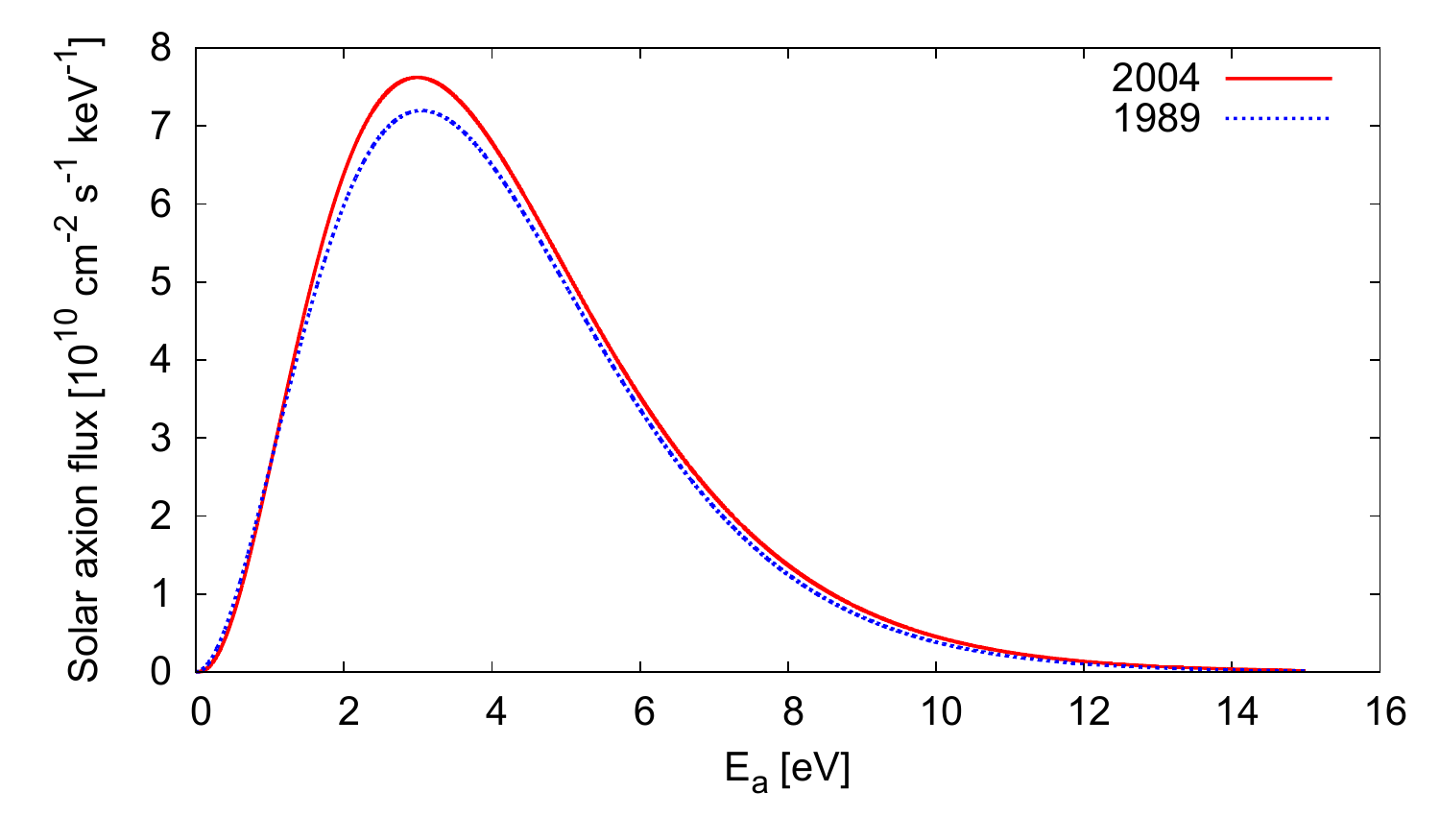}} \par}
\caption{\fontfamily{ptm}\selectfont{\normalsize{Differential solar axion flux spectrum at Earth, for the first calculation in 1989 given at Ref.~\cite{RevModPhys.54.767} and the last calculation in 2004 given in Ref.~\cite{PhysRevLett.92.121301}.  }}}
\label{fi:axionFlux}
\end{figure}

\vspace{0.2cm}
\noindent where $L_{Sun}$ is the solar luminosity and $g_{10} \equiv g_{a\gamma} \cdot 10^{10}$\,GeV. The first calculation of the solar axion flux can be found in~\cite{RevModPhys.54.767}, the last calculations~\cite{PhysRevLett.92.121301} introduce some corrections, however the final results do not defer substantially from the first calculations giving account of the well established Standard Solar model (see Fig.~\ref{fi:axionFlux}).

\vspace{0.2cm}

The differential axion solar flux at Earth, taking into account the mean Sun-Earth distance is well approximated by the relation 

\begin{equation}\label{eq:axionFlux}
\frac{d\Phi_a}{dE_a} =  g^2_{10} \, 6.02 \cdot 10^{10}\, E_a^{2.481}\, e^{-E_a/1.205} \quad \mbox{[{cm}}^{-2}\mbox{s}^{-1}\mbox{keV}^{-1}\mbox{]}
\end{equation}

\vspace{0.2cm}
\noindent where $E_a$ is measured in keV. The maximum axion intensity is reached at $3$\,keV while the average axion energy is $\langle E_a \rangle = 4.2$\,keV

\subsection{Probability of axion conversion.}\label{sc:probConv}

The expected signal at an axion helioscope depends on the number of axions converted to photons at the magnetic region. This conversion is only effective when the polarization of the outcoming photon is parallel to the magnetic field which needs to be transversal to the propagating axion wave~\cite{PhysRevD.37.1237}. The axion state, characterized by the amplitude of the axion field, $a$, propagating in a media along the $z$ axis is given by the relation,

\vspace{0.2cm}

\begin{equation}
i\partial_z
\left( \begin{array}{c}
A\\
a \end{array} \right) =
\left( \begin{array}{cc}
\frac{E_a - m_\gamma^2}{2E_a - i\Gamma/2} & \frac{g_{a\gamma} B}{2}  \\
\frac{g_{a\gamma} B}{2} & \frac{E_a - m_\gamma^2}{E_a} \end{array} \right)
\left( \begin{array}{c}
A\\
a \end{array} \right)
\end{equation}

\vspace{0.2cm}

\noindent where $A$ is the parallel photon component, and $B$ is the component of the magnetic field transversal to the propagation wave. $\Gamma$ is the inverse absorption length for X-rays in the media, and $m_\gamma$ is the effective photon mass which is played by the plasma frequency given by the buffer gas as a function of the number of electrons, which can be expressed in natural units as, 

\begin{equation}\label{eq:mgamma}
m_\gamma = \omega_p = \sqrt{\frac{4\pi\alpha n_e}{m_e}}
\end{equation}

\noindent where $n_e$ is the electron density of the buffer gas which is related with the mass density by the relation

\begin{equation}
n_e = Z\frac{N_A}{W_A} \rho
\end{equation}

\vspace{0.2cm}

\noindent and $Z$ is the corresponding atomic number and $W_A$ is the atomic weight. And allows to rewrite the relation~\ref{eq:mgamma} as a function of the buffer gas density

\begin{equation}\label{eq:mgamma}
m_\gamma \simeq 28.77 \sqrt{\frac{Z}{W_A}\rho\left[ \frac{\mbox{g}}{\mbox{cm}^3} \right]}\,\mbox{eV}
\end{equation}

\vspace{0.2cm}
\noindent where $W_A$ stands for the atomic weight expressed in g/mol.

\vspace{0.2cm}
The conversion probability of axions to photons traveling through a transversal and homogeneous magnetic field $B$ over a total coherence length $L$ was calculated in~\cite{PhysRevD.39.2089} which more general expression can be written as

\begin{equation}\label{eq:probConversion}
P_{a\rightarrow\gamma} = \frac{\left(g_{a\gamma} BL/2\right)^2}{L^2\left( q^2 + \Gamma^2/4 \right)} \left[ 1 + e^{-\Gamma L} - 2e^{-\Gamma L /2} \mbox{cos}\left(qL\right) \right]
\end{equation}

\vspace{0.2cm}
\noindent where $m_\gamma$ has been encoded in the momentum transferred $q$ in the axion-photon interaction which is calculated by the relation

\begin{equation}
q = \left| \frac{m_a^2 - m_\gamma^2}{2E_a} \right|.
\end{equation}

\vspace{0.2cm}
The probability conversion given in relation~\ref{eq:probConversion} can be reformulated and expressed in terms of more suitable experimental units

\begin{equation}
P_{a\rightarrow\gamma} = 1.6278\cdot10^{-17}\left( \frac{BL}{B_{cb}L_{cb}} \right)^2 \left( \frac{ g_{a\gamma} }{10^{-10}\,\mbox{GeV}^{-1}} \right)^2 \mathcal{M}(q,L)
\end{equation}

\vspace{0.2cm}
\noindent where the normalizing factor arises from using the nominal CAST magnetic field intensity $B_{cb} = 8.802$\,T and coherence length $L_{cb} = 9.26$\,m. While $\mathcal{M}$ is a term which quantifies the coherence of the interaction for a given axion mass $m_a$ at a given detection conditions imposed by the buffer medium at the magnetic region which fixes $m_\gamma$ and $\Gamma$,

\begin{equation}\label{eq:Fql}
\mathcal{M}(q,L) = \frac{1}{L^2\left( q^2 + \Gamma^2/4 \right)} \left[ 1 + e^{-\Gamma L} - 2e^{-\Gamma L /2} \mbox{cos}\left(qL\right) \right].
\end{equation}

\vspace{0.2cm}
In the case of the absence of buffer gas ($m_\gamma = 0$ and $\Gamma = 0$) the equation~\ref{eq:Fql} is reduced to the following expression,

\begin{equation}
\mathcal{M}(q,L) = \frac{2}{\left( qL\right)^2} \left[ 1 - \mbox{cos}\left(qL\right) \right].
\end{equation}

\noindent where $q = m_a^2/2E_a$ enhancing the conversion probability for axion masses $m_a \lesssim 0.02$\,eV.

\vspace{0.2cm}

\begin{figure}[!ht]
{\centering \resizebox{0.75\textwidth}{!} {\includegraphics[angle=90]{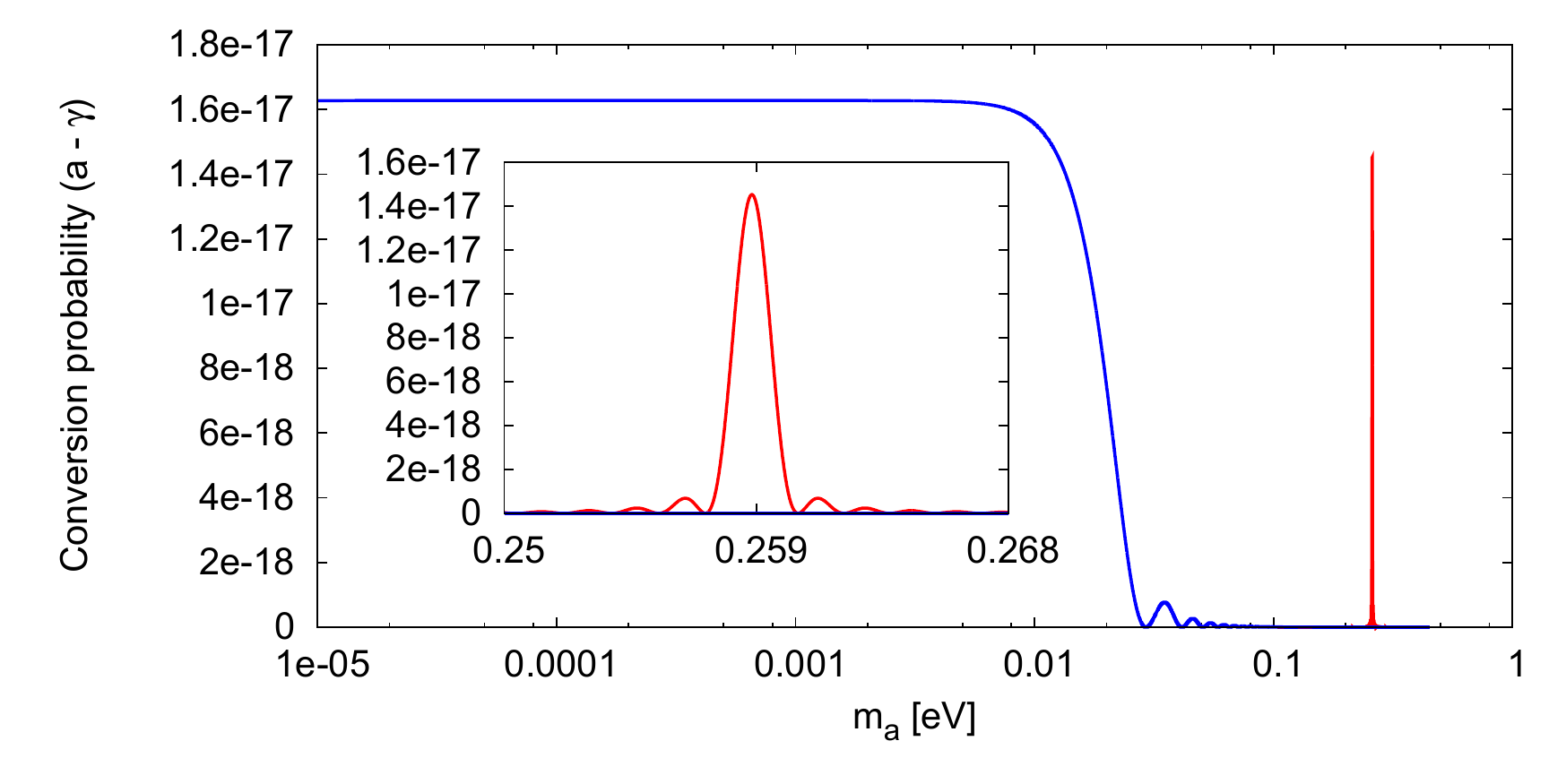}} \par}
\caption{\fontfamily{ptm}\selectfont{\normalsize{ Axion-photon conversion probability, for axions with an energy $E_a = 3$\,keV, in a transversal homogeneous magnetic field of intensity $B_{cb} = 8.802$\,T over a coherence length of $L_{cb} = 9.26$\,m. Two scenarios are presented, vacuum in the magnet bores and magnet bores filled with $^3$He gas at an equivalent pressure $P@1.8K = 6$\,mbar, also a zoom at the coherence mass region with buffer gas inside the bores is shown.   }}}
\label{fi:convProbability}
\end{figure}

The effect of the buffer medium allows to recover the probability conversion for axion masses higher than $m_a \gtrsim 0.02$\,eV in a narrow axion mass range. Fact that is used by helioscope experiment searches, and that allows to cover a full axion mass range by measuring overlapping mass resonances produced by increasing the buffer gas density by small quantities. Thus, this technique for covering wide mass ranges requires long data taking periods due to the short axion mass coverage of a single resonance (i.e. the full width half maximum is about $\simeq2$\,meV at a resonance with maximum at $0.258$\,eV) given by a fixed density in the magnet bores (see Fig.~\ref{fi:convProbability}).

\chapter{The CAST experiment. Technical upgrades and maintenance.}
\label{chap:cast}
\minitoc

\section{The CAST Experiment for Solar Axions Search.}

\begin{figure}
\begin{tabular}{cc}
\includegraphics[width=0.6\textwidth]{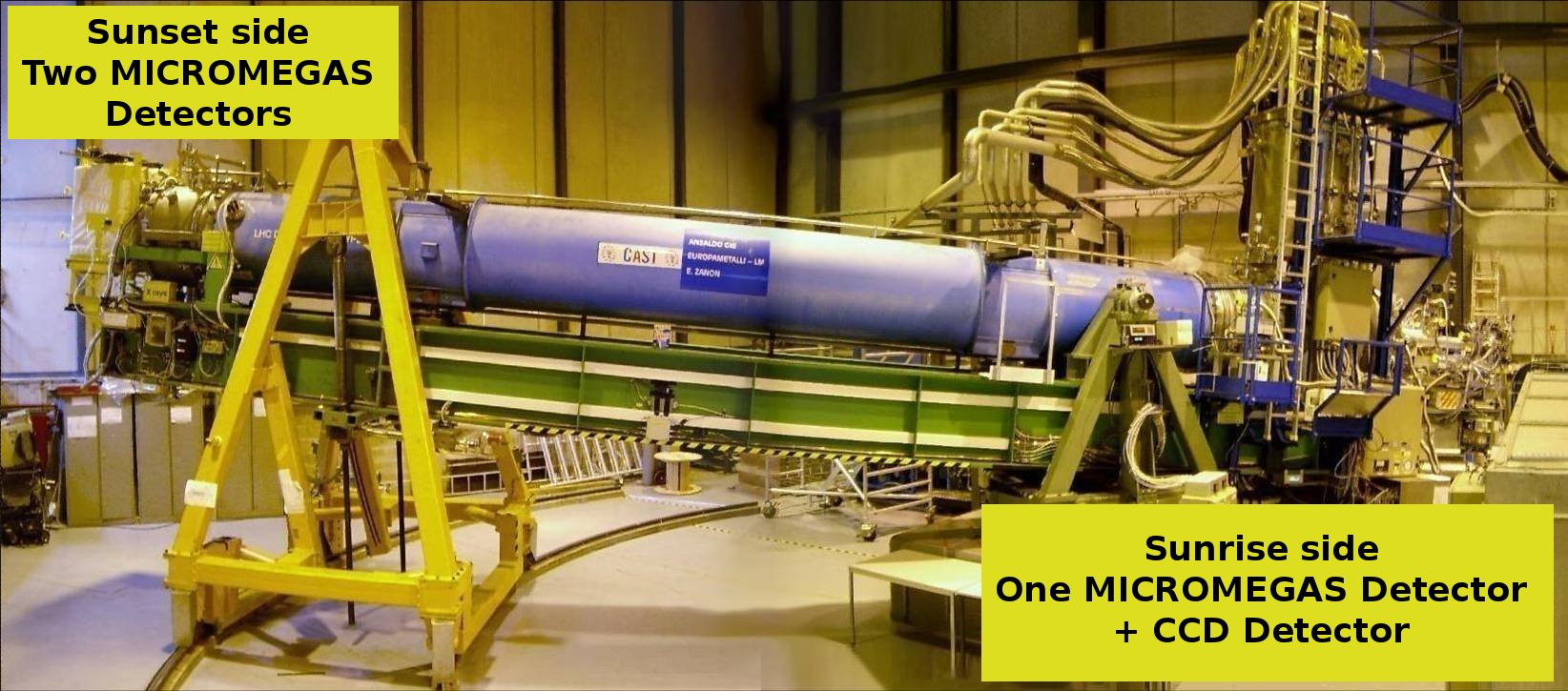} &
\includegraphics[width=0.35\textwidth]{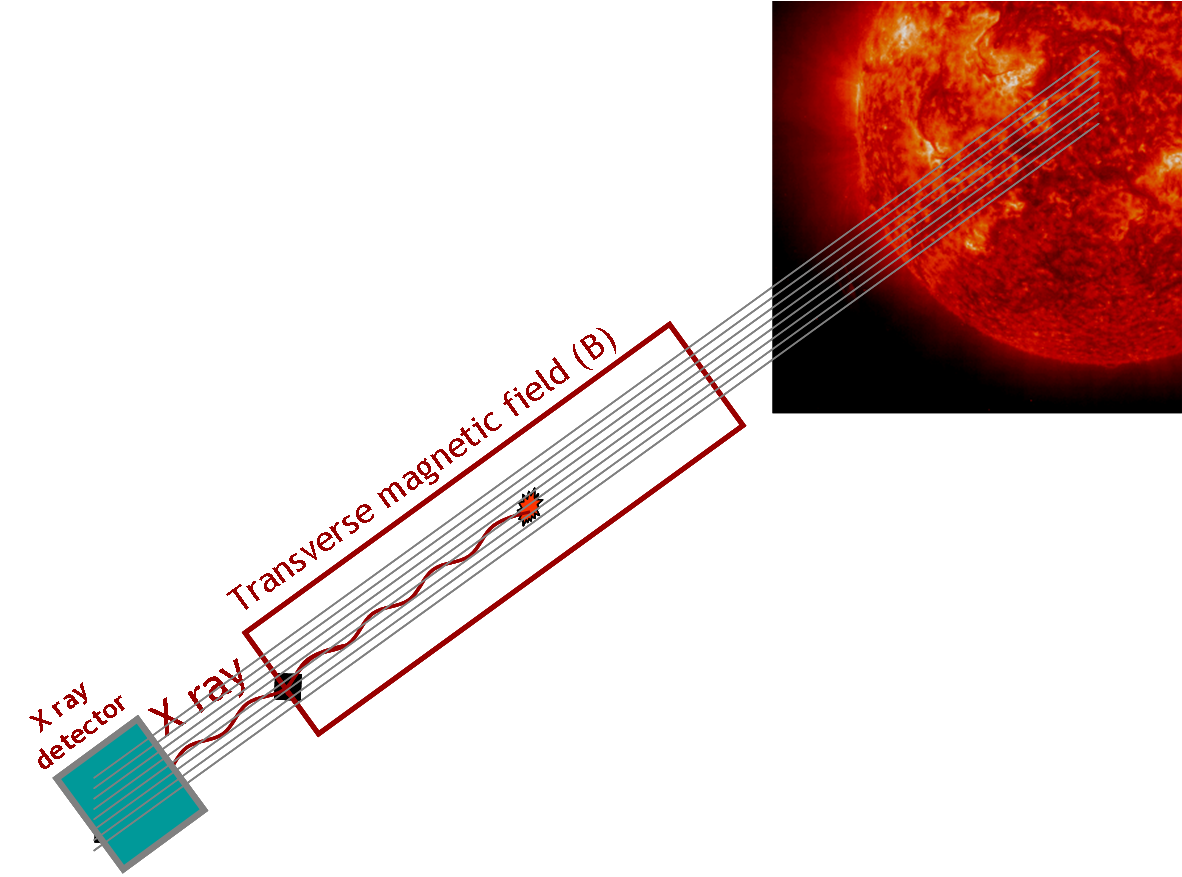} \\
\end{tabular}
\caption{On the left, a picture of the CAST Helioscope running at CERN facilities. The figure shows the general position of the detectors in the experiment, at both ends of the magnet, denominated as sunset and sunrise side. On the right, a sketch of the conversion principle of axions to X-rays.}
\label{fi:CAST}
\end{figure}

Axions could be produced in the Sun via the so-called Primakoff effect. The CERN Axion Solar Telescope (CAST) experiment~\cite{Zioutas:1998cc} uses a Large Hadron Collider (LHC) prototype dipole magnet~\cite{LHCMagnet} (see Fig. \ref{fi:CAST}) providing a magnetic field of about $9$\,T which could reconvert these axions into photons when the magnetic field is transversal to the direction of axions propagation. The magnet aligns with the Sun twice per day (during sunset and sunrise) for about 1.5 hours. The magnet is composed of two independent magnetic bores which ends are covered by X-ray detectors, the magnet bore ends are denominated sunrise and sunset side.

\vspace{.2cm}

The CAST research program required new technical developments and upgrades, the systems had to be re-adapted for the different data taking phases in order to prepare the magnet bores for scanning a new axion mass region. Furthermore, the last \emph{state of art} detectors were implemented during the scheduled upgrading shutdowns in order to improve detector sensitivity and stability (fully described on chapter~\ref{chap:micromegas}). This chapter will describe the main characteristics of the magnet set-up and the required $^3$He system upgrades to fulfill the CAST research program, together with a brief description of the X-ray detectors that have been taking data in the CAST experiment.

\section{The CAST magnet set-up}

The CAST magnet includes a number of additional systems that allow to operate the magnet for the search of solar axions. The magnet implements a cryo system (section~\ref{sc:cryo}) that allows the magnet bores to reach the nominal temperature of $1.8$\,K, in addition, vacuum systems (section~\ref{sc:vacuum}) are installed all around the magnet bore with a double functionality; to insulate the cold parts of the magnet and to increase the transmission of X-rays produced by axions to the detector systems.

\vspace{0.2cm}

\begin{figure}[!ht]
{\centering \resizebox{\textwidth}{!} {\includegraphics{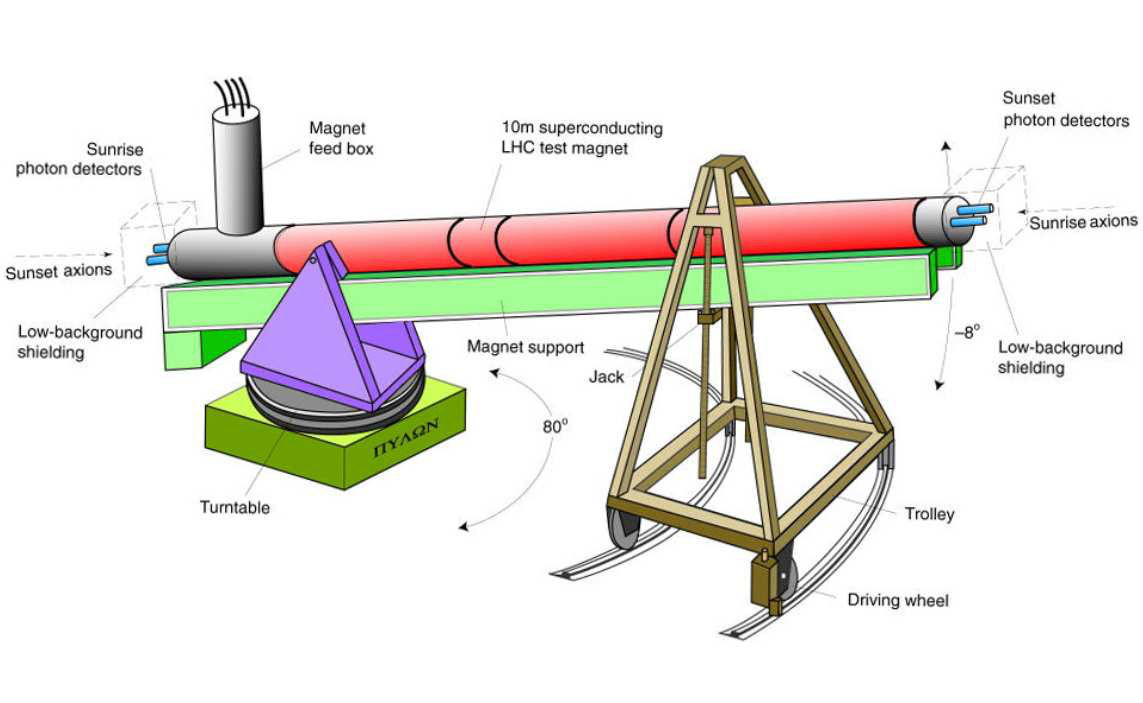}} \par}
\caption{\fontfamily{ptm}\selectfont{\normalsize{ A schematic drawing of the CAST set-up which is supported by a rotating platform.  }}}
\label{fi:magnetSetup}
\end{figure}

The CAST magnet is mounted over a moving structure (see Fig.~\ref{fi:magnetSetup}). The range of movement it is constrained by the magnet weight and size, and its relative position with the floor, and the cooling system which flexible connections allow a reduced movement. The magnet can steer in horizontal movement about $80^\circ$ and about $\pm8^{\circ}$ in vertical movement. The full range of movement of the magnet allows to track the Sun core during the sunrise and the sunset at any period of the year for at least $1.5$\,hours, tracking time that does not change more than $10\%$ due to the relative movement of Sun and Earth during the year. The steering of the magnet it is controlled by a tracking system software (section~\ref{sc:trackingSystem}) using an internal coordinate system. The tracking system is periodically checked in order to assure that the magnet is following the Sun core with the required precision (section~\ref{sc:alignment}).

\vspace{0.2cm}

The monitoring of all the systems related are centralized in a unique acquisition system that allows to continuously monitor the vital constants of the experiment (section~\ref{sc:slowControl}).

\subsection{The CAST Magnet and cooling system}\label{sc:cryo}

\begin{figure}[!ht]
{\centering \resizebox{0.85\textwidth}{!} {\includegraphics{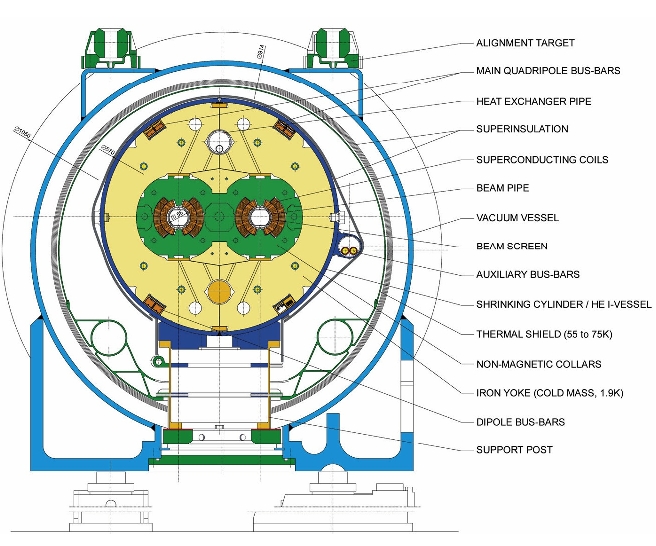}} \par}
\caption{\fontfamily{ptm}\selectfont{\normalsize{ Schematic of the dipole magnet cross-section.}}}
\label{fi:dipoleSchema}
\end{figure}

The CAST magnet is one of the first-generation magnets produced for the LHC. In contrast to the bending magnets presently implemented in the accelerator, it has two straight magnet bores. Each bore has an aperture of $42.5$\,mm, which results in a cross sectional area of $A_{cb} = 14.52$\,cm$^2$. A schematic view of the cross section of this magnet can be seen in figure~\ref{fi:dipoleSchema}. The nominal magnetic field strength over a length of $9.26$\,m is up to $9$\,T. To reach such high magnetic fields, a current of over $13$\,kA is necessary and the magnet has to be operated at a temperature of $1.8$\,K. During a data taking period the magnet is operated at exactly $13$\,kA which provides a magnetic field intensity of $B_{cb} = 8.80$\,T.

\vspace{0.2cm}

The cooling system supplied with liquid helium~\cite{Barth:708949} is implemented in the Magnet Feed Box (MFB) on the top of the magnet at the sunrise side, where the four large section cables driving the high intensity current to the magnet are connected, the cables are protected by a continuous water flow in order to quickly dissipate the heating produced by Joule effect. The fluid helium circuit that cools down the magnet has its end at the Magnet Return Box (MRB). The magnet bores, also denominated as \emph{cold bores}, are isolated from the environmental temperature thanks to the isolation vacuum inside the cryostat and a liquid nitrogen flow that cools down the system to about $77$\,K in a first cooling stage of the system at ambient temperature. The flexible transfer lines were designed to allow the movement of the magnet without an interruption of the helium flow. The whole cryogenic infrastructure was recovered from the Large Electron-Positron collider (LEP) experiment.

\subsection{Vacuum system}\label{sc:vacuum}

The CAST magnet structure is equipped with a system of vacuum pumps in order to evacuate the cold bore. The cryostat vacuum is independently pumped from the vacuum detectors side. In total four gate valves (VT1-4), one at each detector bore end, are installed to separate the magnet from the detectors. During data taking, the gate valves have to be open because their low X-ray transmission. They can be closed automatically to prevent both magnet and detector system from damage in case of failure. However, if there is the need for intervention or problem on one of the detectors, the corresponding valve can be closed and the other detectors can perform as usual. Thus, each detector system implements its own vacuum system which prevents air from going towards the cold magnet bores.

\subsection{The solar tracking system}\label{sc:trackingSystem}

The tracking system consists of a movable platform which is moved by \emph{two} motors, one for horizontal movement and the other for vertical movement. The tracking software is in charge of calculating the rotating motors frequency, or speed, to achieve the desired magnet position. The Sun core is tracked with precision by moving a setup of about $40$\,tons.

\vspace{0.2cm}

\begin{figure}[!ht]
\begin{center}
\begin{tabular}{cc}
{\centering \resizebox{0.48\textwidth}{!} {\includegraphics{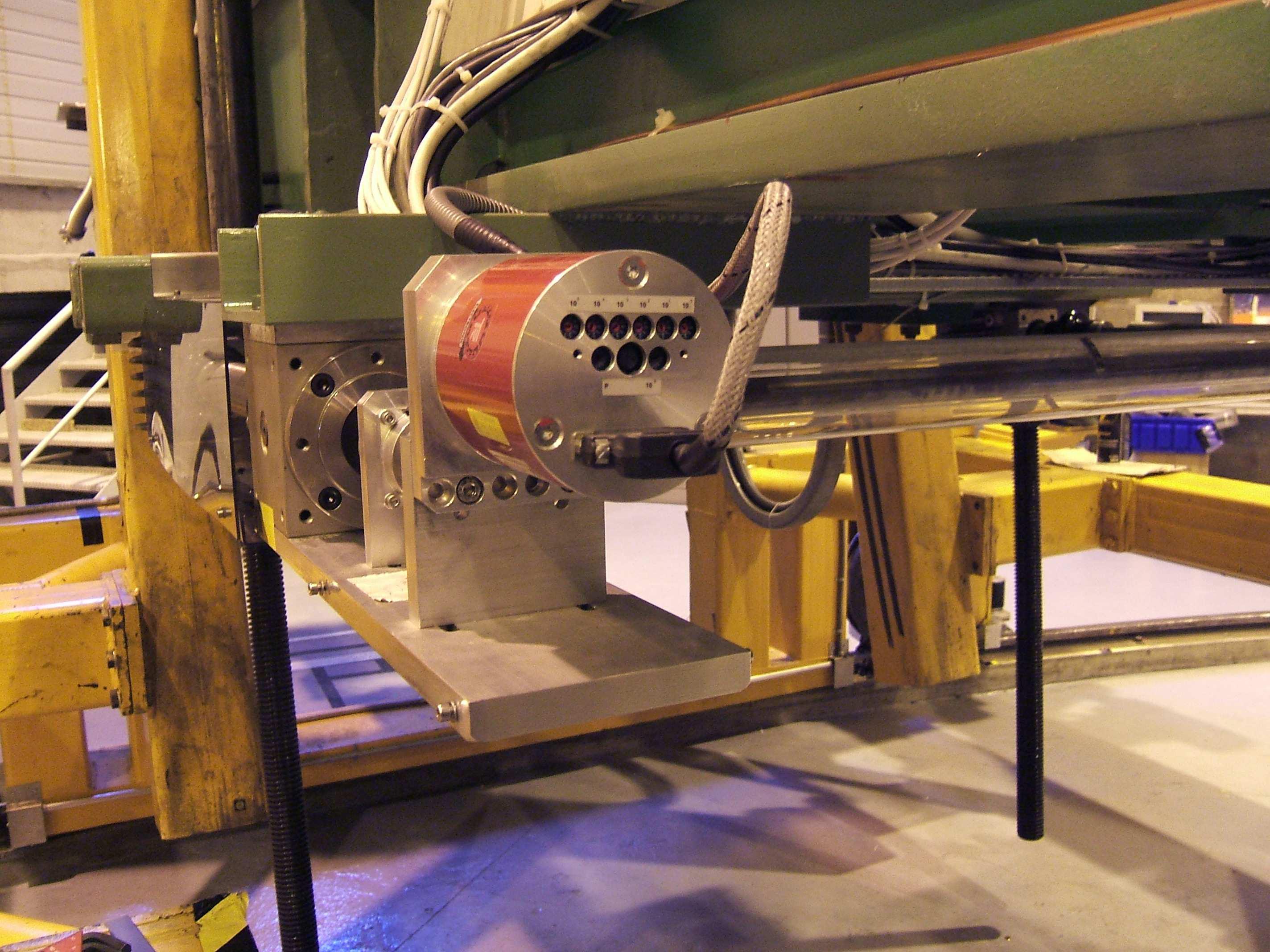}} \par} &
{\centering \resizebox{0.27\textwidth}{!} {\includegraphics{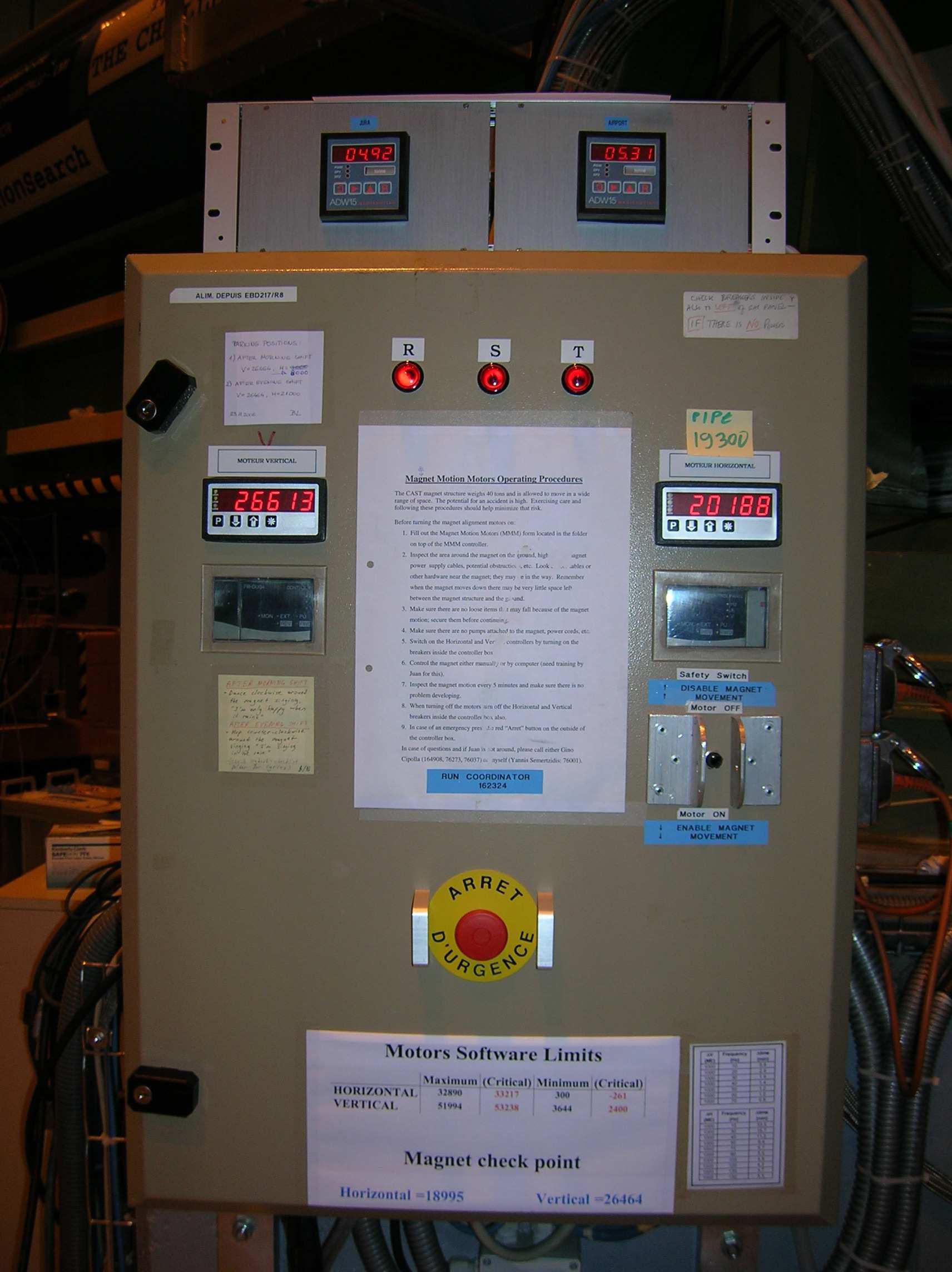}} \par} \\
\end{tabular}
\end{center}
\caption{\fontfamily{ptm}\selectfont{\normalsize{A picture of the motor encoders which correlate the encoder values with angular positions (left), and the encoders control box which allow to move the magnet, manually or by direct communication with the tracking software (right). }}}
\label{fi:encoders}
\end{figure}

The horizontal movement is achieved by using a trolley, where the MRB side of the magnet is sitting, guided along circular rails on the floor of the experimental hall, together with a turntable at the MFB side of the magnet. Additionally the vertical motor is used to rotate \emph{two} lifting jacks which allow for vertical movement. A local reference system consisting of motor encoder-values (see table~\ref{ta:referenceCoordinates}) for these two motors allows to the tracking software to determine the position of the magnet which is used for steering (see Fig.~\ref{fi:encoders}).

\begin{table}[!ht]
\begin{center}
\begin{tabular}{crcrc}
\bf{Ver. encoder}	& \bf{Altitude}	&  \bf{Hor. encoder}	&	\bf{Azimuth}	   &	\bf{Local system} \\
51734           & 7.6$^\circ$       &        184              &           47.19$^\circ$        &     1$^\circ$	\\
45098           & 5.6$^\circ$       &        3639             &           56.19$^\circ$        &     10$^\circ$	\\
38463           & 3.6$^\circ$       &        7477             &           66.19$^\circ$        &     20$^\circ$	\\
31827           & 1.6$^\circ$       &        11316            &           76.19$^\circ$        &     30$^\circ$	\\
26492           & 0$^\circ$         &        15155            &           86.19$^\circ$        &     40$^\circ$	\\
21211           & -1.6$^\circ$      &        18994            &           96.19$^\circ$        &     50$^\circ$	\\
14575           & -3.6$^\circ$      &        22833            &           106.19$^\circ$       &     60$^\circ$	\\
7939            & -5.6$^\circ$      &        26672            &           116.19$^\circ$       &     70$^\circ$	\\
1303            & -7.6$^\circ$      &        30511            &           126.19$^\circ$       &     80$^\circ$	\\
                &              &		33199          	&	  133.19$^\circ$ 	   &     87$^\circ$	\\
\end{tabular}
\end{center}
\caption{\fontfamily{ptm}\selectfont{\normalsize{ Relation between the motor encoder values and the magnet vertical and horizontal position.  }}}
\label{ta:referenceCoordinates}
\end{table}      

\vspace{0.2cm}

\begin{figure}[!ht]
{\centering \resizebox{0.85\textwidth}{!} {\includegraphics{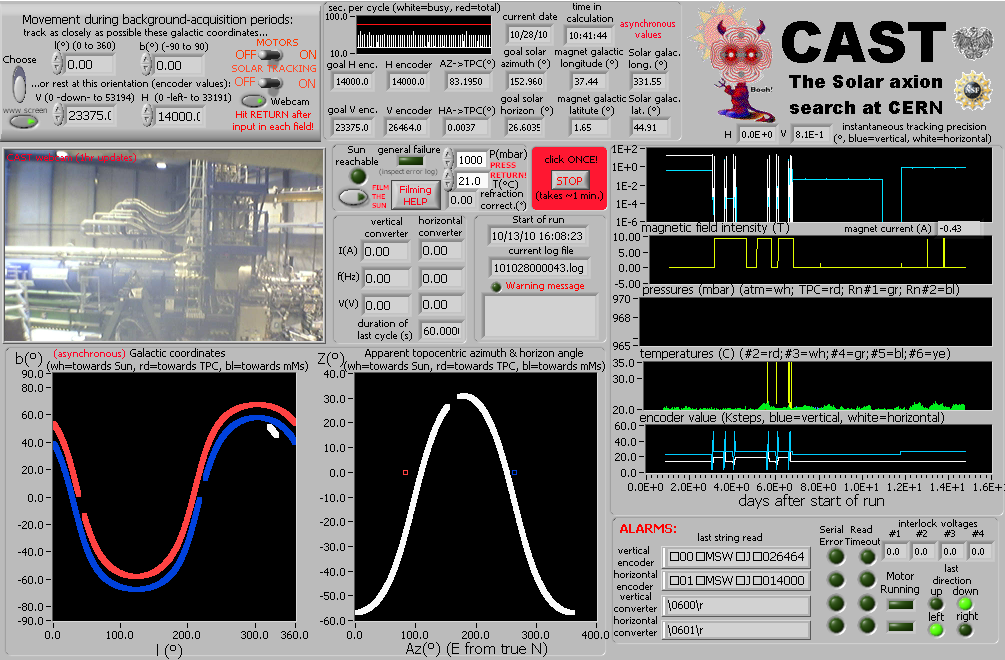}} \par}
\caption{\fontfamily{ptm}\selectfont{\normalsize{Snapshot of the tracking software user interface. }}}
\label{fi:trackingShot}
\end{figure}

The tracking program (see Fig.~\ref{fi:trackingShot}) implements a special mode for solar tracking in which the magnet will automatically follow the Sun when it is reachable, between the vertical range [$-7.2^\circ,7.95^\circ$] and the horizontal angle range [$46.8^\circ,133.1^\circ$]. The program calculates the position of the Sun $1$\,minute in advance (goal position) by using NOVAS, provided by the U.S. Naval Observatory~\cite{NOVAS}, and translates it to the local coordinate system used by the motor encoder-values. The tracking software performs this calculation every minute and compares the actual magnet position to apply the required motors speed to reach the \emph{goal position} within that minute. If the position is not reachable within $1$\,minute the program sets the motors at maximum speed, if the the position is out of range the magnet will move to the parking position manually specified in the program.

The frequency of the motors is set in \emph{two} steps, first horizontal motor and then the vertical motor, which are separated a time interval of $\simeq 1 - 2$\,s. The time used to calculate the Sun position is the mean time between these \emph{two} operations leading to some uncertainty on the coordinates estimation. This uncertainty together with other sources of uncertainty are presented in table~\ref{ta:trackingPrecision}, leading to a tracking precision better than $0.01^\circ$.

\begin{table}
\begin{tabular}{crr}
\bf{Source of error}  		&     \bf{Typical value}	&	 \bf{Maximal value}  \\
			&			&		\\
Astronomical calculations 	&         0.002$^\circ$		& 	     0.006$^\circ$      \\
Uncertainty of coordinates	& $\simeq$ 0.001$^\circ$	&			\\
Clock time 			&	$\simeq 0$$^\circ$	&			\\
Grid measurements		&	0.001$^\circ$		&       < 0.01$^\circ$ 		\\
Interpolation of Grid measurements	&	  0.002$^\circ$	&			\\
Horizontal encoder precision 	& $\simeq$ 0.0014$^\circ$	&			\\
Vertical encoder precision 	&  	0.0003$^\circ$		&			\\
Linearity of motor speed  	& 	< 0.002$^\circ$		&			\\
				&			&			\\
Total				&       < 0.01$^\circ$		&			\\
\end{tabular}
\caption{\fontfamily{ptm}\selectfont{\normalsize{ Summary table of the main error sources in the tracking accuracy. }}}
\label{ta:trackingPrecision}
\end{table}

\subsection{Magnet alignment accuracy.}\label{sc:alignment}

The motor encoding that translates the local coordinate system to the absolute position of the magnet was calibrated before the start of the first CAST data taking in 2002. Since then the magnet alignment has been cross-checked periodically to assure the required tracking accuracy, $1$\,arcmin. \emph{Three} methods are used to perform these measurements; \emph{GRID} measurements, \emph{Laser} position measurements, and \emph{Sun filming}.

\subsubsection{Grid measurements.}

The so called GRID measurements consist in the independent position of the magnet in a set of reference coordinates (GRID) defined during the first calibration of the magnet orientation that cover a wide range of position along the allowed magnet movement range. These measurements are intended to detect any drift in the pointing accuracy with respect to the initial calibration values obtained in 2002.

\begin{figure}[!ht]
{\centering \resizebox{0.85\textwidth}{!} {\includegraphics{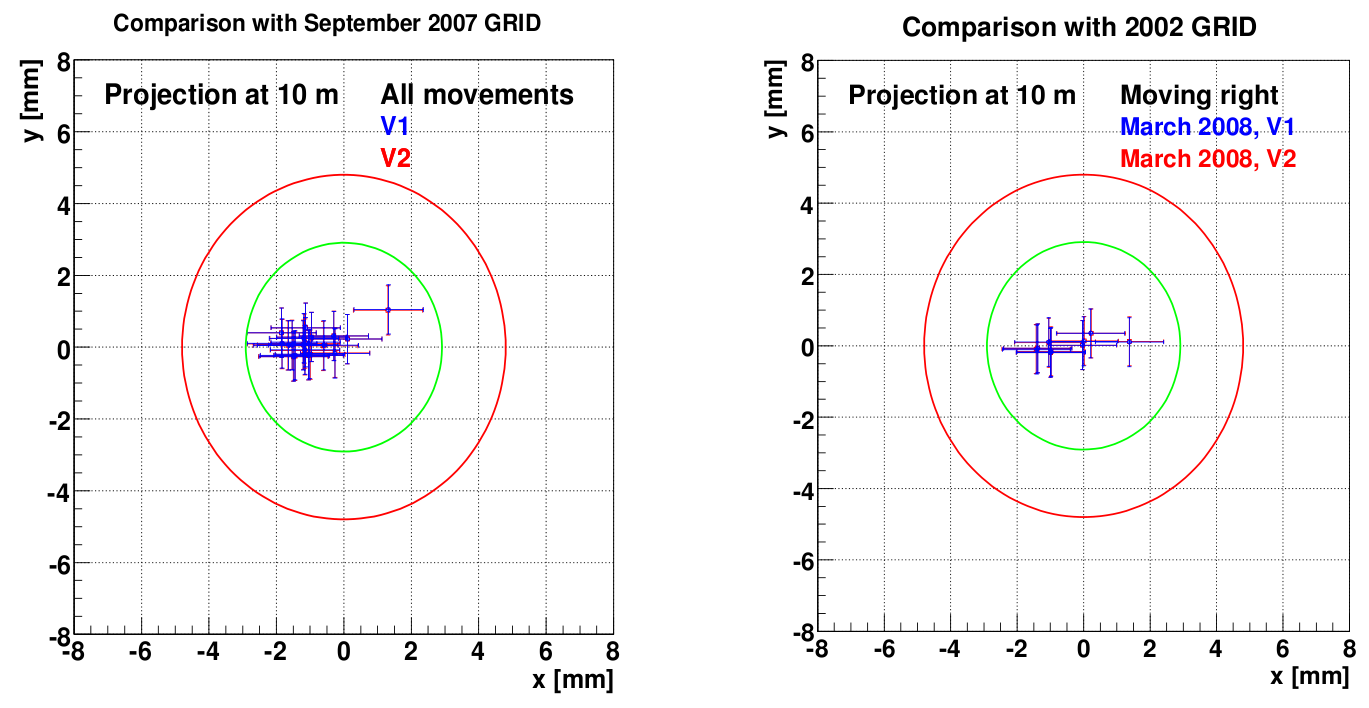}} \par}
\caption{\fontfamily{ptm}\selectfont{\normalsize{GRID measurements carried out in 2008. The relative positions with the original measurements in 2002 (right) and the previous GRID measurements in 2007 (left) are presented for both magnet bores (V1 and V2). The green circle represents a precision of $1$\,arcmin, while the red circle represents the $10\%$ of the Sun radius projected at $10$\,m.  }}}
\label{fi:grid}
\end{figure}

\vspace{0.2cm}
During the GRID measurements carried out in 2008 no significant deviation from the original measurements was observed, all the prefixed positions show a deviation lower than $1$\,arcmin. The GRID measurements are typically represented in a projected plane at $10$\,m in order to compare the coordinates deviation obtained with the required precision and the Sun core (see Fig.~\ref{fi:grid}).

\subsubsection{Laser position measurements.}

In addition, two lasers are placed on the magnet which measure the relative position at some reference points. The horizontal laser points towards the guiding rails where a colored pattern is drawn at the specific reference points, while the vertical laser points to a pattern drawn in the supporting structure. The lasers return the horizontal and vertical encoder values each time a color change is detected in the pattern giving an accurate position which should be reproduced each time the magnet passes through.

\subsubsection{Sun filming.}
The CAST magnet is able to directly observe the Sun twice a year, in March and September. A window was positioned in the CAST host building specifically for this purpose. The Sun filming method allows to perform and independent check of the pointing accuracy of the magnet during tracking movement. Thus, these measurements are able to provide enough information to be comparable with the GRID measurements.

\vspace{0.2cm}

In order to film the Sun a CCD camera is aligned with the magnet axis by using a focusing system and laser that is parallel to the theoretical magnet axis. The tracking system implements a special mode in which the corrections due to refraction of photons into the atmosphere are taken into account for pointing to the Sun, correction which depends on the atmospheric pressure and environment temperature. 

\vspace{0.2cm}

\begin{figure}[!t]
{\centering \resizebox{0.95\textwidth}{!} {\includegraphics{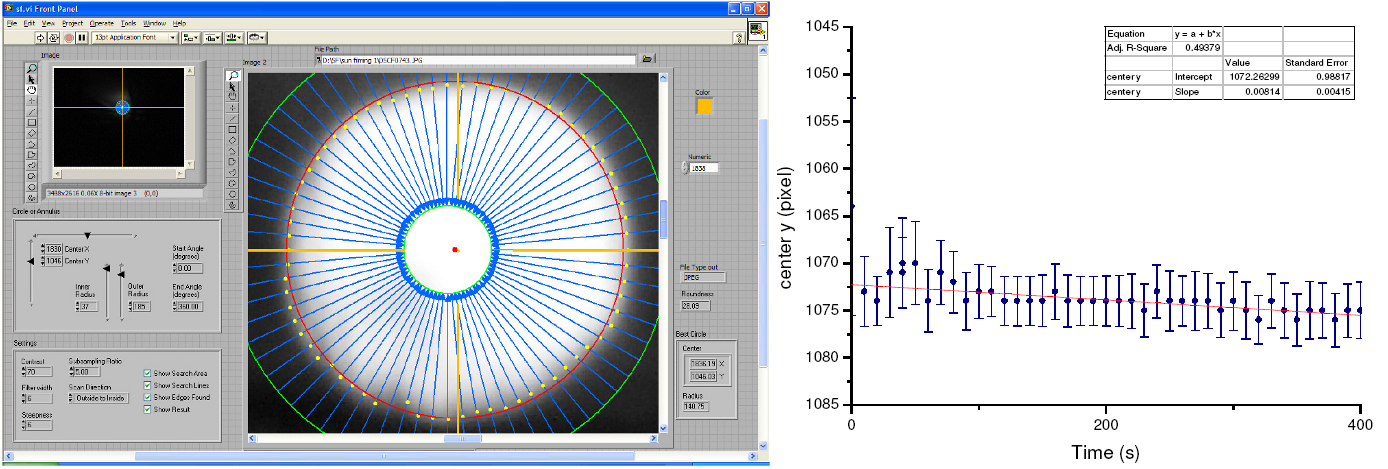}} \par} 
\caption{\fontfamily{ptm}\selectfont{\normalsize{ Sun filming software used to determine the position of the Sun center at the CCD camera (left) and the dynamical evolution of the spot during the Sun filming measurement (right). }}}
\label{fi:sunFilming}
\end{figure}

Actually two independent CCD cameras are installed in order to increase the reliability of these measurements. Figure~\ref{fi:sunFilming} shows the filming software used to determine the position of the Sun center in the CCD camera, together with the dynamical change of this position given in terms of the mean CCD pixel value.

\vspace{0.2cm}

The Sun passes through the window in less than $5$\,minutes, time in which the Sun filming has place. A common problem is that the filming is subject to the weather conditions. However, usually a good measurement is obtained in each of these Sun filming periods which last for $1$\,week.

\subsection{The slow control system.}\label{sc:slowControl}

\begin{figure}[!ht]
{\centering \resizebox{0.85\textwidth}{!} {\includegraphics{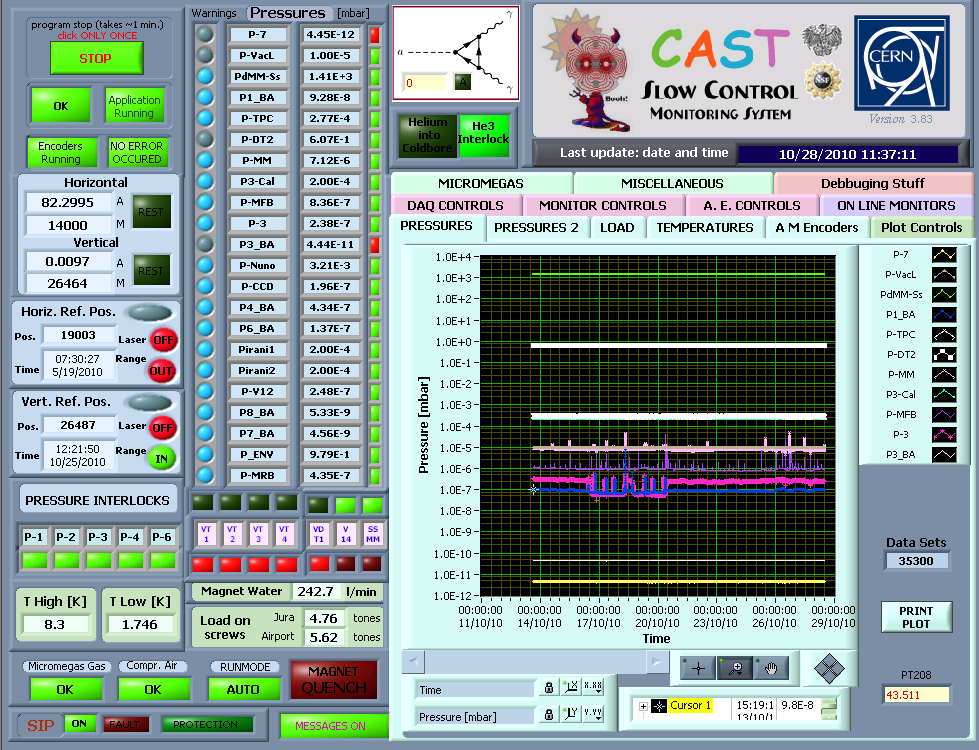}} \par}
\caption{\fontfamily{ptm}\selectfont{\normalsize{Snapshot of the Slow Control system user interface. }}}
\label{fi:scShot}
\end{figure}

The monitoring of the main parameters of each system are recorded in a unique centralized system, denominated slow control. The system implements several National Instruments (NI) cards which allow to acquire analogue and digital signals, and produce output signals as well. A graphical user interface for controlling these cards was developed in LabView (see Fig.~\ref{fi:scShot}) in the early stage of the CAST experiment.

The system implements all the cards functionality allowing to connect or disconnect new sensors and add them to the acquisition software without much code modifications. Thus, the software implements a versioning system which is in charge of auto-formatting the data files produced when new signals are added and/or removed.

\vspace{0.2cm}

The slow control monitors the vacuum pressures of the magnet cryostat and detector systems, as well as, detector parameters, compressed air, magnet temperature, magnet load on screws, magnet position, controls and records the laser reference checkpoints, and keeps track of the magnet valves status and safety systems, between other functionalities.

\vspace{0.2cm}

All the values measured by the slow control system are recorded every minute. An alarm system allows to set threshold values for some of the variables recorded, when a given sensor gives a value above or below a critical value the system sends specific information via mail and SMS to the person in charge of the system. Thus, when an alarm is detected the system enters in fast data taking mode recording the status of the systems every $2$-$3$\,seconds allowing to evaluate later the situation with better accuracy.

\section{The CAST research program.}

The CAST experiment started to take data already by 2003. During its first phase (CAST Phase I) the magnet was operating with vacuum inside the magnet bores, conditions in which the sensitivity of the experiment in the axion-photon coupling $g_{a\gamma}$ was enhanced for axion masses up to $m_a \lesssim 0.02$\,eV. The experiment completed its first phase in 2004. By combining the results of the \emph{three} detectors that took data; CCD~\cite{DonghwaThesis}, TPC~\cite{BertaThesis} and micromegas~\cite{TheopistiThesis} detectors, CAST improved the current limits on the coupling constant~\cite{vacuumCAST},

\begin{equation*}
g_{a\gamma} < 8.8\cdot10^{-11}\,\mbox{GeV}^{-1} \mbox{at 95\%~CL for}\,m_a\leq 0.02\,\mbox{eV}
\end{equation*}

\vspace{0.2cm}
\noindent and providing the best experimental limit over a broad range of axion masses and going beyond the previous limits derived by loss energy arguments and globular-cluster stars.
\vspace{0.2cm}

\begin{figure}[!ht]
{\centering \resizebox{0.8\textwidth}{!} {\includegraphics{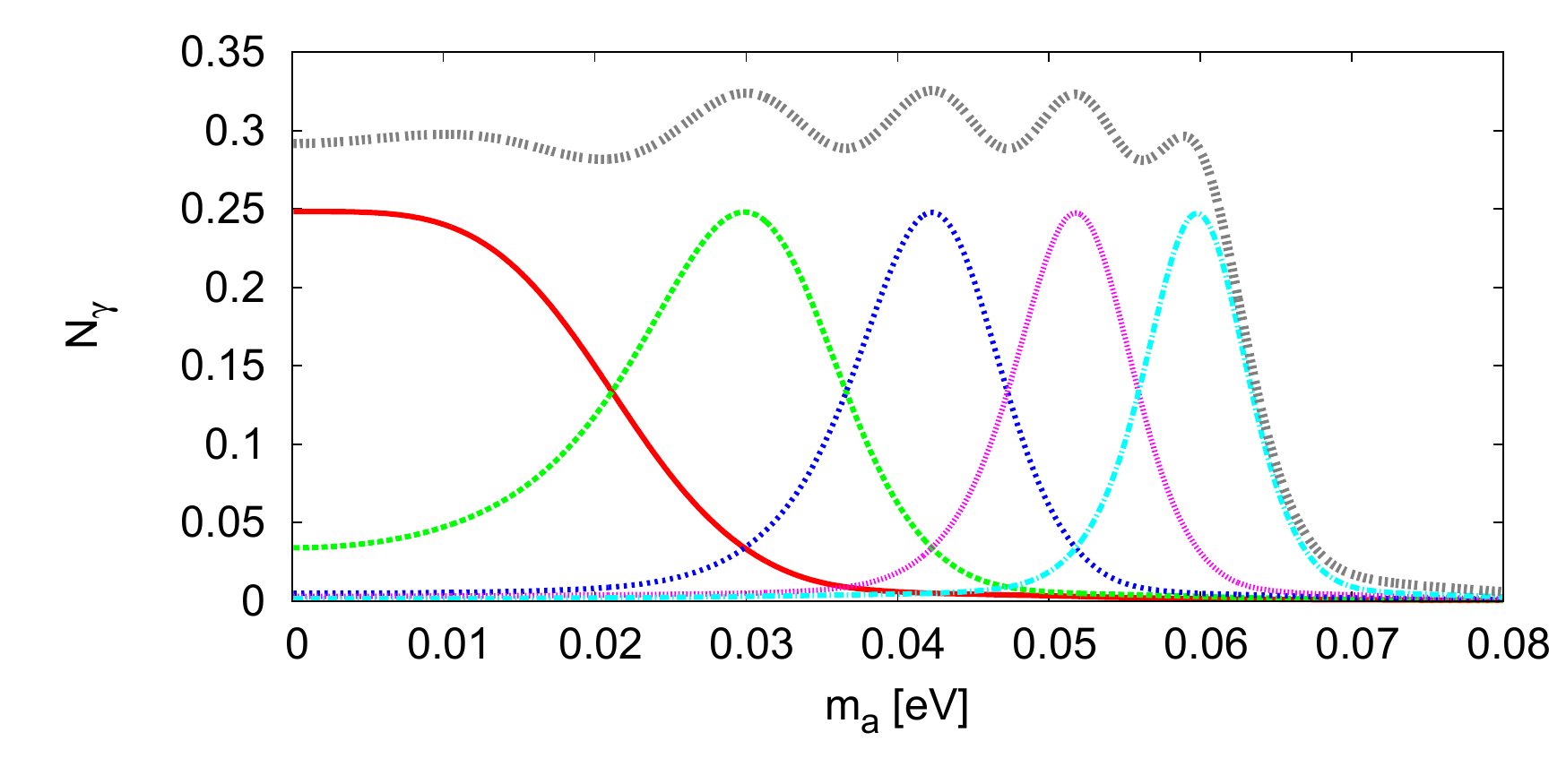}} \par}
\caption{\fontfamily{ptm}\selectfont{\normalsize{Axion mass coverage in terms of the expected number of photons $N_{\gamma}$ at $g_{a\gamma} = 10^{-10}$\,GeV. A tracking in vacuum conditions and the first four density steps are shown, together with the total integrated expected photons due to the steps combination. $N_{\gamma}$ in each step is calculated taking into account one detector (efficiency included) at the nominal magnet operation for a tracking exposure of $5400$\,s.}}}
\label{fi:densitySteps}
\end{figure}

In the second phase (CAST Phase II), CAST extends its sensitivity to higher axion masses by filling the magnet with a refractive buffer gas that maximizes the probability that axions are converted into photons for a narrow mass-range, which will allow CAST to improve the limit on the coupling constant for masses up to about 1\,eV. In order to cover this axion mass region the density of the buffer gas is increased in small steps that are chosen to assure a smooth coverage on the axion coupling limit, directly related with the number of axions that CAST is able to detect in a given axion mass~(see Fig.~\ref{fi:densitySteps}).

\vspace{0.2cm}

\begin{figure}[!ht]
{\centering \resizebox{0.8\textwidth}{!} {\includegraphics{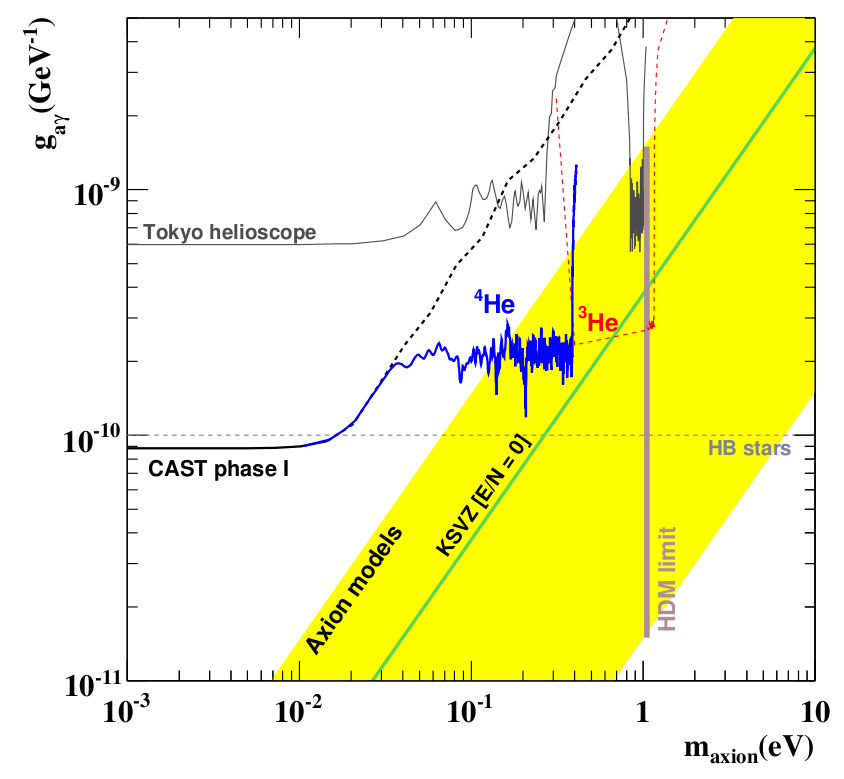}} \par}
\caption{\fontfamily{ptm}\selectfont{\normalsize{Axion-photon coupling limit as a function of the axion mass provided by CAST experiment in the vacuum phase (Phase I) and $^4$He axion mass range coverage. Expected coverage for the $^3$He operating period is also shown. }}}
\label{fi:exclusionHe4}
\end{figure}

In Phase II, the CAST helioscope was first operated with $^4$He as buffer gas and later with $^3$He. $^4$He gas operation finished in 2006 and allowed CAST to scan axion masses up to $0.4$\,eV giving an improved limit for the new scanned region~\cite{He4CAST},

\begin{equation*}
g_{a\gamma} < 2.17\cdot10^{-10}\,\mbox{GeV}^{-1} \mbox{ at 95\%~CL for }0.02\,\mbox{eV} < m_a < 0.4\,\mbox{eV}
\end{equation*}

\noindent providing the best experimental limit in the mass region covered and excluding by first time a favored region by the axion theoretical models (see Fig.~\ref{fi:exclusionHe4}).

\vspace{0.2cm}

$^3$He gas extends the search of axions to higher masses than that masses accessible with $^4$He gas, because the saturation pressure of $^3$He at $1.8$\,K is $135.58$\,mbar compared to the pressure reached with $^4$He at $1.8$\,K, $16.405$\,mbar. The operation with $^3$He will allow the CAST experiment to reach axion masses slightly higher than 1\,eV. The gas system was adapted to work with $^3$He and it required a number of technically challenging upgrades to equip the experiment for the new run.

\vspace{0.2cm}

Masses up to about 0.75\,eV have already been covered by measuring overlapping narrow mass-range settings. CAST continues to complete its program till the end of 2011.


\section{The buffer gas filling system.}

For extending the CAST search to higher axion masses a dedicated system was developed to change the pressure inside the magnet bores in a controlled manner. The system required to precise metering and inserting the gas inside the bores obtaining an accurate pressure step size and a good reproducibility.

\vspace{0.2cm}

The gas must be confined inside the magnetic field region by using four high X-ray transmission windows, installed at the ends of the two magnet bores. Thus, the denominated \emph{cold windows} need to support the pressure difference due to the increasing amount of gas inside the \emph{cold bores}, and to allow a laser to go through for the alignment of the X-ray telescope.

\vspace{0.2cm}

\begin{figure}[!ht]
\begin{center}
\begin{tabular}{cc}
{\centering \resizebox{0.40\textwidth}{!} {\includegraphics{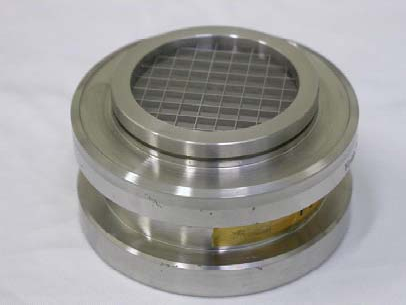}} \par} &
{\centering \resizebox{0.425\textwidth}{!} {\includegraphics{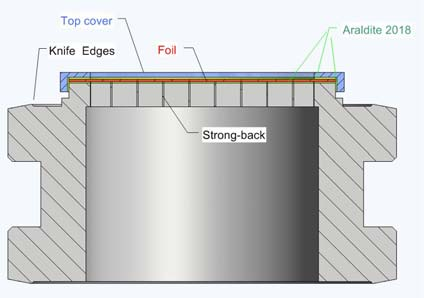}} \par} \\
\end{tabular}
\end{center}
\caption{\fontfamily{ptm}\selectfont{\normalsize{ A picture of one of the four windows installed on the magnet bore (left) and a schematic drawing of the design (right). }}}
\label{fi:coldWindow}
\end{figure}

In order to achieve all these requirements for the containing windows a $15$\,$\mu$m polypropylene foil which provides low helium permeability was used~\cite{Niinikoski:1173057}. The X-ray windows were leak and pressure tested at CERN Cryolab, by applying sudden pressure changes and measuring a leak rate that is in the expected level for the given polypropylene thickness, $< 1\cdot10^{-7}$\,mbar\,l/s. The thin windows are protected from the pressure difference by using a stainless stell grid structure, also denominated \emph{strongback}, that it is attached to the polypropylene layer and is sitting at the vacuum side (see Fig.~\ref{fi:coldWindow}).

\vspace{0.2cm}

Since the X-ray windows are in contact with the helium gas at $1.8$\,K, a heater system is required to keep the windows warm at an approximately constant temperature. The aim of this heating system is to reduce the amount of gas, coming from the vacuum side, that sticks into the \emph{cold windows}. Due to cryogenic restrictions the windows were operating at $120$\,K during the $^4$He phase and at about $80$\,K during the $^3$He period in 2008. Thus, about once every two months the magnet bore is emptied and the heaters are set to maximum power reaching higher temperatures and provoking the degassing of air molecules adsorbed by the polypropylene or frozen (mainly coming from water molecules). This process is denominated windows \emph{bake-out}.

\vspace{0.2cm}

The system used for covering the $^4$He phase mass range consisted basically in a metering volume, immersed in a thermal bath at $36^\circ$C, which was re-filled with helium and directly emptied to the magnet bore. The system allowed to accurately insert small amounts of gas by measuring the initial and final pressures in the metering volume. This system was used to cover the first $160$ density steps, allowing to cover axion masses up to $0.4$\,eV. However, in order to achieve higher pressures inside the bores, and to increase the functionality of the system, a complete new design had to be built for the $^3$He mass coverage.

\section{The $^3$He system upgrades.}\label{sc:He3System}

The experience acquired during the $^4$He phase helped to build a more sophisticated, flexible and hermetically-sealed gas system designed to precisely measure quantities of $^3$He inserted into the cold bore~\cite{Zioutas:987860}.

\vspace{0.2cm}

The metering system was upgraded to adapt to the new data taking period (section~\ref{sc:meteringHe3}). One of the main specifications of the system is to avoid the loss of the valuable $^3$He gas which amount is limited for the remaining running period of the experiment. Thus the thin X-ray windows have to be protected during a magnet quench (section~\ref{sc:expansionHe3}). The new system also implemented a purging system in order to purify the $^3$He gas in case of contamination coming from the pumping system or a possible exposure to air (section~\ref{sc:purgingHe3}). Each part of the system implements remote controlled valves, flowmeters and sensors which allow to monitor the system and to programmatically operate it~(section~\ref{sc:PLCHe3}).

\vspace{0.2cm}

\begin{figure}[!ht]
{\centering \resizebox{0.95\textwidth}{!} {\includegraphics{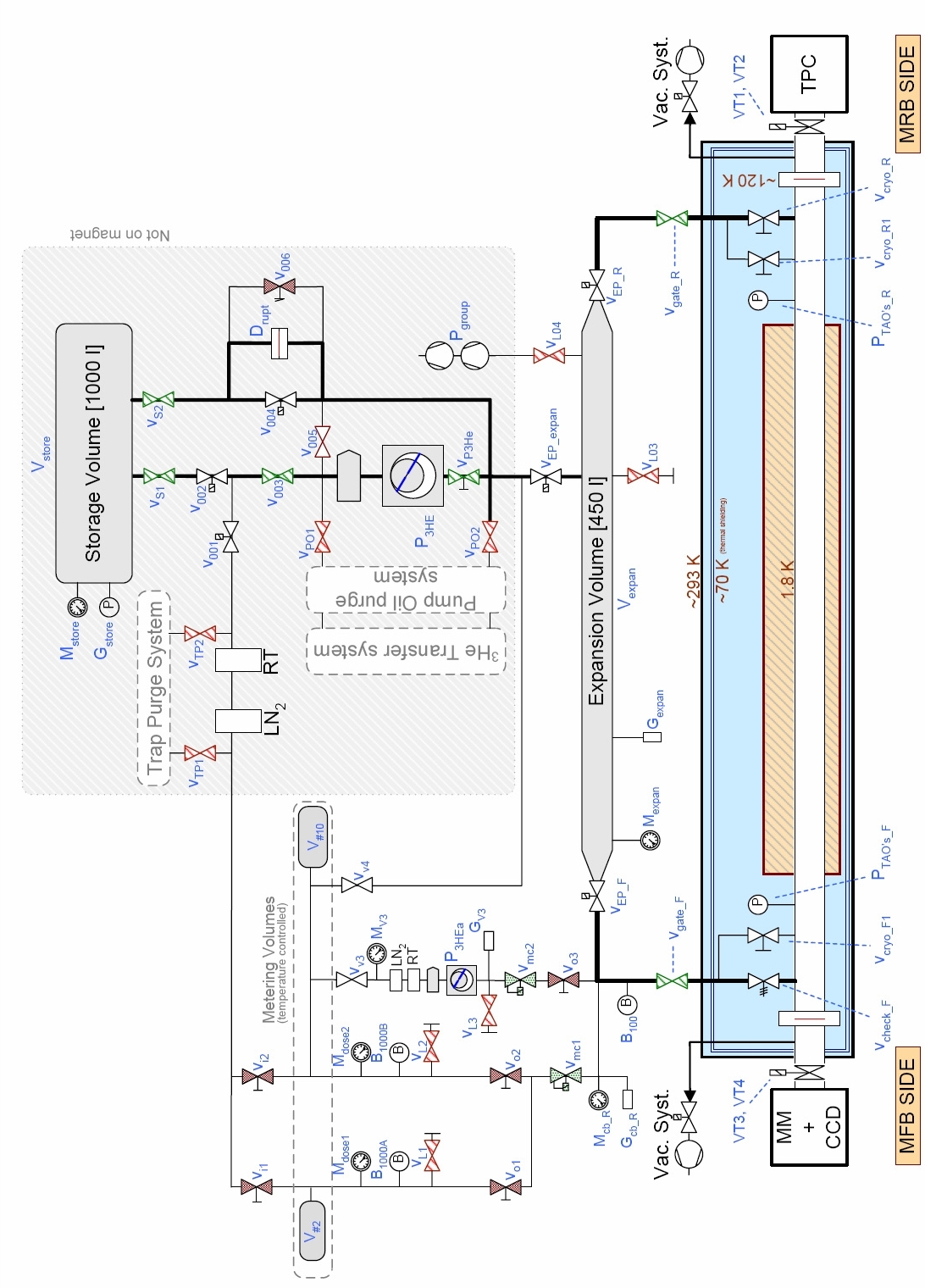}} \par}
\caption{\fontfamily{ptm}\selectfont{\normalsize{Schematic of the $^3$He filling and recovery system connected to the magnet bores. The expansion volume is directly connected to the magnet bore pipes, on grey background is shown the purging and recovery system, and on top the metering system. X-ray windows are placed at the region of $120$\,K in this drawing. }}}
\label{fi:magnetSchema}
\end{figure}

An schematic of all the systems that are involved and its connection to the CAST magnet is shown in figure~\ref{fi:magnetSchema}, where the main parts of the system can be distinguished; expansion volume, storage volume, metering system and purging system.

\subsection{Metering system.}\label{sc:meteringHe3}

\begin{figure}[!t]
\begin{tabular}{cc}
{\centering \resizebox{0.45\textwidth}{!} {\includegraphics{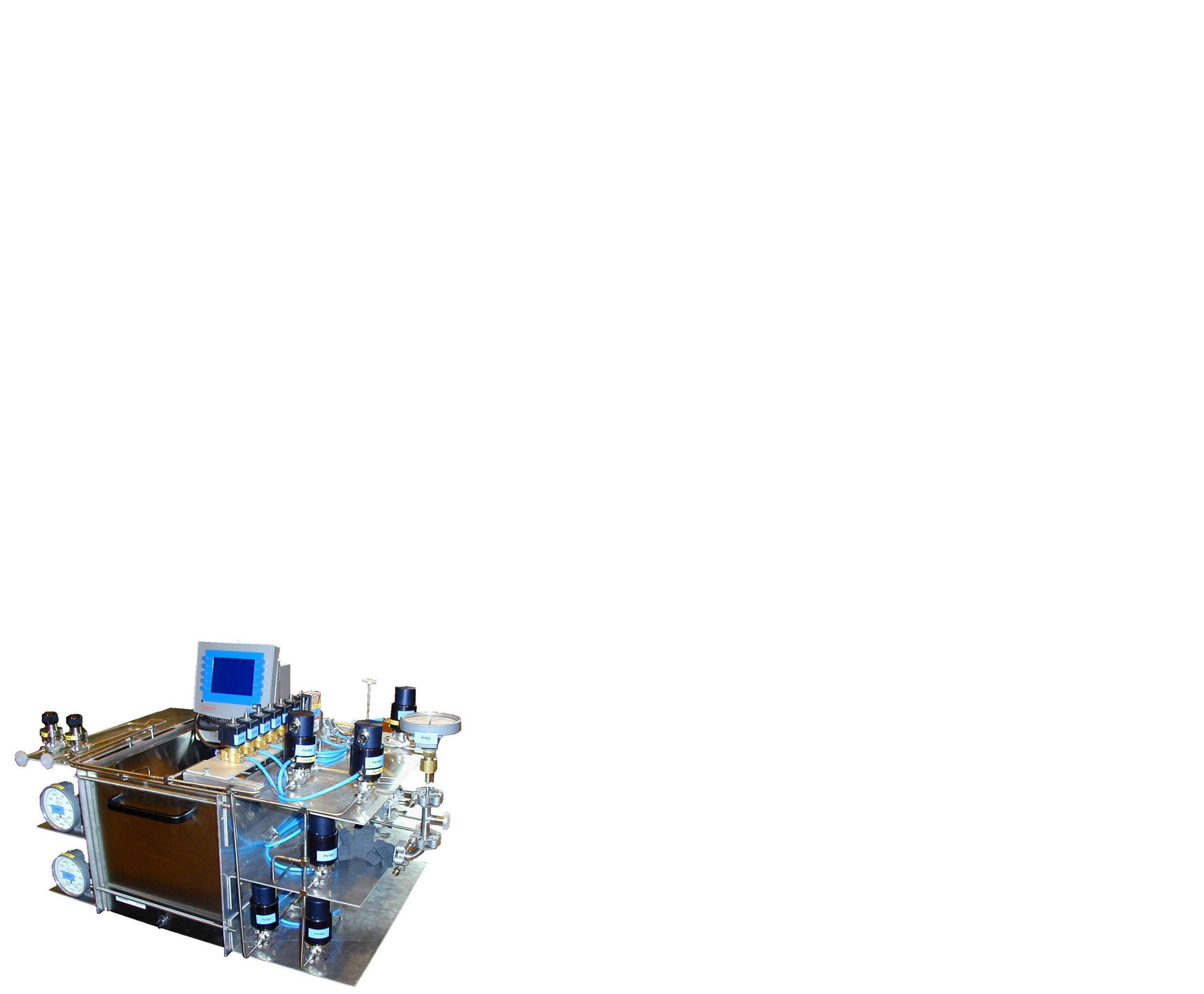}} \par} &
{\centering \resizebox{0.52\textwidth}{!} {\includegraphics{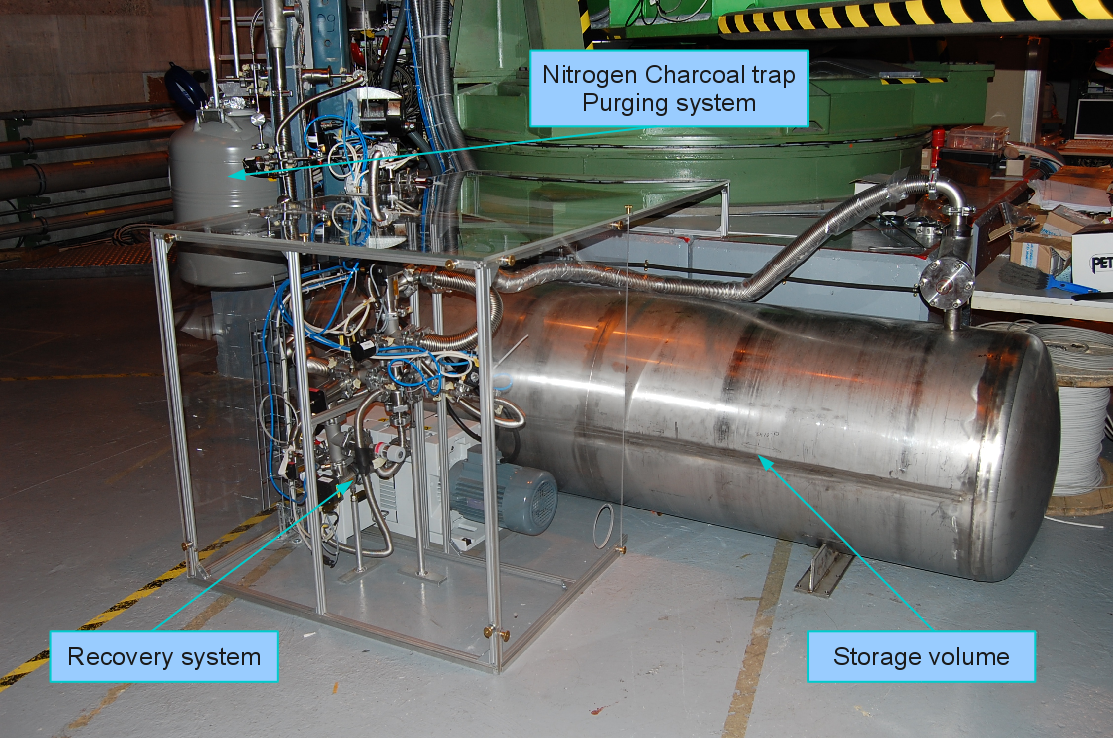}} \par} \\
\end{tabular}
\caption{\fontfamily{ptm}\selectfont{\normalsize{ A picture of the metering system where it is observed the thermal bath, remote controlled valves, and sensors (left), and a picture of the storage volume together with the recovery system and part of the purging system (right).    }}}
\label{fi:meteringVolumes}
\end{figure}

\emph{Two} metering volumes of different capacities, and cylindrical shape, were implemented in the new system (see Fig.~\ref{fi:meteringVolumes}). The smaller volume ($MV2$) contains $1.63$\,liters and allows a total increase in the magnet bores of about $2$ density steps, the second and larger volume ($MV10$) contains $8.58$\,liters allowing to increase the cold bore density by about $10$ steps. The thermal bath temperature is controlled by an embedded system that, between other functionalities, alerts in case the water level decreases.

\vspace{0.2cm}
The accurate metering process when the magnet bore is empty takes several days and increases with the pressure to reach, time that is directly affecting to the assigned days for a data taking period. The metering volume $MV10$ was added to the system in order to reduce the time required to refill the magnet from vacuum, reducing the time spent in this process to $2$-$3$ days. The magnet bore is emptied at least once every $2$ months due to the scheduled windows \emph{bake-out}, but it can also be due to a magnet quench (described on section~\ref{sc:expansionHe3}).

The cold bore filling it is automated through remote controlled pneumatic-valves. The pipework connecting the metering volumes to the magnet implements high accuracy mass-flow controllers that would allow to slowly increase the density in the bores at a constant rate.

\subsection{Purging system.}\label{sc:purgingHe3}

The metering volumes are filled with the available gas inside the storage volume (see Fig.~\ref{fi:meteringVolumes}). The pipework leading to the metering volumes implements $2$ charcoal traps which purify the $^3$He. The first one ($RT$) is at room temperature and traps oil and water vapors, while the second one ($LN_2$) is at liquid nitrogen temperature at about $77$\,K releasing the helium gas from the remaining impurities. The charcoal traps need to be periodically purged in order to maintain its absorption efficiency. The charcoal trap at room temperature can be directly replaced by a new one, while the second one can be warmed up in order to purify it with a neutral gas.

\subsection{Expansion volume and recovery.}\label{sc:expansionHe3}

One of the main challenges was to implement a safety system to protect the thin X-ray windows containing the gas in case of magnet quench. A magnet quench is an embedded protection system of any super-conducting magnet. There is always a possibility that during the operation of the magnet a very small region of the super-conducting coil abandons its super-conducting state. In that case, this small region that has acquired some resistance would absorb all the heating dissipated due to the Joule effect and could be damaged. In order to avoid this damage the magnet safety system increases the temperature of all the super-conducting coil, which will loss its super-conducting properties, allowing the heat dissipation to be distributed homogeneously along the whole coil surface. During this process the temperature of the magnet raises rapidly by a factor of $14$\, in the first $3$ seconds and by a factor of $21$ in the next $200$ seconds, reaching a temperature of about $\lesssim 40$\,K. This temperature increase heats up the magnet bore $^3$He gas inside, increasing the pressure and putting in risk the integrity of the windows. A safety system in the new $^3$He phase with magnet bore pressures reaching values above $100$\,mbar at $1.8$\,K was mandatory.

\vspace{0.2cm}

\begin{figure}[!ht]
{\centering \resizebox{0.95\textwidth}{!} {\includegraphics{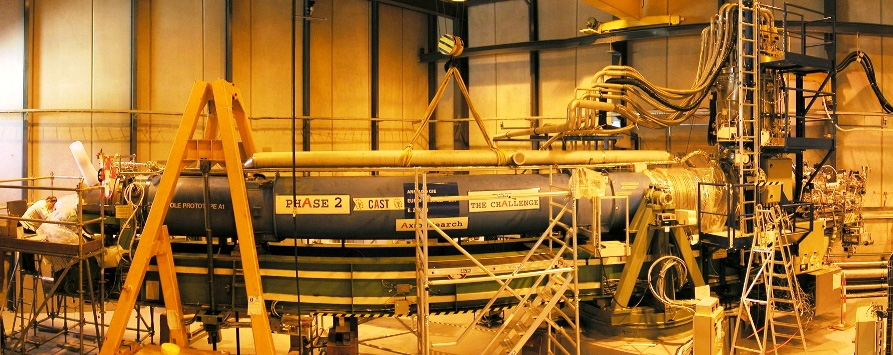}} \par} 
\caption{\fontfamily{ptm}\selectfont{\normalsize{ A picture taken during the installation of the expansion volume on top of the CAST magnet.  }}}
\label{fi:expansionVolume}
\end{figure}

The cold bore is connected to an expansion volume through safety valves, to rapidly vent the $^3$He from the cold bore in case the pressure increases rapidly. The safety valves directly communicate with the quench alarm interlock and will quickly open if it becomes active. The expansion volume (see Fig.~\ref{fi:expansionVolume}) has a storage capacity of $450$\,liters, which in the worst scenario with $140$\,mbar at $1.8$\,K inside the cold bore constrains the maximum pressure due to the increasing magnet temperature to less than $1100$\,mbar, including a safety factor of $1.2$.

\vspace{0.2cm}

When the safety valves are open, the gas distributes between the cold-bore and the expansion volume, the helium is recovered back to the \emph{storage volume} by using a powerful pumping system specially designed to reduce the oil contamination coming from the pump. Then the precision refilling of the magnet can take place again.

\vspace{0.2cm}

In the new running phase, with $^3$He in the cold bores, the vacuum pumps connected to the CAST magnet have to implement an extra safety valve. These safety valves would interrupt all the pumping systems in case a leak of $^3$He to the vacuum side would be detected(i.e. by the rupture of one of the X-ray windows) and would avoid to pump the $^3$He to the air, allowing to recover the gas back to the storage volume following the process previously described.

\subsection{The PLC system.}\label{sc:PLCHe3}

\begin{figure}[!ht]
{\centering \resizebox{\textwidth}{!} {\includegraphics{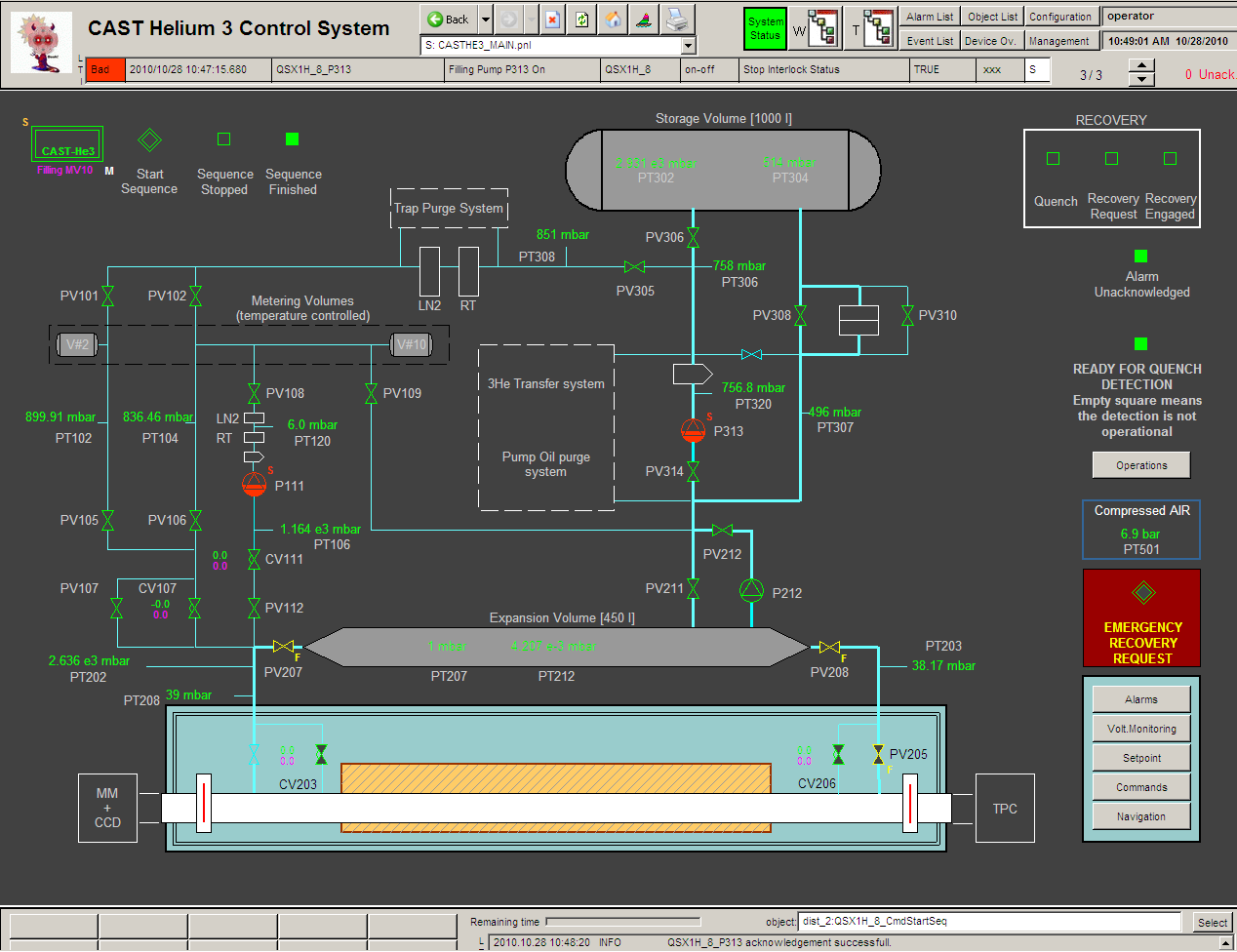}} \par}
\caption{\fontfamily{ptm}\selectfont{\normalsize{ A snapshot of the PLC system graphical user interface. }}}
\label{fi:he3PLCShot}
\end{figure}

A Programmable Logic Controller (PLC) is used for monitoring the status of valves, pressure sensors and flowmeters. A graphical interface (see Fig.~\ref{fi:he3PLCShot}) is used to supervise the system which allows to plot the evolution of its different key parameters. Thus, in the advanced mode, the user can manipulate the status of valves in order to intervene in the operation of the gas through the system.

\vspace{0.2cm}

A relatively simple interface can be used to automate the different tasks that can be performed with the new system. Like the gas recovery to the storage volume, the refilling of the metering volumes, or the filling of the magnet bore during a tracking which must be performed by any standard user of the system without a deep knowledge of the complete system.

\subsection{Functionality of the $^3$He system.}\label{sc:fillingSchema}

The more sophisticated system was designed from the beginning to allow the possibility of different cold bore filling scenarios. The new design was thought to increase the density scanning possibilities of the CAST experiment.

\vspace{0.2cm}

One of the main advantages of the system resides in the possibility of subtracting a metered quantity of gas inside the cold-bore, allowing to move in the scanning axion mass back and forward with much more flexibility. This, together with the possibility of increasing the magnet bore density gradually by controlling the inserting gas flow at an accurate rate, provides the necessary tools to design any possible scanning pattern.

\vspace{0.2cm}

\begin{figure}[!ht]
{\centering \resizebox{0.77\textwidth}{!} {\includegraphics{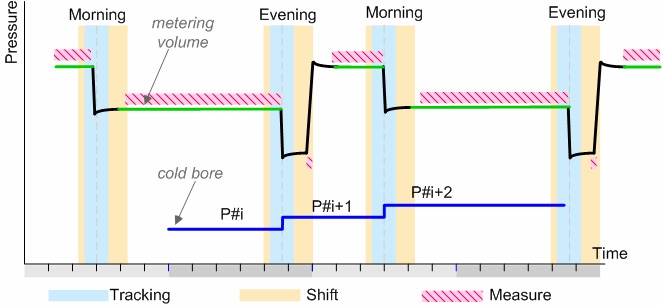}} \par}
\caption{\fontfamily{ptm}\selectfont{\normalsize{ Cold bore filling scheme used in the $^3$He phase, where the increasing density steps during morning and evening trackings ($P_i$, $P_{i+1}$ and $P_{i+2}$) are plotted together with the evolution of pressure in the metering volumes. }}}
\label{fi:fillingSchema}
\end{figure}

The most attractive scenarios were presented to the CAST Collaboration. Between all those scenarios, the filling pattern chosen to increase the magnet bore density during the data taking is shown in figure~\ref{fi:fillingSchema}. In this filling scenario, the density inside the cold bores is increased in the middle of each tracking by $1$ density setting.
\vspace{0.2cm}

This scheme was chosen in order to accelerate the axion mass scanning by covering $2$ density steps per day, however reducing the exposure time in each step to the half which affects to the final sensitivity of the axion-photon coupling. Anyway, this scheme was chosen in order to fulfill the CAST research program for masses up to about $1$\,eV in the assigned time.

\vspace{0.2cm}

The versatility of the new system in removing and inserting metered amounts of gas outside and inside the bores allowed to use a signal triggering protocol which evaluates the significance of the number of tracking counts measured with respect to the mean background of the detectors for a given density step. If the signal significance of all the detectors combined exceeds the threshold imposed by the background detector statistics, the density step is considered to be a \emph{candidate}. The candidate density is measured again in the following days, allowing CAST to detect a signal that would be close to the sensitivity limit of the experiment.

\section{X-ray detectors.}

Three type of X-ray detectors have taken data during CAST data taking periods; a Charge-Coupled Device (CCD), a Time Projection Chamber (TPC), and several MICRO pattern GAseous Structure (MICROMEGAS) detectors.

\vspace{0.2cm}

Currently, only two types are mounted on either end to detect photons from axion-to-photon conversion: three MICROMEGAS detectors and the CCD detector. During the data taking periods corresponding to Phase I and the $^4$He coverage in Phase II a TPC chamber was taking data at the sunset side, covering both magnet apertures. The TPC was replaced by the latest micromegas technology available which provided higher background discrimination capabilities than the TPC chamber (chapter~\ref{chap:micromegas}). The sunrise side of the experiment has been covered, during the full data taking history of CAST, by a micromegas detector and by a CCD detector tuned to work in the required X-ray energy range. For the CCD an X-ray focusing device is used to improve the signal-to-background ratio significantly.

\subsection{The X-ray telescope and the pn-CCD detector.}

\begin{figure}[!ht]
{\centering \resizebox{0.90\textwidth}{!} {\includegraphics{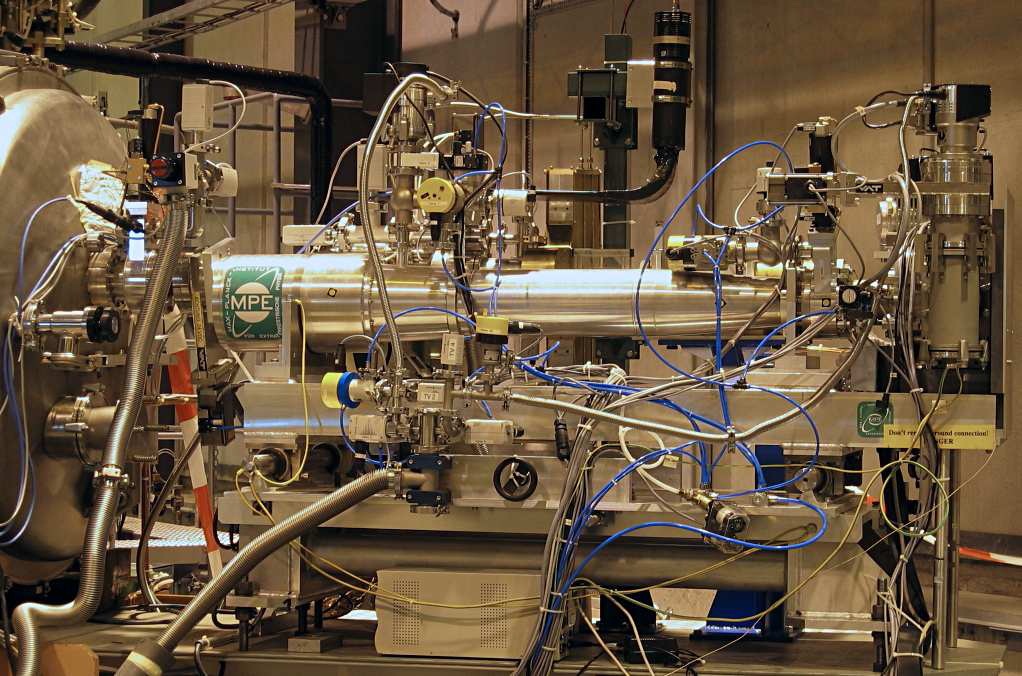}} \par}
\caption{\fontfamily{ptm}\selectfont{\normalsize{ A picture of the X-ray telescope connected to the end of the magnet bore which is on the left, and focuses on the pn-CCD detector which is aligned with the X-ray telescope and the magnet bore pipe, on the right. Electron-pneumatic Valves and vacuum lines are also observed. }}}
\label{fi:telescope}
\end{figure}

An X-ray telescope device~\cite{2005astro.ph.11390K,2007NJPh....9..169K} is placed at the end of one of the magnet bores in the sunrise side of the experiment focusing the expected signal in a small region of a pn-CCD chip specially designed to cover the X-ray energy range $1$-$10$\,keV. The delicate telescope system is kept in vacuum\footnote{During periods when the system is not operating, the X-ray telescope is kept in clean nitrogen for protecting the shells.} at a pressure below $10^{-5}$\,mbar in order to avoid contamination and adsorption on its reflective mirror surface which would result in a degradation of the telescope system efficiency. The system has additional gate valves which separate the magnet from the mirror optics, and the optics from the CCD detector accordingly in order to insulate the mirrors system atmosphere from the other systems as a safety measure. An independent vacuum system for the X-ray telescope and CCD detector is remotely controlled through an electron-pneumatic valves system (see Fig.~\ref{fi:telescope}).

\vspace{0.4cm}
\noindent {\bf The X-ray mirror system.}
\vspace{0.4cm}

\begin{figure}[!ht]
\begin{center}
{\centering \resizebox{0.95\textwidth}{!} {\includegraphics{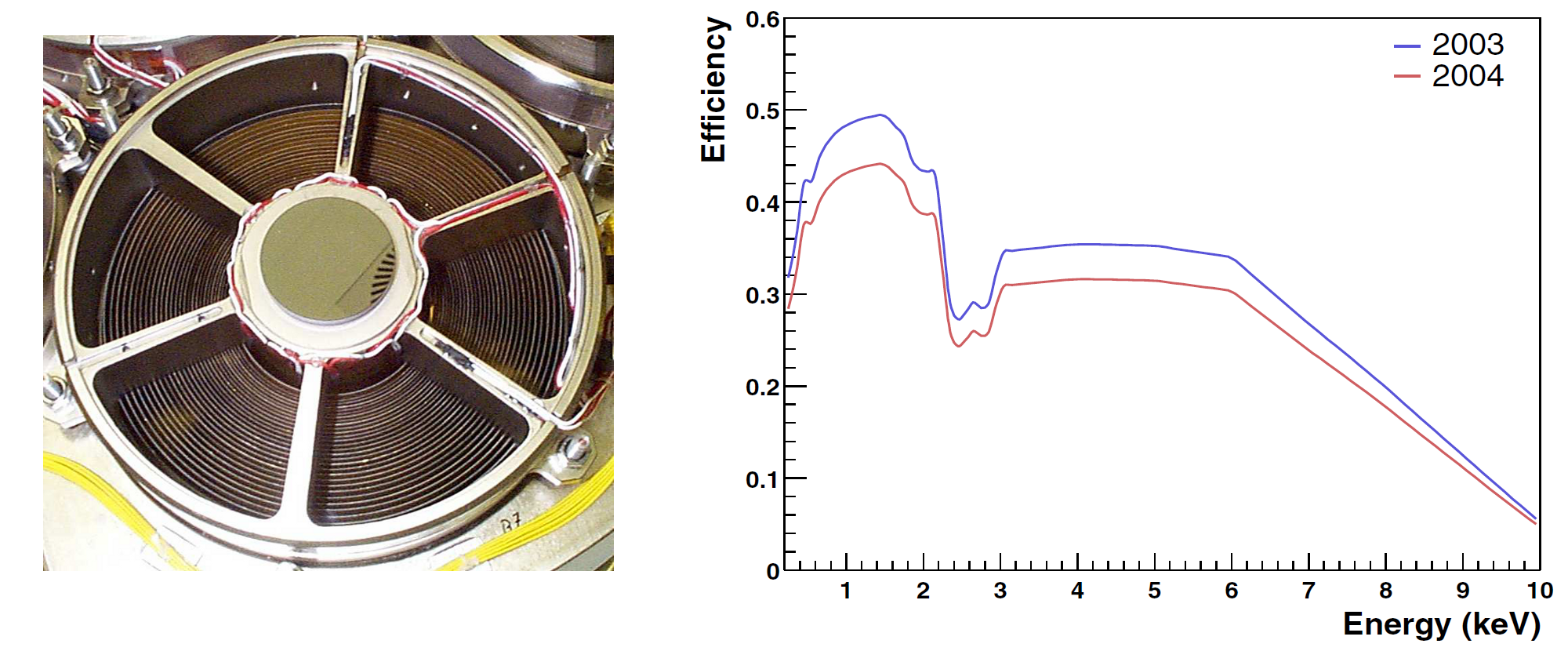}} \par} 
\end{center}
\caption{\fontfamily{ptm}\selectfont{\normalsize{ The front view of the X-ray mirror shells (left) and the telescope system efficiency as a function of the photon energy (right). The reduced efficiency by $11\%$ respect to the 2003 data it is due to the realignment of the telescope necessary for a better centering of the spot on the CCD sensitive area. }}}
\label{fi:xraytelescope}
\end{figure}

The Wolter I type X-ray mirror telescope installed in CAST is a prototype of the X-ray satellite mission ABRIXAS~\cite{ABRIXAS}, which finished in 1999. It consists of a combination of $27$ nested and gold coated parabolic and hyperbolic mirror shells with a focal length of $1600$\,mm. The maximum aperture of the outermost shell is $163$\,mm while the smallest shell has a diameter of only $76$\,mm. The front side of the X-ray mirrors shell is divided into \emph{six} sectors (see Fig.~\ref{fi:xraytelescope}) from which only one is used given that the magnet bore diameter $43$\,mm is smaller than the sector size. The telescope efficiency of each of these sectors was fully characterized at PANTER facilities~\cite{springerlink}, and the sector which presented better performance was chosen.

\vspace{0.2cm}

The overall performance of the X-ray optics depends on two parameters, the transmission efficiency (see Fig.~\ref{fi:xraytelescope}) and the spot size on the pn-CCD chip. The use of a telescope mirror system entails a loss in signal efficiency which is counteracted by the increased significance of a smaller spot size on the detector, which concentrates a signal distributed in an area of $1452$\,mm$^2$ to a spot of about $9$\,mm$^2$, magnifying the signal to background ratio by more than a factor $100$. 

\vspace{0.4cm}
\noindent {\bf The pn-CCD detector.}
\vspace{0.4cm}

\begin{figure}[!ht]
\begin{tabular}{cc}
{\centering \resizebox{0.35\textwidth}{!} {\includegraphics{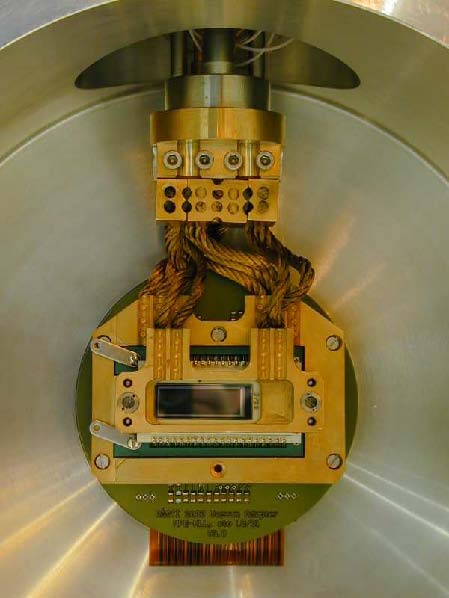}} \par} &
{\centering \resizebox{0.6\textwidth}{!} {\includegraphics{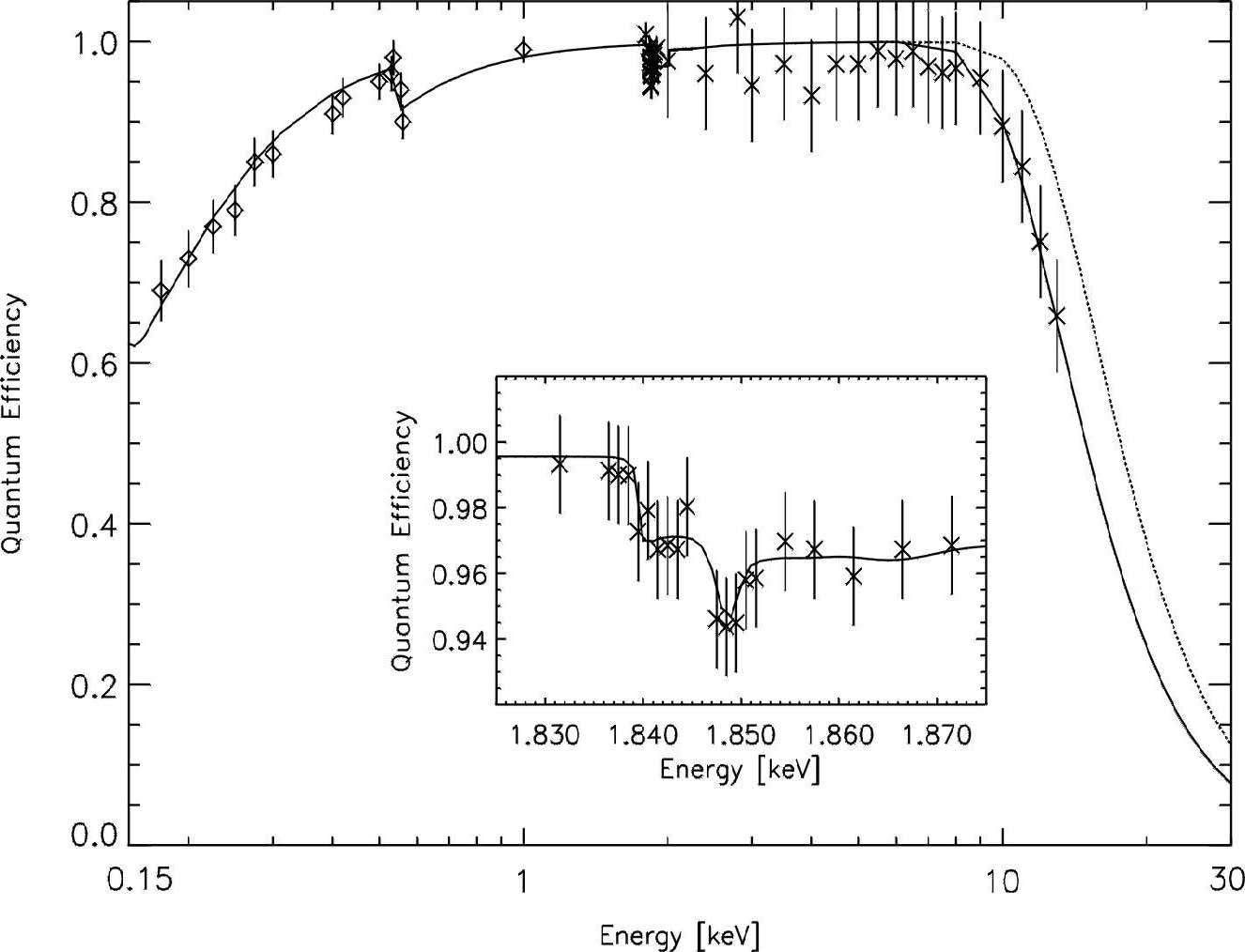}} \par} \\
\end{tabular}
\caption{\fontfamily{ptm}\selectfont{\normalsize{ A picture of the pn-CCD detector installed in the vacuum flange (left) and the quantum efficiency of the pn-CCD detector as measured by the XMM-NEWTON device (right).  }}}
\label{fi:pnCCDchip}
\end{figure}

A CCD detector is placed at the focal plane of the X-ray optics (see Fig.~\ref{fi:pnCCDchip}). The pn-CCD detector for CAST is a prototype, $280$\,$\mu$m thick, developed for the ESA's XMM-NEWTON mission~\cite{pnCCD}. It has a sensitive area of $2.88$\,cm$^2$ divided into $200 \times 64$\,pixels, each pixel covering a region of $150\times150$\,$\mu$m. The effective diameter of the axion signal coming from the Sun core in the CCD chip is $19$\,pixels$=2.83$\,mm. The detector operates at a temperature of $-130^\circ$C which is kept stable over time by using a Stirling cooling system. A major advantage of this type of solid state X-ray detector is the high quantum efficiency close to unity (see Fig.~\ref{fi:pnCCDchip}) in the energy range of interest due to its very thin ($20$\,nm) entrance window at the backside of the chip, allowing to operate the detector in vacuum without additional window.

The good spatial resolution of the CCD chip allows to perform some pattern recognition techniques which are used to discriminate cosmic events such as muons, and other ionizing processes. The remaining background leads to a level of $\left( 7.5\pm0.2 \right)\cdot 10^{-5}$cm$^{-2}$s$^{-1}$keV$^{-1}$ in the axion sensitive range from $1$\,keV to $7$\,keV, which gives a total mean value of $0.15$\,counts during a full CAST solar tracking.

\vspace{0.2cm}

The background level achieved by the CCD detector is partially due to internal radiation coming from the materials close to the detector. Different samples of the detector materials surrounding the detector were measured at the Canfranc Underground Laboratory of the University of Zaragoza. These activity measurements were introduced into a \emph{Geant4} Montecarlo simulation in order to estimate the different contributions to the background, taking into account the geometry of the different parts of the detector system~\cite{2007APh....28..205C}. According to the results of the simulations, the contribution of natural radioactivity coming from the materials surrounding the detector could account for at most $\lesssim 33\%$ of the observed background level while about $\approx 50\%$ of the background is induced by environmental gammas. $^{222}Rn$ is usually one of the strongest sources of natural radioactivity, however, sice the detector is operated in vacuum, the radon contribution is not important at the actual level of sensitivity.

\subsection{The TPC detector.}

The development of the Multi Wire Proportional Chamber (MWPC) by G. Charpak in 1968~\cite{Charpak1968262} allowed for the development of a new type of detectors based on the ionization of a gaseous medium. His work showed that a set of equidistant anode wires could be used as independent proportional counters providing the ionization chamber with position detection capabilities. In the MWPC, the ionized gas charges in the medium are collected by a wired readout plane where the charges are drifted thanks to the high intensity field applied.

\vspace{0.2cm}

\begin{figure}[!ht]
\begin{center}
{\centering \resizebox{0.85\textwidth}{!} {\includegraphics{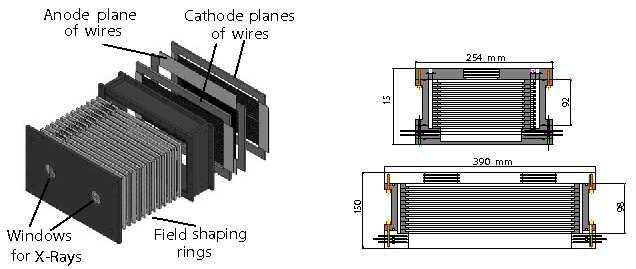}} \par} 
\end{center}
\caption{\fontfamily{ptm}\selectfont{\normalsize{ A schematic of the TPC design at CAST experiment.  }}}
\label{fi:TPCschema}
\end{figure}

The TPC~\cite{Fancher1979383} derives from the MWPC device, while in the MWPC there is only amplification field, in the TPC are distinguished the drift field region and the amplification field region. Thus, the amplification region it is built between anode and cathode wires which are transversally distributed conferring a 2-dimensional readout to the detector (see Fig.~\ref{fi:TPCschema}).

\begin{figure}[!ht]
\begin{center}
{\centering \resizebox{0.85\textwidth}{!} {\includegraphics{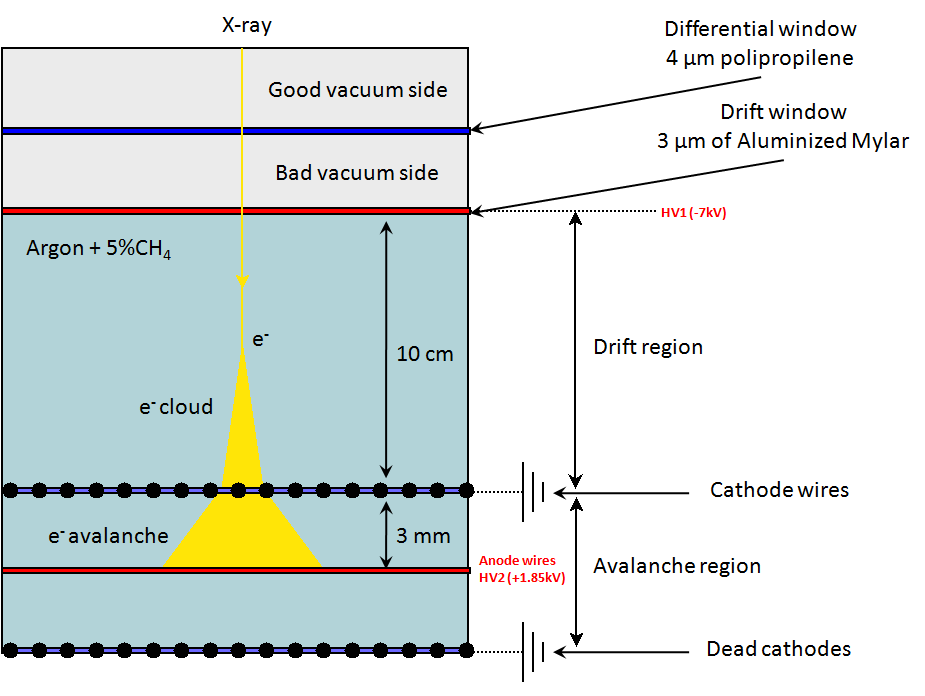}} \par} 
\end{center}
\caption{\fontfamily{ptm}\selectfont{\normalsize{ TPC functional scheme concept and set-up of the TPC detector operating in CAST.  }}}
\label{fi:TPCfunctionalschema}
\end{figure}

In CAST experiment a photon coming from an axion conversion would travel through a vacuum buffer space, entering in the conversion-drift region, where it generates a photoelectron via the photoelectric effect. The photoelectron travels a short distance during which it creates ionization electrons. The electrons drift in a field of about $700$\,V/cm, until they reach the cathode wires and enter the amplification region where a strong field of about $5$\,kV$\cdot$cm$^{-1}$ causes an avalanche magnifying the signal (see Fig.~\ref{fi:TPCfunctionalschema}).

\vspace{0.2cm}

The CAST TPC detector~\cite{1367-2630-9-6-171} has a conversion volume of $10\times15\times30$\,cm$^3$ filled with a mixture of argon with 5\% CH$_4$ gas at atmospheric pressure. The dimensions of the TPC section $15\times30$\,cm$^2$ allow to cover both magnet bores at the sunset side of the experiment. In the front of the detector two windows, consisting of a very thin mylar foil of about $3$\,$\mu$m thick stretched on a metallic strongback, are present in order to allow for X-rays coming from axion-photon conversion to pass-through. The entire chamber is made of $1.7$\,cm thick low radioactivity Plexiglas, except for the electrodes, the the screws and the windows.

\vspace{0.2cm}

The thin mylar windows serve as connection between the detector and the magnet bore pipe which is in vacuum. The requirement of such thin windows for reducing the signal loss, due to the X-ray transmission relation with thickness, allows that some molecules present in the gas mixture inside the detector chamber diffuse through them due to the pressure difference between the chamber and the vacuum side, and disturbing the vacuum side of the CAST magnet bore pipes. To face this problem, a \emph{differential pumping} system is necessary. The vacuum next to the detector is splitten in \emph{two} regions (\emph{bad vacuum} side at about $\lesssim 10^{-3}$\,mbar and \emph{good vacuum} side at about $\lesssim 10^{-6}$\,mbar) by means of a thin ($4$\,$\mu$m) polypropylene window. These volumes are pumped independently. Then, the effect of gas molecules coming from the detector towards the \emph{good vacuum} side is diminished due to the lower diffusion at the polypropylene window, since the pressure difference is much lower. The TPC chamber connection to the magnet and the TPC set-up at the CAST experiment is shown in figure~\ref{fi:TPCfunctionalschema}.

\vspace{0.2cm}
A shielding around the TPC was designed to reduce the background level of the detector and to reduce positional background systematics mainly due to the contamination of the walls at the CAST host building, and the distance to them at the different magnet positions. The design of the shielding was supported by background measurements at different settings and by \emph{Geant4} simulations~\cite{1367-2630-9-7-208}. The resulting shielding design was a compromise between the background reduction of the different components of the shielding and the technical constrains coming from the weight and size available for placing the shielding.

\vspace{0.2cm}

From the inner most part, the shielding is composed by a copper box, $5$\,mm thick, which reduces electronic noise and low energy X-rays, lead bricks, $2.5$\,cm thick, which reduce the low and medium energy environmental $\gamma$ radiation, a cadmium layer, $1$\,mm thick, to absorb thermal neutrons which are slowed down by the surrounding polyethylene pieces with a total thickness of $22.5$\,cm. Thus, a PVC bag tightly covers the whole shielding allowing to reproduce a clean nitrogen atmosphere and to purge possible radon abundance around the detector. After the installation of the shielding the background reduction achieved compared to the previous data showed a reduction factor between $2.5$ and $4$, furthermore the previously observed discrepancies of background levels at different positions were vanished due to the shielding reduction of $\gamma$ sources coming from the walls. The typical background rate of the detector after shielding reduction was about $20$-$60$\,$\cdot$h$^{-1}$ at the energy range $3$-$7$\,keV.

\vspace{0.2cm}

The TPC detector was the first detector which took data at the CAST experiment, and it covered the vacuum Phase I and the $^4$He Phase II periods of the CAST data taking program. However, the TPC was replaced by a new detector system (described on chapter~\ref{chap:micromegas}) composed by \emph{two} Micromegas detectors which showed improved capabilities in background discrimination respect to the TPC.

\subsection{The Micromegas detector.}

The micromegas detector~\cite{Giomataris:299159} emerged to overcome the technical limitation of the wired TPC chambers which were reaching its wiring spacing limit. The micropattern detector consists of conductive strips printed on a board, this technology allows to obtain shorter spacing in the order of $< 300$\,$\mu$m providing a higher spatial resolution, able to reach values lower than $60$\,$\mu$m at the right conditions, moreover improving the modest rate capabilities of conventional MWPC of about $103$\,s$^{-1}$mm$^{-2}$. Moreover, the fact that the strips readout is built over a flat surface allowed to built shorter amplification gaps in the order of $50$\,$\mu$m to $100$\,$\mu$m which provides excellent gain properties~\cite{Giomataris1998239}. The most robust detector structure together with the thinner amplification gap confers the detector an excellent energy resolution, below $12\%$\,FWHM at $6$\,keV~\cite{2009NIMPA.608..259D}, placing the micromegas technology as one of the best candidates for use in multipurpose applications~\cite{Charpak200226} which moreover require an accurate particle identification.

\vspace{0.2cm}

The higher rate capabilities of micropattern detectors, $10^{7}$\,s$^{-1}$mm$^{-2}$, makes them suitable to operate in applications where high-rate beams are present~\cite{Charpak199847,Delbart:772376}. Thus, micromegas detectors are consolidating in several experiments, from a neutron detector at the n-TOF beam facility~\cite{Andriamonje:643275} to a muon tracker for the COMPASS\footnote{COmmon Muon and Proton Apparatus for Structure and Spectroscopy} experiment~\cite{Thers2001133}. Research and development is undergoing to introduce these detectors in future applications, as a large muon track detector for the s-LHC\footnote{super-Large Hadron Collider}~\cite{1748-0221-4-12-P12015} or as a promising candidate for studying neutrino physics at NEXT\footnote{Neutrino Experiment with a Xenon TPC}~\cite{1475-7516-2010-10-010}.


\vspace{0.2cm}

\begin{figure}[!hb]
\begin{center}
{\centering \resizebox{0.75\textwidth}{!} {\includegraphics{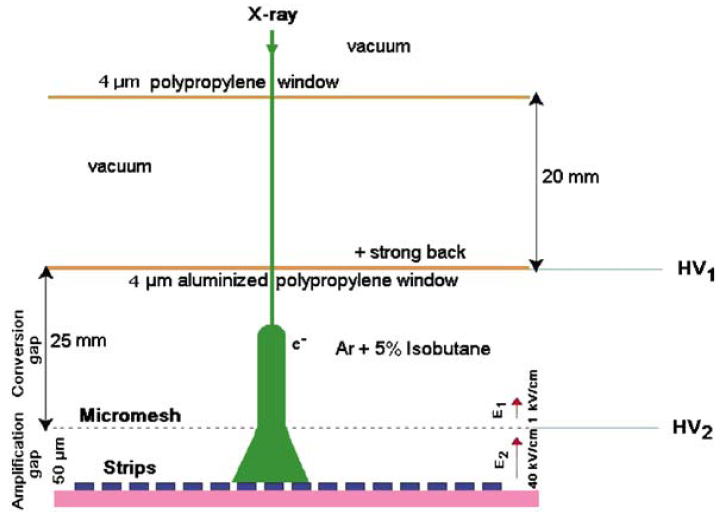}} \par}
\end{center}
\caption{\fontfamily{ptm}\selectfont{\normalsize{ Micromegas detection principle and set-up at the CAST experiment. }}}
\label{mm}
\end{figure}

The micromegas detector at CAST~\cite{Abbon:2007ug} consists of a sensitive area slightly higher than the expected cold-bore signal region of about $15$\,cm$^2$. The micromegas detection principle together with the set-up at the CAST experiment is shown at figure~\ref{mm}.

\begin{figure}[!ht]
\begin{center}
\begin{tabular}{cc}
{\centering \resizebox{0.45\textwidth}{!} {\includegraphics{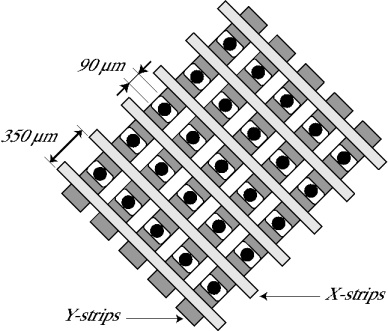}} \par} &
{\centering \resizebox{0.45\textwidth}{!} {\includegraphics{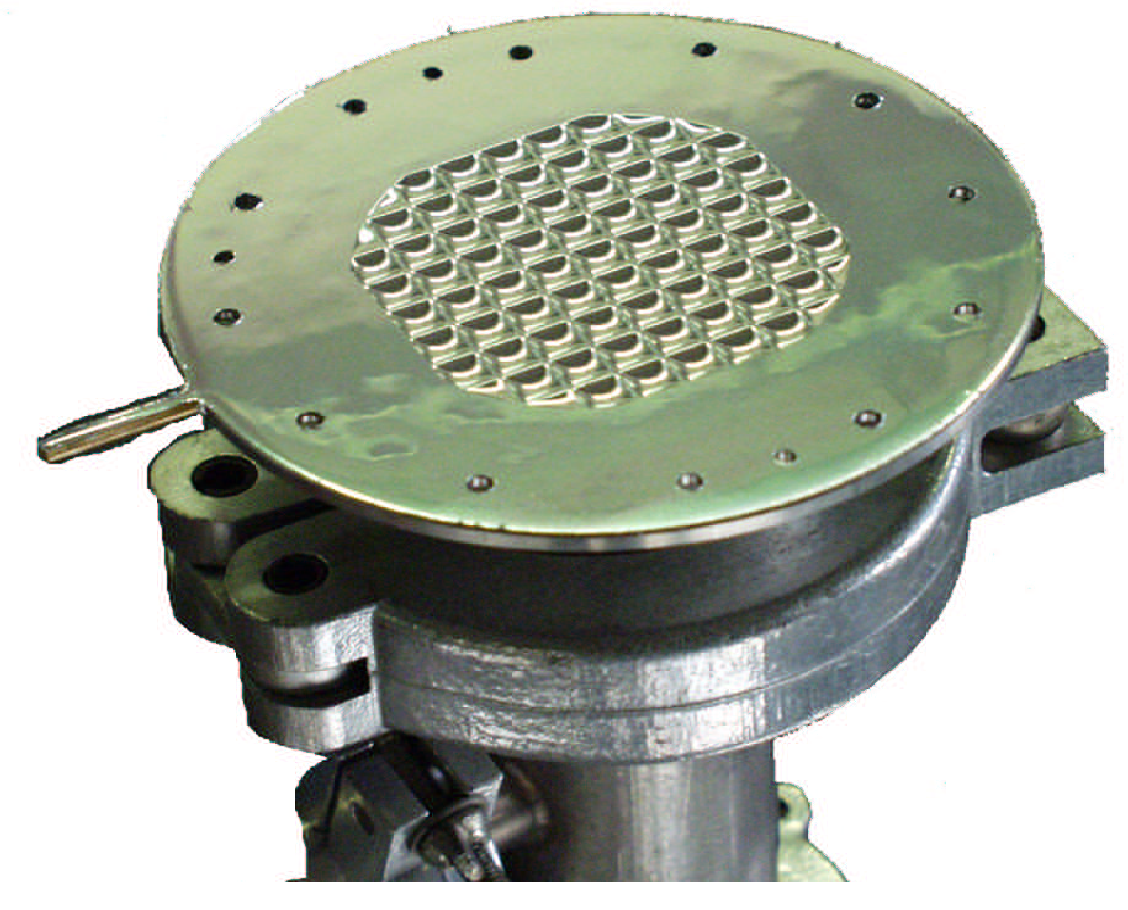}} \par} \\
\end{tabular}
\end{center}
\caption{\fontfamily{ptm}\selectfont{\normalsize{X-Y strip pattern design (left) and a picture of the aluminized polypropylene drift window (right). }}}
\label{fi:castStripsReadout}
\end{figure}

The amplification gap is delimited by a thin micromesh conductive grid which structure depends on the micromegas detector type (detailed on chapter~\ref{chap:micromegas}). The gap homogeneity it is achieved by placing precision insulating pillars, or similar, between the readout plane and the mesh. The amplification field is reached by applying a high voltage at an electrode in contact with the micromesh (usually denominated as $V_m$ or $HV2$) while the strips plane remains grounded. The short amplification distance and the surface homogeneity allows to apply amplification fields of about $40$\,kV/cm. The 2-dimensional pattern of the strips plane (see Fig.~\ref{fi:castStripsReadout}) provides the detector with positional sensing by measuring the total induced charge in each of the strips (detailed on chapter~\ref{chap:rawdata}). A low radioactivity and cylindrical Plexiglas chamber defining a drift distance of $2.5$-$3$\,cm tightly encloses the micromegas readout in order to circulate argon mixtures minimizing external contamination. A thin $4$\,$\mu$m aluminized polypropylene window (see Fig.~\ref{fi:castStripsReadout}) with a high X-ray transmission is placed on top the Plexiglas chamber, where the high drift voltage is applied (usually denominated as $V_d$ or $HV1$) serves at the same time as connection to the vacuum side of the magnet bore pipes. As in the case of the TPC chamber, the gas contained inside the chamber diffuses through the thin drift window towards the vacuum side of the magnet. In order to minimize this effect, a \emph{differential pumping} system was implemented, separating the vacuum next to the detector by means of a $4$\,$\mu$m polypropylene window (see Fig.~\ref{mm}).

During the last years, two new type of micromegas have been under development for the CAST experiment. These new type of micromegas detectors have been taking data in the last running period, thus the detector systems were upgraded substantially during the shutdown period. These new detectors and the upgrades of the detector systems are described in chapter~\ref{chap:micromegas}.



\chapter{New Micromegas detectors and detector systems.}
\label{chap:micromegas}
\minitoc

\section{Introduction}

The mandatory shutdown in 2007 to adapt the CAST magnet to the challenging $^3$He Phase was used by the Micromegas group to upgrade the detector system at the sunrise Micromegas line and to replace the TPC chamber, covering the sunset side in the previous data taking periods, by \emph{two} Micromegas detectors.

\vspace{0.2cm}

The conventional Micromegas detectors taking data in the sunrise line before 2007 showed good discrimination capabilities for the selection of X-ray events. The lower background levels reached and improved energy resolution respect to the TPC taking data in the sunset side pushed the efforts from the group  towards the development of a new detector system equipped with a new branch of Micromegas detectors.

\vspace{0.2cm}

In general, the upgrades carried out during this period were motivated by the possibilities to increase the performance of \emph{three} of the \emph{four} detection lines at the CAST experiment; increasing the detectors efficiency, reaching lower background levels and introducing a new branch of Micromegas detectors with improved stability.

\section{The Micromegas detectors efficiency.}\label{sc:mmEfficiency}

The Micromegas detector chamber was calibrated at Panter X-ray facility in Munich~\cite{springerlink} by 2002. The efficiency measurements obtained at different energies in the range of interest showed a good agreement with the results obtained with a \emph{Geant4}~\cite{Agostinelli} simulation where the main detector contributions were taken into account~\cite{TheopistiThesis}; the drift window made of $4$\,$\mu$m of aluminized polipropilene (the thickness of deposited layer of aluminum was below $50$\,nm which is the final value considered for the \emph{Geant4} simulation), the detector chamber filled with Ar+5\% iC$_4$H$_{10}$ including a drift distance of $3$\,cm and the $4$\,$\mu$m polipropilene differential vacuum window. The effect of the drift window strongback was not considered in the simulation and the reduction effect on the efficiency was estimated by fitting the efficiency measurements and the results from the simulation data, leading to a loss in efficiency of~$5$\,\%.

\vspace{0.2cm}

Thus, in the new CAST phase with buffer gas inside the bores, the effect of cold windows containing the gas had to be included in the detectors efficiency. The transmission of cold windows for X-rays was also calibrated at Panter X-ray facility~\cite{PANTERwindows}. The expected energy profile measured was compatible with the values provided by NIST data base~\cite{nlavn} for a $14$\,$\mu$m polipropilene layer. The effect of the cold windows strong-back was determined by fitting the NIST data at this conditions with the obtained measurements in Panter, leading to an overall reduction in efficiency of $12.6$\,\%.

\vspace{0.2cm}

The \emph{Geant4} code produced at that time has been taken as reference in order to extrapolate the changes produced at the detector systems. The most significative change introduced in 2007 was the overpressured detector chamber (described on section~\ref{sc:monitoringSystem}), that improved the detector efficiency by more than $9$\,\% with respect to the overall efficiency in the previous data taking period, with $^4$He in the magnet bores, when the detector chamber was operating at atmospheric pressure. Table~\ref{ta:efficiencyMM} shows the overall detector efficiency in the range 2-7\,keV by taking into account different contributions; differential vacuum window (DW) and cold windows (CW), together with the overall efficiency when the detector was taking data at atmospheric pressure.

\begin{table}[ht!]
\begin{center}
\begin{tabular}{ccc}
\hline
	&	&	\\
\bf{Pressure [atm]} & \bf{Simulation} & \bf{Efficiency (2-7\,keV)} \\
	&	&	\\
1.4 & Detector & 75 \% \\
1.4 & Detector + DW & 74 \% \\
1.4 & Detector + DW + CW & 60 \% \\
1.0 & Detector + DW + CW & 55 \% \\
	&	&	\\
\hline
\end{tabular}
\end{center}
\caption{\fontfamily{ptm}\selectfont{\normalsize{Overall efficiency of the Micromegas detector including different contributions; intrinsic detector efficiency (at an argon pressure of 1.4\,bar), absorption of cold windows containing the $^3$He buffer gas and the differential vacuum window. }}}
\label{ta:efficiencyMM}
\end{table}

\vspace{0.2cm}

Thus, in order to reduce the diffusion of argon gas circulating inside the overpressured chamber towards the vacuum side, a new set of drift windows was built of aluminized Mylar of $5$\,$\mu$m thickness. The simulations for the $^3$He Phase take into account the new drift window thickness and material.

\vspace{0.2cm}

\begin{figure}[!h]
{\centering \resizebox{0.98\textwidth}{!} {\includegraphics{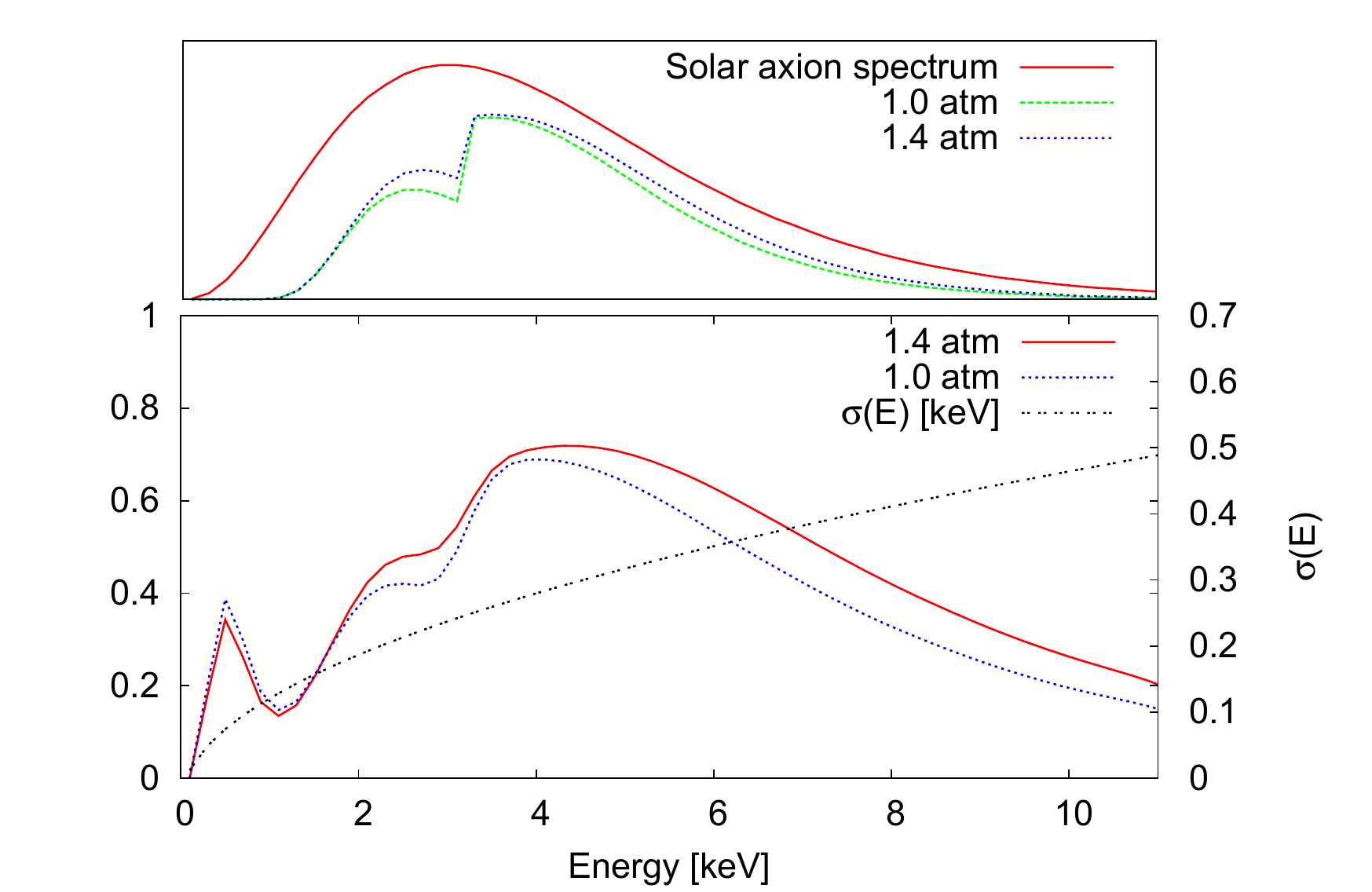}} \par}
\caption{\fontfamily{ptm}\selectfont{\normalsize{ Monte-Carlo axion spectrum generated in \emph{Geant4} and the simulated response of the detector taken into account the full detection system; cold window, differential window and drift window. \emph{Two} scenarios are presented, detector at atmospheric pressure and detector with overpressure. }}}
\label{fi:efficiencyMM}
\end{figure}

The final efficiency calculated in order to obtain the expected axion counts in the detector was performed by including a Monte-Carlo axion solar flux spectrum inside the \emph{Geant4} simulation obtaining a more accurate response of the detector to the expected signal (see Fig.~\ref{fi:efficiencyMM}). The detector response to the axion signal is also smoothed taking into account the energy resolution of the detector by convoluting the efficiency with a Gaussian function, convolution that can be described by the following expression,

\begin{equation}
\epsilon'(E') = \int \frac{\epsilon(E)}{\sqrt{2\pi}\sigma(E)} exp\left( -\frac{1}{2} \frac{(E-E')^2}{\sigma^2(E)} \right) dE
\end{equation}

\vspace{0.2cm}

\noindent where $\epsilon(E)$ is the \emph{Geant4} simulated efficiency and $\sigma(E) \propto \sqrt E$ is introduced by using the expected energy resolution of 15\,\% (FWHM) at $5.96$\,keV. The final efficiency $\epsilon'(E)$ including the energy resolution of the detector it is represented in figure~\ref{fi:efficiencyMM}.

\section{The need for low background detectors.}

\vspace{-0.2cm}

\begin{figure}[h!]
\begin{tabular}{ccc}
	&
	&
\includegraphics[angle=270,width=0.84\textwidth]{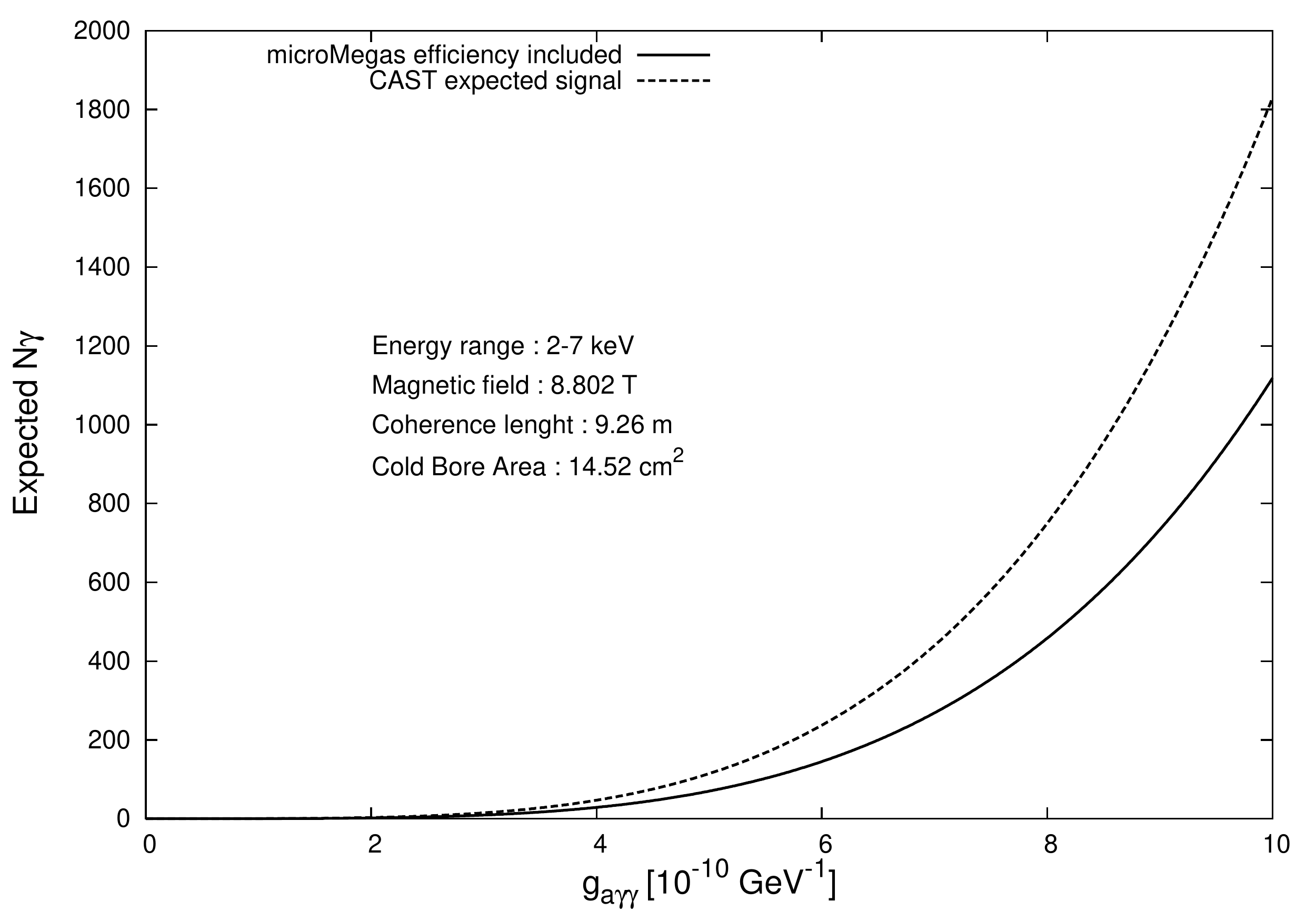}	\\
\end{tabular}

\caption{\fontfamily{ptm}\selectfont{\normalsize{ Integrated number of photons coming from axions expected during half of the tracking period (2700\,s) as a function of the axion-photon coupling constant, $g_{a\gamma}$. }}}
\label{fi:NgammaGa}
\end{figure}

The sensitivity of the CAST experiment in the axion-photon coupling determines the capability of the experiment to find the axion, usually denoted as discovery potential. The sensitivity of CAST is related with the number of counts that are detectable (see Fig.~\ref{fi:NgammaGa}) over the detectors background and is directly dependent on the detector parameters, and therefore their performance, by the relation

\begin{equation}
g_{a\gamma} \propto  \frac{ b^{1/8} } { t^{1/8} \epsilon^{1/4} }
\label{eq:g}
\end{equation}

\vspace{0.1cm}

\noindent where $b$ is the X-ray background rate of the detector in the expected signal range, $t$ is the total exposure time, and $\epsilon$ is the system overall efficiency, that includes the detector efficiency, as well as the transmission of the windows that are in front of the detector.


\vspace{0.2cm}

Once the main parameters of the experiment are fixed (magnetic field intensity and field coherence length) an improvement in discovery potential can only be achieved by an improved performance of the detector systems; by increasing the detectors efficiency or by lowering the background level of the detector (composed of any ionizing process similar to an X-ray pattern). Since the detector efficiency is limited by the gas properties and the X-rays transmission through the detection system, a further increase on efficiency would not be significative. It is reasonable to say that the only possibility of increasing the discovery potential of the experiment is to reduce the detectors background level.

Background reduction is achieved in Micromegas detectors by building the detectors using low radiation materials as kapton and Plexiglas, optimizing the shielding protecting the detector from external radiation (using lead, copper, Plexiglas and cadmium), and by exploiting the signal information that provides the Micromegas readout for rejection of non X-ray events.

\section{New Micromegas detectors for the CAST experiment.}

\begin{figure}[!b]
\begin{tabular}{cc}
{\centering \resizebox{0.47\textwidth}{!} {\includegraphics{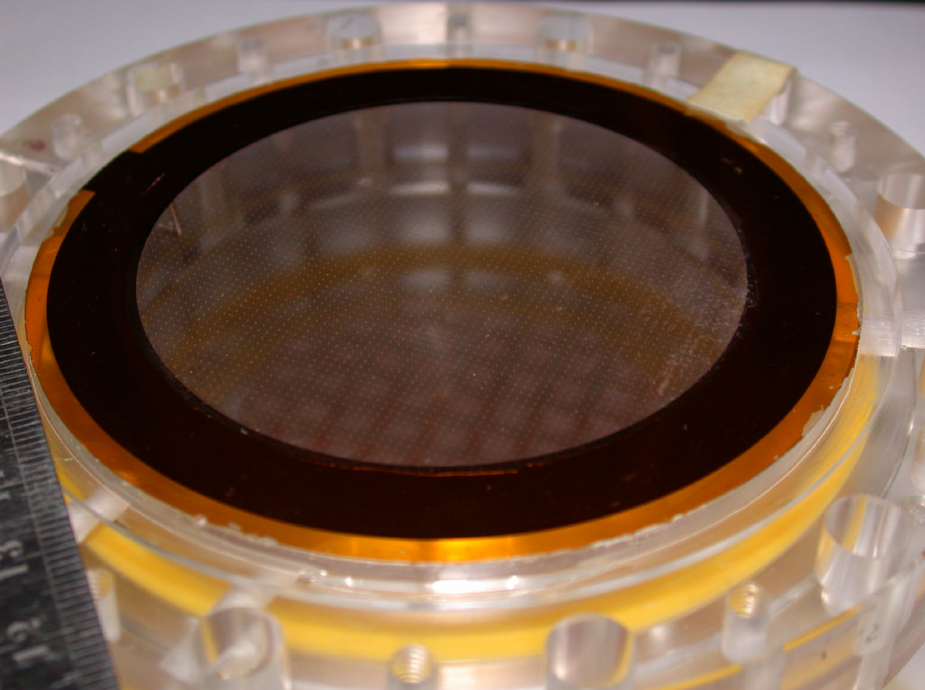}} \par} &
{\centering \resizebox{0.47\textwidth}{!} {\includegraphics{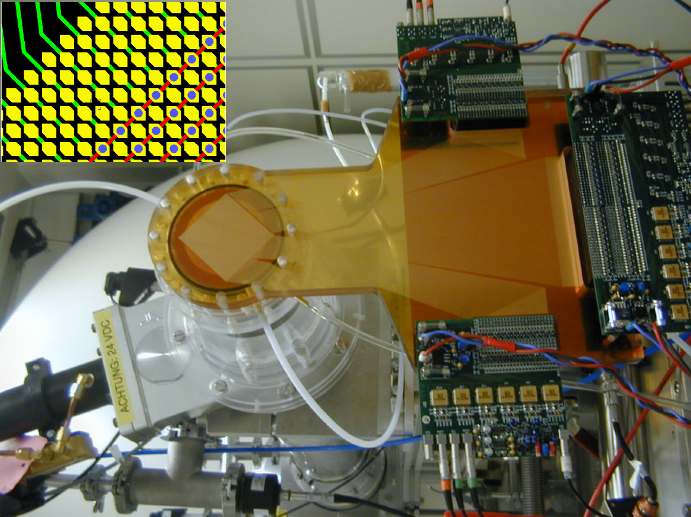}} \par} \\
\end{tabular}
\caption{\fontfamily{ptm}\selectfont{\normalsize{ Mesh (left) and strip readout (right) pictures of the first conventional Micromegas design (V-branch detector) that took data at CAST experiment. The picture was taken at Panter facilities where the detector efficiency was measured. }}}
\label{fi:meshAndStripsVbranch}
\end{figure}

The first Micromegas detectors produced for CAST experiment (denominated V-branch) were built by using a well established \emph{conventional} fabrication process~\cite{Giomataris:299159}. In these detectors (see Fig.~\ref{fi:meshAndStripsVbranch}), the micromesh, $3$\,$\mu$m thick, is made of nickel using the electroforming technique. A grid frame of quartz fibers or other insulating material was mounted on the strips surface composing a precise structure, of about $100$\,$\mu$m thickness, where the micromesh is placed. At the Micromegas conventional technology the micromesh was placed over the strips readout by mechanical means, stretching it by using some screws that exert some pressure at the mesh boundaries. The uniformity of the mesh at the full active area is reached by the electric field forces involved when the high voltage is applied.

During the last years two new generations of Micromegas technologies have been developed; bulk~\cite{2006NIMPA.560..405G} and microbulk \cite{1748-0221-5-02-P02001}. These new technologies are characterized by the fact that the mesh and the readout strips form one single entity.

\subsection{Bulk technology.}

Bulk detectors arose as a promising technology due to the possibility of building larger surface Micromegas detectors~\cite{Anvar2009415} providing a good amplification gap uniformity. In bulk detectors the electroformed mesh is substituted by a stainless steel woven mesh of $30$\,$\mu$m thickness that is placed on top of the readout strips by insulating pillars, produced by a simple process based on PCB technology.

\vspace{0.2cm}

The advantage of the new woven mesh used resides in the possibility to easily stretch and handle it, making it possible to perform a more complex manipulation during the fabrication process. This type of mesh is commercially available at large areas (2x40 m$^2$) and different materials (Au, Cu, Fe, Ni and Ti).

\vspace{0.2cm}

\begin{figure}[!hb]
{\centering \resizebox{0.8\textwidth}{!} {\includegraphics{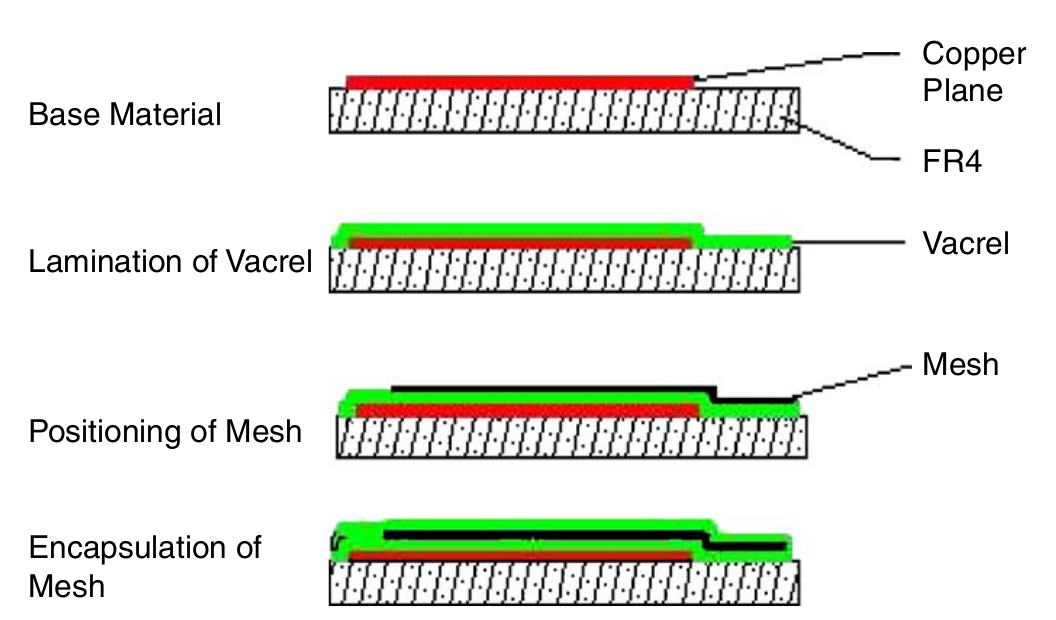}} \par}
\caption{\fontfamily{ptm}\selectfont{\normalsize{ Schematic of the mesh positioning process in bulk detectors. }}}
\label{fi:bulkFabrication}
\end{figure}

During the fabrication process, the anode copper plane where the strips have been printed is placed over a substrate (i.e. FR4), a photoresistive material of the desired amplification gap thickness is deposited over the anode strips together with the woven mesh on top of it, and afterwards encapsulating the mesh under another photoresistive film (see Fig.~\ref{fi:bulkFabrication}).

\begin{figure}[ht!]
\begin{tabular}{cccccccc}
	&
	&
	&
\includegraphics[width=0.33\textwidth]{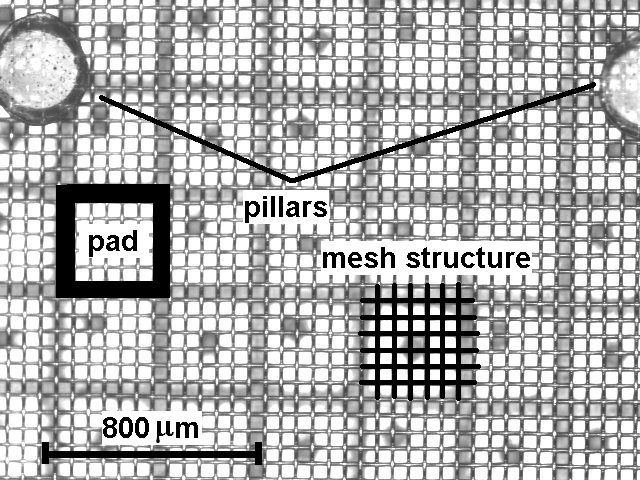} &
	&
	&
	&
\includegraphics[width=0.33\textwidth]{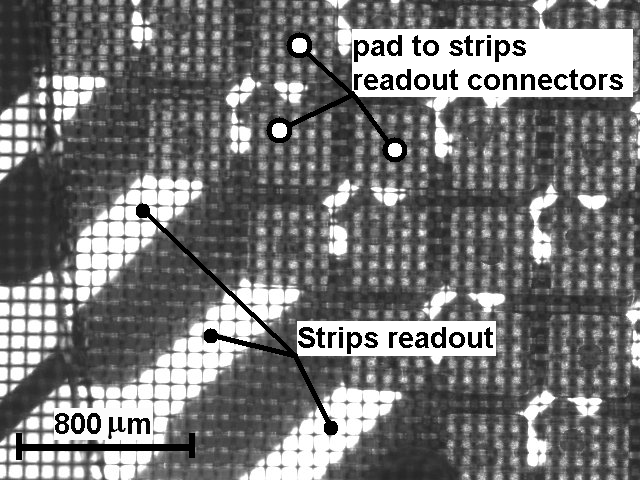} \\
\end{tabular}
\caption{\fontfamily{ptm}\selectfont{\normalsize{ Two microscope pictures from a Bulk type detector. In the first one, the microscope is focusing in the mesh, the ground pads that collect the electrons and pillars supporting the mesh can also be observed. On the second picture readout strips interconnecting the pads are visualized.  }}}
\label{fi:newmmBulk}
\end{figure}

These layers are laminated together at high temperature making up a single entity. The photoresistive film which defines the amplification gap is etched by using a photolithographic method producing the pillars pattern. The pillars have a cylindrical shape of $300$\,$\mu$m and are distributed at equidistant intervals of $2$\,mm (see Fig.~\ref{fi:newmmBulk}).

\vspace{0.2cm}

Bulk detectors are robust, easy to construct and their amplification gap is quite uniform even at large surfaces. They have an acceptable energy resolution only limited by the thicker mesh, in nominal operation these detectors can reach up to 18\% FWHM at $5.9$\,keV. The bigger amplification gap distance produces low noise detectors due to its lower capacitance, but at the same time makes the gain to be more sensitive to pressure variations.

\subsection{Microbulk technology.}

On the other hand, microbulk detectors are manufactured by high accuracy photo-lithography techniques on copper-clad kapton foils resulting in a mesh of $5$\,$\mu$m of copper supported by kapton pillars.

\vspace{0.2cm}

\begin{figure}[ht!]
\begin{tabular}{cccccccc}
	&
	&
	&
\includegraphics[width=0.33\textwidth]{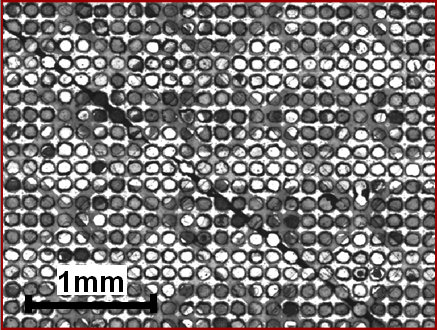} &
	&
	&
	&
\includegraphics[width=0.33\textwidth]{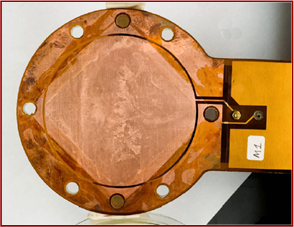} \\
\end{tabular}

\caption{\fontfamily{ptm}\selectfont{\normalsize{ A microscope picture from a Microbulk detector where the mesh holes are visible, and a picture of the full active area of the detector installed in a PCB board. }}}
\label{fi:newmmMicrobulk}
\end{figure}

\begin{figure}[!t]
{\centering \resizebox{0.98\textwidth}{!} {\includegraphics{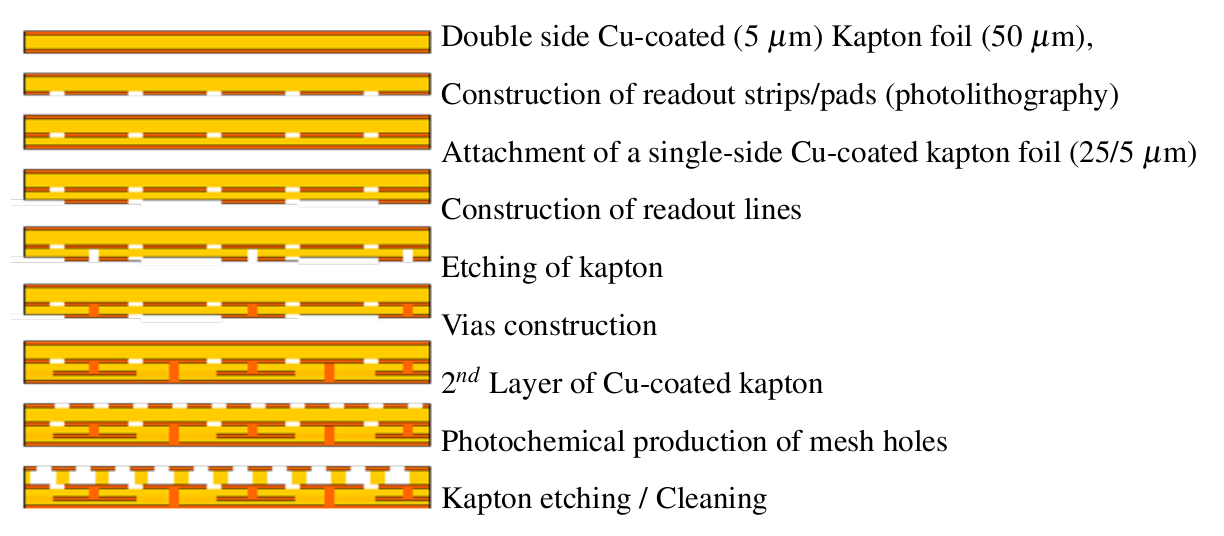}} \par}
\caption{\fontfamily{ptm}\selectfont{\normalsize{ Schematic of fabrication of the readout pattern in microbulk detectors.  }}}
\label{fi:microbulkFabrication}
\end{figure}

The pads pattern and the strips readout are directly built on the copper-clad kapton foil. A photoresistive film is deposited and masked with the desired pattern by using UV light, then the copper can be removed by a standard lithographic process. A first double side copper-clad kapton foil of $50$\,$\mu$m thickness defines the amplification gap. The 2-dimensional readout is achieved by the attachment of additional single side copper kapton foils which are etched in corcondance with the readout pattern. The mesh grid is produced by etching the top copper layer creating a holes pattern (see Fig.~\ref{fi:newmmMicrobulk}), and then the amplification gap space is created removing the kapton by using a photochemical process that with the proper exposure time generates the desired kapton pillars pattern. Figure~\ref{fi:microbulkFabrication} describes the microbulk fabrication process.

\vspace{0.2cm}

These detectors present the highest homogeneity on the amplification gap, given by the accurate thickness of the kapton foil. This novel technology has lead to the best energy resolution ever reached with a micropattern gaseous detector~\cite{2009NIMPA.608..259D}, which are able to reach energy resolutions of $11$\% FWHM at $5.9$\,keV thanks to the great amplification gap homogeneity and its thinner copper mesh. They present higher noise due to the shorter amplification gap but they are less sensitive to pressure variations than bulk detectors.

\subsection{Evolution of Micromegas detectors in the CAST Experiment.}

\begin{figure}[ht!]
\includegraphics[width=1.0\textwidth]{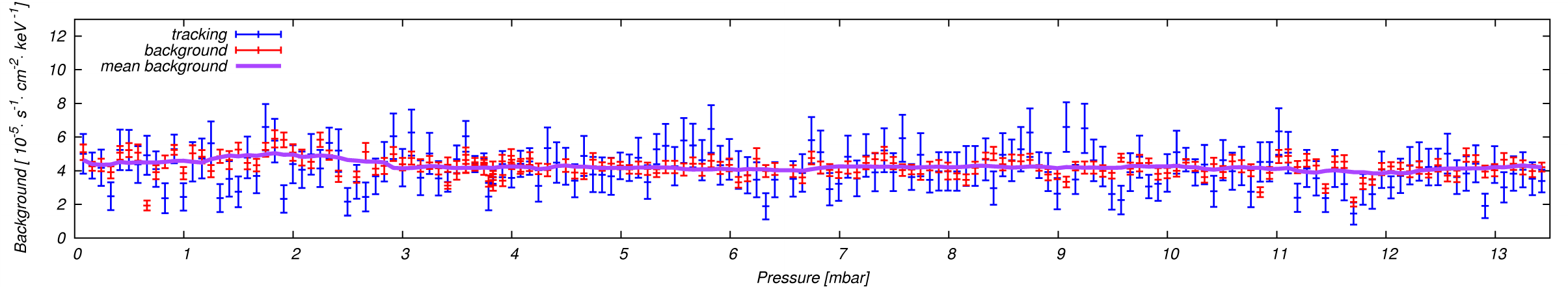}
\caption{\fontfamily{ptm}\selectfont{\normalsize{ Rate evolution for background (magnet not aligned with the Sun) and tracking (magnet aligned with the Sun) settings during the $^4$He data-taking phase in 2005 and 2006. The settings are described as a function of $^4$He pressures inside the cold bore, that define each of the axion mass sensitivity regions. This pressure is always increasing during the completion of the $^4$He Phase. }}}
\label{fi:he4Rate}
\end{figure}

\begin{figure}[h!]
\begin{center}
\begin{tabular}{cccc}
\includegraphics[width=0.45\textwidth]{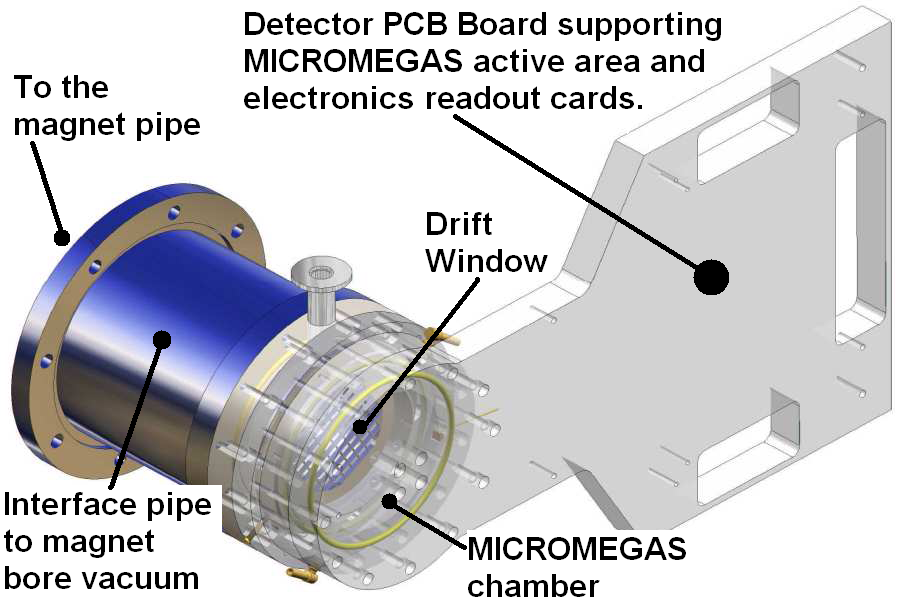} &
 & & 
\includegraphics[width=0.4\textwidth]{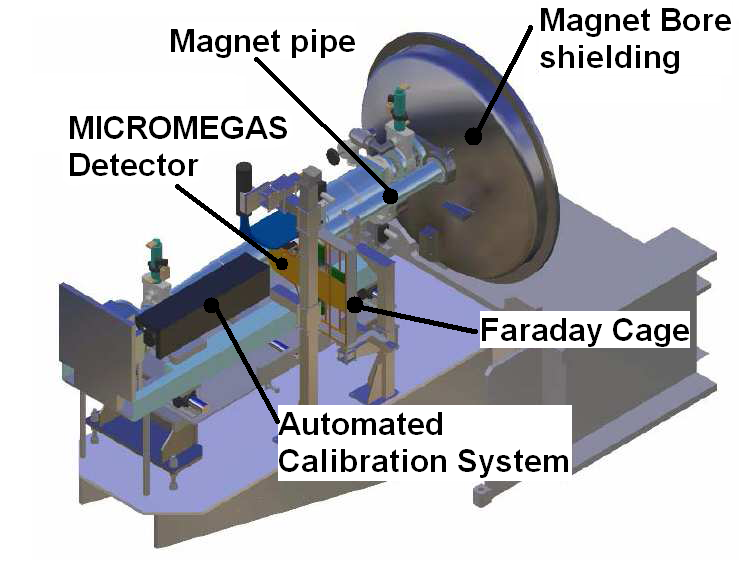} \\ 
\end{tabular}
\end{center}
\caption{\fontfamily{ptm}\selectfont{\normalsize{ Drawings from the Micromegas detection line connected to the magnet bore as it was before the interventions and modifications carried out during 2007. On the left, a zoomed drawing of the detector chamber connection to the system. On the right, the complete system as it was installed during the previous data-taking period. }}}
\label{fi:oldSunrise}

\end{figure}

Since the start of CAST data-taking in 2003, conventional Micromegas detectors (stand alone mesh positioned on top of the anode plane) were taking data in the sunrise tracking side of the magnet. This type of detector showed good performance and stability during the previous data-taking periods before 2007. At the $^4$He buffer gas phase, pressures up to about $13$\,mbar inside the magnet bores were covered (see Fig.~\ref{fi:he4Rate}), using the detection system presented in figure~\ref{fi:oldSunrise}. 

\vspace{0.2cm}

Several detectors, based on the two new technologies described, were produced at CERN facilities and operated in CAST since the beginning of 2008. These new and more versatile technologies replaced the previous conventional technology.

\vspace{0.2cm}

The full range of detector branches produced for CAST during its life time is summarized at table~\ref{ta:micromegasBranches}. The first conventional detectors (named V-branch) were the bigger active area detectors used in CAST ($45$\,cm$^2$ versus the $14.5$\,cm$^2$ region required to fully cover the signal bore area). A new branch of conventional Micromegas detectors of smaller size (T-branch) was produced in order to be used with an under development X-ray focusing device that could be installed in the new Sunrise line described on section~\ref{sc:newLine}. However, the performance of the new optics was not high enough for increasing the CAST discovery potential due to an unexpected reduced transmission efficiency~\cite{Andriamonje:1065408}. A T-branch detector (T3) took some data at the end of 2007 but it had to be finally replaced by a bulk detector (B-branch) able to cover the full signal area for the start of data taking in 2008.

\vspace{0.2cm}


\begin{table}[ht!]
\begin{center}
\begin{tabular}{cccccc}
{\small{Branch}} &	 {\small{Technology}}	&	 {\small{Pitch [$\mu$m]}} &	 {\small{Area [cm$^2$]}}	&	  {\small{\# Strips}}	&	 {\small{Amplification gap [$\mu$m]}}	\\
	&	&	&	&	&	\\
\hline
	&	&	&	&	&	\\
V	&	conventional	&	350		&	45		&	192 x 192	&	50	\\
	&	&	&	&	&	\\
T	&	conventional	&	350		&	11		&	96 x 96		&	50	\\
	&	&	&	&	&	\\
B	&	bulk		&	550		&	36		&	106 x 106	&	128	\\
	&	&	&	&	&	\\
M	&	microbulk	&	550		&	36		&	106 x 106	&	50	\\
\end{tabular}
\caption{\fontfamily{ptm}\selectfont{\normalsize{ The different branches of Micromegas detectors that have taken data at CAST experiment.}}}
\label{ta:micromegasBranches}
\end{center}
\end{table}

The newest microbulk detectors (M-branch) produced have been finally consolidated and is considered as the best option versus the bulk technology for the CAST experiment due to reasons that will be described on chapter~\ref{chap:gLimitHe3}.

\section{The new sunrise Micromegas detection line.}\label{sc:newLine}

\begin{figure}[h!]
\begin{tabular}{cc}
\includegraphics[width=0.73\textwidth]{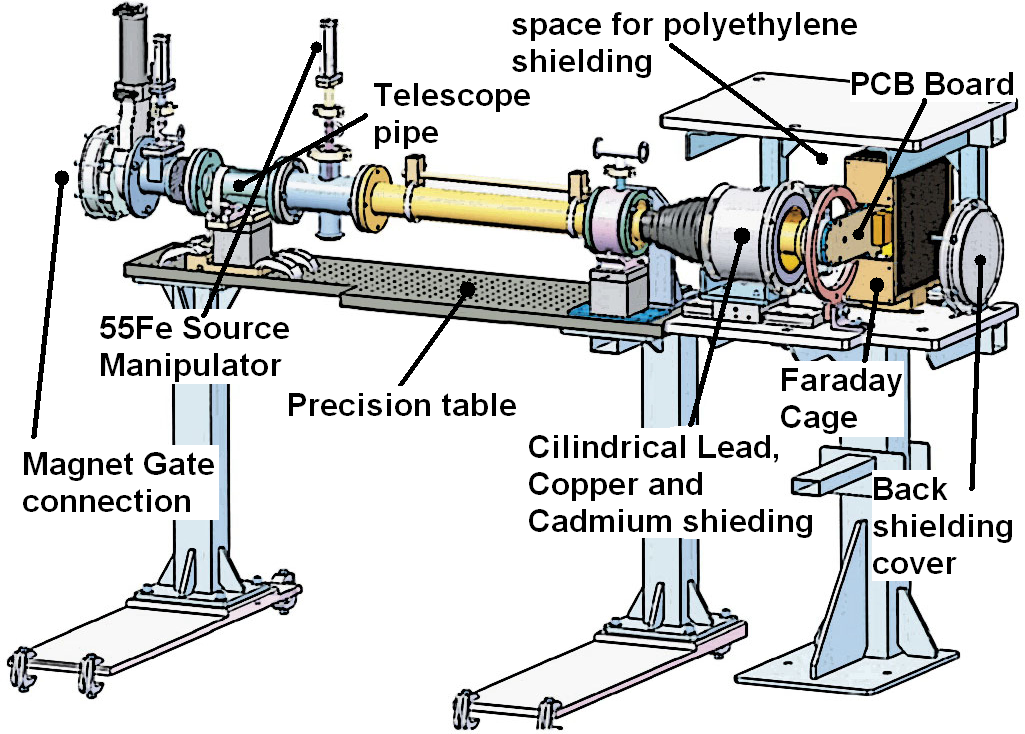} &
\includegraphics[width=0.23\textwidth]{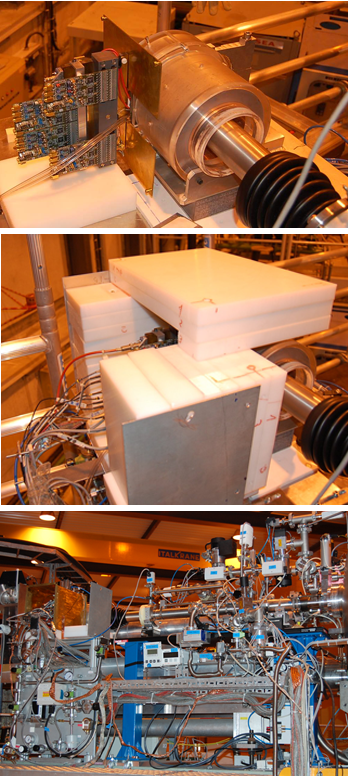} \\
\end{tabular}

\caption{\fontfamily{ptm}\selectfont{\normalsize{ On the left, a sketch of the new Micromegas line in the sunrise side, the line is complemented with vacuum, control and safety systems. On the right, some pictures taken during the detector installation where some parts of the shielding can be observed. The vacuum pumps, displays and sensors, and other mechanical devices are visualized at the bottom picture.}}}
\label{fi:newSunrise}

\end{figure}

In 2007 a new Micromegas-magnet connection line was designed, built and installed in the sunrise side. The new sunrise line (see Fig.~\ref{fi:newSunrise}) was built considering the possibility of placing an X-ray focusing device, that could improve the signal to noise ratio~\cite{Andriamonje:1065408}. The system was designed to fulfill the vacuum requirements for the connection to the magnet bores and several control systems for vacuum and precise measurements of detector gas pressure and flow were implemented to the system.

\vspace{0.2cm}

The new detection line, that replaced the previous design shown in figure~\ref{fi:oldSunrise}, was mounted over an optics table that implements proper alignment mechanisms. The new table allocates extra space for the implementation of a new dedicated shielding composed of lead, copper, cadmium and Plexiglas, while the remaining available space was optimized to complete the shielding with polyethylene blocks. The new shielding reduced the detector background more than a factor~3 (value measured at the final location of the detector in the CAST experiment, see section~\ref{sc:shielding}).

\vspace{0.2cm}

Another important feature of the new system is the circulation of a flux of clean nitrogen inside the inner shielding surrounding the detector in order to reduce the presence of radon near the detector chamber. 

\vspace{0.2cm}

Moreover, the new sunrise line has incorporated an automated calibration system (described on section~\ref{sc:calibrationSystem}) that allows to place the $^{55}$Fe source inside the vacuum illuminating the front of the detector (see Fig.~\ref{fi:newSunrise}). While in the previous system the source was placed at the back and X-rays were passing through perforated holes on the PCB board from the back of detector chamber (see Fig.~\ref{fi:oldSunrise}).

\vspace{0.2cm}

Also a new panel was implemented in order to monitor and control the detector gas pressure and flow, and vacuum systems (section~\ref{sc:monitoringSystem}).

\subsection{The effect of shielding in background reduction.}\label{sc:shielding}

The main population of triggering events in CAST conditions are cosmic muons that are easily discriminated due to the easily detectable long ionizing tracks in the detector. Another source is coming from environmental radiation like gammas and neutrons, that are hardly to discriminate, which can be reduced by using a passive shielding. In CAST the new cylindrical shielding design consists of $2.5$\,cm of archaeological lead, the outer part of the shielding implements a cadmium foil of $2$\,mm to absorb thermal neutrons, and in the inner part there is a $5$\,mm thick radiopure copper layer, also serving as faraday cage. These layers are encapsulated in Plexiglas. The remaining available space was filled with polyethylene varying its thickness up to $25$\,cm.

\vspace{0.2cm}
After the installation of the sunrise line at the middle of 2007, some background measurements took place with the T3 detector, first prototype that took data at the new CAST sunrise line. \emph{Two} sets of measurements were taken; with full shielding coverage, lead cylindrical shielding as it would be during data taking, and half covered detector by opening and displacing the lead shielding towards the magnet bore side leaving the body of the detector naked.

\vspace{0.2cm}

\begin{figure}[!ht]
{\centering \resizebox{1.01\textwidth}{!} {\includegraphics{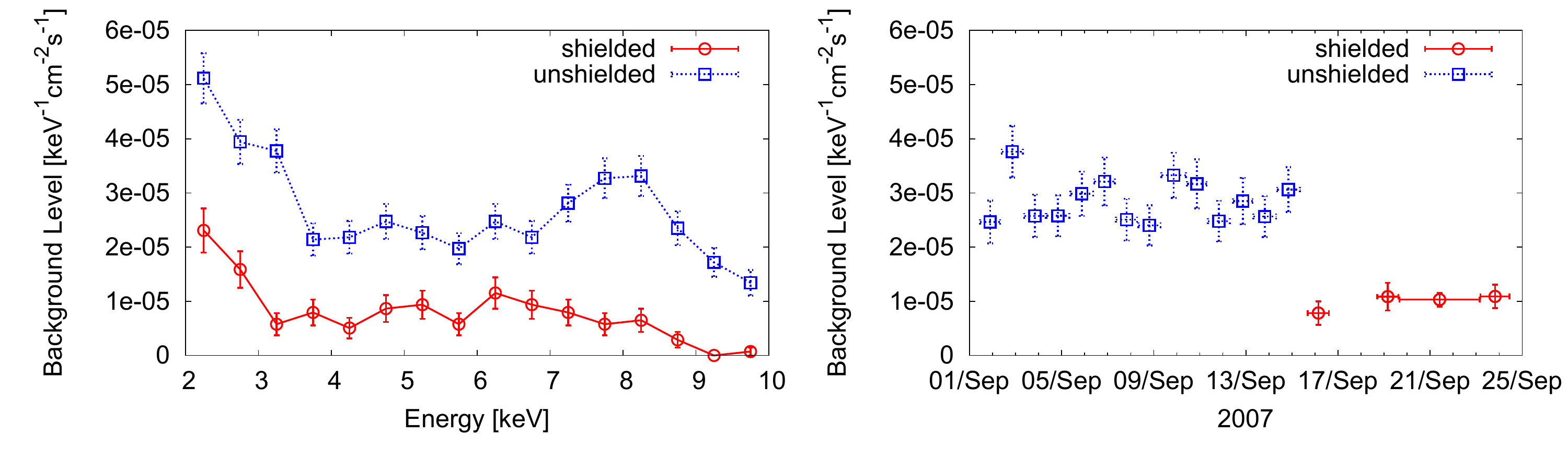}} \par}
\caption{\fontfamily{ptm}\selectfont{\normalsize{Background energy spectrum (left) for the \emph{two} sets of measurements taken with the T3 detector, and final background rate evolution (right) at the integrated energy range 2-7\,keV where the most intense axion signal is expected. }}}
\label{fi:shieldingEffect}
\end{figure}

The effect of the shielding was clearly observed (see Fig.~\ref{fi:shieldingEffect}) after the background discrimination applied to the detector data (as detailed on chapter~\ref{chap:discrimination}). A direct comparison between the background level of these \emph{two} set of measurements lead to a background reduction of about a factor $3$, and a higher reduction respect to the background level ($\gtrsim 4\cdot10^{-5}$\,keV$^{-1}$cm$^{-2}$s$^{-1}$) obtained in the previous data taking periods (see Fig.~\ref{fi:he4Rate}).

\subsection{The calibration system.}\label{sc:calibrationSystem}

Micromegas detectors are calibrated at least once everyday during data taking periods after its corresponding tracking. X-ray calibrations are used to monitor the detector gain, to have a daily crosscheck of the detector main parameters stability (described on chapter~\ref{chap:rawdata}), and to apply a background selection (described on chapter~\ref{chap:discrimination}) to the corresponding background data. The source is moved by a remotely controlled mechanism that places the source in a region visible to the detector (calibration position) or hides the source from it (garage position).

\begin{figure}[!b]
{\centering \resizebox{\textwidth}{!} {\includegraphics{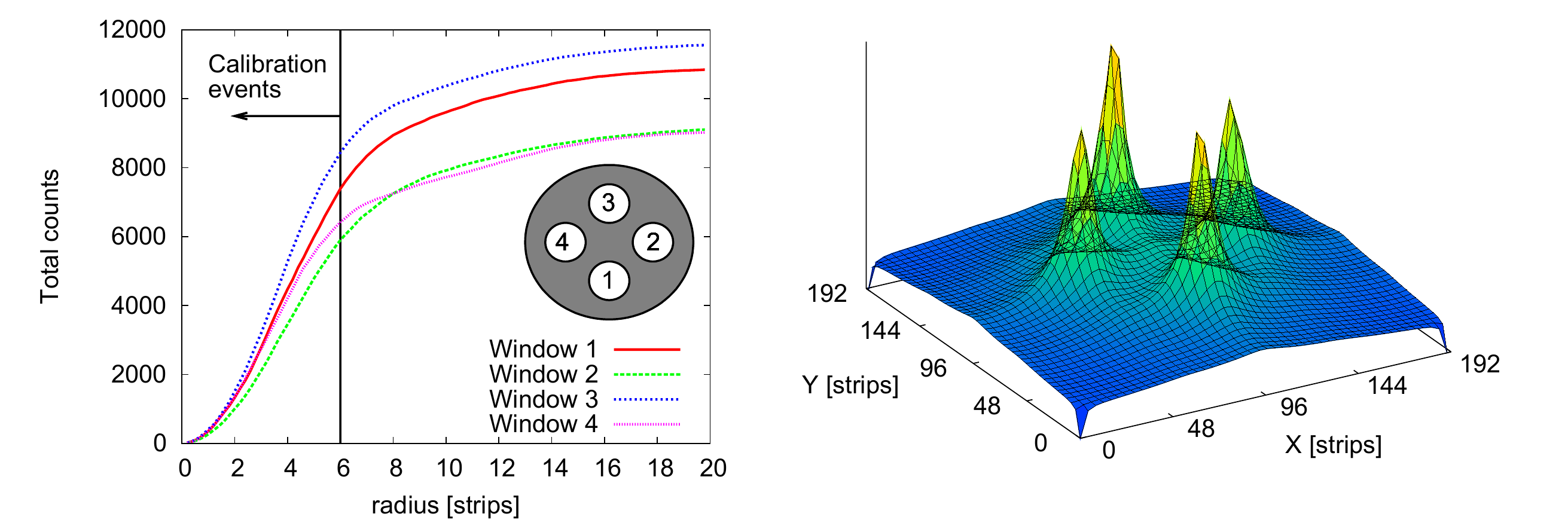}} \par}
\caption{\fontfamily{ptm}\selectfont{\normalsize{On the left, a plot representing the total amount of events inside circular fiducial regions of growing radius, which are centered in the mean position of each calibration spot. On the right, a 3-dimensional map of the distribution of hits position for a calibration run where the effect of the \emph{four} calibration windows is observed.  }}}
\label{fi:calWindows}
\end{figure}

\vspace{0.2cm}

Before the installation of the sunrise line the detector was calibrated from the back, through \emph{four} small cavities perforated at the Plexiglas cover in order to increase the $^{55}$Fe X-ray flux reaching the drift volume. The calibration map obtained had as a result \emph{four} active spots sensitive to X-rays coming from the source (see~Fig.~\ref{fi:calWindows}). The triggering rate during that calibration runs was not much higher ($8$\,Hz) than the acquisition rate during background runs ($1$\,Hz), due to the fact that X-rays had to go through the remaining Plexiglas material and the strips readout plane, together with the low activity of the source. It was expected a non negligible amount of natural background mixed with the X-ray events produced by the radiative source. At those periods, in order to increase the significance of X-ray events produced by the source over those produced by natural background it was performed a fiducial cut to the calibration acquired data, taking into account only the events that had place inside the active spots. In order to determine the fiducial regions to be taken into account in the definition of the X-ray properties (described on chapter~\ref{chap:rawdata}), the mean positions of each of the \emph{four} active calibration regions are calculated. Then, the density of events in each spot inside a circular region around its center position is obtained as a function of the radius (see~Fig.~\ref{fi:calWindows}), only the regions where the event density is higher than the natural background regions are considered, taking for each calibration about $80$\,\% from the $25,000$ events recorded in a calibration run.


\vspace{0.2cm}

\begin{figure}[!ht]
\begin{tabular}{cc}
\resizebox{0.98\textwidth}{!} {\includegraphics{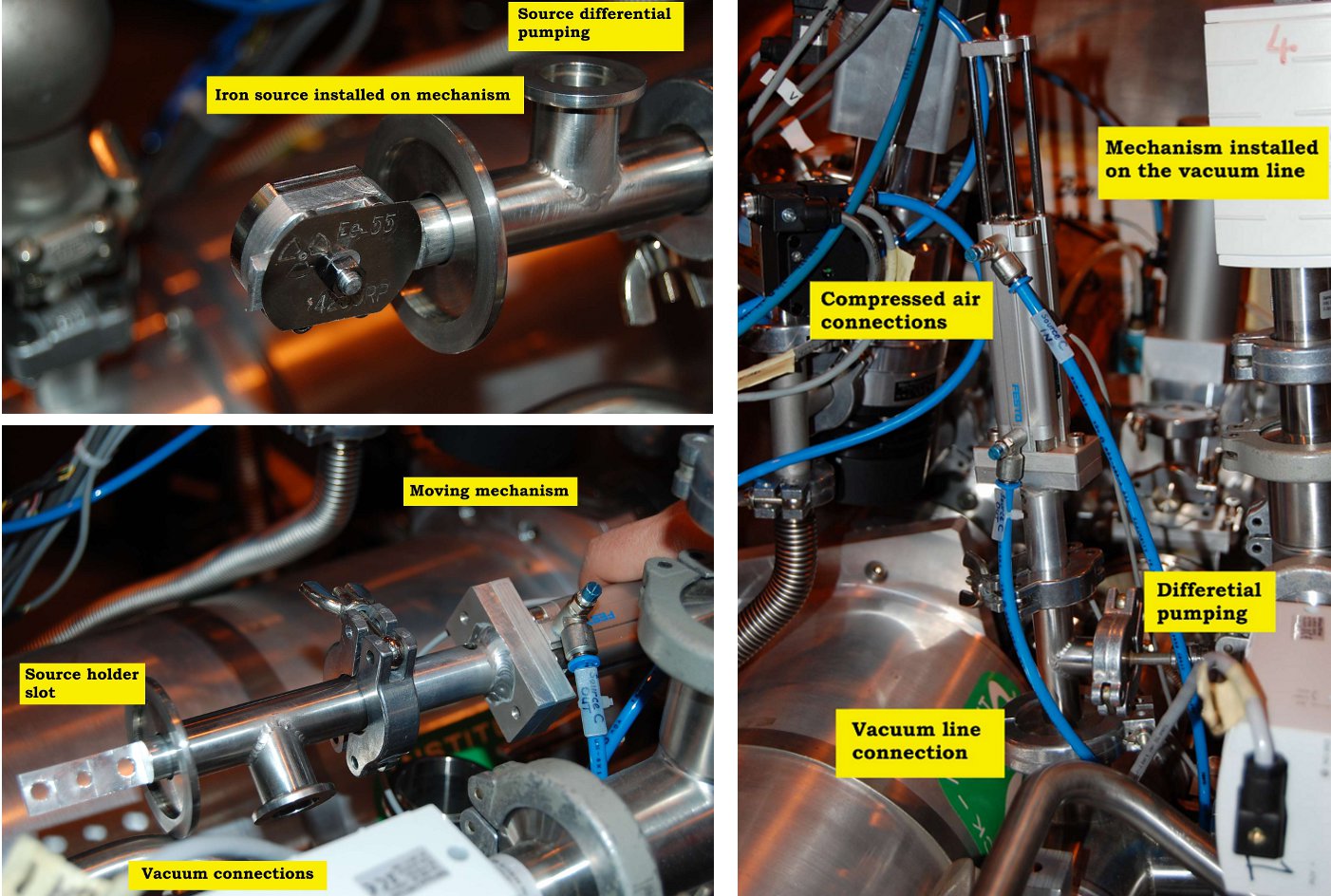} \par} 
\end{tabular}
\caption{\fontfamily{ptm}\selectfont{\normalsize{Descriptive pictures of the manipulator connected to the sunrise vacuum line. }}}
\label{fi:calManipulator}
\end{figure}

The new sunrise line installed in 2007 incorporated an auxiliar vacuum connection which allowed to incorporate a compressed-air controlled source manipulator to the system (see Fig.~\ref{fi:calManipulator}). The mechanism holds the source inside the vacuum side and is able to place it at the middle of the vacuum pipe (calibration position) or to hide it inside the auxiliary vacuum flange (garage position). The source was installed inside a holder provided by a high transmission beryllium window and it was certified at CERN facilities to operate in vacuum assuring the integrity of the source during the data taking period~\cite{vaccuumNotes}. The manipulator incorporates a differential vacuum connection in order to reduce the effect on the vacuum pressure due to the movement of the internal mechanism.

\vspace{0.2cm}

Furthermore, the $^{55}$Fe source used before 2007 was replaced by a higher activity source (about $\sim80$\,MBq at the time of purchase) that increased the calibration rate ($\sim 100$\,Hz) of the detector, limited by the electronic acquisition dead time. The higher X-ray rate favored the increase of statistics collected by the detector during calibration runs to $75,000$\,events due to the shorter periods of time required. 

\vspace{0.2cm}

\begin{figure}[!ht]
\begin{tabular}{ccc}
\resizebox{0.95\textwidth}{!} {\includegraphics{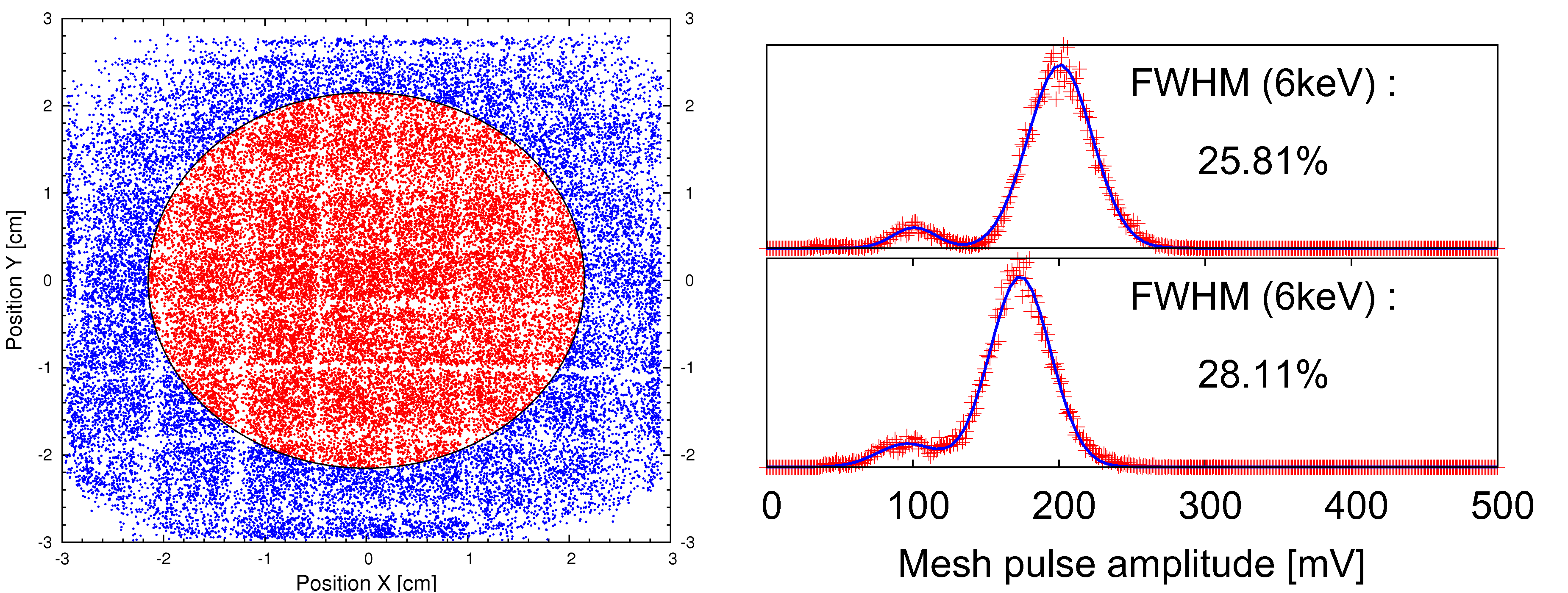} \par} &
	\\
\end{tabular}
\caption{\fontfamily{ptm}\selectfont{\normalsize{ Calibration events hitmap (left) produced by a frontal calibration with the new calibration system. The squared pattern reflects the effect of the drift window strong-back. The circle encloses the region which corresponds with the signal sensitive detector region, and $^{55}$Fe spectra (right) generated with the new line configuration for a bulk detector (bottom) and a microbulk detector (top).   }}}
\label{fi:calHitmapFront}
\end{figure}

Another advantage of the new calibration system resides in the higher similitude to the expected axion signal increasing the reliability of the X-ray signal detection, which interaction probability inside the chamber is higher as closer to the drift window. The source is placed at about $1$\,m of the detector allowing to produce an almost parallel beam that illuminates the full area of the detector (see Fig.~\ref{fi:calHitmapFront}). Thus the daily obtention of the full active hitmap area allows a better monitoring of the performance of the detector and the possibility to apply positional gain corrections.

\subsection{The pressure monitoring and control systems.}\label{sc:monitoringSystem}

\begin{figure}[!ht]
\begin{center}
\begin{tabular}{cc}
\resizebox{0.45\textwidth}{!} { \includegraphics[width=8cm]{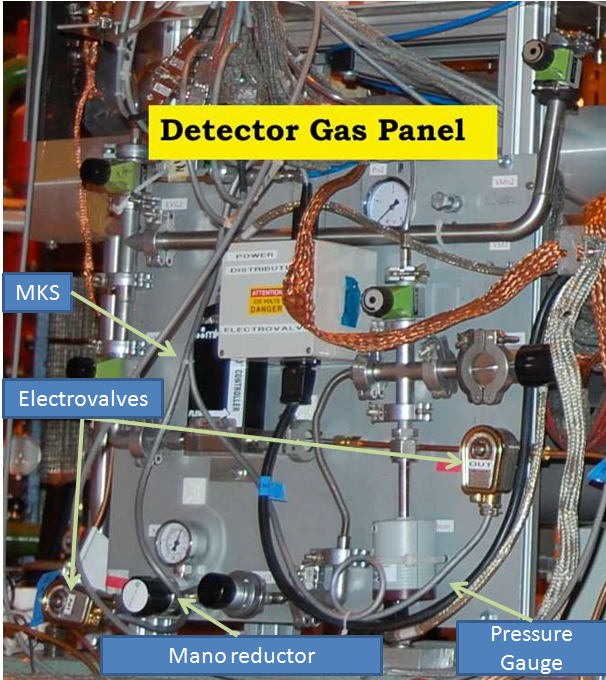} \par} &
\resizebox{0.43\textwidth}{!} { \includegraphics[width=8cm]{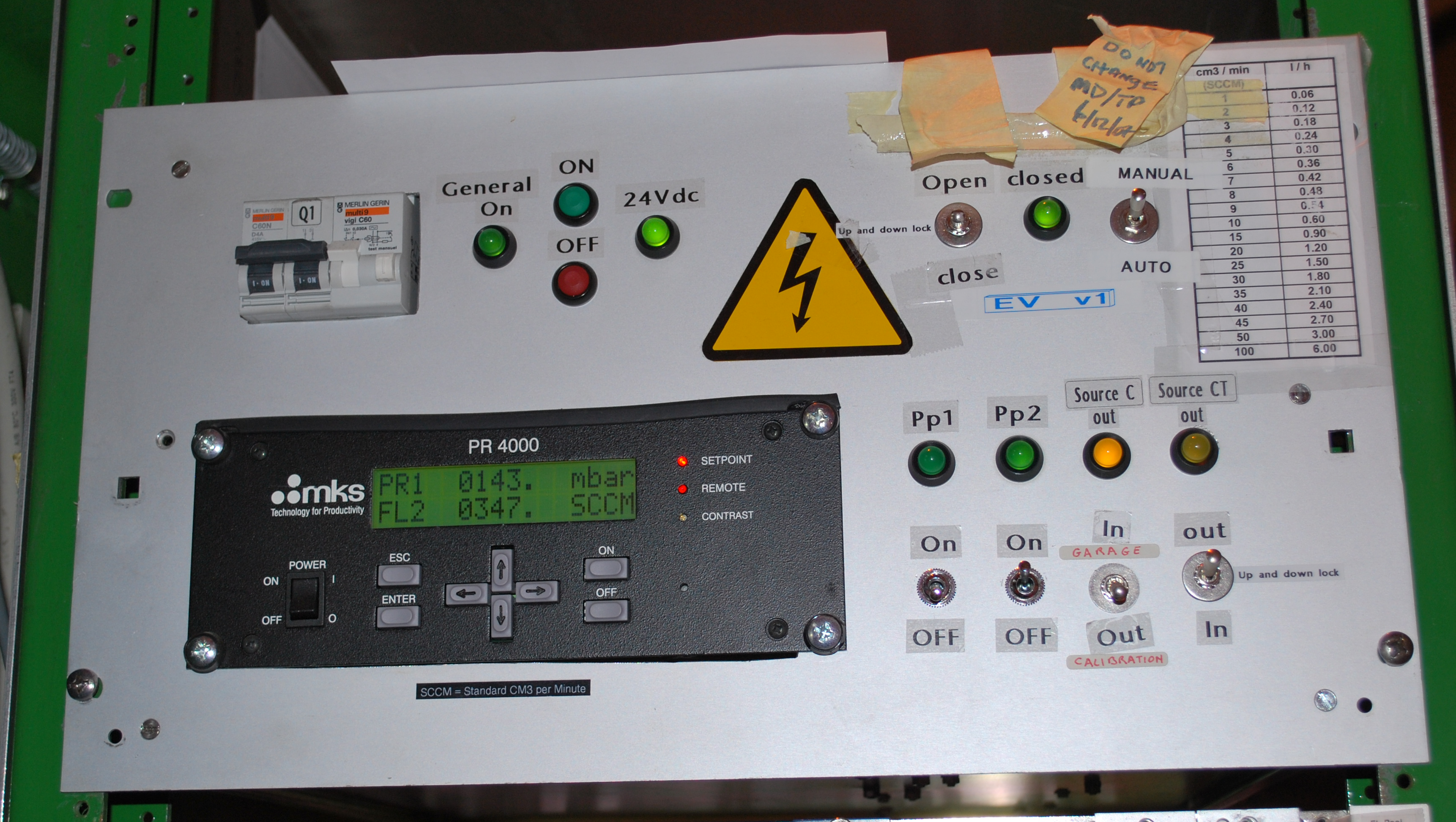} \par} \\
\end{tabular}
\end{center}
\caption{\fontfamily{ptm}\selectfont{\normalsize{ A picture of the detector gas panel where the main components are shown (left) and the sunrise micromegas control box (right).  }}}
\label{fi:gasPanel}
\end{figure}

The new sunrise line assembly at the magnet bore end, together with the experience acquired in the previous data taking periods, allowed the implementation of new monitoring and control systems. A new \emph{control box} and \emph{gas panel} were designed for this purpose (see Fig.~\ref{fi:gasPanel}).

\vspace{0.2cm}

The \emph{gas panel} implements a manometer capable to regulate the detector gas pressure before the chamber and a high accuracy mass flow controller (MKS), commanded through the control box, that fixes the gas output flux (during the operation of the detector the flux was set to $2$\,l/h). A sensor placed just before the MKS is used to monitor the pressure at the detector chamber that after the installation of the line was always higher than $1400$\,mbar during the detector operation. Thus, the gas panel implements \emph{two} electrovalves that insulate the detector and gas panel lines that will close in case of power cut for safety. During the data taking period before 2007 the electrovalves were also connected to the fire alarm service since the gas used was Argon + 5\% iC$_4$H$_{10}$ that was considerated flammable versus the lower concentration of isobutane used afterwards (2.3\%). In addition to the safety function of the electrovalves they are also used to check the leak tightness of the system after any intervention, as the connection of a new detector to the line.

\vspace{0.2cm}

The \emph{control box} implements all the functionality of the monitoring and control devices connected to the line; \emph{vacuum pumps}, \emph{bypass valve}, \emph{source manipulator control} and pressure sensor and flow control monitoring access.

\begin{figure}[!ht]
{\centering \resizebox{0.88\textwidth}{!} {\includegraphics{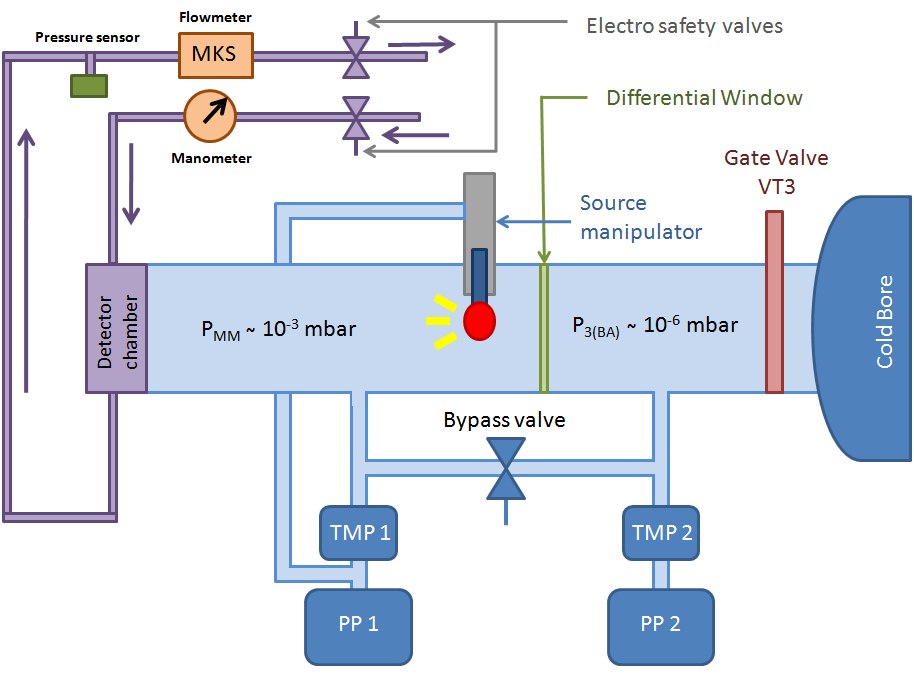}} \par}
\caption{\fontfamily{ptm}\selectfont{\normalsize{Schematic of sunrise detection line vacuum pumping system described in the text. Source manipulator holding the $^{55}$Fe source is also pumped by using PP1 to reduce the effect on vacuum while moving the source. Moreover, a simplified version of the gas flow and pressure sensors, and valves, respect to the detector chamber are shown in the drawing.  }}}
\label{fi:vacuumSunrise}
\end{figure}

\vspace{0.2cm}

Figure~\ref{fi:vacuumSunrise} shows an schematic of the full sunrise line vacuum and detector system. In this system can be distinguished two different vacuum volumes separated by the $4$\,$\mu$m polypropylene differential window\footnote{The \emph{differential window} function is to reduce the amount of gas going to the bores coming from the diffusion of the detector drift window, allowing for high X-ray transmission. This is achieved by using a \emph{differential pumping} system, which working principle is described in chapter~\ref{chap:cast}.}; the volume next to the detector, denominated \emph{bad vacuum side} (measured by $P_{mm}$ sensor), and the volume before the magnet gate valve VT3, denominated \emph{good vacuum side} (measured by $P_{3(BA)}$ sensor). Both volumes are pumped by independent vacuum primary pumps (PP) and turbo molecular pumps (TMP), which are labeled in the drawing as PP1, PP2, TMP1 and TMP2. The \emph{differential pumping} is controlled through the \emph{bypass valve}, which is the only physical connection between the volumes split by the \emph{differential window}. When the \emph{bypass valve} is closed the vacuum reached at the \emph{good vacuum side} is at least 3 orders of magnitude below the vacuum pressure next to the detector, at the \emph{bad vacuum side}. The \emph{bypass valve} it is required to protect the thin polypropylene differential window that supports no more than $10$\,mbar difference of pressure, and it can only be closed when the volumes have reached the nominal vacuum operation levels. Furthermore, a protection system prevents to open the VT3 gate valve if the vacuum reached is not good enough. In data taking conditions, if the pressure at the good vacuum side ($P_{3(BA)}$) rises above $10^{-6}$\,mbar, the gate valve VT3  is closed automatically to protect the cold bore windows and the \emph{bypass valve} is open in order to protect the \emph{differential window}.

\section{Sunset Micromegas installation.}

\begin{figure}[ht!]
\begin{center}
\begin{tabular}{ccc}
\includegraphics[width=0.3\textwidth]{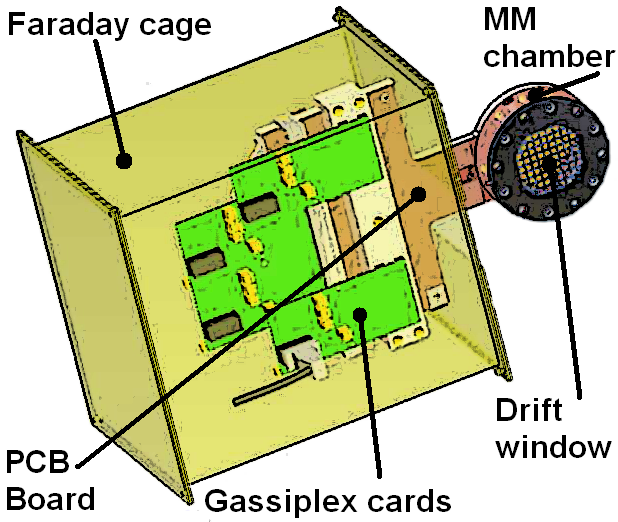} &

\includegraphics[width=0.3\textwidth]{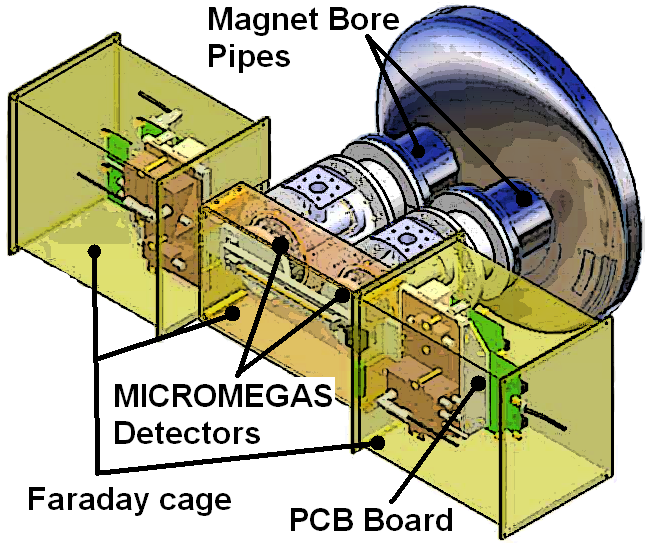} &

\includegraphics[width=0.31\textwidth]{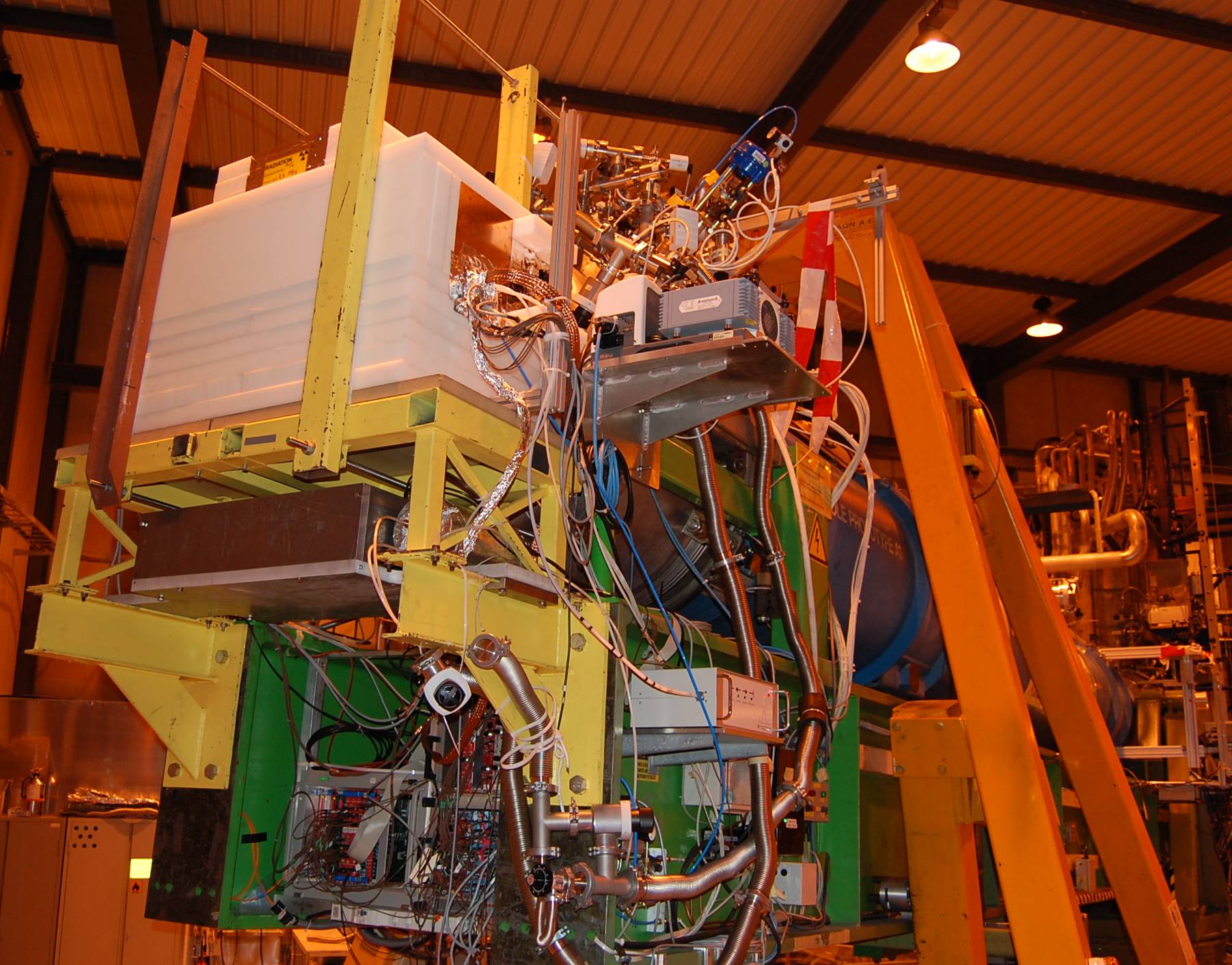} \\

\end{tabular}

\caption{\fontfamily{ptm}\selectfont{\normalsize{On the left, drawings from the new Micromegas detectors system designed for the sunset side and the connection to the magnet bores. On the right, final installation of the shielding covering the sunset side detectors.}}}
\label{fi:Sunset}

\end{center}
\end{figure}

During 2007 and 2008 a new system was designed and built to assemble two new Micromegas detectors on the sunset tracking side. The design of the sunset detectors system (see Fig.~\ref{fi:Sunset}) was restricted by the space available at the sunset side, and it was thought to best fit the previous TPC shielding installed at that side. The new design implemented a new Faraday box scheme specially built for the combined micromegas detector system, which included a calibration system from the back of the detector where both detectors were sharing the same $^{55}Fe$ source. The source was driven inside the Faraday box by a rail mechanism and pushed by a towrope attached to the source. The source was moved right and left to the center of the detectors back (calibration position) and hidden in a small lead shielding placed between the detectors.

\vspace{0.2cm}

\begin{figure}[!ht]
{\centering \resizebox{0.98\textwidth}{!} {\includegraphics[angle=270]{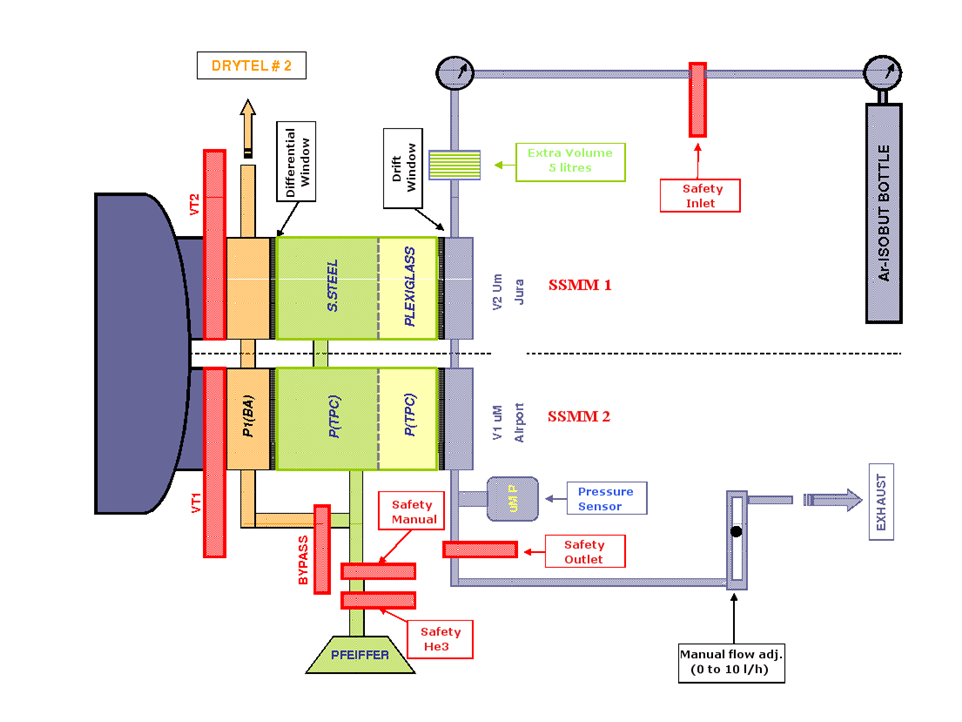}} \par}
\caption{\fontfamily{ptm}\selectfont{\normalsize{Schematic of the micromegas detectors connected to the sunset detection lines described in the text. The VT1 line connection (left) shows the pressure sensors involved in each volume, the VT2 connection line (right) describes the bore material, stainless stell for the vacuum pipes, and Plexiglas for the detector chamber. The schematic of the detector gas flow system with sensors and electrovalves is also shown.  }}}
\label{fi:vacuumSunset}
\end{figure}

Figure~\ref{fi:vacuumSunset} describes the new sunset micromegas vacuum and detector system at the sunset side. The system was developed following the same vacuum scheme described on section~\ref{sc:monitoringSystem}, including a \emph{differential window} per vacuum connection line (corresponding to VT1 and VT2 gate valves) that splits the \emph{bad vacuum} side next to the detectors, measured by $P_{TPC}$ sensor, from the \emph{good vacuum} side, measured by $P_{1(BA)}$ sensor. The \emph{good vacuum side} and \emph{bad vacuum side} volumes are common for both detector lines. The \emph{good vacuum side} is pumped by a Drytel pump which implements both, a turbo molecular pump and a primary pump (DRYTEL\#2), while the \emph{bad vacuum side} is pumped by a Pfeiffer pump.

\vspace{0.2cm}

The same gas line was used to feed both detectors, interconnected one after the other. The absence of a high accuracy flowmeter in the sunset side caused the pressure in the detectors to fluctuate more than expected due to the abrupt adjustments of the pressure mano-reductor connected to the line preceding the detectors, the problem was solved by installing a $5$\,liters reservoir volume in the line that smoothed the pressure changes in the system. The gas line implements also pressure and flow sensors, and electrosafety valves that can be electrically controlled in case of electrical cut or a bad vacuum level reached in the \emph{bad vacuum side}.

\chapter{Rawdata analysis and characterization of Micromegas detectors.}
\label{chap:rawdata}
\minitoc

\section{Introduction}

\begin{figure}[ht!]
\begin{center}
\includegraphics[width=0.95\textwidth]{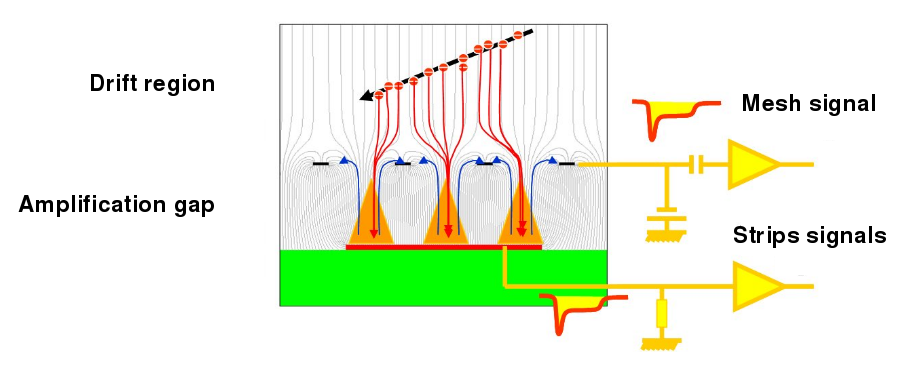} 

\caption{\fontfamily{ptm}\selectfont{\normalsize{Functional schema of the Micromegas detection principle and the readout used by Micromegas detectors in CAST. The mesh signal is processed by a Timing Amplifier shaper (ORTEC 741) while the charge collected by the strips is integrated by using \emph{Gassiplex} chips. }}}
\label{fi:readout}
\end{center}
\end{figure}

Physical ionizing processes are recorded by the electronic acquisition set-up. The data gathered for each event includes information of the \emph{induced charge evolution} in the mesh structure and the integrated \emph{spatial charge} collected by each strip (see Fig.~\ref{fi:readout}).

\vspace{0.2cm}

The information provided by the signal recorded on the mesh and the spatial charge collection in the strips readout makes it possible to differentiate events that are most likely to be an X-ray from any other events. The background is mainly composed by cosmic rays, such as muons and neutrons. Muons are easily discriminated due to the long ionizing tracks produced inside the chamber. Electronic noise is easily rejected from physical processes having place in the detector chamber since real events produce localized deposits of charge in the strips plane.

\vspace{0.2cm}

The Micromegas readout being used in CAST offers enough information to provide the detector with good rejection capabilities making it suitable to distinguish different kind of events by using pattern recognition algorithms.

\vspace{0.2cm}

In order to use this information to obtain the processes we are interested in, it is required to define a minimum set of parameters that best describe the event properties. This event parameterization is usually carried out before applying any background discrimination analysis. The rawdata analysis to be described will be focused in obtaining signals that are most likely to be an X-ray.

\section{Micromegas readout and acquisition set-up}\label{sc:readout}

The \emph{mesh signal} is generated by slow ions produced in the \emph{amplification gap} drifting back towards the mesh structure. This signal has a double functionality; it gives the information on the charge evolution (related with the shape and distribution in the z-axis of the electron cloud created in the \emph{drift region}) and it is used to generate the main \emph{trigger} signal to activate the acquisition and the digitalization of signals.

\vspace{-0.2cm}

Before being digitized, the mesh signal is preamplified using a \emph{Canberra 2004} preamplifier and afterwards shaped by a timing amplifier (ORTEC 471) that allows to adjust the amplification gain and several timing parameters to adjust the signal filtering and integration that affects the final shape of the recorded events. 

\vspace{0.3cm}

The mesh signal output after the Timing Amplifier module is digitized using a \emph{VME MATACQ} (CAEN V1720) module. The signal is acquired at $1$\,GHz sampling frequency in a $2.5$\,$\mu$s window. A discriminator module allows to generate a digital trigger signal when the output of the Timing Amplifier exceeds a given threshold voltage chosen to be slightly above the background electronic noise.

The \emph{strips signals} are induced by the electron avalanche produced in the \emph{amplification gap}. The XY-pattern of the strips allows to determine the position and distribution of the electron cloud projected in the Z-plane.

\vspace{0.3cm}

The charge induced and collected by the strips pattern is integrated and processed by \emph{Gassiplex} cards. A \emph{Gassiplex} card implements 92 pin connectors that are directly connected to each of the detector strips. It is composed by 6 \emph{Gassiplex} chips~\cite{Santiard:272783}, each of them implementing 16 analog memories able to store the integrated voltage of each strip. 

\begin{figure}[ht!]
\begin{tabular}{cc}
\includegraphics[width=0.48\textwidth]{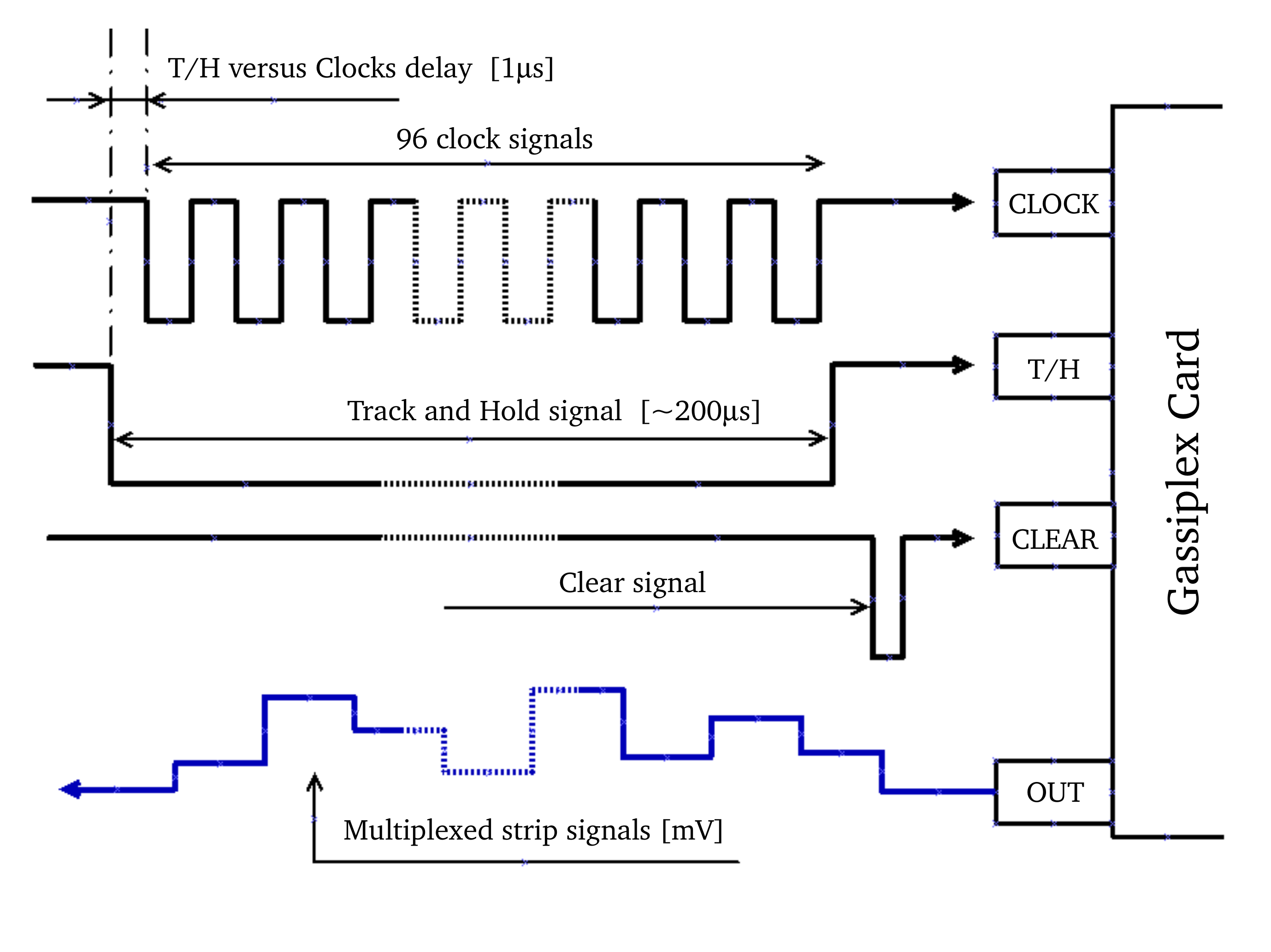} &
\includegraphics[width=0.48\textwidth]{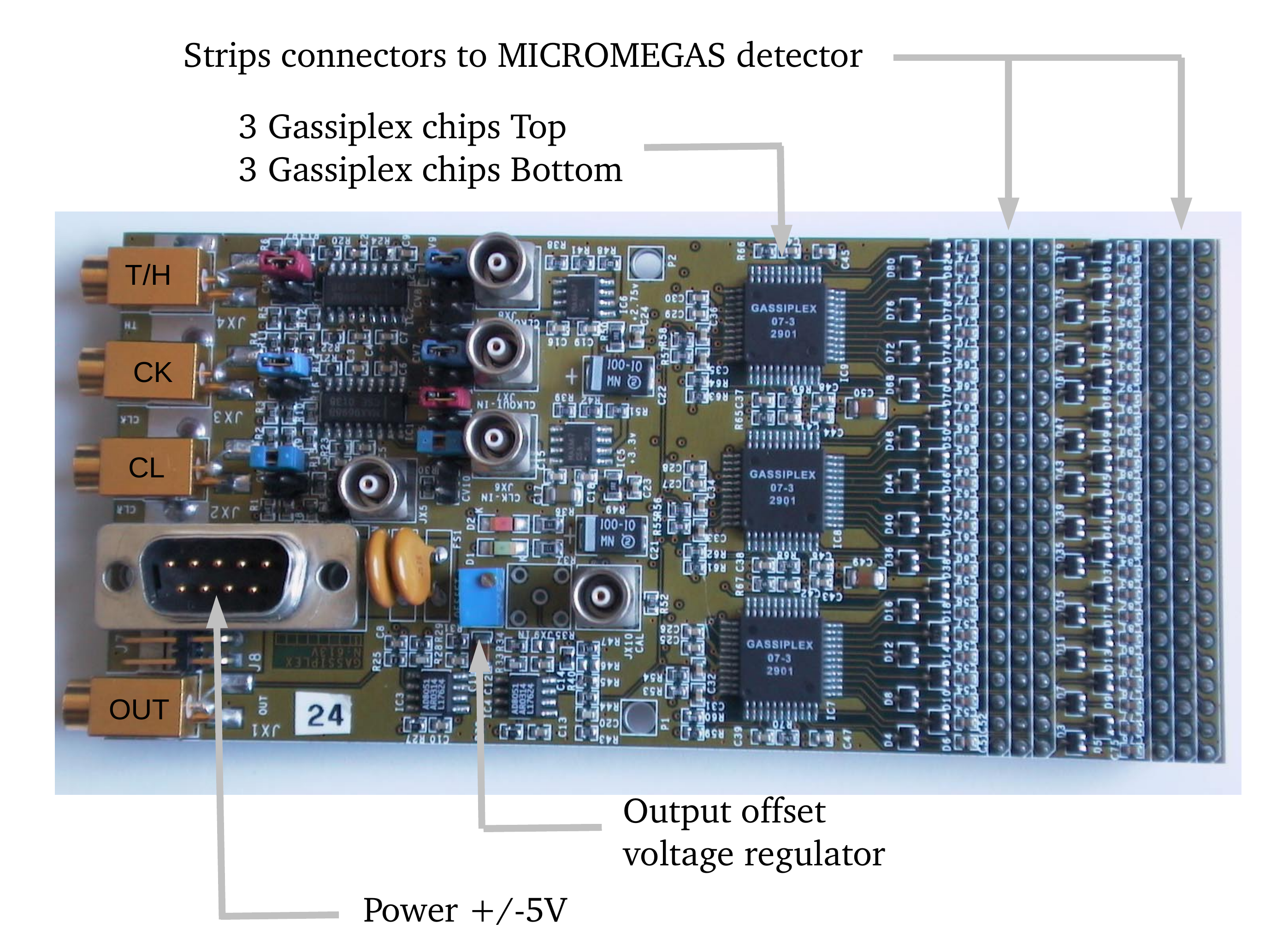} \\
\end{tabular}

\caption{\fontfamily{ptm}\selectfont{\normalsize{On the left, a schematic of the signals going into and coming out the Gassiplex cards. On the right, a picture of a Gassiplex card where Gassiplex chips, input/output connectors and strip pins are observed. The output signals offset can be tuned using a voltage regulator implemented in the card. }}}
\label{fi:gassiSignals}
\end{figure}

The \emph{Gassiplex} card is accessed from an auxiliar electronic system in order to operate it and obtain the measured strip values. Three inputs allow to start the Gassiplex card acquisition and obtain the data: \emph{Track \& Hold} (T/H), \emph{Clock} (CK) and \emph{Clear} (CL) (see Fig. \ref{fi:gassiSignals}). The T/H initializes the integration process until the first CK signal becomes Low and freezes the analog memories in the Gassiplex chips. The strip signals come out multiplexed through the \emph{Output} connector giving the value of the following strip after each CK signal becomes High. After all the CK signals have been sent the CL signal resets the analog memories for recording the next event. This process is started each time a trigger is detected by the mesh signal. Finally, the multiplexed output signal is digitalized by a VME V550 module from CAEN.

\vspace{0.3cm}

\begin{figure}[ht!]
\begin{center}
\includegraphics[width=1.0\textwidth]{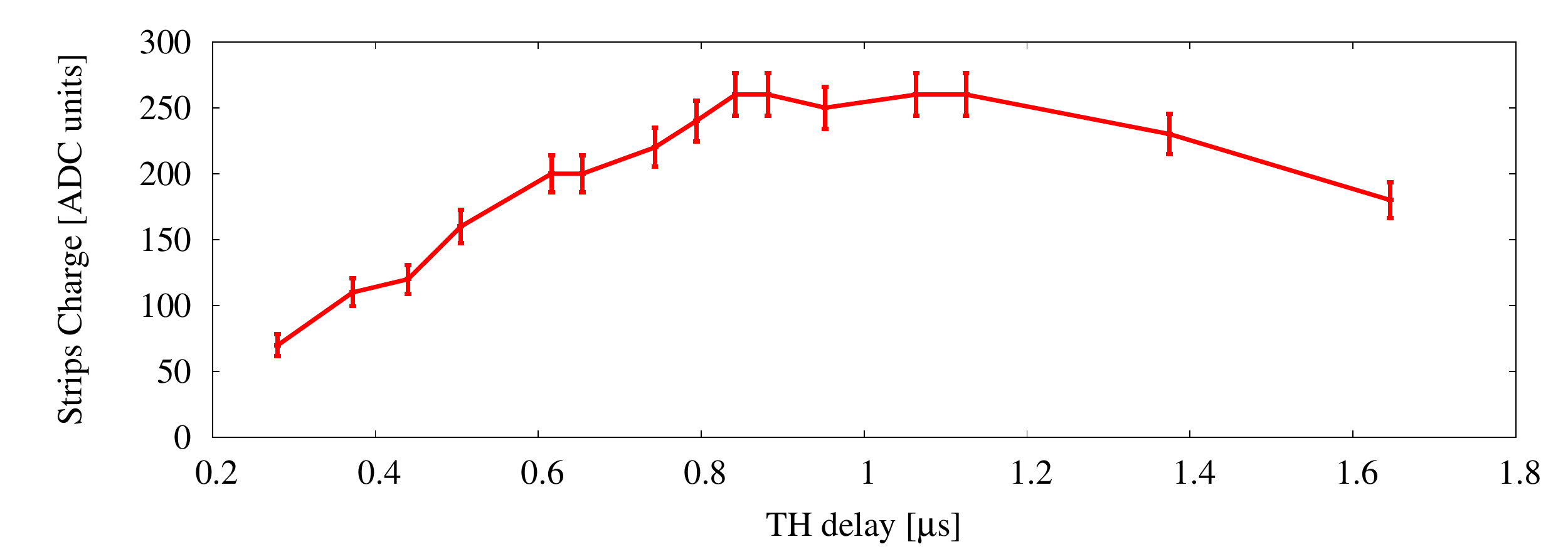} 
\end{center}

\caption{\fontfamily{ptm}\selectfont{\normalsize{Mean strips charge collected during a calibration run for different T/H signal delays. The curve gives the optimum integration time for 1\,$\mu$s delay. }}}
\label{fi:THdelay}
\end{figure}

The \emph{Gassiplex} card can be optimized to maximize the S/N ratio. This optimization is performed by adjusting the time delay between the first CK signal and the T/H signal, related with the chips integration time. In order to find the best timing for these signals it has been measured the total charge collected, for the same running conditions of a Micromegas detector, at different delays giving an optimum at $1$\,$\mu$s delay (see Fig.~\ref{fi:THdelay}).

\vspace{0.2cm}

The Micromegas detectors taking data in CAST require several \emph{Gassiplex} cards connected. In general, at least one card is needed for each strip axis, X~and~Y. If the detector has more than 92 strips per axis, it is required an extra \emph{Gassiplex} card per axis.

\vspace{0.2cm}

The full acquisition of one event requires a \emph{dead time} of about $10$\,ms, during this period the trigger signal is vetoed and no new events can be acquired. The typical background rate is around $1$\,Hz leading to 1\% of the total acquisition time lost.

\section{Positional Cluster Analysis.}

\begin{figure}[ht!]
\begin{center}
\includegraphics[width=1.0\textwidth]{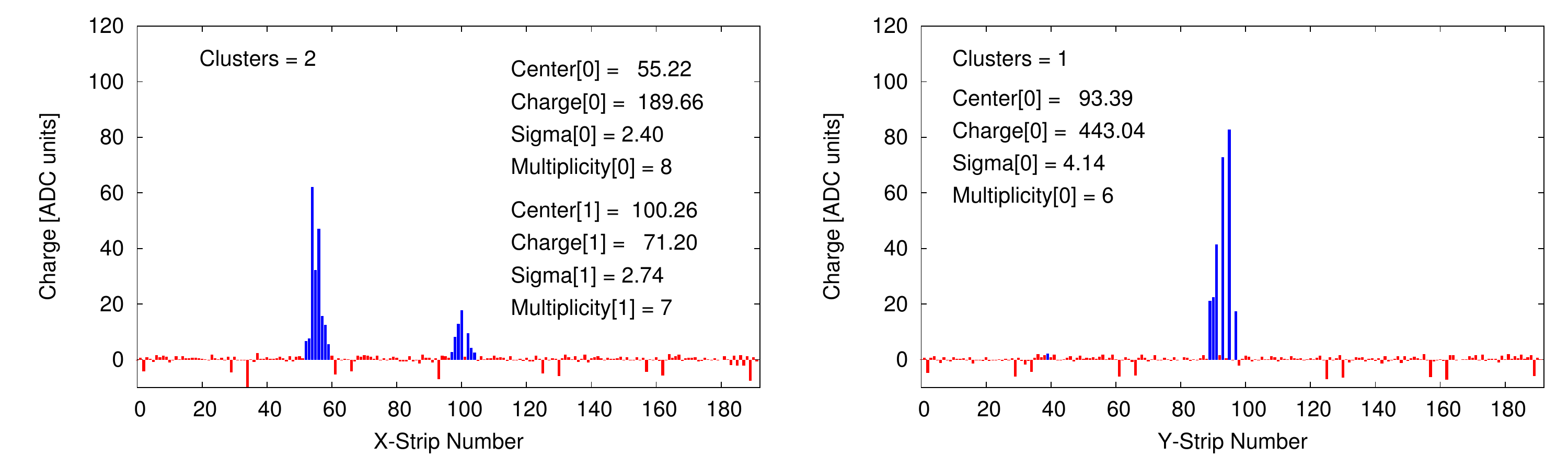} 
\end{center}

\caption{\fontfamily{ptm}\selectfont{\normalsize{Strips charge collection after removing pedestal noise from Gassiplex card rawdata. On the left, an event containing two clusters of charge (non X-ray like event). On the right, a mono-cluster event (X-ray like event) where several inactive strips are present inside the cluster. }}}
\label{fi:stripClusters}
\end{figure}

The charge measured in each strip is used to identify localized charge depositions, denominated clusters, that allow to parameterize the recorded signals in a minimum set of parameters (see Fig.~\ref{fi:stripClusters}).

\vspace{0.3cm}

A cluster of charge is defined by at least two consecutive active strips, we consider as active strips the ones that have collected a charge level higher than the measured noise level.

\vspace{0.3cm}

The noise level of each strip or \emph{pedestal} it is obtained from background events, rejecting from each event the strips that have been activated. The pedestal mean value,~$p_i$, and its fluctuation,~$\sigma_{p_i}$~is obtained for each strip,~$i$. The condition of activation for a real event strip value,~$c_i$, it is expressed as

\begin{equation}
\hat{c_i} = c_i - p_i > 3 \sigma_{p_i}
\end{equation}

\noindent where, in principle, the $\sigma$ factor chosen implies that less than 0.27\% of the strips that are not really active will be detected as active.

\vspace{0.3cm}

The information provided by the strips allows to determine the number of clusters in each axis and their properties. The \emph{position}, \emph{size}, \emph{shape} and \emph{energy} of each of these clusters defines the event and it can be used in a posterior analysis for \emph{detector characterization} and \emph{background selection}.

\vspace{0.3cm}

The main parameters obtained for each cluster can be summarized in the following list.

\begin{itemize}

\item {\bf Cluster charge $\left(\hat{c} = \sum_i \hat{c_i}\right)$ :} Addition of the strips charge that define the cluster. Proportional to the number of electrons that generated the cluster, and to the ionization energy deposited by the originating process.

\item {\bf Cluster position  $\left(\mu = 1/\hat{c} \cdot \sum_i i\cdot\hat{c_i}\right)$ :} Strip value weighted with the charge measured in each strip. This parameter describes the mean cluster position.

\item{\bf Cluster sigma $\left( \sigma^2 = 1/\hat{c} \cdot \sum_i \hat{c_i} \left( i - \mu \right)^2 \right)$ :} Describes the cluster size weighted with the charge detected in each strip.

\item{\bf Cluster skew $\left( \gamma = 1/\hat{c} \cdot \sum_i \hat{c_i} \left( \frac{i - \mu }{\sigma}\right)^3 \right)$ :} Describes the cluster asymmetry in terms of the third standardized moment. Skew values closer to zero mean a more symmetric cluster.

\item{\bf Cluster multiplicity :} Describes the cluster size in terms of the number of strips that have been activated inside the cluster.

\end{itemize}

X-ray events produce in general a unique cluster signal (since they generally interact by photoelectric process), inducing some charge in X-strips and some charge in~Y-strips. In a preliminary analysis it is possible to reject up to about 90\% of the background events by imposing that the event should contain at least a cluster and no more than one cluster in each axis.

\vspace{0.2cm}

In order to apply this preliminary cut, one must assure that X-ray events are not rejected. After calibrating the detector with an X-ray source, some detectors show a rejection factor higher than the expected natural background. This effect could be due to partial \emph{pile up} events which are not completely inside the strips charge integration window or crosstalk, that will produce one or several low energy fake clusters. This effect is commonly observed in high activity sources generating a high rate in the detector, where \emph{pile up} events produce an event that is not observed in the mesh acquisition window but generate a small cluster in the strips readout. In order to recover the acceptance for X-ray events, under some conditions, it is required to introduce a new parameter, the \emph{cluster balance}.

\vspace{0.2cm}

The \emph{cluster balance} allows to redefine events that contain several clusters into a mono-cluster event. It plays a role when an event contains several clusters, but one of the clusters is carrying most of the total clusters charge. The parameter defines the minimum percentage of the total clusters charge that one cluster must carry in order to be defined as a mono-cluster event.

\vspace{0.2cm}

The \emph{cluster balance} value should be kept as high as possible in order to preserve only the events that produced a clear main cluster, but low enough to reach the expected level of acceptance with an X-ray source. Typical values for \emph{cluster balance} are higher than 85\%. Depending on the detector and its running conditions it is enough to accept only pure mono-cluster events by setting the \emph{cluster balance} parameter to 100\%.


\subsection{Cluster correction.}

As mentioned before, a cluster is defined when at least two consecutive strips are found active. A cluster \emph{starts} whenever two consecutive active strips are detected, and it \emph{ends} whenever three non-active consecutive strips are found.  The definition that describes the \emph{ending} of a cluster allows to have clusters with one or two consecutive strips non-active inside the cluster definition (see Fig.~\ref{fi:stripClusters}). 

\vspace{0.2cm}

A cluster is defined this way in order to avoid that \emph{strip defects}, \emph{dead strips}, \emph{cross talk}, or the strips particular \emph{response} at the measuring time, splits in two clusters a physical mono-cluster process.

\vspace{0.2cm}

\begin{figure}[ht!]
\begin{center}
\includegraphics[width=1.0\textwidth]{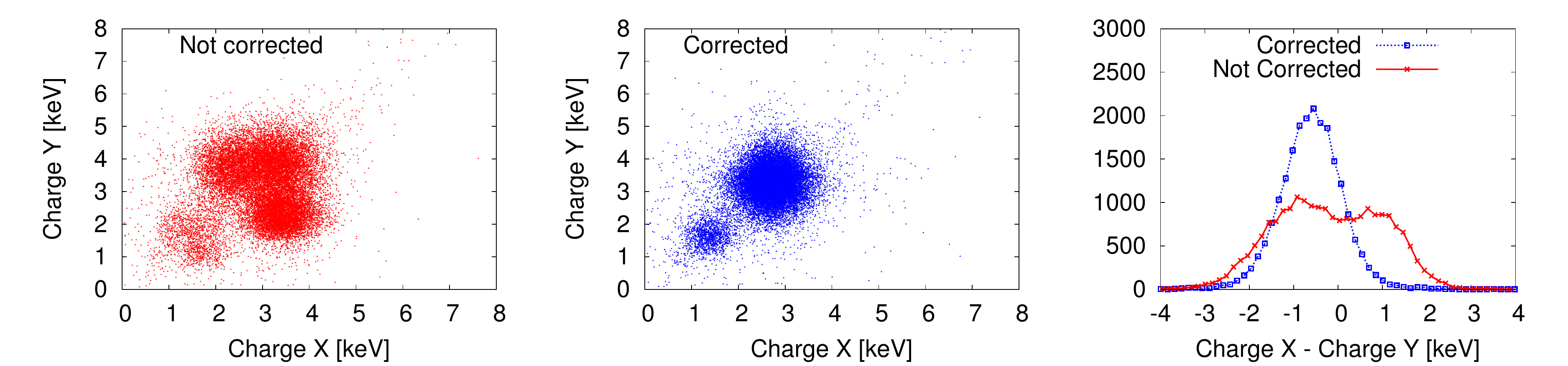} 
\end{center}

\caption{\fontfamily{ptm}\selectfont{\normalsize{On the left, the distribution map of \emph{chargeX} and \emph{chargeY}, generated by an $^{55}$Fe X-ray source, for a detector with missing strips charge. On the middle, the same distribution map after applying the \emph{dead strips} virtual charge correction. On the right, the effect of the correction in the deviation of \emph{chargeX} and \emph{chargeY} parameters.}}}
\label{fi:clusterCorrection}
\end{figure}

For a mono-cluster process it is expected that the X-strip plane collects as much charge as the Y-strips plane, these two parameters should be highly correlated. It has been observed in some Micromegas detectors (conventional type) that this correlation is distorted, due to some missing charge in one axis but not in the other axis (see Fig.~\ref{fi:clusterCorrection}).

\vspace{0.2cm}

 In order to recover the expected correlation between these two parameters, \emph{chargeX} and \emph{chargeY}, it is required to introduce some \emph{virtual} charge for the strips that have been detected inactive inside the cluster. This \emph{virtual} charge is calculated using the information provided by the surrounding strips. For a isolated dead strip it can be written as,

\begin{equation}
\hat{c}_{i+1} = \frac{\hat{c}_i + \hat{c}_{i+2}}{2}
\end{equation}

\vspace{0.2cm}

\noindent and for two consecutive dead strips inside a cluster it can be written as,

\begin{equation}
\hat{c}_{i+1} = \frac{2\hat{c}_i + \hat{c}_{i+3}}{3} \quad\quad \hat{c}_{i+2} = \frac{\hat{c}_i + 2\hat{c}_{i+3}}{3}
\end{equation}

\vspace{0.2cm}

\noindent where the \emph{inactive strips} charge is associated with the pondered value of the delimiting active strips.

\vspace{0.2cm}

Figure \ref{fi:clusterCorrection} shows the result of applying these corrections. The deviation between X-charge and Y-charge it is used as a potential discriminant for non X-ray events. In figure~\ref{fi:clusterCorrection} it is observed how this deviation recovers a Gaussian shape. The correction allows to recover a strips data set that otherwise would have been useless. 

\vspace{0.2cm}

An offset is observed in the deviation between \emph{chargeX} and \emph{chargeY}, related with the fact that the \emph{top} strips plane always receive a slightly higher charge than the \emph{bottom} plane. Thus the \emph{virtual} charge added could be emphasizing this effect. The mean shifting value can be taken into account and it will not affect the background discrimination analysis as long as the deviation is gaussian distributed.

\section{Pulse shape analysis.}

The pulse parameterization is intended to describe the full pulse signal with a minimum amount of parameters. These few parameters allow to define the pulse shape and introduce them in a posterior discrimination analysis, or to reproduce post-processed pulses coming from simulated parameters.


\subsection{Standard pulse shape parameters.}\label{sc:pulseStdParameters}

\begin{figure}[!ht]\begin{center}

\includegraphics[width=0.6\textwidth]{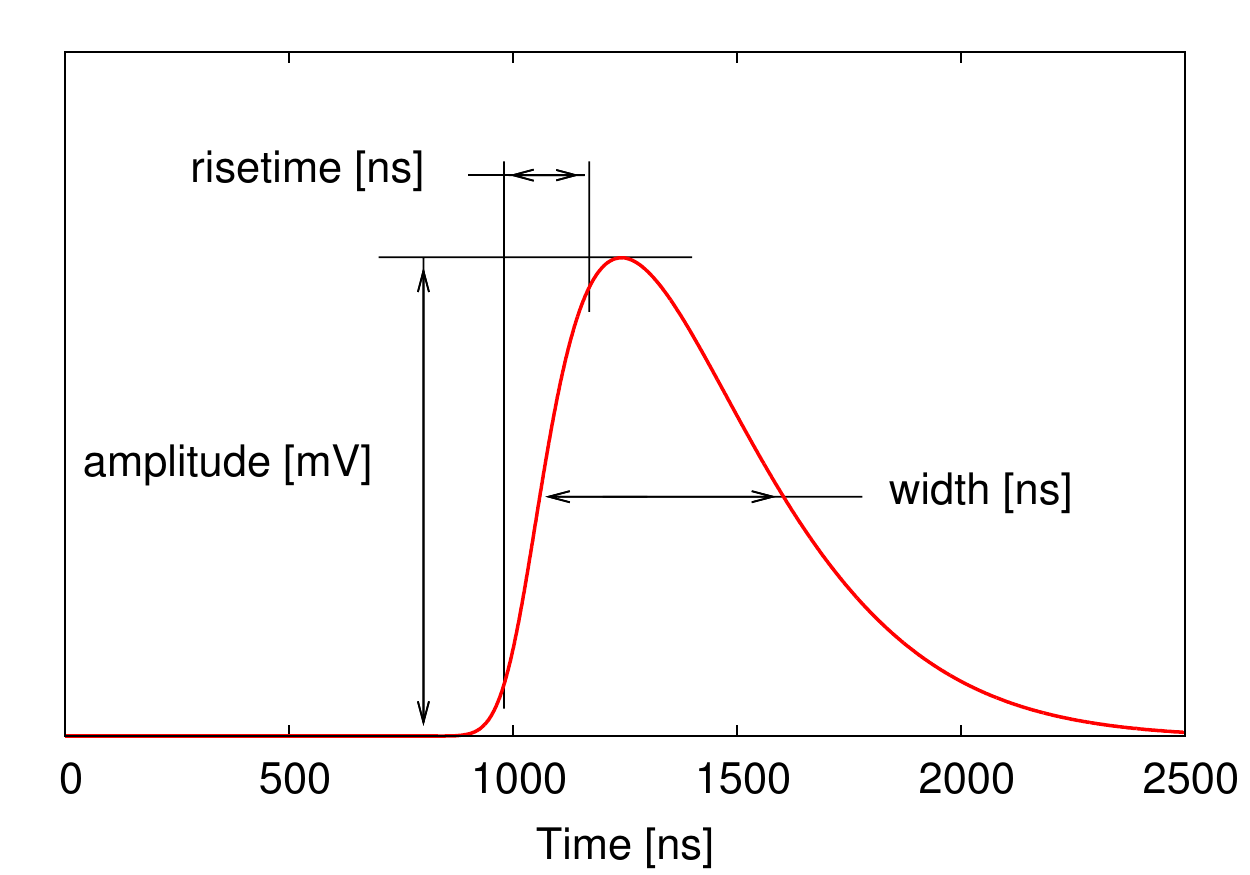}
\caption{\fontfamily{ptm}\selectfont{\normalsize{Standard pulse signal parameters definition.}}}
\label{fi:pulseParams}
\end{center}\end{figure}

The Micromegas mesh signal gives energy and temporal information from the triggering event. The event energy is well described by the amplitude and area of the generated pulse. Two parameters are commonly used in order to describe the pulse shape, \emph{risetime} and \emph{width}. In figure \ref{fi:pulseParams} are shown these parameters for a pulse generated by an x-ray event.

The pulse parameter values are slightly dependent on the definition used, the main standard parameters definition used for the rawdata analysis carried out in this thesis can be summarized in the following list.

\begin{itemize}

\item {\bf Pulse baseline :} Baseline voltage offset calculated with the first 100 pulse data points.

\item {\bf Pulse center :} Time at which the pulse reaches the maximum value.

\item {\bf Pulse amplitude :} The value at the \emph{pulse center} after subtracting the \emph{pulse baseline} value.

\item {\bf Pulse integral :} The pulse area enclosed between the \emph{starting pulse time} and the \emph{ending pulse time}. The starting and ending pulse times are defined when the pulse reaches 15\% of the amplitude value to the left and the right of the \emph{pulse center}.

\item {\bf Pulse mean center :} The addition of time values between the \emph{pulse starting time} and the \emph{pulse ending time} weighted with the pulse amplitude at the given times and normalized with the \emph{pulse integral}.

\item {\bf Pulse risetime :} Time length between the \emph{starting pulse time} and the time at which the pulse reaches 85\% of the \emph{pulse amplitude}.

\item {\bf Pulse width :} Time length between the points where the pulse reaches 50\% of the \emph{pulse amplitude}.

\end{itemize}

Before defining these parameters the pulse is preprocessed, some frequencies are subtracted using FFT analysis in order to reduce fluctuations and to smooth the pulse shape (see section~\ref{sc:frequential}). This preprocessing allows a more accurate definition of pulse parameters in noisy pulses.

\begin{figure}[!ht]\begin{center}

\includegraphics[width=1.0\textwidth]{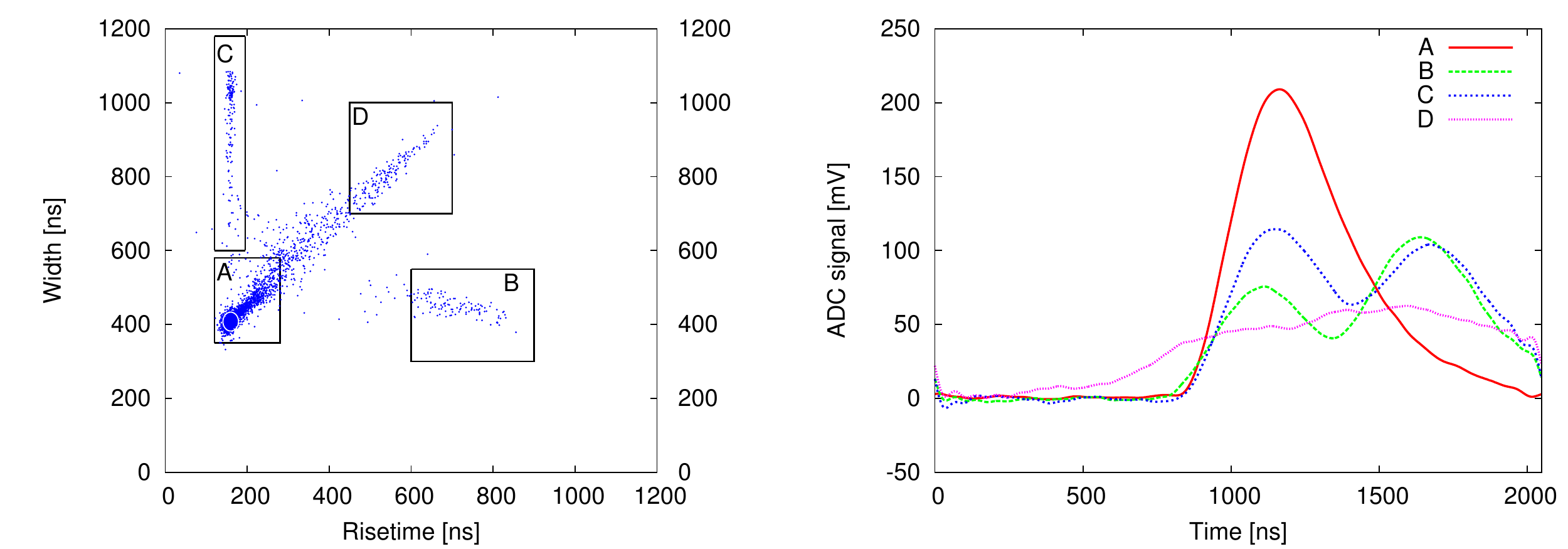}
\caption{\fontfamily{ptm}\selectfont{\normalsize{On the left, \emph{pulse risetime} and \emph{pulse width} parameter space for an $^{55}$Fe source. Events type A are X-ray events, the red circle places the region where the highest density of events is found. Events type B and C are mainly events that have overlapped, pile up events. The low source rate allowed to capture a non negligible population of cosmic rays, type D events. On the right, a mesh pulse extracted from each of the event type regions.}}}
\label{fi:rtVsWdAndPulses}
\end{center}\end{figure}

The parameters that define the pulse shape allow to classify different kind of events by drawing its parameter space (see Fig. \ref{fi:rtVsWdAndPulses}). Pile up events usually generate fake pulse parameters throwing them out the X-ray mean values. The rejection of this type of events in a background run is included in the acquisition \emph{dead time}, due to the low probability of occurrence during background measurements.

\subsection{A new fitting method for pulse shape analysis.}\label{sc:fitting}

\begin{figure}[!ht]\begin{center}
\includegraphics[width=0.4\textwidth]{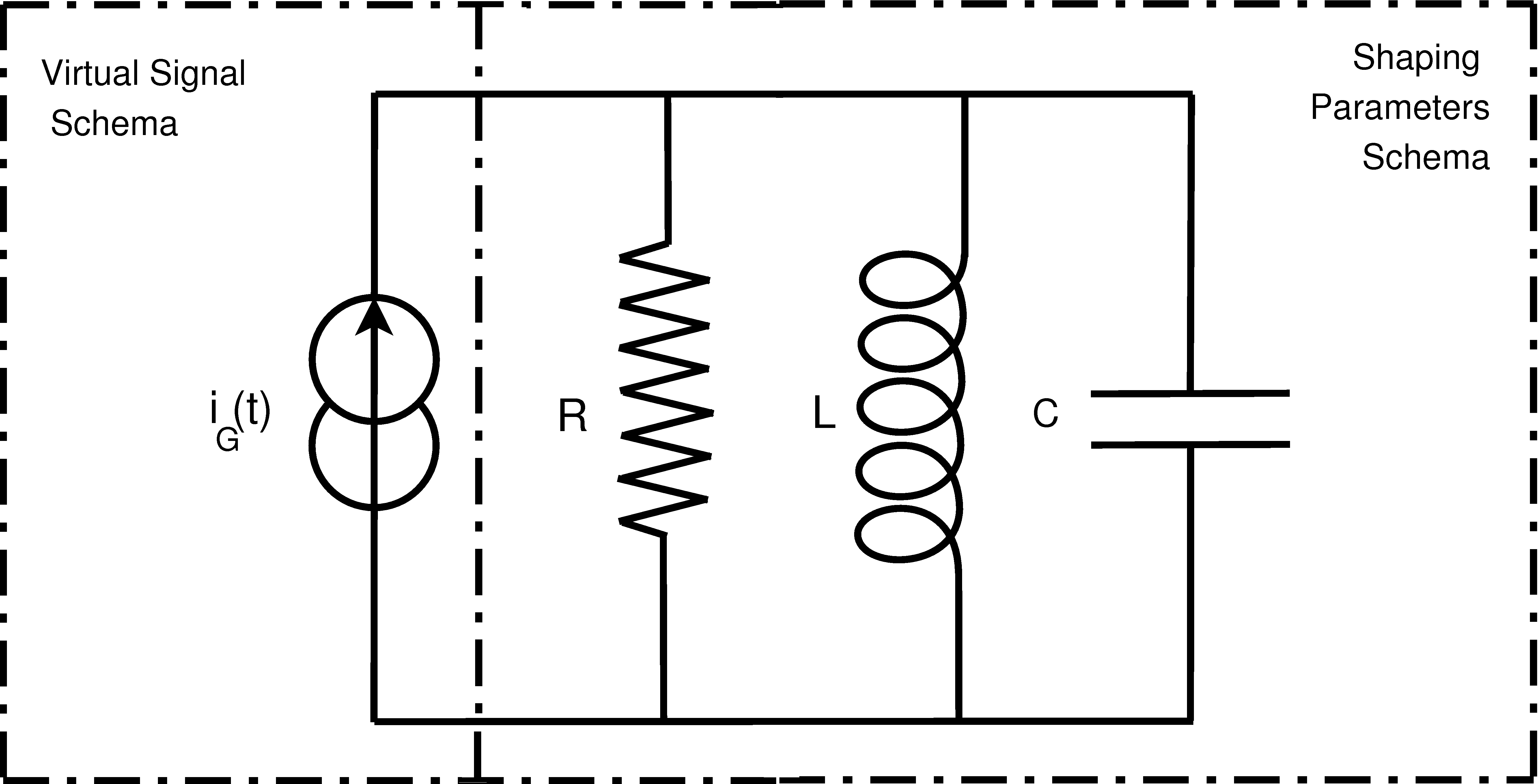}
\caption{\fontfamily{ptm}\selectfont{\normalsize{Pulse shaping schema.}}}
\label{fi:LCschema}
\end{center}\end{figure}

A method for pulse fitting was developed following the equations from an RLC circuit (see Fig.~\ref{fi:LCschema}). This circuit is described by a second order differential equation,

\begin{equation}\label{eq:RC}
\frac{1}{w_C} \frac{d^2 V(t)}{dt^2} + \frac{dV(t)}{dt} + w_L V(t) = \frac{d\left(R\cdot i_g(t)\right)}{dt}
\end{equation}

\vspace{0.2cm}

\noindent and involves two characteristical times, $w_C$ and $w_L$, allowing to adjust the growing and decaying pulse shape. $V(t)$ is the output voltage measured in the mesh.

\vspace{0.2cm}

The generating input signal, $di_g(t)/dt$, has been chosen to be gaussian shaped (see equation \ref{eq:iG}) assuming point-like energy deposition and an ideal gaussian distributed electrons cloud. The gaussian description defines the triggering event with a minimum amount of parameters; reference time, $t_g$, signal intensity, $A_g$ and temporal signal width, $\tau_g$,

\begin{equation}\label{eq:iG}
\frac{d\left( R \cdot i_g(t)\right)}{dt} = \frac{A_g}{\tau_g^2} \cdot e^{ -\frac{\left(t-t_g\right)^2}{2{\tau_g^2}} } 
\end{equation}

\noindent which amplitude is normalized by $\tau_g^2$ based on goodness of fit obtained with different approaches of expression~(\ref{eq:iG}) to experimental pulses acquired with the CAST micromegas electronic set-up. Small variations to this equation lead to fitting curves with slightly higher errors and that do not reproduce as adequately the shape of the pulse.

\vspace{0.2cm}

The resistive parameter in this fitting model, $R$, could give account for the different amplifier factors by the electronic set-up, however this parameter is mixed with the input signal amplitude and it lacks of interest to be determined for the following analysis. The parameter can be considered inside $A_g$, $A_g' \equiv A_g/R$.

\begin{figure}[!ht]\begin{center}
\includegraphics[width=0.85\textwidth]{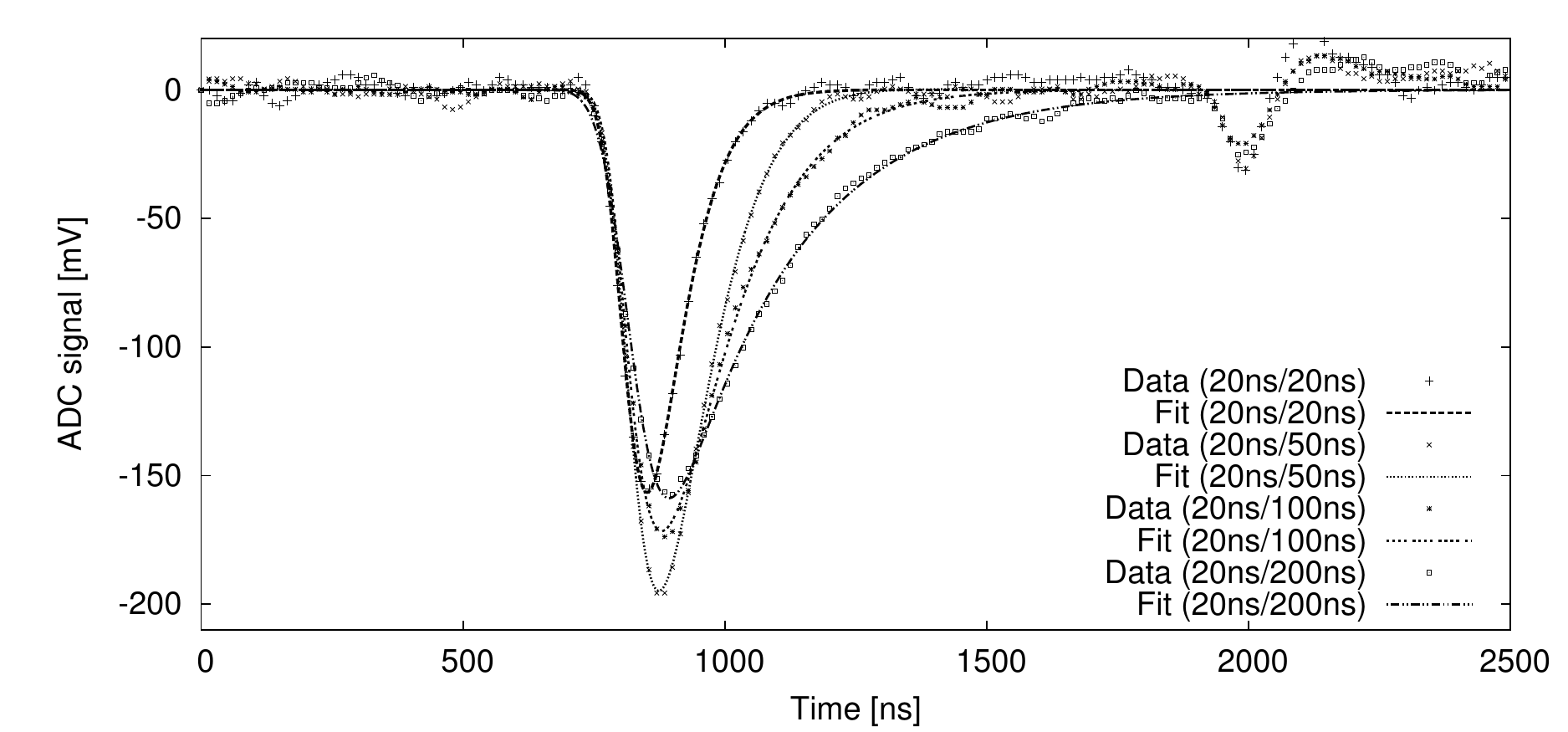}
\caption{\fontfamily{ptm}\selectfont{\normalsize{ Mesh signal pulses recorded with a Micromegas microbulk detector for different Timing Amplifier settings (int/diff), together with the fitted curve obtained. The peak appearing in the pulse tail is peak-up noise that appears due to interferences with the VME acquisition, and lacks physical meaning. }}}
\label{fi:PulsesFits}
\end{center}\end{figure}

The parameters that define the mesh signal in equation \ref{eq:RC} is obtained for each mesh pulse by performing a Monte-Carlo fitting, varying the parameters and finding the ones that minimize the distance between the experimental mesh pulse and $V(t)$ (see Fig.~\ref{fi:PulsesFits}). The fitting allows to determine all the parameters that define the pulse shape in equation (\ref{eq:RC}).

\begin{figure}[!ht]\begin{center}
\includegraphics[width=1.0\textwidth]{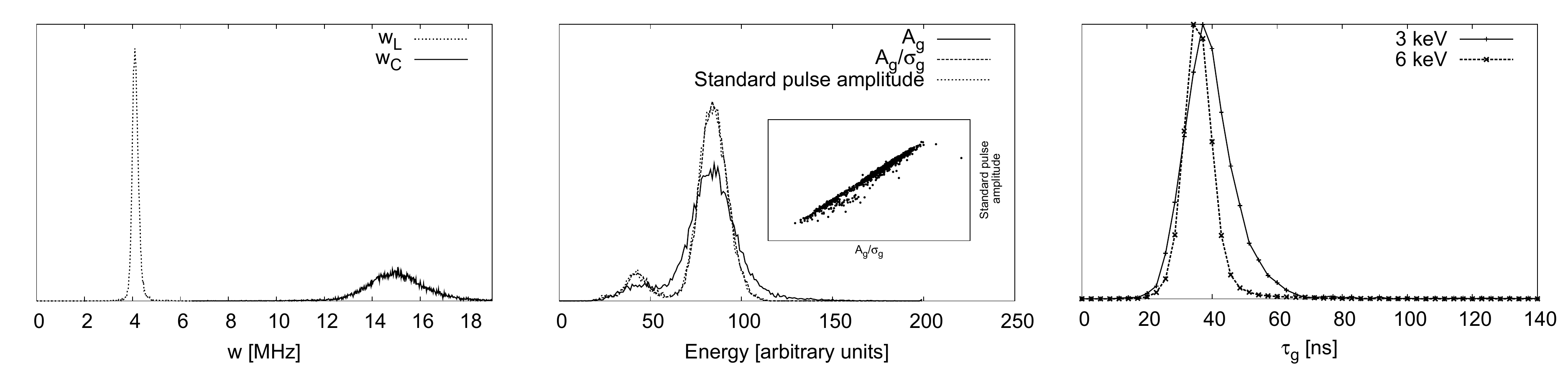}
\caption{\fontfamily{ptm}\selectfont{\normalsize{On the left, distribution of $w_C$ and $w_L$ fitted parameters for the events generated with an $^{55}$Fe source. On the middle, $^{55}$Fe spectrum reconstruction with the input signal parameters and the standard pulse amplitude. Standard pulse amplitude correlation with $A_g/\tau_g$ is also shown. On the right, distribution of $\tau_g$ for 3\,keV and 6\,keV events, introducing an arbitrary normalization.}}}
\label{fi:rcAnalysis}
\end{center}\end{figure}

\vspace{0.2cm}
This method has been used to parameterize pulses for different Timing Amplifier settings (see Fig. \ref{fi:PulsesFits}). The method allows to describe the pulse shape by using \emph{five} parameters. \emph{Two} of those parameters are related with the electronic set-up, $w_C$ and $w_L$, and their values only depend on the settings imposed in the Timing Amplifier module (see Fig.~\ref{fi:rcAnalysis}). The other \emph{three} parameters, $A_g'$, $\tau_g$ and $t_g$, are the few remaining parameters that contain physical information from the event.
 
\vspace{0.2cm}

The relation~(\ref{eq:iG}) intends to be related to the original electron cloud created in the drift region. The number of electrons $n_e$ in the electron cloud is directly related with the energy $\epsilon$ of the originating event, which is given by

\begin{equation}
\epsilon \propto n_e = \sqrt{8\pi^3} \, \rho_c \, \sigma_l \, \sigma_t^2
\end{equation}

\noindent where $\rho_c$ is the electron energy density at the cloud center, $\sigma_l$ the longitudinal diffusion, and $\sigma_t$ the transversal diffusion. Thus, the density of an ideal electron cloud integrated at the mesh plane can be described by the following relation as a function of the arrival time, assuming the electron avalanche time negligible compared to the drift time,

\begin{equation}\label{eq:cloud}
\lambda(t) = \frac{n_e}{\sqrt{8\pi} \sigma_l }\, e^{-\frac{\left(t-t_o\right)^2}{2(\sigma_l/v_d)^2}}
\end{equation}

\vspace{0.2cm}

\noindent where $v_d$ is the drift velocity, used to transform the spatial electron cloud in its temporal evolution in the mesh structure. 

Based on the assumption that the gaussian parameter $\tau_g$ is related with the longitudinal diffusion $\tau_g = \sigma_l/v_d$, the analogy between the generating input signal, $di_g/dt$, and the relation~(\ref{eq:cloud}) allows us to describe the energy of the event as

\begin{equation}
\epsilon \propto n_e = \frac{\sqrt{8\pi}\,A_g\,v_d}{\tau_g} \propto \frac{A_g}{\tau_g}
\end{equation}

\vspace{0.2cm}
\noindent indicating that the right parameterization of the energy in the parameter set introduced in the model is given by $A_g/\tau_g$. Figure~\ref{fi:rcAnalysis} shows the $^{55}$Fe spectrum obtained by the parameter $A_g$ and $A_g/\tau_g$, where it is observed the improved energy definition by normalizing $A_g$ with $\tau_g$. Moreover, the parameter $A_g/\tau_g$ shows a good correlation with the standard pulse amplitude described on section~\ref{sc:pulseStdParameters}, which gives account of the good agreement between both methods for the obtention of the event energy.





\subsection{Fitting multicluster events.}

The equation~(\ref{eq:RC}) can be generalized to reproduce pulses containing more than one cloud of charge. The left side of the equation reproduces the behavior of the electronics in the incoming signal, producing the final shape that is measured in the mesh. The equation can be extended by adding as many new signals as necessary to the right side of the equation,

\begin{equation}\label{eq:RCext}
\left[ \frac{1}{w_C} \frac{d^2}{dt^2} + \frac{d}{dt} + w_L  \right] V(t)  = \sum_n \frac{d\left(R\cdot i^n_g(t)\right)}{dt}
\end{equation}

\noindent where $n$ identifies the input signal of each electron cloud.

\vspace{0.2cm}

By applying the same fitting method it is possible to parameterize more complex events by sequentially adding extra signal terms to the expression~(\ref{eq:RCext}) till the error is low enough, allowing to differentiate and identify independent electron clouds, and parameterize them.

\begin{figure}[!ht]\begin{center}
\includegraphics[width=1.0\textwidth]{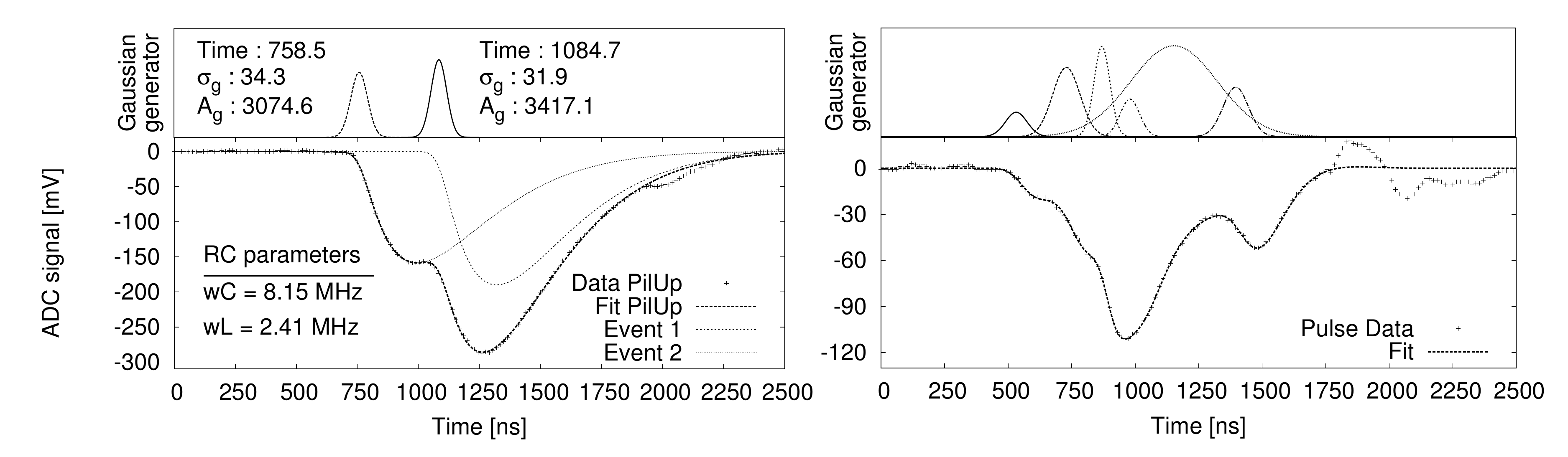}
\caption{\fontfamily{ptm}\selectfont{\normalsize{On the left, the fitting of a double cluster pile up event extracted from an $^{55}$Fe calibration run. On the right, a more complex event structure extracted from a cosmic run. }}}
\label{fi:pilupevent}
\end{center}\end{figure}

\vspace{0.2cm}
In order to illustrate the result of the method for multi-cluster events it has been chosen a pile up event coming from an $^{55}$Fe calibration run, and a much more complex event structure coming from a cosmic run (see Fig.~\ref{fi:pilupevent}).

\subsection{Frequential analysis.}\label{sc:frequential}

The relation~\ref{eq:RCext} can be re-expressed in the frequential domain by applying the \emph{Fourier transform}, giving a theoretical description of the pulse shape in terms of its characteristical frequencies. The resulting algebraic expression can be re-written to describe the pulse shape as a function the frequency $V(w)$, 

\begin{equation}\label{eq:Vw}
V(w) = \frac{w_C}{w_C w + j \left(w_L w_C - w^2\right)} \sum_{i} A_i \sigma_i e^{-j t_i w} e^{ - \frac{w^2 \sigma_i^2}{2} }
\end{equation}

\begin{figure}[!htb]\begin{center}
\includegraphics[width=1.0\textwidth]{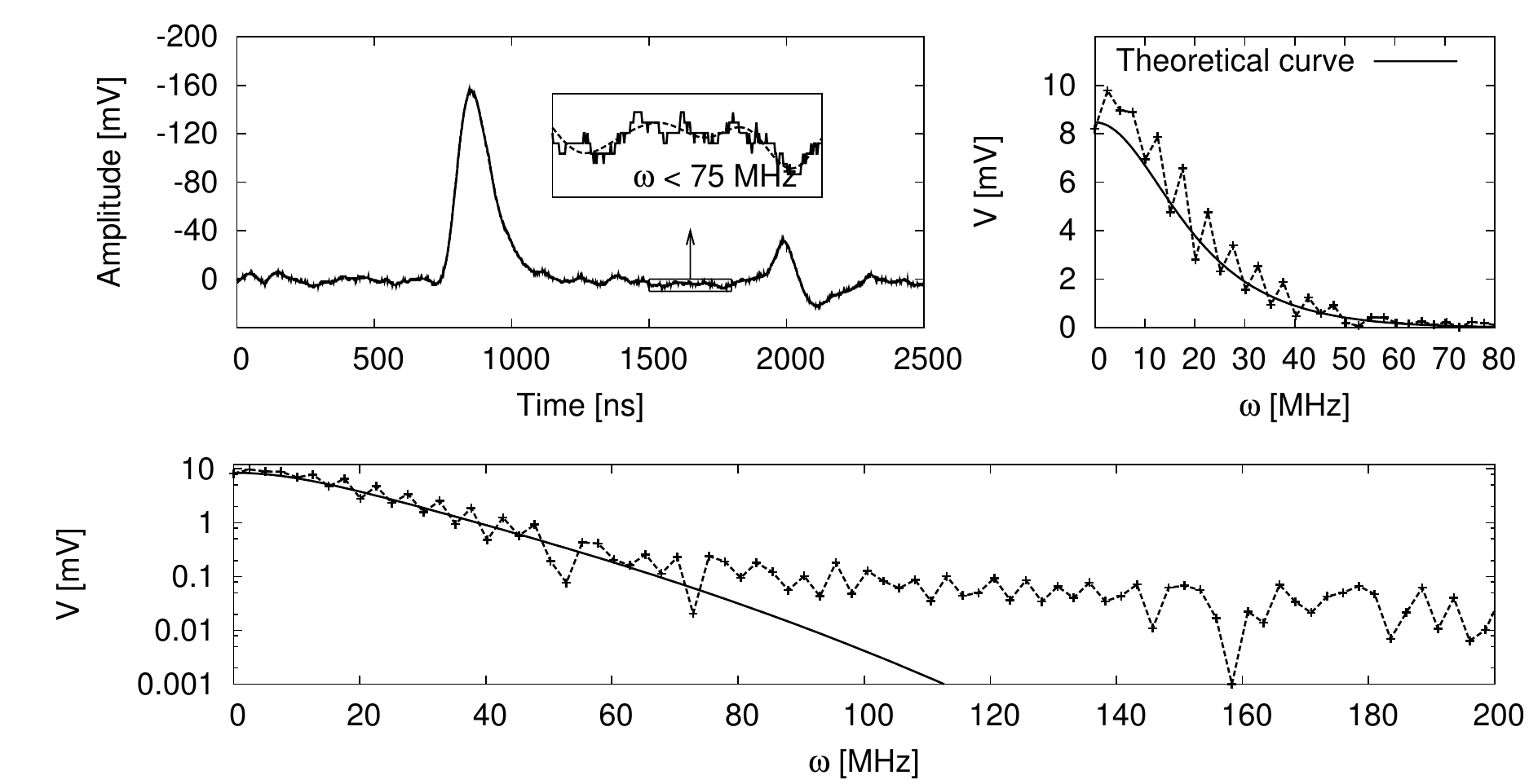}
\caption{\fontfamily{ptm}\selectfont{\normalsize{Temporal mesh pulse signal data together with the results of smoothing the noise fluctuations after applying the cut frequency $w_t < 75$\,MHz (left-top), and pulse mesh signal in the frequential domain, where the FFT data obtained is represented with dashed line, and the theorical curve with continous line. The signal amplitude is presented in linear scale (top-right) and logarithmic scale (bottom).}}}
\label{fi:fft}
\end{center}\end{figure}

\vspace{0.2cm}
\noindent where the modulus of this complex expression gives the contribution of each frequency to the signal. A single X-ray pulse event has been chosen to show the properties of its frequential shape (see Fig.~\ref{fi:fft}). The frequential data from the measured pulse is obtained by applying a Fast Fourier Transform (FFT) method implemented in the c library \emph{fftw3}~\cite{FFTW05}, which can be directely compared with the theoretical description given in relation~(\ref{eq:Vw}). In the frequential domain is observed how the lower frequencies contribute more to define the pulse shape. The higher frequencies are related with the baseline noise fluctuations, this represention allows to define a frequency threshold that should be always lower than the cut frequency $w_t = \sqrt{w_C w_L}$ to include only the main frequencies that are involved, thus allowing to smooth the pulse shape fluctuations.

\section{Characterization of Micromegas detectors.}

Before being finally installed in CAST, the new Micromegas detectors built are tested at CEA Saclay. The performance tests carried out in Micromegas detectors intend to assure the quality of the data provided by the detector in terms of energy resolution, active detection area, stability and background discrimination capabilities.  

\vspace{0.2cm}

\begin{figure}[!ht]\begin{center}
\begin{tabular}{cc}
\includegraphics[width=0.455\textwidth]{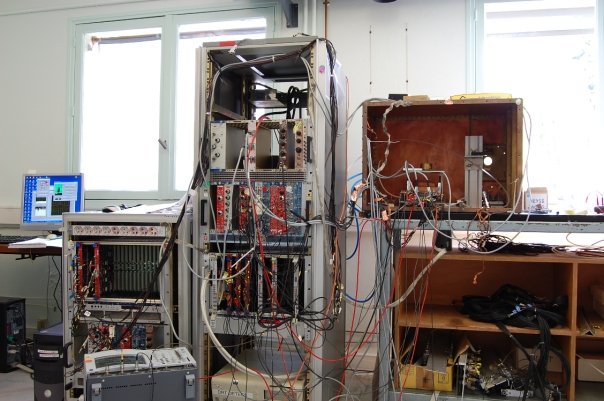} &
\includegraphics[width=0.455\textwidth]{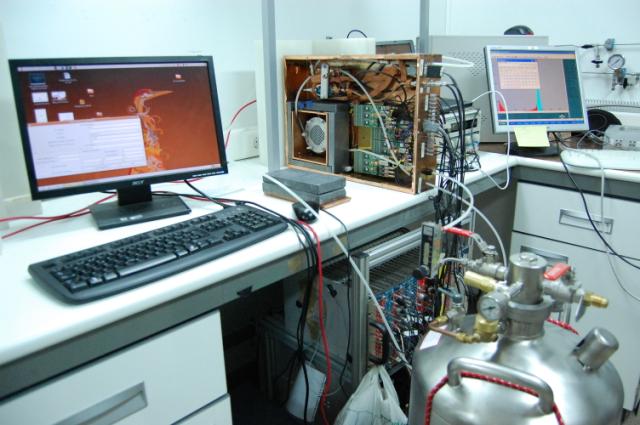} \\
\end{tabular}
\caption{\fontfamily{ptm}\selectfont{\normalsize{On the left, Micromegas acquisition set-up at CEA Saclay. On the right, Micromegas set-up at Zaragoza.}}}
\label{fi:saclaySetup}
\end{center}\end{figure}

Figure~\ref{fi:saclaySetup} shows the Micromegas acquisition systems at CEA Saclay and at Zaragoza laboratory. These systems have been developed keeping the same acquisition standards as the detectors that are actually taking data in the CAST experiment. A full characterization of a bulk Micromegas detector was carried out at Saclay allowing to study the behavior of the detector under different conditions, isobutane concentration, pressure and, mesh and drift voltages applied (section~\ref{sc:B4}). Later on, a more concise characterization of a microbulk type detector obtained at Zaragoza laboratory will be also presented (section~\ref{sc:M13}).

\vspace{-0.4cm}
Moreover, a full characterization of the electronic set-up for different Timing Amplifier settings is given by using the method described on section~\ref{sc:fitting}.

\subsection{Characterization of pulse acquisition shapes.}\label{sc:pulseChar}

The fitting method described on section~\ref{sc:fitting} has been used to carry out a full characterization of the electronic set-up related with the mesh pulse acquisition. The parameters $w_L$ and $w_C$, together with the input signal parameters, from expression~(\ref{eq:RC}) have been obtained for all the possible timing combinations that offers the timing amplifier module (ORTEC 471). In order to characterize these settings it was used a bulk detector (B4) at the nominal conditions of operation in the CAST experiment. The parameters in each of the measurements conditions are obtained by calculating the mean value of these parameters, which keep a Gaussian distribution, for X-ray events coming from an $^{55}$Fe source.

\vspace{0.2cm}

For each of the timing settings combination was taken more than one measurement in which the relation of coarse and fine gain of the ORTEC module was modified, trying to preserve the same amplifier gain. Tables~\ref{ta:wCwLtable} and~\ref{ta:gaussianParameters} show the resulting parameter values obtained for the timing amplifier frequencies $w_C$ and $w_L$, and the input signal parameters obtained in each calibration run. At table~\ref{ta:wCwLtable} it is noticed how the $w_C$ and $w_L$ parameters depend only on the imposed setting at the amplifier module. The input signal parameters given at table~\ref{ta:gaussianParameters} show how the input signal amplitude depends more strongly with the amplifier gain settings, while the signal width $\tau_g$ keeps a more constant value than the original \emph{pulse width} parameter indicating a better corelation with the diffusion of the originating physical event due to the correlation reduction with the electronic parameters achieved with the new fitting method, in spite of the variations observed at different settings that might be due to the information loss related with the signal filtering process, since higher timing settings lead to a shorter cut frequency $w_t$, it is expected that the shorter timing amplifier values lead to a more accurate value of the original signal width.

\begin{figure}[!ht]\begin{center}
\includegraphics[width=0.95\textwidth]{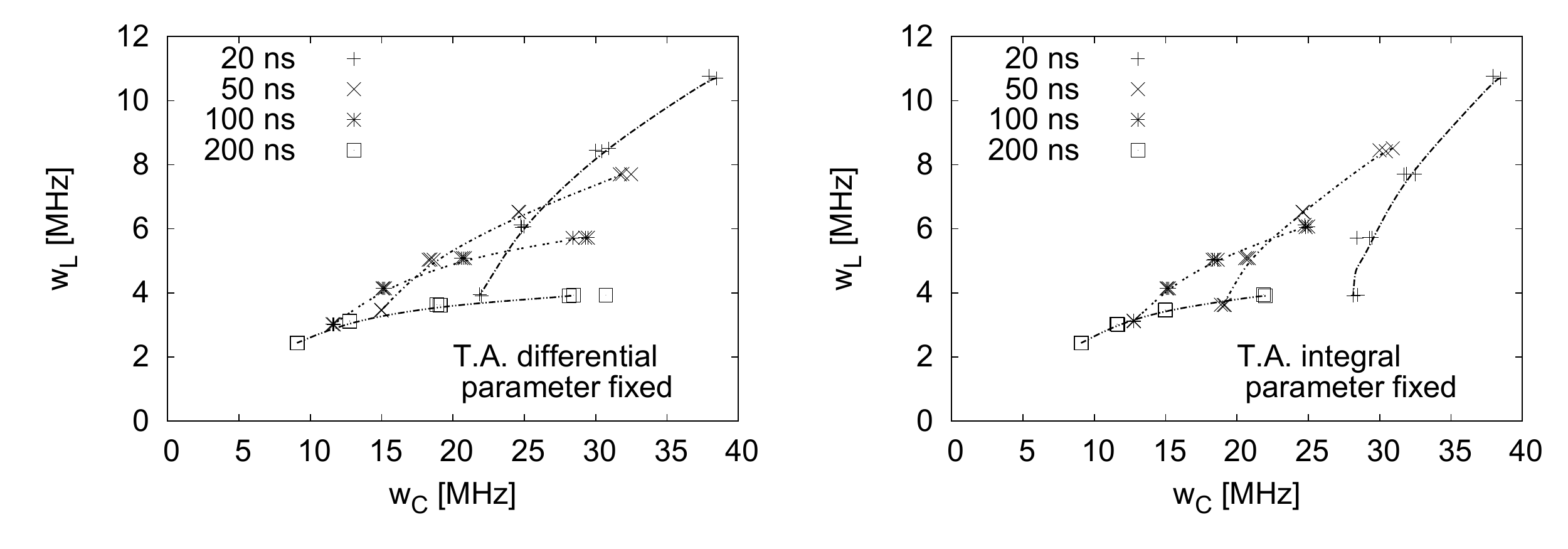}
\caption{\fontfamily{ptm}\selectfont{\normalsize{wC and wL mean values distribution curves for different amplifier settings.}}}
\label{fi:wCwLcurves}
\end{center}\end{figure}

Tables~\ref{ta:wCtable} and~\ref{ta:wLtable} summarize the mean $w_C$ and $w_L$ values obtained for each timing amplifier settings combination shown in figure~\ref{fi:wCwLcurves}.

\vspace{-0.2cm} 

\vspace{-0.8cm}

\begin{table}[!hb]
\begin{center}
\begin{tabular}{c|cccc}
Int \ Dif & 20 & 50 & 100 & 200 \\
\hline
20  & 38.20 & 32.02 & 29.04 & 29.11 \\
50  & 30.45 & 24.62 & 20.71 & 19.03 \\
100 & 24.85 & 18.46 & 15.14 & 12.76 \\
200 & 21.93 & 14.98 & 11.61 & 9.09 \\
\end{tabular}
\caption{\fontfamily{ptm}\selectfont{\normalsize{Values of $w_C$[MHz] for different Timing Amplifier settings.}}}
\label{ta:wCtable}
\end{center}
\end{table}

\vspace{-0.8cm}

\begin{table}[!hb]
\begin{center}
\begin{tabular}{c|cccc}
Int \ Dif & 20 & 50 & 100 & 200 \\
\hline
20  & 10.73 & 7.70 & 5.71 & 3.92 \\
50  &  8.46 & 6.52 & 5.08 & 3.62 \\
100 &  6.08 & 5.03 & 4.14 & 3.12 \\
200 &  3.92 & 3.46 & 3.02 & 2.44 \\
\end{tabular}
\caption{\fontfamily{ptm}\selectfont{\normalsize{Values of $w_L$[MHz] for different Timing Amplifier settings.}}}
\label{ta:wLtable}
\end{center}
\end{table}

\begin{table}
\begin{center}
\begin{tabular}{c|cccc|cccc}
 & \multicolumn{4}{|c|}{Timing amplifier settings} & \multicolumn{4}{|c}{Timing amplifier parameters} \\
\hline
Run&	Coarse&	Fine& 	Int&	Dif&	$w_C$ & $\sigma_{w_C}$ &	$w_L$ &  $\sigma_{w_L}$ \\
\hline
\hline
15464	&	6	&7.3	&20	&20	&37.93& 7.35	&10.77&	0.77 \\
15466	&	20	&6.8	&20	&20	&38.46& 6.79	&10.70&	0.70 \\
\hline
15467	&	6	&7.3	&50	&20	&30.92& 7.65	&08.51&	0.53 \\
15468	&	10	&7.3	&50	&20	&30.44& 4.81	&08.42&	0.49 \\
15469	&	20	&6.0	&50	&20	&29.99& 4.64	&08.45&	0.44 \\
\hline
15470	&	6	&7.3	&100	&20	&24.74& 4.07	&06.12&	0.30 \\
15471	&	10	&7.3	&100	&20	&24.99& 4.02	&06.07&	0.34 \\
15472	&	20	&6.5	&100	&20	&24.82& 3.91	&06.05&	0.33 \\
\hline
15473	&	10	&7.3	&200	&20	&21.86& 4.15	&03.95&	0.21 \\
15474	&	20	&7.3	&200	&20	&22.00& 3.22	&03.90&	0.17 \\
\hline
15475	&	6	&6.0	&20	&50	&31.70& 5.14	&07.70&	0.39 \\
15476	&	10	&5.0	&20	&50	&31.88& 4.96	&07.71&	0.38 \\
15477	&	20	&3.0	&20	&50	&32.49& 5.03	&07.70&	0.34 \\
\hline
15478	&	6	&6.5	&50	&50	&24.64& 3.84	&06.52&	0.31 \\
15479	&	10	&5.7	&50	&50	&24.59& 3.82	&06.53&	0.29 \\
\hline
15481	&	6	&6.8	&100	&50	&18.42& 2.41	&05.04&	0.20 \\
15482	&	10	&6.0	&100	&50	&18.65& 2.38 	&05.03&	0.19 \\
15483	&	20	&4.3	&100	&50	&18.30& 2.33	&05.03&	0.20 \\
\hline
15484	&	6	&7.3	&200	&50	&14.98& 1.76	&03.47&	0.17 \\
15485	&	10	&6.5	&200	&50	&14.97& 1.52	&03.46&	0.14 \\
15486	&	20	&5.5	&200	&50	&15.00& 1.69	&03.45&	0.16 \\
\hline
15487	&	6	&5.5	&20	&100	&28.39& 4.76	&05.71&	0.23 \\
15488	&	10	&4.0	&20	&100	&29.27& 4.94	&05.72&	0.23 \\
15489	&	20	&1.0	&20	&100	&29.46& 4.96	&05.72&	0.22 \\
\hline
15506	&	6	&5.5	&50	&100	&20.57& 2.91	&05.08&	0.20 \\
15507	&	10	&4.5	&50	&100	&20.71& 2.64	&05.08&	0.20 \\
15508	&	20	&2	&50	&100	&20.84& 2.62	&05.08&	0.20 \\
\hline
15509	&	6	&6.0	&100	&100	&15.06& 1.54	&04.14&	0.17 \\
15510	&	10	&5.0	&100	&100	&15.13& 1.52	&04.14&	0.16 \\
15511	&	20	&3.0	&100	&100	&15.23& 1.53	&04.14&	0.16 \\	
\hline
15512	&	6	&6.5	&200	&100	&11.55& 1.03	&03.02&	0.17 \\
15513	&	10	&6.0	&200	&100	&11.65& 0.96	&03.01&	0.14 \\
15514	&	20	&4.0	&200	&100	&11.62& 0.98	&03.02&	0.14 \\
\hline
15490	&	4	&6.0	&20	&200	&28.15& 5.37	&03.91&	0.16 \\
15491	&	6	&5.0	&20	&200	&28.46& 5.48	&03.93&	0.17 \\
15492	&	10	&3.5	&20	&200	&30.72& 6.67	&03.93&	0.14 \\
\hline
15493	&	4	&5.5	&50	&200	&19.07& 3.22	&03.62&	0.17 \\
15494	&	6	&4.5	&50	&200	&19.15& 3.05	&03.62&	0.16 \\
15495	&	10	&2.5	&50	&200	&18.88& 2.81	&03.63&	0.16 \\
\hline
15496	&	4	&6.5	&100	&200	&12.77& 1.29	&03.11&	0.17 \\
15497	&	6	&5.0	&100	&200	&12.77& 1.25	&03.12&	0.16 \\
15498	&	10	&3.0	&100	&200	&12.75& 1.27	&03.13&	0.17 \\
\hline
15499	&	4	&6.5	&200	&200	&09.09& 0.71	&02.43&	0.16 \\
15500	&	6	&6.0	&200	&200	&09.10& 0.69	&02.44&	0.16 \\
15501	&	10	&4.5	&200	&200	&09.08& 0.71	&02.44&	0.17 \\
\end{tabular}
\caption{\fontfamily{ptm}\selectfont{\normalsize{Values of $w_C$ and $w_L$ for different time settings.}}}
\label{ta:wCwLtable}
\end{center}
\end{table}

\begin{table}
\begin{center}
\begin{tabular}{c|cccc|cccccc}
 & \multicolumn{4}{|c|}{Timing amplifier settings} & \multicolumn{6}{|c}{Input signal parameters} \\
\hline
Run&	Coarse&	Fine& 	Int&	Dif&	$A'_g$&	$\sigma_{A_g}$ & $\tau_g$&	$\sigma_{\tau_g}$ &	$t_o$& $\sigma_{t_o}$  \\
\hline
\hline
15464	&	6	&7.3	&20	&20	&3077	&698	&31.67	&5.18	&797.1	&8.5	\\
15466	&	20	&6.8	&20	&20	&3971	&877	&31.89	&4.93	&796.3	&8.8	\\
\hline
15467	&	6	&7.3	&50	&20	&2573	&554	&33.87	&5.45	&778.9	&11.	\\
15468	&	10	&7.3	&50	&20	&4069	&896	&34.30	&5.35	&793.4	&10.	\\
15469	&	20	&6.0	&50	&20	&3751	&871	&34.41	&6.15	&791.2	&10.	\\
\hline
15470	&	6	&7.3	&100	&20	&1929	&432	&35.16	&6.09	&761.8	&13.	\\
15471	&	10	&7.3	&100	&20	&3021	&652	&35.38	&5.85	&781.3	&12.	\\
15472	&	20	&6.5	&100	&20	&3573	&795	&35.91	&6.15	&786.3	&12.	\\
\hline
15473	&	10	&7.3	&200	&20	&2055	&471	&36.01	&6.80	&762.3	&16.	\\
15474	&	20	&7.3	&200	&20	&3820	&839	&37.05	&4.81	&786.4	&12.	\\
\hline
15475	&	6	&6.0	&20	&50	&3349	&705	&33.27	&4.69	&790.2	&9.4	\\
15476	&	10	&5.0	&20	&50	&3476	&753	&33.18	&4.63	&791.4	&10.1	\\
15477	&	20	&3.0	&20	&50	&3930	&867	&33.48	&4.42	&794.8	&9.3	\\
\hline
15478	&	6	&6.5	&50	&50	&4137	&894	&36.45	&4.94	&791.5	&10.7	\\
15479	&	10	&5.7	&50	&50	&4068	&864	&36.19	&4.90	&791.4	&9.7	\\
\hline
15481	&	6	&6.8	&100	&50	&3814	&819	&36.90	&5.12	&783.9	&11.0	\\
15482	&	10	&6.0	&100	&50	&3881	&829	&37.22	&5.10	&784.7	&11.3	\\
15483	&	20	&4.3	&100	&50	&3698	&791	&36.93	&4.99	&782.7	&11.3	\\
\hline
15484	&	6	&7.3	&200	&50	&3411	&724	&38.13	&5.44	&774.5	&12.6	\\
15485	&	10	&6.5	&200	&50	&3507	&752	&38.61	&4.97	&775.2	&12.7	\\
15486	&	20	&5.5	&200	&50	&3840	&807	&38.75	&5.31	&778.9	&11.9	\\
\hline
15487	&	6	&5.5	&20	&100	&3128	&685	&33.51	&4.85	&790.2	&9.9	\\
15488	&	10	&4.0	&20	&100	&3254	&710	&33.79	&4.83	&791.5	&9.5	\\
15489	&	20	&1.0	&20	&100	&3443	&753	&33.82	&4.76	&793.0	&9.2	\\
\hline
15506	&	6	&5.5	&50	&100	&3044	&661	&35.59	&5.08	&781.8	&11.6	\\
15507	&	10	&4.5	&50	&100	&3573	&772	&35.86	&4.86	&787.1	&10.8	\\
15508	&	20	&2	&50	&100	&3833	&832	&36.22	&4.80	&789.1	&11.7	\\
\hline
15509	&	6	&6.0	&100	&100	&3345	&720	&37.53	&5.21	&777.3	&12.8	\\
15510	&	10	&5.0	&100	&100	&3542	&752	&37.71	&5.02	&779.4	&12.2	\\
15511	&	20	&3.0	&100	&100	&4020	&853	&38.18	&4.90	&783.7	&11.8	\\	
\hline
15512	&	6	&6.5	&200	&100	&3222	&675	&39.13	&5.52	&767.5	&14.5	\\
15513	&	10	&6.0	&200	&100	&3957	&826	&39.70	&5.27	&775.8	&13.2	\\
15514	&	20	&4.0	&200	&100	&3694	&772	&39.74	&5.33	&773.1	&13.6	\\
\hline
15490	&	4	&6.0	&20	&200	&3316	&738	&34.91	&5.12	&790.9	&9.9	\\
15491	&	6	&5.0	&20	&200	&3316	&729	&35.06	&5.11	&791.1	&10.2	\\
15492	&	10	&3.5	&20	&200	&3455	&729	&36.46	&4.67	&792.5	&10.4	\\
\hline
15493	&	4	&5.5	&50	&200	&2464	&556	&37.24	&5.56	&772.8	&13.1	\\
15494	&	6	&4.5	&50	&200	&2797	&605	&37.42	&5.41	&776.4	&12.5	\\
15495	&	10	&2.5	&50	&200	&2874	&605	&37.15	&5.20	&777.2	&12.3	\\
\hline
15496	&	4	&6.5	&100	&200	&2004	&674	&38.85	&6.24	&767.0	&14.8	\\
15497	&	6	&5.0	&100	&200	&2970	&604	&38.79	&5.31	&766.5	&13.9	\\
15498	&	10	&3.0	&100	&200	&2953	&606	&38.86	&5.34	&766.3	&14.1	\\
\hline
15499	&	4	&6.5	&200	&200	&3482	&727	&40.82	&5.94	&760.9	&15.9	\\
15500	&	6	&6.0	&200	&200	&3597	&759	&40.92	&5.92	&763.3	&15.4	\\
15501	&	10	&4.5	&200	&200	&3413	&715	&40.94	&5.93	&760.3	&15.8	\\
\end{tabular}
\caption{\fontfamily{ptm}\selectfont{\normalsize{Gaussian parameters.}}}
\label{ta:gaussianParameters}
\end{center}
\end{table}

\clearpage

\subsection{Characterization of a bulk type detector.}\label{sc:B4}

A bulk type detector (B4) was fully characterized as a function of the main parameters related with the operating conditions of the detector; gas pressure, isobutane concentration, and mesh and drift voltages applied. These measurements intend to be a reference for the operating conditions of a bulk type detector, typical voltages required and effect of gas properties in the main parameters of the detector readout; mesh pulse and cluster properties.

\vspace{0.2cm}

The parameters description shown in this section correspond to the ones obtained from the main peak ($5.9$\,keV) and scape peak of an $^{55}$Fe source. It must be noticed that the error bars in \emph{risetime}, \emph{width}, \emph{multiplicity} and \emph{cluster size} plots represent the width of the distribution.

\subsubsection{Drift voltage measurements}
The detector was characterized for different drift voltages for a given mesh voltage of $V_m = 350$\,V, a fixed timing amplifier settings (x2/7/100ns/100ns\footnote{coarse gain/fine gain/differential timing/integration timing}) and gas argon+$2$\,\% iC$_4$H$_{10}$ at a pressure of $1.4$\,bar, which are the typical running conditions used in CAST bulk detectors. 

\vspace{0.2cm}

\begin{figure}[!ht]\begin{center}
\begin{tabular}{cc}
\includegraphics[angle=270,width=6.7cm]{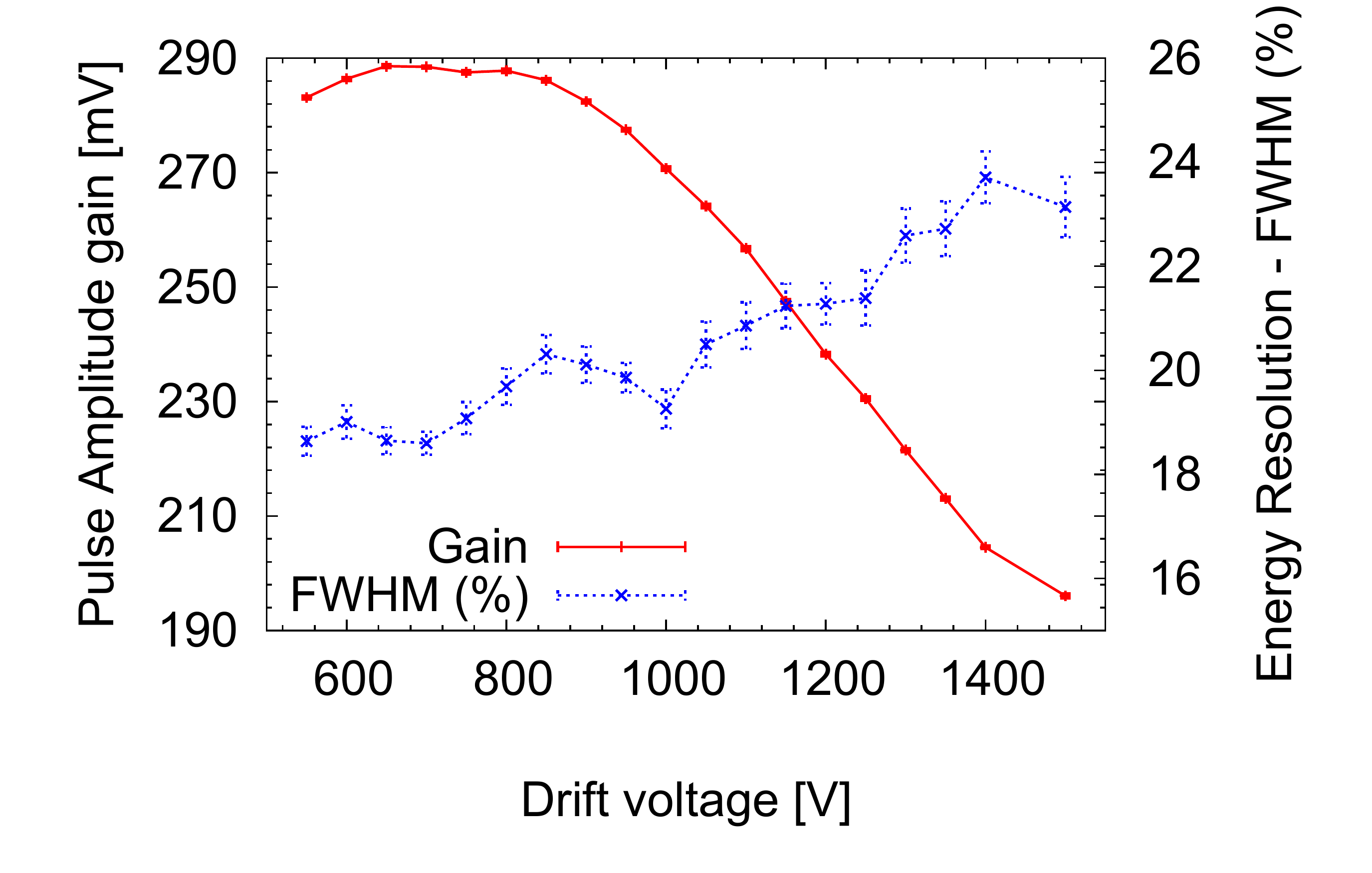} &
\includegraphics[angle=270,width=6.7cm]{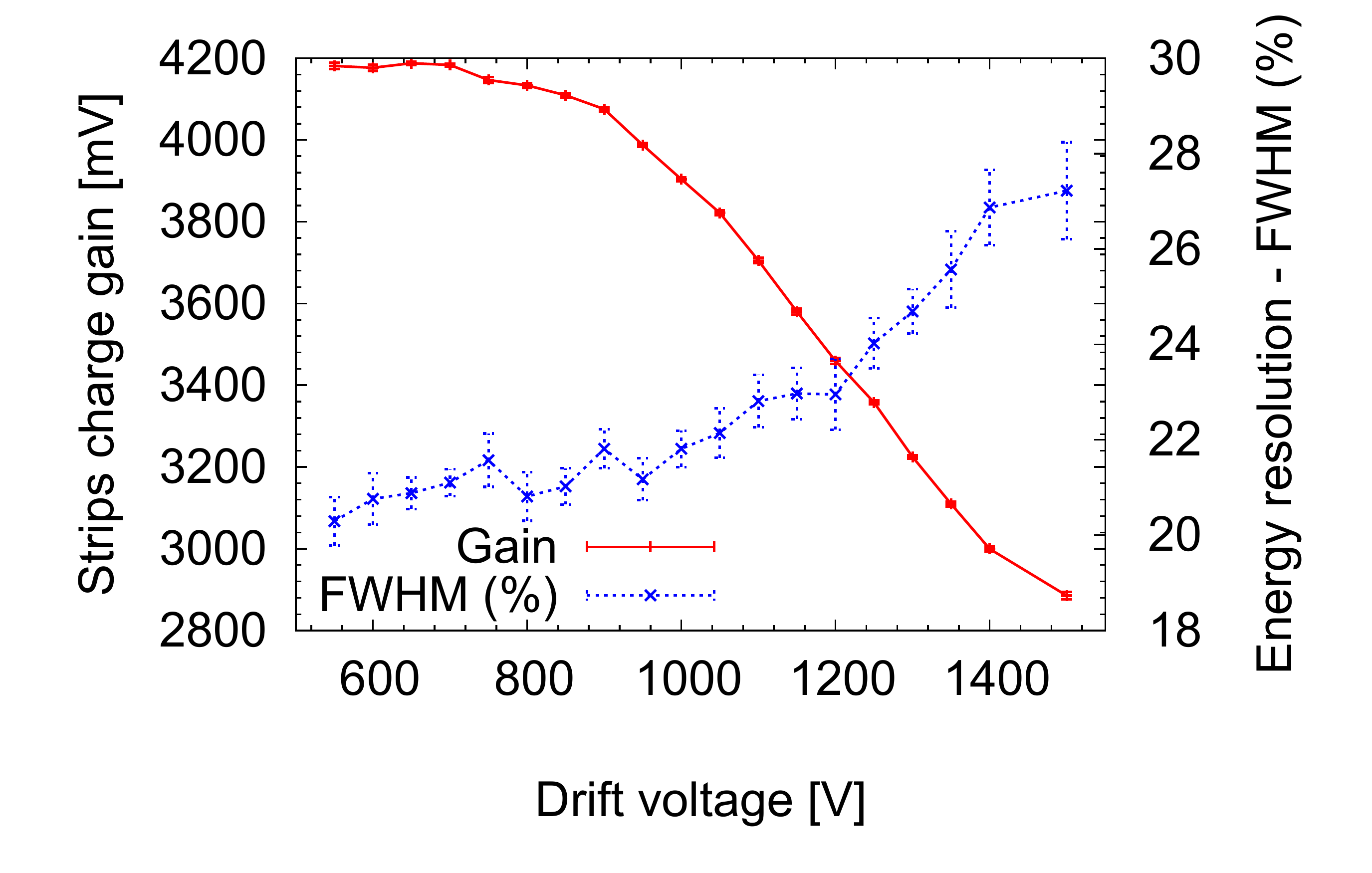} \\
\end{tabular}

\vspace{-0.2cm}
\caption{\fontfamily{ptm}\selectfont{\normalsize{ Relative gain and energy resolution obtained from pulse amplitude and strips charge at the $5.9$\,keV peak for different drift voltage settings. }}}
\label{Drift21}
\end{center}\end{figure}

The effect on the gain due to the sequentially increased drift voltage is presented in figure~\ref{Drift21}, together with the energy resolution represented at the auxiliary axis. The gain curve describes the typical loss of transparency behavior related with the lower drift/amplification field ratio that reduces the number of electrons that pass through the mesh to the amplification region. Fact that is directly related with the worsening of the energy resolution at higher drift voltages, having a bigger effect on the strips collected energy due to the independent electronic noise added by each strip composing the cluster.

\vspace{-0.6cm}

\vspace{-0.6cm}

\begin{figure}[!ht]\begin{center}
\includegraphics[angle=270,width=6.7cm]{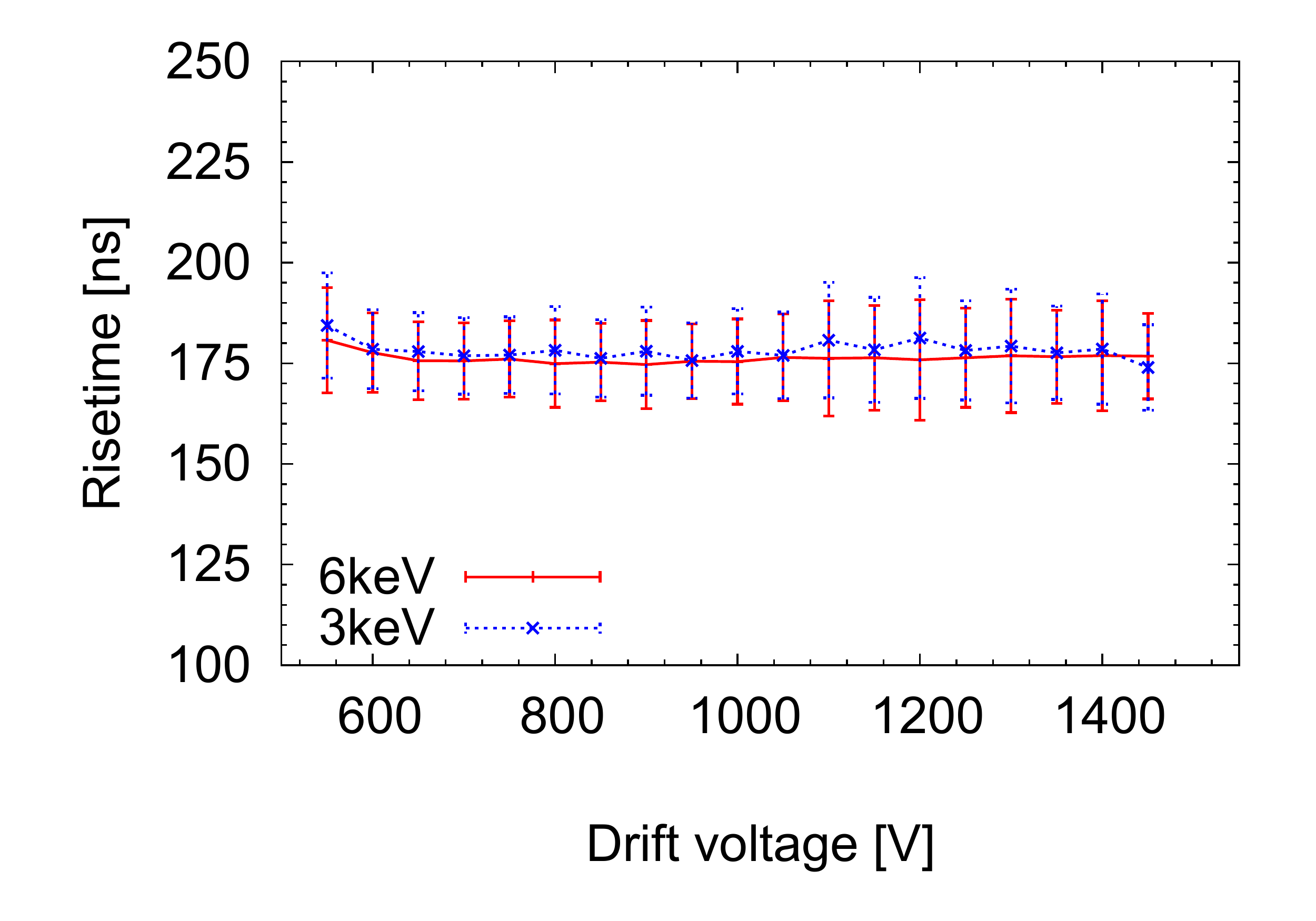}
\includegraphics[angle=270,width=6.7cm]{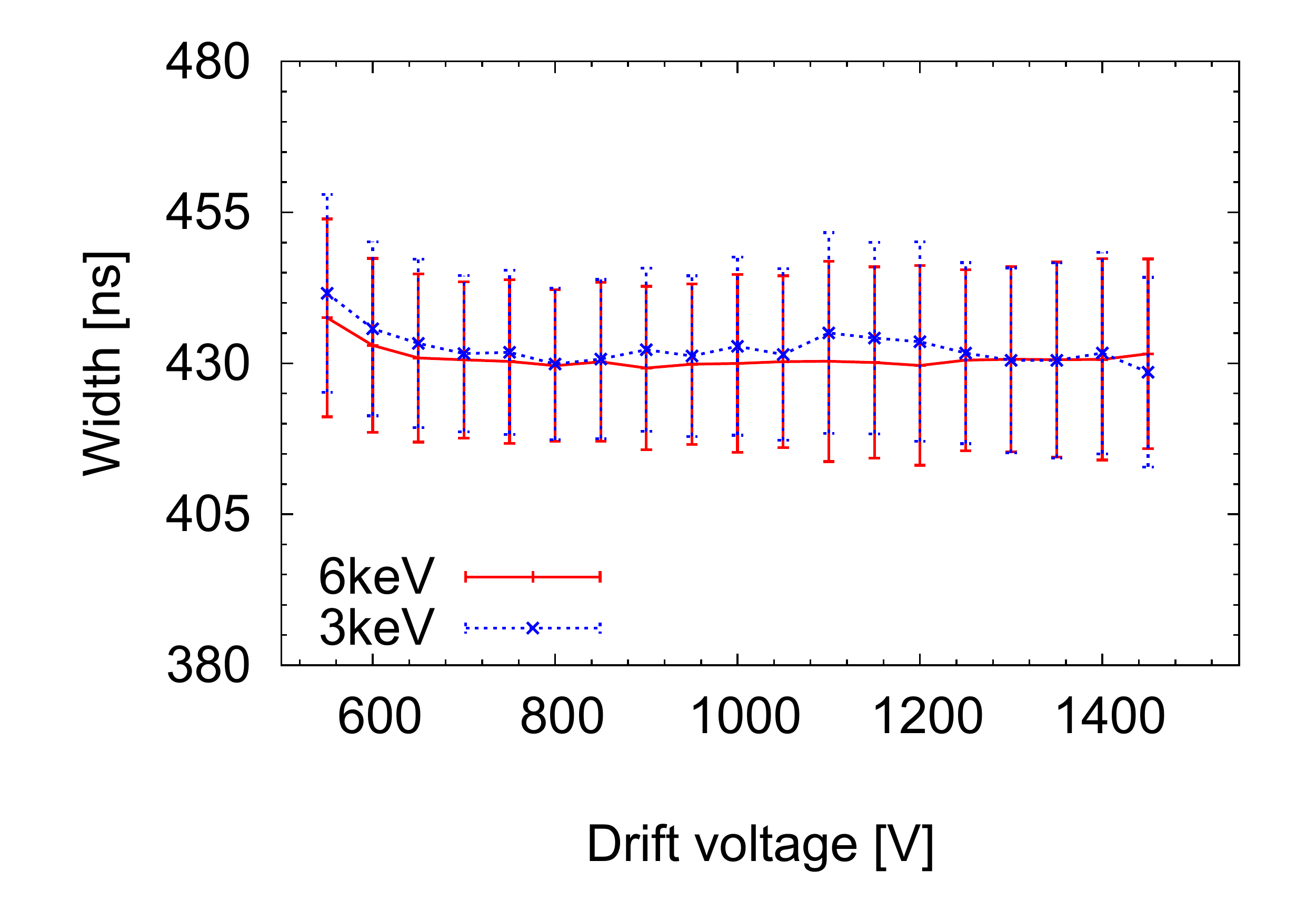}

\caption{\fontfamily{ptm}\selectfont{\normalsize{  Pulse risetime and width evolution at different drift voltage settings. }}}
\label{Drift22}
\end{center}\end{figure}

\begin{figure}[!ht]\begin{center}
\begin{tabular}{cc}
\includegraphics[angle=270,width=6.45cm]{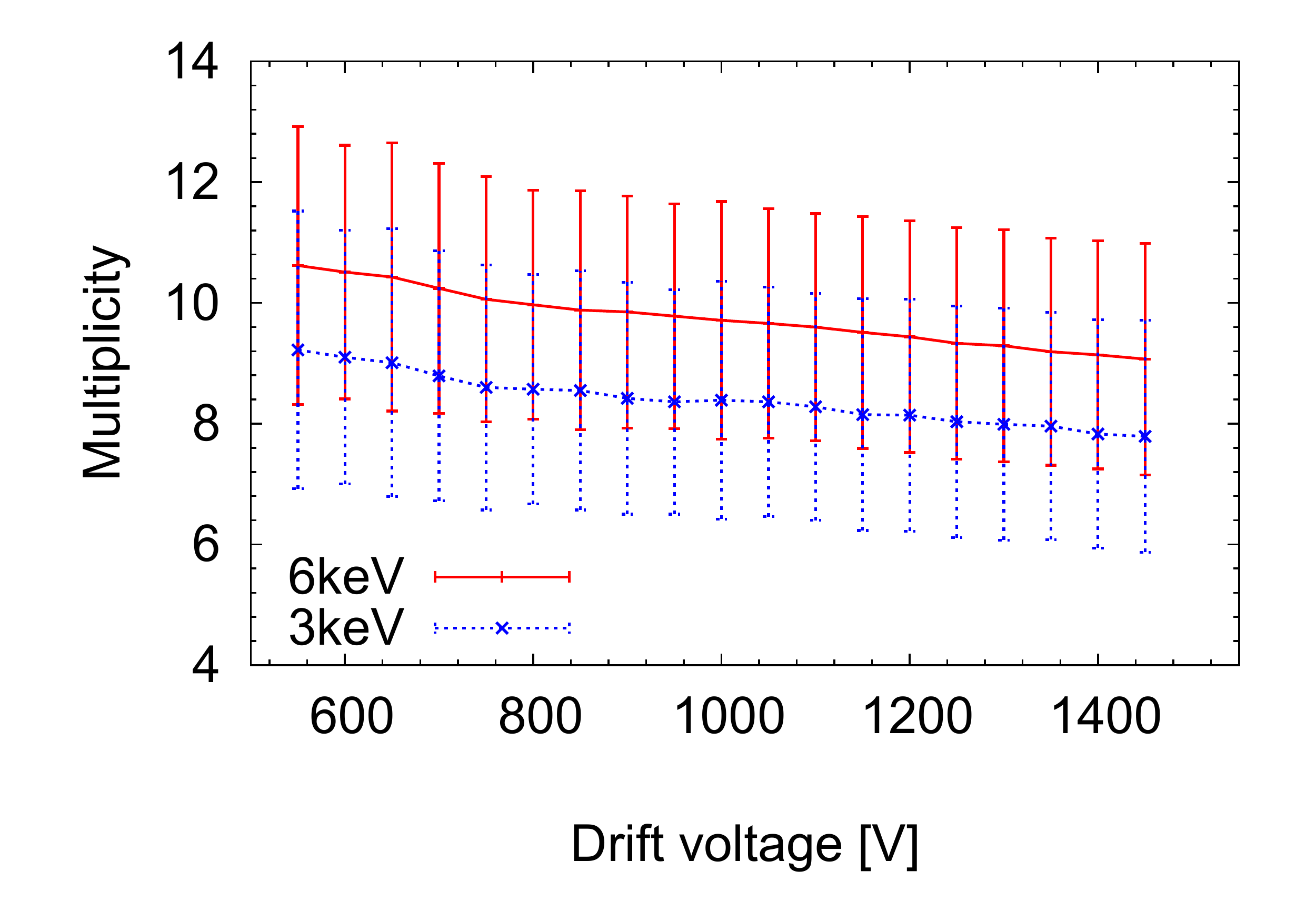} &
\includegraphics[angle=270,width=6.45cm]{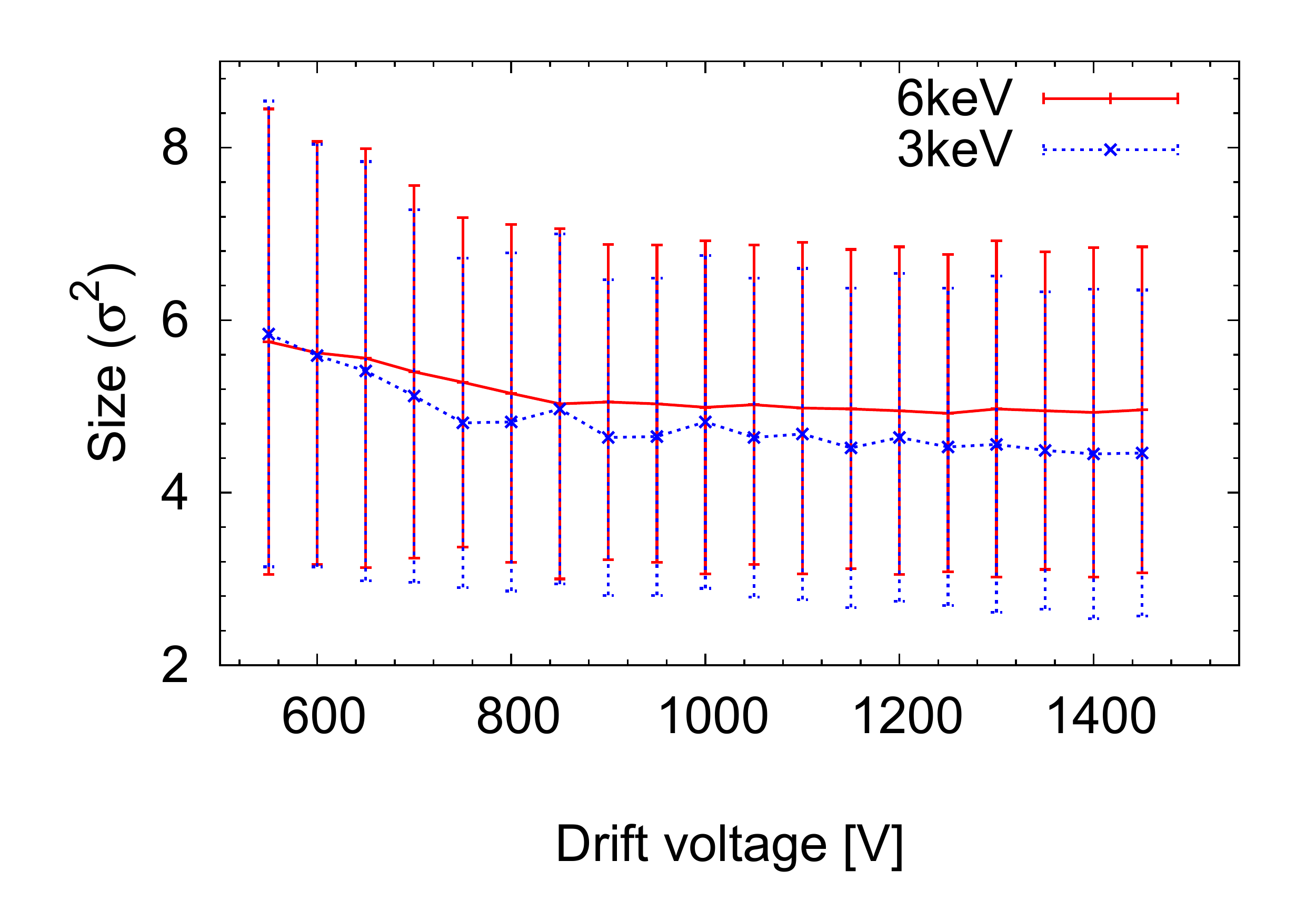} \\
\end{tabular}

\caption{\fontfamily{ptm}\selectfont{\normalsize{ Multiplicity and size in strip units as a function of the drift voltage.}}}
\label{Drift23}
\end{center}\end{figure}

\begin{figure}[!ht]\begin{center}
\includegraphics[width=0.96\textwidth]{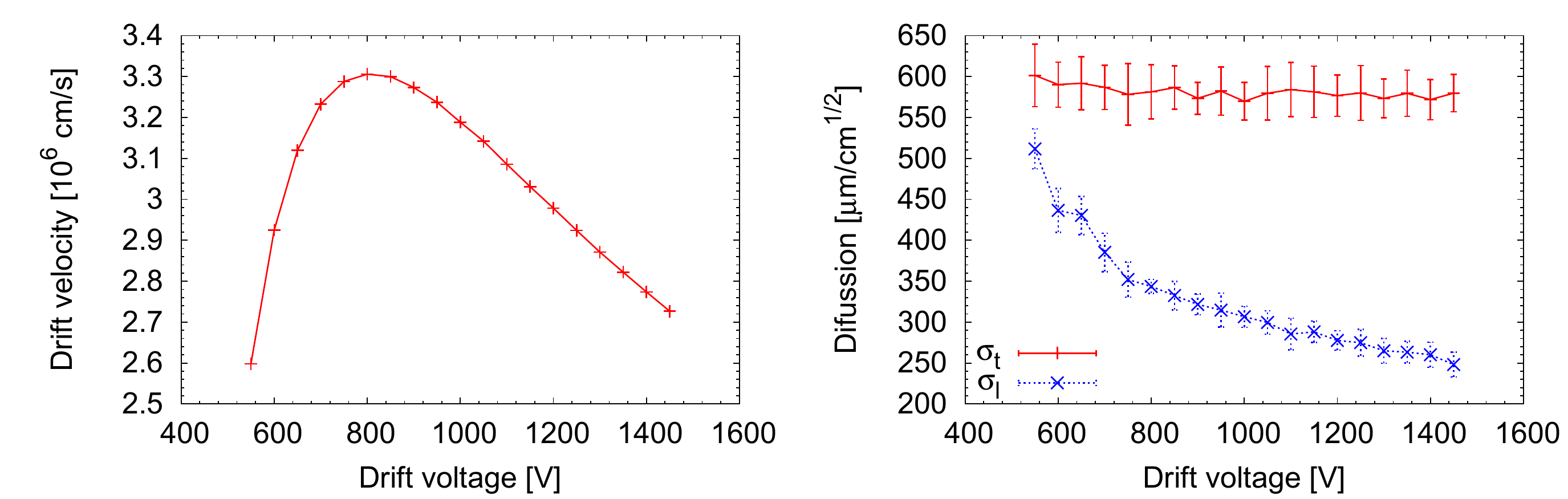}
\caption{\fontfamily{ptm}\selectfont{\normalsize{ Drift velocity (left) and longitudinal (dashed line) and transversal (continous line) difussion (right) for the corresponding drift measured settings (obtained with Magboltz~\cite{Biagi:1018382}).  }}}
\label{fi:diffusionDrift}
\end{center}\end{figure}

Figures~\ref{Drift22}~and~\ref{Drift23} show the effect of the increasing drift voltage on \emph{pulse risetime}, \emph{pulse width}, \emph{cluster multiplicity} and \emph{cluster size}. The pulse parameters do not show a big dependency on drift field except for the extremely low drift voltage applied (below $650$\,V) where the effect on these parameters coming from a higher longitudinal charge diffusion and smaller drift velocity (see Fig.~\ref{fi:diffusionDrift}) can be observed. The \emph{cluster multiplicity} decreases as the drift field increases keeping a close correlation with the reduced gain due to transparency loss, since fewer charges are reaching the strips readout the number of strips becoming active is lower. However, the cluster size, which only depends on the shape of the cluster, is not much affected by the lower gains and keeps a better correlation with the transversal charge diffusion which is lower as the drift field increases.


\vspace{-0.2cm}

\vspace{-1.8cm}

\vspace{-1.4cm}



\subsubsection{Mesh voltage measurements}
Detector behavior at different mesh voltages was also analyzed. The gas pressure for these runs was $1.4$\,bar and $2.5$\,\% of isobutane concentration. The drift voltage was kept constant during these measurements at $V_d = 1043$\,V.

\vspace{0.2cm}

\begin{figure}[!ht]\begin{center}
\begin{tabular}{cc}
\includegraphics[angle=270,width=0.48\textwidth]{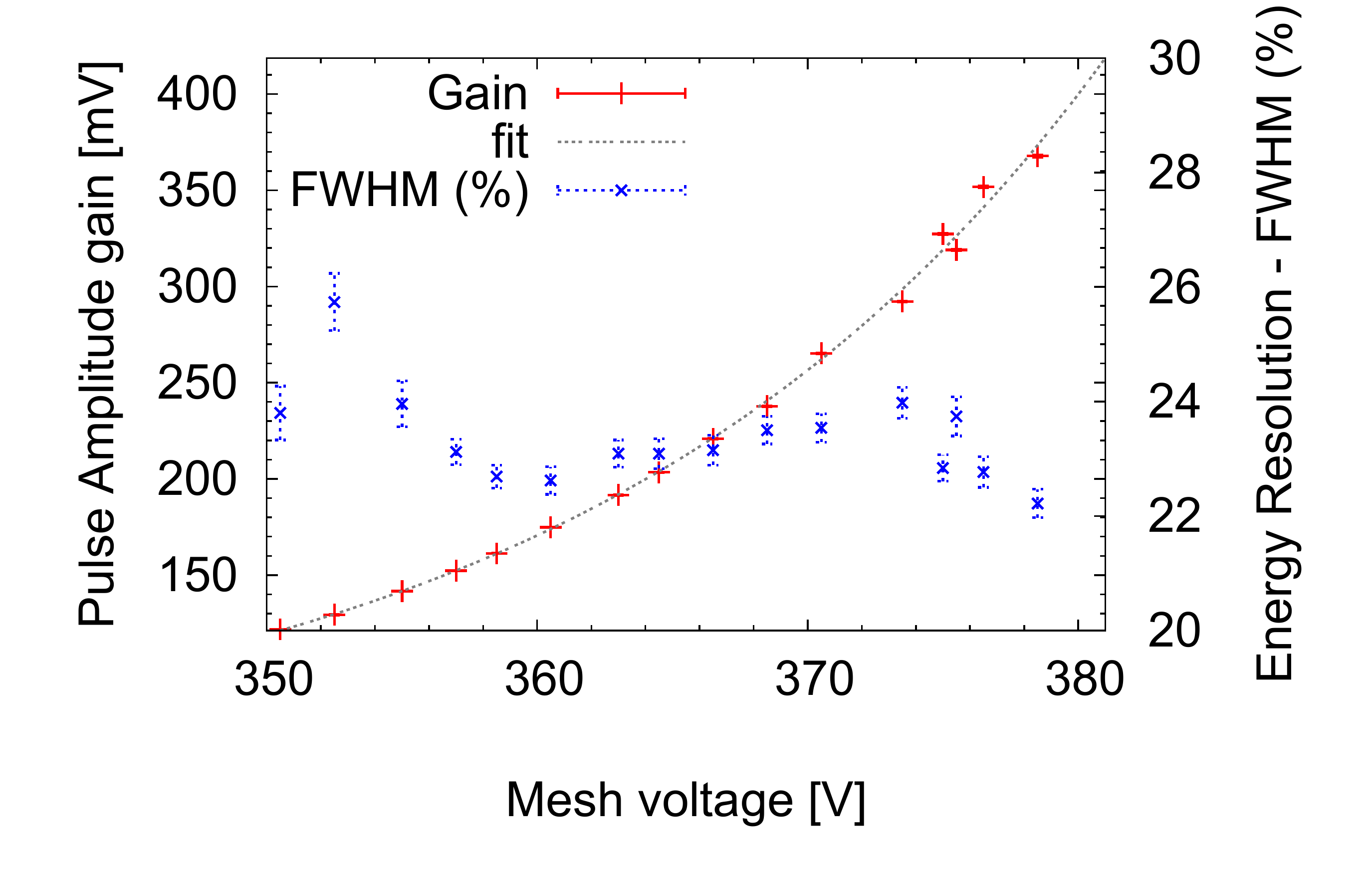} &
\includegraphics[angle=270,width=0.48\textwidth]{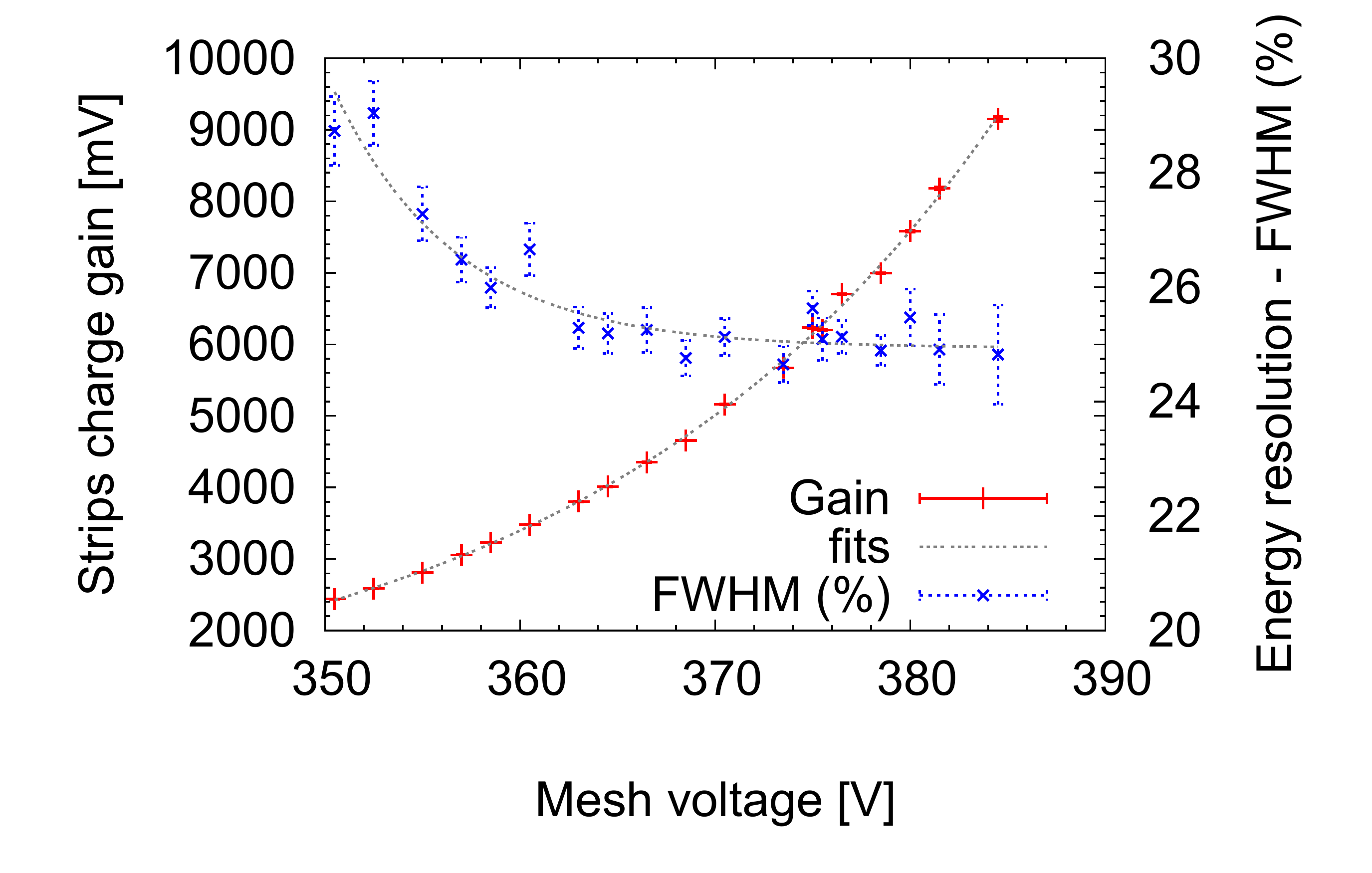} \\
\end{tabular}

\caption{\fontfamily{ptm}\selectfont{\normalsize{ Evolution of relative gain and energy resolution at different mesh voltage settings. The energy is obtained from pulse amplitude and the strips clusters charge collected. }}}
\label{Mesh1}
\end{center}\end{figure}

The relative gain signal measured with the mesh pulse and strips charge shows an expected exponential dependency with the mesh voltage (see Fig.~\ref{Mesh1}). In order to parameterize the detector gain relative to the main parameters providing the event energy as a function of the applied mesh voltage $V_m$ the following expression has been used

\begin{equation}
g(V_m) = g_{ref} + \alpha \cdot \left( exp( (V_m - V_{ref})/\gamma ) - 1 \right)
\end{equation}

\vspace{0.2cm}

\noindent where $g_{ref}$ is the parameters gain at the reference voltage $V_{ref} = 350$\,V. The obtained parameters to this expression are presented in table~\ref{ta:meshFits} using as reference the $5.9$\,keV peak.

\begin{table}[!hbt]
\begin{center}
\begin{tabular}{c|ccc}
Source  & 	{\bf$g_{ref}$}	&	$\alpha$ [V]	&  $\gamma$[V] \\
	&	&	&	\\
\hline
	&	&	&	\\
Pulse amplitude & 119.2\,mV (2.7\%) & 76.7824 (17.4\%) & 19.5056 (9.1\%) \\
	&	&	&	\\
Pulse integral  & 52991\,mV$\cdot$ns (2.7\%) & 30425.0 (18.0\%) & 18.8862 (9.3\%) \\
	&	&	&	\\
Cluster signal  & 2385.4\,mV (1.7\%) & 1695.22 (141.4\%) & 21.3941 (4.2\%) \\
\end{tabular}

\vspace{0.2cm}
\caption{\fontfamily{ptm}\selectfont{\normalsize{Relation of fitting parameters for \emph{pulse amplitude} energy, \emph{pulse integral} energy and \emph{strips clusters} energy.}}}
\label{ta:meshFits}
\end{center}
\end{table}

An improvement in the energy resolution is observed as the mesh voltage is higher, fact that is specially strong in the case of the energy obtained from the strips cluster, related with the benefit of the increased signal to noise due to the higher amount of electrons inducing charges in each strip. The improvement on the mesh is not so important since it already provides a good time integrated signal that is much higher than the noise.

\vspace{0.2cm}

\begin{figure}[!ht]\begin{center}
\includegraphics[angle=270,width=0.49\textwidth]{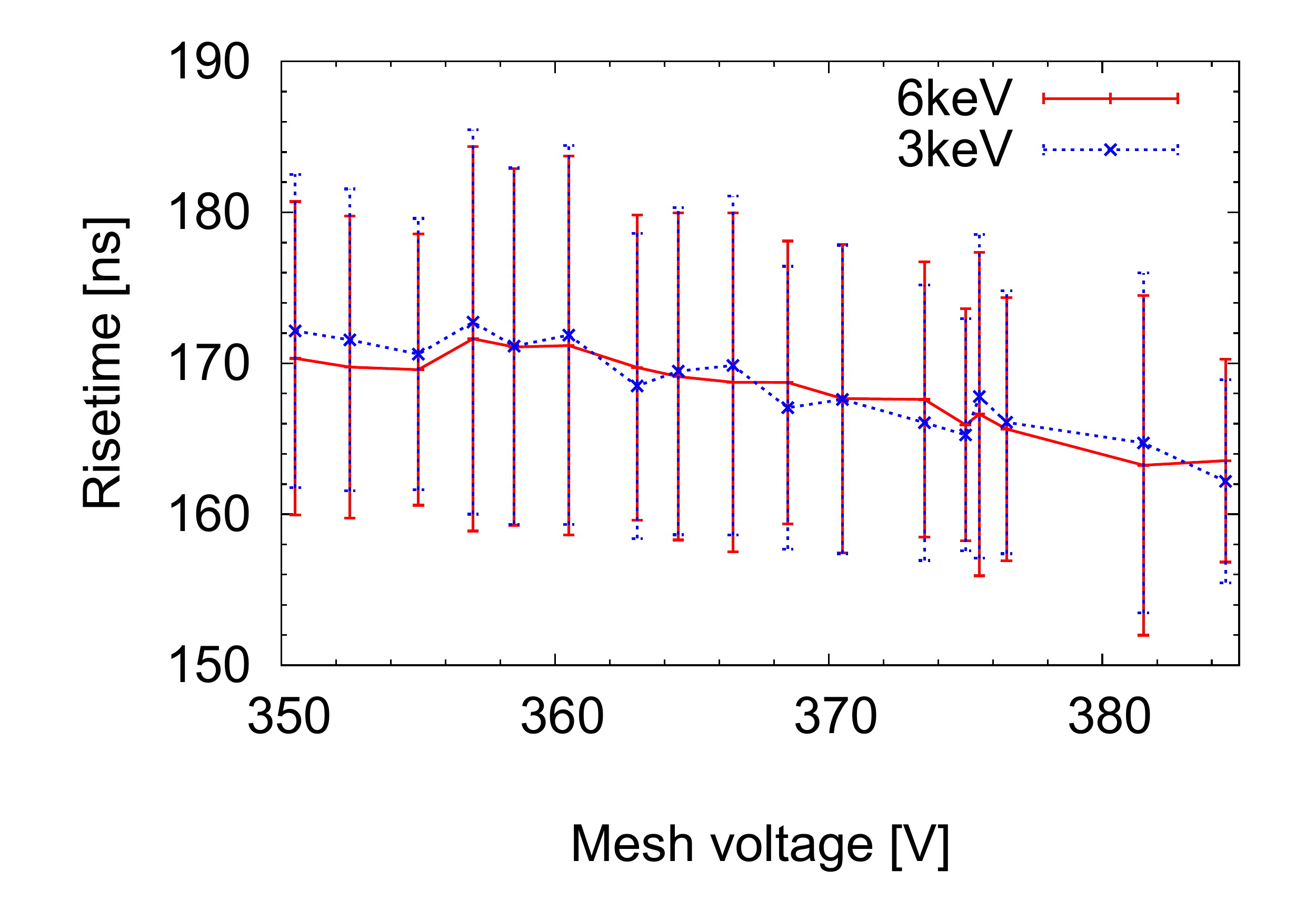}
\includegraphics[angle=270,width=0.49\textwidth]{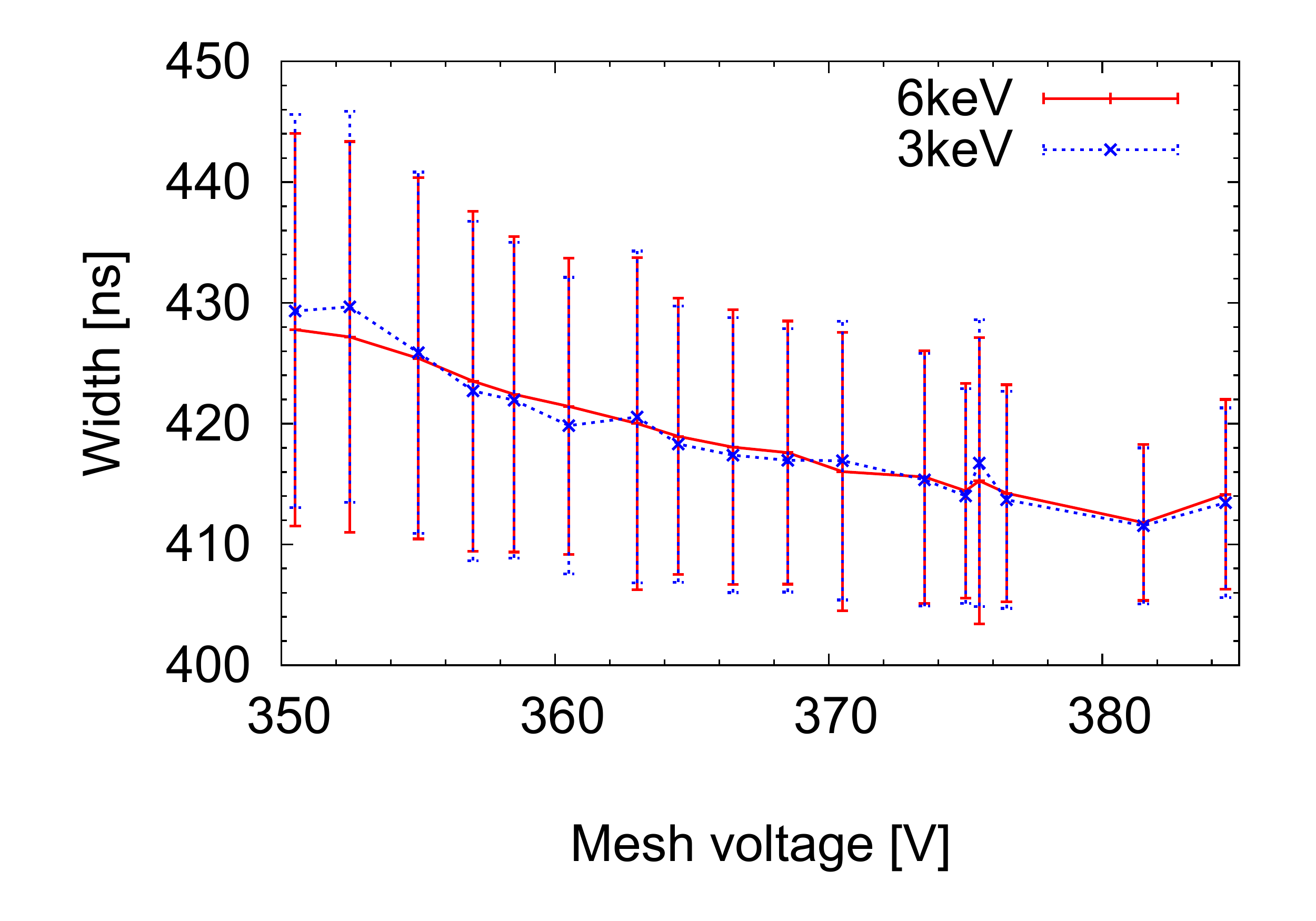}

\caption{\fontfamily{ptm}\selectfont{\normalsize{ Pulse risetime and width evolution at different mesh voltage settings. }}}
\label{Mesh2}
\end{center}\end{figure}

\begin{figure}[!ht]\begin{center}
\includegraphics[angle=270,width=0.49\textwidth]{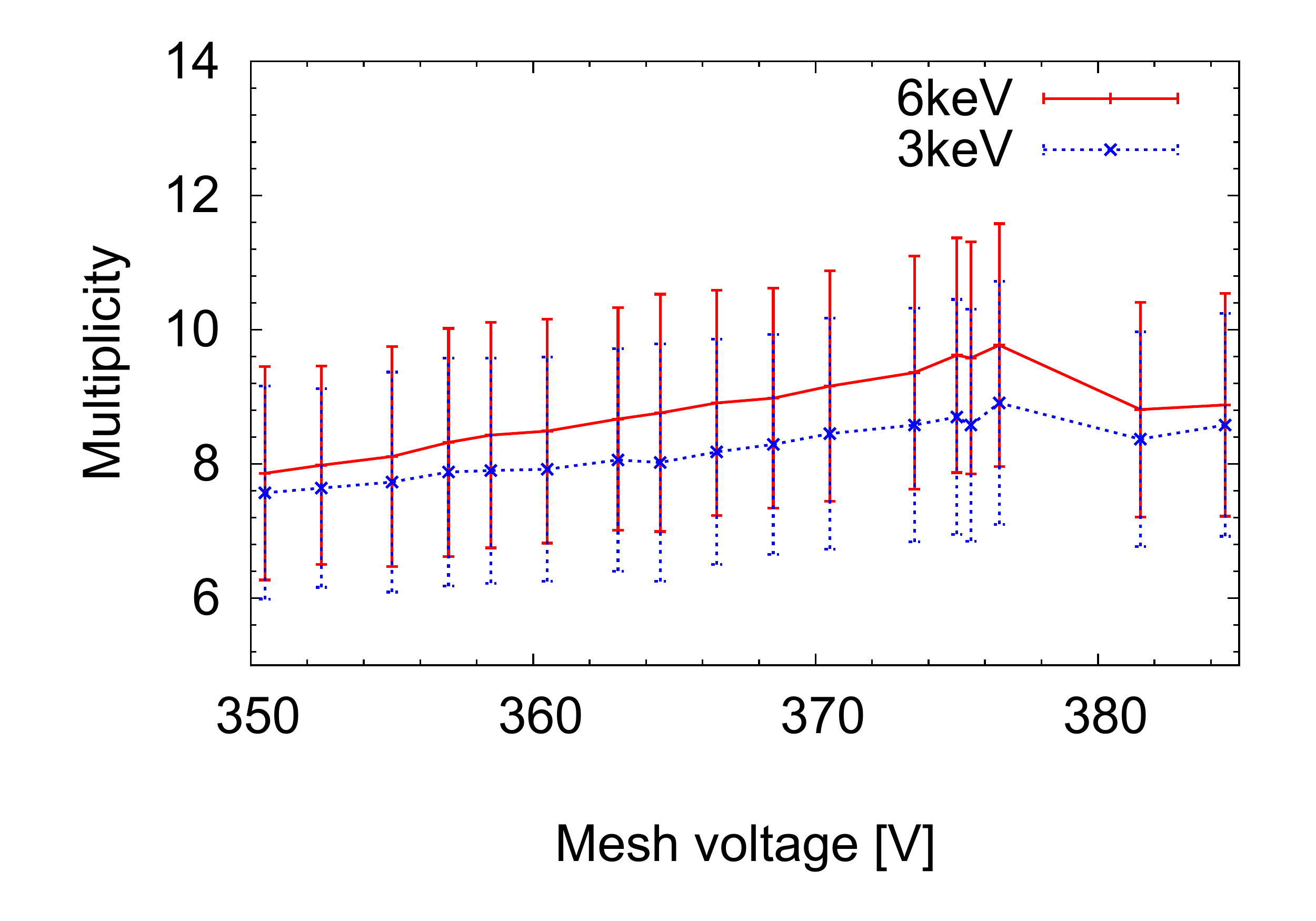}
\includegraphics[angle=270,width=0.49\textwidth]{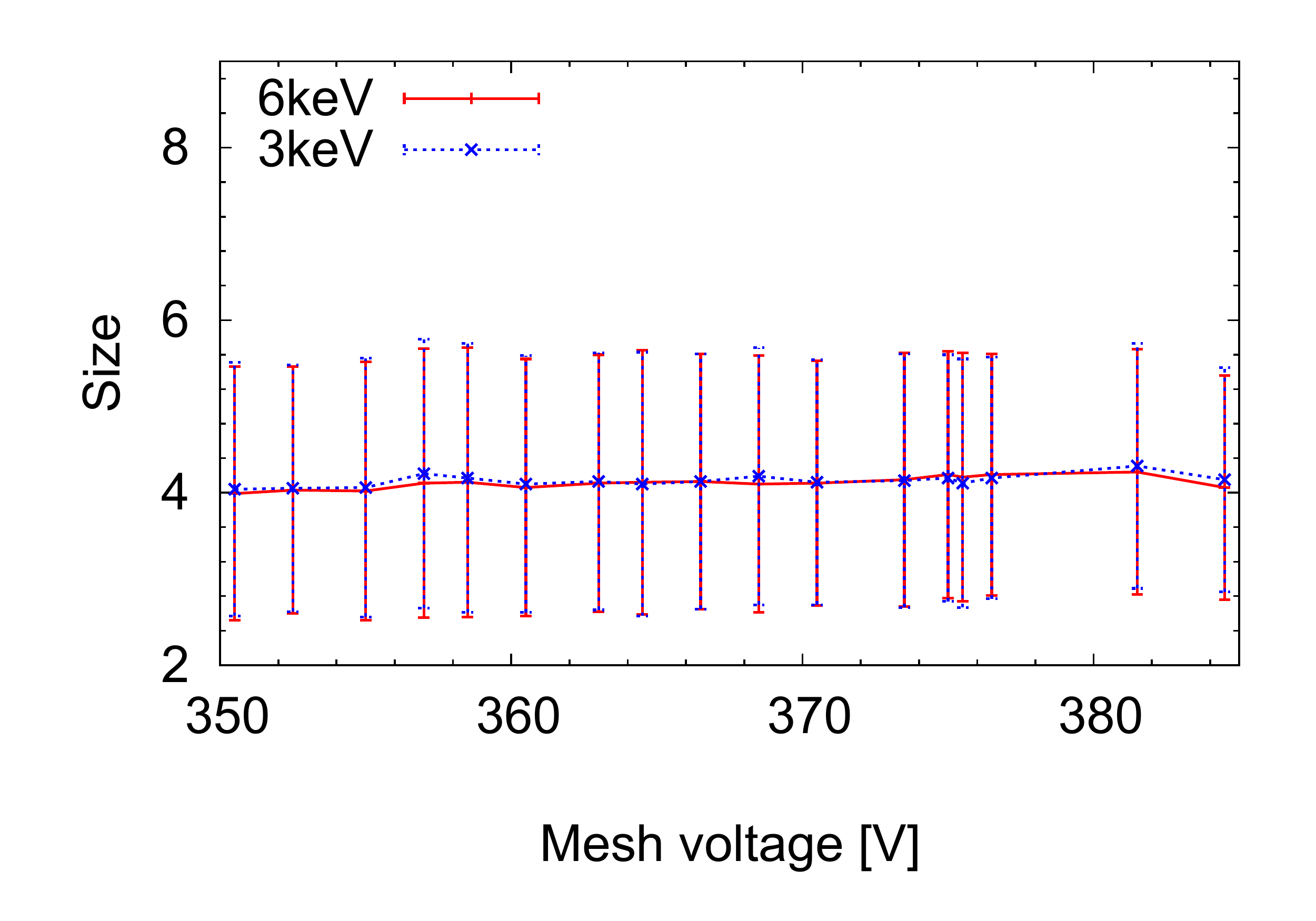}

\caption{\fontfamily{ptm}\selectfont{\normalsize{ Multiplicity and cluster size ($\sigma^2$) evolution in strip units as a function of the mesh voltage applied, for $3$\,keV and $6$\,keV events. }}}
\label{Mesh3}
\end{center}\end{figure}

Figure~\ref{Mesh2} shows the pulse parameters dependency on the mesh voltage applied at these measurements. The diffusion effect should be negligible due to the small decrease of drift field due to the mesh voltage increase. The lower risetime and width values obtained for higher amplification fields are necessarily related with the faster arrival of ions to the mesh plane. Figure~\ref{Mesh3} shows the main strip parameters, \emph{cluster multiplicity} and \emph{cluster size}, where it is observed the growing tendency of multiplicity with the increased amplification field due to the higher amount of charges involved. However, the \emph{cluster size} remains constant since the diffusion of charges it is not affected notably.

%

\subsubsection{Isobutane concentration measurements}

\begin{figure}[!ht]\begin{center}
\includegraphics[angle=270,width=0.7\textwidth]{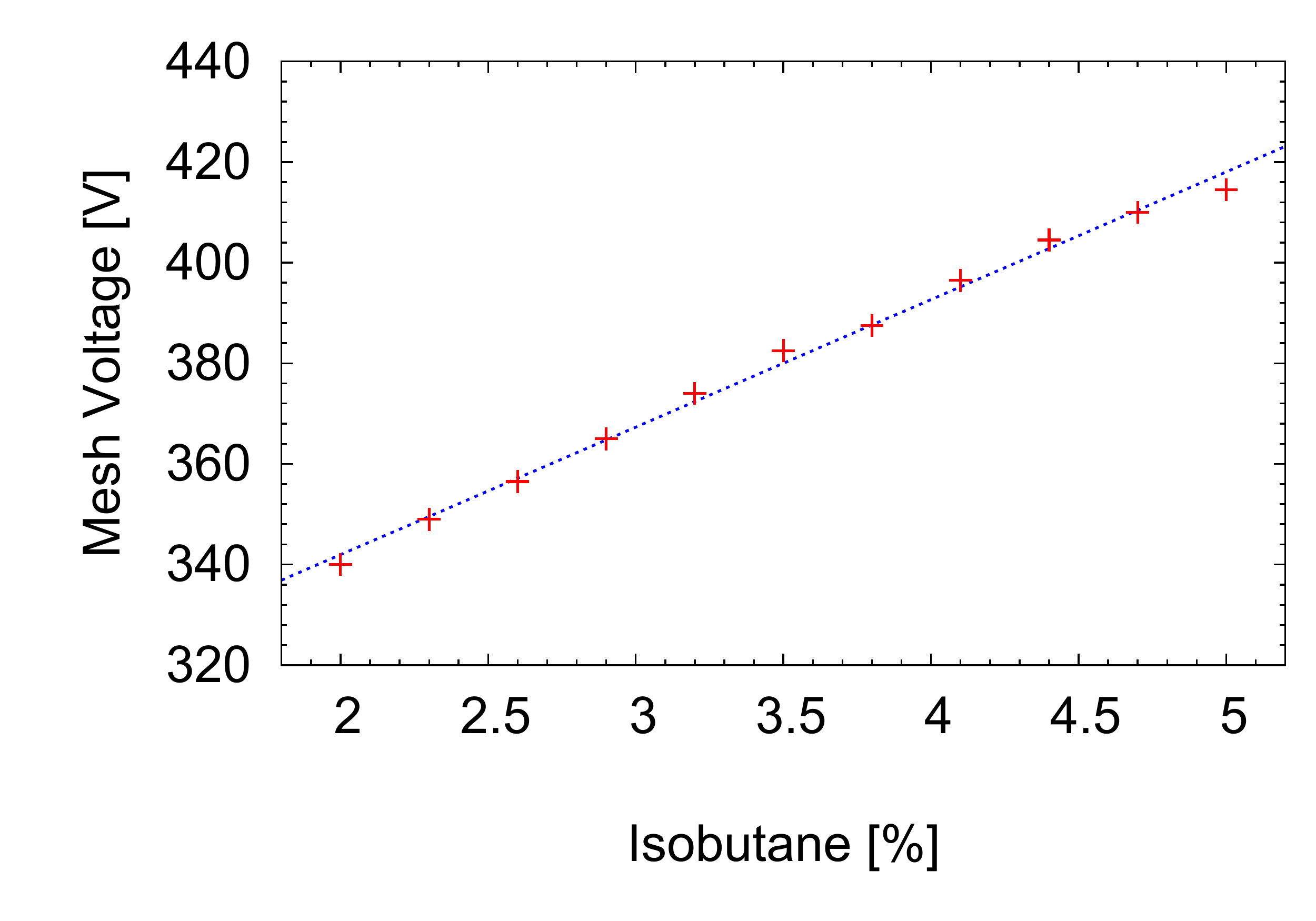}
\caption{\fontfamily{ptm}\selectfont{\normalsize{ Evolution of the mesh voltage needed during isobutane concentration measurements for keeping a constant gain.   }}}
\label{IsoVm}
\end{center}\end{figure}

Measurements at different Ar+iC$_4$H$_{10}$ concentrations were taken at a pressure of $1.4$\,bar keeping the drift field constant at $E_{drift} = 212$\,V/cm. The mesh voltage was readjusted during these measurements in order to keep always the same relative gain using as reference the mesh pulse amplitude, the timing amplifier settings remained untouched at x2/7/100ns/200ns to maintain the mesh signal to noise ratio. Figure~\ref{IsoVm} shows the required mesh voltage to keep a constant gain for the different argon-isobutane mixtures showing a linear dependency in isobutane concentration. The required voltage increase in order to keep a constant gain was $\Delta V = 25.4$\,V per $1$\,\% isobutane concentration increase.

\begin{figure}[!ht]\begin{center}
\includegraphics[angle=270,width=0.48\textwidth]{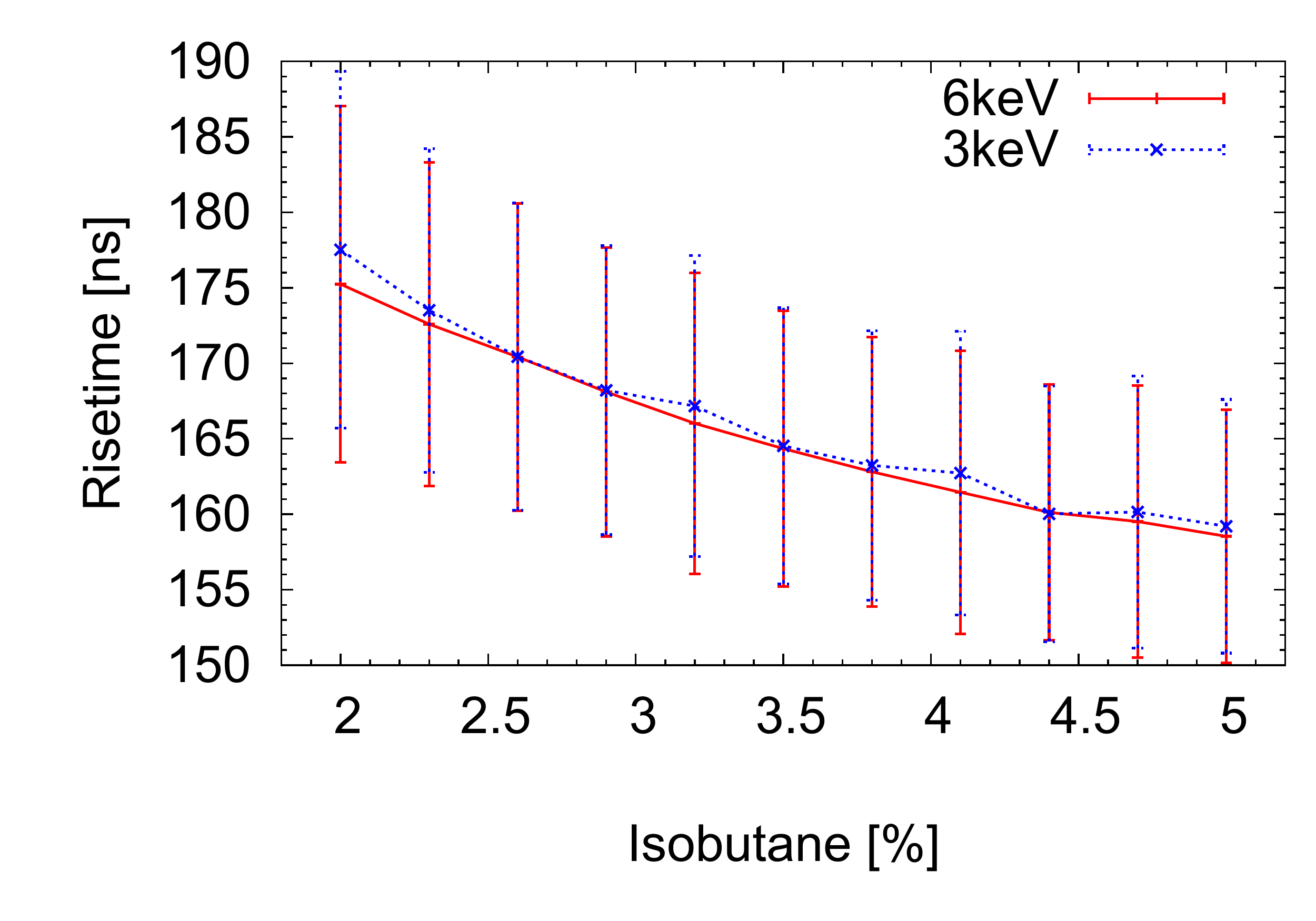}
\includegraphics[angle=270,width=0.48\textwidth]{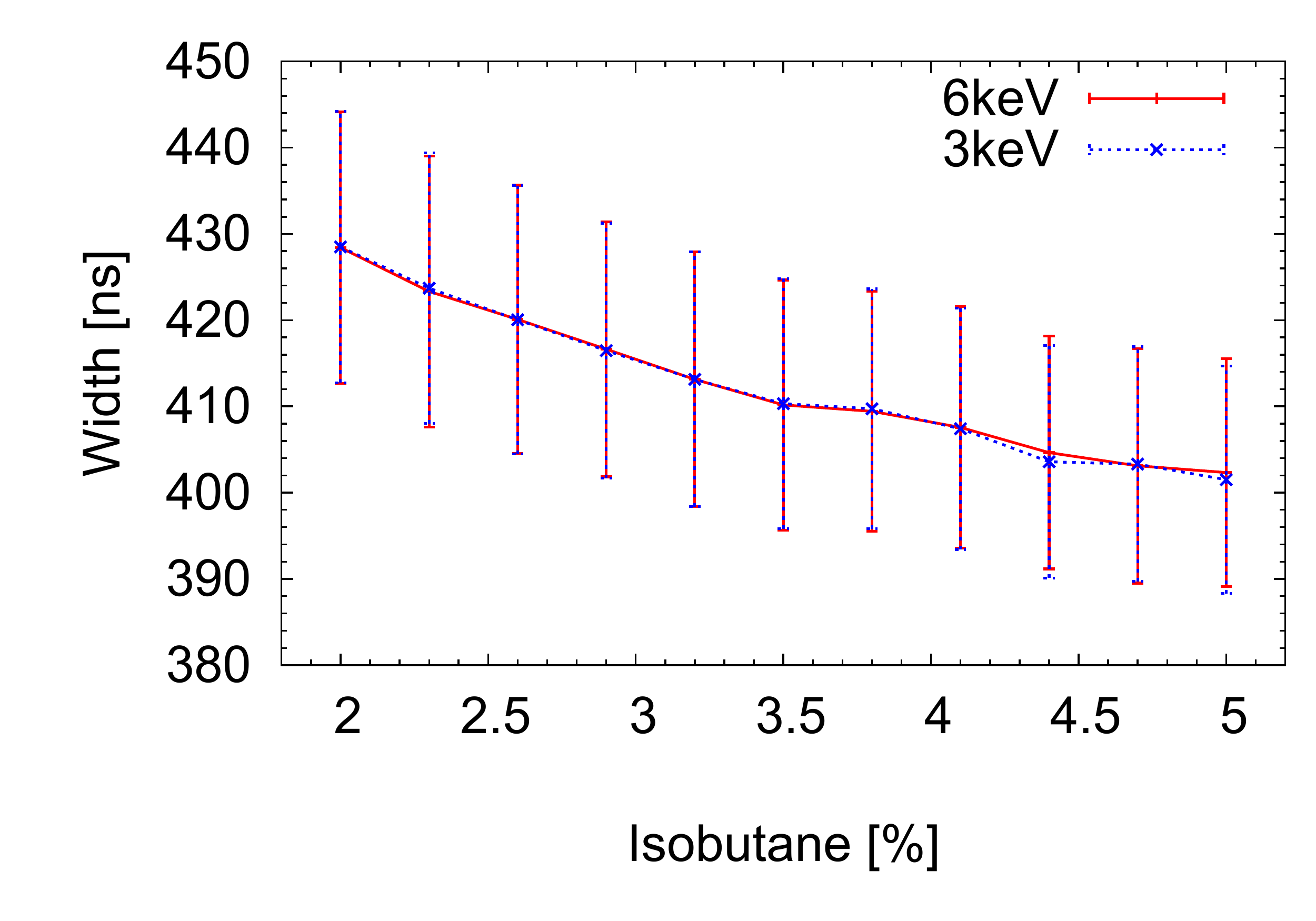}

\caption{\fontfamily{ptm}\selectfont{\normalsize{ Pulse risetime and width evolution at different isobutane concentrations, for 3\,keV and 6\,keV. }}}
\label{Iso2}
\end{center}\end{figure}

\begin{figure}[!ht]
\includegraphics[width=\textwidth]{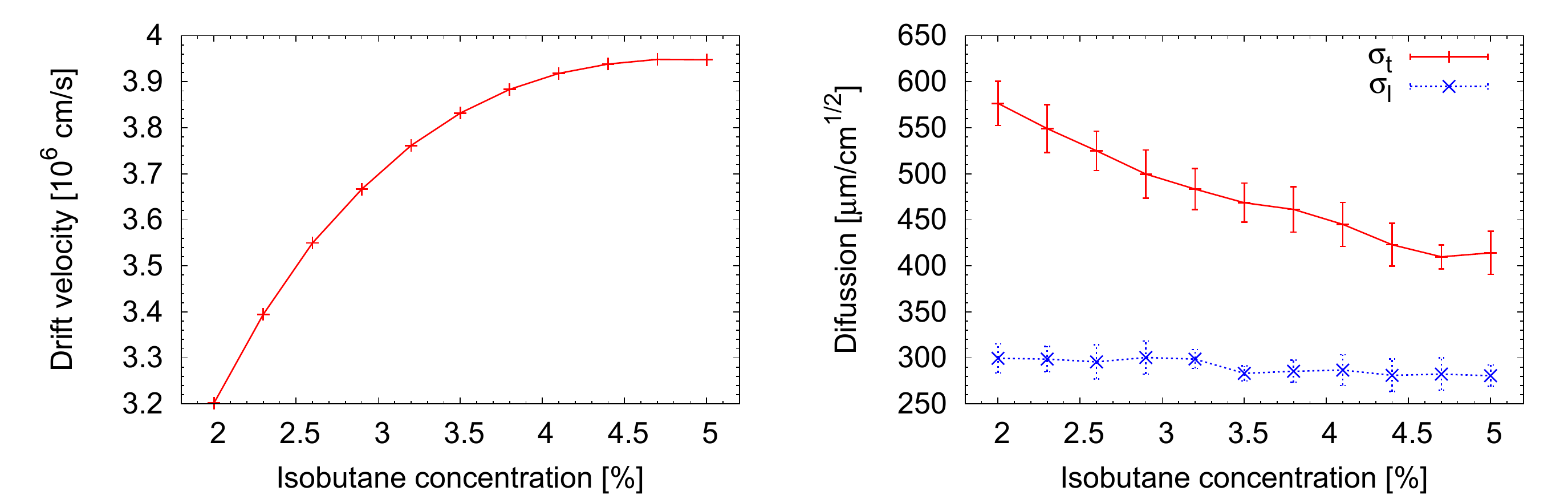}
\caption{\fontfamily{ptm}\selectfont{\normalsize{ Drift velocity (left) and longitudinal (represented with blue dashed line) and transversal (represented with red continous line) diffusion (right) for the different isobutane concentrations (data generated with Magboltz). }}}
\label{fi:diffusionIsobutane}
\end{figure}

In these measurements the \emph{pulse risetime} and \emph{pulse width} decrease for higher isobutane concentrations (see Fig.~\ref{Iso2}). Since the longitudinal diffusion does not change significatively for different isobutane concentrations, and the drift velocity increases for higher isobutane concentrations (see Fig.~\ref{fi:diffusionIsobutane}) this effect is probably associated to the higher amplification field required at higher isobutane concentrations, and the higher mobility of ions drifting back to the mesh, thus reducing the amplification process time required. 

\begin{figure}[!ht]\begin{center}
\begin{tabular}{cc}
\includegraphics[angle=270,width=0.47\textwidth]{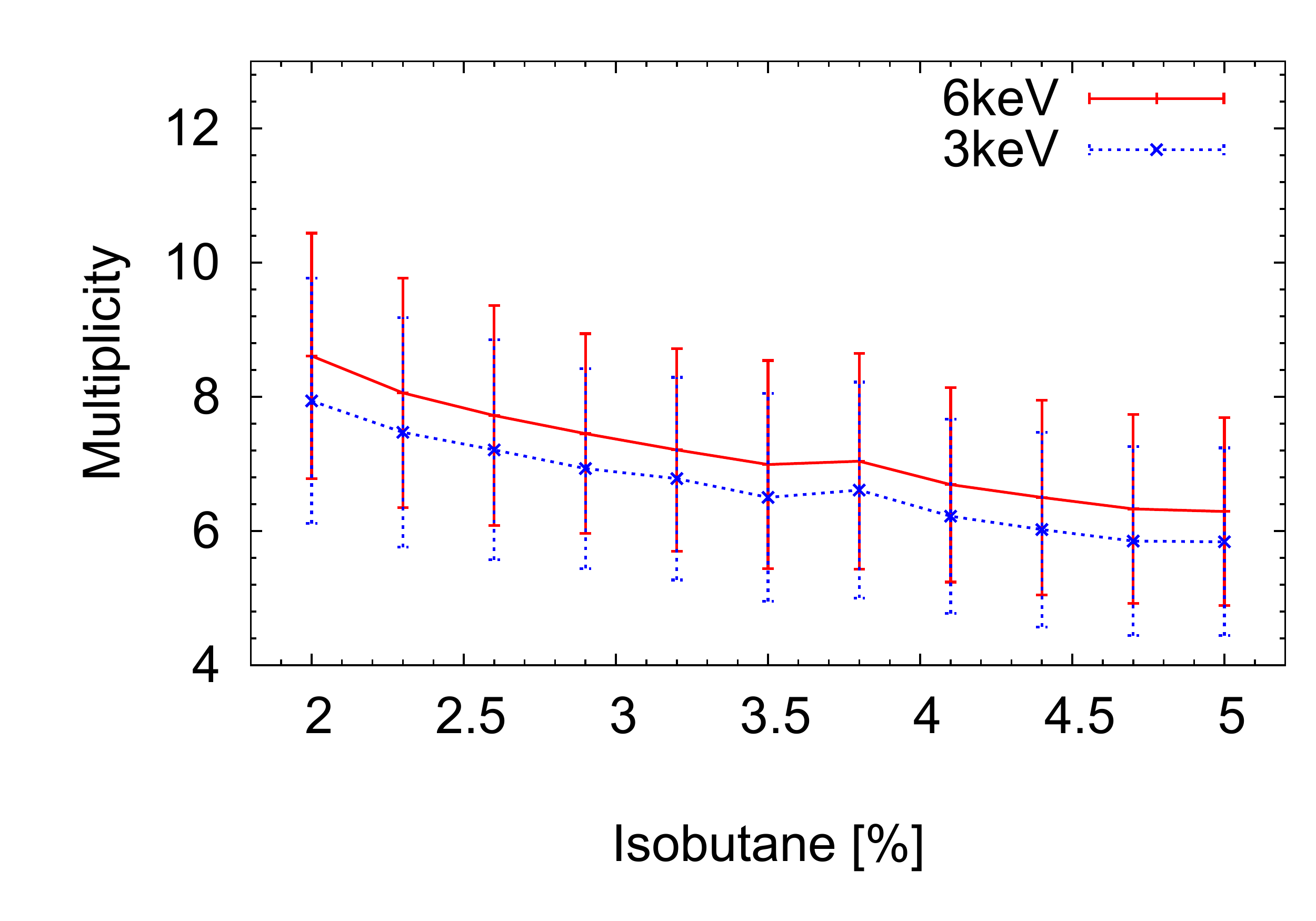} &
\includegraphics[angle=270,width=0.47\textwidth]{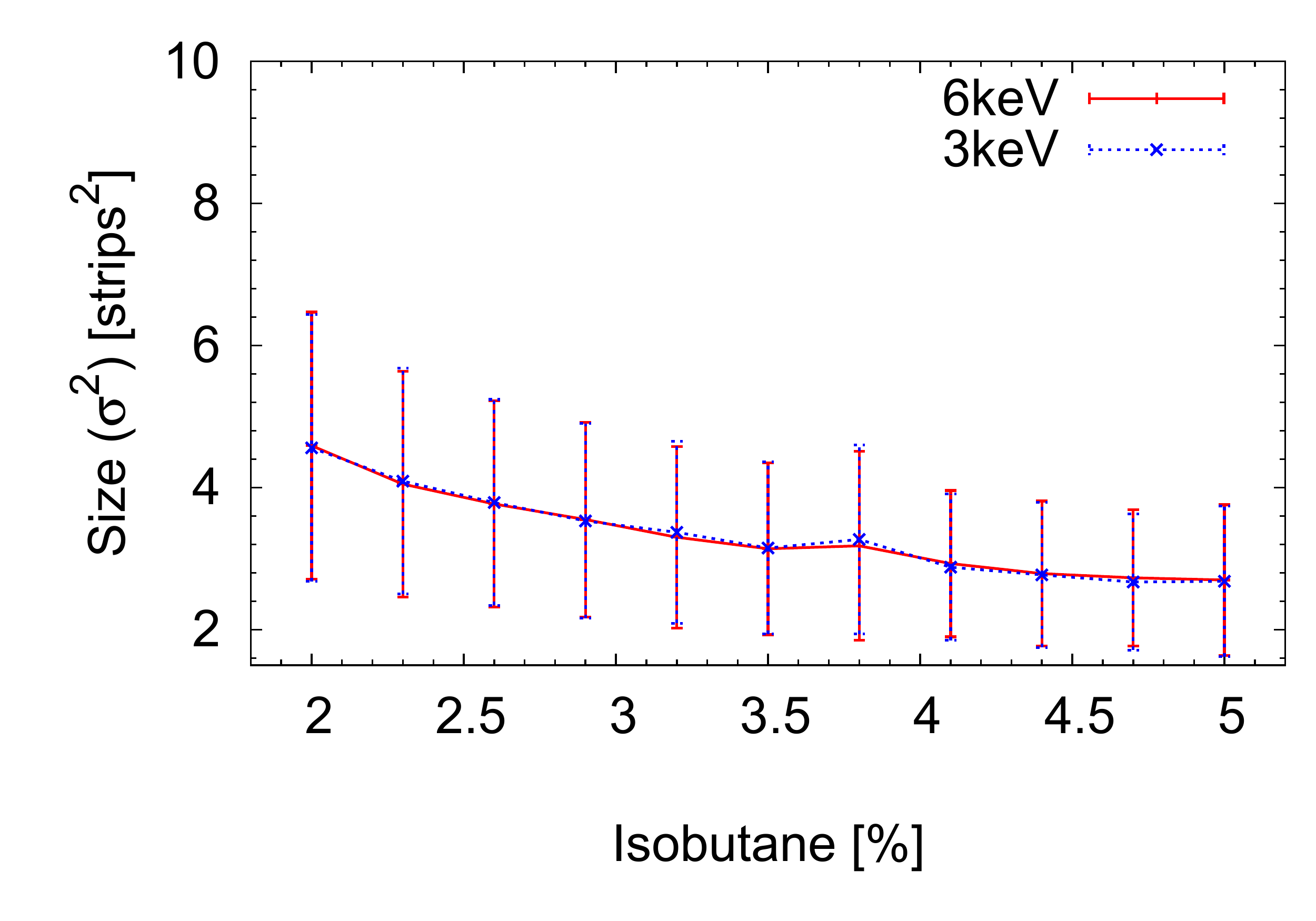} \\
\end{tabular}

\caption{\fontfamily{ptm}\selectfont{\normalsize{ Cluster multiplicity and cluster size evolution as a function of the isobutane concentrations measured, for 3\,keV and 6\,keV events. }}}
\label{Iso3}
\end{center}\end{figure}

\vspace{0.2cm}
During these measurements the \emph{cluster multiplicity} reproduces also the effect of the transversal diffusion since the gain remained constant and the clusters measure a comparable amount of electrons, thus reducing systematic effects on the multiplicity. Furthermore a close correlation between the \emph{cluster multiplicity} and the \emph{cluster size} as a function of the isobutane concentration is observed (see Fig.~\ref{Iso3}).


%

\subsubsection{Pressure evolution measurements}
\begin{figure}[!h]
\begin{center}
\includegraphics[angle=270,width=0.7\textwidth]{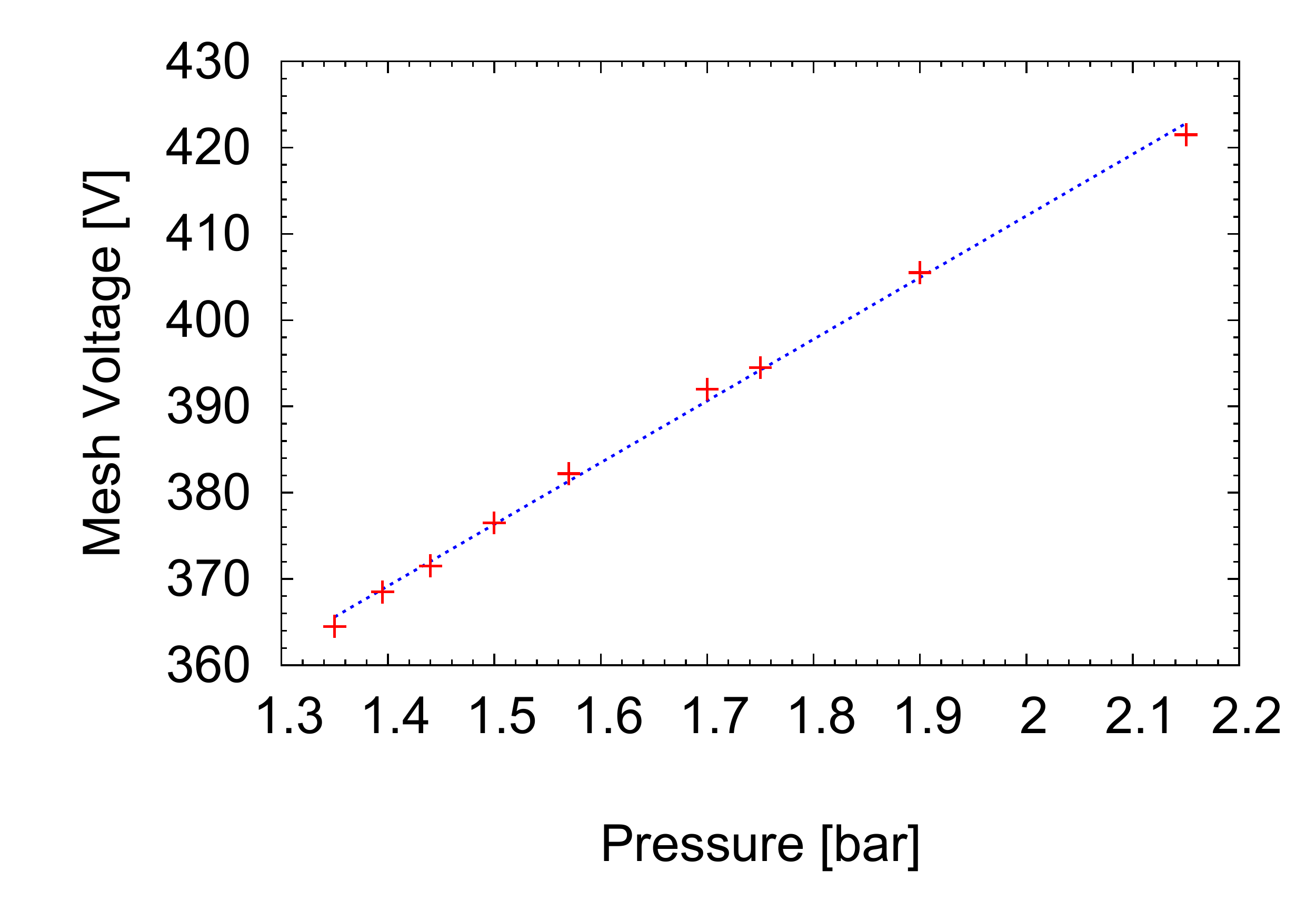}
\end{center}
\caption{\fontfamily{ptm}\selectfont{\normalsize{ Evolution of the mesh voltage needed during pressure measurements for keeping a constant gain.}}}
\label{PrVm}
\end{figure}

Measurements at different pressures were carried out with argon+2.5\%iC$_4$H$_{10}$ gas at a fixed drift voltage $V_d = 1041$\,V. The timing amplifier settings remained untouched x2/6/100ns/200ns, and the mesh voltage was adjusted to keep a constant gain by following the same arguments as in the isobutane measurements. The required voltage increase (see Fig.~\ref{PrVm}) to keep a constant gain was obtained for $\Delta V/\Delta P = 71.5$\,V/mbar, starting at a mesh voltage of $V_m = 340.5$\,V at $1$\,bar.

\vspace{0.2cm}

\begin{figure}[!ht]\begin{center}
\includegraphics[angle=270,width=0.49\textwidth]{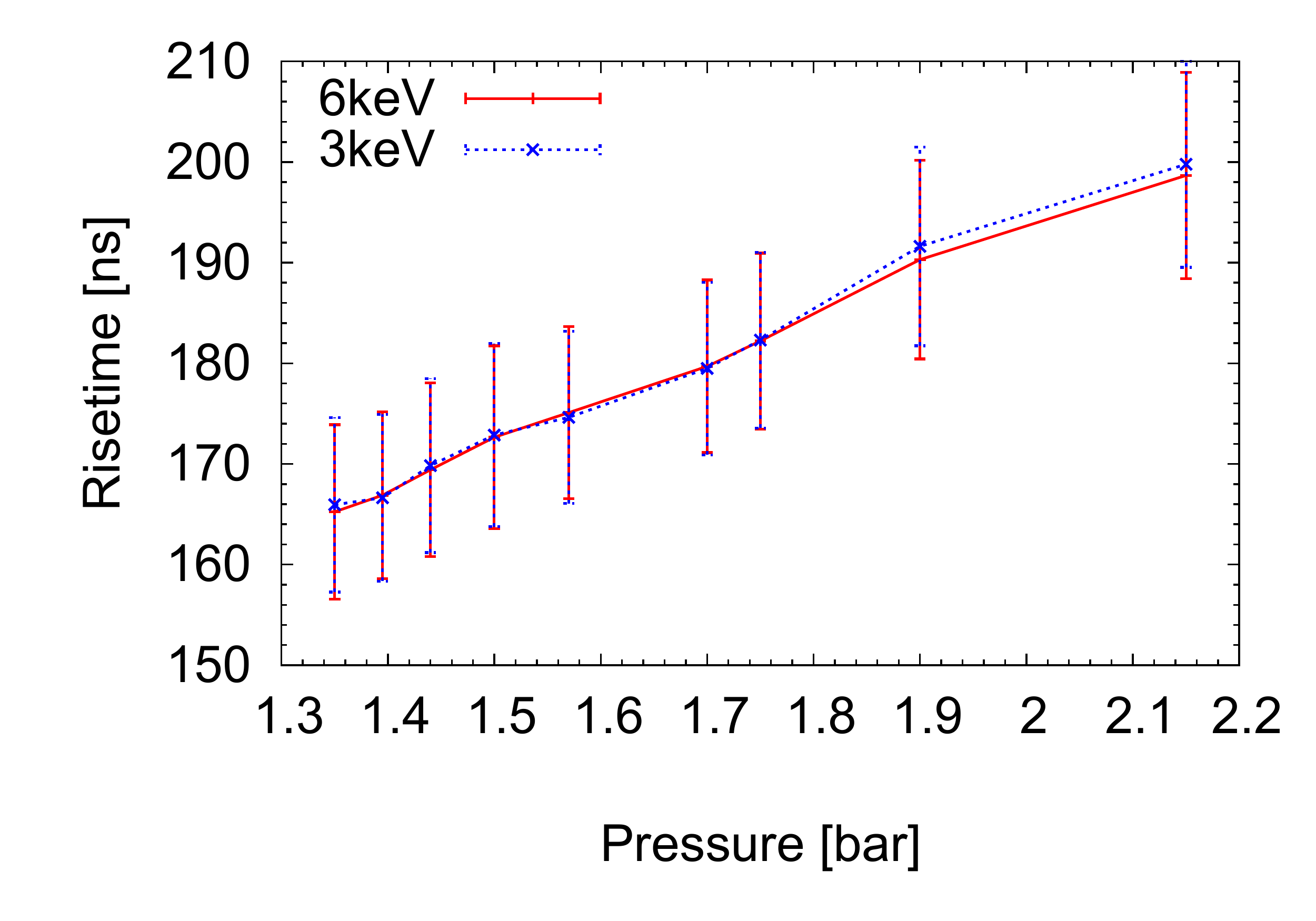}
\includegraphics[angle=270,width=0.49\textwidth]{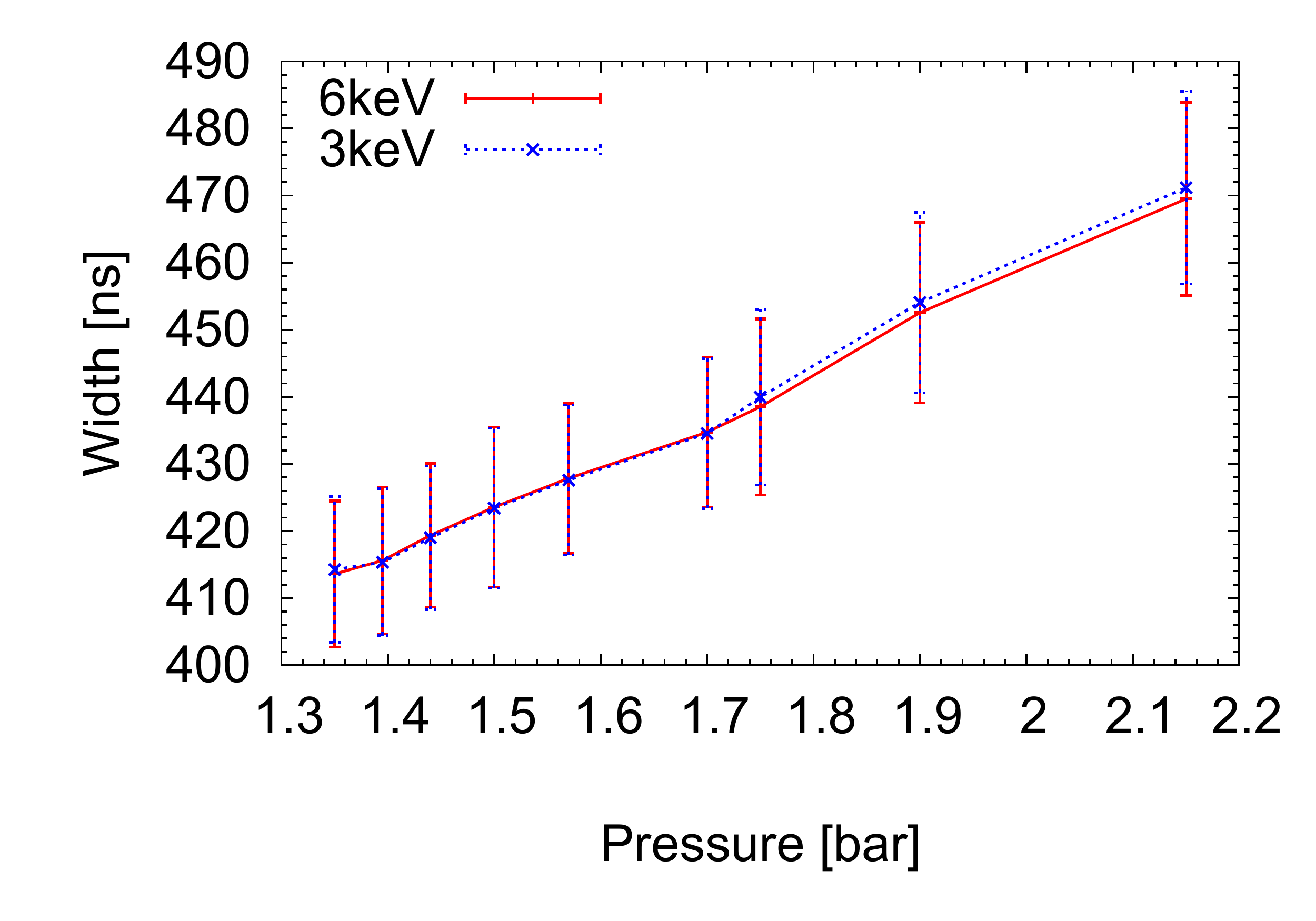}
\caption{\fontfamily{ptm}\selectfont{\normalsize{ Pulse risetime and width evolution at different detector pressure settings. }}}
\label{Pressure22}
\end{center}\end{figure}

\begin{figure}[!ht]\begin{center}
\begin{tabular}{ccc}
\includegraphics[angle=270,width=0.49\textwidth]{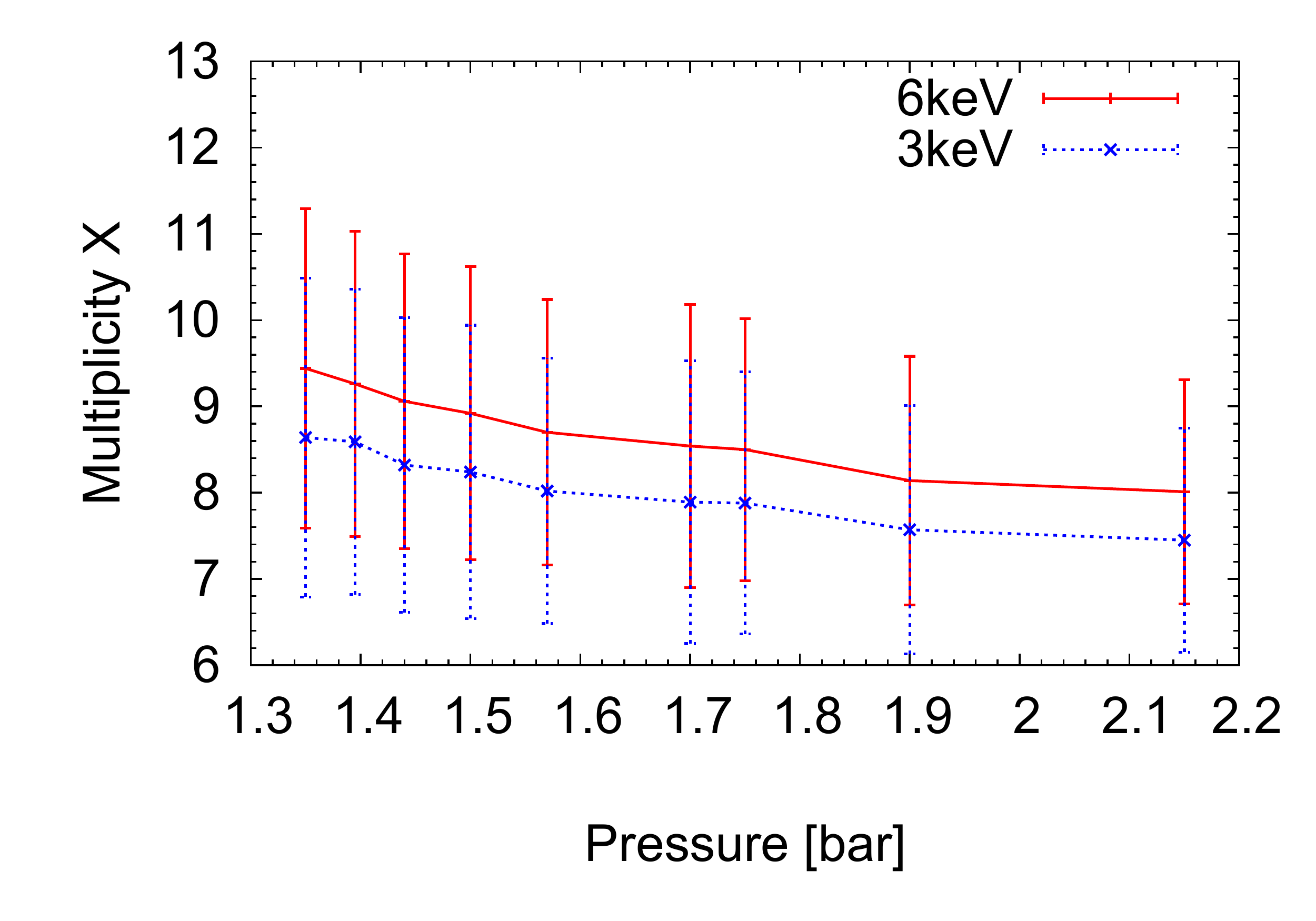} &
\includegraphics[angle=270,width=0.49\textwidth]{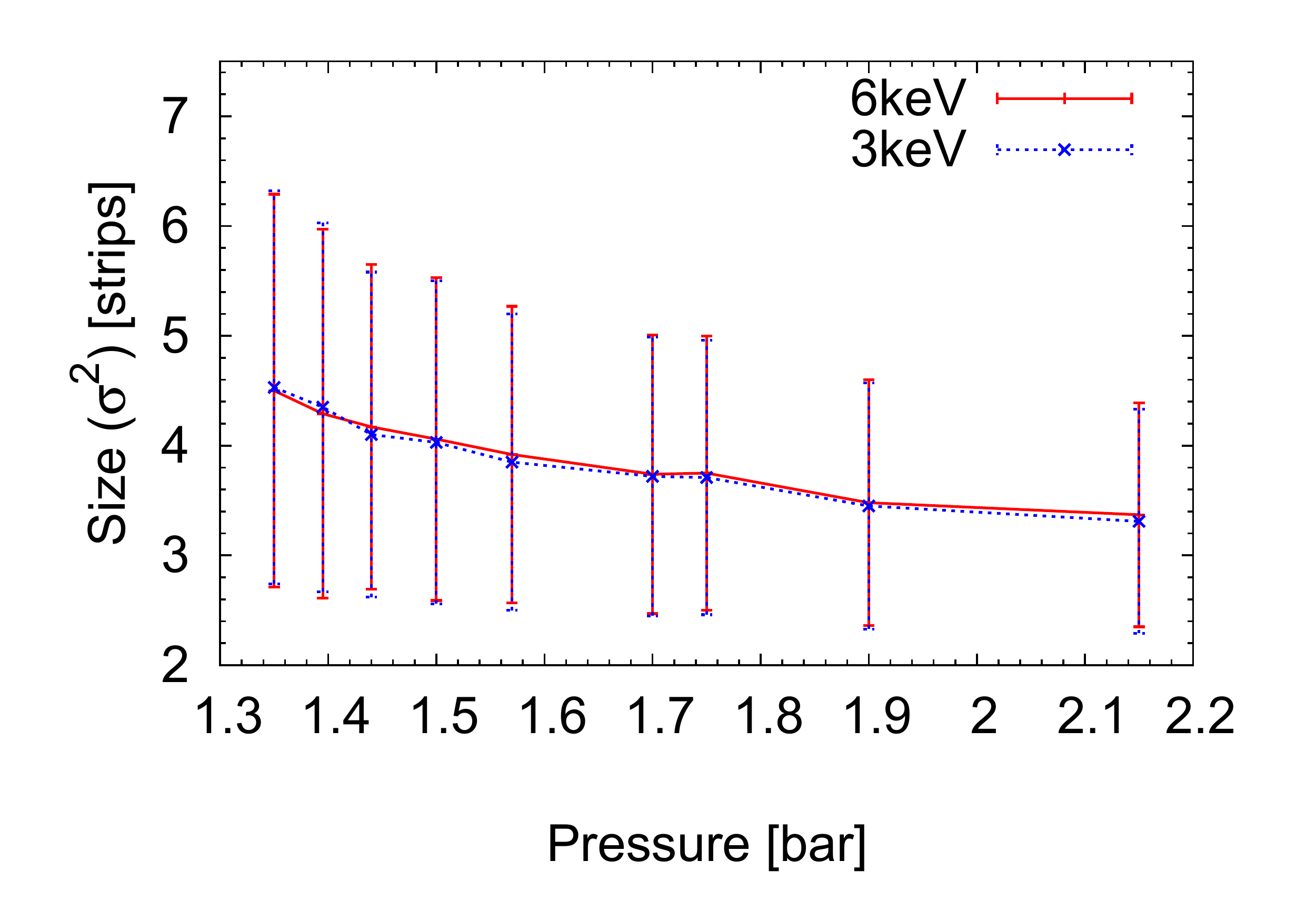} &
\end{tabular}

\caption{\fontfamily{ptm}\selectfont{\normalsize{ Cluster multiplicity and cluster size evolution with detector pressure, for $3$\,keV and $6$\,keV events. }}}
\label{Pressure23}
\end{center}\end{figure}

\begin{figure}[!ht]\begin{center}
\includegraphics[width=\textwidth]{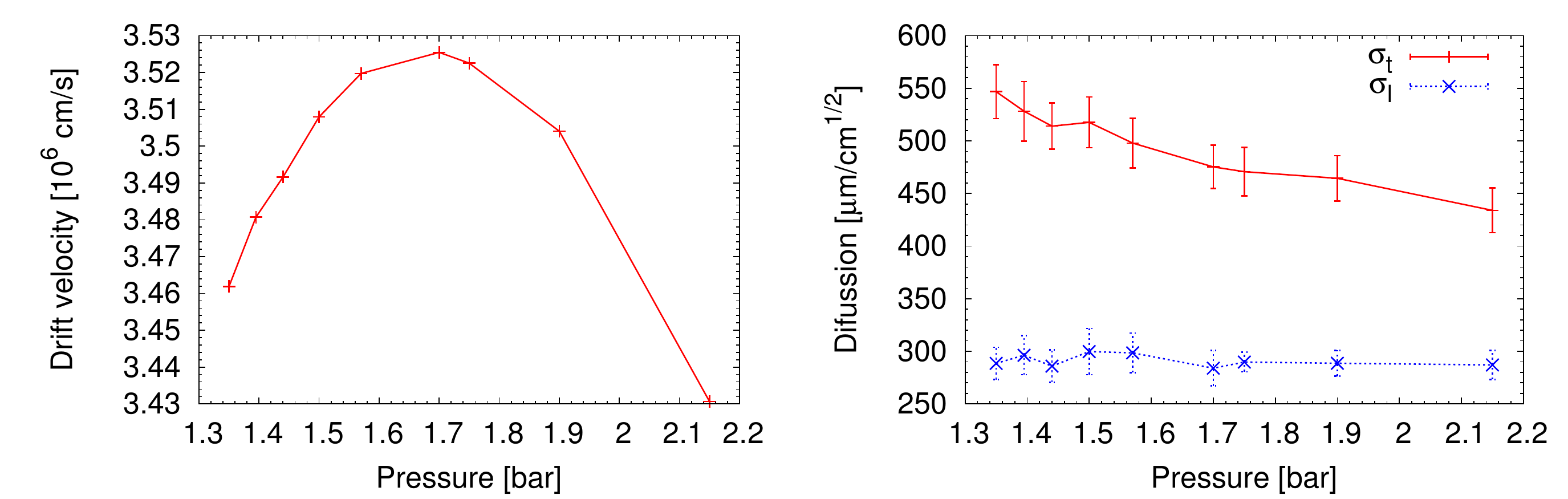}
\caption{\fontfamily{ptm}\selectfont{\normalsize{  Drift velocity (left) and longitudinal (represented with blue dashed line) and transversal (represented with red continous line) diffusion (right) for the different pressure measurements (data generated with Magboltz for the corresponding pressures and drift fields).  }}}
\label{fi:diffusionPressure}
\end{center}\end{figure}

The \emph{pulse risetime} and \emph{pulse width} (see Fig.~\ref{Pressure22}) are higher as the pressure in the detector chamber increases, probably this fact being related with the shorter mean free path of ions making them to move slowly, and increasing the avalanche time process. In spite of the increased mesh pulse timing, the effect on diffusion due to the pressure increase is attenuated as observed in the \emph{cluster multiplicity} and \emph{cluster size} (see Fig.~\ref{Pressure23}), in agreement with the diffusion decreasing tendency obtained with Magboltz~(see Fig.~\ref{fi:diffusionPressure}).

%
%

%
%
\clearpage

\subsection{Characterization of a microbulk type detector.}\label{sc:M13}

A more concise characterization of a microbulk type detector was carried out in the new acquisition set-up built in Zaragoza shown in figure~\ref{fi:saclaySetup}. The main purpose for the development of this additional detector set-up was focused in background measurements at different conditions, which include background measurements at the \emph{Canfranc Underground Laboratory}. The characterization measurements carried out were intended to be a reference for the operation of the detector with argon+2\%iC$_4$H$_{10}$ gas operating at $1.4$\,bar which are the actual conditions of the Micromegas detectors taking data in CAST.

\vspace{0.2cm} 

\begin{figure}[!ht]\begin{center}
\begin{tabular}{ccc}
\includegraphics[width=0.47\textwidth]{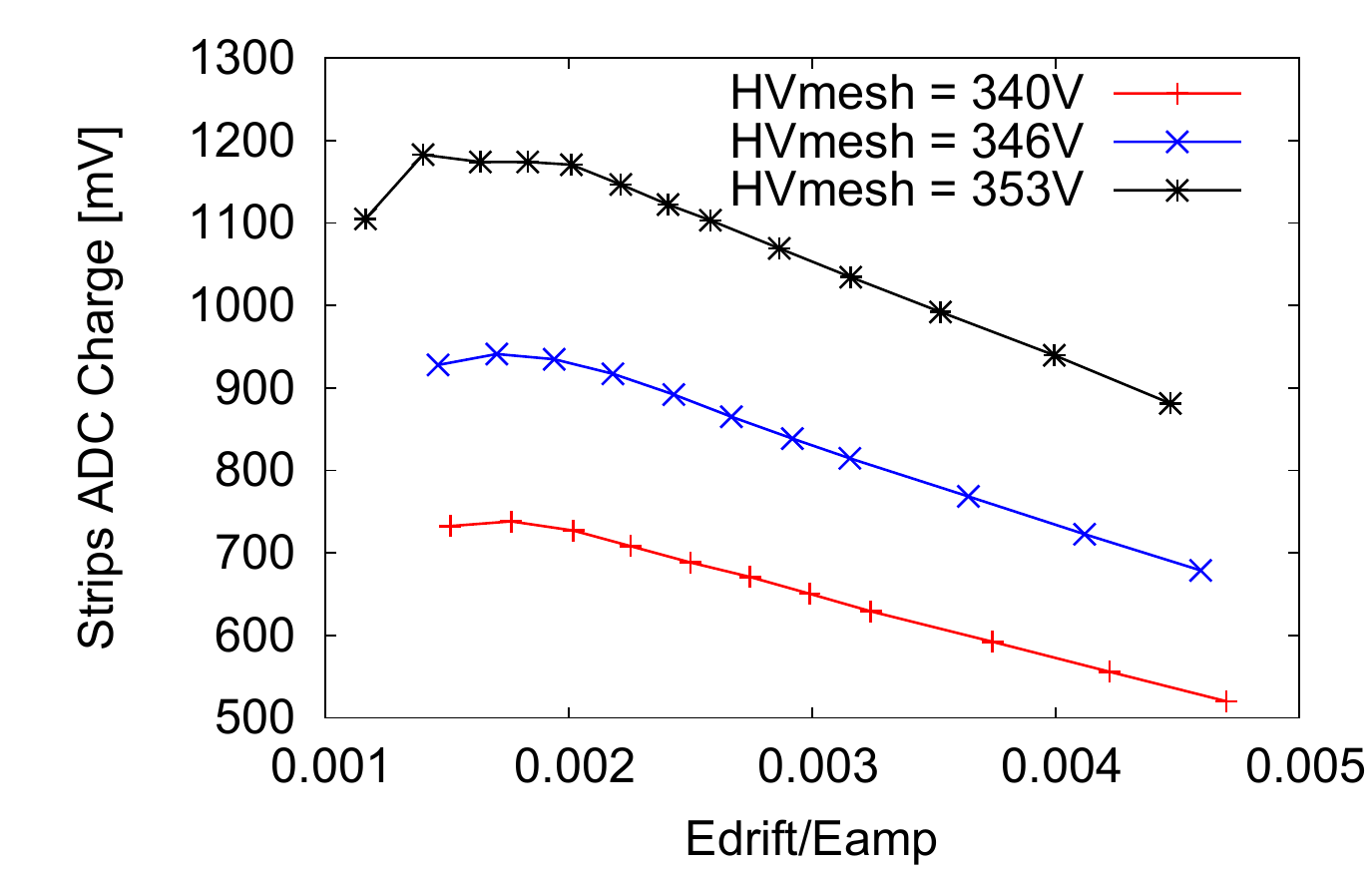} &
\includegraphics[width=0.47\textwidth]{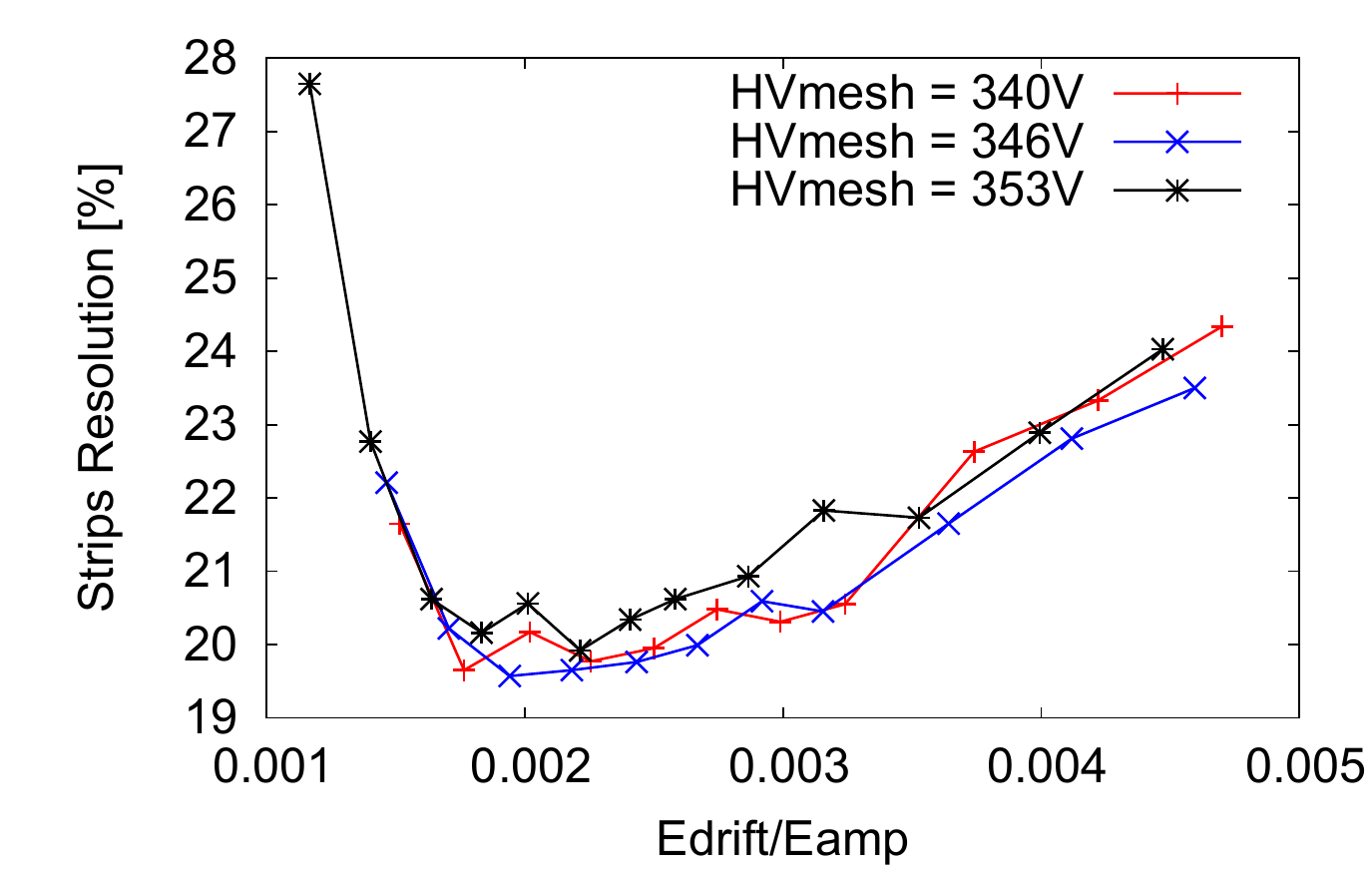} \\
\end{tabular}

\caption{\fontfamily{ptm}\selectfont{\normalsize{Relative gain as a function of the ratio between the drift field $E_{drift}$ and the amplification field $E_{amp}$.}}}
\label{fi:GainM13}
\end{center}\end{figure}

Calibrations with an $^{55}$Fe source were taken characterizing the main detector properties for $5.9$\,keV X-ray events. A full range of drift fields was measured for a set of \emph{three} different amplification fields (applying a mesh voltage of $340$\,V, $346$\,V and $353$\,V). The gain curves obtained are shown in figure~\ref{fi:GainM13} where a short plateau is observed at the lower drift fields measured, the best energy resolution it is reached for these same values. It must be noticed that the energy resolution presented is obtained from the strip clusters, always leading to a worse energy resolution than the mesh \emph{pulse amplitude}) giving account of the improvement on energy resolution versus the bulk type detectors.

\begin{figure}[!ht]\begin{center}
\includegraphics[width=0.9\textwidth]{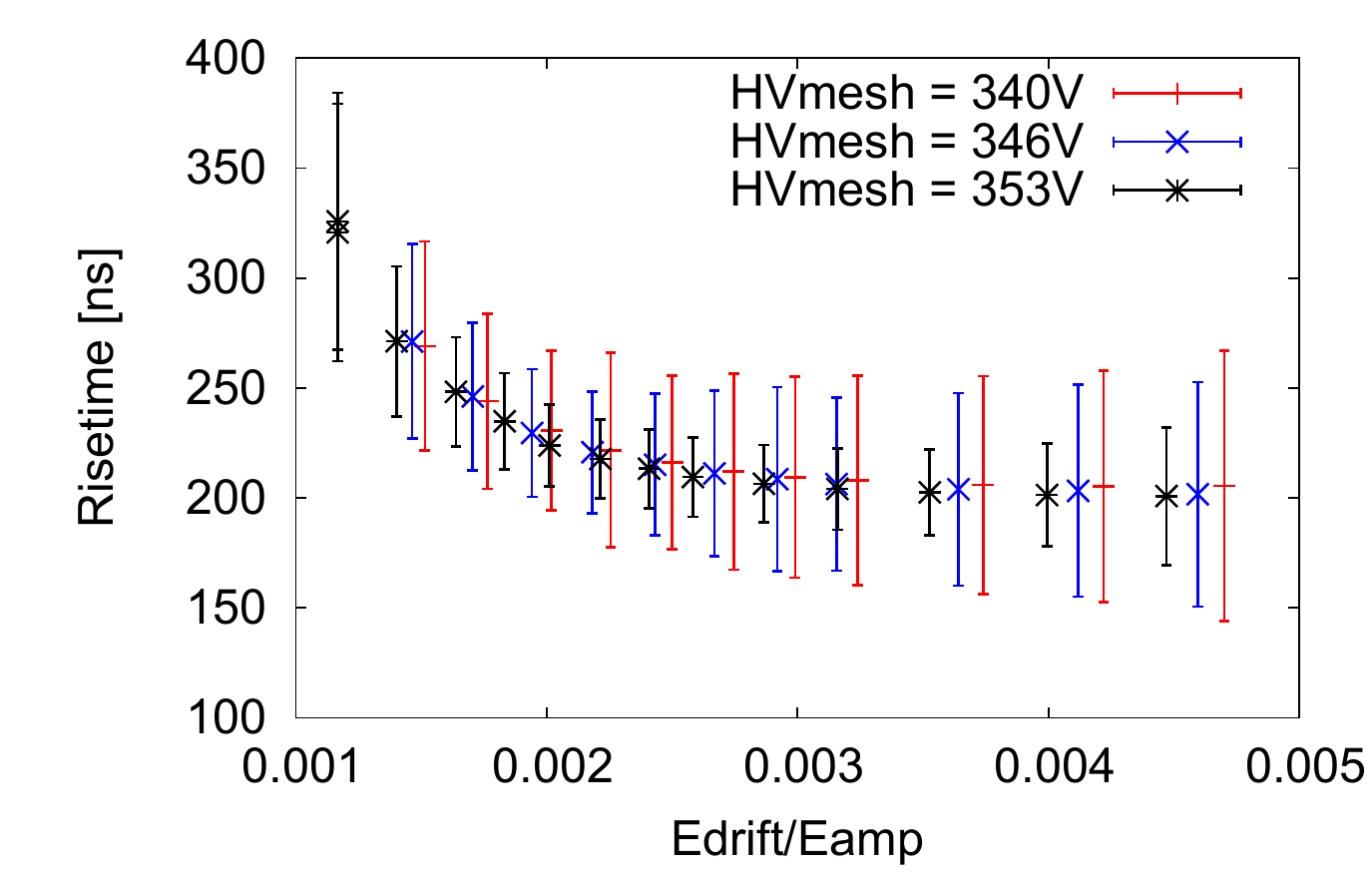} 
\caption{\fontfamily{ptm}\selectfont{\normalsize{ Pulse risetime as a function of the ratio between the drift field $E_{drift}$ and the amplification field $E_{amp}$. }}}
\label{fi:parametersM13}
\end{center}\end{figure}

\begin{figure}[!ht]\begin{center}
\includegraphics[width=0.9\textwidth]{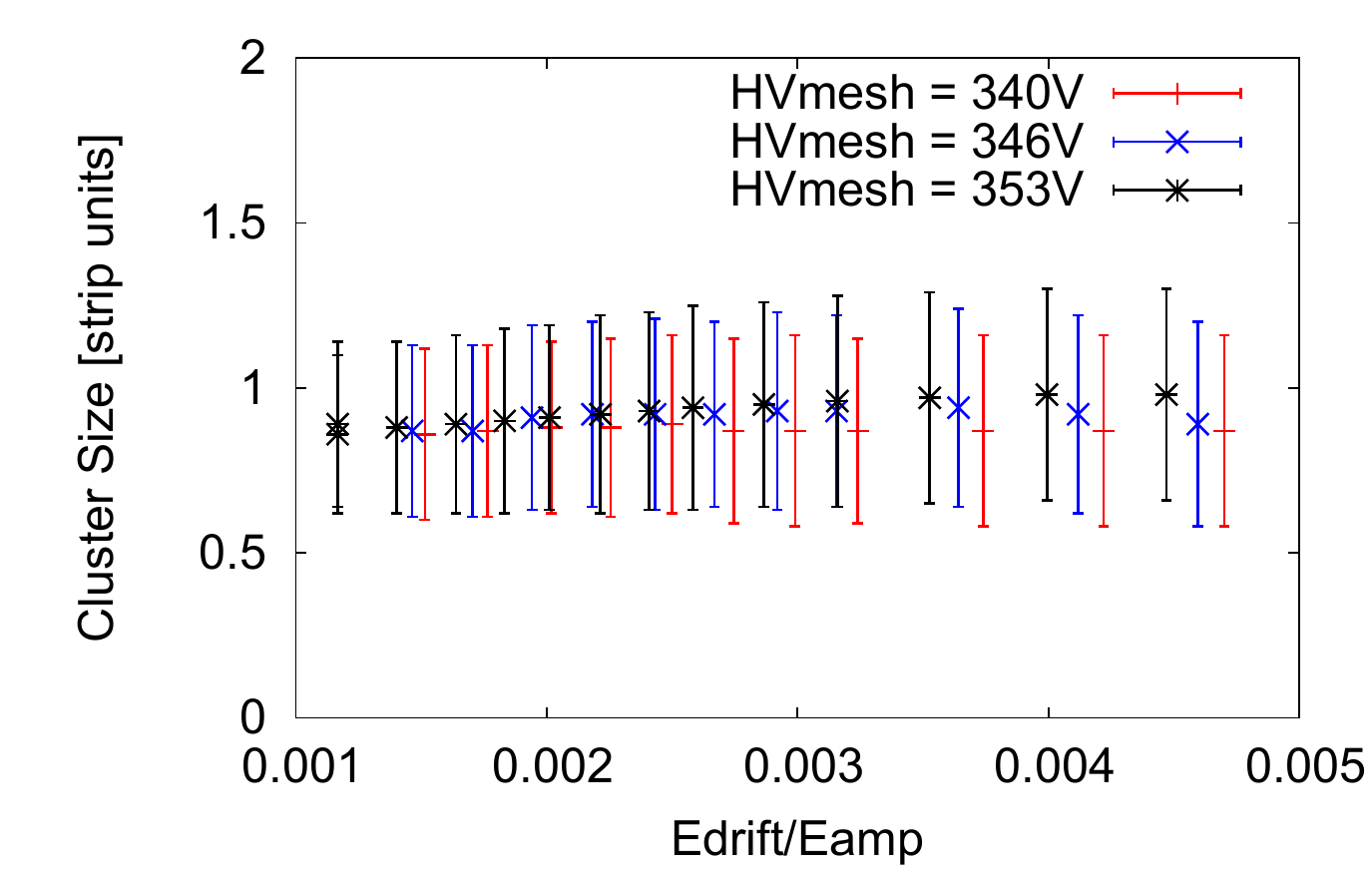} 

\caption{\fontfamily{ptm}\selectfont{\normalsize{ Cluster size as a function of the ratio between the drift field $E_{drift}$ and the amplification field $E_{amp}$. }}}
\label{fi:parametersM13_2}
\end{center}\end{figure}

The \emph{pulse risetime} presents an expected dependency with the drift field related with the arrival of charges to the mesh electrode, and the mean \emph{cluster size} is observed to be relatively constant in these measurements having a mean value much lower than the expected in bulk type detectors (see Fig.~\ref{fi:parametersM13} and Fig.~\ref{fi:parametersM13_2}).

\clearpage
\section{X-ray distributions.}
\begin{figure}[!ht]\begin{center}
\includegraphics[angle=270,width=0.65\textwidth]{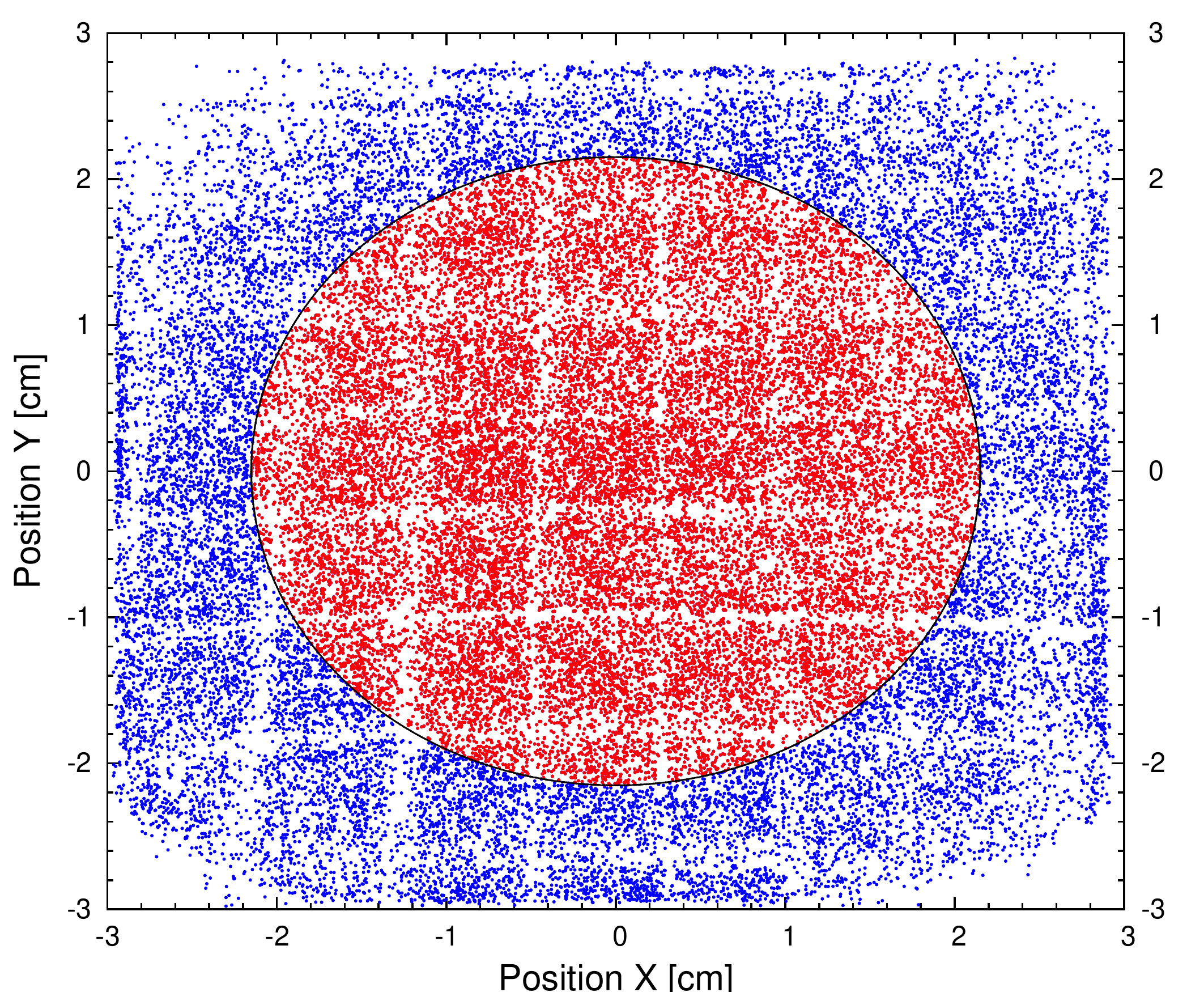} 
\includegraphics[angle=270,width=0.65\textwidth]{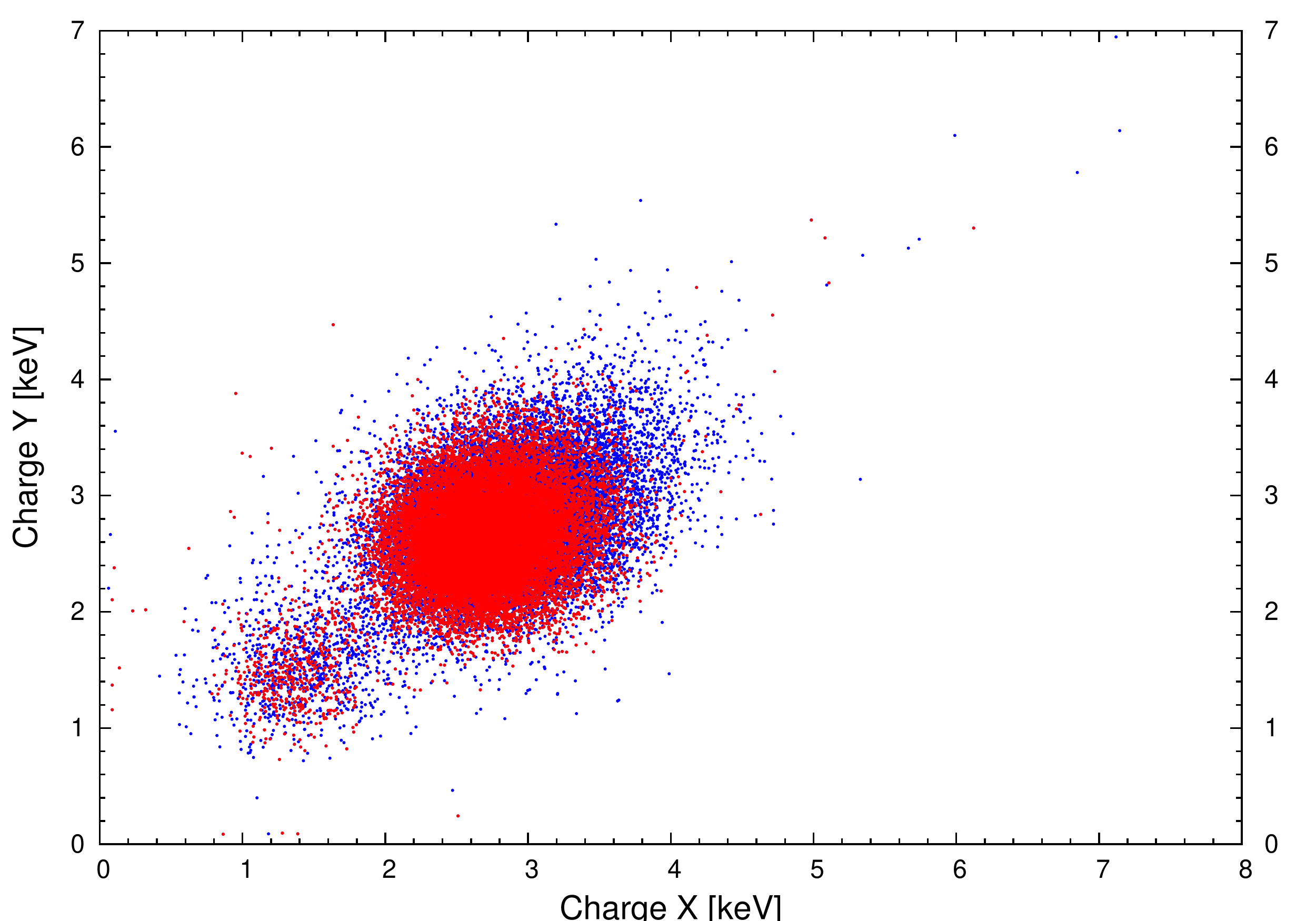} 
\caption{\fontfamily{ptm}\selectfont{\normalsize{ On top, hitmap distribution for a frontal calibration run, shadowed regions produced by drift window strongback are observed. Events hitting the expected CAST coldbore (red) region inside the detector are drawn inside the circle. On the bottom, charge independently collected by the strips in X and Y axis showing the good energy balance for X-ray events.  }}}
\label{hitmap}
\end{center}\end{figure}

\begin{figure}[!ht]\begin{center}
\includegraphics[angle=270,width=0.89\textwidth]{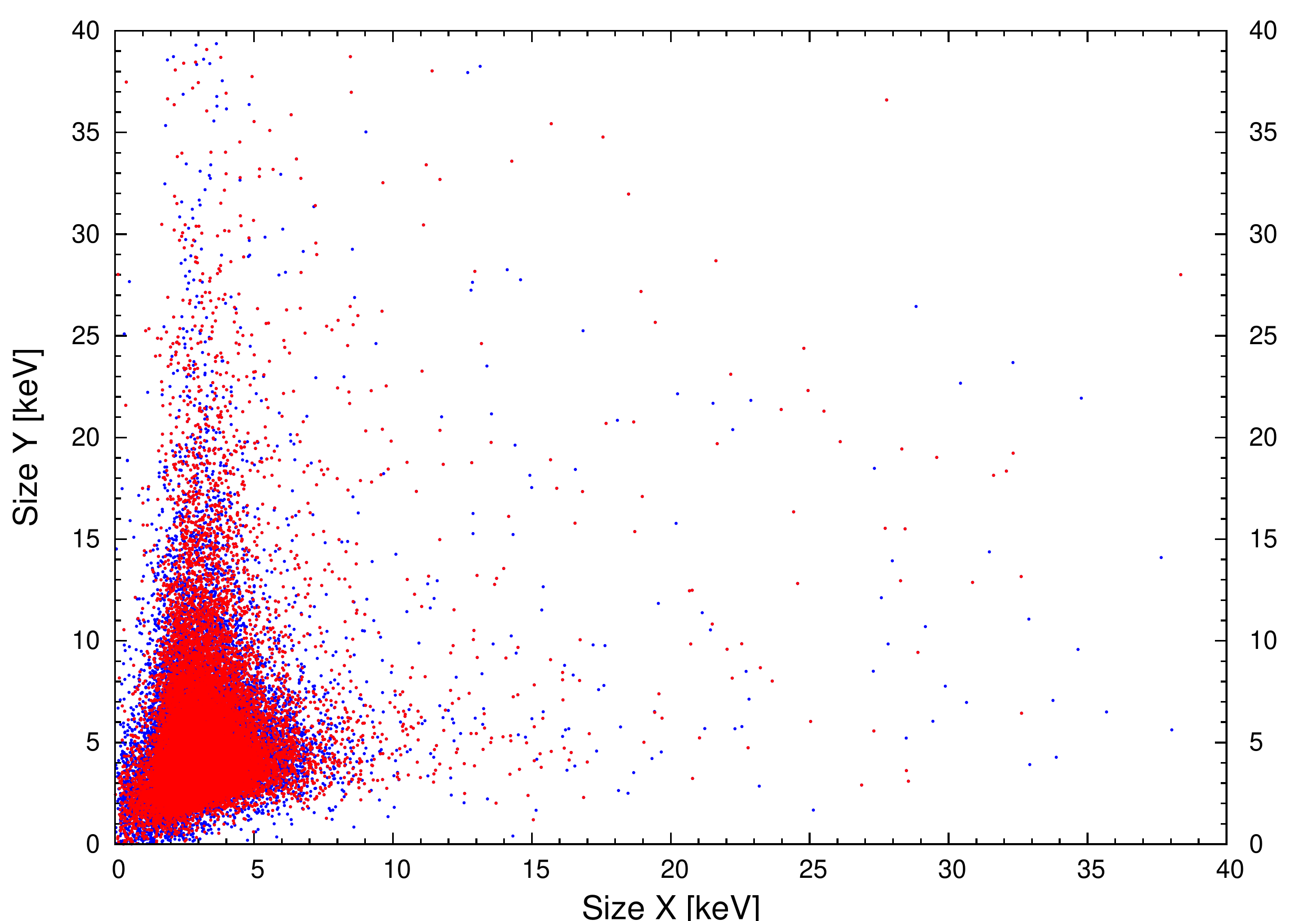}
\includegraphics[angle=270,width=0.89\textwidth]{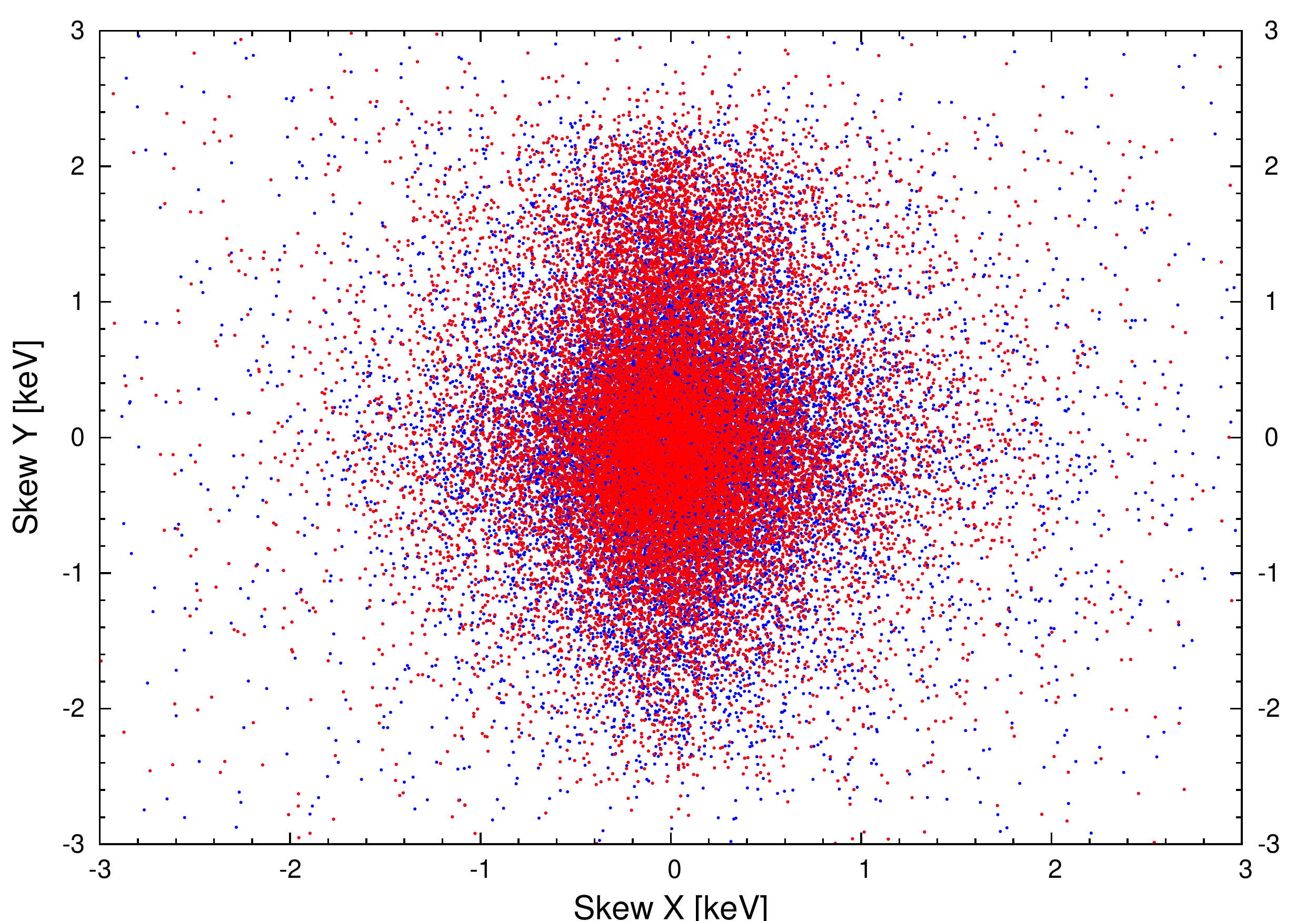}

\caption{\fontfamily{ptm}\selectfont{\normalsize{ On top, the \emph{cluster size} for X and Y clusters showing the typical \emph{cluster size} distribution for X-ray events. On the bottom, typical \emph{cluster asymmetry} distribution quantified with the \emph{cluster skew} parameter.  }}}
\label{size}
\end{center}\end{figure}

\begin{figure}[!ht]\begin{center}
\includegraphics[width=0.89\textwidth]{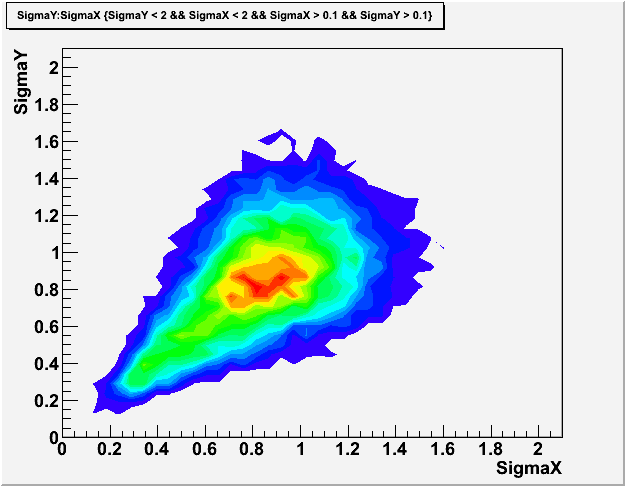}

\includegraphics[width=0.89\textwidth]{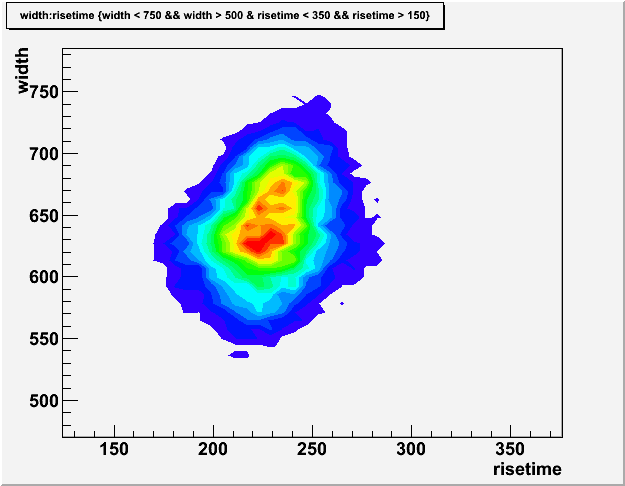}

\caption{ \fontfamily{ptm}\selectfont{\normalsize{Contour maps of the main population of events for the cluster size in X and Y (top) and the pulse parameters (bottom).  }}}
\label{contourRTvsWD}
\end{center}\end{figure}

This section presents the typical bidimensional distributions for the main Micromegas readout parameters, calibration hitmap and strips charge distribution (See Fig.~\ref{hitmap}), \emph{pulse risetime} and \emph{pulse width} distribution with the event energy (see Fig.~\ref{size}), X versus Y \emph{cluster size} and \emph{cluster skew} (see Fig.~\ref{size}), and contour map from the main population of events in the \emph{cluster size} X and Y distribution, and pulse parameters (see Fig.~\ref{contourRTvsWD}).

%

\begin{figure}[!ht]\begin{center}
\includegraphics[angle=270,width=0.89\textwidth]{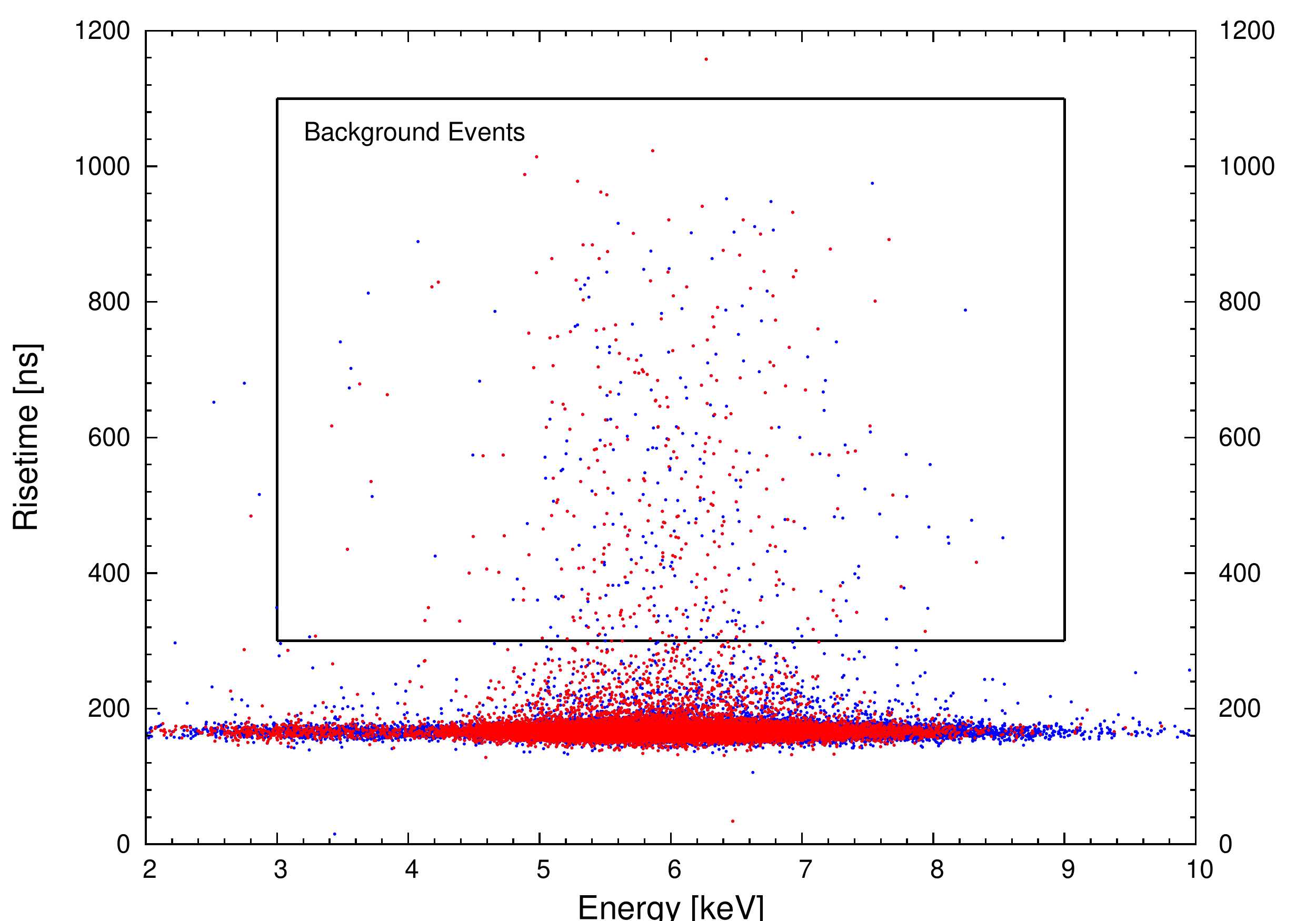}
\includegraphics[angle=270,width=0.89\textwidth]{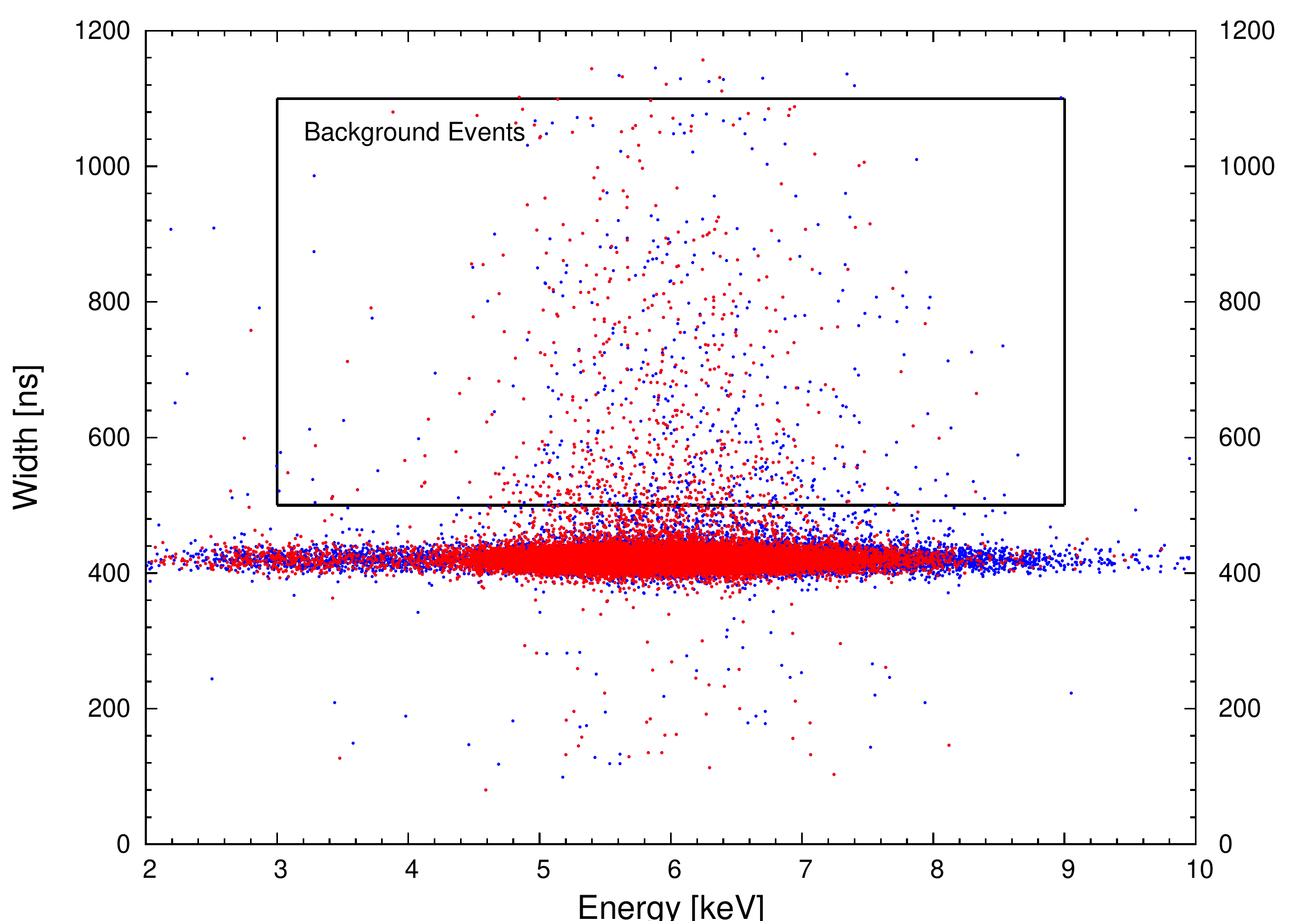}

\caption{\fontfamily{ptm}\selectfont{\normalsize{ Pulse parameters distribution as a function of the energy showing the non dependency of these parameters with the event energy. A non negligible amount of background events is mixed with the calibration run.  }}}
\label{pulseVsE}
\end{center}\end{figure}

\chapter{Background discrimination.}
\label{chap:discrimination}
\minitoc

\section{Introduction}

The selection of X-ray events between all the ionizing processes that have place in the detector chamber is based on a statistical discrimination analysis. The nature of the statistical analysis implies that the selection of X-ray events is not deterministic. The statistical method allows to lay down some selection rules by delimiting a region in the parameter space, or \emph{selection volume}, defined by the information collected by the detector readout, where X-ray events are likely to be found.



\vspace{0.2cm}

An X-ray source is used to determine the typical values of mesh pulse parameters and cluster properties for X-ray events in order to define the \emph{selection volume}. In particular, an $^{55}$Fe source is used since its signal is in the range of interest of the expected axion signal (2-7\,keV).

\vspace{0.2cm}

This chapter describes different methods studied to perform the desired X-ray selection. The \emph{selection volume} is chosen by each method trying to maximize the number of calibration accepted events (usually described in terms of \emph{software efficiency}) and minimize the number of selected background events. The way the \emph{selection volume} is defined depends on the method used.

\vspace{0.2cm}

The final set of background events selected must be understood as the most likely population of events to be an X-ray event. The presence of any other kind of events mixed with the real X-ray population cannot be determined. However, the method or rules that minimize the total population selected preserving the \emph{software efficiency} will lead to a more pure selection.

\section{Discriminants definition.}\label{sc:discriminants}

The parameters obtained in the rawdata analysis of the detector are used to built \emph{discriminant observables} that are finally used for background rejection. Thus, in order to introduce them in some statistical selection methods they are preferred to be Gaussian distributed and to be as less energy dependent as possible.

\vspace{0.2cm}

The most significant \emph{observables} that can be built with the Micromegas readout used for discrimination and selection of X-ray events, together with its typical or expected value for X-ray events, are given in the following list.

\begin{itemize}

\item {\bf 0. Pulse risetime ($rt$) : } A localized deposit of charge tends to give lowest \emph{risetime} values, usually limited by the electronic set-up. Typical mean value 80-160\,ns, depending on detector amplification gap and electronic set-up.

\item {\bf 1. Pulse width ($wd$) : } A localized deposit of charge tends to give the lowest \emph{width} values. Typical mean value 150-300\,ns. As the \emph{risetime} parameter, it is related with the amplification gap and the electronic set-up.

\item {\bf 2. Cluster charge balance $\left( \left(c_x - c_y\right)/\left(c_x+c_y\right) \right)$ : } Deviation of charge collected between X-strips plane and Y-strips plane. Expected mean value zero.

\item {\bf 3. Pulse risetime versus pulse width ($rt/wd$) : } \emph{risetime} versus \emph{width} ratio. Gives pulse temporal shape information. Typical value 0.5-0.7.

\item {\bf 4. Energy balance 1 $\left( \left(E_a - E_i\right)/\left(E_a+E_i\right) \right)$ : } Deviation between energy measured by the mesh \emph{pulse amplitude} and the mesh \emph{pulse integral}. Expected mean value zero.

\item {\bf 5. Energy balance 2 $\left( \left(E_a - E_c\right)/\left(E_a+E_c\right) \right)$ : } Deviation between energy measured by the mesh \emph{pulse amplitude} and the strips \emph{cluster charge}. Expected mean value zero.

\item {\bf 6. Energy balance 3 $\left( \left(E_i - E_c\right)/\left(E_i+E_c\right) \right)$ : } Deviation between energy measured by the mesh \emph{pulse integral} and the strips \emph{cluster charge}. Expected mean value zero.

\item {\bf 7. Cluster size balance $\left( \left(\sigma_x - \sigma_y\right)/\left(\sigma_x+\sigma_y\right) \right)$ : } Deviation between the X-strips and Y-strips cluster size. Expected value zero.

\item {\bf 8. Cluster multiplicity balance $\left( \left(m_x - m_y\right)/\left(m_x+m_y\right) \right)$ : } Deviation between the X-strips and Y-strips cluster multiplicity. Expected value zero.

\item {\bf 9. Skew $\left( \gamma_x \cdot \gamma_y \right)$ : } X and Y combined cluster asymmetry. Expected value zero.

\item {\bf 10. Multiplicity $\left( m_x \cdot m_y \right)$ : } X and Y combined cluster multiplicity. Expected value $\sim35$.

\item {\bf 11. Size $\left( \sigma_x \cdot \sigma_y \right)$ : } X and Y combined cluster size, or cluster area. Expected value $\sim5$.

\item {\bf 12. Pulse center $\left( t_o \right)$ : } Highest peak amplitude time inside the acquisition window. The trigger defines a constant peak position time for X-ray events. Typical value 800-1200\,ns.

\item {\bf 13. Pulse center mean $\left( \bar{t}_o \right)$ : } Mean expected time value including pulse amplitude weighting. Typical value 700-1100\,ns.

\end{itemize}

The background rejection is performed by using a subset of observables from the list. The number given at each \emph{discriminant} inside the list is used to create an identifier of the combination of several discriminants. In order to introduce different observable combinations in a discrimination analysis each combination must be associated to a unique number. This association has been created by using a 14 bit digital word where each bit relates to the presence or not of an observable in the analysis. The bit position for each observable is the one given in the observables list.

\vspace{0.2cm}

\begin{figure}[!ht]
{\centering \resizebox{0.9\textwidth}{!} {\includegraphics{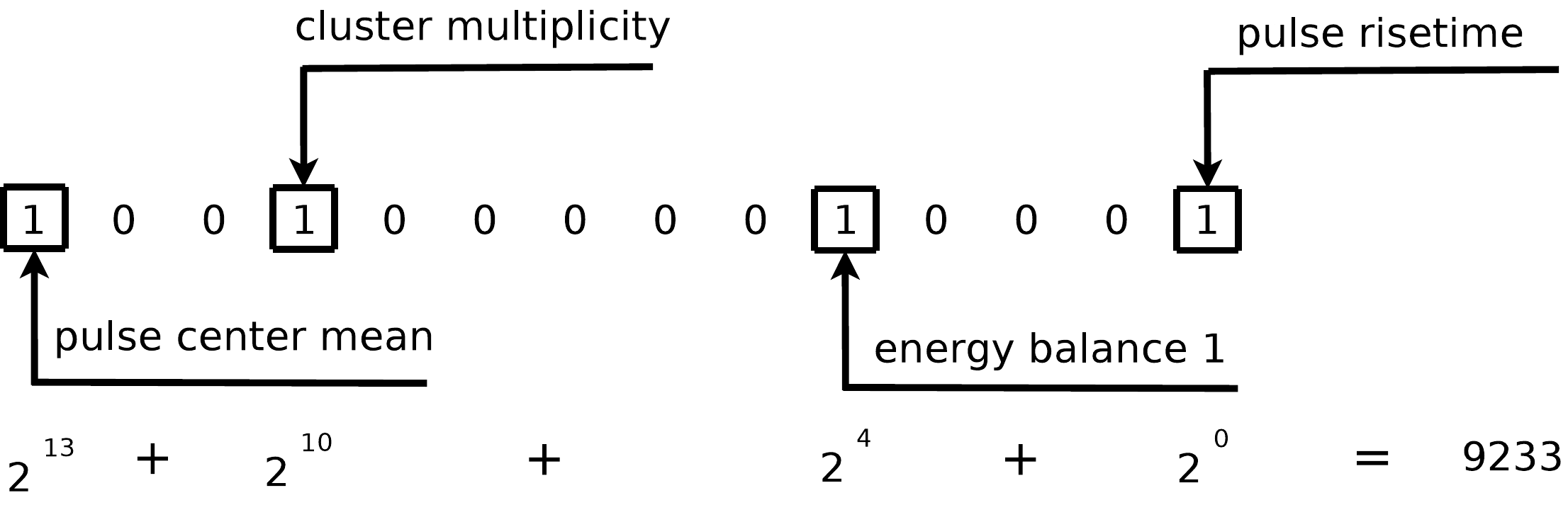}} \par}
\caption{\fontfamily{ptm}\selectfont{\normalsize{Codification of an observables subset combination. }}}
\label{fi:bitCombinations}
\end{figure}

For example, the observables combination that involves \emph{pulse risetime}, \emph{energy balance 1}, \emph{multiplicity} and \emph{pulse center mean} is the combination 9233~(see Fig.~\ref{fi:bitCombinations}).

\vspace{0.2cm}

The many \emph{observables} described in the list are chosen in order to evaluate them in an iterative analysis and to find an optimum \emph{observables} combination to be used for discrimination.

\subsection{Optimum discriminants selection.}\label{sc:optimumdiscriminants}

Some of the \emph{observables} defined are clearly correlated and introduce redundant information when they are put together in a statistical analysis, also known in the literature as \emph{statistical noise}~\cite{Towers:2002hz}. One of the most important steps in the discrimination analysis is to determine the subset of observables that will lead to the optimum X-ray selection, understanding by optimum the best background rejection preserving the same \emph{software efficiency}.

\vspace{0.2cm}

The best and worst combinations can be obtained after analyzing all the possible combinations of at least \emph{two} observables. This study has been performed using the multivariate method described in section~\ref{sc:multivariate}. 

\vspace{0.2cm}

\begin{figure}[!ht]
{\centering \resizebox{\textwidth}{!} {\includegraphics{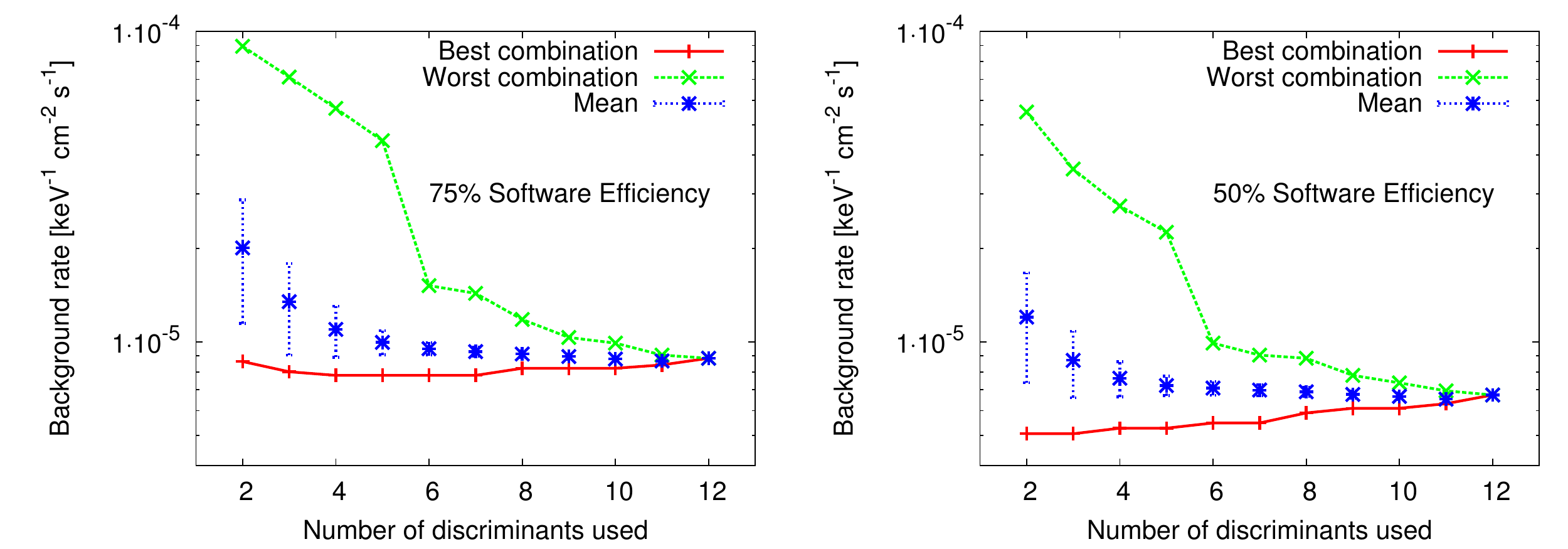}} \par}
\caption{\fontfamily{ptm}\selectfont{\normalsize{The resulting background as a function of the number of observables used in the discrimination analysis for software efficiencies of 75\% and 50\%. The background level is shown for the worst and best combinations, also the mean value from all the combinations that correspond to the given number of observables. }}}
\label{fi:multStudies}
\end{figure}

Figure~\ref{fi:multStudies} summarizes the results obtained after analyzing a set of background runs with all the possible observable combinations as a function of the number of observables used. At first, it is noticed that the combinations that involve a higher number of observables do \emph{not} lead to the best rejection. This fact is clearly observed for the analyzed combinations at 50\% software efficiency. 

It is observed that the combinations leading to the worst background selection are also the ones involving a fewer amount of observables. This result implies that in order to obtain a good background rejection with a minimum amount of observables the election of these few observables is critical. However, this is not the case for combinations involving a higher amount of observables which election to be introduced in the analysis is not so relevant, leading always to a reasonable background reduction.

\vspace{0.2cm}

The best and worst combinations involving \emph{two} and \emph{three} observables are shown in tables \ref{ta:twoBestWorst} and \ref{ta:threeBestWorst}. From these tables is induced that the \emph{best} observables to be used in a discrimination analysis for X-ray selection are \emph{pulse risetime}, \emph{pulse width} and \emph{multiplicity}, while the worst ones are \emph{charge balance}, \emph{multiplicity balance} and \emph{skew}.

\begin{table}[ht!]
\begin{center}
\vspace{0.3cm}
\begin{tabular}{c|c|c|c}
& id & Observable 1 & Observable 2 \\
\hline
\multirow{2}{*}{Best} & 129 & Risetime & Cluster size balance \\
 & 516 & Risetime/Width & Cluster size balance \\
\hline
\multirow{2}{*}{Worst} & 516  & Charge balance & Skew \\
 & 260 & Charge balance & Multiplicity balance \\
\hline
\end{tabular}
\caption{ Best and worst combinations involving \emph{two} observables. }
\label{ta:twoBestWorst}
\end{center}
\end{table}

\begin{table}[ht!]
\begin{center}
\vspace{0.3cm}
\begin{tabular}{c|c|c|c|c}
& id & Observable 1 & Observable 2 & Observable 3 \\
\hline
\multirow{2}{*}{Best} & 1033 & Risetime/Width & Risetime & Multiplicity \\
 & 1034 & Risetime/Width & Width & Multiplicity \\
 & 1027 & Risetime & Width & Multiplicity \\
\hline
\multirow{2}{*}{Worst} & 772  & Charge balance & Multiplicity balance & Skew \\
 & 580 & Charge balance & Energy balance 3 & Skew \\
 & 324 & Charge balance & Energy balance 3 & Multiplicity balance \\
\hline
\end{tabular}
\caption{ Best and worst combinations involving \emph{three} observables. }
\label{ta:threeBestWorst}
\end{center}
\end{table}

The observables that produce the best combinations are those that are strongly related with point-like charge depositions, higher \emph{risetime}, \emph{width} and \emph{multiplicities} are expected for non point-like events. On the other hand, the worst combinations are the ones that do not contribute to distinguish point-like charge depositions between all the possible ionizing processes. Many ionizing processes are going to produce a similar \emph{charge balance} and \emph{multiplicity balance} since there is no physical reason to produce more charge in one strips plane than in the other. The \emph{skew} and/or cluster symmetry rejection potential is masked by the effect of the diffusion of charges in the gas, X-ray patterns generated by those parameters have comparable values to any other process having place in the chamber.

\section{The discrimination methods.}\label{sc:methods}

Each particular combination of observables generated with the list provided at section~\ref{sc:optimumdiscriminants} defines a parameter space where each event is located at a given coordinates fixed by its pattern vector $\tilde{X}_i$, which components are related with the chosen observables. The particular statistical methods applied to the data are based on different ways to define the \emph{selection volume} where mainly X-rays are likely to be found. Up to now, three different methods have been studied and applied to the background discrimination of CAST Micromegas detectors; \emph{sequential analysis}, \emph{modified multivariate analysis} and \emph{neural networks}.

\vspace{0.2cm}

In the \emph{sequential} method, the \emph{selection volume} is chosen by inspecting the data and parameterizing basic geometrical shapes that enclose the main population of X-ray events in a projected plane of the defined parameters space. The cuts are applied to each defined observable one after the other at different projections of the parameter space. There are advanced ways to define the shapes of cuts that will have effect on the calibration and background events. This method was used in previous data analysis of detectors in CAST~\cite{TheopistiThesis}, however the method has been substituted by more powerful and efficient methods in the analysis of CAST Micromegas data, and it will not be discussed in detail.

\vspace{0.2cm}

The \emph{multivariate analysis} defines the \emph{selection volume} by weighting the standard deviation of each discriminant, and takes into account the correlations between them in order to define the optimum multidimensional ellipsoid that will enclose the X-ray events. The discriminants used in this analysis are expected to be Gaussian distributed. The covariance matrix obtained with a known population of X-ray events defines the shape and rotation of the ellipsoid.

\vspace{0.2cm}

The \emph{neural network} method used goes a step further in the definition of the \emph{selection volume}. The neural network is divided in correlated cells, where each cell represents an irregular volume that covers a particular region of the parameter space. The size and position of the particular volume defined by each cell is determined by training the neural network with background data. Each cell and their neighbors will get specialized in recognizing an event with a given input pattern $x_i$. By using events coming from an X-ray calibration source the cells that recognize X-ray events are revealed, and the background events that are associated to that cells are the resulting background selection.

\subsection{Modified multivariate method.}\label{sc:multivariate}

The multivariate method used in Micromegas detectors for extracting the X-ray data was first introduced in CAST by \emph{K. Kousouris}. This multivariate method uses the standard multivariate theory for defining the \emph{selection volume}, and introduces a energy correction factor for assuring the imposed software efficiency in the expected signal energy range. 

\subsubsection{Theory.}

A set of $N$ discriminant observables, $x_i$, defined using the ones described at section~\ref{sc:discriminants}, which are non fully correlated and follow Gaussian distributions with mean values $\mu_i$, are used to define the joint probability density function (pdf) expressed as,

\begin{equation}
f(\tilde{X}) = \frac{1}{\left( 2 \pi \right)^{N/2} |\rho|^{1/2}} exp\left( -\frac{1}{2} \tilde{X}^\mathrm{T} \rho^{-1} \tilde{X} \right)
\end{equation}

\vspace{0.1cm}

\noindent where $\tilde{X}_i = \frac{1}{\sigma_i} \left( x_i - \mu_i \right)$ and $\rho$ is the correlation matrix, related with the covariance matrix $V_{ij}$ and the standard deviation $\sigma_i$ of the observable $x_i$, by the expression

\begin{equation}
\rho_{ij} = \frac{1}{\sigma_i \sigma_j} V_{ij}
\end{equation}

\vspace{0.1cm}

\noindent which non diagonal elements define the correlations between the different discriminant observables chosen.

\vspace{0.2cm}

The expression $q = \tilde{X}^\mathrm{T} \rho^{-1} \tilde{X}$ can be reformulated after a rotation of the vector $\tilde{X}$ with an orthogonal transformation $U$ ($\tilde{X} = UY$ and $U^\mathrm{T} = U^{-1}$). With a suitable choice of $U$ the matrix $\Lambda = U^\mathrm{T} \rho^{-1}U$ is diagonal with eigenvalues $\lambda_i$ which are at the same time eigenvalues of $\rho^{-1}$ allowing us to write,

\begin{equation}
q = \sum_i \lambda_i Y_i^2 \quad \Longleftrightarrow \quad \sum_i \frac{Y_i^2}{\sqrt{q/\lambda_i}^2} = 1
\end{equation}

\begin{figure}[t]
\begin{center}
\begin{tabular}{c}
\includegraphics[width=0.28\textwidth]{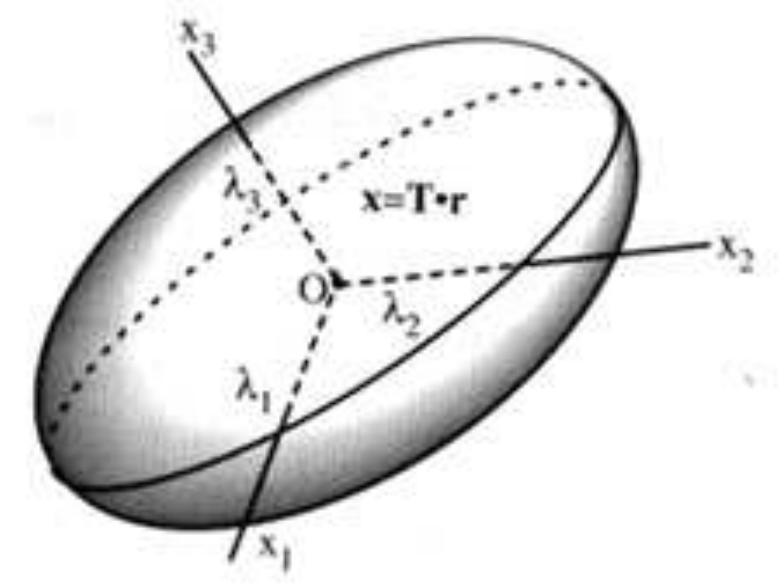} 
\end{tabular}
\end{center}
\caption{\fontfamily{ptm}\selectfont{\normalsize{Schema from a standard multivariate analysis where an acceptance elliptical volume is defined by weighting the standard deviation of each parameter.}}}
\label{fi:ellipsoid}
\end{figure}

\vspace{0.1cm}

\noindent relation that defines an N-dimensional ellipsoid (see~Fig.~\ref{fi:ellipsoid}), for every value of $q$, with axis half lengths $\sqrt{(q/\lambda_i)}$. Thus, the pdf of the quantity $q$ is,
\vspace{0.1cm}

\begin{equation}
f_q(q) = \frac{1}{ 2^{N/2}\, \Gamma(N/2) } q^{\frac{N}{2}-1}\,e^{-q/2}
\end{equation}

\vspace{0.1cm}

\noindent where the quantity $q$ follows a $\chi^2$ distribution with $N$ degrees of freedom.

\vspace{0.2cm}

The importance of the quantity $q$ relies in the fact that all the information of the $N$ discriminant observables is projected in a single quantity which may be used as an efficient selection rule.

\subsubsection{Applying the multivariate cut.}

The quantity $q$ is calculated for each calibration event by using the expression $q = \tilde{X}^\mathrm{T} \rho^{-1} \tilde{X}$, where the correlation matrix is precalculated using the most representative $^{55}$Fe events from the main peak at $5.96$\,keV. The discriminant observables coming from an $^{55}$Fe source follow a Gaussian distribution and reproduce the theoretical distribution of $f_q(q)$~(see~Fig.~\ref{fi:multDists}).

\vspace{0.2cm}

In other words, the quantity $q$ can be understood as the distance between the given event pattern, $\tilde{X}$, coming from a background or a calibration run, and the center of the ellipsoid, sitting at the origin by definition. This distance is calculated by using the metric given by the correlation matrix, $\rho$, that relates the same $q$-value to all the events placed in the same ellipsoid surface, appropriately rotated to suit the eigenvectors of $\rho$. 

\vspace{0.2cm}

In such way, event patterns that are far away from the $^{55}$Fe source observables mean values will have a higher $q$-value than events that are close to the X-ray patterns mean value. The method allows then to set a $q$-limit value, $q_L$, that sets the frontier between X-ray like events and other natural background events.

\vspace{0.2cm}

Once the $q$-distribution of $^{55}$Fe events is known,  the value of $q_L$ is determined by choosing a given acceptance, or \emph{software efficiency}, $\epsilon'$ of X-rays in the method.

\begin{equation}
\epsilon' = \int_{q<q_L} f_q(q) dq
\end{equation}

\vspace{0.1cm}

The same $q_L$ calculated to obtain a given acceptance for X-rays, together with its corresponding correlation matrix, it is used to determine the background events that are most like to be an X-ray, that ones that satisfy $q < q_L$.

\subsubsection{Energy correction.}

The covariance matrix is calculated by using only $6$\,keV peak events since they are the most abundant and representative in the energy range of interest. However, X-ray events at a different energy would lead to a different correlation matrix given that the discriminants variance $\sigma_i$ depends on the energy. The metric of the space defined by $\rho$ depends on the energy, and the same $q_L$ leads to a different $\epsilon'$ for different X-ray energies.

\vspace{0.2cm}

In order to preserve the same software efficiency, in the energy range we are interested on, the $q$-value of each event is calculated with the correlation matrix obtained at $E_o = 6$\,keV, $q(E) = \tilde{X}(E)^\mathrm{T} \rho^{-1}(E_o)\tilde{X}(E)$. Then $q(E)$ is transformed using the following the expression

\begin{equation}
\tilde{q} = \left( \frac{E}{E_o} \right)^a q(E)
\end{equation}

\noindent that rises from the fact that the variance follows a power law, $\sigma(E)\propto E^{-a/2}$. The parameter $a$ must be determined by demanding the energy invariance of the $\tilde{q}$ distribution.

\vspace{0.2cm}

\begin{figure}[!ht]
{\centering \resizebox{\textwidth}{!} {\includegraphics{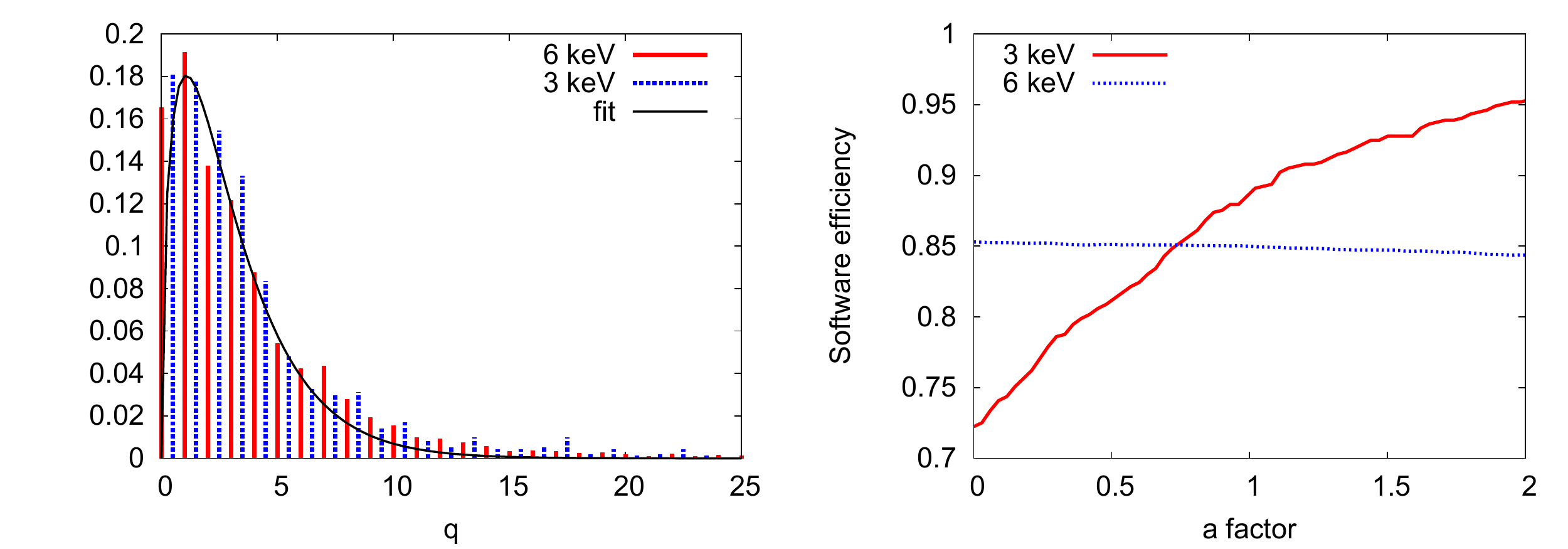}} \par}
\caption{\fontfamily{ptm}\selectfont{\normalsize{ On the left, $q$-distribution for $6$\,keV and $3$\,keV events (already corrected) corresponding to the $^{55}$Fe main and scape peak. The discriminants used to generate this distribution are \emph{risetime}, \emph{cluster size balance} and \emph{multiplicity}. The distribution obtained for $6$\,keV events is fitted with the theoretical expression of $f_q(q)$, obtaining $N=3.08\pm0.33$. On the right, software efficiency value, $\epsilon'$, as a function of $a$ parameter, leading to a correction parameter, $a=0.72$. After applying the correction, the events at $3$\,keV recover from a $72\%$ efficiency without correction to the imposed $85\%$ efficiency.  }}}
\label{fi:multDists}
\end{figure}

In practice, the parameter $a$ is determined by calculating the efficiencies at $3$\,keV and $6$\,keV, and determining the $a$ factor that makes them equal. The value of $a$ is usually close to $1$ (see Fig.~\ref{fi:multDists}).

\subsection{SOFM method.}

An alternative method based on a branch of \emph{neural networks} was developed for the selection of X-ray events in Micromegas detectors.

\vspace{0.2cm}

\begin{figure}[!ht]
{\centering \resizebox{0.5\textwidth}{!} {\includegraphics{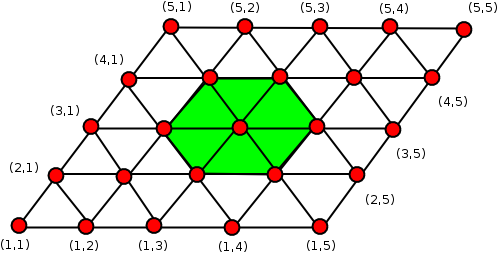}} \par}
\caption{\fontfamily{ptm}\selectfont{\normalsize{Hexagonal network topology. The neurons, or nodes, are described by a bidimensional coordinate system. }}}
\label{fi:topomap}
\end{figure}

The Self Organized Feature Maps (SOFM)~\cite{1402-4896-39-1-027} is based on the definition of a topological map, this map describes a network of interconnected nodes, or neurons. Each node keeps a correlation with its nearby neighbors as a function of the distance between them in the topological space. For simplicity, these topological networks are chosen to be squared or hexagonal, in our case we have chosen an hexagonal network given the higher directional independency that provides (see~Fig.~\ref{fi:topomap}).

\vspace{0.2cm}

\vspace{0.2cm}

Once the topological space is defined by means of a coordinate system, each neuron is associated to a k-dimensional \emph{weights} vector. The weights vector of each neuron will be used to identify the neuron with a given type of event.


\vspace{0.2cm}

In order to use the network as a method for selecting X-ray events, the network needs to be \emph{trained}. When the network is trained, events with similar characteristics are grouped together in the network topological space. The weights vector is updated during the network training in order to obtain this result. X-ray events should be identified by neurons that are inside a neighborhood.

\vspace{0.2cm}

This idea has been used in order to distinguish an X-ray event from any other kind of event produced in the chamber. The network is trained using background events, mainly composed by cosmic rays, electronic noise, and other processes we are interested to discard. Once the network has been trained, it is ready to identify or classify events with different characteristics which are defined by its pattern vector~$\tilde{X}_i$.

\vspace{0.2cm}

At the end of the training, or learning process, only a subset of neurons will be activated when an X-ray pattern is presented to the net. The neurons that identify the X-ray events are determined by using the pattern vectors generated with an $^{55}$Fe source. 

\vspace{0.2cm}

When the X-ray cells have been determined, it is possible to obtain the X-ray background from the data. In order to optimize the rejection of non X-ray events, only the most significative neurons, the ones that identify a higher amount of X-ray events, are used for background selection. This allows to increase the rejection power of the network by discarding some of the neurons that identified a few percentage of X-ray events by introducing a given \emph{software efficiency} to the method.

\subsubsection{Network training.}

The information provided by each Micromegas event is parameterized by a \emph{pattern} vector, $x^n$, using some of the discriminants defined in section \ref{sc:discriminants}, where $n$ identifies the event process. The \emph{pattern} vector components are renormalized using the variance obtained from the set of events that will be used to train the network, $x^n_i/\sigma^n_i$, so each component has an equivalent contribution to the metric space defined by them.

\vspace{0.2cm}

When a new event pattern is shown to the network, each neuron evaluates the distance of its weights vector with the event pattern vector. The neuron that is closest to this pattern vector is denominated the \emph{winner} neuron, and it becomes active.

\vspace{0.2cm}

The training of the neural network is an iterative process. In each iteration, all the event patterns from a background data set are shown to the network. In order to obtain a positive result in the training of the network the number of event patterns available must be as high as possible.

\vspace{0.2cm}

In each iteration, when a new event pattern, $x^n$, is presented to the network, the weights vector of the winner neuron, $w_i$, and their neighbors are updated in such way that their value is closer to the event pattern vector.

\vspace{0.2cm}

The weights vector is updated using the following relation,

\begin{equation}
w_{ik}(t+1) = w_{ik}(t) + \alpha(t) * h(d_{ig}, R(t)) \cdot ( x^n_k - w_{ik} ) 
\end{equation}

\noindent where $i$ denotes the neuron to be updated, $k$ the weights vector component and $t$ the iteration number. The function $\alpha(t) < 1$ is a parameter denominated \emph{learning rate}, and its value decreases as the iteration number increases. The function $h(d_{ig}, R(t))$, or neighborhood function, defines the topological network region, centered in the winner neuron, where the weights update will have effect. A wide type of neighbors functions can be used in this type of analysis; for simplicity, in this study it has been used a Gaussian shaped function 


\begin{equation}
h(d_{ig}, R(t)) = exp(-d_{ig}/R(t) )
\end{equation}

\noindent where $R(t)$ is the denominated neighborhood radius, that in the same way as $\alpha(t)$ decreases with the iteration number, and $d_{ig}$ is the topological distance between the winner neuron, $n_g$, and the neuron that updates its weight, $n_i$. The updating weights algorithm is iterated over all the neurons in the net, but only the closest cells to the winner neuron have a substantial change of its weight vector.

\vspace{0.2cm}

The evolution of $R(t)$ used in the weights updating process is calculated as a function of the iteration number using the following expressions,

\begin{equation}
R(t) = R_o + \frac{I_t}{I_f} (R_f - R_o)
\end{equation}

\noindent where $I_t$ is the iteration number and $I_f$ is the total number of iterations to be used to complete the learning process. The neighborhood radius varies from $R_o$ to $R_f < R_o$ during the learning process resulting in lower influence to the neighbor neurons when the training is reaching the end.

\vspace{0.2cm}

The learning rate function is in charge of smoothing the influence of the event pattern presented to the network in the weights updating process. It is calculated as follows

\begin{equation}
\alpha(t) = \alpha_o + \frac{I_t}{I_f} (\alpha_f - \alpha_o)
\end{equation}

\noindent where $\alpha(t)$ evolves from $\alpha_o$ to $\alpha_f < \alpha_o$ during the learning process.

\vspace{0.2cm}

\begin{figure}[hb!]
\begin{center}
\begin{tabular}{c}
\includegraphics[width=\textwidth]{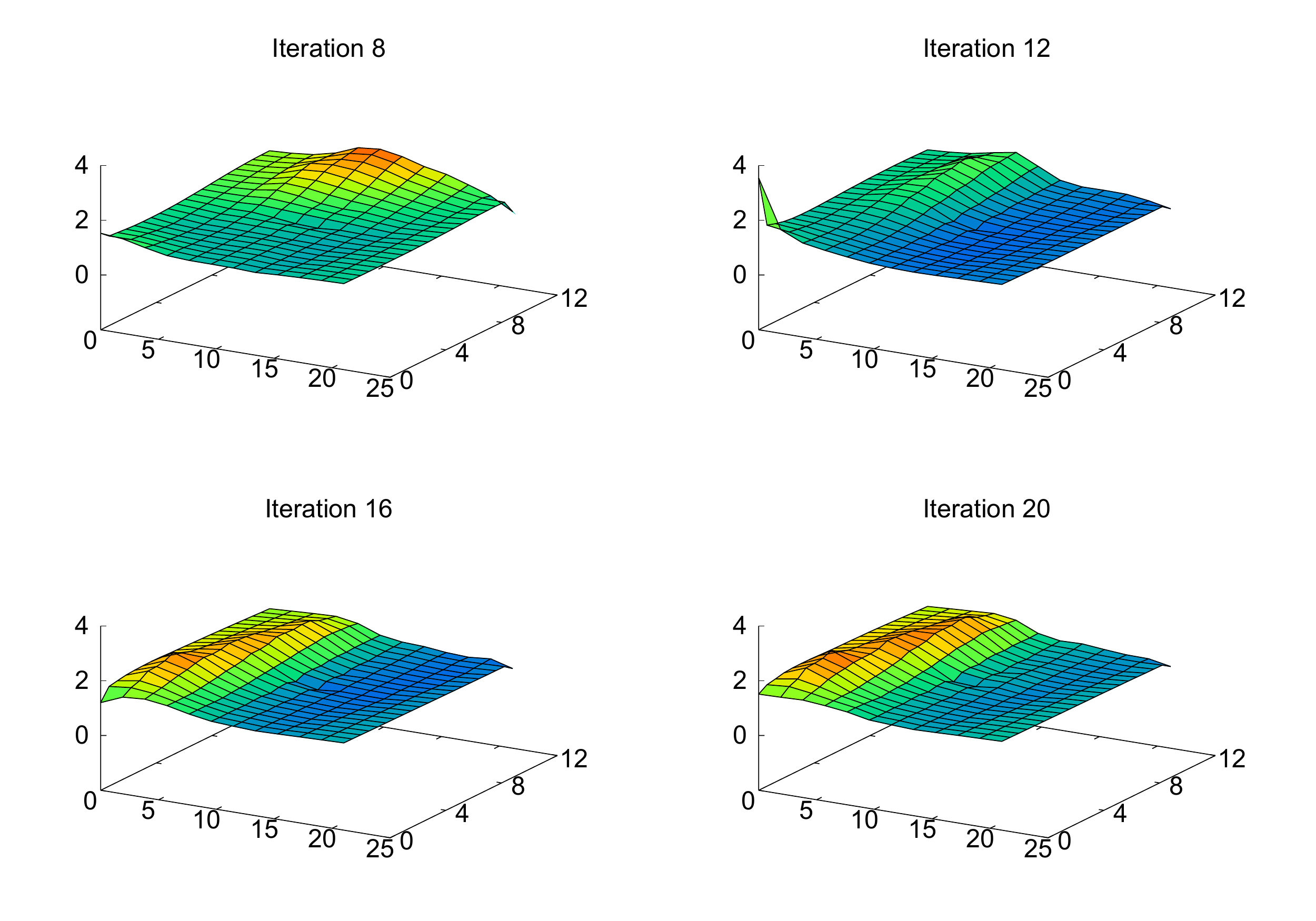} 
\end{tabular}
\end{center}

\caption{\fontfamily{ptm}\selectfont{\normalsize{Evolution of pulse risetime weights component during the learning process in the topological network space, for a network of 20x20 nodes with hexagonal distribution.}}}
\label{fi:rtWeightEvolution}

\end{figure}

The initial and final values are arbitrary parameters and are chosen empirically, testing different values, and using the ones that better suit to the right weights network evolution. Figure~\ref{fi:rtWeightEvolution} represents the evolution of the \emph{pulse risetime} weights vector component at the topological network during the network training process.

\subsubsection{Testing the SOFM method.}\label{sc:TestingSOFM}

In order to test the method a given set of data from the $V$-branch Micromegas detector was used. During the network trainning were used a total of 20~iterations, showing in each iteration all the background patterns selected; around $23,000$\,events taken from about 276~hours of background data. 

\vspace{0.2cm}

The weights vectors of each neuron are initially set with a random value uniformly distributed around the mean value of the patterns to be shown for each component. During the trainning, the weights vector of each neuron evolves to the typical values of a given type of event present in the background, keeping some correlation with its neighborhood (see~Fig.~\ref{fi:rtWeightEvolution}).

\vspace{0.2cm}

During the trainning process, the values for the neighborhood radius used were $R_o = 5$ and $R_f = 1$, and for the learning rate $\alpha_o = 1$ and $\alpha_f = 0.01$.

\vspace{0.2cm}

Once the network has been trained the values of the weights vectors are not updated anymore, remain fixed. The known X-ray event patterns coming from an $^{55}$Fe source are then given to the network in order to determine which cells have specialized in detecting X-ray events, which are the ones that have a weights vector comparable to the X-ray pattern.

\vspace{0.2cm}

\begin{figure}[!ht]
{\centering \resizebox{\textwidth}{!} {\includegraphics{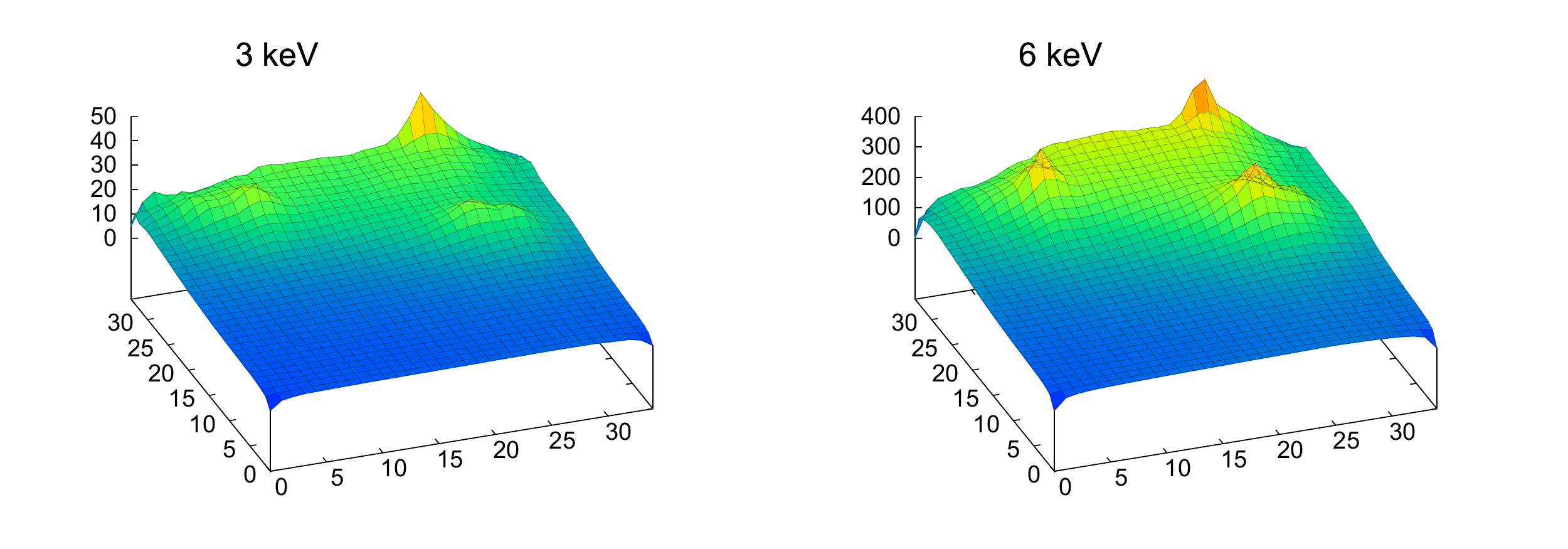}} \par}
\caption{\fontfamily{ptm}\selectfont{\normalsize{A representation of the frequency that neurons in the topological map are activated in response to calibration events coming from an $^{55}$Fe source. On the left plot, neurons activated by events coming from the escape peak, and on the right plot, events coming from the main peak.}}}
\label{fi:calNeuronHitmap}
\end{figure}

In order to obtain the X-ray calibration map of the neural network were used around 230,000 events coming from about 13 hours of different calibration runs. Each event is associated with its corresponding winner neuron, counting the number of times that each neuron becomes active is obtained the relevance of that neuron in detecting X-ray like events (see Fig.~\ref{fi:calNeuronHitmap}).

\vspace{0.2cm}

\begin{figure}[!ht]
{\centering \resizebox{\textwidth}{!} {\includegraphics{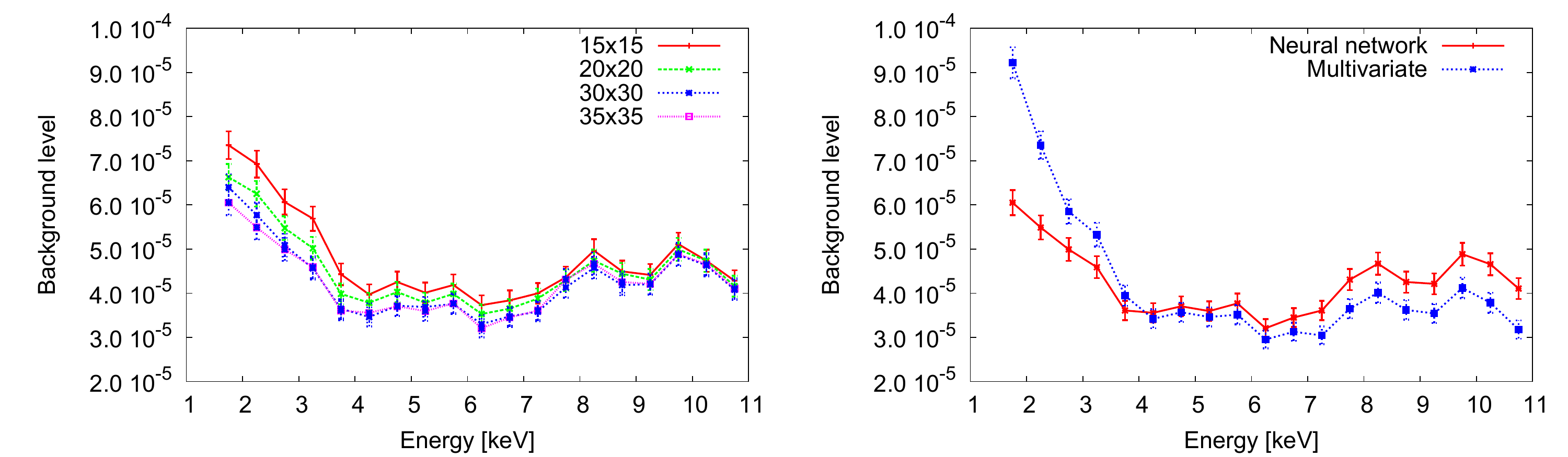}} \par}
\caption{\fontfamily{ptm}\selectfont{\normalsize{On the left, X-ray background spectra obtained for different network sizes. A major reduction is observed at low energies as the network size is increased. On the right, the comparison between the background spectrum obtained with the SOFM method and the multivariate method described. }}}
\label{fi:NNspectras}
\end{figure}

The optimum network size was obtained for a net of 35x35 neurons. The network size definition problem is usually one of the most important parameters that affect to the result obtained in SOFM methods. In our case, it was observed an improved background rejection as the network was larger, but without further improvement after a network size of 30x30 (see Fig.~\ref{fi:NNspectras}). Probably the size is bounded by the statistics available.

\section{Comparison between the methods described.}

\begin{figure}[!ht]
{\centering \resizebox{0.9\textwidth}{!} {\includegraphics{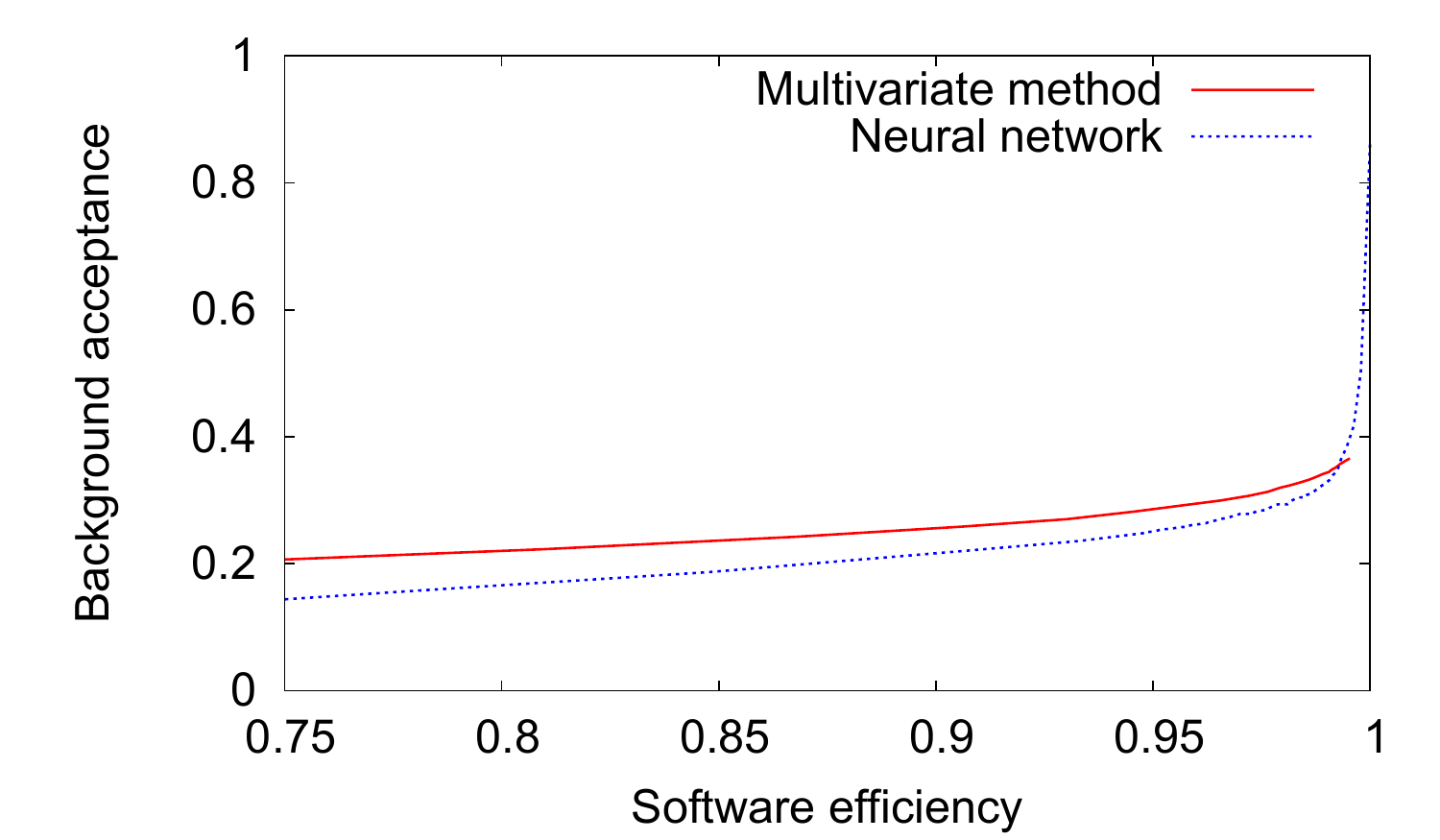}} \par}
\caption{\fontfamily{ptm}\selectfont{\normalsize{ Comparison between the background acceptance as a function of the software efficiency for the discrimination methods studied. }}}
\label{fi:efficiency}
\end{figure}

The modified multivariate method and the SOFM method have been compared by analyzing the background data set used to test the SOFM method (section~\ref{sc:TestingSOFM}) at different software efficiencies for both methods. The rejection power at each efficiency is described by the percentage of background events accepted after applying the selection rules imposed by each method (see Fig.~\ref{fi:efficiency}).

The final background reduction obtained with the SOFM method improves the one reached by the modified multivariate method for any given software efficiency. This fact is associated to a better definition of the \emph{selection volume} that can be provided by the neural network method, which in practice is defining a \emph{selection volume} for each X-ray network cell.

\section{Optimum software efficiency.}\label{sc:optimumeff}

The background reduction carried out through the event selection leads to a software efficiency loss by rejecting a minimal fraction of calibration events. However the lost in X-ray detection efficiency affects directly to the sensitivity of the experiment. It is required to ponderate the benefits in sensitivity due to background rejection with the disadvantage of losing some expected axion signal. 

\vspace{0.2cm}

The sensitivity of the experiment can be quantified by the axion-photon coupling value $g_{a\gamma}$ that is able to reach. An approach to the limit on the measurable coupling was given in expression~\ref{eq:g}, written as a function of the X-ray background level and efficiency of the detector.

\vspace{0.2cm}

The value for software efficiency can be easily chosen by modifying the size of the \emph{selection volume} in a given method, in order to modify its acceptance. This allows to analyse a sufficient amount of background data obtaining the background reduction at different software efficiency values.

\vspace{0.2cm}

The optimum software efficiency loss to be used in the Micromegas background data can be obtained by minimizing the expression \ref{eq:g}. That is the same as maximizing the relation $\epsilon/\sqrt{b}$, related with the \emph{detector discovery potential}. 

\vspace{0.2cm}

\begin{figure}[!ht]
{\centering \resizebox{0.95\textwidth}{!} {\includegraphics{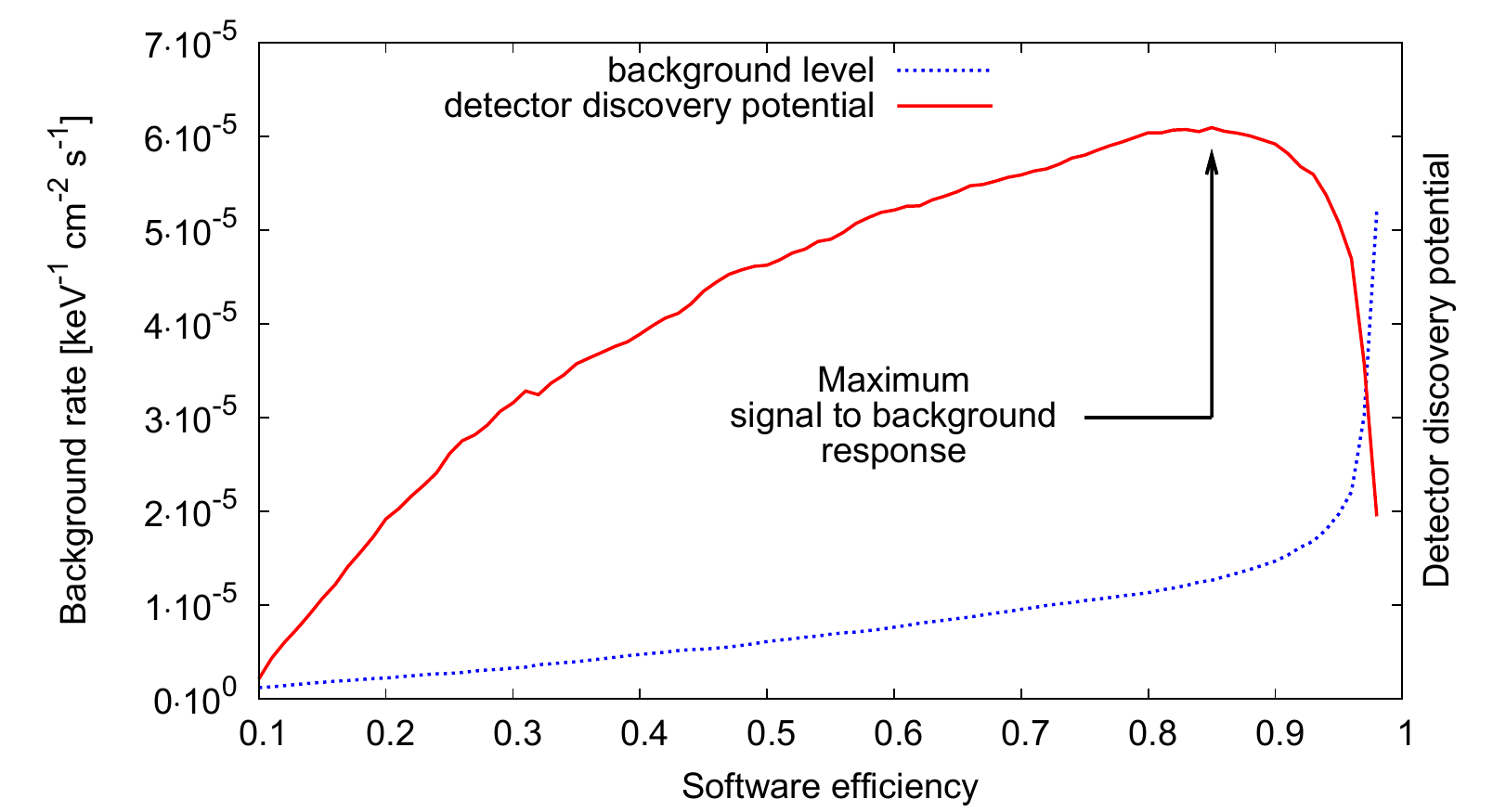}} \par}
\caption{\fontfamily{ptm}\selectfont{\normalsize{Background analysis extracted from a data set of a Bulk detector taking data in the Micromegas Sunrise CAST line. The plot is obtained using 386~hours of background data. In dashed blue line, the background level of the detector as a function of the software efficiency. In red, the equivalent detector discovery potential giving the maximum signal response at a software efficiency of 85~\%.}}}
\label{fi:optimumEff}
\end{figure}

Figure~\ref{fi:optimumEff} presents the background level reached at each imposed software efficiency. As we observe in this plot, the background rejection is intense up to a efficiency of 90\%. However, as the software efficiency is decreased the background level of the detector evolves to a linear decrease with the signal efficiency, which reduces the signal to background potential in the coupling sensitivity. The maximum value for \emph{detector discovery potential} as a function of the software efficiency determines the optimum software efficiency to be used in the CAST experiment.

This analysis was performed independently for each Micromegas detector running in the Micromegas Sunrise line of the CAST experiment. Bulk and microbulk technologies, running in the $^3He$ Phase, presented an optimum efficiency at around 85\%, while the conventional Micromegas technology running in the $^4He$ Phase showed an optimum signal to background ratio at 92\% of software efficiency.




\section{Effect of discrimination in background data.}

This section presents the background reduction on the overall rawdata acquired, the Microbulk detector taking data in the Sunrise line is used as a representative of Micromegas detectors which show similar event patterns. The background events shown correspond to those events that have passed the preliminar cuts described on chapter \ref{chap:rawdata}, which are those where a unique cluster has been detected in X and Y strip planes.

\vspace{0.2cm}

The background rejection effect on the energy spectrum is presented in figure~\ref{fi:bckCutsrtVsWd}. It is noticed how the background selection provides to the energy spectrum a more defined structure. The overall background acceptance in the 2-7\,keV range is around 5\% for Bulk and Microbulk detectors, while it was about 20\% for conventional detectors (see Fig.~\ref{fi:optimumEff}), fact that is mainly attributed to the presence of the new shielding installed at the time these detectors started to operate in CAST.

\begin{figure}[!h]
\begin{center}
\begin{tabular}{cc}
{\centering \resizebox{0.48\textwidth}{!} {\includegraphics{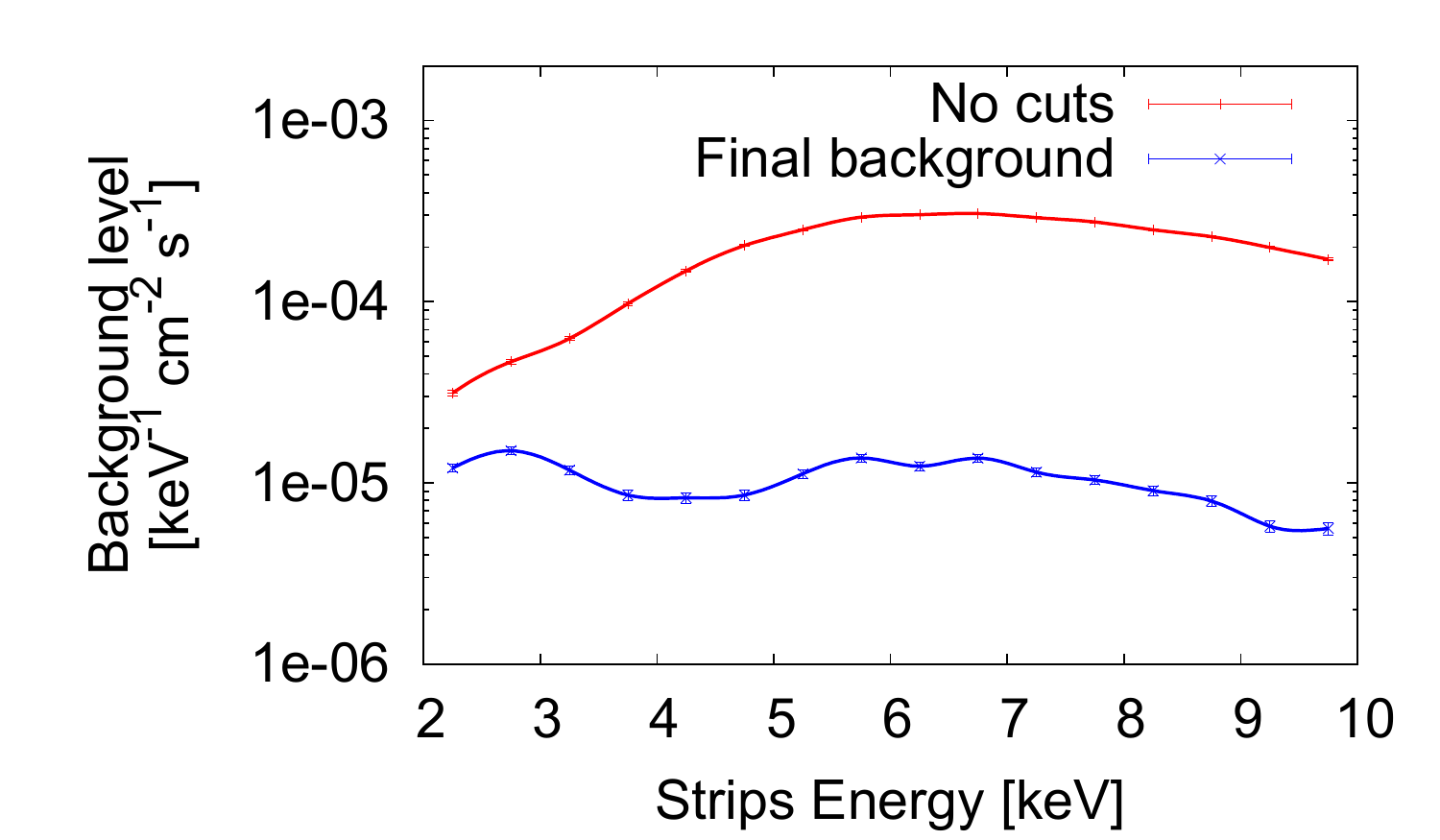}} \par} &
{\centering \resizebox{0.48\textwidth}{!} {\includegraphics{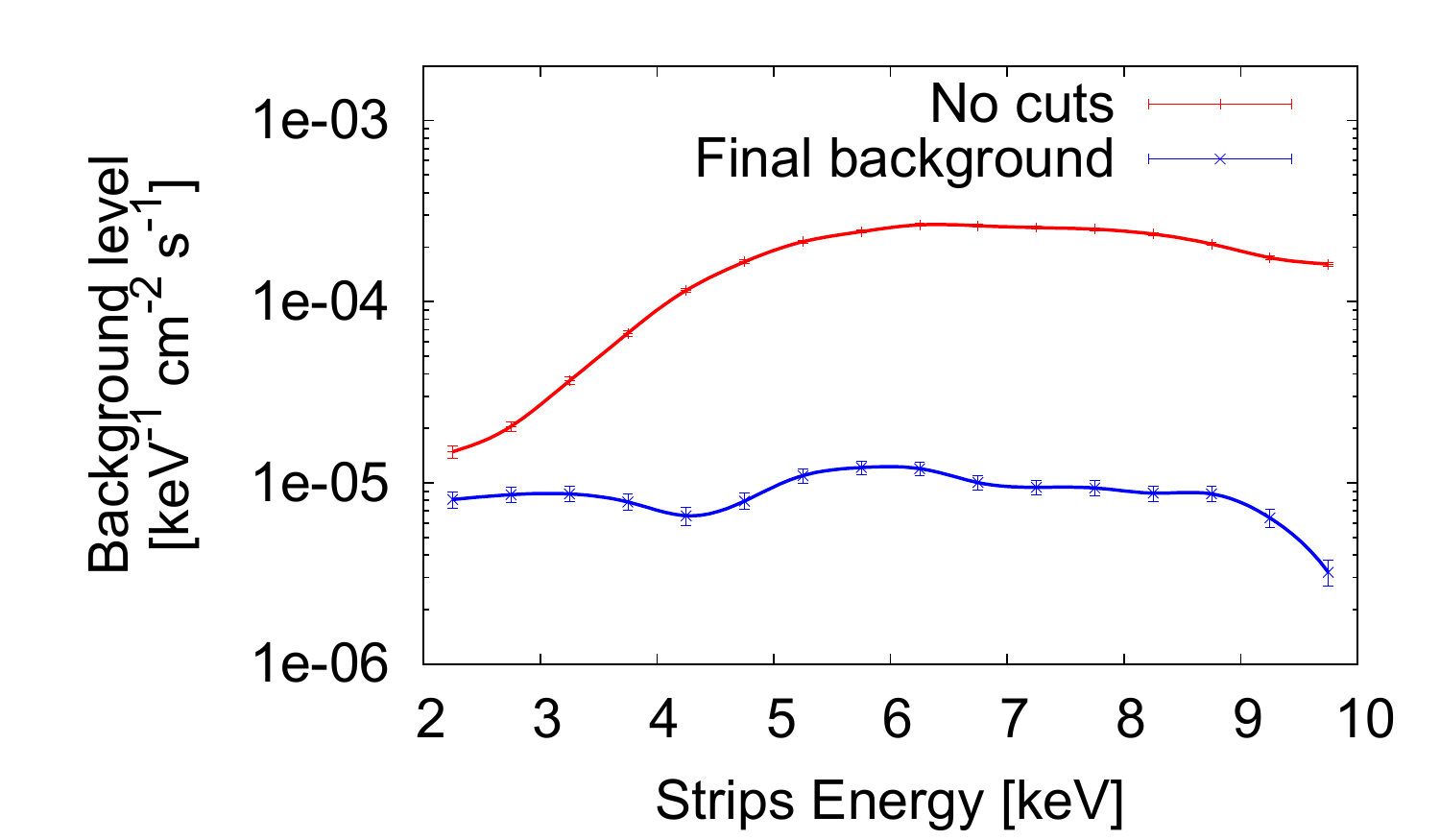}} \par} \\
\end{tabular}
\caption{\fontfamily{ptm}\selectfont{\normalsize{ Effect of discrimination on background energy spectrum for a Bulk detector (left) and a Microbulk detector (right).   }}}
\label{fi:bckCutsrtVsWd}
\end{center}
\end{figure}

\begin{figure}[!h]
\begin{tabular}{cc}
\resizebox{0.5\textwidth}{!} {\includegraphics{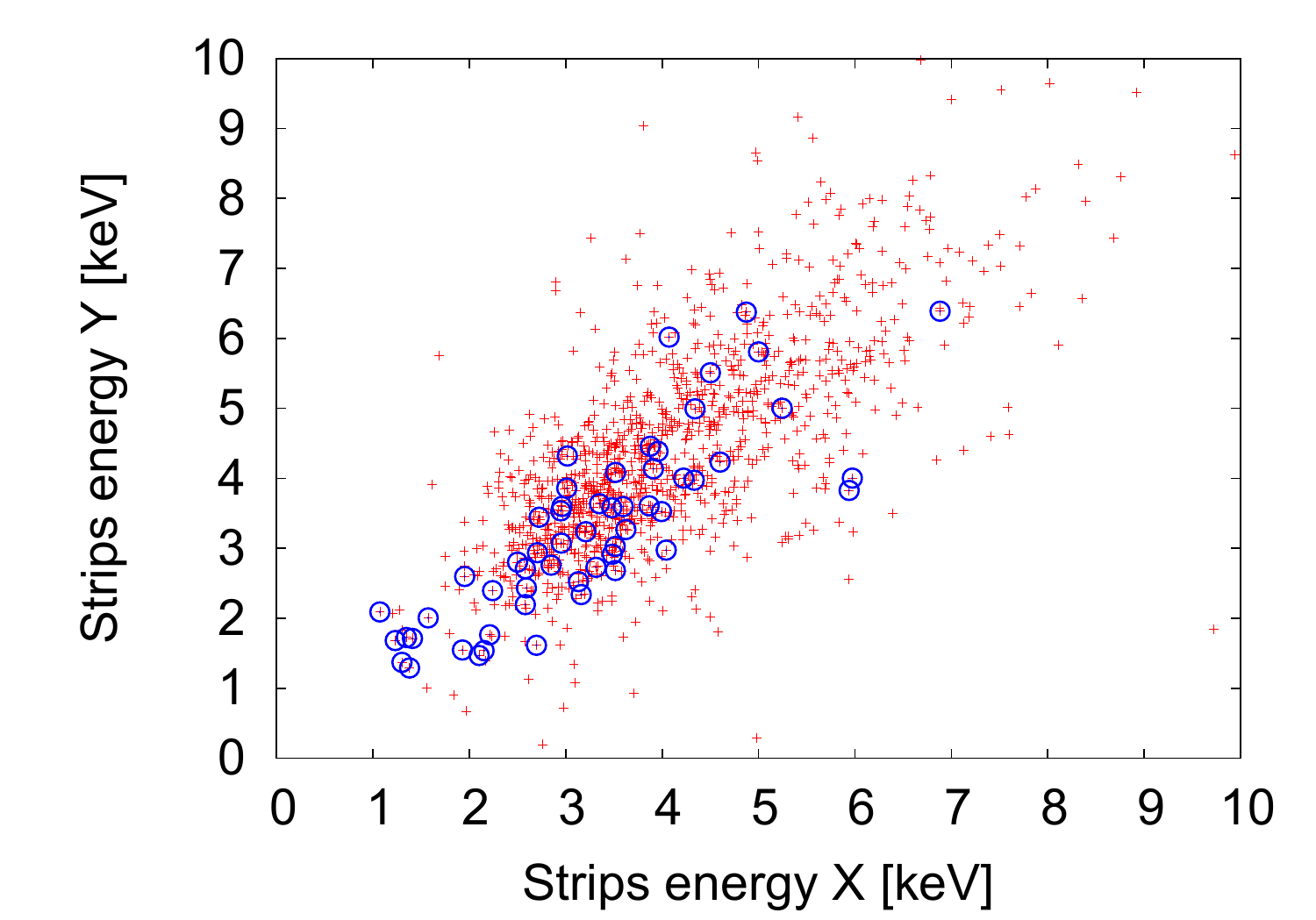}} &
\resizebox{0.5\textwidth}{!} {\includegraphics{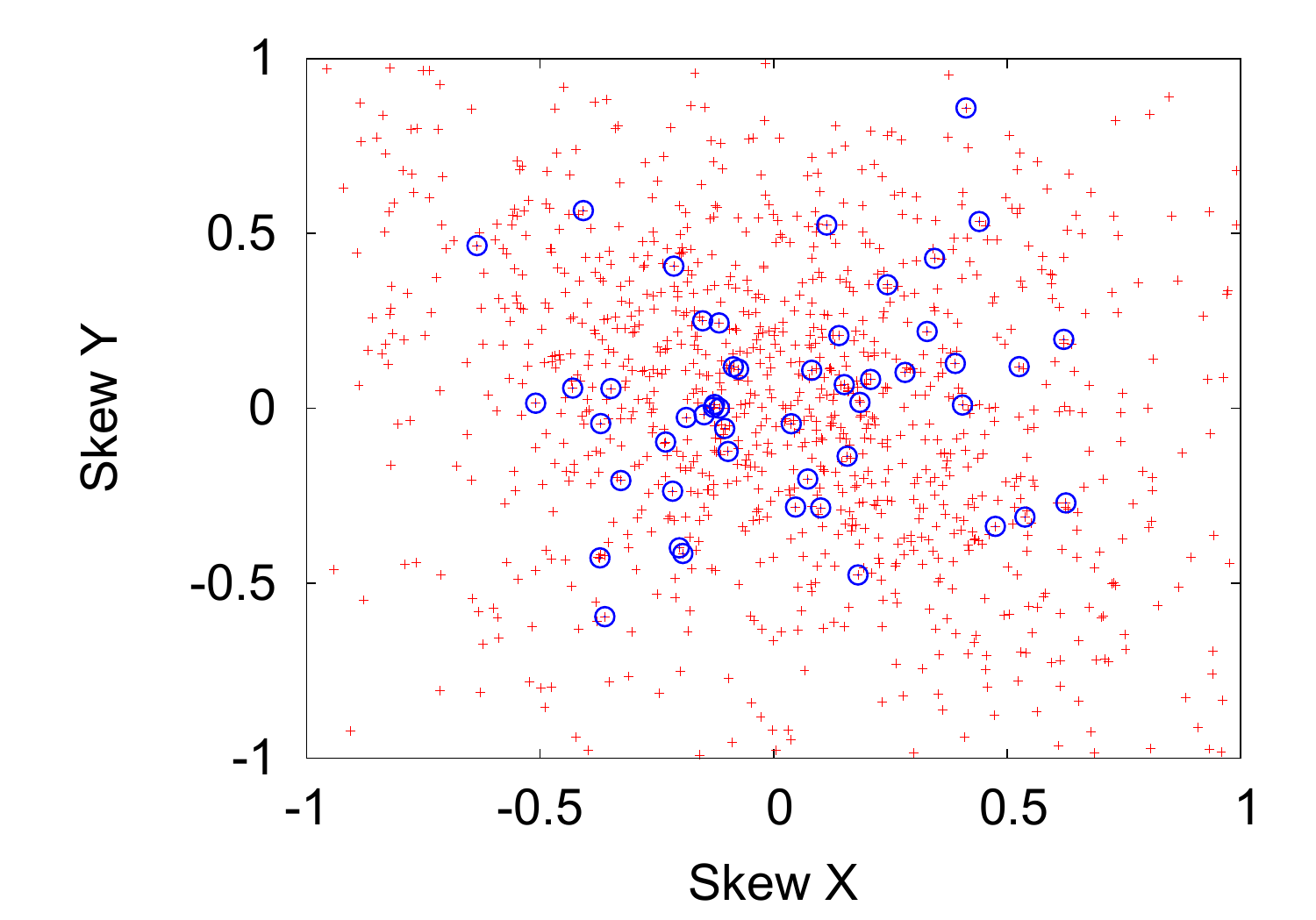}} \\
\end{tabular}
\caption{\fontfamily{ptm}\selectfont{\normalsize{ Background distribution maps for discriminants related with cluster assymmetries. Blue circles identify the background selected events likely to be an X-ray. }}}
\label{fi:bckCutsChargeSkew}
\end{figure}

\begin{figure}[!h]
\begin{tabular}{cc}
{\centering \resizebox{0.5\textwidth}{!} {\includegraphics{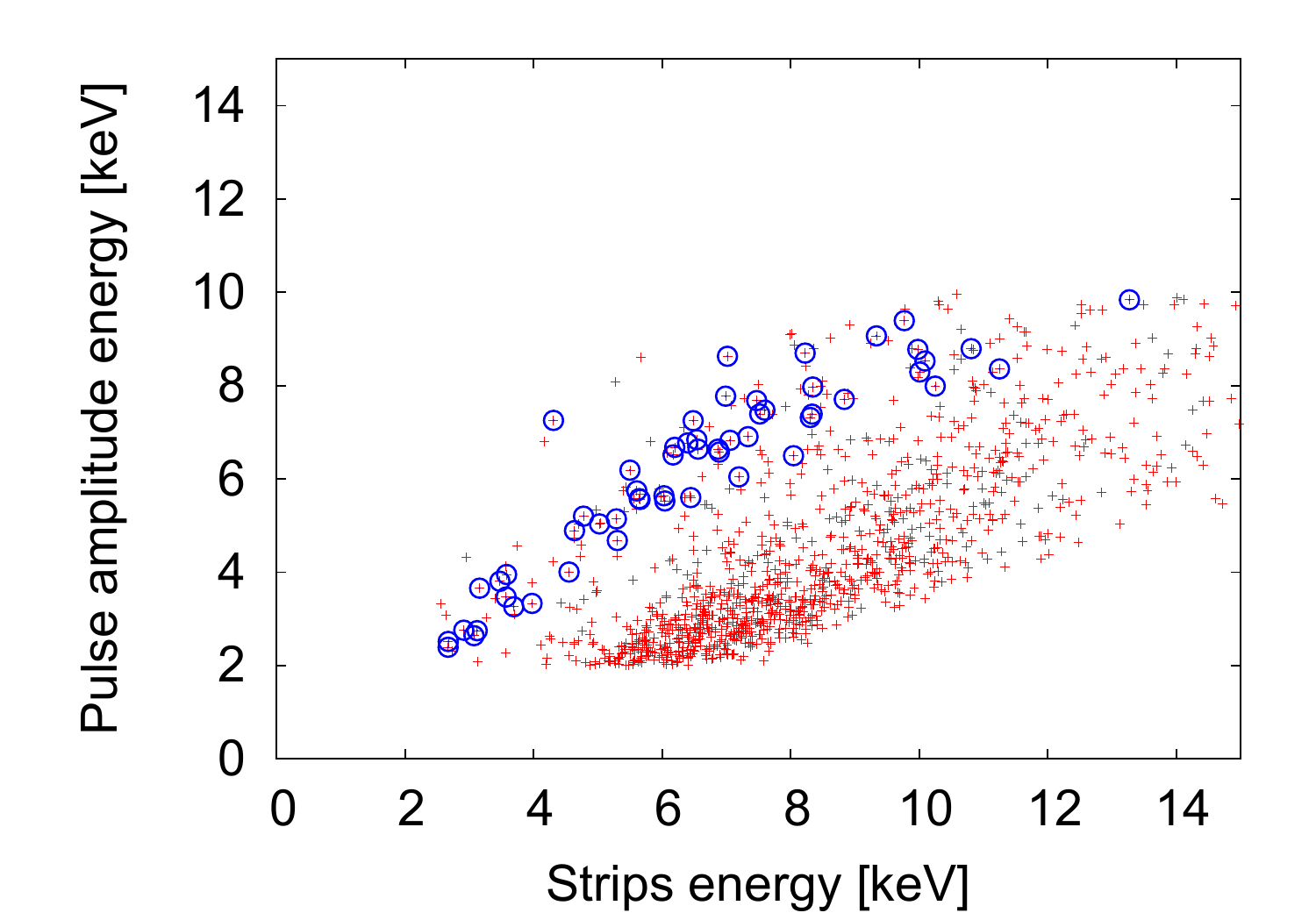}} \par} &
{\centering \resizebox{0.5\textwidth}{!} {\includegraphics{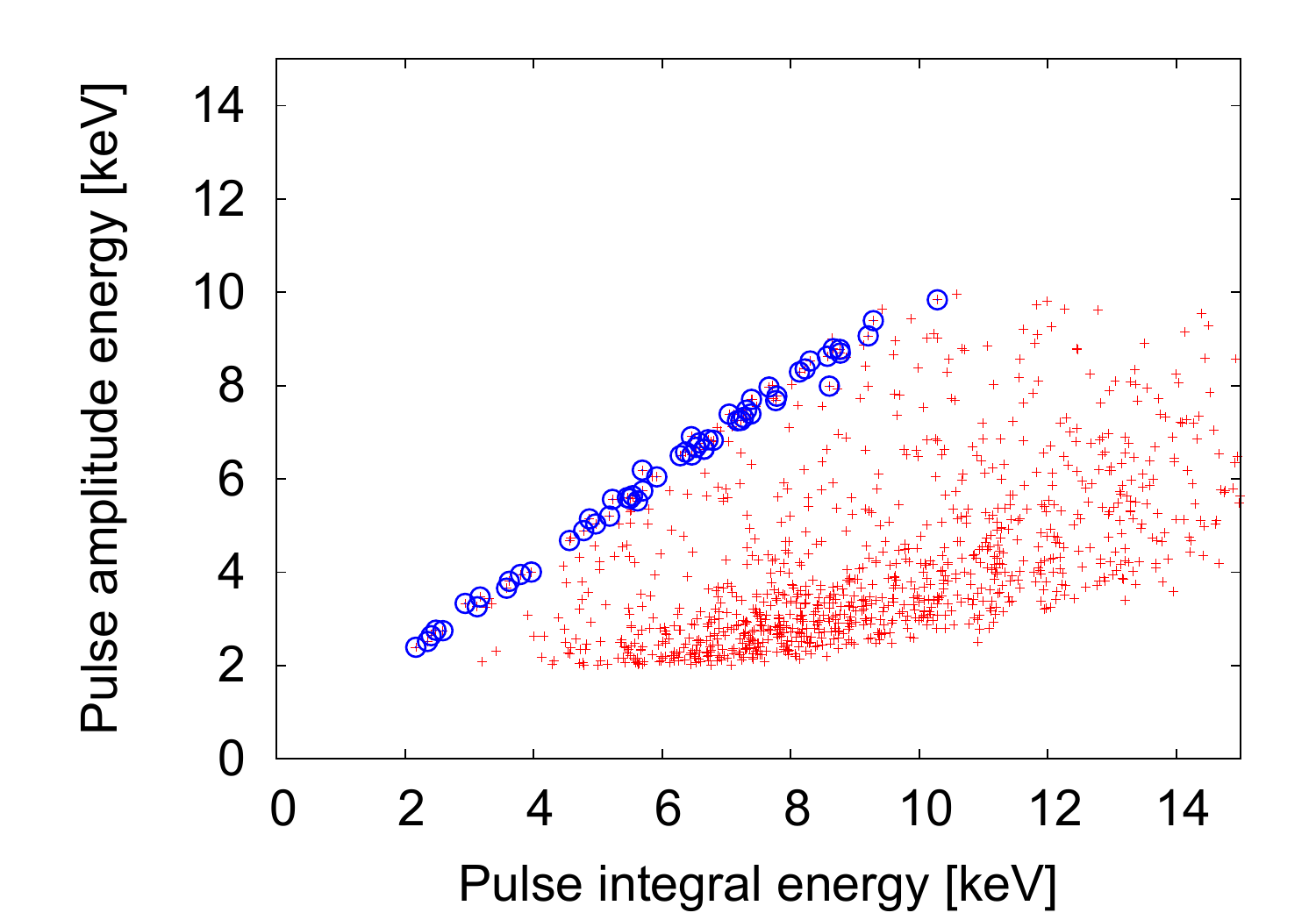}} \par} \\
\end{tabular}
\caption{\fontfamily{ptm}\selectfont{\normalsize{ Energy distribution maps from the different event energy definitions for background events. Blue circles identify the background selected events likely to be an X-ray. }}}
\label{fi:bckCutsEnHeightVsEnIntegral}
\end{figure}

\begin{figure}[!h]
\begin{tabular}{cc}
{\centering \resizebox{0.5\textwidth}{!} {\includegraphics{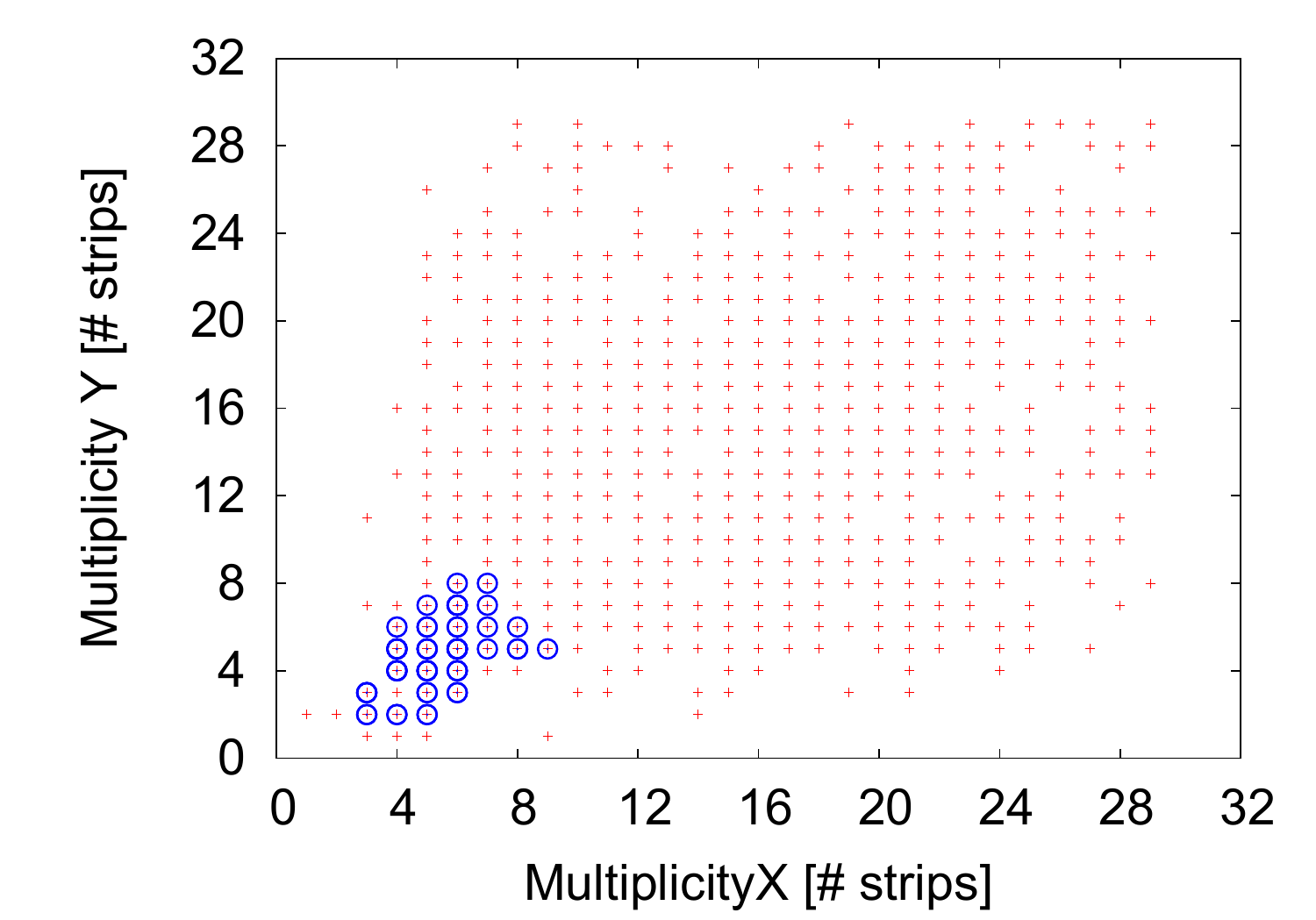}} \par}
{\centering \resizebox{0.5\textwidth}{!} {\includegraphics{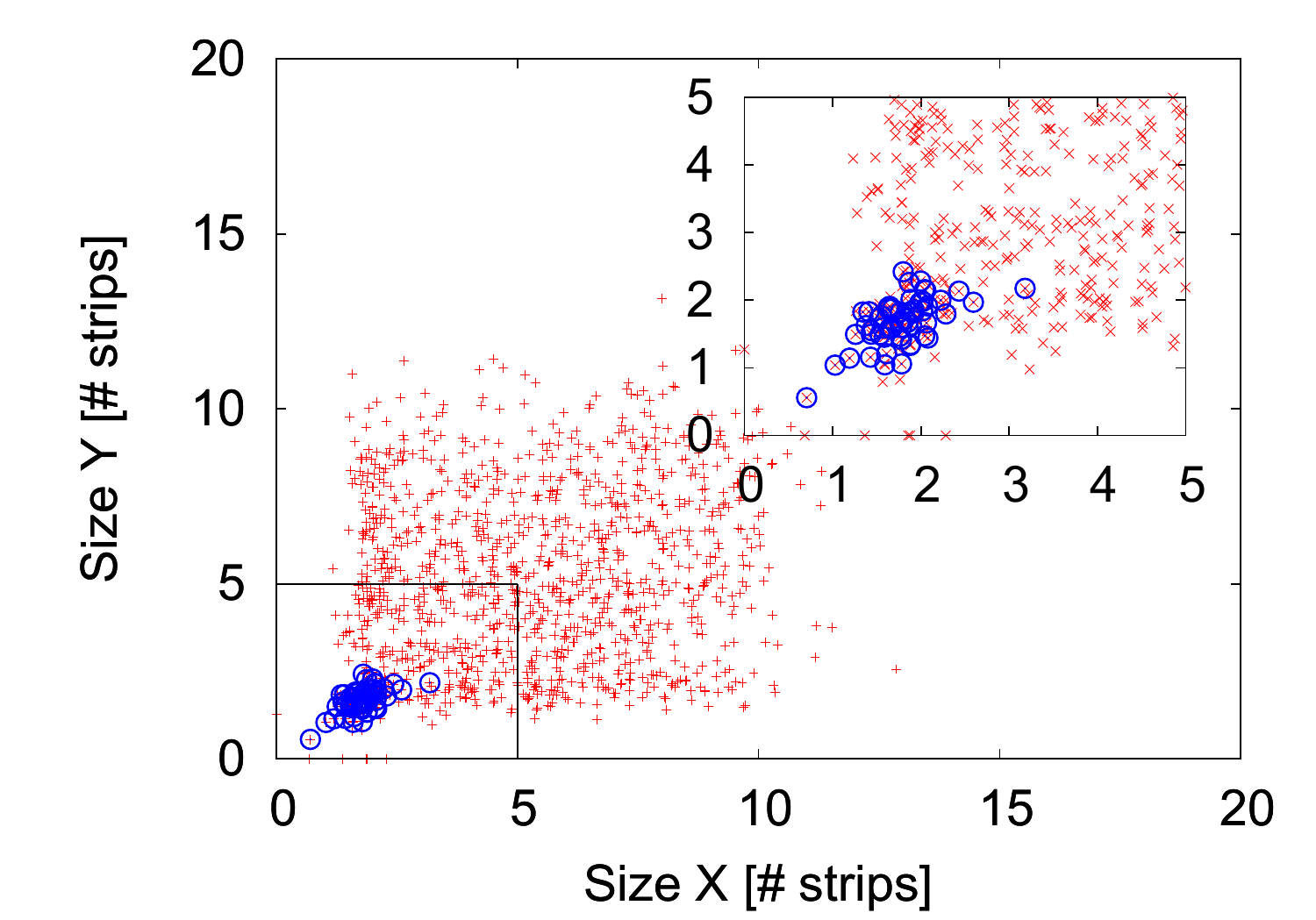}} \par}
\end{tabular}
\caption{\fontfamily{ptm}\selectfont{\normalsize{ Background distribution maps for discriminants related with cluster size. Blue circles identify the background selected events likely to be an X-ray.  }}}
\label{fi:bckCutsMultiplicity}
\end{figure}

Figures~\ref{fi:bckCutsrtVsWd}, \ref{fi:bckCutsChargeSkew}, \ref{fi:bckCutsEnHeightVsEnIntegral}, \ref{fi:bckCutsMultiplicity} show the background distributions map of the main parameters used to calculate the \emph{discriminant observables} defined on section~\ref{sc:discriminants}, where the final selected background events are highlighted.

\vspace{0.2cm}

The \emph{pulse risetime} and \emph{pulse width} (see Fig.~\ref{fi:bckCutsrtVsWd}) correspond to the set of best observables to be used in an X-ray selection analysis (as detailed on section \ref{sc:optimumdiscriminants}). The discrimination potential of these observables is due to the fact that many events present higher \emph{risetime} and \emph{width} than the typical values for X-ray events.

\begin{figure}[!h]
{\centering \resizebox{0.75\textwidth}{!} {\includegraphics{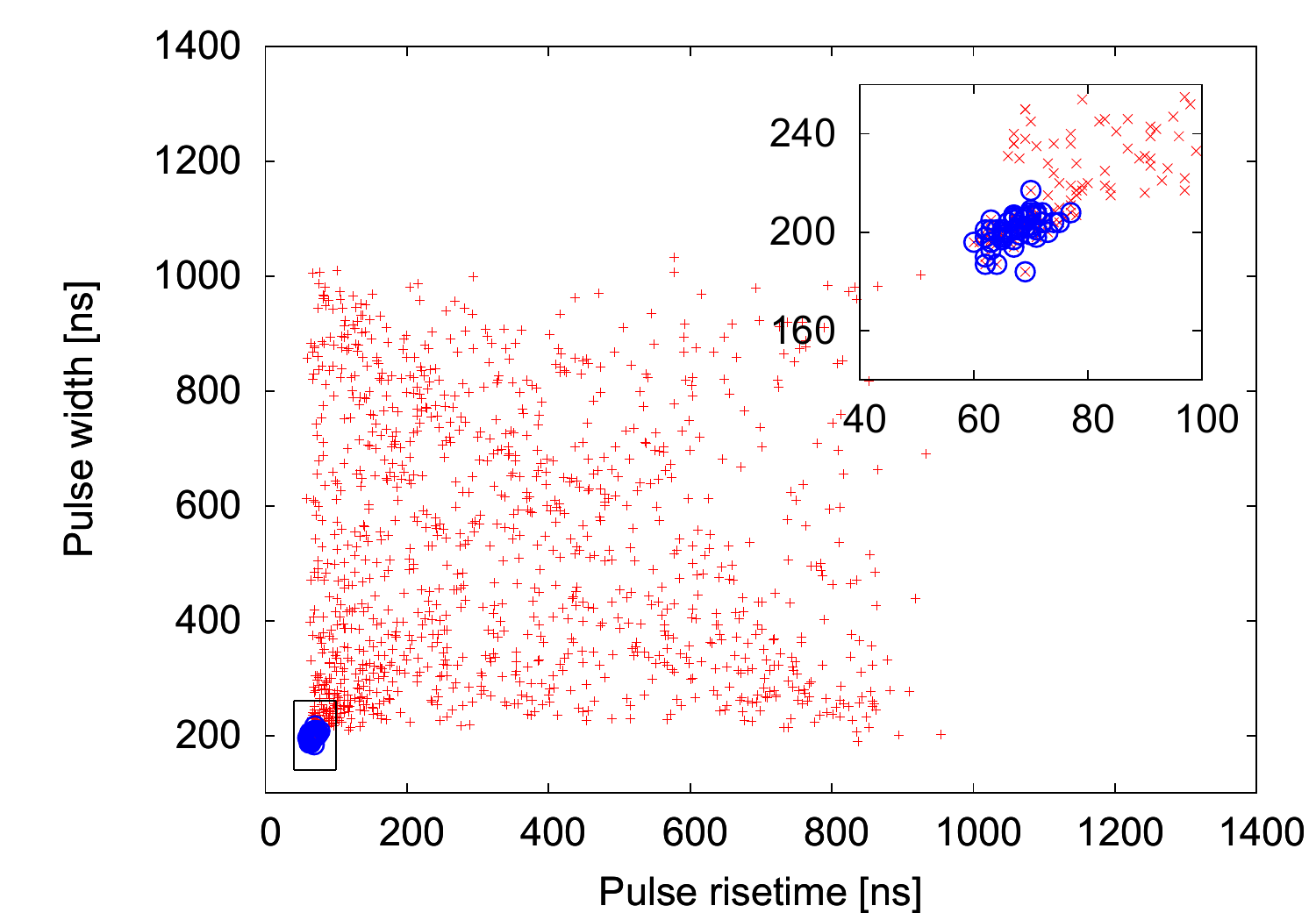}} \par}
\caption{\fontfamily{ptm}\selectfont{\normalsize{ Background distribution map for the discriminants related with the pulse shape. Blue circles identify the background selected events likely to be an X-ray. }}}
\label{fi:bckCutsrtVsWd}
\end{figure}

\vspace{0.2cm}

The \emph{cluster skew} and \emph{cluster charge balance} (see Fig.~\ref{fi:bckCutsChargeSkew}) were presented in section~\ref{sc:optimumdiscriminants} as the worst observables to be introduced in the analysis for increasing the rejection factor of Micromegas detectors. It is observed how the final background events selected, which correspond with short \emph{pulse risetime} events, cannot be distinguished between the events that form the total background at the \emph{skew} and \emph{charge balance} distributions.

\vspace{0.2cm}

The distribution maps corresponding to the different ways to obtain the energy of the event (see Fig.~\ref{fi:bckCutsEnHeightVsEnIntegral}) show that a big population of events are pile up events that are clearly identified by the difference in energies described by the \emph{pulse integral} and the \emph{pulse amplitude}. In this case, the right parameter describing the energy of the event is either the \emph{cluster energy} or the \emph{pulse integral}. However, for localized point-like depositions as X-rays the \emph{pulse amplitude} constitutes a good reference of the event energy. Small energy deviations between the different definitions of event energy, even when the \emph{energy balance} is not introduced in the analysis, represents a good indicator of the goodness of the analysis in the selection of X-ray events. 

The size of the event represented by the clusters size and multiplicity (see Fig.~\ref{fi:bckCutsMultiplicity}) constitute another good candidates for X-ray discrimination since main population of background events detected provide a higher cluster size and multiplicity than the expected low values for X-ray events.

These results show the high capability of selection of particular events as X-rays with micromegas detectors. The low background levels achieved are related with the fact that the main population of background events acquired are easily rejected by using the temporal and spatial information provided with the detector.

\chapter{The leak problem during 2008.}
\label{chap:leak}
\minitoc


\section{Leak problem description.}

During the data taking period in 2008 the CAST magnet was operating with the presence of a leak in the Helium system. The effect of the leak was observed for the first time at the end of 2008. Unfortunately, the leak was present in the system since the first time the cryostat was closed and cooled down at the end of 2007.

\vspace{0.2cm}

\begin{figure}[!hb]\begin{center}
\includegraphics[width=\textwidth]{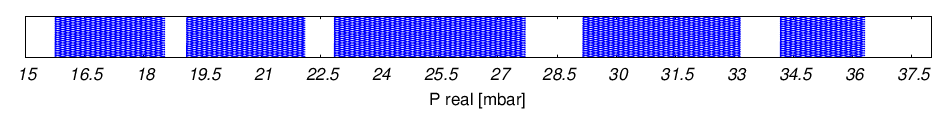}
\caption{An schematic of the \emph{gaps} produced during 2008 data taking period due to several breaks in which the cold bores where emptied and the magnet was re-filled to the point where we should be in the case a leak would not be present.}
\label{fi:gapSchema}
\end{center}\end{figure}

The leak affected the desired density that was expected to be covered for each pressure step during that year. The presence of the leak was actually shifting the desired density towards lower values. Whenever there was a break or small shutdown within the data taking period (due to a magnet quench or a cold windows bake-out) the magnet was emptied and re-filled back (see table \ref{ta:fillingStarts}) to the last density step scanned. These cold bore re-fillings produced several non covered axion mass \emph{gaps} due to the fact that the density step in which we actually stopped was lower than we expected since the effect of the leak was still not recognized. A preliminary analysis of the available data gave the coverage observed in figure \ref{fi:gapSchema}.

\vspace{0.2cm}

\begin{table}[!ht]
\begin{center}
\begin{tabular}{cccc}
\bf{Date} & \bf{Time} \\
\hline
	&	\\
\vspace{0.1cm}
Tue,2008-04-08 & 15:26 \\
\vspace{0.1cm}
Mon,2008-05-05 & 09:37 \\
\vspace{0.1cm}
Mon,2008-06-02 & 16:43 \\
\vspace{0.1cm}
Thu,2008-07-03 & 09:46 \\
\vspace{0.1cm}
Thu,2008-08-07 & 08:46 \\
\vspace{0.1cm}
Thu,2008-10-16 & 17:13 \\
\end{tabular}
\end{center}
\vspace{-0.3cm}
\caption{Starting times when the CAST magnet was re-filled with gas before of each data taking sub-period in 2008.}
\label{ta:fillingStarts}
\end{table}

\begin{figure}[!ht]\begin{center}
\includegraphics[width=\textwidth]{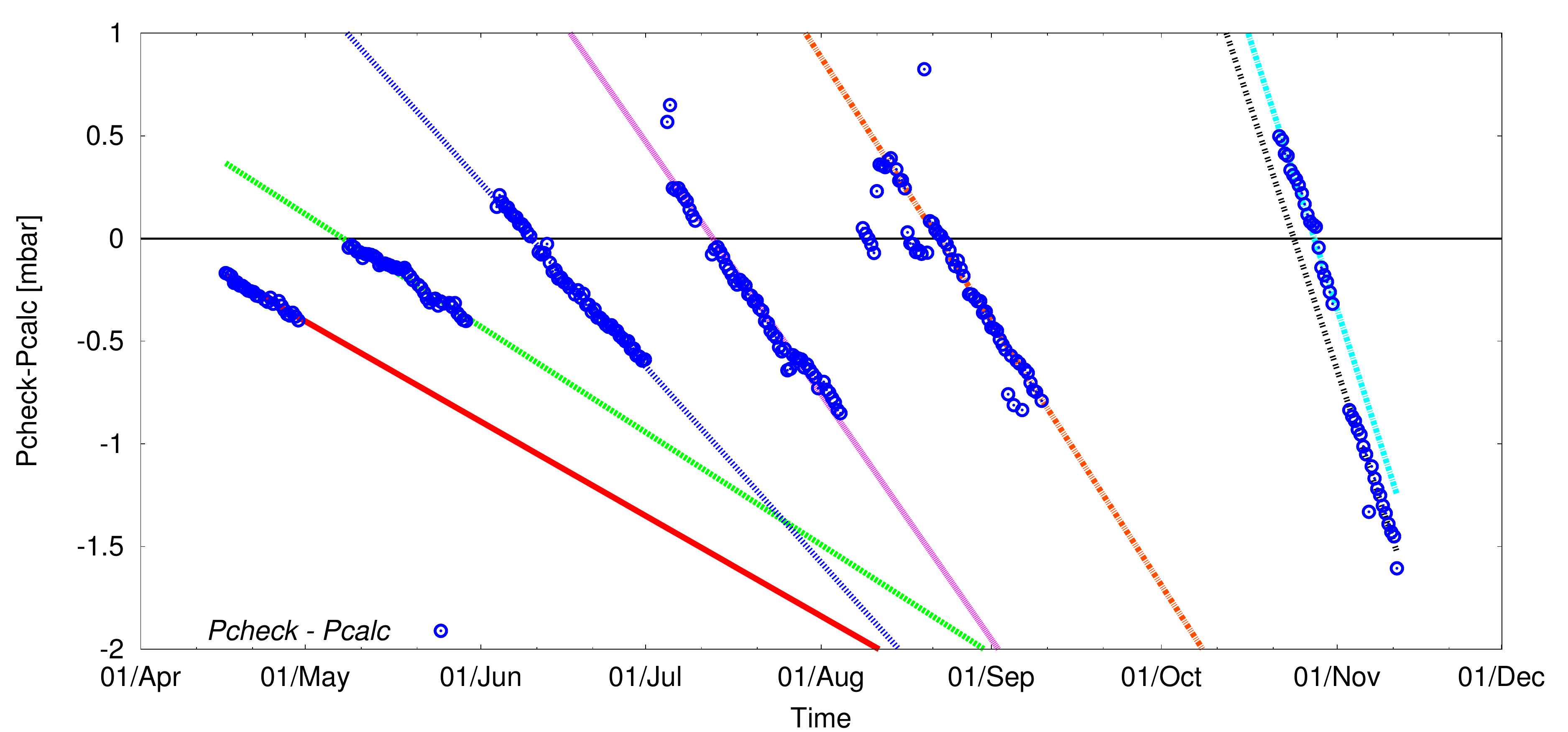}
\vspace{-0.5cm}
\caption{Leak evolution in the different periods after refilling back the cold bore and the corresponding linear fittings. $P_{check}$ is the value of $P_{cb}$ re-normalized by $1.8$\,K$/T_m$, being $T_m$ the measured temperature of the Helium bath surrounding the cold bore gas at the magnetic region. The y-axis defines the deviation of the measured pressure $P_{check}$ from the equivalent calculated pressure, $P_{calc}$. }
\label{fi:leakfits}
\end{center}\end{figure}

The first indication of leak was observed in the value of the pressure sensor, $P_{cb}$, that is inserted inside the pipes containing the $^3$He gas. The value before emptying and after refilling the cold bores was too much biased to be considered as a possible systematic error. After analyzing the evolution of $P_{cb}$ during 2008 data taking period it was concluded that the leak was present during the whole period (see figure \ref{fi:leakfits}). In this plot, the effect of the leak is observed to increase as a function of time, fact that is directly related with the increasing amount of gas inside the magnet.

\vspace{0.4cm}

Figure \ref{fi:leakfits} shows the six data taking sub-periods with starting times given by table \ref{ta:fillingStarts}. The data set chosen to draw this plot is the value of cold bore pressure measured by $P_{cb}$ an hour after the end of each tracking run. Points that are found out of the main tendency are points when the magnet conditions were not stable (due to magnet movement, cold windows bake-out, etc).

\vspace{0.4cm}

The last period presents an offset at the beginning of November, this offset is known to be produced by the effect of the delimiting cold bore window temperatures on the gas distribution, affecting to the value measured by $P_{cb}$. The heaters power provided to the windows was decreased at that time due to a problem with cryogenics resulting in a reduction of the windows temperature from around $70$\,K to $60$\,K. 

\vspace{0.2cm}
The offsets observed at the beginning of each period are mainly related with the fact that $P_{calc}$ does not include corrections with the windows temperature evolution during the data taking period. 


\section{Direct leak measurements }

\begin{figure}[!ht]\begin{center}
\includegraphics[width=0.9\textwidth]{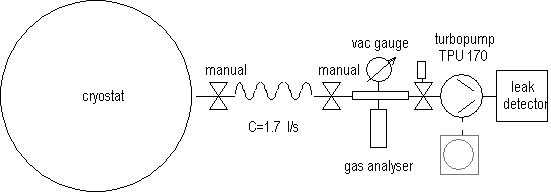}
\caption{ Schematic of the system connected to the CAST magnet cryostat for direct leak detection.}
\label{fi:LeakSearchSchema}
\end{center}\end{figure}

After the suspicion of the leak, on \emph{28th of November, 2008} a mobile pump station was connected to the cryostat (see figure \ref{fi:LeakSearchSchema}) via a manual valve, MFB side, and a long flexible pipe. The station is equiped with a residual quadrupole gas analyzer and a Helium leak detector.\footnote{ The measurements presented in this section are the result of a small part of the work carried out by \emph{Jean-Michel Laurent}, from the TE/VSC group at CERN, for the CAST experiment.}

\vspace{0.2cm}

Before opening the cryostat valve, only the Hydrogen peak ($2\cdot10^{-12}\,$A) was measurable by the gas analyzer. Peaks 3 and 4 were recorded before and after opening the valve (see table \ref{ta:leakGasAnalyser}). The calibration factor (QME120) for $^4$He is $6.4\cdot10^{-5}\,$A/mbar, the corresponding value for $^3$He is unknown and assumed the same. Thus the corrected $^3$He partial pressure at the analyzer level is then $5.5\cdot10^{-8}\,$mbar. 

\vspace{0.2cm}

Given that the flow rate for Helium should be between 100 l/s and 150 l/s for the pump mobile station the corresponding $^3$He flux going to the pump was between $5.5\cdot10^{-6}\,$mbar$\cdot$l/s and $8.2\cdot10^{-6}\,$mbar$\cdot$l/s, to be compared to the $7.8\cdot10^{-6}\,$mbar$\cdot$l/s measured by the (uncalibrated) leak detector.

\begin{table}[!ht]
\begin{center}
\begin{tabular}{c|cccc}
        &  \multicolumn{2}{c}{\bf{Leak detector} [$mbar \cdot l / s$]} &  \multicolumn{2}{|c}{\bf{RGA Gas analyzer} [A]} \\
\hline
		&				&				&		&		\\
\bf{valve}	&	$^3He$	&	$^4He$	&	Peak 3	&	Peak 4	\\
\hline
		&				&				&		&		\\
\bf{Closed}	&	$7.9\cdot10^{-9}$	&	$3.3\cdot10^{-10}$	&	0	&	0	\\
		&				&				&		&		\\
\bf{Open}	&	$7.8\cdot10^{-6}$	&	$9\cdot10^{-7}$		&	$3.5\cdot10^{-12}$	&	$0.4\cdot10^{-12}$	\\
\end{tabular}
\end{center}
\caption{First values obtained with the mobile pumping system with cold bores at $35\,$mbar. }
\label{ta:leakGasAnalyser}
\end{table}

According to the size of the two turbomolecular pumps connected to the cryostat ($2\,$x$\,700\,$l/s) and the low conductance of the mobile station flexible pipe (1.7 l/s for $^3$He) the gas flux to the said pumping station is lower than $0.3\%$ the total flux escaping from the cryostat. A first estimation for the amount of $^3$He gas coming out the cryostat is fixed at around $2\cdot10^{-3}\,$mbar$\cdot$l/s, measured at ambient temperature. 

\vspace{0.2cm}

\begin{figure}[!ht]\begin{center}
\includegraphics[width=\textwidth]{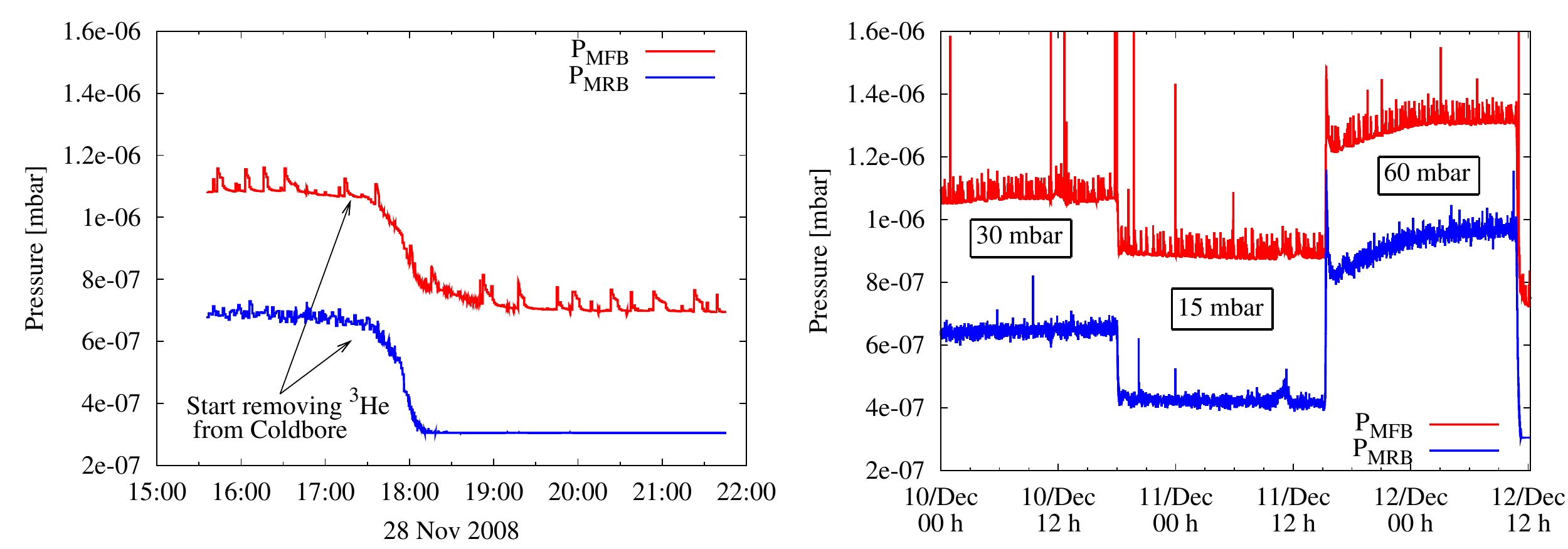}
\vspace{-0.3cm}
\caption{ On the left, evolution of the pressures in the cryostat isolation vacuum after emptying the cold bores. On the right, the effect of different pressures inside the cold bores ( $30\,$mbar, $15\,$mbar, $60\,$mbar) to the behavior of the pressures in the isolation vacuum due to the presence of the leak. }
\label{fi:LeakVacuumPressures}
\end{center}\end{figure}

The same day the $^3$He gas was removed from the cold bores; the peak 3 measured by the gas analyzer decreased from $4\cdot10^{-12}\,$A to $5.3\cdot10^{-13}\,$A, while the signal of the leak detector decreased from $8.1\cdot10^{-6}\,$mbar$\cdot$l/s to $7\cdot10^{-7}$mbar$\cdot$l/s. These variations were also observed on both cryostat pressures, $P_{MFB}$ and $P_{MRB}$ (see figure \ref{fi:LeakVacuumPressures}).

\vspace{0.3cm}

The same measurements were performed for different amounts of gas inside the cold bores, allowing to characterize the leak for different values of pressure. In the following days, on \emph{2nd and 4th of December} the cold bores were filled to $10\,$mbar, $30\,$mbar and $60\,$mbar. The results of these measurements at the mentioned cold bore pressures are summarized in table \ref{ta:leakGasAnalyserSummary}, where a proportional dependency between the leak rate and the pressure inside the cold bore is observed.

\vspace{0.2cm}

Furthermore, on \emph{10th of December}, the cold bore was filled to $30\,$mbar, emptied to $15\,$mbar and re-filled again to $60\,$mbar giving a clear evidence of the effect of the leak in the vacuum pressures measured in the cryostat as it is observed in figure \ref{fi:LeakVacuumPressures}.

\begin{table}[!ht]
\begin{center}
\begin{tabular}{ccc}
$P_{cb}$ [mbar]        &  \bf{Gas analyzer} [mbar$\cdot$l/s]	&	\bf{Leak detector} [mbar$\cdot$l/s] \\
\hline
	&				&		\\
\vspace{0.3cm}
0	&	0			&	$7.4\cdot10^{-9}$	\\
\vspace{0.3cm}
10	&	$4.0\cdot10^{-4}$	&	$7.0\cdot10^{-4}$	\\
\vspace{0.3cm}
30	&	$1.2\cdot10^{-3}$	&	$1.9\cdot10^{-3}$	\\
\vspace{0.3cm}
35	&	$1.4\cdot10^{-3}$	&	$2.1\cdot10^{-3}$	\\
\vspace{0.3cm}
60	&	$2.5\cdot10^{-3}$	&	$3.5\cdot10^{-3}$	\\
\end{tabular}
\end{center}
\caption{Summary for the leak values estimations obtained with the mobile pumping station equipment. Note that the gas analyzer calibration is the mean value published for $^4$He peak. It varies from one instrument to the other and depends on the ion source and resolution settings of each instrument. The Helium leak detector is factory calibrated for $^3$He, and not checked in situ.}
\label{ta:leakGasAnalyserSummary}
\end{table}

\vspace{-0.5cm}


\section{Search for $^3$He leak.}

Once the leak was characterized, on \emph{January-February 2009}, the MFB side of the cryostat was opened without breaking the CAST vacuum connected to the Helium system, pumped with the two Drytel groups and a mobile pumping station implementing a leak detector, in order to search for the $^3$He leak.

\vspace{0.2cm}

\begin{figure}[!h]\begin{center}
\includegraphics[width=0.8\textwidth]{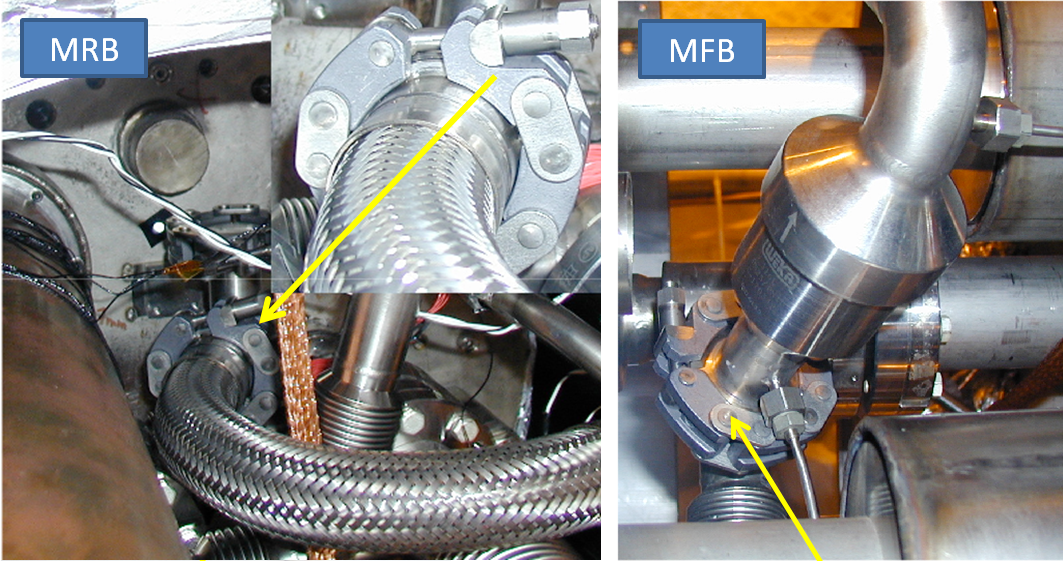}
\caption{Pictures of the flanges where the leaks were found on MRB and MFB sides.}
\label{fi:LeakPhotos}
\end{center}\end{figure}

\begin{figure}[!ht]\begin{center}
\includegraphics[width=0.8\textwidth]{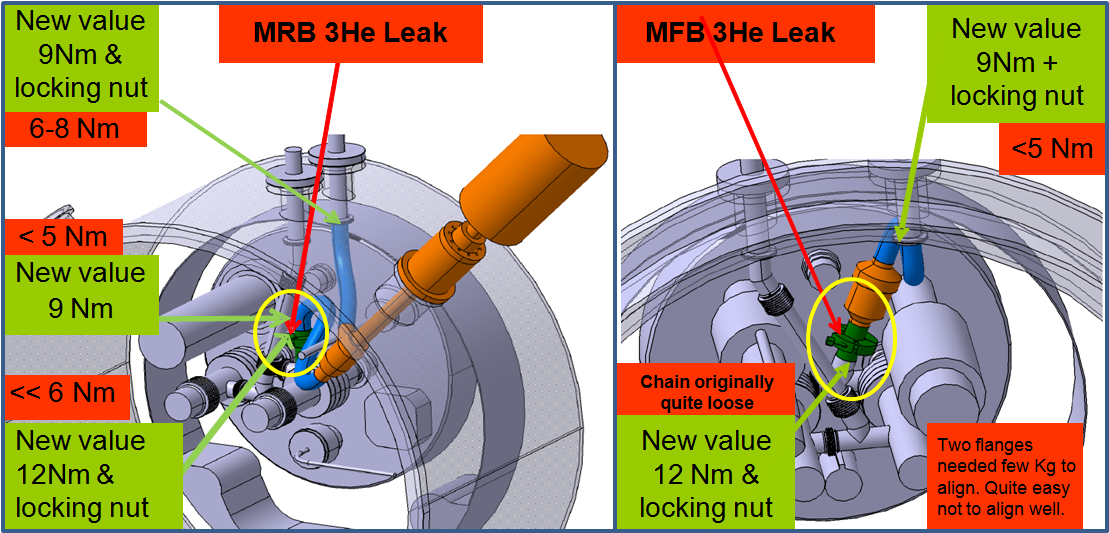}
\caption{ Schematic of the interventions carried out in the magnet bore ends connections in order to cure the leak and to re-assure some of the junctions present in the Helium system.}
\label{fi:LeakDescription}
\end{center}\end{figure}

A first leak was found on the Helium system inside the cryostat on the MFB side on a pair of DN40 KF flanges with an Helicoflex seal and closed by a Cefilac chain clamp (see figures \ref{fi:LeakPhotos} and \ref{fi:LeakDescription}).

\vspace{0.2cm}

The measurements with the Helium leak detector allowed to estimate that the leak found in the MFB side was around $1\cdot10^{-3}\,$mbar$\cdot$l/s (taking into account the splitting of the gas from the leak between the pumps and the leak detector). The chain clamp was found loose and tightening to nominal torque cured the leak.

\vspace{0.2cm}

The second leak was found on the Helium circuit inside the cryostat on a pair of KF flanges with Helicoflex seal and Cefilac chain clamp connecting the flexible and the tee welded to the cold bores (see figures \ref{fi:LeakPhotos} and \ref{fi:LeakDescription}).

\vspace{0.2cm}

The magnitude of the leak in this side was estimated to be greater than $5\cdot10^{-5}\,$mbar$\cdot$l/s. The clamp was tightened to torque $6\,$N$\cdot$m and the leak decreased to the $10^{-9}\,$mbar$\cdot$l/s range. The remaining vacuum pressures in the system decreased after these interventions. After further tightening at $7\,$N$\cdot$m there was not detectable leak in the $10^{-9}\,$mbar$\cdot$l/s range (no leak greater than $2\cdot10^{-10}\,$mbar$\cdot$l/s). All the other junctions were checked and re-assured (see figure \ref{fi:LeakDescription}).

\vspace{0.2cm}

\vspace{0.2cm}

In \emph{2007}, when the cryostat was enclosed the system was completely leak tested. It is believed that the leak appeared just after or during the cooling down of the system; the dilatation coefficient of flanges (two flanges were found not completely aligned and were supporting too much stress) and magnet vibrations due to pumping systems and magnet movement might have been the cause of the beginning of the leak. The new torque values applied to the junctions and the re-alignment of flanges were done at the beginning of \emph{2009} in order to avoid this problem to appear in the future.


\vspace{-0.2cm}
\section{Further tests to estimate the number of moles lost.}

The estimation of the leak described in the previous sections characterizes the leak in its final status. However, the leak could have had an evolution during the data taking phase in 2008, and in order to reconstruct the effect of the leak in each pressure setting during 2008 further measurements were mandatory.

\vspace{0.2cm}

Moreover, the leak estimations performed with the gas analyzer and the leak detector are based in several assumptions that might affect the accuracy in the leak determination (flexible pipe capacitance, pumping speeds, unknown conversion factor for $^3$He in leak measuring instrumentation, etc).

\vspace{0.2cm}

The intention of the tests that will be described in the following sections is to characterize the behavior of the pressure inside the cold bore\footnote{$P_{cb}$ is measured outside the magnet cryostat at ambient temperature at a volume connected to the cold bore and monitored by B100 sensor (see Fig.~\ref{fi:magnetSchema})} $P_{cb}$, as a function of the amount of moles introduced in the absence of a leak. The characterization will allow to determine the amount of gas lost in each pressure setting by direct comparison with data available for 2008, when the leak was present.

\vspace{0.2cm}

The basic idea consisted on finding a curve that could describe the measured pressure $P_{check} = P_{check}(n_T)$ as a function of the total number of moles introduced in the system, where $n_T$ is the total number of moles introduced in the system\footnote{$P_{check}$ is the measured pressure, $P_{cb}$, normalized to $1.8\,$K corrected with the measured magnet temperature, $T_{cb}$. This pressure was tagged from the very beginning as \emph{check} since it gives a control pressure that smoothes the fluctuations of $P_{cb}$ with the temperature of the magnet}. A set of measurements of $P_{check}$ for different amounts of gas inside the cold bore would allow to correct the systematics in the data shown in figure~\ref{fi:leakfits} and to obtain an estimation of the leak by using the pressure measured inside the cold bore at each step.


\subsection{Filling measurements description.}
\label{sc:filling}

These measurements were carried out on \emph{June-July, 2009} when the system was again ready for data taking, seeking to recreate the same magnet conditions as in the data taking period covered in 2008.

\vspace{0.2cm}

During these tests were obtained two set of measurements at different heating powers ($3.6$\,W and $7.2$\,W) in order to compare the behavior of $P_{check}$ in two different conditions. Each set of measurements would be completed by filling the cold bore in steps of about $3$\,mbar up to about $40$\,mbar (value slightly higher than the desired pressure at the end of 2008). In each step the value of the pressure was increased every 8 hours allowing the system to stabilize and to study such stabilization.

\vspace{0.2cm}

For filling the cold bore it was used the metering volume \emph{MV10} (described on section~\ref{sc:meteringHe3}). The pressures recorded in the metering volume were pondered in periods of 2-3 minutes of stability in order to calculate the amount of moles introduced in the system. Tables \ref{ta:MV10_1} and \ref{ta:MV10_2}, provided inside the appendix \ref{chap:appendix1}, show these pondered values together with the temperature of the thermal bath before and after the fillings. In table \ref{ta:moles} it is presented the amount of moles introduced in the system for each pressure setting.

\vspace{0.2cm}

Thus, table \ref{ta:moles} shows the calculation of the number of moles assuming a constant temperature for the thermal bath of $36$\,$^\circ$C and the number of moles calculated using the measured temperature in the thermal bath, showing the reduced effect of thermal bath temperature in the precision of the system and the propagation in several number of consecutive fillings. The direct comparison between the moles inserted in the corresponding steps at $3.6$\,W and $7.2$\,W gives account for the high reproducibility on the total number of moles introduced in the system.

\vspace{0.2cm}

\begin{figure}[!ht]\begin{center}
\begin{tabular}{c}
\includegraphics[width=0.92\textwidth]{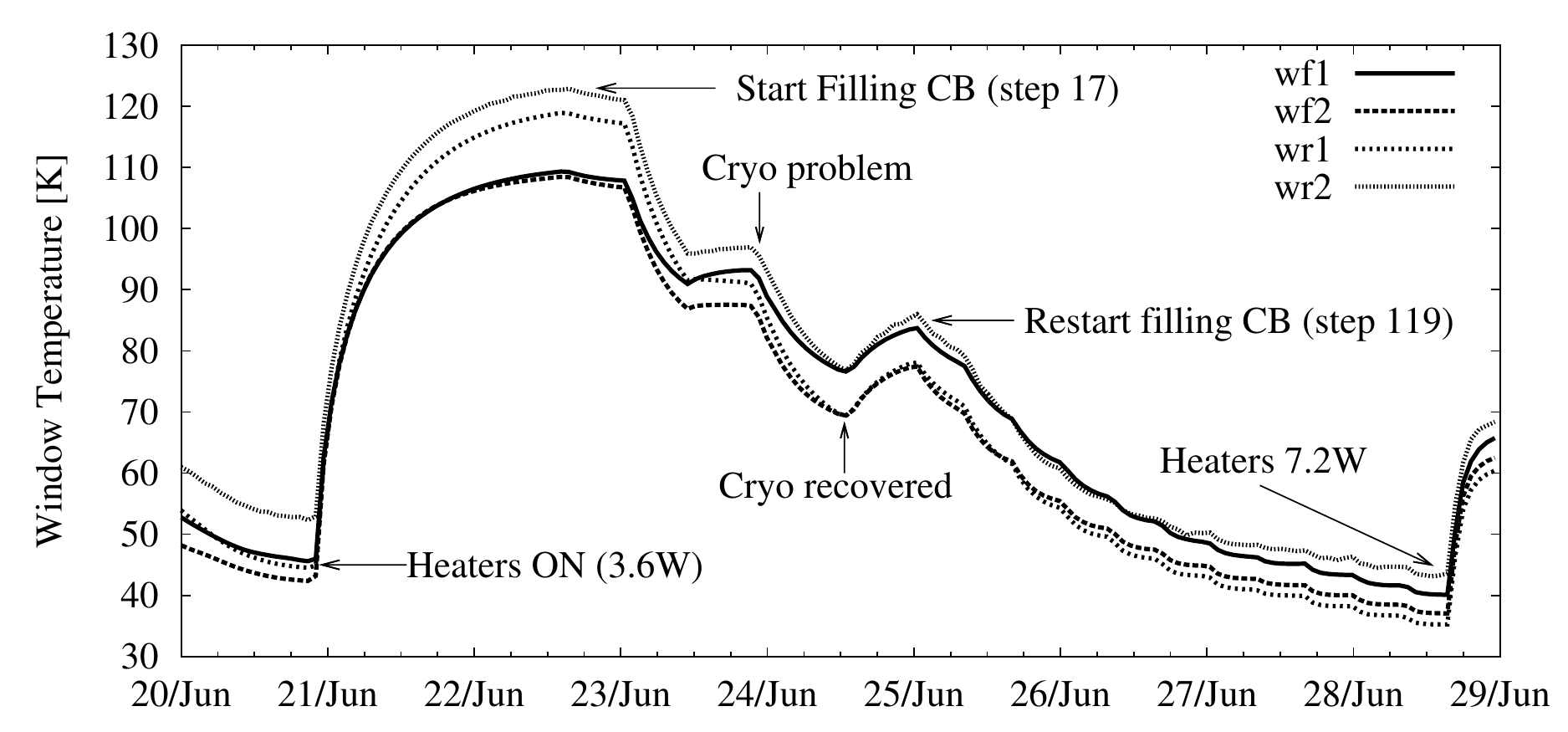}\\ 
\includegraphics[width=0.92\textwidth]{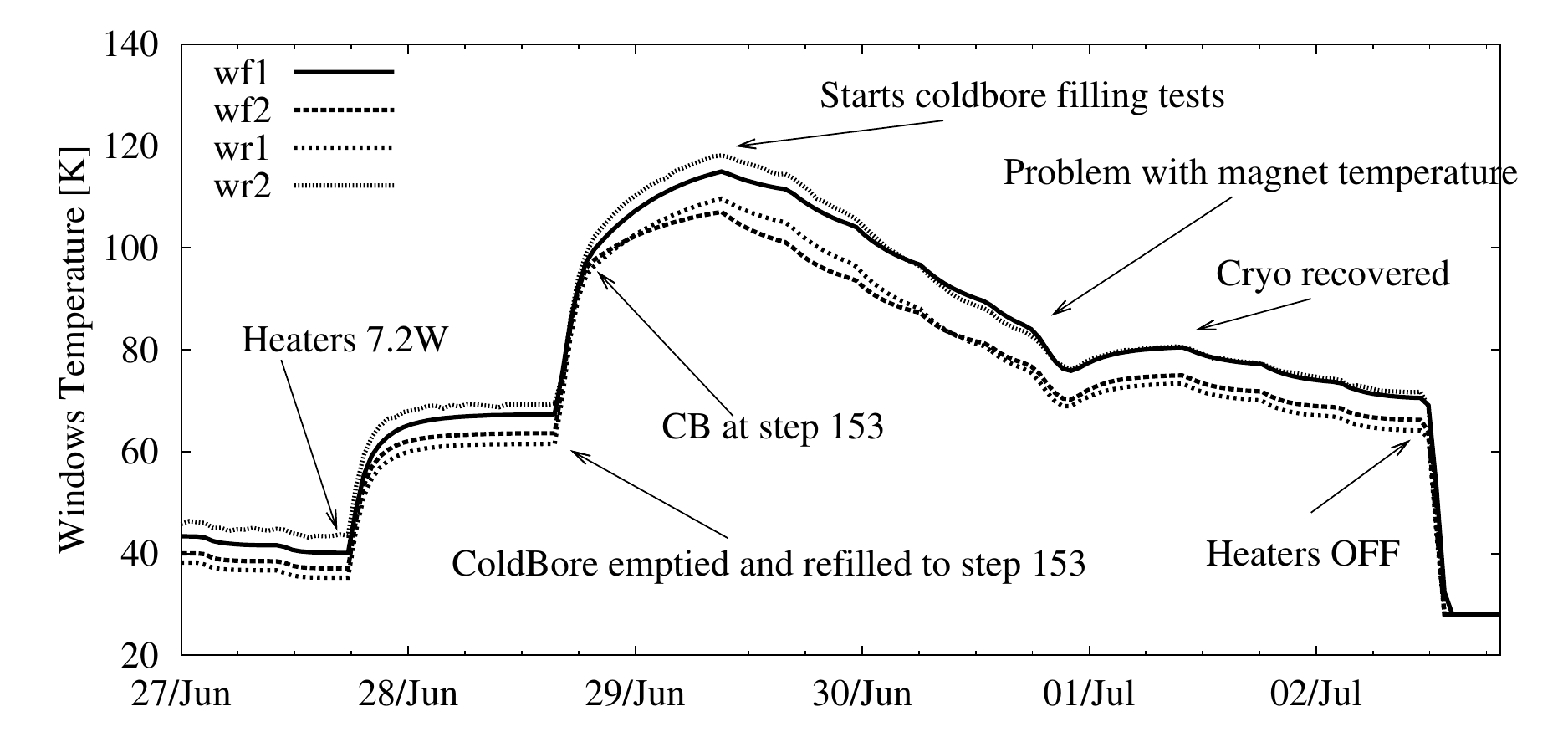} \\
\end{tabular}
\caption{Window temperature evolution during the $^3$He tests performed at the end of June and beginning of July. The graph on the top shows the windows temperature evolution for a constant heating power of $3.6$\,W, and on the bottom plot for a constant heating power of 7.2~W. Both plots show how the windows temperature is affected by filling the cold bore with $^3$He gas, by changing the heating power to the windows and by problems suffered with the cryo system during these tests.}
\label{fi:LeakTestsHistory}
\end{center}\end{figure}

Figure~\ref{fi:LeakTestsHistory} summarizes the tests carried out in terms of the evolution of four different temperature sensors that are placed on each of the cold windows. Changes in heating power and in the amount of gas inside the cold bore have a relevant influence on the temperature of the windows. The wavy decreasing shape of the temperature of the windows corresponds exactly with the $3$\,mbar periodic re-filling of the cold bore. The increasing gas amount inside the cold bore rises the heat transfer between gas and cold windows thus decreasing the temperature of the windows.

\vspace{0.2cm}

\vspace{0.2cm}

During these tests there were some incidences that affected directly to the temperature of the windows, but after each recovery the system came back to the expected nominal values. In the set of measurements corresponding to $3.6$\,W (see Fig.~\ref{fi:LeakTestsHistory}), the temperatures in the windows behaved in an unexpected way, compared to the general trend. Before the "Cryo Problem", during the second cold bore filling, the temperatures of the windows started to increase when the expected behavior was to decrease due to the increased amount of gas. 

\vspace{0.2cm}

Moreover, another unexplained phenomenum was observed, the temperature probes $wf1$ and $wr1$ seem to exchange themselves, being $wf1$ closer to $wr2$ and $wr1$ closer to $wf2$. The same configuration remained till the end of these tests. It might be that the system was in a kind of unstable equilibrium state, and after the recovering of the "Cryo problem" it went to a stable equilibrium state. These effects are related with critical thermodynamical values reached during the evolution of the system that do not interfere in the cold bore density range we are interested in.

\begin{figure}[!ht]\begin{center}
\includegraphics[width=0.82\textwidth]{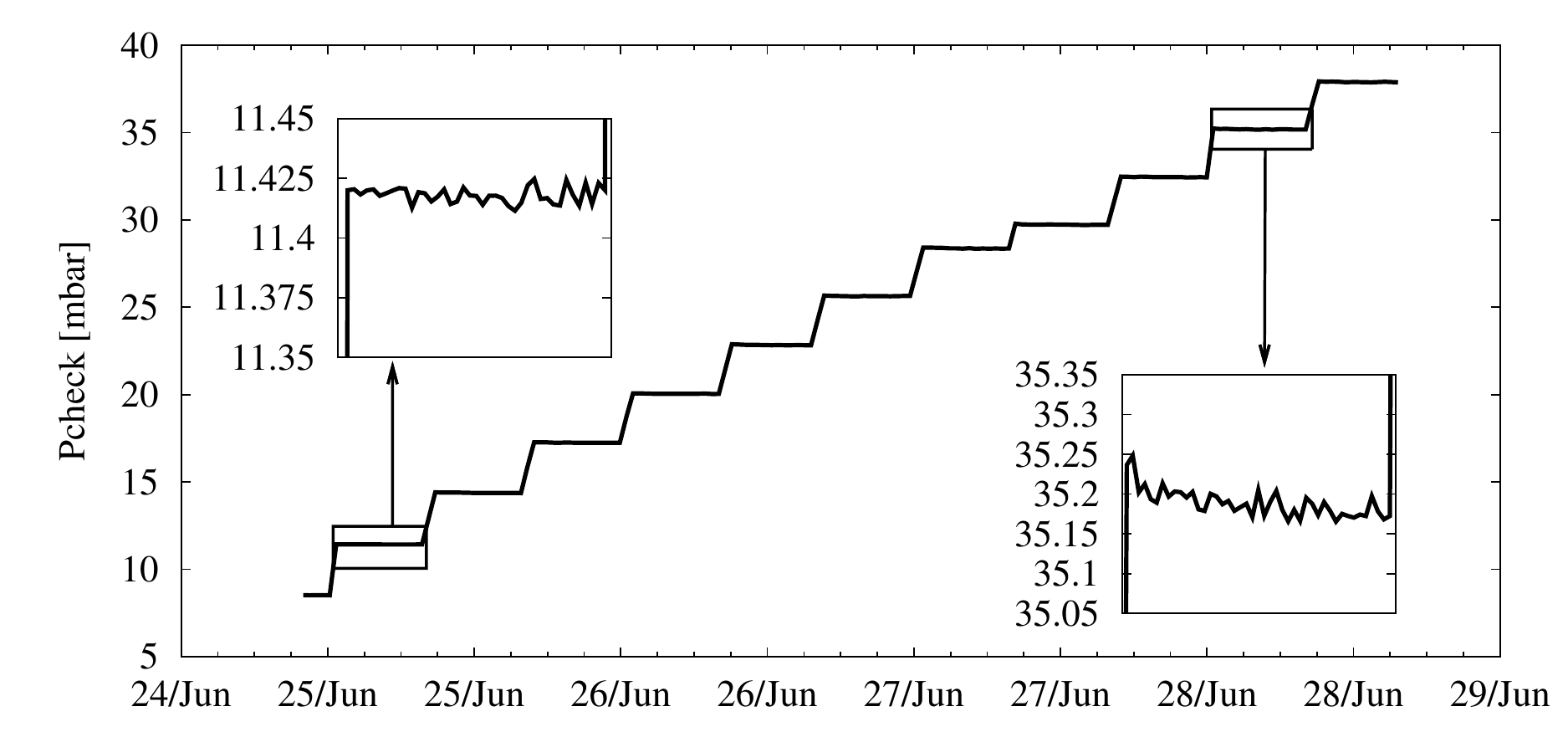}
\caption{ Plot representing the measured pressure, already corrected with the temperature of the magnet. The different steps observed in the plot correspond to the 8 hours stabilization time of the system. In the last steps it is already observed a dependency with the windows temperature.}
\label{fi:LeakFillingSteps}
\end{center}\end{figure}

\begin{figure}[!hb]\begin{center}
\includegraphics[width=0.92\textwidth]{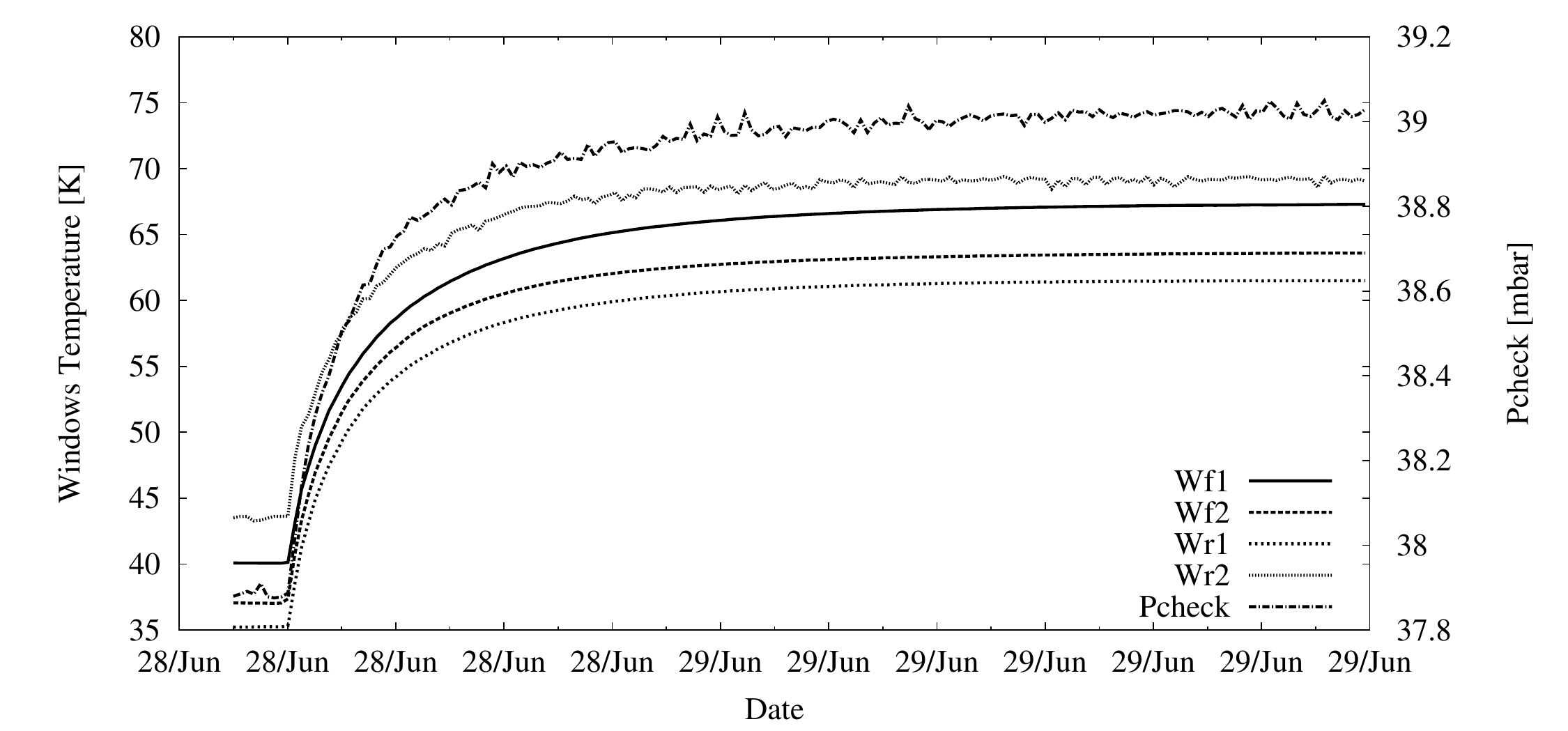} 
\caption{ This plot represents the change of temperature due to increasing the heating power from $3.6$\,W to $7.2$\,W in each of the four cold bore window sensors, together with the change in $P_{check}$ plotted in the auxiliary y-axis. It is appreciated how the evolution of $P_{check}$ keeps a close correlation with the temperature of the windows.}
\label{fi:LeakWindowTemperatures}
\end{center}\end{figure}

The pressure steps measured in conditions of stability (see Fig.~\ref{fi:LeakFillingSteps}) including 8 hours time evolution measurement are finally the ones presented in table \ref{ta:steps}. These are the steps to be considered for this study, corresponding with the density range covered during 2008. There are also two transient periods when the temperature of the windows changes due to the change of heating power with a constant amount of gas inside the cold bore (see Fig.~\ref{fi:LeakWindowTemperatures}). These sets of data will be also useful for investigating the gas behavior.

\vspace{0.2cm}

\begin{table}[h]
\begin{center}
\begin{tabular}{c|cccccccccc}
\hline
 &  &  &  &  &  &  &  &  &  &  \\
\bf{3.6\,W} & 136 & 170 & 203 & 235 & 267 & 295 & 321 & 360 & 386 & 412 \\
 &  &  &  &  &  &  &  &  &  &  \\
\hline
 &  &  &  &  &  &  &  &  &  &  \\
\bf{7.2\,W} & 152 & 187 & 219 & 251 & 282 & 308 & 334 & 360 & 386 & \\
 &  &  &  &  &  &  &  &  &  &  \\
\hline

\end{tabular}
\caption{Pressure steps measured in conditions of stability and with long evolution time (around 8 hours).}
\label{ta:steps}
\end{center}
\end{table}


\subsection{Windows temperature stabilization time.}

A first result from the tests carried out is the characterization of the windows temperature evolution at \emph{two} different heating powers as a function of the amount of moles inside the magnet.

\vspace{0.2cm}

\begin{figure}[!ht]\begin{center}
\begin{tabular}{cc}
\includegraphics[width=0.5\textwidth]{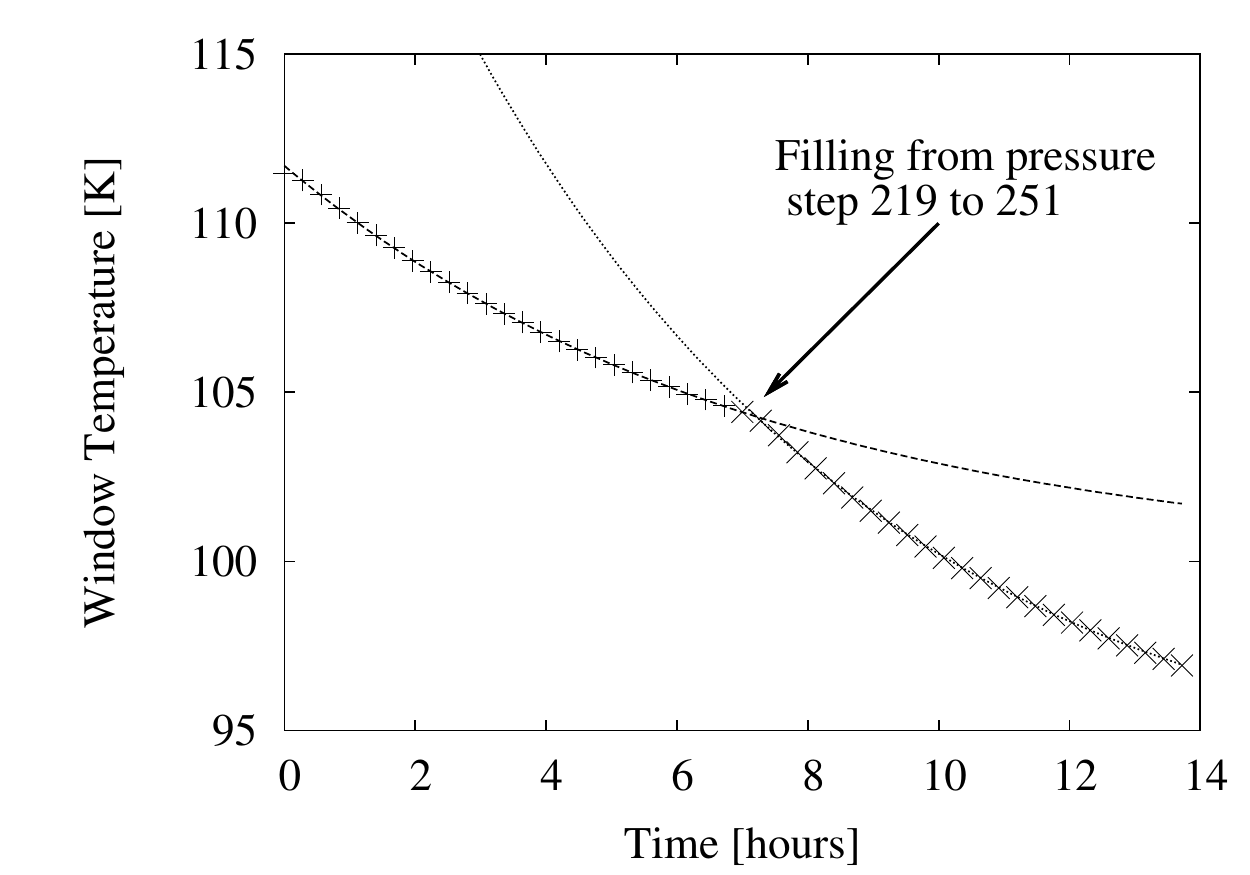} & \includegraphics[width=0.45\textwidth]{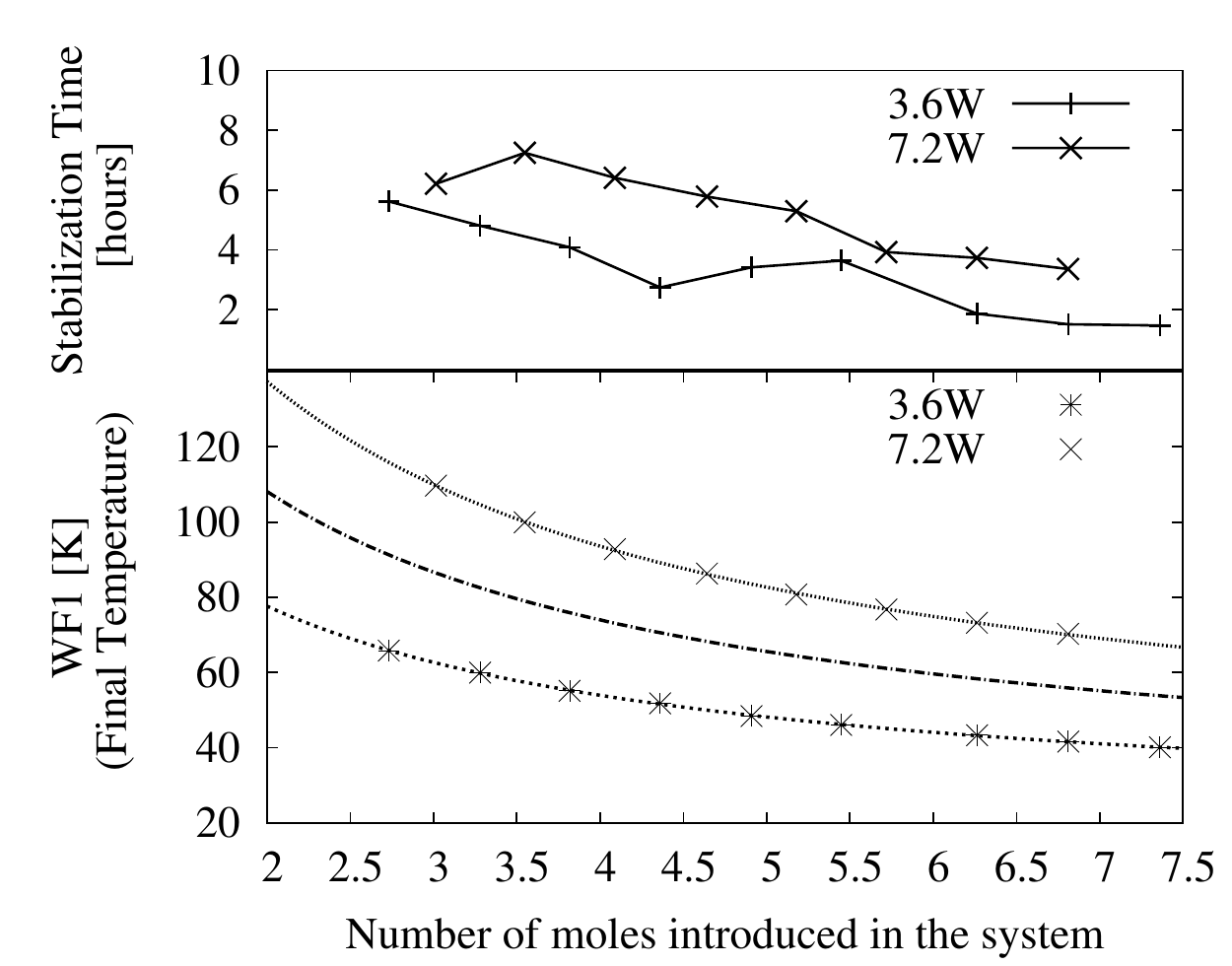} \\
\end{tabular}
\caption{On the left, the change of tendency in the windows temperature after re-filling the cold bore during the 7.2~W series (the stabilization period during those measurements was fixed in 7 hours). On the right, the \emph{final stabilization temperature} and \emph{stabilization time} versus the amount of moles inside the magnet, obtained from the exponential fit parameters of the evolution of windows temperature for each re-filling.}
\label{fi:LeaktemperatureFits}
\end{center}\end{figure}

The windows temperature follows a characteristic exponential evolution with time that changes tendency after each re-filling of the cold bore (see Fig.~\ref{fi:LeaktemperatureFits}). Each of the stabilization periods after re-filling the cold bore has been fitted to an exponential function and the two main parameters, \emph{final stabilization temperature} and \emph{stabilization time} have been obtained. The evolution of these two parameters as a function of the number of moles is also shown in figure \ref{fi:LeaktemperatureFits}, where it is observed how the \emph{stabilization time} becomes shorter as a higher gas density is inside the cold bores.

\vspace{0.2cm}

The final stabilization temperature can be described by a rational function $T_{st}(n_T) = a + b/(n_T+c)$ as a function of the number of moles inserted, for both heating power measurement series. This description allows to determine the equilibrium temperature that the windows will reach due to the re-filling of the magnet to higher densities. According to the sensor $wf1$, the value obtained is 20.27\,K for a heating power of 3.6~W, and $28.43$\,K for a heating power of 7.2\,W.

\subsection{A model for data description.}
To obtain any useful information from the tests performed it is needed to introduce a simplified model that will describe the data sets measured. It was corroborated during these tests that the measured pressure $P_{check}$ depends on and keeps a close correlation with the windows temperature. It is required to introduce a clear dependency with the windows temperature.

\vspace{0.2cm}

In order to be able to handle the data one must introduce a model that is simple enough and still is a good approach to the reality. The following temperature distribution is chosen to describe the primary effects observed during these tests,

\begin{equation}\label{eq:Tz}
T(z) = T_m + \left(T_w - T_m\right) \cdot \mbox{exp}( - z/z_o(n) )
\end{equation}

\vspace{0.2cm}

\noindent where $T_w$ is the temperature on the windows, and $z$ is the coordinate system along the magnet axis with reference in the windows, $T(z=0) = T_w$. The parameter $z_o(n)$ can be considered as the heat penetration length. This parameter is expected to be increasing with the amount of gas inside the magnet bores.

\vspace{0.2cm}

It is expected that the effects of the cold windows are only appreciable in their neighborhood, the expected values for $z_o$ should be small compared to the bore pipe length, $L_m$.

\vspace{0.2cm}

Some formulation needs to be developed to find a relation that describes the pressure in the system as a function of the amount of moles inserted and the temperatures of the windows. We must assume that the pressure in the system is constant, so we are assuming there are no gravitational or convective effects in our model.







\vspace{0.2cm}

Then, the measured pressure $P_{check}$ can be described (detailed in appendix~\ref{app:Pcheck}) as a function of the windows temperature $T_w$, the total number of moles introduced $n_T$, and the penetration lenght $z_o(n)$

\begin{equation}\label{eq:PmVsTw}
P_m = P_m^o (T_m) \frac{1}{ 1 - \frac{z_o(n)}{L_m/2} \mbox{log}(T_w/T_m) }
\end{equation}

\vspace{0.2cm}

\noindent where $P_m^o$ is the reference pressure of the system when the whole volume is at $T_m$,

\begin{equation}\label{eq:Pmo}
P_m^o (T_m) = \frac{R T_m n_T}{V_{cb}}
\end{equation}

\vspace{0.2cm}

\noindent and $T_m$ is the measured temperature in the cold bore center.

\subsection{Finding the parameter $z_o(n)$.}

$z_o(n)$ remains as a free parameter in the relation (\ref{eq:PmVsTw}) and needs to be determined by using the data sets available. This section describes the technical details to obtain such parameter.

\vspace{0.2cm}

The theoretical relation (\ref{eq:Pmo}) assumes that all the moles inserted in the system will go inside the volume defined by the cold bore pipes, $V_{cb}$. Experimentally this is not true since the system is composed by some extra pipework that increases the volume where the gas will be distributed, also denominated as \emph{dead volumes}. We need to reformulate as follows,

\begin{equation}
P_m^o (T_m)^* = \frac{R T_m n_T^*}{V_{cb}} = \gamma P_m^o (T_m) 
\end{equation}

\noindent where $n_T^*$ denotes the fraction of moles that are contained inside the cold bore volume $V_{cb}$ by the addition of to the dead volumes. It seems a good approach that the ratio between moles going to the \emph{dead volumes} and moles going to the cold bore volume given by the parameter $\gamma$ remains constant as long as the temperature ratio between both volumes does not change dramatically.

\vspace{0.2cm}

Now $P_m^o(T_m)^*$ is a quantity that is closer to the value given by $P_{cb}$ and that is related with $P_{check}$ as follows,

\vspace{0.2cm}

\begin{equation}
P^o_{check} = \gamma P_m^o (1.8\,\mbox{K})
\end{equation}

\vspace{0.2cm}

\noindent the relation \ref{eq:PmVsTw} needs to be rewritten to describe the experimental value $P_{check}$,

\vspace{0.2cm}

\begin{equation}\label{eq:PcheckFinal}
P_{check} =\frac{1.8\,\mbox{K}}{T_m} P_m^o(T_m)^* = \gamma P_m^o(1.8\,\mbox{K}) \frac{1}{ 1 - \frac{z_o(n)}{L_m/2} \mbox{log}(T_w/T_m) }
\end{equation}

\vspace{0.2cm}

\noindent expression that can be directly used for fitting the evolution of pressure versus windows temperature using the experimental data available. The parameter $T_w/T_m$ should also be renormalized with the experimental value of $T_m$ for each value of $P_{check}$, however, given that the effect is almost negligible it will be considered constant. 

\vspace{0.2cm}

The final fitting it is performed using three free parameters, $\gamma$, $\beta$ and $\sigma$,

\vspace{0.2cm}

\begin{equation}\label{eq:fit}
P_{check} = \gamma  \frac{ P_m^o(1.8\,\mbox{K}) }{ 1 - \beta \mbox{log}(T_w-\sigma) }
\end{equation}

\vspace{0.2cm}

\noindent where $\sigma$ is a parameter that is introduced to assume a possible systematic on the real temperature of the cold window surface respect to the value that is measuring the sensor. The value for $P_m^o(1.8\,\mbox{K})$ it is calculated for each pressure setting by using the amount of moles transferred given by the metering volumes, values presented in table \ref{ta:moles}. The parameter $\beta$ it is related with the parameter $z_o(n)$ by the following expression,

\vspace{0.2cm}

\begin{equation}\label{eq:zo2beta}
\beta = \frac{ \frac{z_o(n)}{L_m/2} }{ 1 - \frac{z_o(n)}{L_m/2} \mbox{log}(T_m) } \qquad \frac{z_o(n)}{L_m/2} = \frac{\beta}{1+\beta \mbox{log}(T_m)}
\end{equation}

\vspace{0.1cm}

\noindent that it is used to simplify the fitting and obtain afterwards the parameter $z_o(n)$.

\vspace{0.2cm}

The first data sets used to check the model are the pressure settings 360, 386 and 412. The reason is that for these steps there are available stabilization time measurements at $3.6\,$W and $7.2\,$W settings, and cooling and heating transient measurements, meaning a long range fitting as a function of the windows temperature is viable. The resulting fitting parameters from expression~(\ref{eq:fit}) are summarized in table~\ref{ta:fitParams}.

\begin{table}[ht]
\begin{center}
\begin{tabular}{cccc}
\vspace{0.3cm}
\bf{Step}    &	$\gamma$	& 	$\beta$	& 	$\sigma\,$[K]	 \\
\hline
	&	&	&	\\
\vspace{0.3cm}
\bf{360}	&	0.921975 (0.11\%) &  0.032497 (0.81\%) & 17.607 (1.534\%) \\			
\vspace{0.3cm}
\bf{386}	&	0.924297 (0.10\%) &  0.032280 (0.59\%) & 19.025 (1.228\%) \\
\vspace{0.3cm}
\bf{412}	&	0.923 (*)	  &  0.032093 (0.005\%) & 18.811 (0.04\%) \\  
\end{tabular}
\caption{Parameter values obtained for steps 360, 386 and 412 (see Fig.~\ref{fi:LeakPTFullFits}). The parameter $\gamma$ in step 412 is forced to be $\gamma = 0.923$ due to technical problems finding an appropriate solution for the fit.}
\label{ta:fitParams}
\end{center}
\end{table}

\begin{figure}[!ht]\begin{center}
\begin{tabular}{cc}
\includegraphics[width=0.48\textwidth]{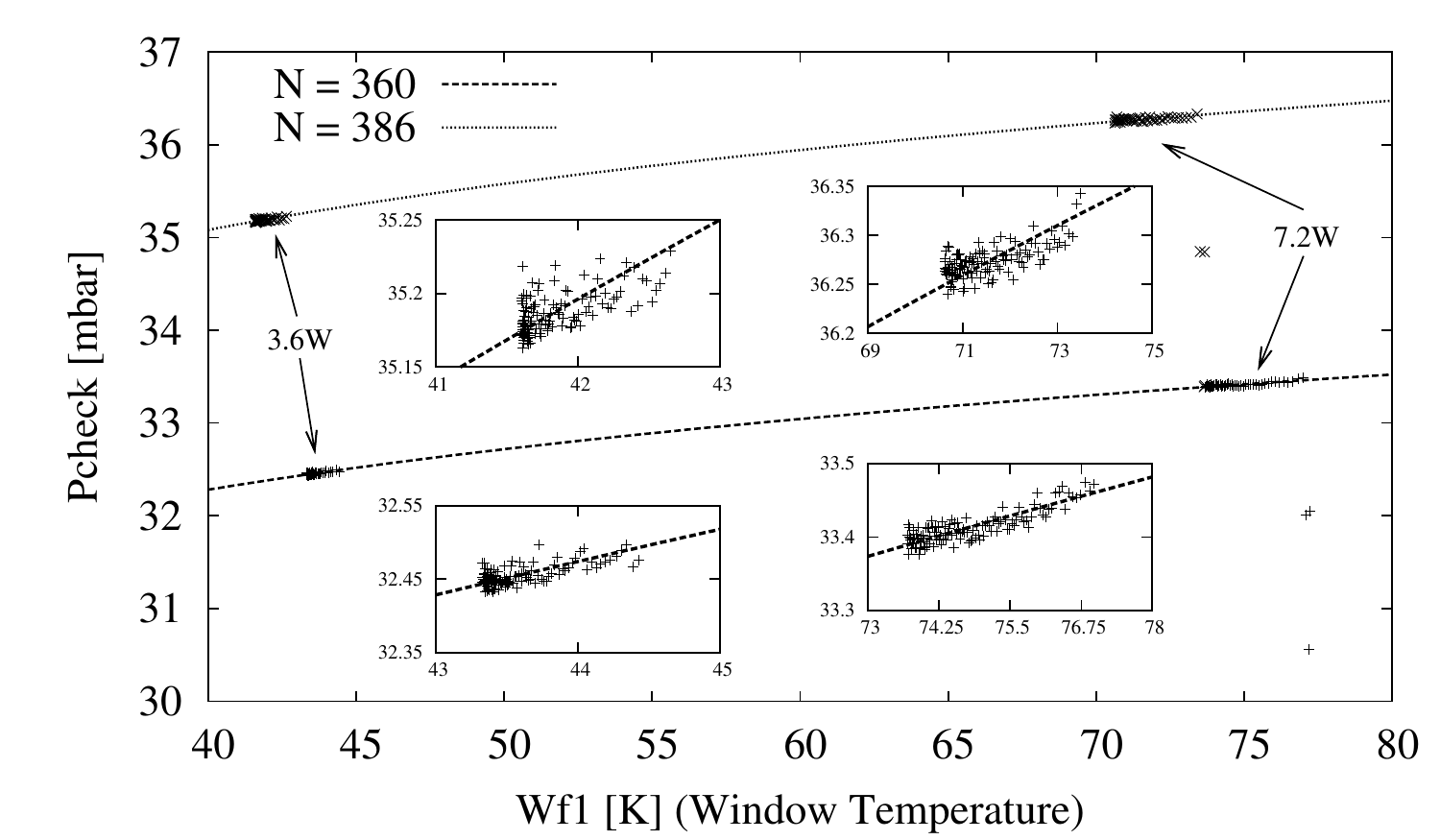} & \includegraphics[width=0.48\textwidth]{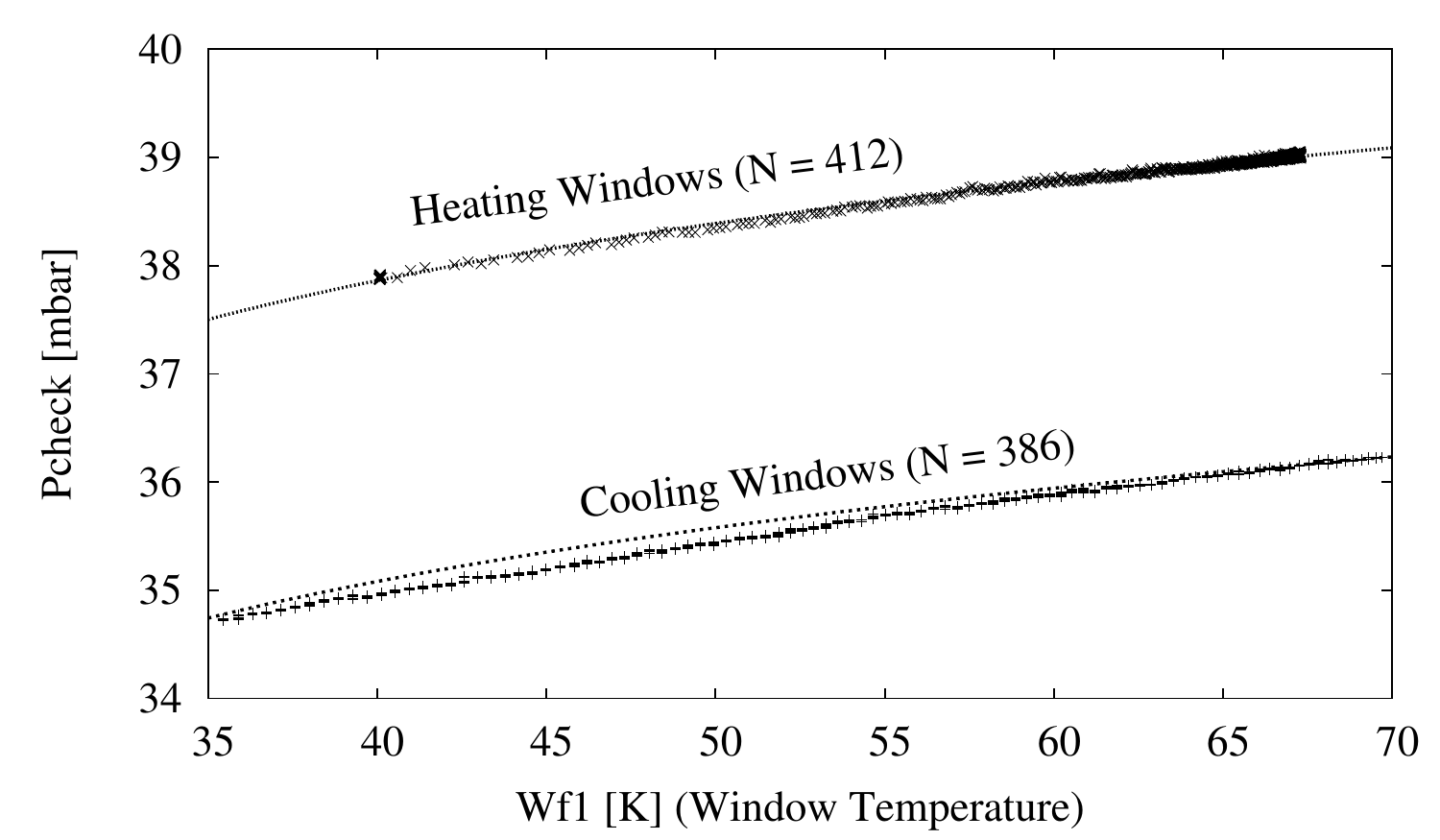} \\
\end{tabular}
\caption{On the left, $P_{check}$ is plotted versus the window temperature, $wf1$, for two pressure settings measured with long stabilization time at 3.6~W and 7.2~W. The curves are obtained using the data model described. On the right are plotted the transients due to a change in heating power. Pressure step 412 is fitted, and the curve that describes the pressure step 386 transient was obtained from the rest of data available for that step. }
\label{fi:LeakPTFullFits}
\end{center}\end{figure}

The resulting fits are shown in figure \ref{fi:LeakPTFullFits}, where it is observed a good agreement between the experimental data and the described model. Only, the period of windows cooling shows a small discrepancy in the evolution of $P_{check}$ versus the window temperature, $wf1$, this could be due to the fast cooling process after switching off the heaters (see Fig.~\ref{fi:LeakTestsHistory}). The experimental value seems to be matching the value predicted by the model after the temperature rate change decreases.

\begin{figure}[!htb]\begin{center}
\begin{tabular}{cc}
\includegraphics[width=0.48\textwidth]{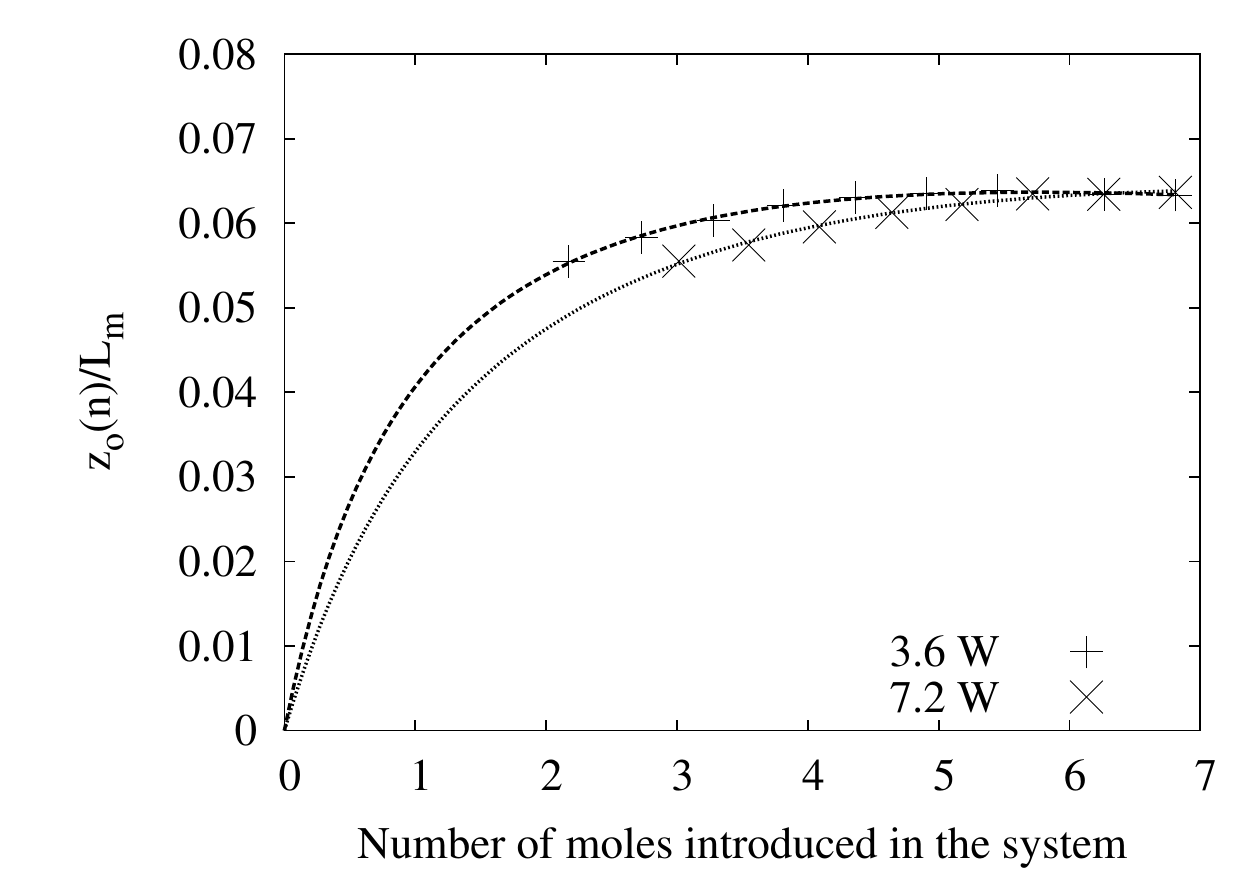} & \includegraphics[width=0.48\textwidth]{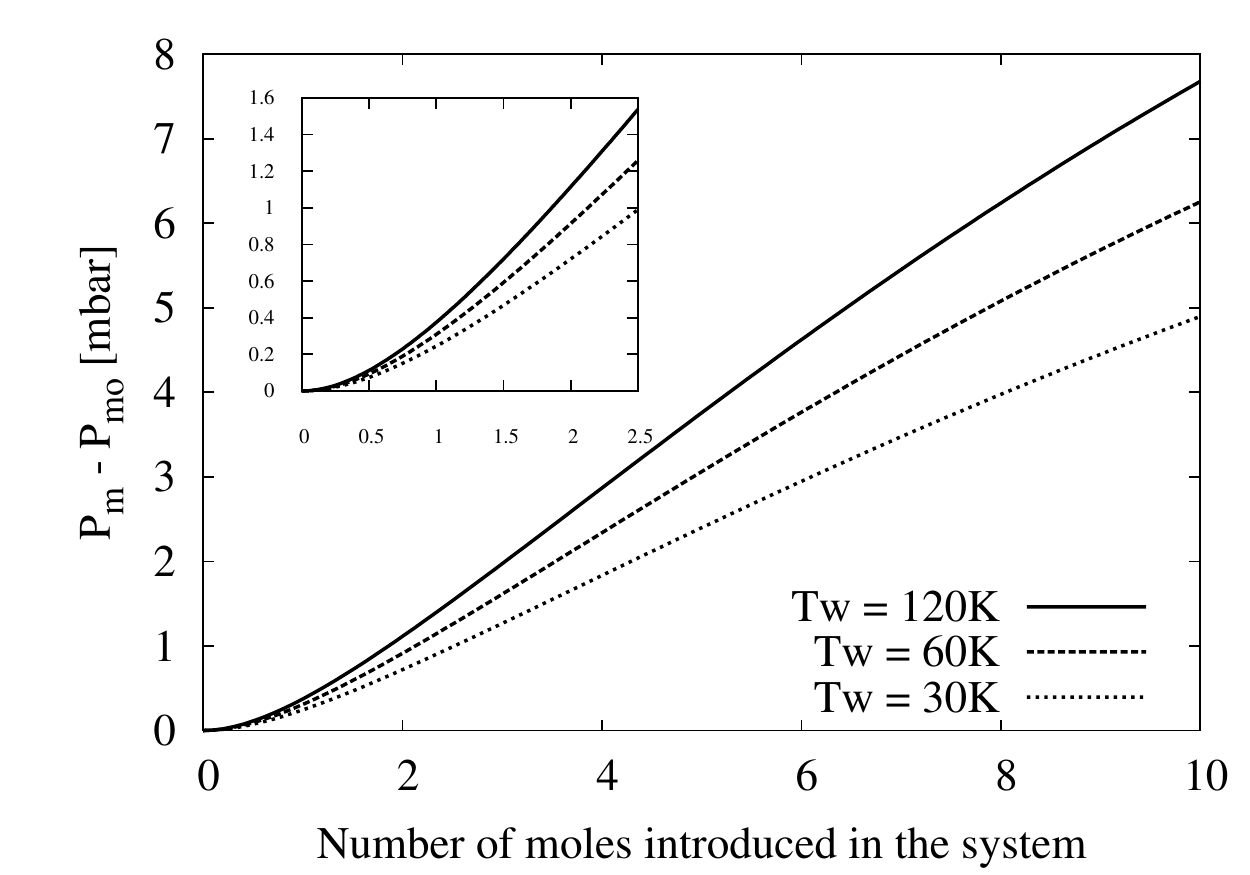} \\
\end{tabular}
\caption{ On the left, parameterization of the parameter $z_o(n)$. On the right plot, the difference in pressure between the equivalent pressure corrected with the windows temperature and the parameter $z_o(n)$ obtained, and the equivalent pressure with the whole magnet at $T_w = 1.8\,\mbox{K}$.  }
\label{fi:LeakzParameter}
\end{center}\end{figure}

\vspace{0.2cm}

For the remaining pressure settings, the measured stabilization times are not in coincidence for measurements at 3.6~W and 7.2~W, see table \ref{ta:steps}. This supposes a problem because the data set is defined in a short range of window temperature, when a wider range of windows temperature would be desirable in order to obtain more reliable values for the fitting parameters.

\begin{table}[ht]
\begin{center}
\begin{tabular}{ccccc}
\vspace{0.2cm}
    &	$a$	& 	$b$	& 	$c$  &     $d$	 \\
\hline
    &		&		&		&	\\
\bf{3.6~W}	&	-0.000514896 & 0.0210818 & 0.952314 & 1.07004 \\
    &		&		&		&	\\
\bf{7.2~W}	&	-0.000348953 & 0.0169092 & 0.722037 & 1.28596 \\
\end{tabular}
\vspace{0.1cm}
\caption{ Parameters that describe $z_o(n)$ as a function of the number of moles following the expression~\ref{eq:zn}. }
\label{ta:zParams}
\end{center}
\end{table}

In order to make the fitting of the rest of data available easier, the parameters $\gamma$ and $\sigma$ have been fixed. The parameter values $\gamma = 0.923$ and $\sigma = 18.0\,\mbox{K}$ have been taken as safe parameters. The only remaining free parameter, $\beta$, is obtained by fitting the rest of pressure settings present in table \ref{ta:steps}.

\vspace{0.2cm}

Using the relation (\ref{eq:zo2beta}), we find the experimental values for the description of $z_o(n)$ that are presented in figure \ref{fi:LeakzParameter}. The discrepancies found between the data sets for 3.6~W and 7.2~W might be due to the fact that the data range for the fit is not wide enough to find accurate parameters, as it was mentioned before. A rational function, 

\begin{equation}\label{eq:zn}
\frac{z_o(n)}{L_m/2} = n \frac{a n + b}{c n + d}
\end{equation}

\vspace{0.2cm}

\noindent has been chosen for a parameterized description of the function $z_o(n)$. In figure \ref{fi:LeakzParameter} it is plotted the parameterized curve, and in table \ref{ta:zParams} are presented the values obtained. These values should be considered as valid only for the range from 2 to 7\,moles, since there is no data analyzed for the pressure settings corresponding to $^4$He Phase. The range from $0$ to $7$ moles is extrapolated by arguing that the heat transfer must be zero in absence of gas.

\vspace{0.2cm}

In figure \ref{fi:LeakzParameter} it is also presented the increased pressure in the magnet compared to the pressure in case of windows temperature $T_w = 1.8\,\mbox{K}$. It is worth to mention that this increase of pressure is directly related with the density inside the cold bore, since relation (\ref{eq:PmVsTw}) must be satisfied all along the magnet axis.


\subsection{Determining the leak.}

After finding $z_o(n)$, the measured pressure, $P_{check}$, can be described by the expression (\ref{eq:PcheckFinal}) as a function of the number of moles introduced in the system and the temperature of the cold windows. In order to find the real number of moles inside the magnet we can now observe the behavior of $P_{check}$ as a function of the temperature of the windows during data taking periods and compare it with the results obtained in the previous section.

\vspace{0.2cm}

If now we use the fact that all the parameters that define $P_{check}$ are already known, the relation (\ref{eq:PcheckFinal}) can be used to fit the data from any period where the magnet is in stable conditions (not moving). Introducing $z_o(n)$ parameterized, as described in relation (\ref{eq:zn}), in the equation (\ref{eq:PcheckFinal}), and letting $n$ be the only free parameter, the fitting to the data will return directly the value of the real amount of gas inside the magnet in moles.

\begin{table}[ht]
\begin{center}
\begin{tabular}{cccc}
\bf{Step}    &	\bf{187}	& 	\bf{223}	& 	\bf{268}   \\
\hline
\bf{Date}	&	\bf{26/April}	&	\bf{27/May}	&	\bf{30/June} \\
\hline
			&		&		&		\\
\bf{Moles in absence of leak} &  3.03095	&	3.63566	&	4.40297	\\
\bf{Moles calculated}	 &  2.95692	&	3.53880	&	4.22656 \\
			 &			&			&		\\
\hline
\bf{Moles lost}		 &  0.07403	&	0.09686	&	0.17641 \\
\hline
			 &			&			&		\\
\bf{Mean leak rate [moles/day]}	 &  4.05e-3	&	5.38e-3	&	6.78e-3	\\
\bf{Mean leak rate [mbar$\cdot$l/s]}	 &  1.17e-3	&	1.55e-3	&	1.96e-3	\\
			 &			&			&		\\
\bf{Step}			 &	  \bf{324}   	&	\bf{374}		&	\bf{412}	 \\
\hline
\bf{Date}			 &    	\bf{3/August}  	&	\bf{9/September}	&	\bf{13/November} \\
\hline
			&		&		&		\\
\bf{Moles in absence of leak} &	5.54347		&	6.59905		&	7.41042	\\
\bf{Moles calculated}	 &	5.32831		&	6.36736		&	7.00110 \\	
			 &			&			&		\\
\hline
\bf{Moles lost}	         &     	0.21516		&	0.23169		&	0.40932 \\
\hline
			 &			&			&		\\
\bf{Mean leak rate [moles/day]}  &	6.94e-3		&	7.02e-3		&	14.7e-3 \\		
\bf{Mean leak rate [mbar$\cdot$l/s]}	 &  2.00e-3	&	2.03e-3		&	4.24e-3	\\
\end{tabular}
\caption{ Summary of moles lost due to $^3$He leak. The mean leak rate is calculated taking into account the first gas re-filling days given in table \ref{ta:fillingStarts}. Leak rate is normalized to standard leak units [mbar$\cdot$l/s] at an environmental temperature of $300\,$K. }
\label{ta:lostmoles}
\end{center}
\end{table}

\begin{figure}[!htb]\begin{center}
\begin{tabular}{cc}
\includegraphics[width=0.49\textwidth]{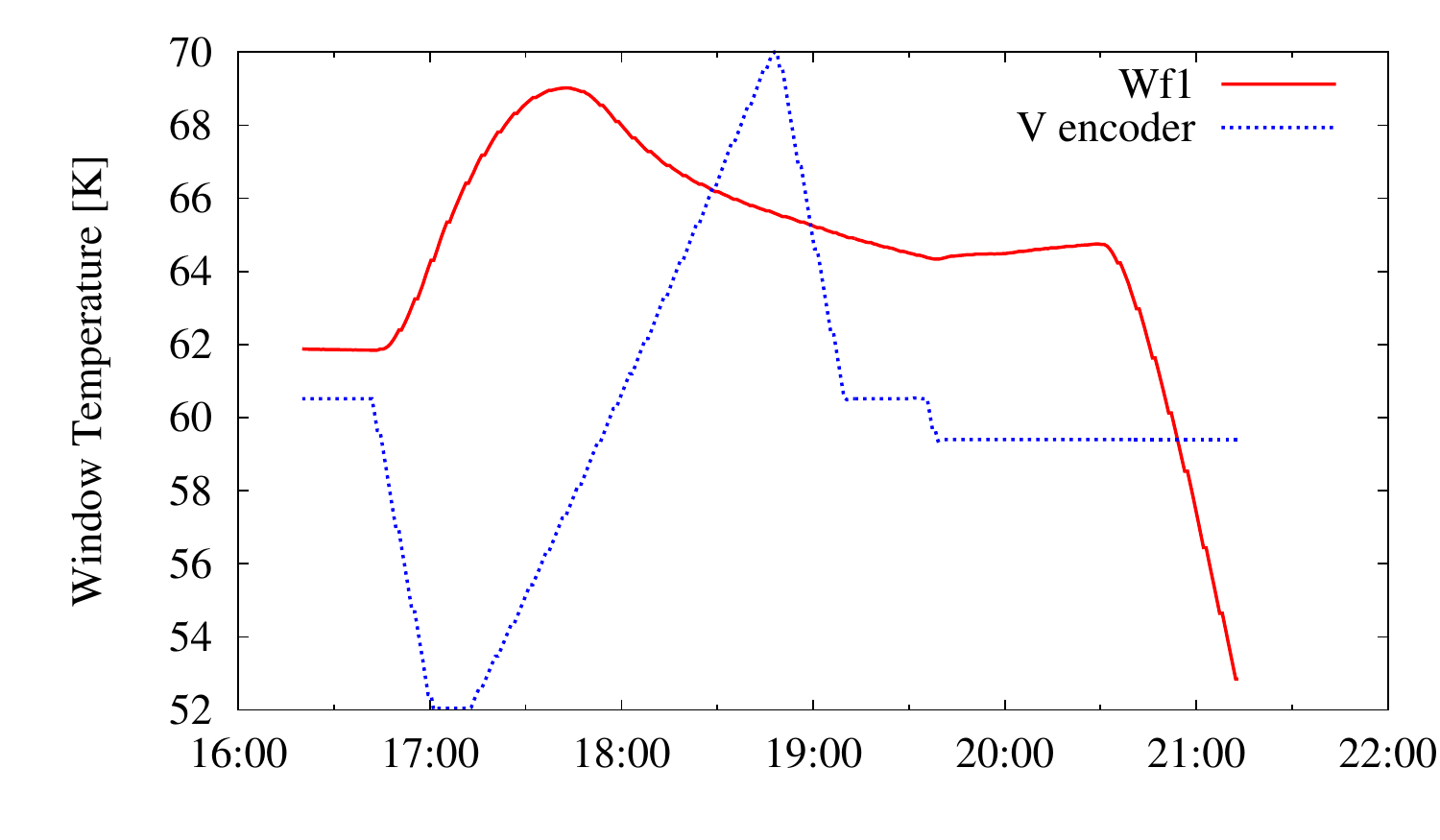} & \includegraphics[width=0.49\textwidth]{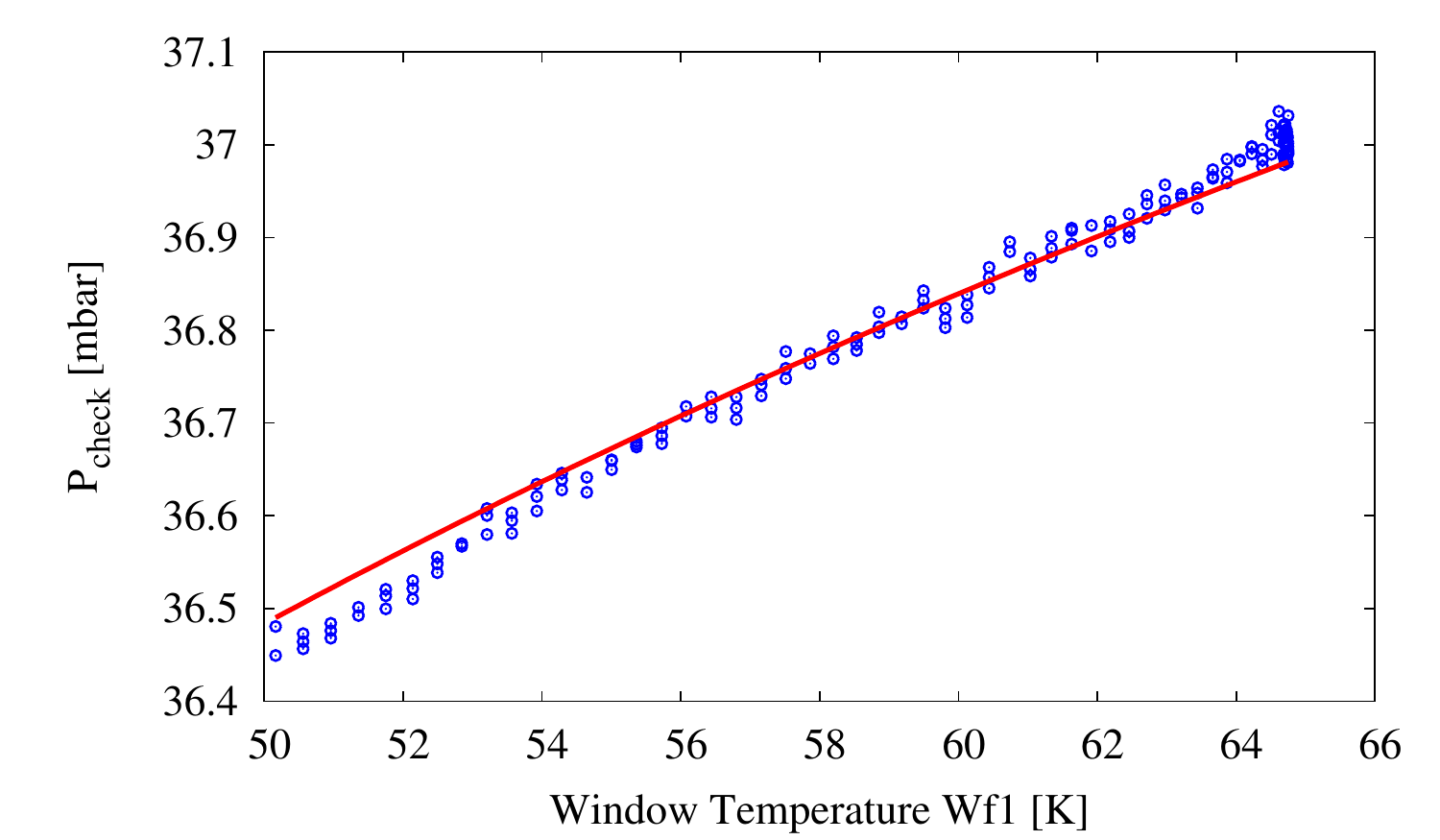} \\
\end{tabular}
\caption{ On the left, last tracking from 2008 data, the magnet movement it is represented by plotting the vertical angle encoder. On the right the dependency of $P_{check}$ with the windows temperature after sometime that the magnet has stopped his movement. The data it is fitted following expression (\ref{eq:PcheckFinal}), obtaining a value of 7.0011~moles, versus the 7.4104~moles expected for that setting. }
\label{fi:LeakdataTakingFit}
\end{center}\end{figure}

This technique for finding the number of moles inside the magnet has been applied to the last pressure settings before emptying the cold bore, in order to know the total amount of moles lost in absolute terms for each sub-period of data taking. Table \ref{ta:lostmoles} shows the obtained moles inside the magnet in those last pressure steps. By taking into account the time since the last re-filling of the cold bore it is estimated the mean leak rate. In figure \ref{fi:LeakdataTakingFit} is shown the last tracking of 2008, corresponding to the last pressure step, $N = 412$.

\vspace{0.2cm}

The calculation for the moles in the last pressure steps of each subperiod allows to obtain an estimation of the overall amount of gas lost in each of these periods. In order to determine the evolution of the leak day by day, it could be done by performing the calculation of the number of moles for each pressure step during the data taking periods. 



\section{Towards a description of the axion mass coverage.}

The measurements obtained in the determination of the leak concluded that the observed behavior in the measured pressure inside the cold bore is closely related to the inner cold bore density. The good agreement between the theoretical model and the experimental data obtained in static conditions during the filling tests manifests the physical relation between them.

\vspace{0.2cm}

After these measurements it was concluded that the determination of the leak was not enough to describe the densities that CAST had scanned during 2008. At the higher pressures covered during the $^3$He phase, the density in the magnetic region cannot be directly determined by the amount of gas inserted in the cold-bore, but it is just a boundary parameter in the description of the gas distribution. 

\vspace{0.2cm}

The reason resides in several factors that might affect the density profile during the tracking movement as it will be argumented in section~\ref{sc:trackingDensity}. The only solution to obtain an accurate measurement of the density in the magnetic region relays in the understanding of the measured pressure quantity $P_{check}$, and in describing the nature of its evolution in relation to the known parameters in the CAST magnet; window temperatures, inner magnet temperature and total amount of gas inside the cold-bore.

\vspace{0.2cm}

A better understanding of the gas behavior can only be reached by performing Computational Computational Fluid Dynamics (CFD) simulations (section~\ref{sc:CFDsimulations}) due to the extreme complexity of the whole system that involves convection and a real gas equation of state dependent on the temperature of the gas that is not homogeneously distributed on the system.

\vspace{0.2cm}

Preliminary, but not concluding results have already been obtained by CFD simulations. This section describes an experimental approach to the description of the density in the inner cold bore by using the value of the measured pressure.

\subsection{Effects on the magnet cold bore density during tracking.}\label{sc:trackingDensity}

\begin{figure}[!ht]
{\centering \resizebox{1.0\textwidth}{!} {\includegraphics{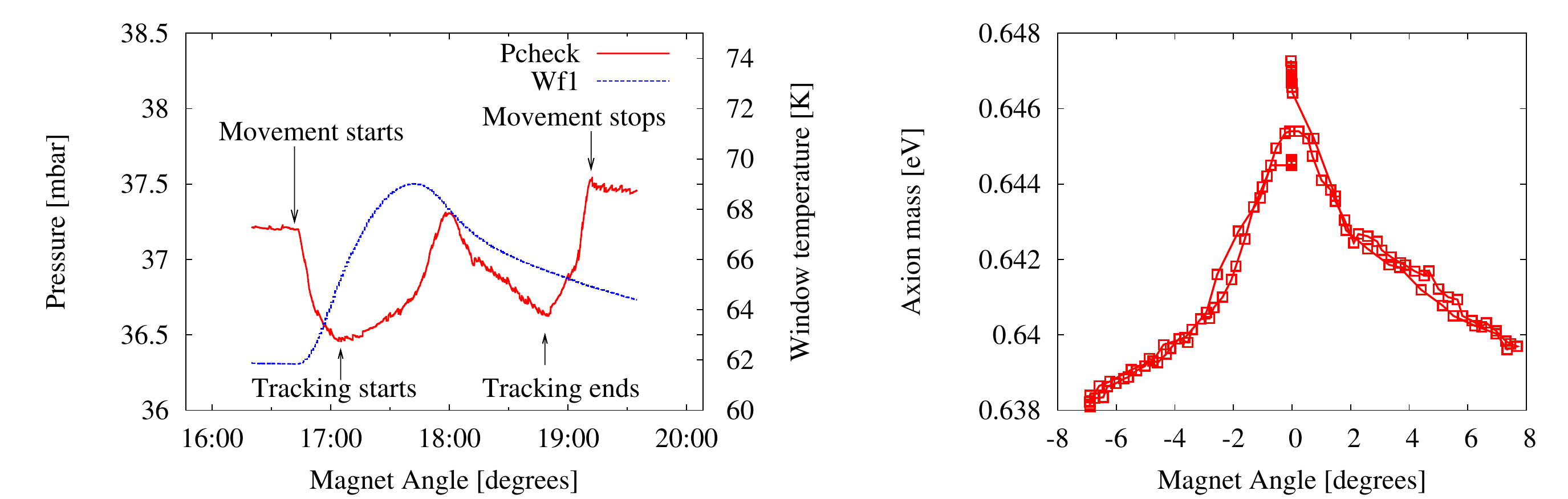}} \par}
\caption{\fontfamily{ptm}\selectfont{\normalsize{ On the left, evolution of $P_{check}$ during the last tracking in 2008, together with the temperature evolution from one of the four cold bore windows, represented in the auxiliary scale. On the right, $P_{check}$ as a function of the magnet vertical angle.}}}
\label{fi:pcheckTracking}
\end{figure}

The pressure of the magnet varies due to the vertical movement of the magnet. Figure~\ref{fi:pcheckTracking} represents the variations of pressure during the tracking movement. The cold-bore windows temperatures change during tracking due to the increased or reduced heat transfer to the windows coming from gas convection and depend on the inclination of the magnet. The pressure change is too high to be attributed to the change on windows temperature during the tracking. The pressure change due to the effect of windows temperatures can be immediately observed after the magnet movement stops, as it is observed in figure~\ref{fi:pcheckTracking}.

The effect on the pressure of the system as measured by the $P_{check}$ sensor is not dependent on the magnet speed , that it is slowly moving during the Sun tracking movement and fast when the magnet moves to the position where the tracking starts at $-8^\circ$, or moving to the parking position, coming back to the horizontal position. Figure~\ref{fi:pcheckTracking} shows the high reproducibility in pressure in spite of the \emph{two} different magnet speeds covered in the full vertical range from about $-8^\circ$ to $8^\circ$, that gives account of the short stabilization time required during the vertical movement.

\vspace{0.2cm}

The pressure offset observed between $-8^\circ$ and $8^\circ$ is mainly due to the gas cold bore re-filling that has place in the middle of the tracking, at about $0^\circ$.

\vspace{0.2cm}

Eventhough the effect on the measured pressure during the magnet movement is not completely described and no concluding remarks can be made about the phenomena involved, the observed effect on pressure might be mainly associated with convection\footnote{Convective effects are related with gravitational effects which affects in different ways as the magnet is placed at different vertical positions.} which depends on the absolute angle. The lower pressure measured at higher angles must be related to the fact that gas convection currents are changing the density profile squeezing the regions of lower density (producing an expansive effect) making the density in the magnetic region to decrease.

\subsection{Effect of hydrostatic pressure.}\label{sc:hydrostatic}

The fact that $P_{check}$ is measured in a pressure sensor that is placed $1\,$m above the cold bore axis\footnote{The axis is defined assuming the magnet is placed in horizontal position.} introduces a systematic in the pressure defined at the magnet axis. In order to estimate the offset produced between the cold bore pressure and $P_{check}$ we must use the following expression

\begin{equation}
\frac{dP}{dz} = - \rho g = -\frac{N_{^3He}\cdot g}{R\cdot T(z)} P
\label{eq:HydroEq}
\end{equation}

\noindent and choosing for simplicity a linear density distribution with boundary conditions $T(z=0) = T_w$ and $T(z=1\,\mbox{m}) = T_{env}$ we obtain the relation between the cold bore axis pressure and the sensor pressure,

\begin{equation}
P_{check} = P_{axis} \cdot exp \left[ -\frac{N_{^3He}\cdot g}{R\cdot T_{env}} \int_0^1 \left( 1-\frac{T_{env}}{T_w} \right) z + \frac{T_{env}}{T_w}  dz \right]
\end{equation}

\noindent where $T_{env}$ is the environmental temperature where it is placed the pressure sensor, and $T_w$ is the temperature of the cold windows.

\vspace{0.2cm}

Considering $T_{env} = 300\,$K and $T_w = 50\,$K the offset produced in $P_{check}$ is not bigger than $0.004\%$. However, this error does not have influence in the estimation of the number of moles for the leak estimation since the offset is always present in the measurements. 

\vspace{0.2cm}

The main systematic error due to the hydrostatic pressure effect comes from the possibility of measuring $P_{check}$ at different windows temperatures. Assuming that the fluctuations of $T_{env}$ are negligible compared to the fluctuation of $T_w$, and considering that the biggest effect on the hydrostatic pressure offset is given by $T_w$, we can express the estimation of the systematic error in $P_{axis}$ as a function of $T_w$ for two different measurements of $P_{check}$ leading to the same experimental value,

\vspace{0.2cm}

\begin{equation}
P_{axis}^{T_{w_1}} = P_{axis}^{T_{w_2}}\cdot exp \left[ -\frac{N_{^3He} \cdot g}{2R}  \left( \frac{1}{T_{w_2}} - \frac{1}{T_{w_1}} \right)  \right]
\label{eq:systHydro}
\end{equation}

\vspace{0.2cm}

\noindent where $T_{w_1}$ and $T_{w_2}$ are the temperatures of the windows at the time of measurement of $P_{check}$. From expression \ref{eq:systHydro} it is deduced that lower window temperatures lead to a higher deviation from the real value of $P_{axis}$. The temperature of the windows in the measurements described in section \ref{sc:filling} were never bellow $40\,$K, if we consider the extreme case that the windows would be at $T_w^1 = 20\,$K and $T_{w_2} \to \infty$ the ratio between both $P_{axis}$ values does not differ more than $0.01\%$.

\vspace{0.2cm}

The deviation effect from the real pressure inside the cold bore respect to the pressure measured in the sensor can be considered negligible in horizontal position. The maximum deviation of the value measured in the pressure sensor and the pressure at the magnetic region, at the maximum vertical position of the magnet ($8^\circ$), considering a constant density ($\rho \simeq 0.77$\,mg$/$cm$^3$, corresponding to the last tracking from 2008) along the cold bore pipe it is estimated a maximum deviation from the real value of $0.08$\,mbar.

\subsection{Systematic effect on the measured magnet temperature.}\label{sc:MagnetTemperature}

A source of error in the obtention of the cold bore density by using the value of the measured pressure arises from the fact that the magnet temperature $T_m$ used to normalize $P_{cb}$ to obtain the $P_{check}$ parameter it is not the temperature of the gas but the temperature of the Helium bath surrounding the magnetic region.

\vspace{0.2cm}

By assuming a constant temperature difference $\phi$ between the real gas temperature and the measured temperature, the effect on the re-normalization is described by

\begin{equation}
\frac{P_{check}}{P_{cb}\cdot1.8\,\mbox{K}/(T_m + \phi)} = \frac{T_m + \phi}{T_m}
\end{equation}

\vspace{0.2cm}

\noindent that leads to a difference in the absolute value of $P_{check}$, around $0.28\%$ for $\phi = 5$\,mK and $2.91\%$ for $\phi = 50$\,mK.

\vspace{0.2cm}

\begin{figure}[!ht]
{\centering \resizebox{1.0\textwidth}{!} {\includegraphics{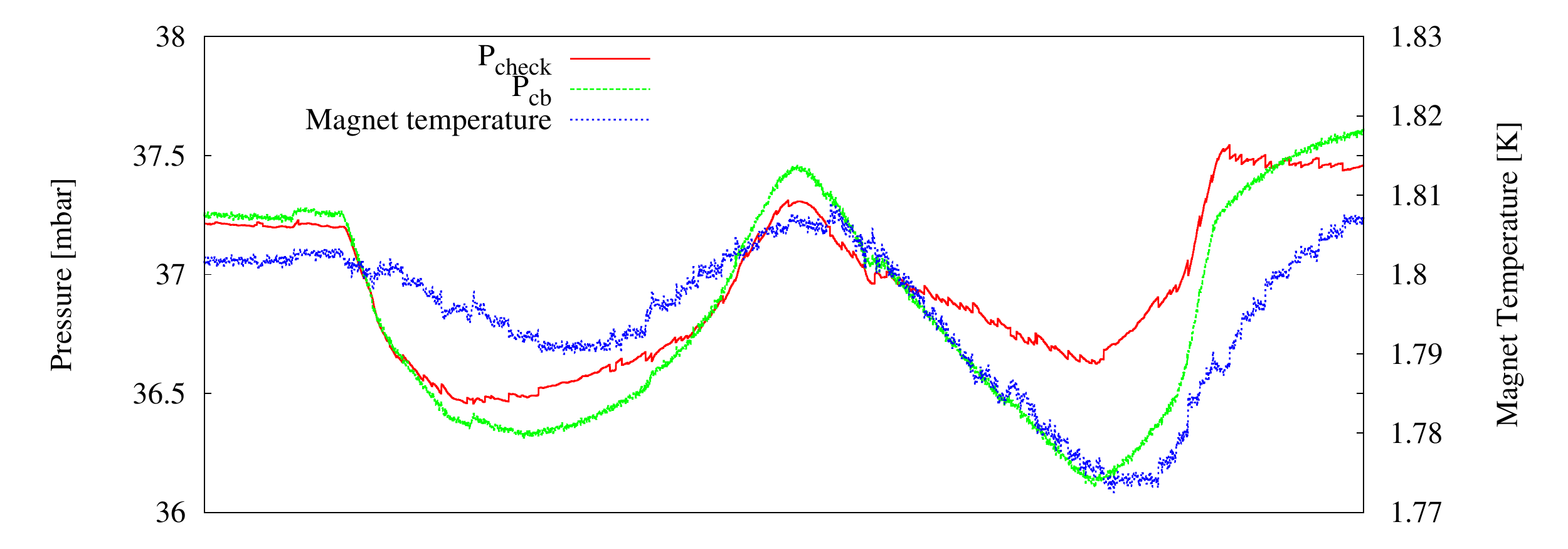}} \par}
\caption{\fontfamily{ptm}\selectfont{\normalsize{ Magnet temperature evolution during tracking together with the measured pressure $P_{cb}$ and the renormalized pressure at $1.8$\,K, $P_{check}$.  }}}
\label{fi:Tmagnet}
\end{figure}

Figure~\ref{fi:Tmagnet} shows the magnet temperature evolution during the tracking movement for the same tracking as in figure~\ref{fi:pcheckTracking}. The shape of the magnet temperature could lead to believe that the effects on $P_{check}$ during tracking are coming from a systematic produced by a variation of the parameter $\phi$ during tracking, however after the magnet movement stops $P_{check}$ starts decreasing following the change on windows temperature, eventhough the temperature of the magnet continuous changing. This fact, together with the high correlation with the magnet vertical angular position observed in figure~\ref{fi:pcheckTracking}, reinforces the hypothesis that the main tendency on $P_{check}$ it is ruled by convection effects and is due to real changes on density in the magnetic region of the magnet.

\subsection{Van der Walls corrections.}\label{sc:VDW}

The higher densities and the low temperatures inside the magnet make necessary to introduce second order corrections to the ideal gas equation of state in order to relate the measured pressure with the inner density of the magnet. These corrections start to be considerable in the relation between density and pressure at the new scanning ranges in the $^3$He Phase.

\vspace{0.2cm}

An extension to the ideal gas equation of state~\cite{HurlyHelium,HarveyHelium,0026-1394-46-5-017} is given by the virial expansion

\begin{equation}
\frac{p}{\rho RT} = 1 + B(T)\rho + C(T)\rho^2 + ...,
\end{equation}

\vspace{0.2cm}

\noindent where $p$ is the pressure, $\rho$ is the molar density, $R$ the molar gas constant, and $T$ the absolute temperature. $B(T)$ is the second virial coefficient, representing the lowest order deviation from ideal-gas behavior. As the density increases, the contribution from the third virial coefficient $C(T)$ becomes significant. The second virial coefficient is calculated in~\cite{HurlyHelium} for $^4$He and $^3$He gas where values for $T = 1.8$\,K are provided, $B(1.8\,\mbox{K}) = -145.295$\,cm$^3\cdot$mol$^{-1}$ and $dB(1.8\,\mbox{K})/dT = 75.286$\,cm$^3\cdot$mol$^{-1}\cdot$K$^{-1}$. No value is found in the literature for the third virial coefficient at $T=1.8$\,K, in spite that some calculations exist at higher temperatures~\cite{HarveyHelium}. The final equation of state used to obtain the density of the cold bore at the magnetic region is given by following relation,

\begin{equation}\label{eq:stateVDW}
P_{check} = \rho R T_o + B(1.8\,\mbox{K}) \rho^2 R T_o
\end{equation}

\begin{figure}[!ht]
{\centering \resizebox{0.7\textwidth}{!} {\includegraphics{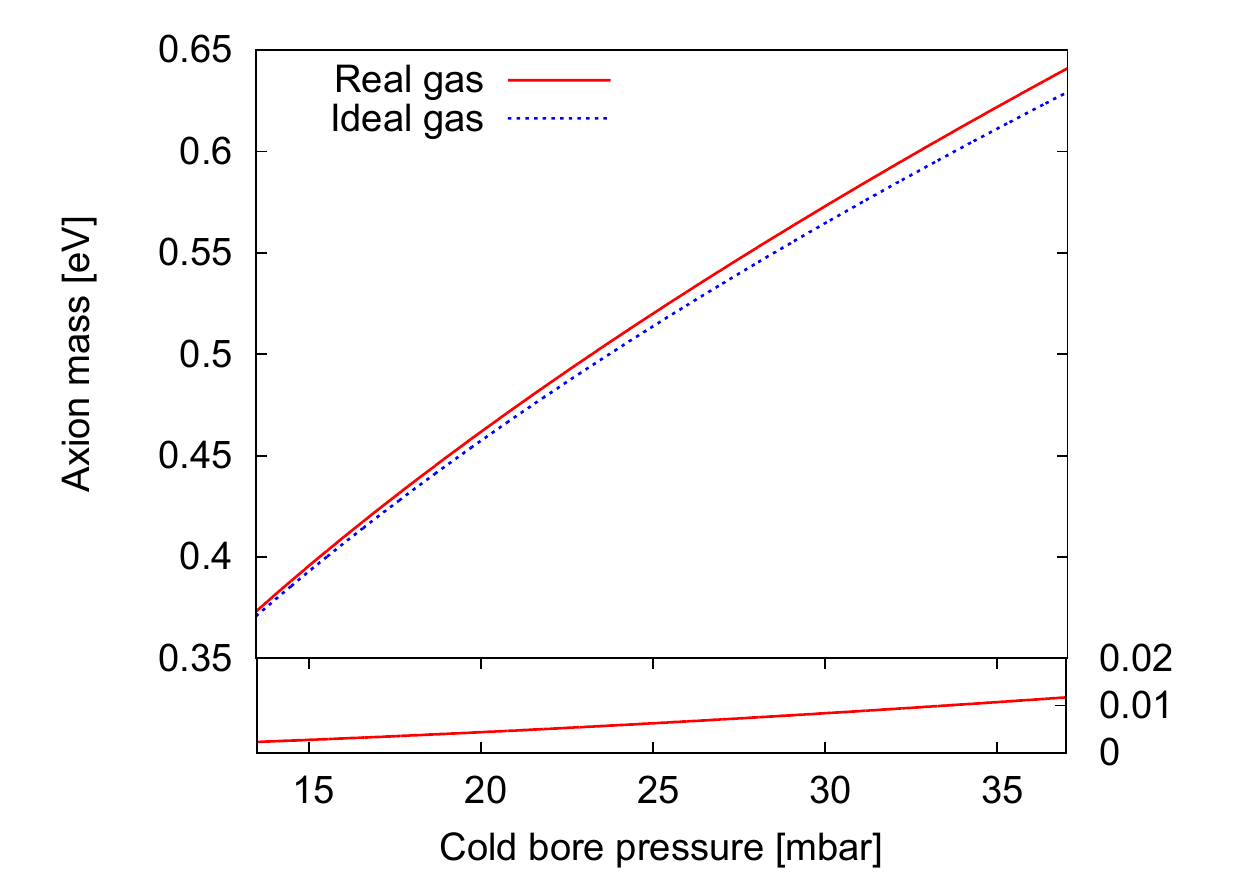}} \par}
\caption{\fontfamily{ptm}\selectfont{\normalsize{ Axion mass calculation as a function of the cold bore pressure by using a real gas model or an ideal gas model. The bottom plot shows the divergence on axion mass given between models.  }}}
\label{fi:AxionMassVDW}
\end{figure}

\vspace{0.2cm}

\noindent where $T_o = 1.8$\,K is the renormalization temperature used to obtain $P_{check}$. Figure~\ref{fi:AxionMassVDW} represents the axion mass, obtained from the cold bore density by relation~XX, as a function of the cold bore pressure in the case of a real gas approach given in relation~\ref{eq:stateVDW} and the obtained with an ideal gas.

\vspace{0.2cm}

The effect of changes in magnet temperature to the pressure of the system due to the dependency of the second virial coefficient $B(T)$ with the temperature can be calculated by using the following expression.

\begin{equation}
\Delta P_{check} = \rho^2 \frac{dB(1.8\,\mbox{K})}{dT} RT_o ( T_{m} - T_o )
\end{equation}

\vspace{0.2cm}

The last pressure measured in 2008, $P \leq 38$\,mbar, corresponds to a density $\rho \leq 0.77$\,mg$/$cm$^3$. The maximum change of magnet temperature during tracking is around $\Delta T = 40$\,mK, leading to a maximum change in pressure $\Delta P_{check} \simeq 0.03$\,mbar. Value that cannot explain by itself the effects observed in the pressure evolution during tracking movement.

\subsection{CFD simulations for gas behavior characterization.}\label{sc:CFDsimulations}

The complexity of the system has pushed the efforts from the CAST collaboration towards a full description of the system by using Computational Fluid Dynamics (CFD) software. All the effects described in the previous sections, hydrostatic pressure, real equation of state, window temperature effects and convection, can be combined together in a simulation in order to obtain a more accurate relation between the measured pressure and the density in the magnetic region.

\vspace{0.2cm}

\begin{figure}[!htb]\begin{center}
\begin{tabular}{cc}
\includegraphics[width=0.5\textwidth]{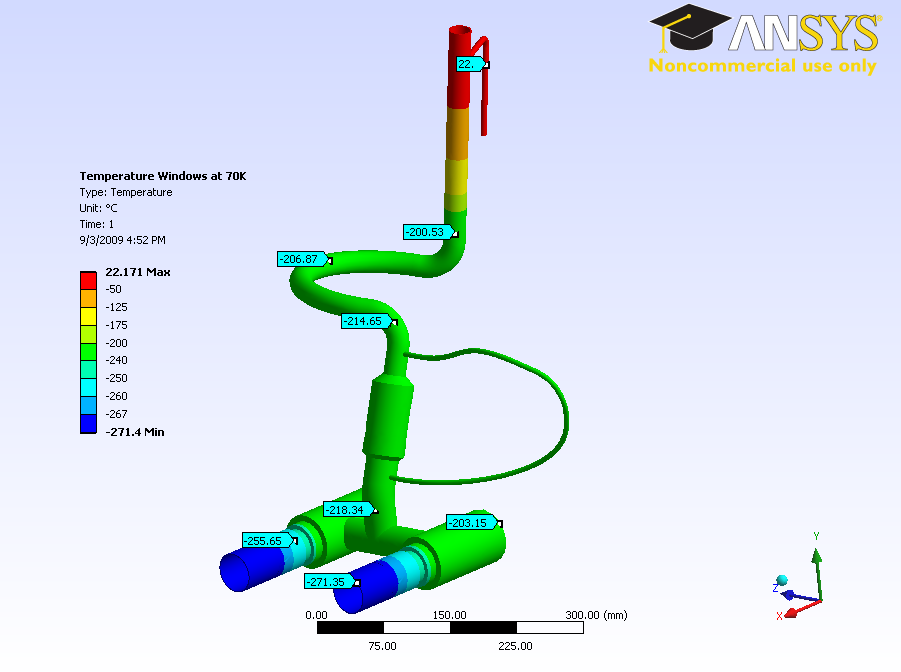} & \includegraphics[width=0.5\textwidth]{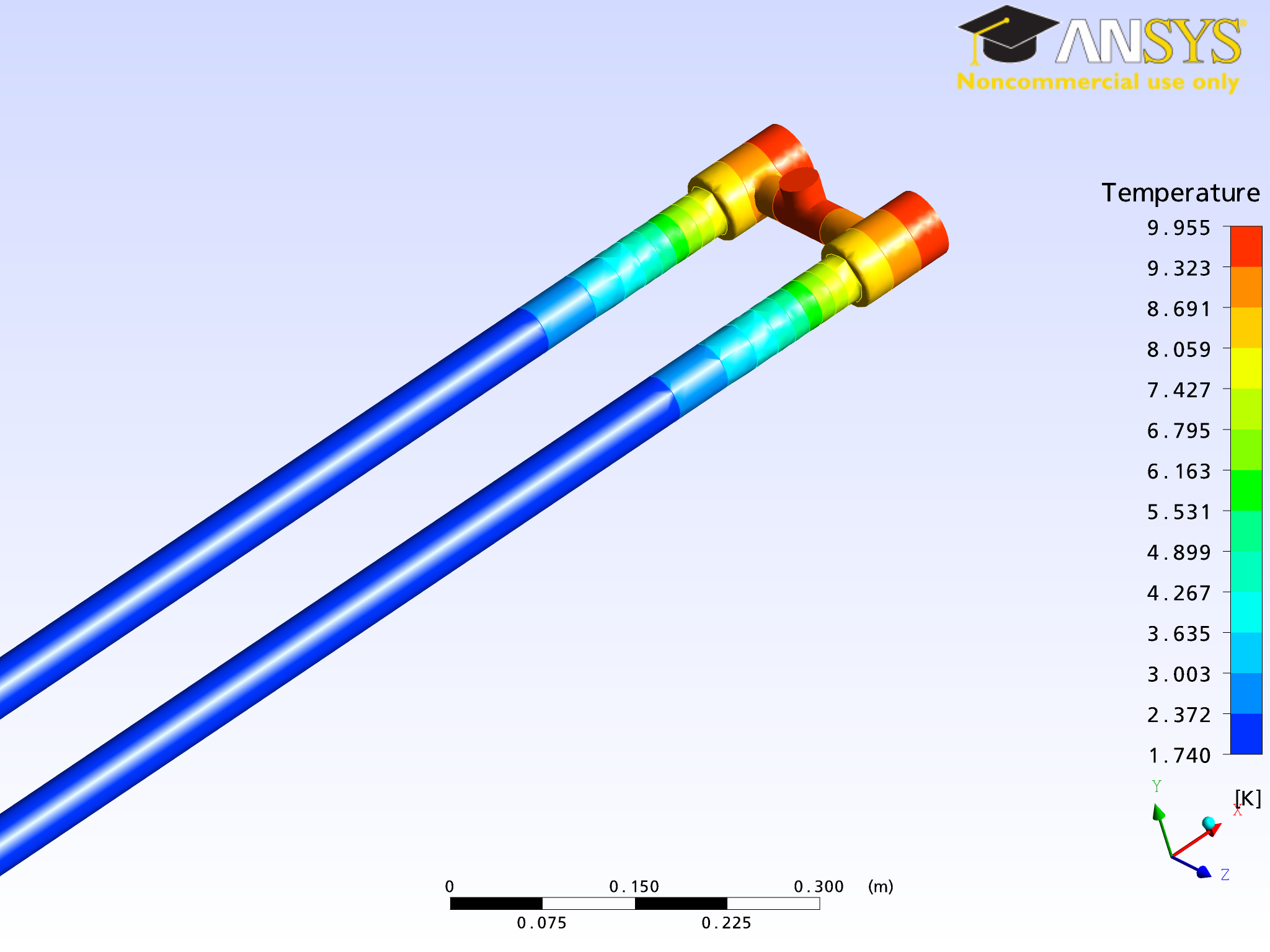} \\
\end{tabular}
\caption{\fontfamily{ptm}\selectfont{\normalsize{ On the left, a CAD drawing coming from a simulation of the temperature distribution of the connecting pipework from the measured pressure sensor location, on top at environment temperature, to the cold bore region. On the right, cold bore pipes temperature distribution. }}}
\label{fi:LeakSimulations}
\end{center}\end{figure}

The inner geometry of the CAST magnet where the gas is distributed has been reproduced with the maximum detail (see~Fig.~\ref{fi:LeakSimulations}) in order to describe the gas distribution and to fully understand the behavior of the gas inside the cold bore pipes and the effect of the different gas distributions on the pressure sensor at different configurations.

\vspace{0.2cm}

The computation inside the CAST magnet geometry can take into account all the available measured values as input parameters for the simulation, such as number of moles, measured pressure, temperature of the windows and the different vertical positions of the magnet that have place during tracking movement.

\vspace{0.2cm}

It is expected from simulations to reproduce the effects that are observed during tracking in order to separate the contributions to the changes in the pressure sensor coming from real density changes and the other secondary effects, due to changes on magnet temperature and hydrostatic pressure. In summary, CFD simulations will provide

\begin{itemize}

\item Correction of hydrostatic pressure.
\item Inclusion of convection and gravitational effects in the model.
\item A more accurate gas temperature distribution.
\item Real equation of state applied to the non constant temperature distribution along the pipes.
\item Calculation of gas behavior at different angles and simulation of tracking.

\end{itemize}

\vspace{0.2cm}

\begin{figure}[!htb]
{\centering \resizebox{1.0\textwidth}{!} {\includegraphics[angle=90]{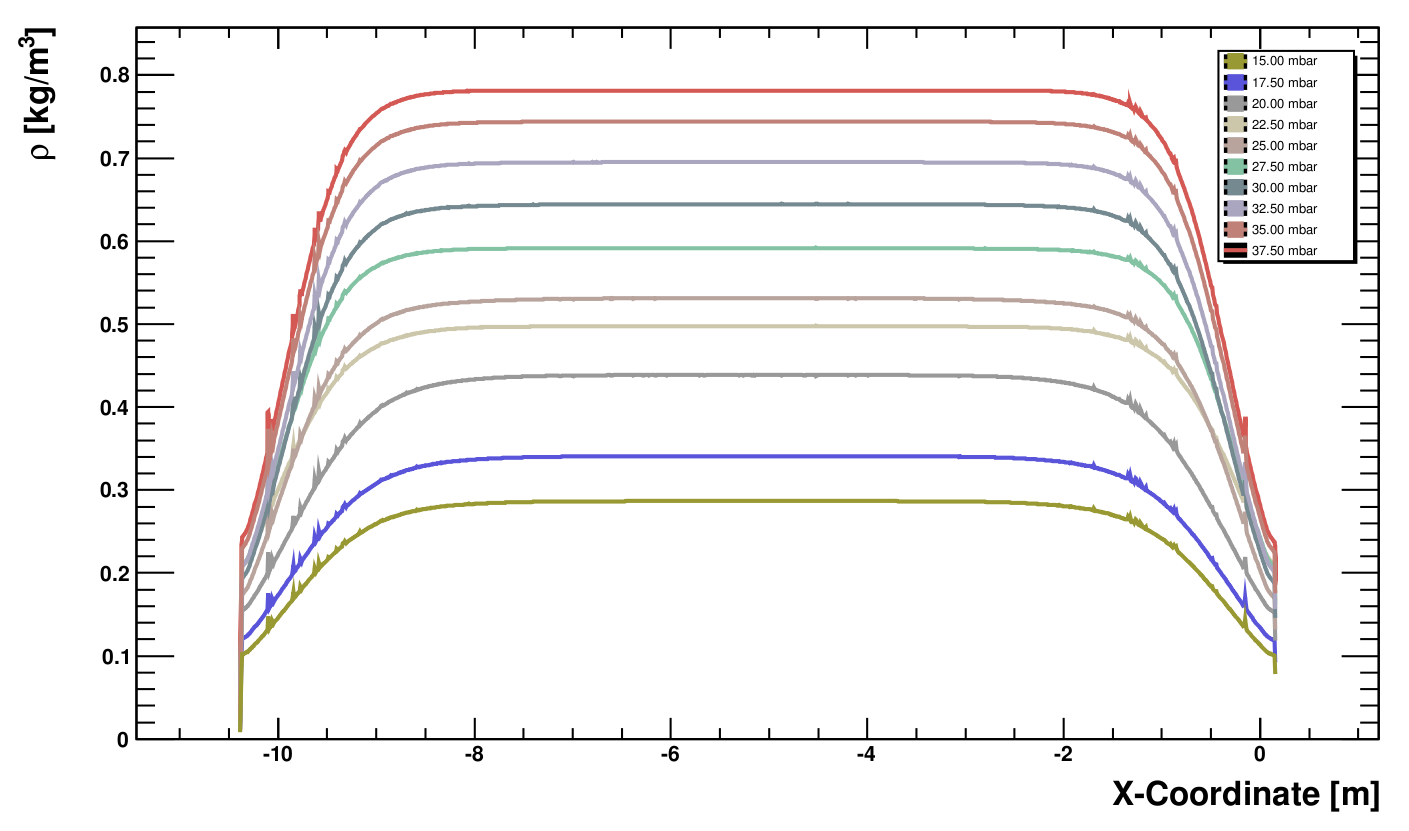}} \par}
\caption{\fontfamily{ptm}\selectfont{\normalsize{ Density profile along the magnet axis for the simulated pressure scanning carried out in 2008. The magnetic region limits are defined between $-0.49$\,m and $-9.74$\,m.  }}}
\label{fi:cbProfiles}
\end{figure}

Preliminary but not concluding results have already been obtained by running the first simulations for the corresponding 2008 data period (see~Fig.~\ref{fi:cbProfiles}). These first results obtain a homogeneous density profile length shorter than that defined the magnetic region of the CAST magnet. The maximum length loss along the 2008 data taking period is always below $\Delta L \leq 2$\,m, fact that will be necessarily taken into account for the calculation of the axion-photon coupling limit.

\vspace{0.2cm}

Simulations will be able to reduce the systematic errors coming from the description of a real gas and the hydrostatic pressure and other second order effects. However, the determination of the axion mass covered by CAST magnet will anyway relay on the measured pressure $P_{check}$, errors on the measurement of $P_{check}$ cannot be bypassed. 

\vspace{0.2cm}

Adding the errors on the systematics presented in sections~\ref{sc:hydrostatic}, \ref{sc:MagnetTemperature} and \ref{sc:VDW}, are higher than the errors coming from the pressure sensor. The error on the axion mass being covered by assuming the presence of these systematic effects it is estimated to be not more than $m_a \lesssim 1.5$\,meV, value to be improved by the on going simulations.

\chapter{The sunrise Micromegas tracking and background data in 2008.}
\label{chap:gLimitHe3}
\minitoc

\section{Introduction.}

After a short commissioning phase at the end of 2007, when the new $^3$He filling system described in section \ref{sc:He3System} was tested together with the new vacuum and detectors systems described in chapter~\ref{chap:micromegas}, the CAST experiment was ready for a new data taking period to fulfill its program and extend the sensitivity to axion masses higher than $0.38$\,eV. During the commissioning period, standard procedures were established for the tasks to carry on during the long coming data taking phase.

\vspace{0.2cm}

This chapter summarizes this data taking period, presents the final background and tracking analysis of the data measured in 2008 and fixes a first upper-limit on the axion-photon coupling in the new axion mass range explored.

\section{Detectors density steps coverage}

During 2008, three different detectors took data in the Micromegas sunrise line of CAST. The bulk Micromegas detectors (named B2 and B4) showed weakness when exposed to tracking movement vibrations. Mechanical stress provoked the degrading of the two Bulk detectors that took data in the Sunrise line the first density steps corresponding to the $^3$He Phase.

\vspace{0.2cm}

The first Bulk detector installed (B2) started malfunctioning after some time of operation. The detector was substituted by another Bulk detector (B4) that after some time showed also some degradation. More details are given in section~\ref{sc:Bulkproblem}.

\vspace{0.2cm}

The programmed September shutdown was used to replace the last Bulk detector by a Microbulk technology detector (M10). Since then, all the Micromegas detectors used in CAST have been built in this technology, since it has been proven to be the most performing in CAST conditions.

\vspace{0.2cm}

During 2008 data taking, the density settings from N160 to N411 were covered. Each of the detectors mentioned covered a density range given by the CAST settings shown in table~\ref{ta:Ncoverage}.

\begin{table}[bh!]
\begin{center}
\begin{tabular}{c|cc}
\small{Detector} & \small{Initial step} & \small{Final step} \\
\hline
\small{B2}	&  \small{N160} & \small{N269} \\
\small{B4}	&  \small{N270} & \small{N375} \\
\small{M10} 	&  \small{N374} & \small{N411} \\
\end{tabular}
\caption{Steps coverage for each Micromegas detector taking data in 2008.}
\label{ta:Ncoverage}
\end{center}
\end{table}

\section{CAST incidences affecting tracking data.}

The restart of data taking corresponding to 2008, after a long shutdown, was a period of instabilities where the CAST experiment had to overcome many challenging puzzles related with tracking system, detectors, vacuum systems and $^3$He system, including the leak problem described in chapter~\ref{chap:leak} that affects to the expected axion mass coverage during this period. The technical problems that will be described in this section affected the resulting data taking efficiency concerning the Sunrise Micromegas detection line. 

\subsection{Tracking system failure.}\label{sc:trackingproblem}

\begin{figure}[!ht]
{\centering \resizebox{1.0\textwidth}{!} {\includegraphics{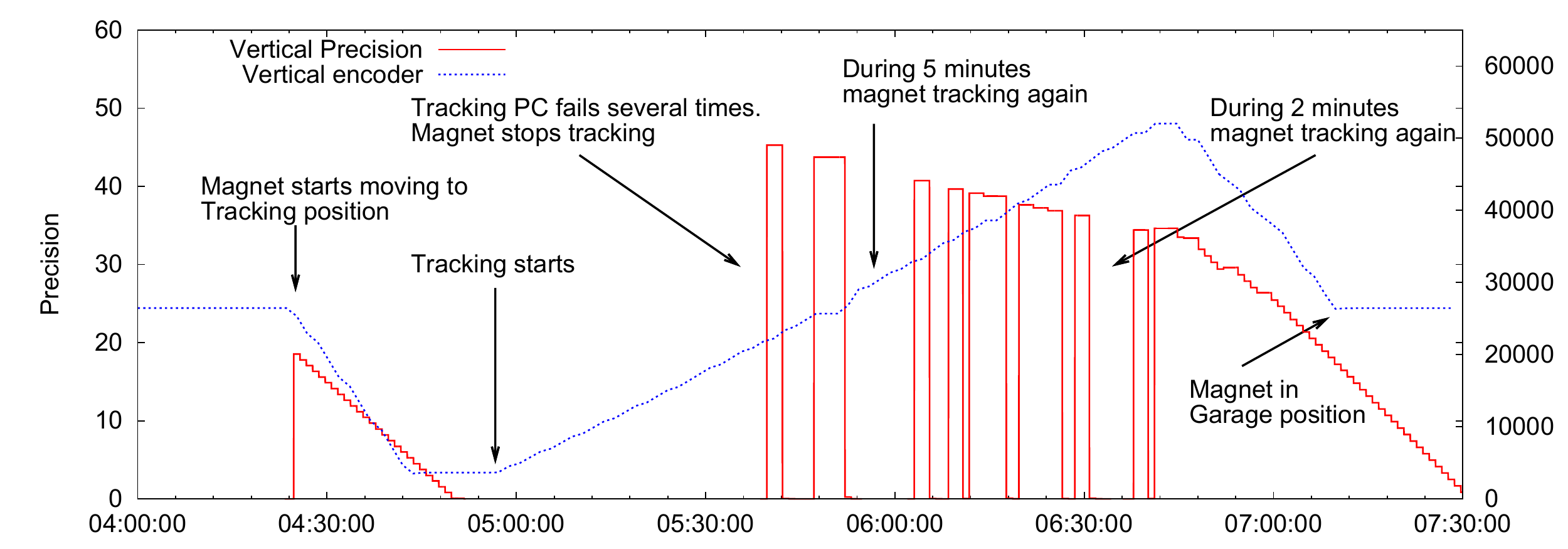}} \par}
\caption{\fontfamily{ptm}\selectfont{\normalsize{Morning tracking corresponding to settings N252/N253, on 21$^{st}$ of June. In red, the vertical tracking precision value during the magnet movement. The periods when the magnet stopped are revealed by the periods where the precision rises. In blue (dashed), tracking vertical encoder that describes the vertical movement; the magnet goes to the lowest position, stays for the start of Sun tracking and increases gradually its vertical value to follow the Sun.  }}}
\label{fi:trackingProblem}
\end{figure}

At the end of June, when the data taking was running smoothly, the magnet stopped its movement suddenly in the middle of a solar tracking. When the tracking system software was restarted the magnet was returning to the expected Sun position recovering the way lost. This process was repeating several times during tracking (see~Fig.~\ref{fi:trackingProblem}) leading to uncomplete tracking coverage for the affected settings.

The unexpected failure, not well understood at the beginning, was attributed to network and/or software issues due to suspicious time correlations with system antivirus update activity and the fact that the application restart was solving temporally the problem. However, after a couple of tracking days the problem was persisting stopping randomly during movement. The system hardware was replaced by a newer one where network access was limited. Moreover, all the cabling and connections related with the communication with the motor encoders box were replaced by newer ones and the communication cables were shortened.

\vspace{0.2cm}

Finally the problem was solved concluding that it was coming from noise affecting the communication between tracking system and hardware and/or aging connectors. The increasing number of systems connected during the shutdown in 2007, together with the increasing activity of the LHC\footnote{LHC power converters do operate in the same building as the CAST experiment.}, could account for the extra noise affecting a system that was running without problems since the very beginning of CAST data taking activity.

\vspace{0.2cm}

This tracking technical problem affected the Sunrise density settings N252, N253, N255 and N258, where a reduction of exposure time is present (see~appendix~\ref{chap:appendix4}). From which the most affected settings are N253 and N255.

\subsection{Bulk Micromegas detectors performance.}\label{sc:Bulkproblem}

The first period of 2008 phase was covered by Bulk Micromegas detectors. These detectors were fulfilling the required CAST experiment performance after installation.

\vspace{0.2cm}

\begin{figure}[!ht]
{\centering \resizebox{1.0\textwidth}{!} {\includegraphics{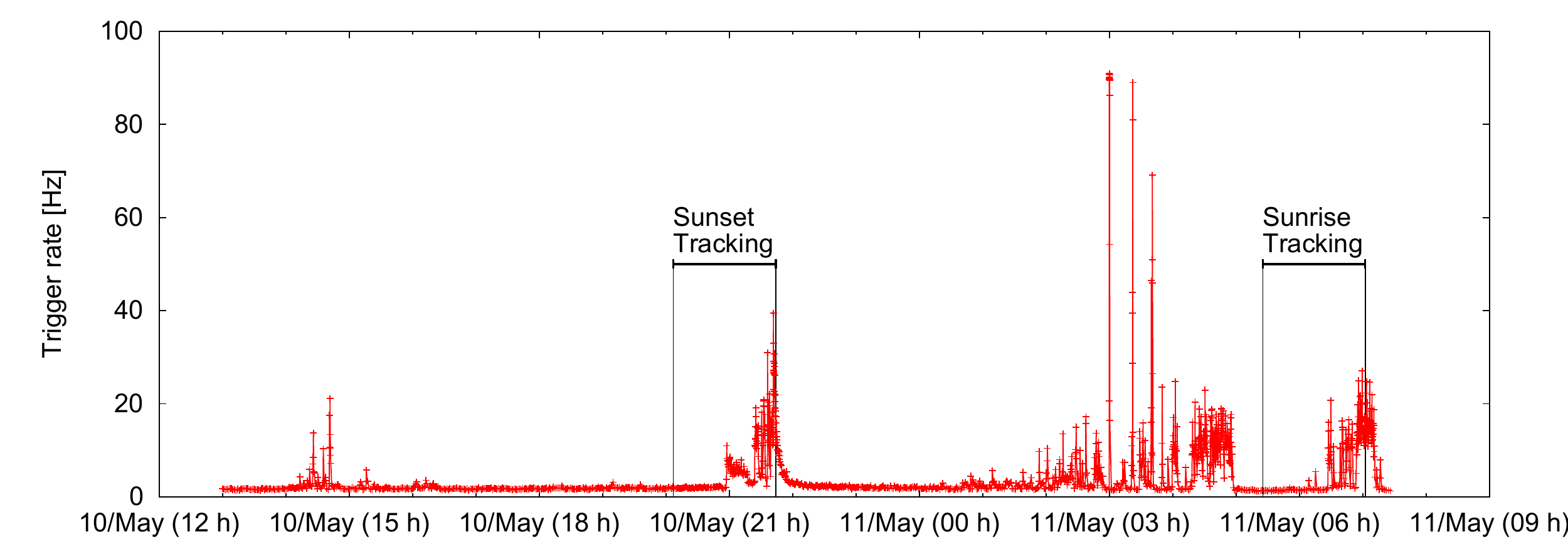}} \par}
\caption{\fontfamily{ptm}\selectfont{\normalsize{The presence of noise and sparks is reflected in the acquisition trigger rate. The effect on the trigger rate can be observed during magnet movement.   }}}
\label{fi:sparksProblem}
\end{figure}

The first detector (B2) started to show sparking processes after the first month of operation. At that time, these processes started to appear from time to time, and frequently during tracking periods (see~Fig.~\ref{fi:sparksProblem}). The problem could be bypassed by making mechanical interventions in the detector system. But after some period the detector developed a short circuit and had to be replaced by a new detector.

\vspace{0.2cm}

The new detector installed (B4) did not show the same sparking problems as the previous detector, however it showed a gradual degrading that was noticed in the energy resolution of the detector (detailed on section~\ref{sc:gain2008}). In spite of the worst energy resolution the background discrimination of the detector was not affected.


\vspace{0.2cm}

The Bulk detector behavior observed and the fact that Microbulk detectors did not show this kind of problems, pointed that the phenomena observed in CAST Bulk detectors was surely due to mechanical stress and vibrations at the CAST experiment data taking conditions. Mechanical stress and vibrations could be related with a mesh displacement or distortion that would explain the start of sparking processes and the non homogeneity of the mesh amplification gap at the detector active area, a fact that it is not affecting to Microbulk detectors since mesh and strips readout are built in a single entity.

\subsection{The VME acquisition problem.}\label{sc:VMEproblem}
During a short period, the acquisition system of the Sunrise Micromegas detector was running with reduced performance. A problem with the \emph{MATACQ} acquisition module (see section~\ref{sc:readout}) in charge of recording the mesh pulse (also named FADC data) was malfunctioning. A new module was available for replacement after few days affecting a total of 8 scanning settings (N181 - N187).

\vspace{0.2cm}

The reduced information recorded during these days for background and tracking data affected directly the background discrimination potential of the detector. The background level achieved by using only the strips data information was increased by more than a factor $3$ (see~Fig.~\ref{fi:bckandmeanBckVsTime}).

\subsection{Sunrise Micromegas data taking efficiency in 2008.}

A total of 252 steps were covered by the CAST experiment in 2008. From which $6$ full steps were lost related with acquisition technical problems and Bulk sparking issues. Another $3$ steps were affected by low tracking exposure time due to the tracking system problem, and $8$ settings were covered with reduced performance due to the acquisition problem described in section~\ref{sc:VMEproblem}.

\vspace{0.2cm}

The Sunrise Micromegas line covered at least $248$ settings ($98.4\%$), from which $235$ settings ($93.2\%$) were covered with full tracking time and performance.

\section{Detector stability and gain history.}\label{sc:gain2008}

\begin{figure}[!ht]
{\centering \resizebox{1.0\textwidth}{!} {\includegraphics{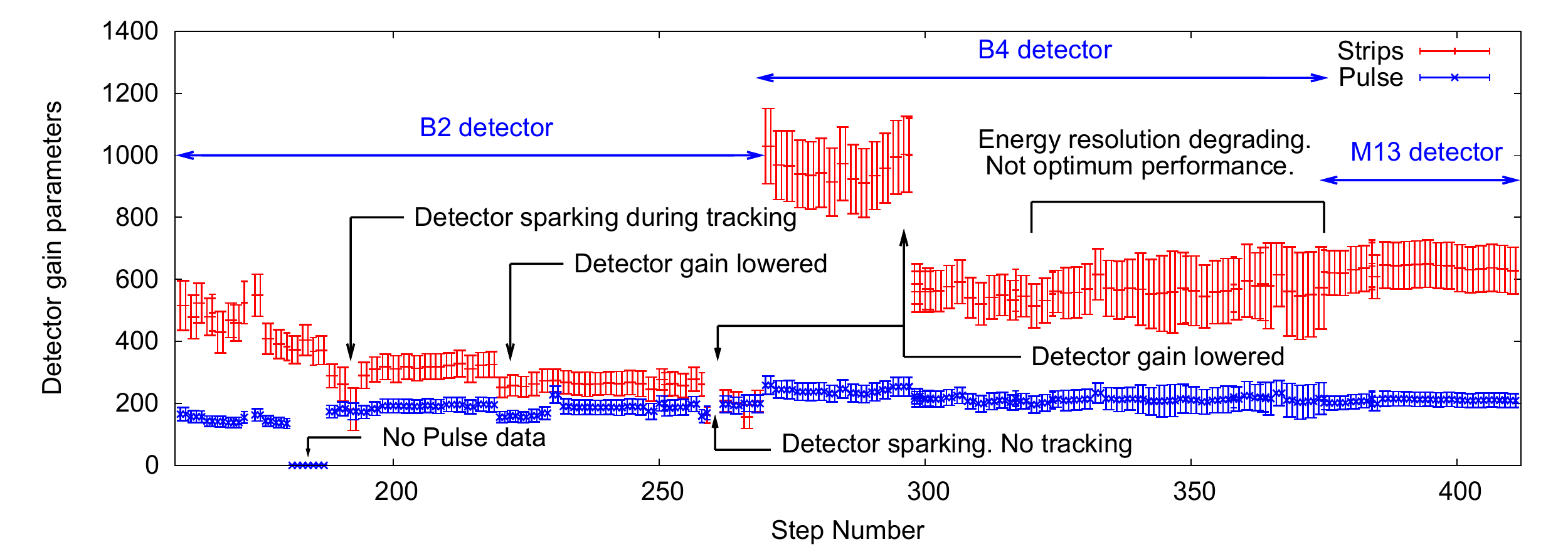}} \par}
\caption{\fontfamily{ptm}\selectfont{\normalsize{ Mesh pulse and strips gain evolution (ADC units in mV) for each of the detectors taking data in 2008 as a function of the step covered, which is always increasing over time. Error bars are proportional to the obtained detector energy resolution at each calibration run. }}}
\label{fi:gain2008}
\end{figure}

Figure~\ref{fi:gain2008} summarizes the gain evolution as a function of the step numbers covered during 2008. At the beginning of the run were observed some gain instabilities with the first Bulk detector (B2). These instabilities might be related with the fact that some changes were having place in the mesh structure (due to mechanical stress and vibrations produced by magnet movement) that was affecting the gain day by day. As a result of these gain fluctuating period the detector started to have sparking processes from time to time as described in section~\ref{sc:Bulkproblem}. A mechanical intervention at the detector PCB support holding it in a better way improved the detector gain stability. Afterwards, the detector gain was intentionally lowered in order to diminish the potential of the detector to become in short-circuit, however after some weeks of operation the detector became suddenly unoperational and had to be replaced by the B4 detector.

The new B4 detector showed a better gain stability. The first days of operation the detector was performing properly. The detector gain was higher than the required and at some point it was decided to reduce the gain by lowering the amplification field, thus showing the detector was operating far away from its operational limit voltage. Anyway, after some weeks of operation, without presence of sparking processes, the detector started to show reduced performance that was reflected in an increasing energy resolution over time represented in figure~\ref{fi:gain2008} by the increasing error bar size.

\vspace{0.2cm} 

The new M10 detector installed during the September shutdown showed a better gain stability and a higher robustness than Bulk detectors, in CAST data taking conditions. 

\vspace{0.2cm}

The effect of gain changes in the final background measured by Micromegas detectors is minimized by the fact that $^{55}Fe$ calibration runs are taken just after the tracking is finished and each calibration it is used to analyze the corresponding background data.

\vspace{0.2cm}

\begin{figure}[!ht]
{\centering \resizebox{1.0\textwidth}{!} {\includegraphics{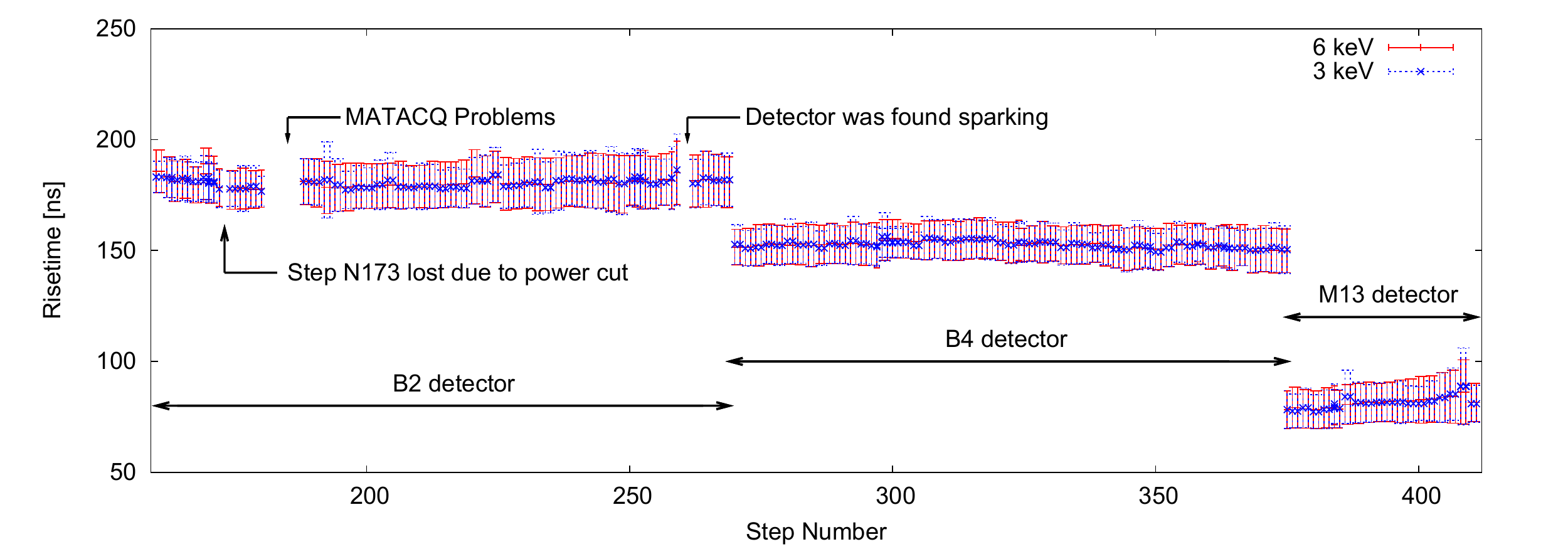}} \par}
\caption{\fontfamily{ptm}\selectfont{\normalsize{ Pulse risetime parameter evolution for each of the detectors taking data in 2008 as a function of the step covered. The pulse risetime is directly related with the electronic settings imposed in each period. Thus, Microbulk detector shows a lower risetime value due to the thinner amplification gap. }}}
\label{fi:meanrt2008}
\end{figure}

\begin{figure}[!ht]
{\centering \resizebox{1.0\textwidth}{!} {\includegraphics{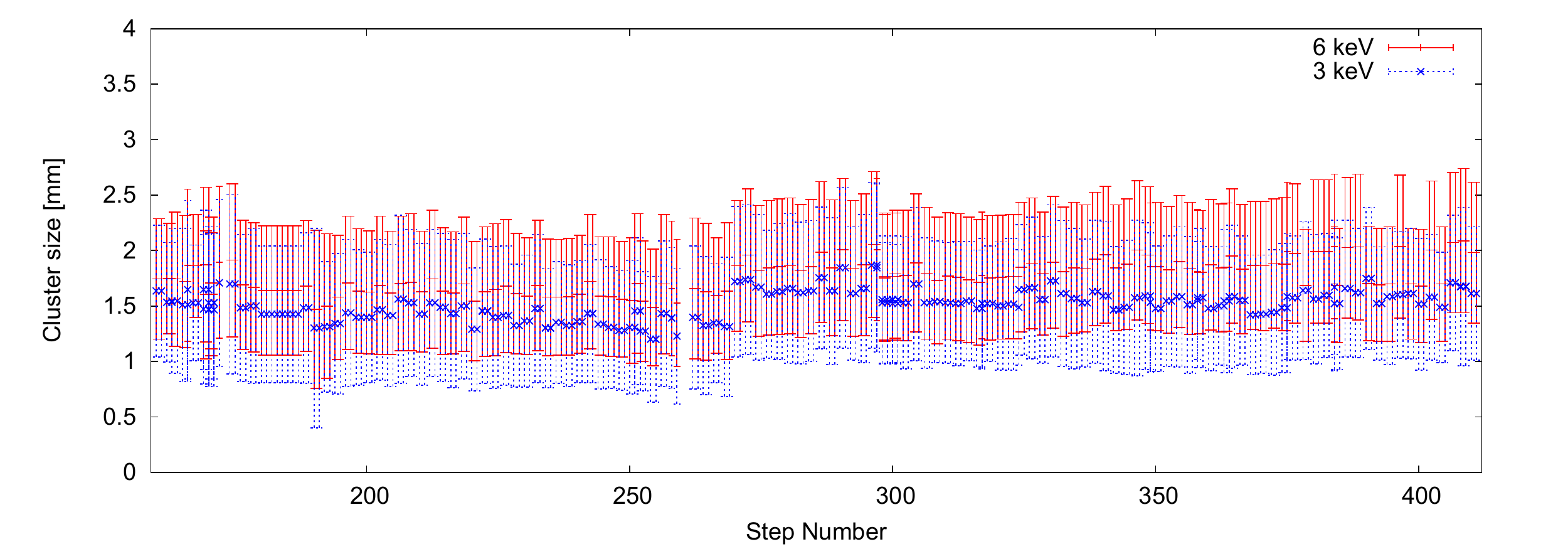}} \par}
\caption{\fontfamily{ptm}\selectfont{\normalsize{ Cluster size parameter evolution in x-plane strips as a function of the steps covered in 2008.  }}}
\label{fi:meansize2008}
\end{figure}

In spite of the gain instabilities observed at some periods, non significative systematic effects are observed in the parameters used for background discrimination. Figures~\ref{fi:meanrt2008}~and~\ref{fi:meansize2008} show the evolution of some of the Micromegas main parameters obtained with the $^{55}Fe$ calibration runs taken during the data taking period in 2008.

Moreover, the events measured at the escape peak and at the main $^{55}Fe$ peak present compatible mean values. The effects of parameter fluctuations on the background discrimination rules is smoothed by the redefinition of background cuts applied by using the values obtained in each calibration run.

\section{Micromegas background 2008 data analysis.}\label{sc:background2008}

The background level of the detectors needs to be defined at each tracking day in order to obtain the coupling constant limit reached by CAST data taking in 2008. In absence of signal, the tracking data should be compatible within the statistical expectation. The rules used to define the mean background level at each density step, together with some systematics studies showing the background and tracking compatibilities and a full summary of the background data collected are presented in this section.

\subsection{Background definition.}\label{sc:backgroundDefinition}

The final background for 2008 data was obtained by applying the modified multivariate method described in chapter~\ref{chap:discrimination}. The cuts efficiency was fixed at $85\%$, chosen by the arguments given at section~\ref{sc:optimumeff} for events coming from both, the main and escape, peaks from the $^{55}Fe$ source.

\vspace{0.2cm}

After several empirical tests, using the arguments given in section~\ref{sc:optimumdiscriminants}, the best discriminants combination that better suited to all three Micromegas detectors taking data in the Sunrise line are the ones given in the following list.

\begin{itemize}
\item Risetime
\item Cluster size balance
\item Cluster multiplicity
\end{itemize}

The pulse width parameter, as described in chapter~\ref{chap:discrimination}, is one of the best candidates to be used for background rejection. However, this parameter had to be removed from the analysis due to an unexpected and intense background reduction period in detector B2 when this parameter is introduced in the analysis. This parameter did not show significative deviations between the different calibration runs taken, however the physical reason for such background reduction is not well understood up to now. This effect could be attributed to a systematic effect affecting background data (described on appendix~\ref{chap:appendix5}) and I decided not to take it into account for the analysis of this data set.

\vspace{0.2cm}

Once the background counts have been selected after applying the X-ray discrimination rules described in chapters~\ref{chap:rawdata}~and~\ref{chap:discrimination}, the background level to be used for each tracking run can be obtained.

\vspace{0.2cm}

Every day a new background file of about $23$\,hours is taken. Background files with recording time lower than $5$\,hours have been rejected in order to avoid fake background data due to testing or detector problems.

\vspace{0.2cm}

In order to take into account possible detector background evolution during the data taking period in 2008 it was considered to take for each tracking the mean background of the surrounding days. For each tracking day $6$ surrounding natural days were taken into account, if available. Tracking data periods have been removed from the background level definition, as well as a fixed time interval of $15$\,minutes around the beginning and the end of each tracking.

\vspace{0.2cm}

\begin{figure}[!ht]
{\centering \resizebox{1.0\textwidth}{!} {\includegraphics{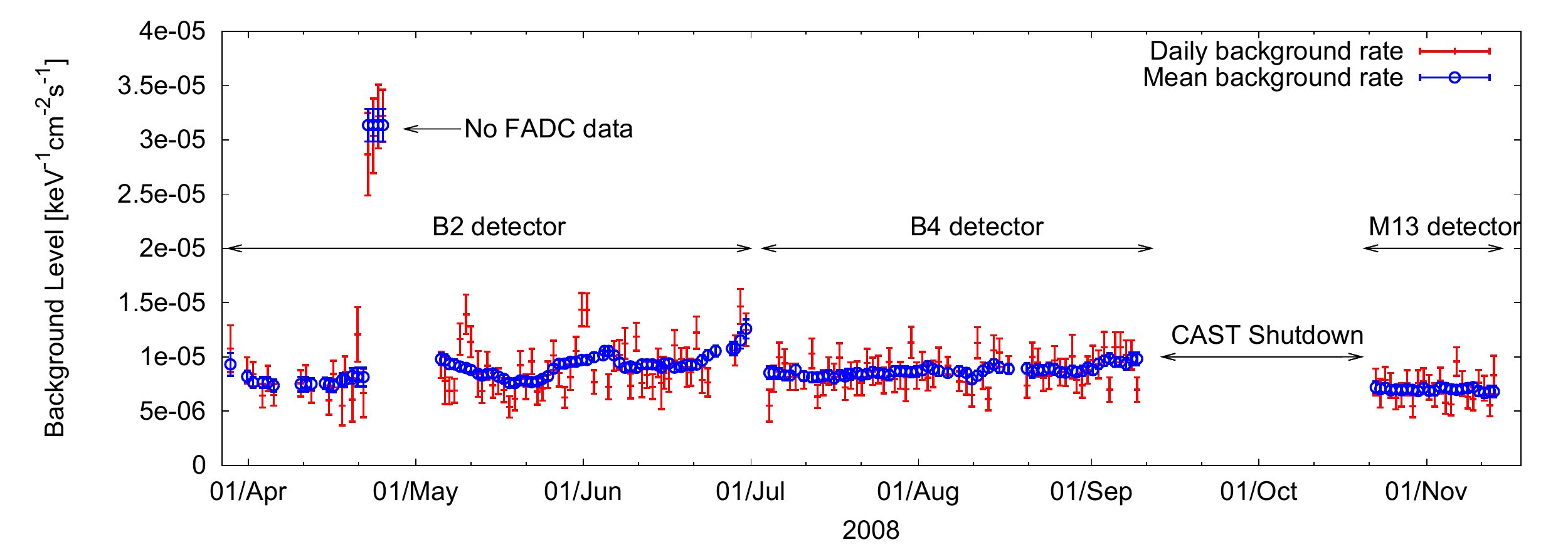}} \par}
\caption{\fontfamily{ptm}\selectfont{\normalsize{Daily background rate during 2008 data taking period together with the mean background rate calculated. }}}
\label{fi:bckandmeanBckVsTime}
\end{figure}

In order to obtain a mean background level at each given tracking day, by using the remaining backround data left after applying the mentioned conditions, the data had to be split into different subsets concerning the periods that different detectors were taking data. In the case of B2 detector, the mean background calculation was split into more sets due to the presence of background days with non full rejection potential (see Fig.\ref{fi:bckandmeanBckVsTime}) as described in section~\ref{sc:Bulkproblem}.

\vspace{0.2cm}

Figure~\ref{fi:bckandmeanBckVsTime} shows the resulting mean background for each day. The B2 detector shows the higher background fluctuations along its data taking period. The periods when the trigger rate was saturating the electronic acquisition ($>50$\,Hz), and/or detector mesh current crossed the threshold of $50$\,nA were removed from the final background data. However, these background fluctuations might be related with the sparking detector state that it was appearing from time to time. Probably the mesh re-stabilization processes that were having place and were also affecting to the gain of the detector caused sparks with a pattern similar to X-rays and could not be rejected by the discrimination analysis.

\subsection{Tracking definition}\label{sc:trackingDefinition}

In order to assure that the detected counts during the CAST tracking correspond to the conditions when the detector was exposed to the possible axion-photon conversion in the magnet some selection rules have been applied. These selection rules are defined by the fullfilling of all the following conditions that define a tracking

\begin{itemize}
\item Internal tracking system determining Solar tracking conditions; Sun reacheable and solar tracking system switched on.
\item Horizontal and vertical magnet position expected by tracking system coordinates $<0.01$\,degrees.
\item Vertical and horizonal movement.
\item Magnetic field on.
\item VT3 gate valve is open.

\end{itemize}

\begin{figure}[!ht]
{\centering \resizebox{1.0\textwidth}{!} {\includegraphics{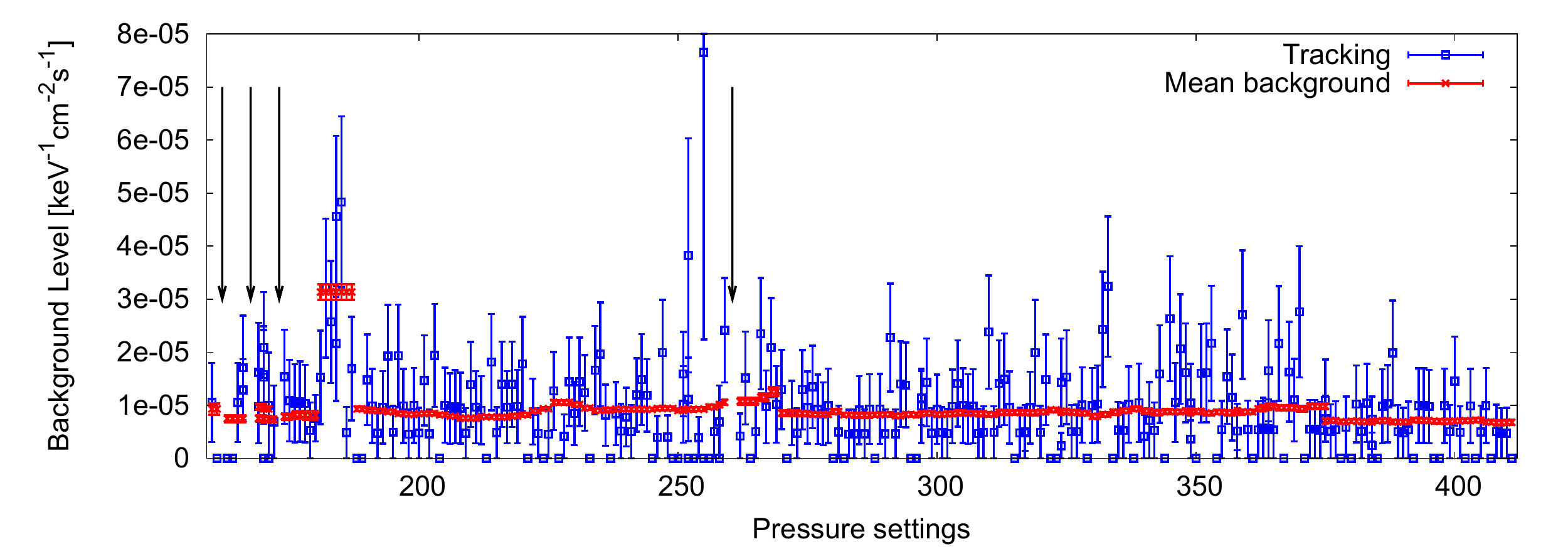}} \par}
\caption{\fontfamily{ptm}\selectfont{\normalsize{Tracking rates and mean background rate defined as a function of the step number measured in 2008. The arrows point the step settings that were missed, the very first ones related with DAQ problems and the middle one related with tracking system problem. }}}
\label{fi:bckandtrackingvssetting}
\end{figure}

The corresponding tracking counts for each tracking were obtained from the resulting background selection by imposing the tracking conditions described. Figure~\ref{fi:bckandtrackingvssetting} represents the mean background level used for each tracking step covered together with the normalized tracking rate. The higher rates observed at around step N250 are due to the shorter exposure time related with the tracking system failures (described on section~\ref{sc:trackingproblem}).

\vspace{0.2cm}

\begin{center}
\begin{figure}[!h]
{\centering \resizebox{0.95\textwidth}{!} {\includegraphics{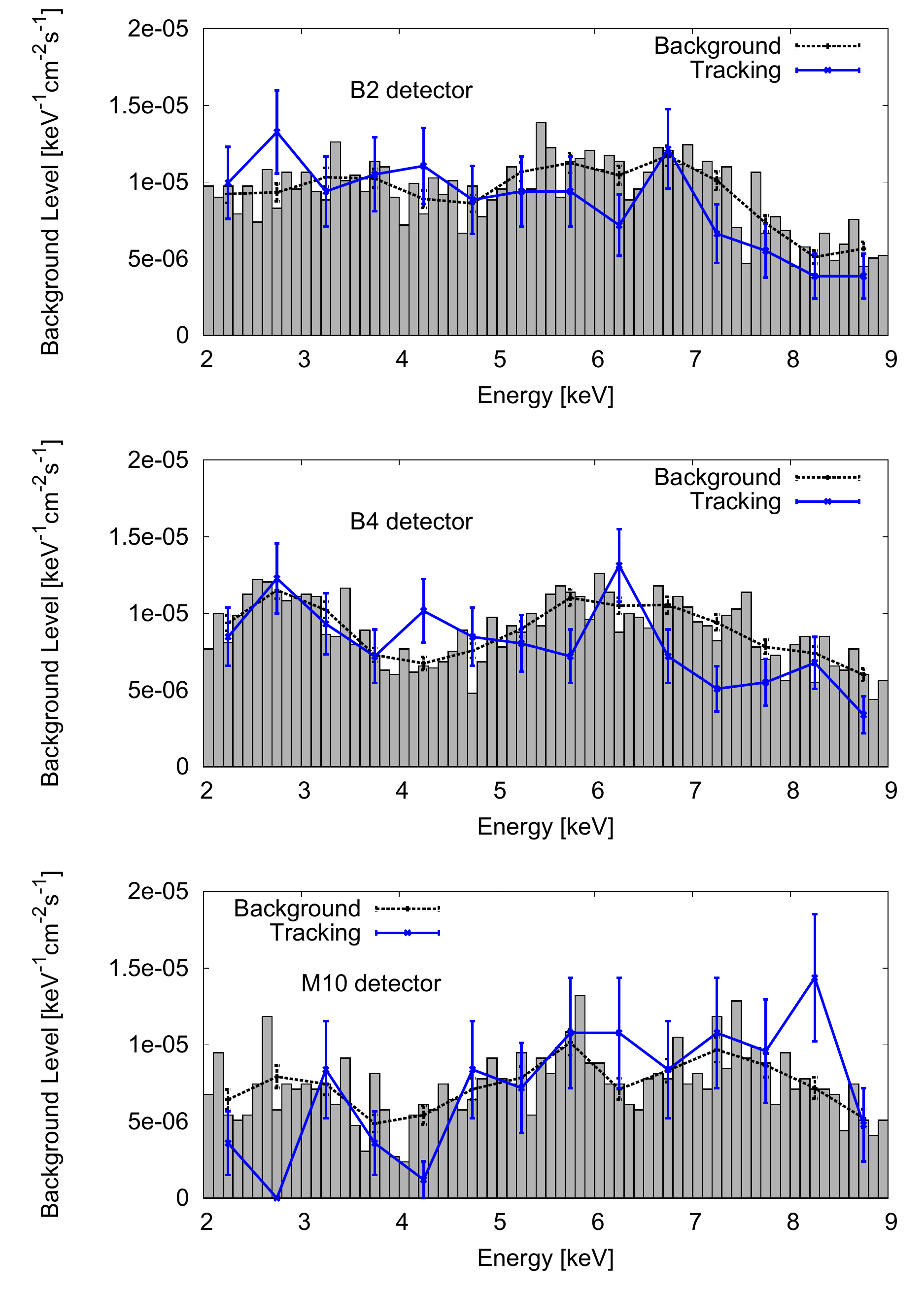}} \par}
\caption{\fontfamily{ptm}\selectfont{\normalsize{ Background and tracking analogue spectrum for each of the detectors taking data in the Sunrise line during 2008. An standard binning of $0.5$\,keV is chosen to compare tracking and background data. Boxes represent the background level for a $0.1$\,keV energy~binning width.  }}}
\label{fi:TrackingBackgroundSpectra2008}
\end{figure}
\end{center}

The overall background and tracking spectra for each of the detectors taking data in 2008 is shown in figure~\ref{fi:TrackingBackgroundSpectra2008}. The B2 detector spectrum shown is integrated over the period after the FADC acquisition problem.

\vspace{-0.8cm}

\subsection{Background and tracking data.}

The total tracking and background data collected by the Sunrise Micromegas detectors during the three data taking sub-periods adds up to $3662.7$\,hours from which some data has been subtracted by the conditions given by the background definition given in section~\ref{sc:backgroundDefinition}. After removal of the corresponding tracking data periods from the total background time the remaining effective data is in total $3284.5$\,hours.

\vspace{0.2cm}

The overall selected background statistics for each Micromegas detector taking data in 2008 is summarized in table~\ref{ta:background2008}.

\vspace{0.2cm}


\begin{table}[hb]
\begin{center}
\begin{tabular}{cccc}
  &  \bf{Time [$h$]} &   \bf{Counts} &   \bf{Level [$cm^{-2}s^{-1}keV^{-1}$]} \\
\hline
  &  &   &  \\
\vspace{0.2cm}
\bf{B2} &  1323.85 &  3078 &   $8.90 \cdot 10^{-6} \pm 0.16 \cdot 10^{-6}$  \\
\vspace{0.2cm}
\bf{B4} &  1395.29 &  3177 &   $8.71 \cdot 10^{-6} \pm 0.15 \cdot 10^{-6}$  \\
\vspace{0.2cm}
\bf{M10} &  565.32 &  1048 &   $7.09 \cdot 10^{-6} \pm 0.21 \cdot 10^{-6}$  \\
\hline
\vspace{0.2cm}
  &  &   &  \\
\bf{Total} &  3284.46 &  7303 &    \\
\end{tabular}
\end{center}
\caption{Overall background statistics for each of the Micromegas detectors taking data during 2008. In B2 detector statistics have been removed $53.20$ background hours corresponding to no FADC data.}
\label{ta:background2008}
\end{table}

The Sunrise Micromegas detectors were exposed to tracking conditions a total of $212.12$\,hours. Table~\ref{ta:tracking2008} summarizes the tracking coverage of each detector together with the tracking counts detected and the normalized tracking level. The tracking statistics in table~\ref{ta:tracking2008} for B2 detector do not include about $5.85$\,hours of tracking data corresponding to non FADC data taking days.

\begin{table}[hb]
\begin{center}
\begin{tabular}{cccc}
 &   \bf{Time} [$h$] &   \bf{Counts} &   \bf{Level [$cm^{-2} s^{-1} keV^{-1}$]}  \\
\hline
  &  &   &  \\
\vspace{0.2cm} 
\bf{B2}  &  84.01   &  201  &   $9.15 \cdot 10^{-6}  \pm 0.65 \cdot 10^{-6} $   \\
\vspace{0.2cm} 
\bf{B4}  &  90.32  &  206  &   $8.73 \cdot 10^{-6}  \pm 0.61 \cdot 10^{-6} $   \\
\vspace{0.2cm} 
\bf{M10}  &  31.95  &  47  &   $5.63 \cdot 10^{-6}  \pm 0.82 \cdot 10^{-6} $   \\
\hline
\vspace{0.2cm}
  &  &   &  \\
\bf{Total}  &  206.27  &  454  &   \\
\end{tabular}
\end{center}
\caption{Overall tracking statistics for each of the Micromegas detectors taking data during 2008. In B2 detector statistics have been removed $5.85$ tracking hours corresponding to no FADC data.}
\label{ta:tracking2008}
\end{table}

\vspace{0.2cm}

Background and tracking levels for each setting in 2008, together with the total detected counts and exposure time in each tracking can be found at appendix~\ref{chap:appendix4}.

\subsection{Background systematic studies for 2008 data.}

Some statistical tests have been performed in order to check the compatibility of tracking and background data after the assumption of the absence of signal for each of the detectors taking data in the Sunrise Micromegas line.

\subsubsection{Bulk detectors}

Bulk detectors show an overall background and tracking compatibility. The mean background and tracking levels are comparable within the poissonian errors calculated, as it is noticed by comparing the values shown in tables~\ref{ta:background2008}~and~\ref{ta:tracking2008}.

\vspace{0.2cm}

In order to cross-check the good statistical behavior of background and tracking data it has been chosen to split the final available background data, as defined in the previous section, in binning periods of $1$\,hour. The distribution of counts in this short period should follow a poissonian distribution.  Only full hours have been taken into account to describe the counts distribution at the given binning time period. The resulting mean from the total number of bins taken into account is $\mu_B = 2.38$ for the B2 detector, and $\mu_B = 2.27$ for the B4 detector.

\vspace{0.2cm}

\begin{figure}[!ht]
{\centering \resizebox{0.75\textwidth}{!} {\includegraphics{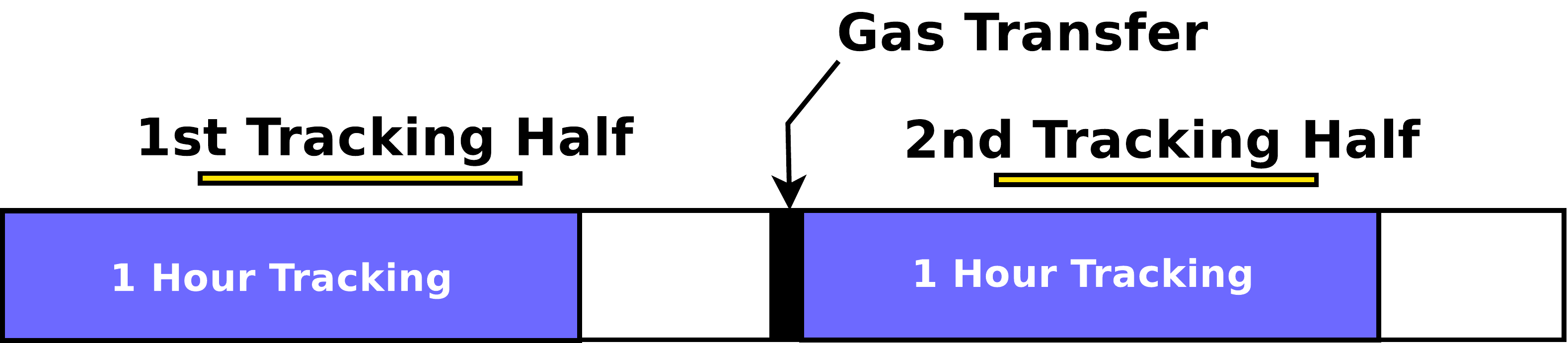}} \par}
\caption{\fontfamily{ptm}\selectfont{\normalsize{Time binning schema used to study the statistics compatibility of tracking and background data. First binning period actually overlaps with the second tracking half since half tracking period is slightly shorter than $1$\,hour, and the second binning period will contain some counts that are not completely in tracking conditions but still contain magnet movement and detector exposed to cold bore magnetic field and buffer gas.  }}}
\label{fi:BulksTrackingDefinition}
\end{figure}

The statistical tracking counts distribution subject of this study is obtained by taking the first hour since the beginning of the tracking and the first hour since the end of the gas transferred to the cold bore in the middle of the tracking~(see~Fig.\ref{fi:BulksTrackingDefinition}). 

\begin{figure}[!ht]
{\centering \resizebox{1.0\textwidth}{!} {\includegraphics{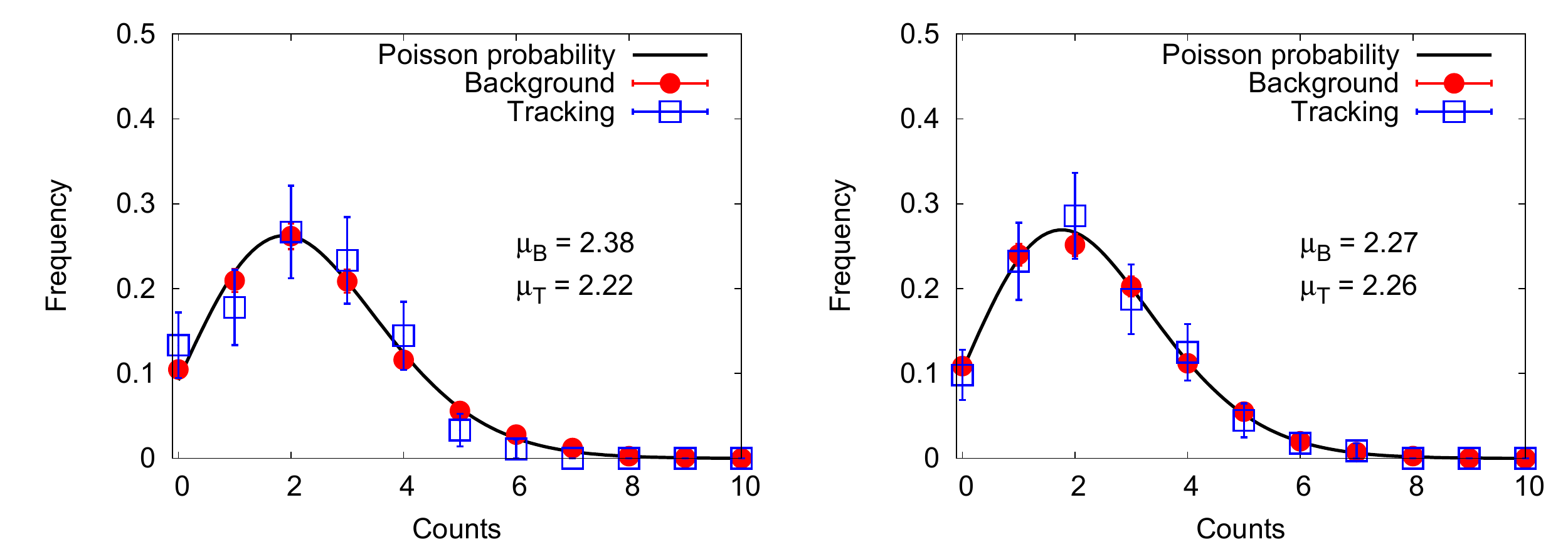}} \par}
\caption{\fontfamily{ptm}\selectfont{\normalsize{Number of counts distribution for a time binning length of 1\,hour. Tracking distribution is obtained from the schema given in figure~\ref{fi:BulksTrackingDefinition}. On the left, \emph{B2} detector, on the right \emph{B4} detector. }}}
\label{fi:BulksTrackingCompatibility}
\end{figure}

The theoretical distribution for the mean background counts obtained in this particular time binning it is drawn in figure~\ref{fi:BulksTrackingCompatibility}, for B2 and B4 detectors. In these plots, the experimental background counts distribution and the tracking counts distribution are also represented. Background distribution shows a perfect agreement of the experimental distribution with the expected theoretical poissonian distribution, and tracking counts distribution is within the expected errors due to the shorter data period available in tracking conditions.

\subsubsection{Microbulk detector}

The data taking period covered with the M10 detector shows a bigger discrepancy between tracking and background levels (see tables~\ref{ta:background2008}~and~\ref{ta:tracking2008}). The background level measured in tracking conditions is lower than the background level defined in non-tracking conditions. 

\vspace{0.2cm}

\begin{figure}[!ht]
{\centering \resizebox{0.75\textwidth}{!} {\includegraphics{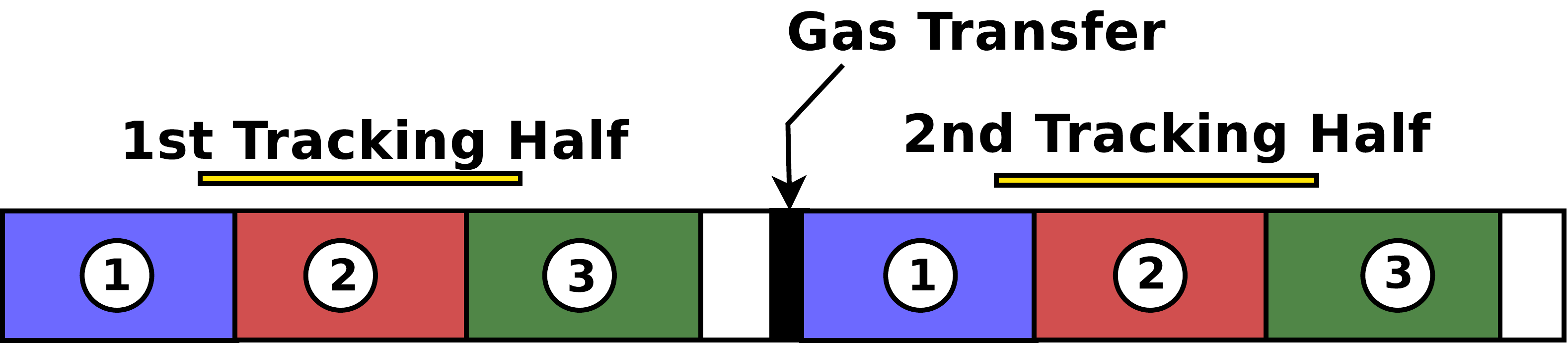}} \par}
\caption{\fontfamily{ptm}\selectfont{\normalsize{ Time binning schema used to study the statistics compatibility of tracking data with background data. }}}
\label{fi:M10TrackingDefinition}
\end{figure}

This systematic effect has been studied by using the same method as for Bulk detectors. In order to investigate the source of this systematic effect, the background from M10 detector has been split in binning sets of $15$\,minutes. The tracking data has been divided in three different groups associated with a relative tracking period of $15$\,minutes inside each tracking run (see~Fig.~\ref{fi:M10TrackingDefinition}).

\begin{figure}[!ht]
\begin{tabular}{cccccc}
\quad \quad \quad \quad \quad \quad  &
\resizebox{0.04\textwidth}{!} {\includegraphics{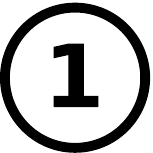}} & 
\quad\quad \quad \quad \quad \quad \quad \quad \quad \quad &
\resizebox{0.04\textwidth}{!} {\includegraphics{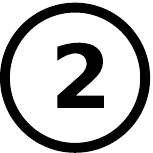}} &
\quad \quad \quad \quad \quad \quad \quad \quad\quad &
\resizebox{0.04\textwidth}{!} {\includegraphics{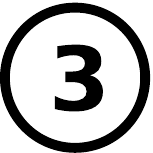}} \\
\end{tabular}

{\centering \resizebox{1.0\textwidth}{!} {\includegraphics{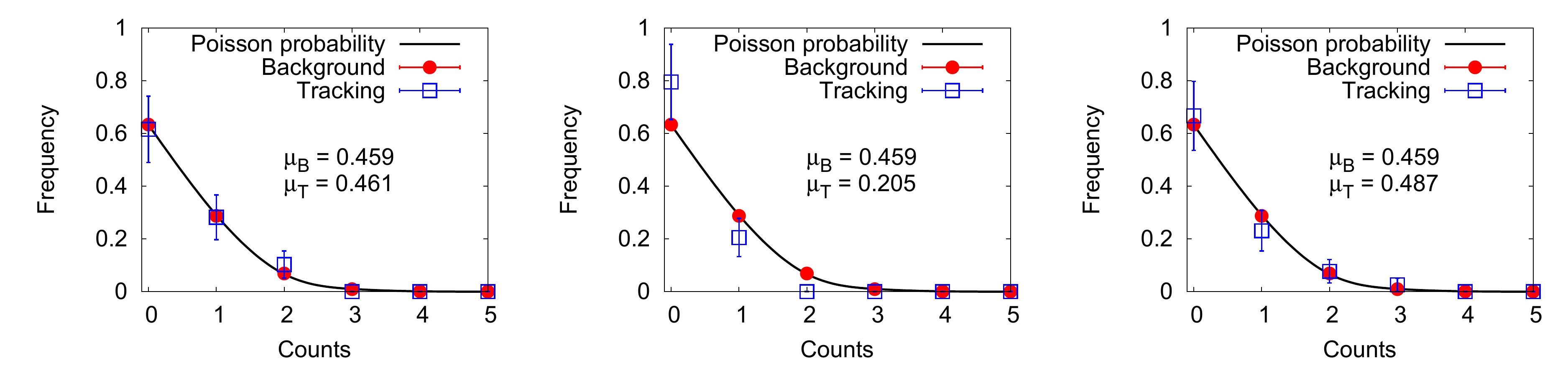}} \par}
\caption{\fontfamily{ptm}\selectfont{\normalsize{Number of counts distribution for a time binning length of 15\,minutes. Tracking distribution is obtained from the schema given in figure~\ref{fi:M10TrackingDefinition}.  }}}
\label{fi:M10TrackingCompatibility}
\end{figure}

The mean of the counts distribution for background data in this particular time binning scheme was $\mu_B = 0.459$. The theoretical poissonian distribution for this mean value is represented in figure~\ref{fi:M10TrackingCompatibility}, where it is observed that the experimental background counts distribution fits perfectly with the theoretical one.

\vspace{0.2cm}

Figure~\ref{fi:M10TrackingCompatibility} shows also the tracking counts distribution in each of the binning groups defined in figure~\ref{fi:M10TrackingDefinition}. It is observed how the first and the third group follow reasonable poissonian distributions and have a mean tracking value compatible with the background mean value. However, the mean tracking value for the second group defined is much lower than the expected background counts in $15$\,minutes. The reason for the lower tracking mean in the second group is coming from the fact that a higher number of bins than expected contained zero counts, and a lower number of bins than expected contained one and two counts.

The lower tracking mean counts obtained in the second group could not be attributed to any systematic effect. The fact that no systematic was found in the first and third binning group reduces the chances to have a systematic effect in the second group. Moreover, the experimental tracking counts distribution defined in the second binning group is still in the statistical error limit. The background and tracking incompatibilities (tracking level about $1.7\sigma$ lower than the background level) observed in tables~\ref{ta:background2008}~and~\ref{ta:tracking2008} are necessarily due to the low statistics collected during the data taking period covered by the M10 detector.

\chapter{A limit on the axion coupling in the eV-mass scale.}
\label{chap:gLimitHe3Final}
\minitoc

\section{Introduction.}

The analysis methodology used to obtain an upper limit in the previous $^4$He Phase it is not accurate enough to be used with the mass scanning carried out during 2008. The results presented in this thesis for the first data set corresponding to $^3$He buffer gas, covered in 2008, are obtained using an unbinned likelihood method.

\vspace{0.2cm}

This section briefly describes the method that lead to an exclusion limit for $^4$He Phase axion mass coverage, and presents the results obtained for 2008 data with the new methodology.

\section{The X-ray expected signal.}

The sensitivity in the axion-photon coupling is directly related with the CAST X-ray expected signal, which is proportional to $g_{a\gamma}^4$ due to the double interaction process required, a factor $g_{a\gamma}^2$ is coming from the Primakoff conversion in the Sun, and the other $g_{a\gamma}^2$ factor is related to the axion-photon conversion in the CAST magnet. The final X-ray signal in CAST is obtained by combining four main contributions 

\begin{itemize}
\item The axion solar flux at Earth ($d\Phi_a/dE_a$)
\item The probability conversion at the CAST magnet ($P_{a\rightarrow\gamma}$)
\item The detectors efficiency ($\epsilon_d$)
\item The attenuation lenght ($\Gamma$)
\end{itemize}

The differential axion solar flux $d\Phi_a/dE_a$ and the conversion probability $P_{a\rightarrow\gamma}$  in the presence of a magnetic field inmersed in a buffer gas were described at section~\ref{sc:detectionDetails}. The solar axion flux is calculated by using the expression~\ref{eq:axionFlux} which only depends on the axion energy $E_a$. The conversion probability at the energy $E_a$ is fixed by the buffer gas density $\rho_{cb}$ by means of the effective photon mass $m_\gamma$ and the attenuation lenght $\Gamma$. The relation~\ref{eq:mgamma} gives $m_\gamma$ as a function of $\rho$ for any buffer gas, by setting the atomic number for helium $Z=2$ and the atomic weight for $^4$He, $W_{He^4} = 4$\,g/mol or $^3$He $W_{He^3} = 3$\,g/mol, $m_\gamma$ can be calculated for any of the corresponding data taking periods with buffer gas in the magnet bores. The X-ray absorption due to the probability of conversion from axion to photon at any point along the coherence lenght $L_{cb}$ is already included in $P_{a\rightarrow\gamma}$, by using expression~\ref{eq:probConversion}.

\vspace{0.2cm}

\begin{figure}[!ht]
{\centering \resizebox{0.95\textwidth}{!} {\includegraphics{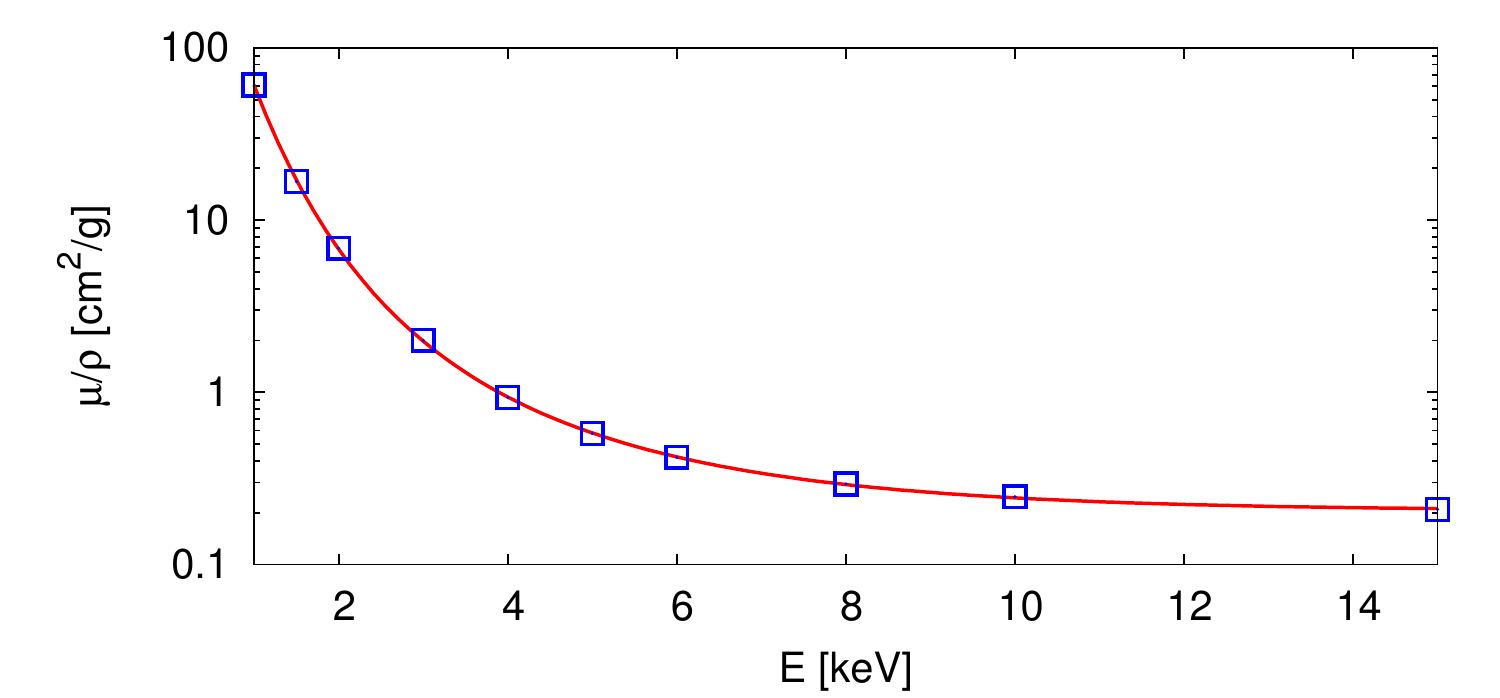}} \par}
\caption{\fontfamily{ptm}\selectfont{\normalsize{ The helium mass absorption coefficient data obtained from NIST database, and the interpolated curve obtained with expression~\ref{eq:absCoeff} }}}
\label{fi:absorptionCoefficient}
\end{figure}

The attenuation lenght $\Gamma$ is obtained by using the total photon mass attenuation coefficient $\mu/\rho$ which is provided by National Institute of Standards and Technology (NIST) database~\cite{nla.cat-vn3823794}. In order to introduce the NIST data into the final calculation, the functional shape of $\mu/\rho$ for helium in the energy range from $1$\,keV to $15$\,keV was found to be well described (see Fig.~\ref{fi:absorptionCoefficient}) by the following expression,

\vspace{-0.2cm}
\begin{equation}\label{eq:absCoeff}
\mbox{log}\left(\mu/\rho\left(E\right)\right) = -1.5832 + 5.9195 \cdot e^{-0.353808\cdot E} + 4.03598 \cdot e^{-0.970557\cdot E}
\end{equation}

\vspace{0.2cm}
\noindent being $E$ the X-ray photon energy expressed in keV, and $\mu/\rho$ in cm$^2$/g. This expression allows to calculate the total attenuation lenght as a function of the energy for any cold bore density $\rho_{cb}$, $\Gamma = \rho_{cb} \cdot \mu/\rho$ 

Finally, the detectors efficiency is introduced in the calculation of the expected X-ray signal. The total micromegas detector efficiency at the CAST running conditions was provided at section~\ref{sc:mmEfficiency}. The different contributions to the expected signal in a tracking are integrated for the corresponding exposure time $t_{tk}$ at the running conditions given by $B_{cb}$, $L_{cb}$ and $\rho_{cb}$, expressed as,

\begin{equation}
S(g_{a\gamma}, m_a, m_\gamma, E) = g_{a\gamma}^4 \int_{t_{tk}} \epsilon_d(E)\,\frac{d\Phi_a}{dE}\,P_{a\rightarrow\gamma}\,A_{cb}\,dt_{tk}~\left[\mbox{keV}^{-1}\right]
\end{equation}

\vspace{0.2cm}
\noindent where the $g_{a\gamma}$ factors in $d\Phi_a/dE_a$ and $P_{a\gamma}$ have been extracted in order to emphasyze the dependency of the expected signal on $g_{a\gamma}^4$.

\section{Coupling limit obtained in $^4$He period coverage.}

The buffer gas effects shown in chapter~\ref{chap:leak} do not apply to $^4$He Phase in the same magnitude due to the lower densities covered during this data taking period. Thus, the effects during the $^4$He period were considered negligible compared to the effects observed in the density range covered in 2008. 

\vspace{0.2cm}

The cold bore density was not increased in the middle of the tracking, as in the following $^3$He phase, but increased between trackings assuring the coverage of a setting by each detector.


\vspace{0.2cm}

In $^4$He phase the likelihood function was defined for each density step $k$, by arguing that a tracking was sensitive to a well defined axion mass resonance given by a constant buffer density during a full tracking run. The likelihood for each density setting (or pressure setting) was defined as $L_{k}$:

\vspace{0.2cm}

\begin{equation}\label{eq:He4Likelihood}
L_k = \frac{1}{L_{0k}} \prod_{i}{} e^{-\mu_{ik}}\frac{\mu_{ik}^{n_{ik}}}{n_{ik}!} 
\end{equation}

\vspace{0.2cm}

\noindent where $L_{0k} = \prod_{i} e^{-n_{ik}}\frac{n_{ik}^{n_{ik}}}{n_{ik}!}$ is the appropriate normalization factor, the index $i$ does reference to a given energy bin, and $\mu_{ik}$ is the expected number of counts in pressure step $k$ and in energy bin $i$. $\mu_{ik}$ is therefore the sum of of the expected background $b_{ik}$ estimated plus the theoretical axion signal $s_{ik}$, which depends on the theory parameters ($m_a$ and $g_{a\gamma}$),

\vspace{0.2cm}

\begin{equation}\label{eq:expected_counts}
\mu_{ik} = b_{ik} + s_{ik}(g_{a\gamma},m_a, \rho_k)
\end{equation}

\vspace{0.2cm}

\noindent where $s_{ik}$ is calculated by using the expected solar axion flux $\frac{d\Phi_a}{dE}$, given at section \ref{sc:axionFlux}, and the conversion probability $P_{a\gamma}$, given at section \ref{sc:probConv}, for the cold bore density, $\rho_k$, and it can be formulated as follows,

\vspace{0.2cm}

\begin{equation}\label{eq:expected}
s_{ik} = \int_{E_i}^{E_i + \Delta E} \frac{d\Phi_a}{dE} P_{a\rightarrow\gamma} \epsilon A_{cb} \Delta{t_k} dE = g_{a\gamma}^4 \int_{E_i}^{E_i + \Delta E} \frac{d n_{a\gamma}}{dE} \Delta t_k dE
\end{equation}

\vspace{0.2cm}

\noindent where $\Delta t_k$ is the total tracking time spent at a the density setting $k$.

\vspace{0.2cm}

The global likelihood of CAST was generated by adding the contributions of all detectors and all the neighbor pressures that might contribute to the axion mass where we are calculating $g_{a\gamma}(m_a)$,

\vspace{0.2cm}

\begin{equation}\label{eq:Lglobal}
L_{m_a}(g_{a\gamma}) = \prod_{detector} \prod_{k} L_k
\end{equation}

\vspace{0.2cm}

In practice, the upper-limit for the coupling constant in a given mass was calculated by integrating the Bayesian probability from zero up to $95\%$ of its area in $g_{a\gamma}^4$ given by the following expression

\vspace{0.2cm}

\begin{equation}\label{eq:Chi}
\frac{\int_0^{g_{a\gamma}^4} e^{-\frac{1}{2} \chi^2 } dg_{a\gamma}^4}{\int_0^{\infty} e^{-\frac{1}{2} \chi^2 } dg_{a\gamma}^4} = 0.95
\end{equation}

\vspace{0.2cm}

\noindent where $\chi^2$ is calculated at each $m_a$ by using the global likelihood defined in equation~\ref{eq:Lglobal}.

\vspace{0.2cm}

\begin{equation}\label{eq:Chi2}
-\frac{1}{2}\chi_{m_a}^2 = \mbox{log}\left( L_{m_a}(g_{a\gamma} ) \right)
\end{equation}

\vspace{0.2cm}

The micromegas background and tracking data that was used to calculate the micromegas contribution to the axion-photon coupling limit in $^4$He Phase (see Fig.~\ref{fi:limitBufferGas}) is presented at appendix~\ref{chap:appendix3}. The micromegas detector obtained an upper-limit on the axion-photon coupling constant of

\vspace{0.2cm}

\begin{equation}
g_{a\gamma} \leq 2.48\cdot10^{-10}\,\mbox{GeV}^{-1}
\end{equation}

\vspace{0.2cm}

\noindent at 95\% C.L. for $m_a \lesssim 0.4$\,eV.

\vspace{0.2cm}

The contributions of the CCD X-ray telescope system~\cite{JuliaThesis,AnnikaThesis}, the TPC detector~\cite{JaimeThesis}, and the micromegas detector taking data in $^4$He Phase were combined setting a typical upper limit on the axion-photon coupling of

\begin{equation}
g_{a\gamma} \leq 2.20 \cdot 10^{-10}\,\mbox{GeV}^{-1}
\end{equation}

\vspace{0.2cm}
\noindent at $95\%$ CL for $m_a \lesssim 0.4$\,eV~\cite{He4CAST}.

\section{Coupling limit calculation for 2008 data.}

The likelihood method applied for the first period covered with $^4$He buffer gas cannot be applied to the mass region covered in 2008. The presence of the leak during 2008 data taking period and the density changes affected by the vertical movement of the magnet (section \ref{sc:trackingDensity}), due to gas convection and window temperature changes, makes not possible to apply the binned likelihood method used in the previous data taking period. The new likelihood scheme is better suited to account for a continuously varying density in the magnet bore.

\vspace{0.2cm}

Moreover, the use of an unbinned likelihood method is motivated by the overall reduction of background achieved by CAST detectors with respect to the ones of the $^4$He Phase data, as well as to the reduced $^3$He density step time described in section XX.

\vspace{0.2cm}

A new likelihood function has been derived from the function used in $^4$He data period that takes into account the detection time of each tracking count into the analysis. The detection time of each tracking count is associated to the conditions of axion detection given by the cold bore density at that particular time. By using the new method the expected axion flux is evaluated at the time the tracking count was detected.

\vspace{0.2cm}

The mathematical equivalence in the obtention of an upper-limit by using one or the other method is described in appendix~\ref{app:likelihood}.

\subsection{Unbinned likelihood method.}

The unbinned likelihood function used to analyse the data corresponding to 2008 data taking period can be described as,

\vspace{0.2cm}

\begin{equation}\label{eq:newLikelihood}
\mbox{log} L \propto -R_T + \sum_i^N \mbox{log} R(t_i,E_i,d_i)
\end{equation}

\vspace{0.2cm}

\noindent where the sum runs over each of the N detected tracking counts and $R(t_i)$ is the event rate expected at the time $t_i$, energy $E_i$ and detector $d_i$ of the event $i$, while $R_T$ is the integrated expected number of counts over all the exposure time, energy and detectors,

\vspace{0.2cm}

\begin{equation}
R(t,E,d) = B_d + S(t,E,d)
\end{equation}

\vspace{0.2cm}

\noindent being $B_d$ the background rate of the detector $d$, and $S(t, E, d)$ is the expected rate from converted axions in the detector $d$ which depends on the axion properties $g_{a\gamma}$ and $m_a$,

\vspace{0.2cm}

\begin{equation}
S(t,E,d) = \frac{d\Phi_a}{dE} P_{a\gamma} \epsilon_d
\end{equation}

\vspace{0.2cm}

\noindent where $d\Phi_a/dE$ is the expected solar axion rate spectrum on Earth (calculated by expression~\ref{eq:axionFlux}), $P_{a\gamma}$ the axion photon conversion probability in the CAST magnet (calculated by expression~\ref{eq:probConversion}), and $\epsilon_d$ the detector efficiency, calculated as described in section~\ref{sc:mmEfficiency}.

\vspace{0.2cm}

The contribution of $B_d$ to the integrated signal $R_T$ does not need to be calculated since this parameter does not depend on $g_{a\gamma}$ and it does not affect the upper limit on $g_{a\gamma}$ to be obtained. The background of the detector has been binned in intervals of $0.5$\,keV, so the expected signal at the detection time $R(t_i,E_i,d_i)$ is necessarily integrated for the energy bin corresponding to $E_i$.

\vspace{0.2cm}

The expression~\ref{eq:newLikelihood} is split in two contributions, the first term takes into account the total exposure and the number of X-ray photons the detectors should have detected for a given coupling, $g_{a\gamma}$, the second term takes into account the detected counts during tracking at the given conditions of detection. This leads to derive a quick estimator of the sensitivity of the experiment in the case of zero counts detected that could be reached by extremely low background detectors,

\vspace{0.2cm}

\begin{equation}
g_{a\gamma} \leq \left( \mbox{log}(20)/N_\gamma \right)^{\frac{1}{4}} \quad [10^{-10} \mbox{GeV}^{-1}]
\end{equation}

\vspace{0.2cm}

\noindent for a $95\%$ CL, where $N_\gamma$ is the total expected axion counts at $g_{a\gamma} = 10^{-10}$\,GeV$^{-1}$.

\subsection{Axion flux estimation for the $^3$He coverage in 2008.}

In chapter~\ref{chap:leak} was shown how the pressure measured in the cold bore can describe perfectly the window temperature changes using an idealized model of the temperature profile. The measured pressure can be directly related with the inner cold bore density except for the necessary corrections to be introduced by CFD simulations, as hydrostatic pressure and real gas corrections due to the magnet temperature global change during tracking.

\vspace{0.2cm}

Up to now, the best approach to the axion mass covered by the CAST magnet in each tracking is the value of the measured cold bore pressure $P_{check}$ that is related with the density in the magnetic region by expression~\ref{eq:stateVDW}. The buffer density is proportional to $m_\gamma$ by relation~\ref{eq:mgamma} which is later introduced in the conversion probability $P_{a\rightarrow\gamma}$ to obtain the contribution of each density to a given axion mass.

\vspace{0.2cm}

\begin{figure}[!ht]
{\centering \resizebox{1.0\textwidth}{!} {\includegraphics{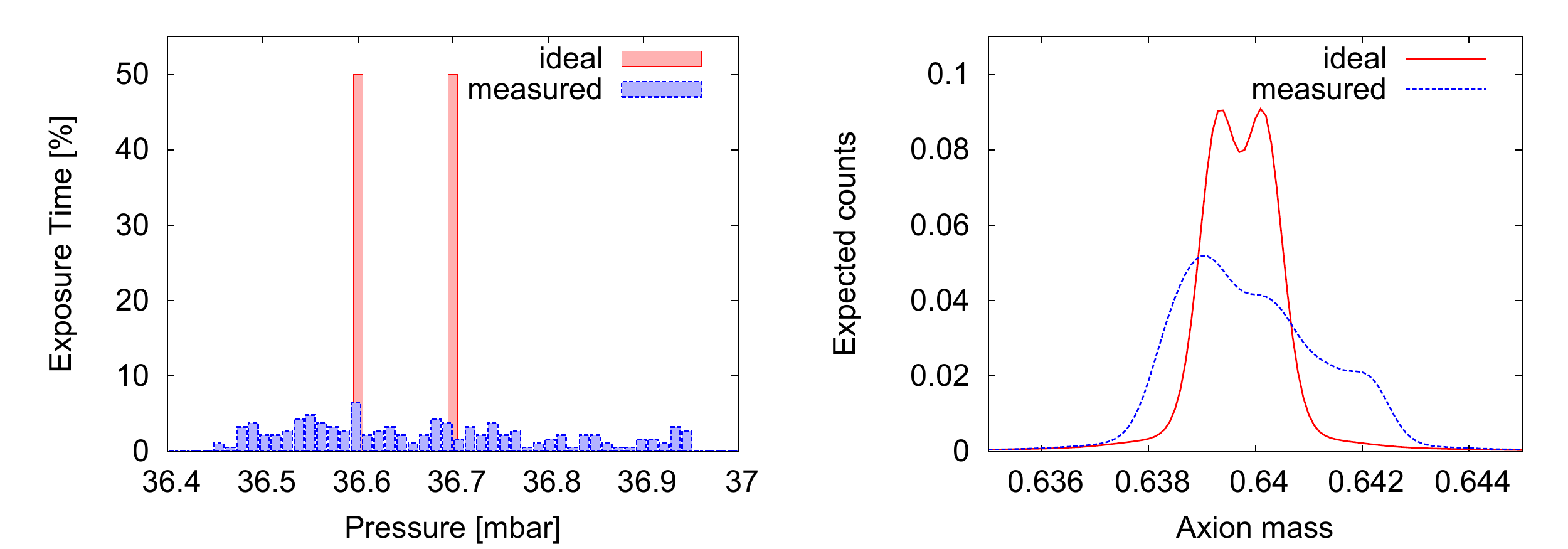}} \par}
\caption{\fontfamily{ptm}\selectfont{\normalsize{ In these figures is represented the last tracking from 2008. On the left, the ideal scanning step, 2 settings per tracking, together with the measured pressure scanning in percentage. On the right, the equivalent axion mass resonance at $g_{a\gamma} = 10^{-10}$GeV$^{-1}$ for the ideal case and for the generated with the measured pressure.   }}}
\label{fi:trackingScan}
\end{figure}

In order to obtain the effective exposure in each axion mass the time spent in each tracking day has been derived as a function of $P_{check}$ and rewritten as a function of the cold bore density $dt/d\rho_{tck}$. The expected signal $S_T$ is calculated as the contribution of the cold bore densities scanned in tracking conditions at a given axion mass, and it can be summarized by the following expression,

\begin{equation}\label{eq:expectedCorrected}
S_T (m_a, g_{a\gamma}) = \int_{\rho_{tck}} \frac{dt}{d\rho} \left[ \int_{E_i}^{E_f} \frac{d\Phi_a}{dE} P_{a\rightarrow\gamma} \epsilon A_{cb} dE \right] d\rho_{tck}
\end{equation}

\vspace{0.2cm}

\noindent where $E_i = 2$\,keV and $E_f = 7$\,keV is the energy range defined for the maximum expected axion signal.

\vspace{0.2cm}

Figure~\ref{fi:trackingScan} shows the pressure scanning for the last tracking of 2008 (same tracking data as the corresponding figure \ref{fi:pcheckTracking}). Ideally two densities should have been measured during that tracking, but instead a range of densities are scanned. The effect on the expected mass resonance is shown in the same figure, leading to a less intense but broader mass range coverage.

\vspace{0.2cm}

\begin{figure}[!ht]
{\centering \resizebox{1.0\textwidth}{!} {\includegraphics{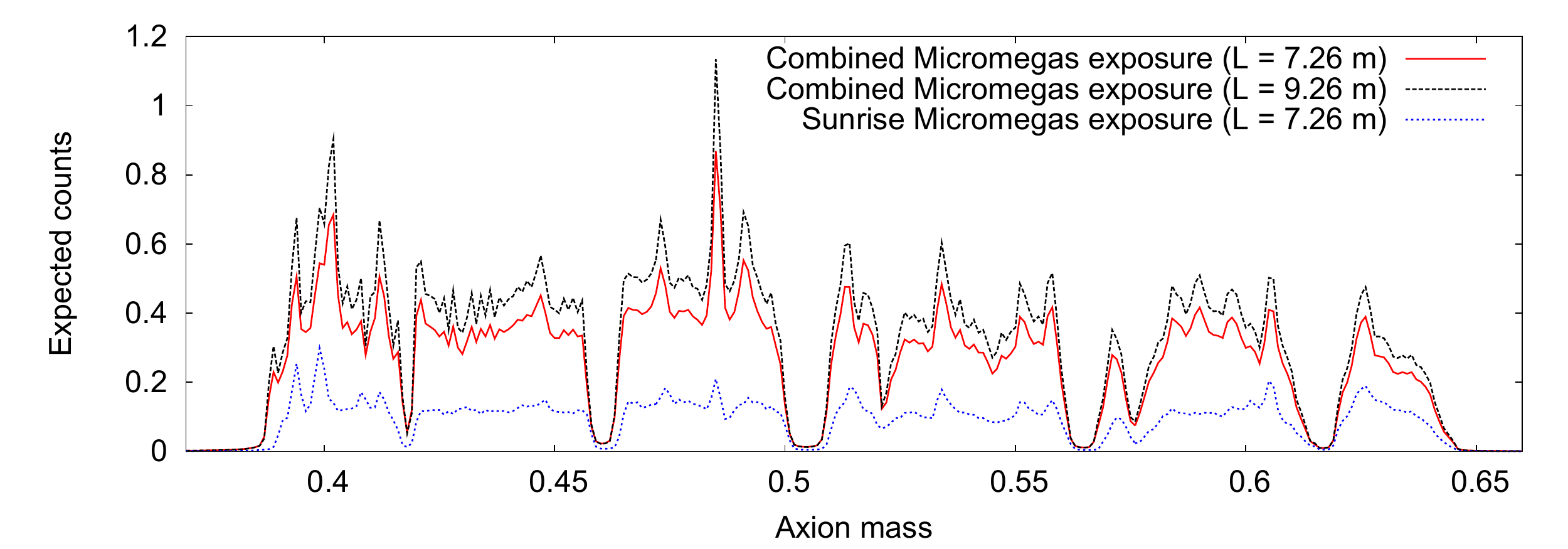}} \par}
\caption{\fontfamily{ptm}\selectfont{\normalsize{ Expected axion counts ($g_{a\gamma} = 10^{-10}$GeV$^{-1}$) for each axion mass after the integration of the full 2008 tracking exposure with micromegas detectors (Sunrise micromegas detector and Sunrise micromegas detector exposure combined with the \emph{two} Sunset micromegas detectors taking data in 2008). The expected axion flux was calculated for different magnetic lengths.  }}}
\label{fi:fullTrackingScan}
\end{figure}

In order to obtain the expected axion flux at a given axion mass $m_a$ the full tracking exposure in 2008 has been integrated. The result is encoded as $S_T(m_a)$ and it will be introduced in expression~\ref{eq:newLikelihood} in order to obtain a limit for each axion mass (see Figure~\ref{fi:fullTrackingScan}).

\subsection{A limit for 2008 mass range scanning.}

The background and tracking data presented in section~\ref{sc:background2008} has been used to define the contribution of the detected counts into the upper-limit coupling calculation. The expected axion rate in the detector $S(t, E, d)$ at each $m_a$ is calculated taking into account the value of the pressure $P_{check}$ at the time of detection and deriving the cold bore density and the effective photon mass $m_\gamma$ in the same way as it was estimated for the calculation of $S_T$. The mean background level of the detector was shown in section~\ref{sc:backgroundDefinition} for each tracking step in 2008. The energy of the detected tracking count is associated with the energy bin of the mean background level obtained (appendix~\ref{chap:appendix4} shows the integrated mean background level defined in the range from 2 to 7\,keV).

\vspace{0.2cm}

\begin{figure}[!ht]
{\centering \resizebox{1.0\textwidth}{!} {\includegraphics{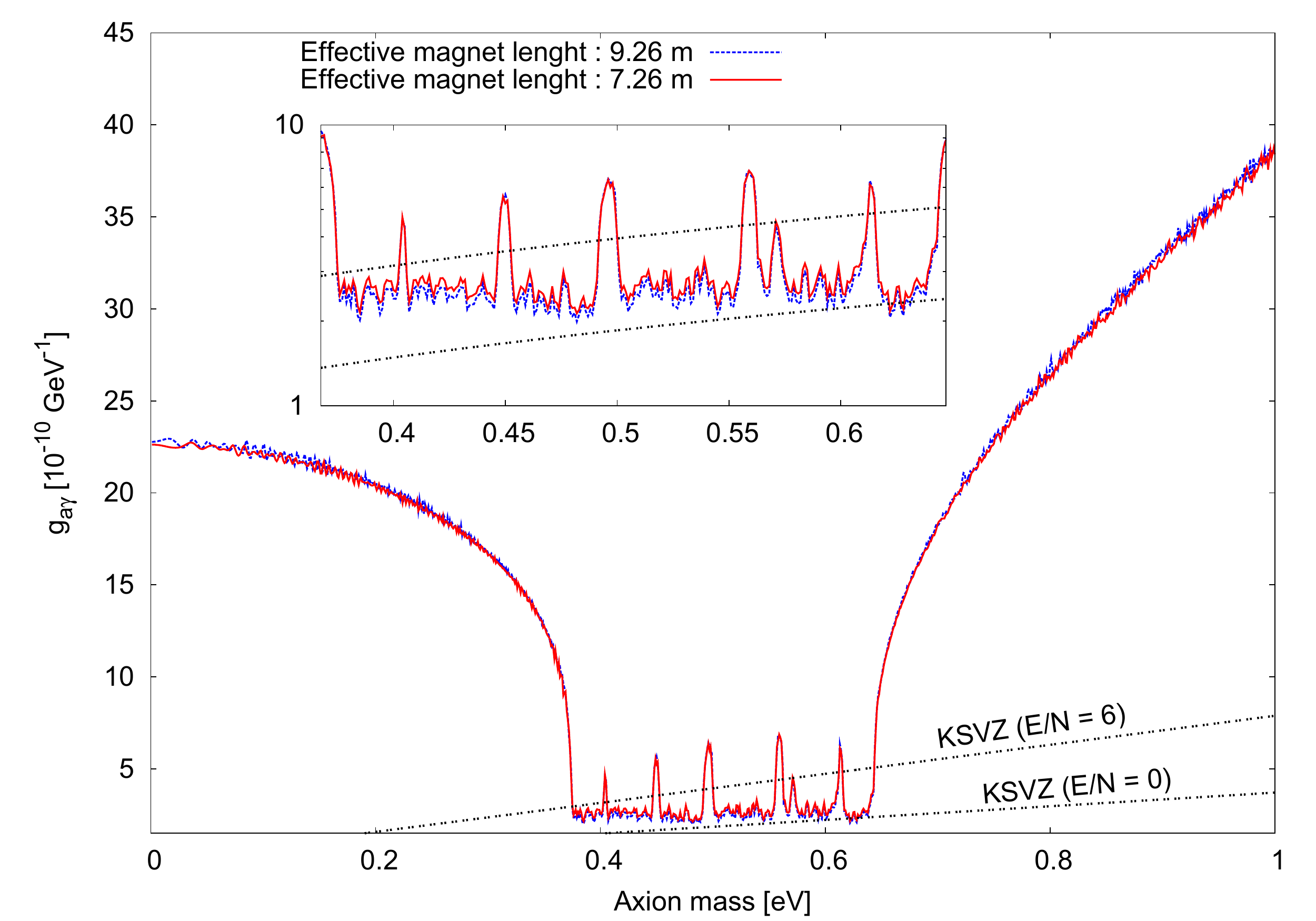}} \par}
\caption{\fontfamily{ptm}\selectfont{\normalsize{ Axion-photon coupling upper-limit for Sunrise micromegas detector in the axion mass covered in 2008. The limit for an effective length of $9.26$\,m and $7.26$\,m is shown.   }}}
\label{fi:limitSunrise}
\end{figure}

The exclusion plot has been calculated for \emph{two} magnetic field effective lengths ($7.26$\,m and $9.26$\,m) due to the uncertainty on the limits of the density homogeneity at the magnetic region. CFD simulations have already shown that in the worst scenario the effective magnet length is reduced by about $2$\,meters. The exclusion limit is calculated at the best ($9.26$\,m) and the worst ($7.26\,m$) scenario in order to delimit the sensitivity reached by CAST by the axion mass scanning carried out in 2008. Figure~\ref{fi:limitSunrise} shows the exclusion limit obtained by the Sunrise micromegas line for these \emph{two} extreme scenarios where the unexplored regions due to the leak problem are clearly observed.

\vspace{0.2cm}

\begin{figure}[!ht]
{\centering \resizebox{1.0\textwidth}{!} {\includegraphics{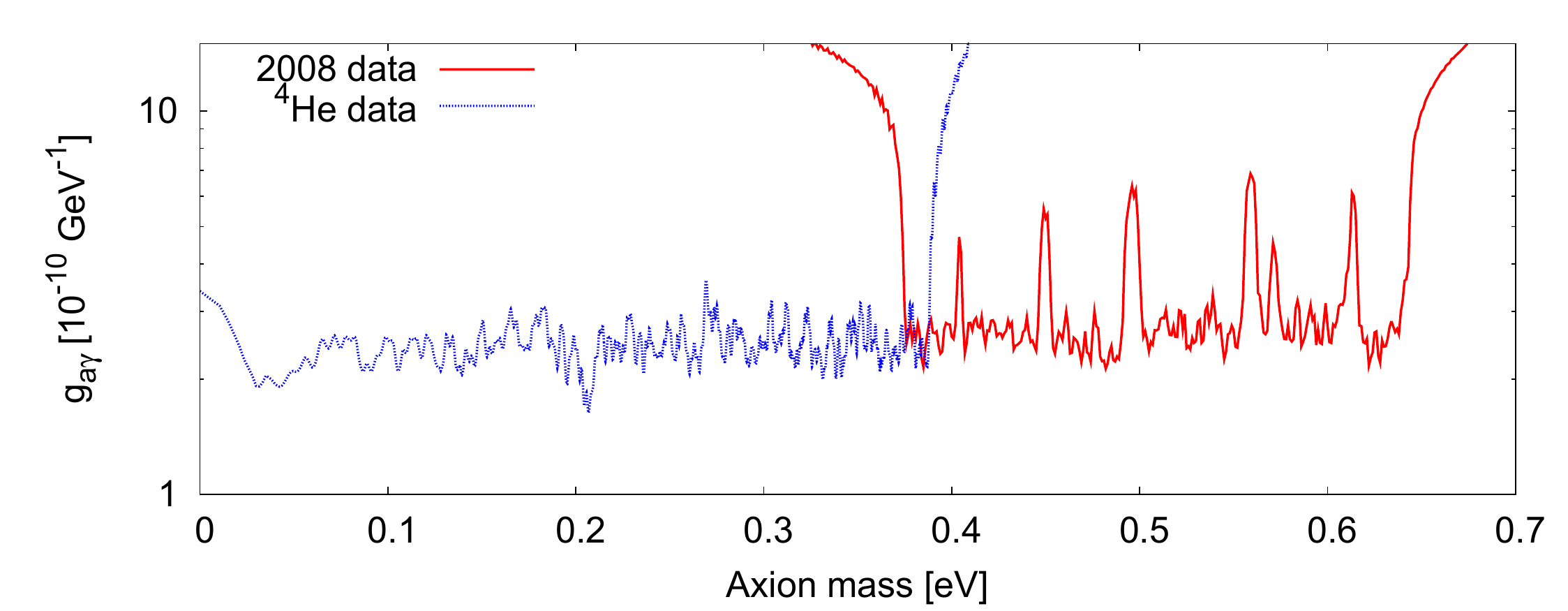}} \par}
\caption{\fontfamily{ptm}\selectfont{\normalsize{ Axion-photon coupling limit for the second phase of CAST, corresponding to $^4$He Phase data taking period in 2005 and 2006, and $^3$He Phase in 2008, reached with the micromegas Sunrise Line.    }}}
\label{fi:limitBufferGas}
\end{figure}

The limit obtained by the Sunrise line during the full data taking period with buffer gas in the magnet bore, including $^4$He coverage, is presented in figure~\ref{fi:limitBufferGas}.

\vspace{0.2cm}

The Sunrise micromegas limit has been combined with the data corresponding to the other \emph{two} micromegas detectors covering the Sunset side of the experiment. The total tracking time exposure in axion detection conditions is $207.80$\,hours and $176.84$\,hours for each Sunset detector. The data is combined by introducing the total expected counts given by each detector as a function of $m_a$ (see Fig.~\ref{fi:fullTrackingScan}) and the total tracking counts detected at the given detection conditions in expression~\ref{eq:newLikelihood}. The results obtained for the combined and sunrise alone upper-limits are shown in table~\ref{ta:limitsTable}, where \emph{two} mass ranges have been chosen for the calculation of the coupling upper-limit value reached by CAST in 2008. The value obtained in the region from $0.39$\,eV to $0.64$\,eV represents the coverage in the full period, including the gaps produced by the effect of the leak, and the second scanning range, $0.52$\,eV to $0.55$\,eV has been chosen as the representative region for the coupling limit reached in the absence of non-scanned mass regions.

\begin{table}[htp!]
\begin{center}
\begin{tabular}{cccc}
& \bf{Effective Length} & \bf{0.52-0.55 eV} & \bf{0.39-0.64 eV} \\
& \bf{[m]} & \bf{[GeV$^{-1}$]} & \bf{[GeV$^{-1}$]} \\
& & & \\
\hline
& & & \\
\multirow{2}{*}{\bf{Sunrise}}	&  $7.26$ & $2.71\cdot10^{-10}$ & $3.01\cdot10^{-10}$ \\
&	$9.26$ &  $2.55\cdot10^{-10}$ & $2.87\cdot10^{-10}$ \\
& & & \\
\hline
& & & \\
\multirow{2}{*}{\bf{Combined}}	&  $7.26$ & $2.21\cdot10^{-10}$ & $2.44\cdot10^{-10}$ \\
&	$9.26$ &  $2.07\cdot10^{-10}$ & $2.32\cdot10^{-10}$ \\
& & & \\
\end{tabular}
\caption{ Summary of axion-photon coupling limits at different configurations.}
\label{ta:limitsTable}
\end{center}
\end{table}

\begin{figure}[!htb]
{\centering \resizebox{1.0\textwidth}{!} {\includegraphics{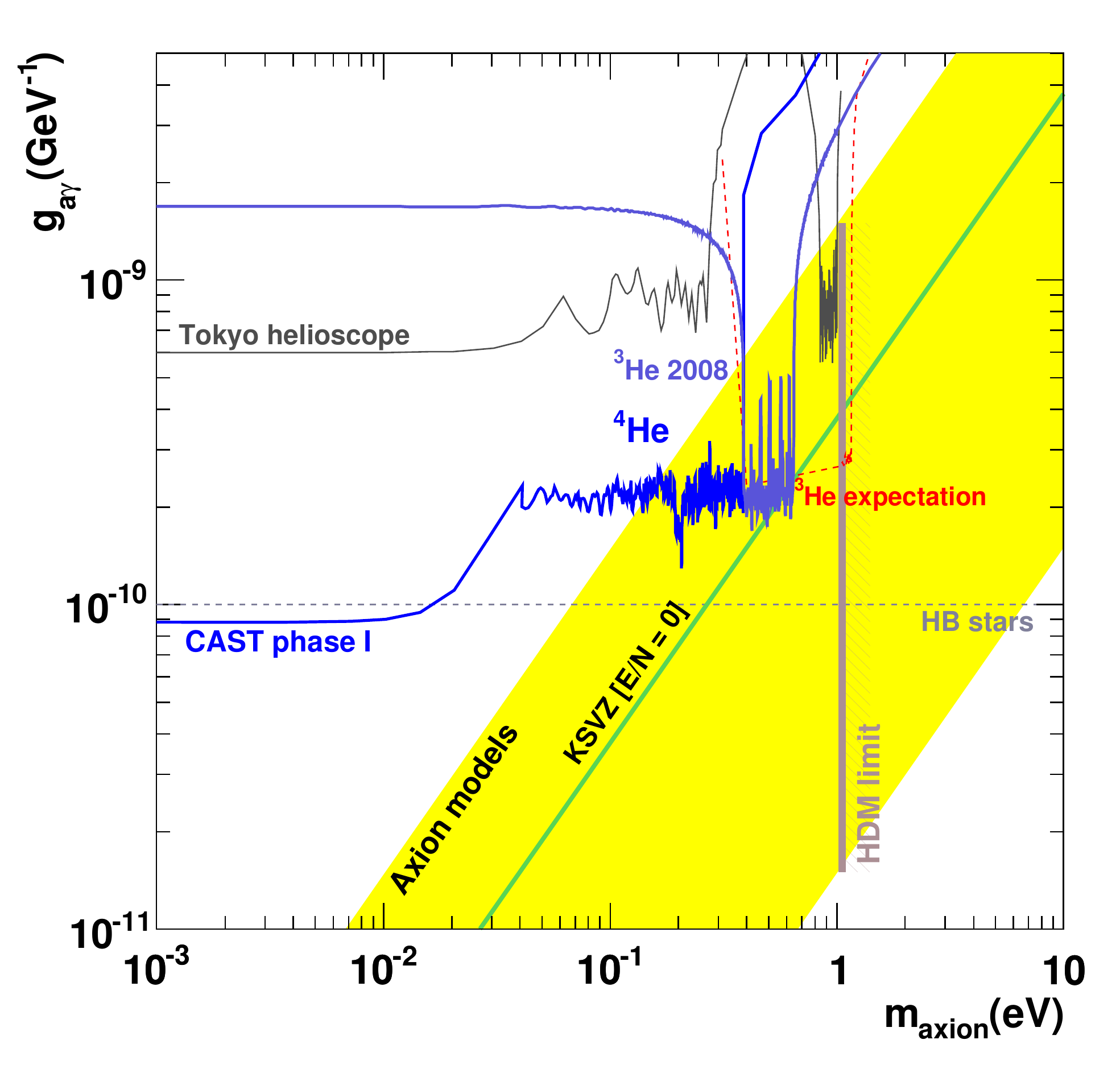}} \par}
\caption{\fontfamily{ptm}\selectfont{\normalsize{ CAST exclusion plot for the axion-photon coupling. In blue it is presented the limit obtained in the previous CAST data taking periods, vacuum Phase I and $^4$He phase. In violet is presented the first axion mass scanned region covered in the new $^3$He phase, obtained by the combination of the three micromegas detection lines taking data in 2008. In red, the expected axion mass coverage during the full $^3$He phase data taking period. The limit obtained in the different phases of Tokyo helioscope are also shown. The theoretical favoured axion models region within the range $0.07 < \left| E/N -1.92 \right| < 7$ is represented by a yellow band, together with other theoretical constrains coming from astrophysical and cosmological arguments, as the HDM limit ($m_a < 1.05$\,eV) and the HB stars limit. }}}
\label{fi:exclusionFinal}
\end{figure}

Figure~\ref{fi:exclusionFinal} shows the most conservative limit, given by using the shorter coherence lenght $L = 7.26$\,m in the calculation, combining the three micromegas lines as the final result obtained in this thesis, together with the combined exclusion limit obtained by the previous CAST axion mass coverage.

\vspace{0.2cm}

We can conclude that the CAST experiment has fixed the best experimental upper-limit to the axion-photon coupling by using the most conservative limit in a new axion mass range

\begin{equation}
g_{a\gamma} \leq 2.44\cdot 10^{-10} GeV^{-1}
\end{equation}

\vspace{0.2cm}

\noindent at $95$\% C.L. from $0.39$\,eV to $0.64$\,eV, to be improved by the $4^{th}$ CAST detection line and the coverage of the unexplored regions produced by the $^3$He leak problem.

\vspace{0.2cm}

The CAST experiment entered by first time in the $g_{a\gamma}-m_a$ parameter space favoured by the theoretical axion models during the $^4$He data taking phase. The sensitivity reached by CAST during the $^3$He data taking phase for the new axion mass coverage has lead to the first experimental limit, obtained by direct axion search, that excludes a region in the eV-mass scale of the KSVZ model with $E/N = 0$. The future inclusion of the revisited density settings which were not covered due to the leak will allow to set the limit on the axion-photon coupling of at least $g_{a\gamma} \lesssim 2.2\cdot10^{-10}$\,GeV$^{-1}$ at 98\,\% C.L. in the axion mass range presented. This sensistivity will allow CAST to exclude the existence of standard KSVZ axions ($E/N = 0$) for masses $m_a \gtrsim 0.6$\,eV. In the near future, the continous axion mass scanning which is actually going on, and it will stablish a new limit of a comparable coupling sensitivity for higher axion masses. If no signal is found CAST experiment will exclude standard KSVZ axions for masses up to about $m_a \lesssim 1$\,eV.

\chapter{Summary and conclusions}
\label{chap:summary}

The axion is a hypothetical particle that emerged as a natural solution to the CP problem in the strong interactions, which seem to not violate CP. The problem arises from the fact that the description of strong interactions given by quantum chromodynamics (QCD) theory does not forbid such violation. Furthermore, since electroweak interactions violate CP it is difficult to understand why strong interactions do not. Several solutions were considered, however the most elegant and natural solution has as an outcome a new hypothetical particle, the axion. The axion was postulated by \emph{H. Quinn} and \emph{R. Peccei} in 1977, based on the previous work of \emph{'t Hoof} and \emph{Weinberg} where they emphasize the need for a CP violating term in the theory.

\vspace{0.1cm}

Different axion models arose, fixing the relation between the axion mass and the interaction strength with ordinary matter. Depending on the model the axion would couple to matter in different ways, defining the interaction rules with each of the existing particles. The coupling with ordinary matter would have some observational implications in astrophysics and cosmology which allows to constrain the axion properties in terms of axion mass and coupling strength. In particular, the existence of axions would play a role in stellar evolution, and the fact that the axion has mass places it in cosmological models as a possible candidate for dark matter, which depending on its mass could account for a dominant contribution to the total dark matter in the Universe.
\vspace{0.1cm}

The search for axions is well motivated as a solution to the strong CP problem and as a possible candidate to dark matter. In addition, axions could also be related with other physical processes of non well understood nature, as the solar corona heating problem or the modulation of the Earth magnetic field.

\vspace{0.1cm}
A common feature of any axion model is the fact that the axion always couples to \emph{two} photons. \emph{P.~Sikivie} introduced in 1983 the basis for a direct experimental search of axions by using the analogy of the coupling to photons with the pion decay to \emph{two} photons via the Primakoff effect. The experimental detection principle is based on applying an intense magnetic field which could provide one of the photons of the interaction allowing to basically convert axions to photons, or vice-versa. Since then, several proposals for experimental axion searches have been carried out, continue and are in development. After 30 years the motivation for the discovery of the axion has not deceased but intensified.

\begin{figure}[!ht]
{\centering \resizebox{0.9\textwidth}{!} {\includegraphics{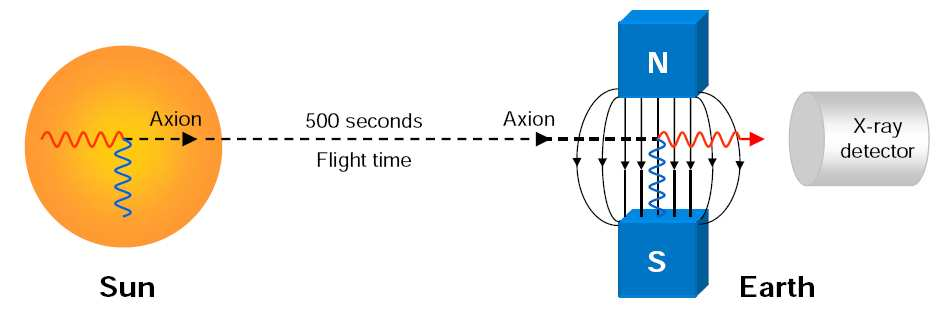}} \par}
\caption{\fontfamily{ptm}\selectfont{\normalsize{ Detection concept for the magnet helioscope idea. Where the Sun is used an intense source of axions. }}}
\label{fi:summaryHelioscopeConcept}
\end{figure}

\vspace{0.1cm}
The axion search experiments can be split into \emph{three} main groups depending on the source of axions. The first possibility are \emph{haloscope searches} which are based on the relic axion cold dark matter (CDM) density which would be gravitationally linked to our galaxy. The second are \emph{laboratory searches} which use the possibility of generating axions by using an intense photon beam traveling through an intense magnetic field, the existence of the axion would then have some effects on the beam properties, or axions could be re-converted back to photons in an second magnetic region. The third are \emph{helioscope searches} which are based in the favorable conditions at the core of the Sun for producing a non negligible amount of axions. The axions generated at the Sun core travel through the space without any obstacle except for the special conditions given by the transversal magnetic field produced at the helioscope where axions can be converted to photons (see Fig.~\ref{fi:summaryHelioscopeConcept}).

\vspace{0.1cm}

\vspace{0.2cm}

\begin{figure}[!ht]
\begin{center}
\begin{tabular}{cc}
{\centering \resizebox{0.44\textwidth}{!} {\includegraphics{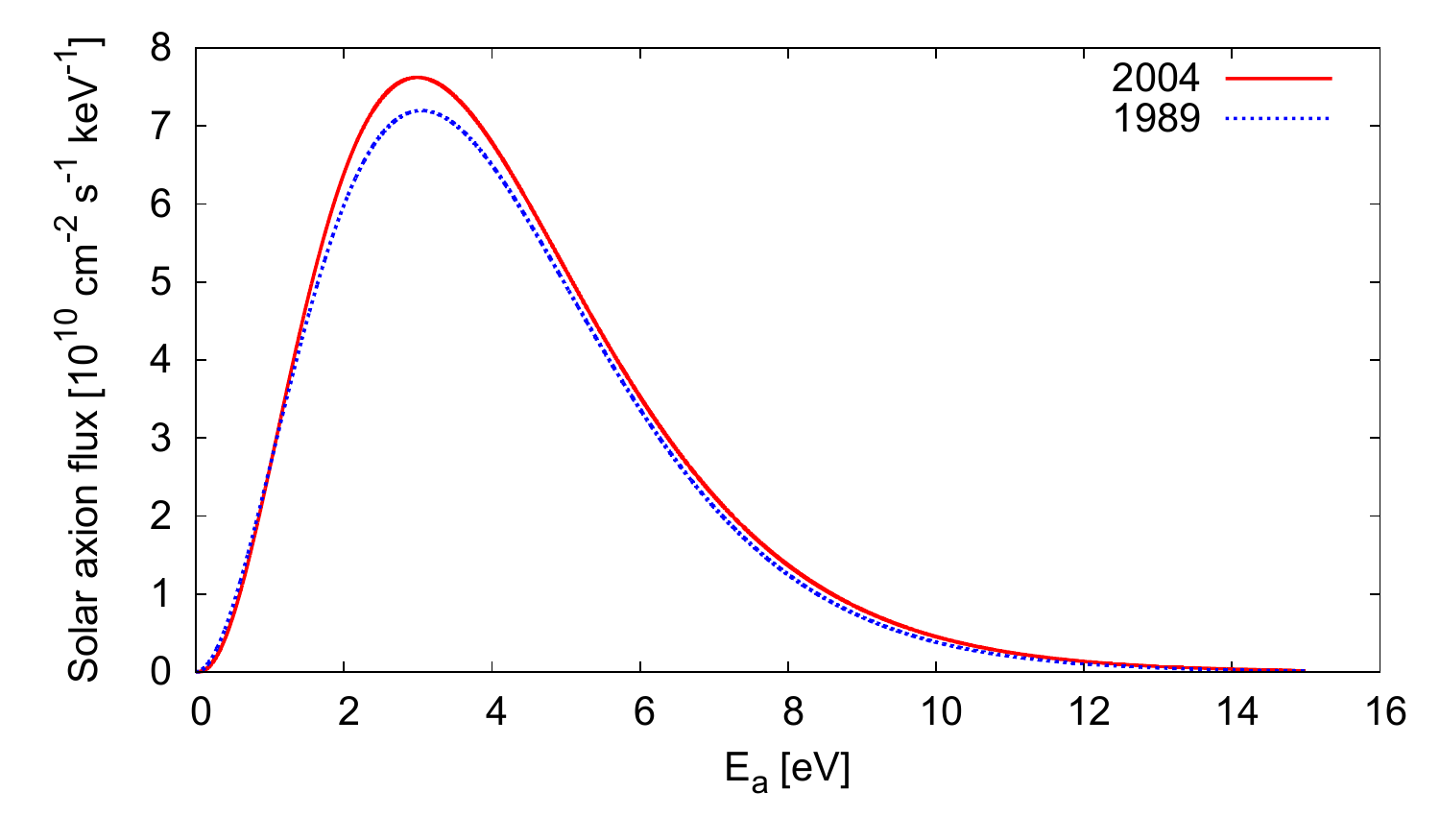}} \par} &
{\centering \resizebox{0.51\textwidth}{!} {\includegraphics{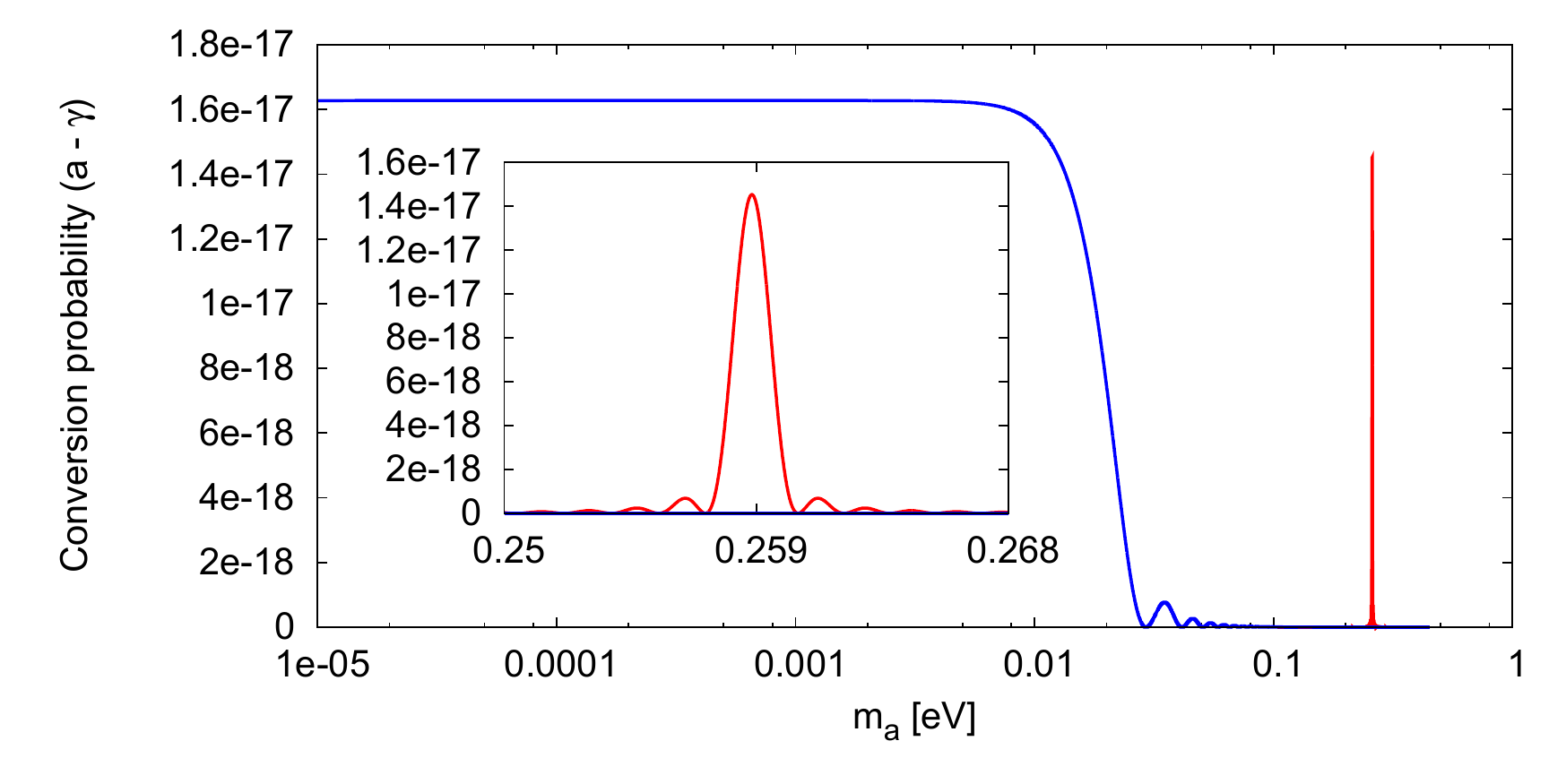}} \par} \\
\end{tabular}
\end{center}
\caption{\fontfamily{ptm}\selectfont{\normalsize{ Expected axion solar flux on Earth (left) and probability conversion of axion to photon in the presence of a transversal magnetic field in CAST conditions (CAST vacuum phase and a resonance given by a fixed density in the magnet bore).  }}}
\label{fi:summaryFluxProb}
\end{figure}

The helioscope search, in which the work presented in this thesis is focused, is based on the well established standard solar model which allows to calculate the expected axion flux generated in the core of the Sun as a function of the axion-photon coupling. The axion solar flux at Earth was calculated by \emph{G. Raffelt} and the probability conversion of axion to photons in the presence of an external magnetic field was given by \emph{K. Bibber} (see Fig.~\ref{fi:summaryFluxProb}). Thus allowing to determine the expected number of converted axions into photons, when the magnet is aligned with the Sun, as a function of the axion-photon coupling, which in absence of signal provides an upper-limit to the axion-photon coupling.

\vspace{0.1cm}

\vspace{0.1cm}

This thesis presents the latest results from CAST (CERN Axion Solar Telescope), an experiment hosted at the European Organization for Nuclear Research (CERN) which is searching for solar axions. CAST consists of a superconducting prototype dipole magnet from the Large Hadron Collider (LHC) which is able to provide an intense magnetic field of almost 9\,T, over a length of $9.26$\,m. The magnet is mounted over a rotating platform which allows to continuously follow the Sun during the sunrise and the sunset, each for about $1.5\,$hours. The expected axion signal mean energy is within the X-ray range thus CAST implements X-ray detectors to cover both magnet sides, corresponding to the sunrise side and the sunset side.

\vspace{0.1cm}

\begin{figure}[!ht]
{\centering \resizebox{0.8\textwidth}{!} {\includegraphics{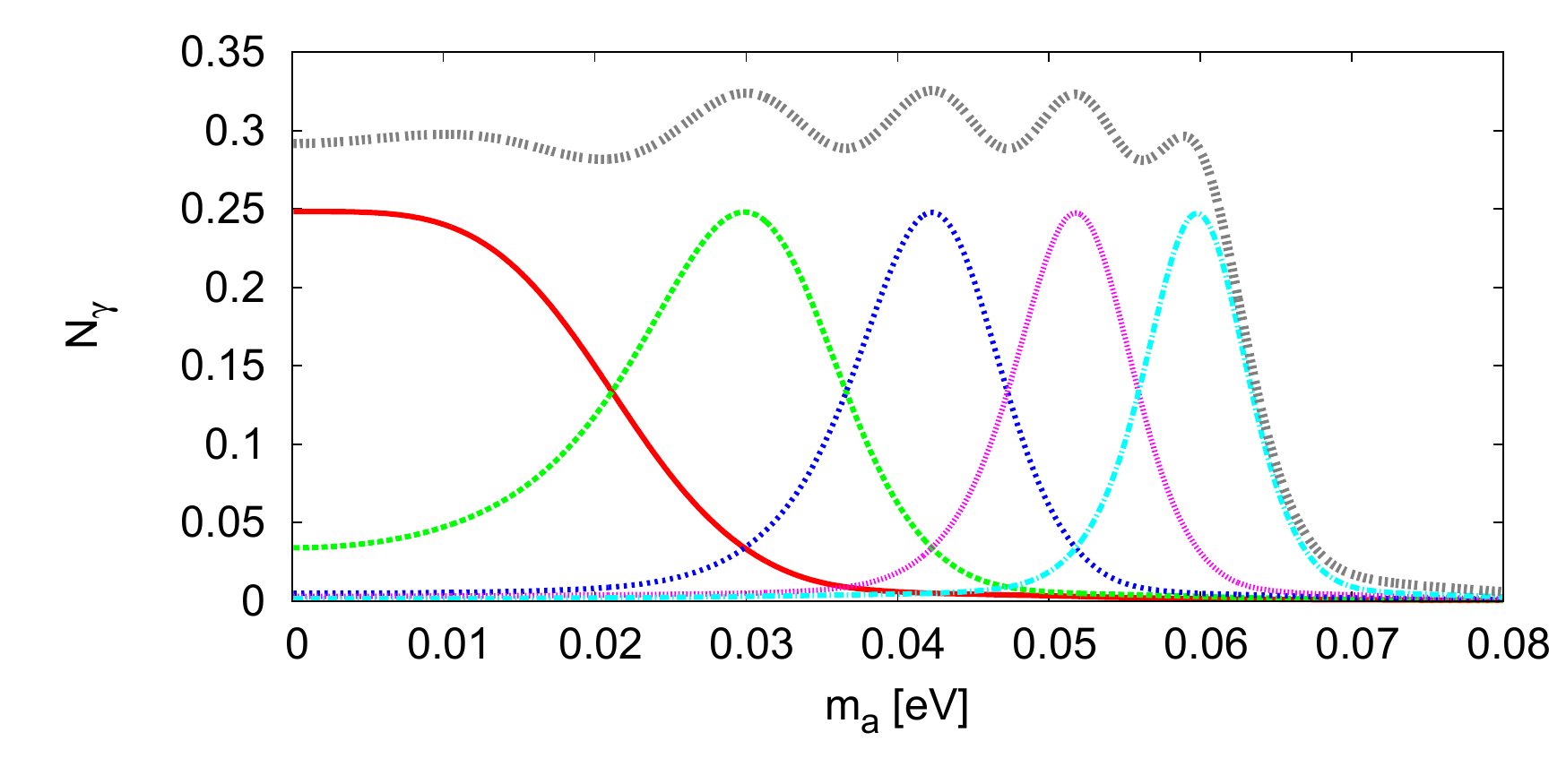}} \par}
\caption{\fontfamily{ptm}\selectfont{\normalsize{ Density steps scanning, in terms of the expected photons for a tracking with one detector at $g_{a\gamma} = 1\cdot10^{-10}$\,GeV$^{-1}$, leading to coverage in a wide axion mass range, first scanning steps are presented in this plot. }}}
\label{fi:summaryNgamma}
\end{figure}

CAST has published the most restrictive axion-photon coupling limit in a wide axion mass range. The CAST experiment started to take data already by 2003. During its first phase (CAST Phase I) the magnet was operating with vacuum inside the magnet bores, conditions in which the sensitivity of the experiment in the axion-photon coupling $g_{a\gamma}$ was enhanced for axion masses up to $m_a \lesssim 0.02$\,eV. The experiment completed its first phase in 2004 improving the current limits on the coupling constant, $g_{a\gamma} < 8.8\cdot10^{-11}\,\mbox{GeV}^{-1} \mbox{at 95\%~CL for}\,m_a\leq 0.02\,\mbox{eV}$. In the second phase (CAST Phase II), the experiment extends its sensitivity to higher axion masses by filling the magnet with a refractive buffer gas that enhances the conversion of axions to photons for a narrow axion mass range which depends on the buffer gas density. The continuous scanning in axion mass is achieved by smoothly increasing the amount of gas inside the magnet bore allowing to be sensitive to axions in a wider axion mass range (see Fig.~\ref{fi:summaryNgamma}).

\begin{figure}[!hb]
{\centering \resizebox{0.8\textwidth}{!} {\includegraphics{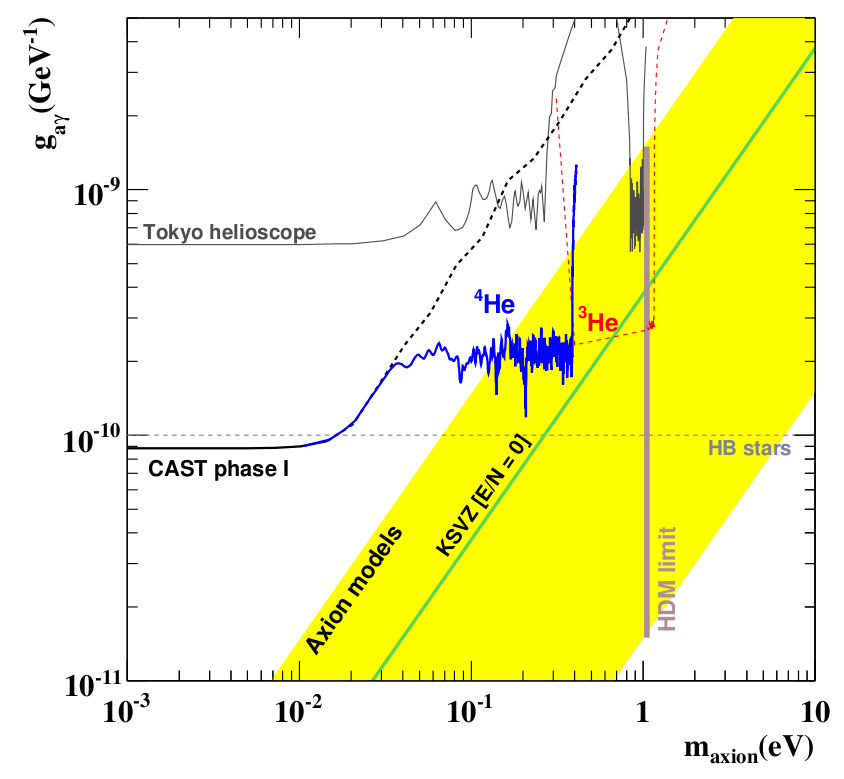}} \par}
\caption{\fontfamily{ptm}\selectfont{\normalsize{Axion-photon coupling limit as a function of the axion mass provided by CAST experiment in the vacuum phase (Phase I) and $^4$He axion mass range coverage. Expected coverage for the $^3$He operating period is also shown. }}}
\label{fi:summaryExclusionHe4}
\end{figure}

The second phase of CAST used $^4$He as buffer gas during the first data taking period which allowed to cover axion masses up to 0.4\,eV. This data taking period finished in 2006 allowing to set a new limit on the axion-photon coupling for the new explored axion mass range of $g_{a\gamma} < 2.17\cdot10^{-10}\,\mbox{GeV}^{-1} \mbox{ at 95\%~CL for }0.02\,\mbox{eV} < m_a < 0.4\,\mbox{eV}$ being the first experiment entering in the axion models favored region for eV-mass scale axions. These results are summarized by representing the excluded $g_{a\gamma}-m_a$ parameter space (see Fig.~\ref{fi:summaryExclusionHe4}) together with other helioscope searches, astrophysical constrains, and favored axion models region.

\vspace{0.1cm}

In order to sweep higher axion masses CAST requires to be operated with $^3$He gas that allows to reach higher gas densities, the saturating pressure of $^3$He above 130\,mbar allows to scan masses up to about 1\,eV. The gas system had to be adapted to work with $^3$He and it required a number of technically challenging upgrades to equip the experiment for the new run.

\vspace{0.1cm}

\begin{figure}[!ht]
\begin{center}
\begin{tabular}{cc}
{\centering \resizebox{0.35\textwidth}{!} {\includegraphics{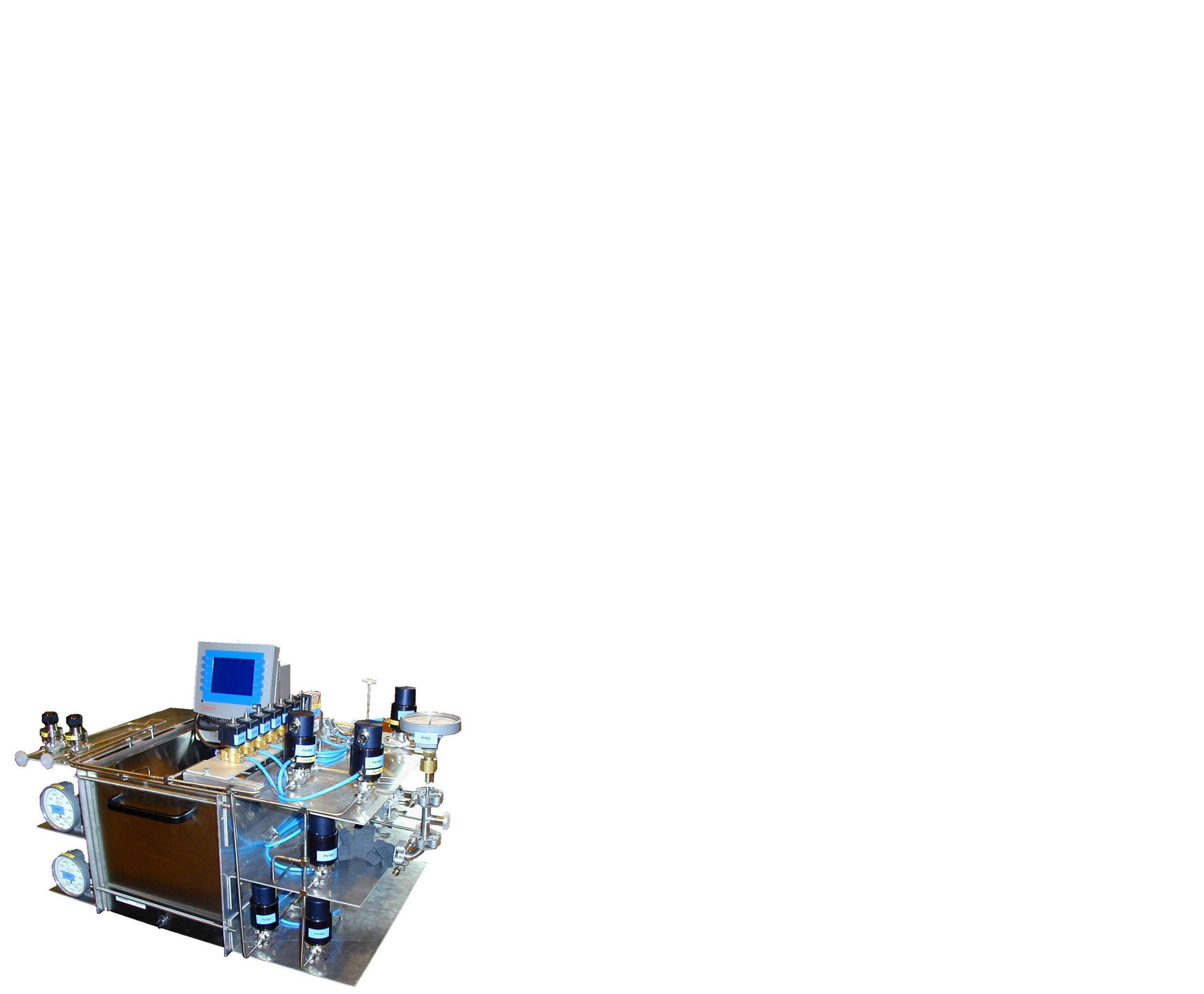}} \par} &
{\centering \resizebox{0.57\textwidth}{!} {\includegraphics{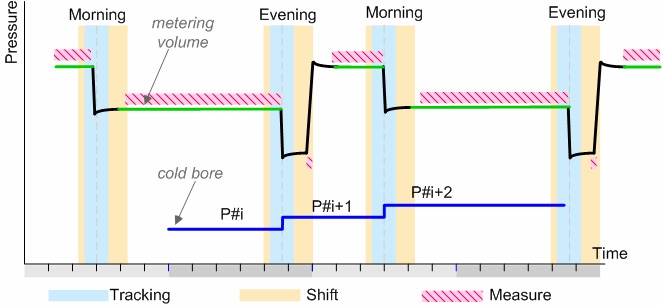}} \par}
\end{tabular}
\end{center}
\caption{\fontfamily{ptm}\selectfont{\normalsize{ A picture of the new metering system developed for the new data taking phase (left), and a plot representing the cold bore filling scheme (right) during the data taking period. }}}
\label{fi:summaryHe3Filling}
\end{figure}

The new $^3$He equipment required to implement a recovery system to avoid the loss of the valuable $^3$He gas in case of a magnet quench, a superconducting magnet protection system which raises the temperature of the magnet thus increasing the pressure of the gas inside the bores. Then, the gas needs to be quickly evacuated, via an expansion volume, in order to protect the thin high X-ray transmission windows containing the gas. Moreover, the required upgrade was used to increase the functionality and reliability of the system. A new metering volumes design was constructed for fastening the magnet bores gas filling process from empty bores state, which requires longer periods of time as the pressure in the bores is higher. Furthermore, the $^3$He pipework connections are controlled by automated electro-pneumatic valves allowing to program the functionality of the system through a Power Line Communications (PLC) system, which in addition allows to monitor an extense distribution of pressure and flow sensors all along the system. The new functionalities of the system include automatic recovery and gas purging system, between others. Furthermore, the new more versatile system allows to define different magnet bore filling schemas, during the data taking periods, which included a slow density ramping during tracking thanks to the accurate mass flow controllers implemented at the metering system connection to the bores. The final bores filling schema carried out during the $^3$He data taking period was the one covering \emph{two} density steps per tracking by increasing the density in the bores to the next step in the middle of the tracking in a short period of time, of about $3$\,minutes (see Fig.~\ref{fi:summaryHe3Filling}).

\vspace{0.1cm}

CAST exploits the availability of the \emph{four} magnet bore apertures by implementing different type of X-ray detector systems reinforcing the detection of a hypothetical signal which should be detected independently at each detection system during sunrise and sunset trackings. The vacuum phase and $^4$He data taking periods were covered by a Time Projection Chamber (TPC) covering both magnet bores in the sunset side, and each bore in the sunrise side being covered by a X-ray telescope focusing in a Charge-Coupled Device (CCD) adapted to operate in the X-ray energy range, and a Micromegas detector.

\vspace{0.1cm}

\begin{figure}[h!]
\begin{tabular}{cc}
\includegraphics[width=0.52\textwidth]{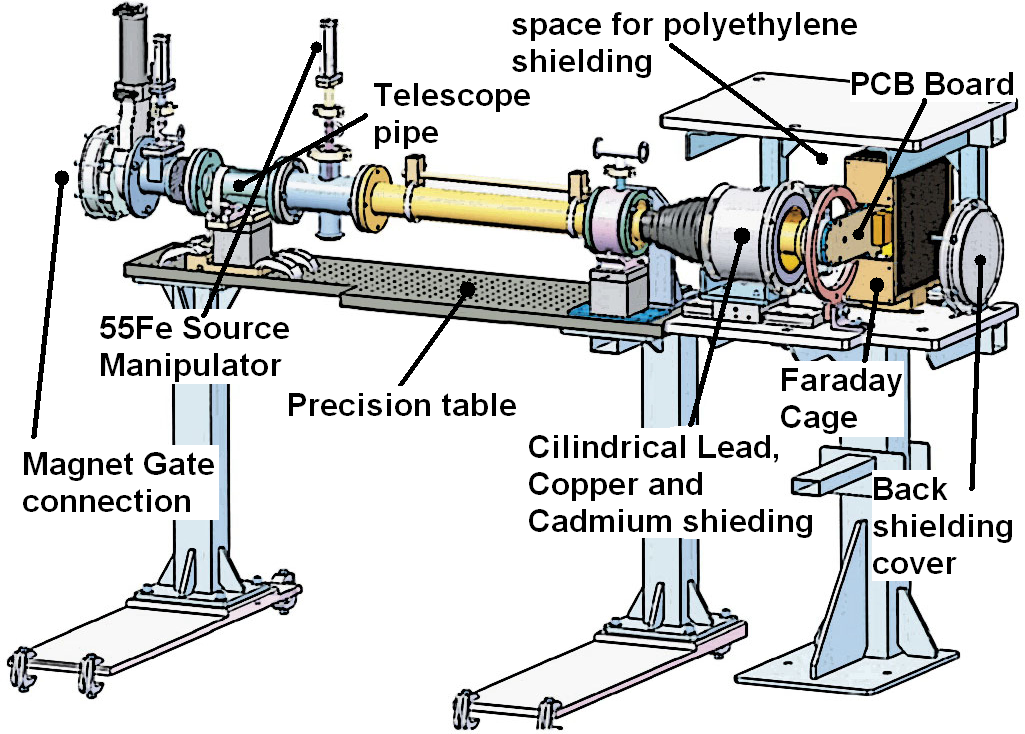} &
\includegraphics[width=0.42\textwidth]{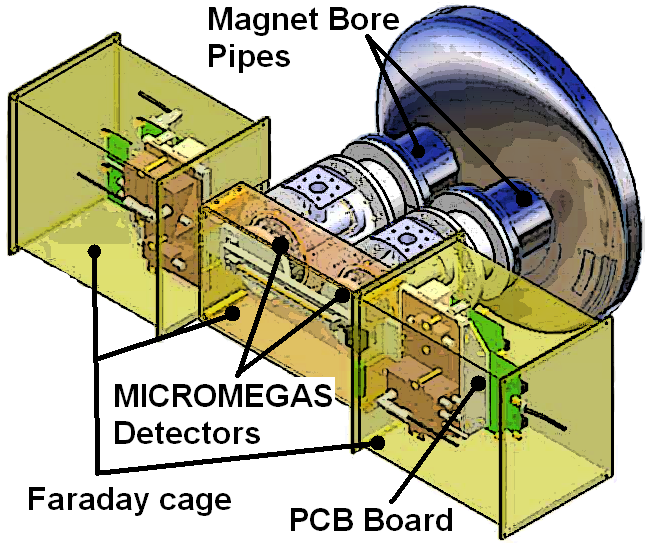} \\
\end{tabular}
\caption{\fontfamily{ptm}\selectfont{\normalsize{ Drawing from the new Micromegas detection line installed at the sunrise side (left) and the new Micromegas detectors system installed at the sunset side (right).  }}}
\label{fi:summaryMicromegasSystems}
\end{figure}

The CAST shutdown in 2007 required to implement the new $^3$He system was used by the Micromegas detector groups, hosted by CEA Saclay and University of Zaragoza, to upgrade the detection systems of CAST. The good performance of Micromegas detectors in terms of stability and lower background pushed the replacement of the TPC taking data at the sunset side by \emph{two} Micromegas detectors, increasing considerably the discovery potential at the sunset side of the experiment. In addition, the Micromegas detection system at the sunrise side was upgraded by installing a new vacuum line, with the capability of holding a future X-ray focusing system, that further improved the monitoring of the main detector parameters. The new sunrise line was particularly designed to allocate a complementary shielding made of lead, copper, cadmium, Plexiglas and polyethylene, that allowed to reduce the final background level of the detector by more than a factor~3. Furthermore, the last Micromegas technologies, bulk and microbulk types, have been developed to operate in CAST. These new technologies have improved the robustness of the previous conventional technology by implementing the 2-dimensional strip readout and mesh in one single entity. The new technologies, bulk and microbulk, are able to reach energy resolutions of about 18\% and 12\% at 6\,keV, respectively. Chapter~\ref{chap:micromegas} fully describes these new technologies and the new systems installed, shown in~Figure~\ref{fi:summaryMicromegasSystems}.

\vspace{0.1cm}

\vspace{0.1cm}

\begin{figure}[!ht]\begin{center}
\begin{tabular}{cc}
\includegraphics[width=0.455\textwidth]{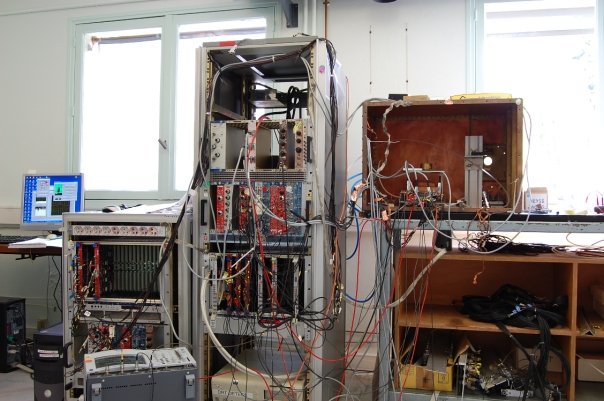} &
\includegraphics[width=0.455\textwidth]{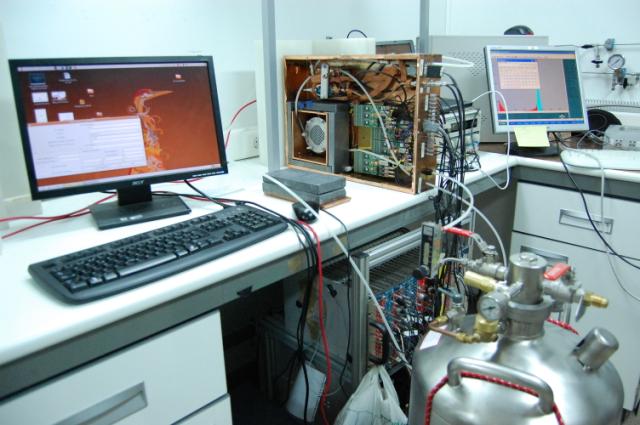} \\
\end{tabular}
\caption{\fontfamily{ptm}\selectfont{\normalsize{On the left, Micromegas acquisition set-up at CEA Saclay. On the right, Micromegas acquisition set-up at Zaragoza.}}}
\label{fi:summaryZaragozaSetup}
\end{center}\end{figure}

The Micromegas detector readout allows to extract useful information that describes the ionizing processes having place at the detector chamber. The time and spatial resolution allow to easily distinguish X-ray events from cosmic muons and electronic noise, which are the main source of triggering events in Micromegas detectors. An extense characterization of these detectors is carried out in CEA Saclay before being finally installed at the CAST magnet bore ends. The chapter~\ref{chap:rawdata} of this thesis fully describes the detector readout and the typical pattern of X-ray event obtained from the parameterization of the temporal signal and spatial information. Moreover, a characterization of a bulk detector has been studied at different conditions; related with pressure, high voltage settings, and isobutane concentration for a Ar+iC$_4$H$_10$ gas mixture. Moreover, a recently built Micromegas acquisition system in Zaragoza (see Fig.~\ref{fi:summaryZaragozaSetup}) was used to perform a basic characterization of a microbulk detector. As well, a model to describe the shape of the temporal mesh pulse was developed and used to characterize different configurations of the electronic set-up concerning the shaping of the signal.

\vspace{0.1cm}

The selection of X-ray events between all the processes recorded by the Micromegas data acquisition is done by applying statistical methods. The readout data of the detector is synthesized to obtain potential discriminant parameters, or observables, which are introduced in the analysis in order to recognize X-ray events. The statistical method allows to reduce significatively the amount of X-ray representative background events by introducing a high acceptance of real X-rays coming from an $^{55}$Fe source. In chapter~\ref{chap:discrimination} are described the potential observables definition and the ones leading to a higher background rejection are presented. \emph{Two} statistical methods are described in this chapter; an modified multivariate analysis and a rejection based on the theory Self-Organized Feature Maps (SOFM) showing the capabilities of background rejection of Micromegas detectors by using different approaches.

\vspace{0.1cm}

\vspace{0.1cm}

The data taking period in 2008 had to deal with an unexpected problem, the presence of a leak on the $^3$He magnet bores circuit that was actually shifting the desired axion mass coverage, thus losing the reference of density inside the bores due to the unknown nature of the leak. Since the effect of the leak was not recognized till the end of the run it produced some uncovered axion mass regions, due to the required emptying and re-filling of the cold bores, which lead to start back at the expected density in the absence of a leak. An extense analysis of the behavior of the pressure inside the $^3$He is presented in chapter~\ref{chap:leak}, which allowed to characterize the leak and to establish the relation between the cold bore density and the measured pressure in the bores, which is continuously monitored during the data taking periods. Moreover, the influence of the delimiting windows temperature on the gas distribution, being a key parameter, was analyzed for different amounts of gas corresponding to the density range covered in 2008. The better understanding of the behavior of the gas inside the bores allowed to realize of necessary real gas corrections, that start to be non negligible at the starting of the new $^3$He density coverage, which showed better agreement with Computational Fluid Dynamics (CFD) calculations than the ideal gas approach.

\vspace{0.1cm}

\begin{figure}[!ht]
{\centering \resizebox{1.0\textwidth}{!} {\includegraphics{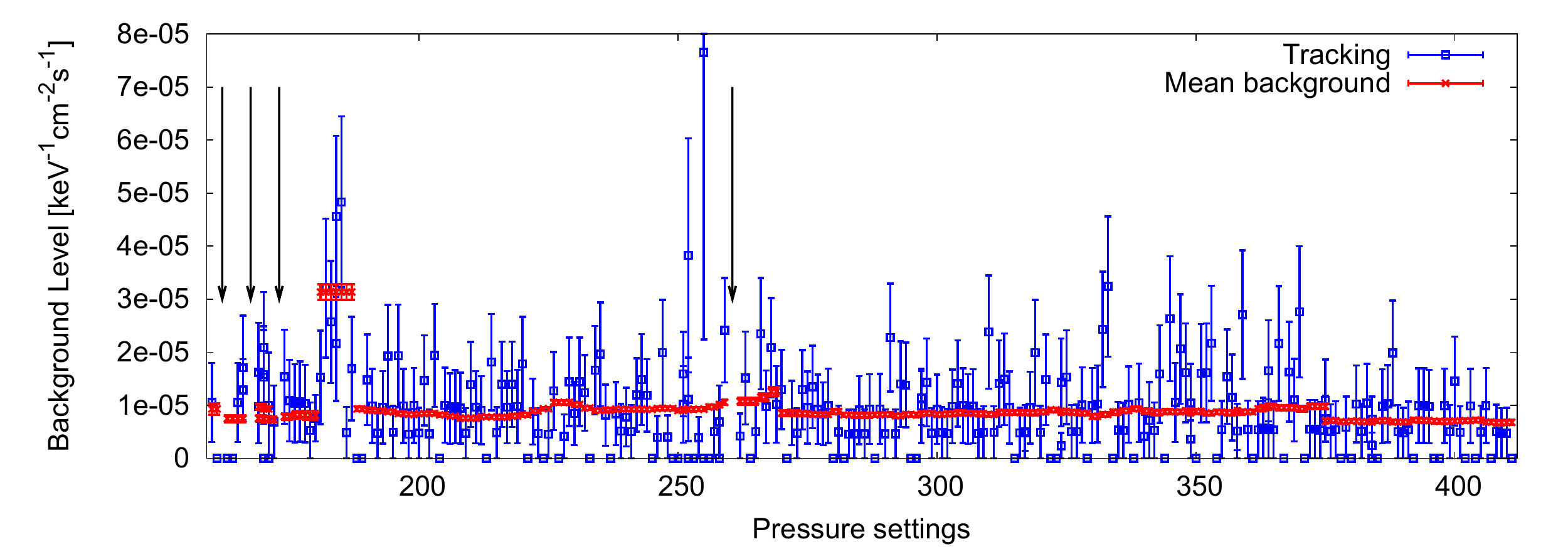}} \par}
\caption{\fontfamily{ptm}\selectfont{\normalsize{Tracking rates and mean background rate defined as a function of the step numbers measured in 2008. }}}
\label{fi:summaryTracking}
\end{figure}

This thesis focuses on the analysis of the background and tracking data of 2008 for the sunrise Micromegas detection line. During 2008 \emph{three} different detectors took data at the sunrise line, since \emph{two} of those detectors had to be replaced due to technical problems. Chapter~\ref{chap:gLimitHe3} describes the coverage of each of those detectors, and the technical issues affecting the data taking period. In addition, this chapter describes the background definition used for each tracking and studies possible systematic effects in background and tracking data. The background, to be compared with the tracking counts, is defined for each tracking day by integrating background data measured in surrounding days. The final tracking counts detected at each cold bore density allow to obtain a limit on the axion-photon coupling in the absence of signal (see Fig.~\ref{fi:summaryTracking}).

\begin{figure}[!ht]
{\centering \resizebox{1.0\textwidth}{!} {\includegraphics{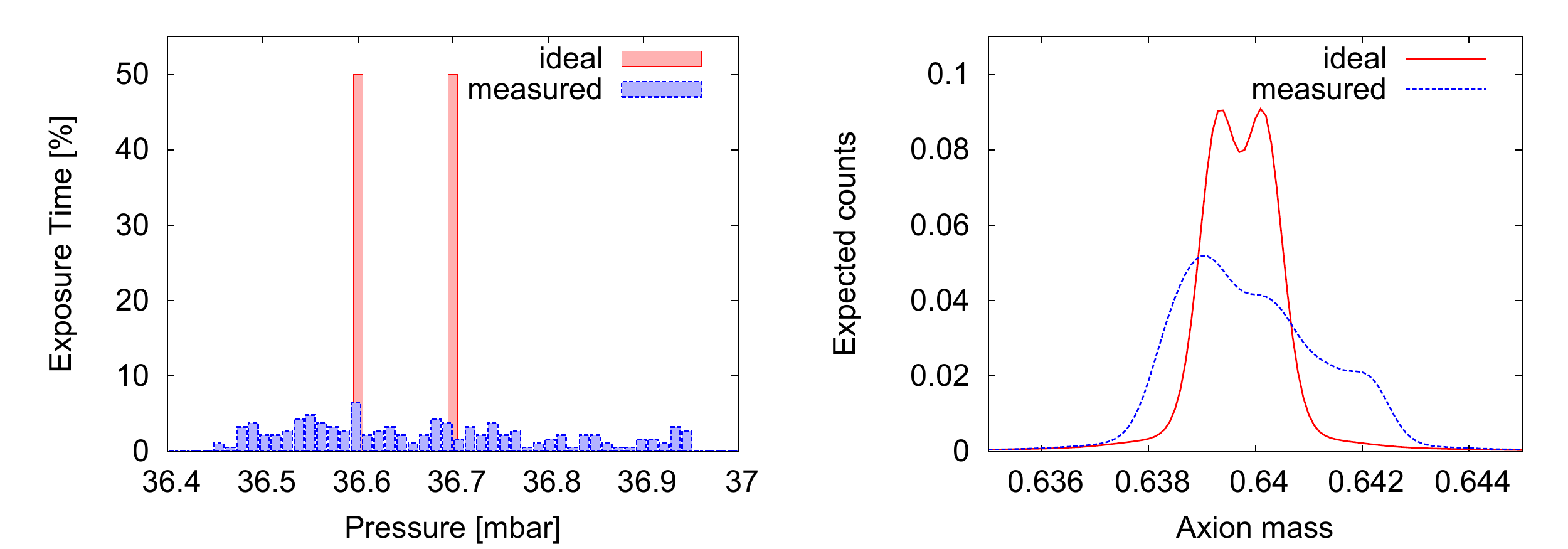}} \par}
\caption{\fontfamily{ptm}\selectfont{\normalsize{ On the left, the ideal scanning step, 2 settings per tracking, together with the measured pressure scanning in percentage for the last tracking in 2008. On the right, the equivalent axion mass resonance at $g_{a\gamma} = 10^{-10}$\,GeV$^{-1}$ for the ideal case and for the generated with the measured pressure.   }}}
\label{fi:summaryTrackingScan}
\end{figure}

The tracking and background data is then introduced in a likelihood method that allows to compute the compatibility of the tracking measured data with the expected background level, the signal is introduced in the likelihood for different values of the axion-photon coupling $g_{a\gamma}$ that allows to obtain $\chi^2(g_{a\gamma})$. In the absence of signal the minimum of $\chi^2$ should be compatible with zero, and an upper limit on $g_{a\gamma}$ can be obtained integrating to a certain confidence level, usually fixed in 95\% C.L. Chapter~\ref{chap:gLimitHe3Final} describes the binned likelihood method used in the previous phases of the experiment to obtain a limit, together with the Micromegas data that lead to a limit in the $^4$He data taking period. The likelihood method in the new $^3$He phase had to be re-adapted in order to take into account the changes in density during tracking movement, due to window temperature changes and gravitational effects on the gas distribution along the magnet bore during tracking. The ideal \emph{two} step coverage per tracking was re-calculated in order to obtain a more realistic axion mass coverage (see Fig.~\ref{fi:summaryTrackingScan}), and the new unbinned likelihood presented allows to take into account the conditions of the bore density at the time of detection of each tracking count.

\begin{figure}[!ht]
{\centering \resizebox{1.0\textwidth}{!} {\includegraphics{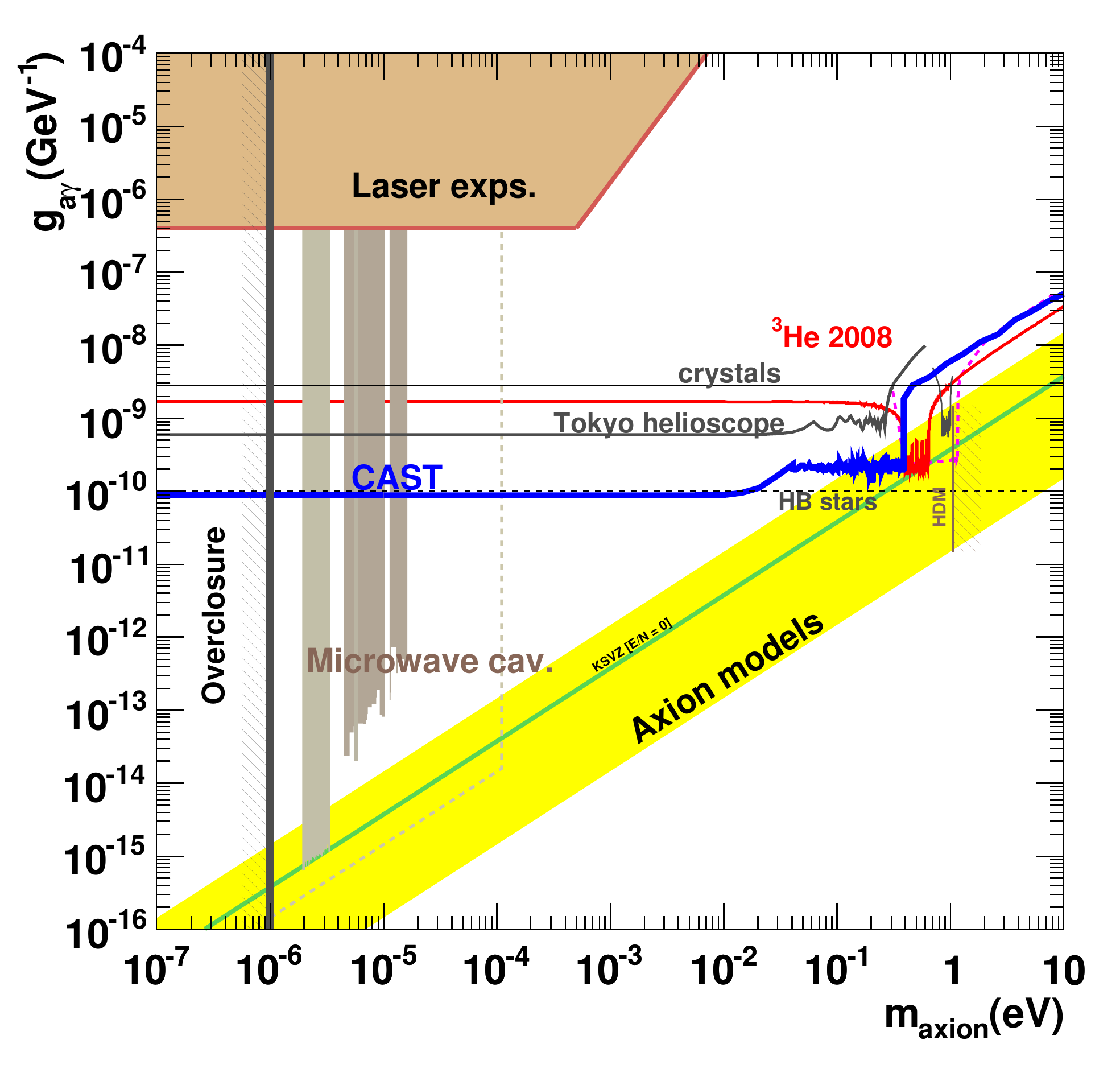}} \par}
\vspace{-0.6cm}
\caption{\fontfamily{ptm}\selectfont{\normalsize{ CAST axion-photon coupling limit obtained in 2008 by combining \emph{three} Micromegas detection lines (red), together with the previous CAST published results (blue). The sensitivity reached in this new axion mass coverage allows to exclude a significant region favored by different axion models, represented with a yellow band. In addition, the CAST experiment crossed the standard KSVZ axions model line by the end of 2008 data run. The most relevant astrophysical limits and cosmological limits are also shown, HB stars from globular clusters ($g_{a\gamma} < 10^{-10}$\,GeV$^{-1}$), cold dark matter and hot dark matter overclossures. Laser based searches, microwave cavity searches, and crystal searches are also represented, together with the exclusion of the Tokyo helioscope recently published results. }}}
\label{fi:summaryExclusionFinalBig}
\end{figure}

\vspace{0.1cm}
The tracking and background data from the sunrise Micromegas line presented in this thesis has been combined with the data from the \emph{two} Micromegas detectors running at the sunset side of the experiment, leading to a more restrictive limit on the axion-photon coupling in the new axion mass coverage, given by the first $^3$He data taking period carried out in 2008. The gas density distribution along the magnet bore causes a negative effect on the effective coherence length, which in the worst scenario would be reduced to $7.26$\,m. Thus, resulting in a conservative limit of $g_{a\gamma} < 2.7\cdot10^{-10}$\,GeV$^{-1}$ given by the sunrise Micromegas detector, and a combined limit of $g_{a\gamma} < 2.2\cdot10^{-10}$\,GeV$^{-1}$, at the mass regions covered during 2008 data taking, 0.39\,eV $< m_a <$ 0.64\,eV, excluding the uncovered regions due to the effect of the leak. The final result on the coupling for the axion-photon interaction in terms of the axion mass is presented in figure~\ref{fi:summaryExclusionFinalBig}, together with the main limits provided by other axion searches.

\vspace{0.1cm}

The results presented in this thesis conclude that the negative effect of the reduction of the magnet coherence length, and the reduced tracking exposure in the new $^3$He phase coverage, by covering \emph{two} settings per tracking, has been counter-balanced by the higher performance introduced by the new Micromegas detectors and detection systems, in terms of lower background level and higher efficiency. Thus placing the detector discovery potential in a level competitive with the X-ray focusing telescope and CCD detection potential. The lower background levels achieved, which placed the detected tracking counts per tracking in few counts per density steps, with a mean rate of about 2 counts per hour, has lead to a combined Micromegas coupling limit at the same level as the combined limit reached in the $^4$He period coverage with \emph{four} detection lines. The future inclusion of the data collected by the CCD detector, and the coverage of the unexplored mass regions which appeared due to the effect of the leak, will push the limit even lower.

\vspace{0.1cm}

In addition, the extended CAST search to higher masses has entered in a region of the $g_{a\gamma}-m_a$ parameter space that has never been explored before, excluding by first time standard KSVZ axions (E/N = 0). Moreover, providing the first experimental limit based in direct axion search excluding standard KSVZ axions for the eV-mass scale. The results presented in this thesis allow to conclude that CAST will exclude standard KSVZ axions for $m_a \gtrsim 0.6$\,eV, when the current CAST physics program reaches its end measuring axion masses close to $1$\,eV, in case no signal is found.

\vspace{0.1cm}

Figure~\ref{fi:summaryExclusionFinalBig} presents the coupling limit obtained for the axion-photon interaction as a final result of the work of this thesis, taking into account \emph{three} Micromegas detection lines with a coherence length of $7.26$\,m. The result is presented with the previous published work of the CAST collaboration during the vacuum and $^4$He data taking phases of the experiment, together with the astrophysical and cosmological bounds and the main axion searches limits.

\chapter{Resumen y conclusiones}
\label{chap:resumen}

El axion es una part\'icula hipot\'etica que apareci\'o como una soluci\'on natural al problema de la simetr\'ia CP de las interactiones fuertes, las cuales parecen no violar CP. El problema surge del hecho de que la descripci\'on de las interacciones dada por la cromodin\'amica cu\'antica (QCD) no prohibe dicha violaci\'on. Adem\'as es dif\'icil de comprender el porqu\'e la interacci\'on fuerte no viola CP cuando \'este no es el caso en la interacci\'on electrod\'ebil. Se han considerado varias soluciones a este problema, sin embargo, la soluci\'on m\'as elegante y natural tiene como resultado una nueva part\'icula que todav\'ia no ha sido observada experimentalmente, el axion. El axion fue postulado por \emph{H. Quinn} y \emph{R. Peccei} en 1977, bas\'andose en los trabajos previos de \emph{'t Hooft} y \emph{Weinberg} donde enfatizan la necesidad de incluir un termino que viole CP en la teor\'ia de las interacciones fuertes.

\vspace{0.1cm}

Diferentes modelos de axion fueron apareciendo en los a$\tilde{\mbox{n}}$os sucesivos, fijando la relaci\'on entre la masa del axion y la fuerza de interacci\'on con la materia ordinaria. Cada modelo describe la manera en la que el axion se acoplar\'ia con la materia, describiendo las reglas de interacci\'on con cada una de las part\'iculas existentes. El acoplo del axion con la materia ordinaria tendr\'ia ciertas implicaciones observacionales en astrof\'isica y cosmolog\'ia, lo que permite restringir las propiedades del axion en t\'erminos de su masa y su constante de acoplo. En particular, la existencia de axiones afectar\'ia en cierta medida en evoluci\'on estelar si su fuerza de interacci\'on fuese demasiado intensa, y el hecho de que el axion tenga masa lo sit\'ua en los modelos cosmol\'ogicos como un posible candidato a materia oscura, el cual, en funci\'on de su masa podr\'ia dar cuenta de una contribuci\'on dominante al total de la materia oscura en el Universo.

\vspace{0.1cm}

La b\'usqueda de axiones esta bien motivada como una soluci\'on al problema CP de la interacci\'on fuerte y como un posible candidato a materia oscura. Adem\'as, los axiones podr\'ian estar relacionados con otros fen\'omenos f\'isicos de naturaleza desconocida, como el problema del recalentamiento de la corona solar, o la modulaci\'on del campo magn\'etico terrestre.

\vspace{0.1cm}
Una caracter\'istica com\'un en cualquier modelo de axiones es que el axion se acopla con \emph{dos} fotones. \emph{P.Sikivie} introdujo en 1983 las bases para la detecci\'on directa de axiones usando para ello la analog\'ia de la interacci\'on del decaimiento del pion en \emph{dos} fotones, descrita por el efecto Primakoff. El principio de detecci\'on experimental se basa en aplicar un campo magn\'etico intenso que podr\'ia aportar uno de los fotones de la interacci\'on, mediante un fot\'on virtual, b\'asicamente permitiendo la conversi\'on directa de axiones a fotones, y viceversa. Desde entonces, varias propuestas de b\'usqueda experimental de axiones se han realizado, contin\'uan y estan en desarrollo. Despu\'es de 30 a$\tilde{\mbox{n}}$ la motivaci\'on para descubrir el axion no ha cesado sino intensificado.

\vspace{0.1cm}

\begin{figure}[!ht]
{\centering \resizebox{0.9\textwidth}{!} {\includegraphics{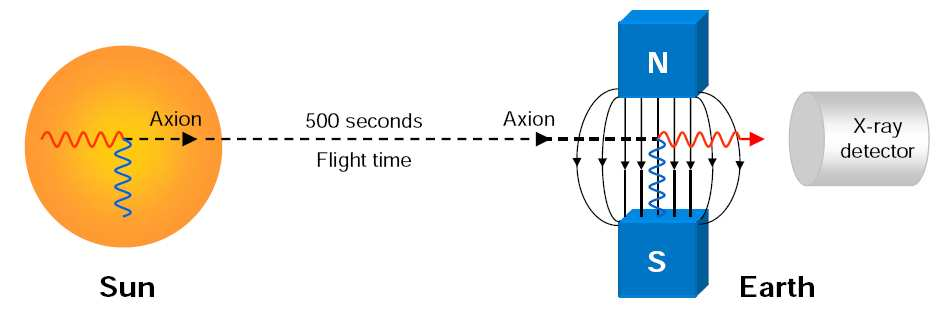}} \par}
\caption{\fontfamily{ptm}\selectfont{\normalsize{ Detection concept for the magnet helioscope idea. Where the Sun is used an intense source of axions. }}}
\label{fi:resumenHelioscopeConcept}
\end{figure}

Los experimentos de b\'usqueda de axiones se pueden dividir principalmente en \emph{tres} grupos dependiendo de la fuente de axiones utilizada. La primera posibilidad son los denominados \emph{"haloscope searches"} que se basan en la posible abundancia de axiones reliquia que contribuir\'ian a la densidad de materia oscura fria (CDM) ligada gravitacionalmente a nuestra galaxia. La segunda son los denominados \emph{"laboratory searches"} que utilizan la posibilidad de generar axiones mediante un intenso haz de fotones viajando a trav\'es de un campo magn\'etico intenso, la existencia del axion producir\'ia efectos observables en las propiedades del haz de fotones, o los axiones generados podr\'ian ser reconvertidos a fotones en una segunda region de campo magn\'etico aislada de la primera. La tercera son los denominados \emph{"helioscope searches"} que se basan en las condicioens favorables en el n\'ucleo del Sol para producir una cantidad suficiente de axiones en funci\'on de su fuerza de interacci\'on. Los axiones generados en el Sol viajar\'ian a trav\'es del espacio sin interaccionar con nada excepto en las condiciones particulares impuestas por el campo magn\'etico producido en el helioscopio y transversal al flujo de axiones, donde los axiones ser\'ian reconvertidos a fotones (ver Fig.~\ref{fi:resumenHelioscopeConcept}).

\vspace{0.2cm}

El trabajo realizado en esta tesis se centra en la b\'usqueda de axiones mediante un experimento basado en \emph{helioscope search}, estos experimentos se basa

\begin{figure}[!ht]
\begin{center}
\begin{tabular}{cc}
{\centering \resizebox{0.44\textwidth}{!} {\includegraphics{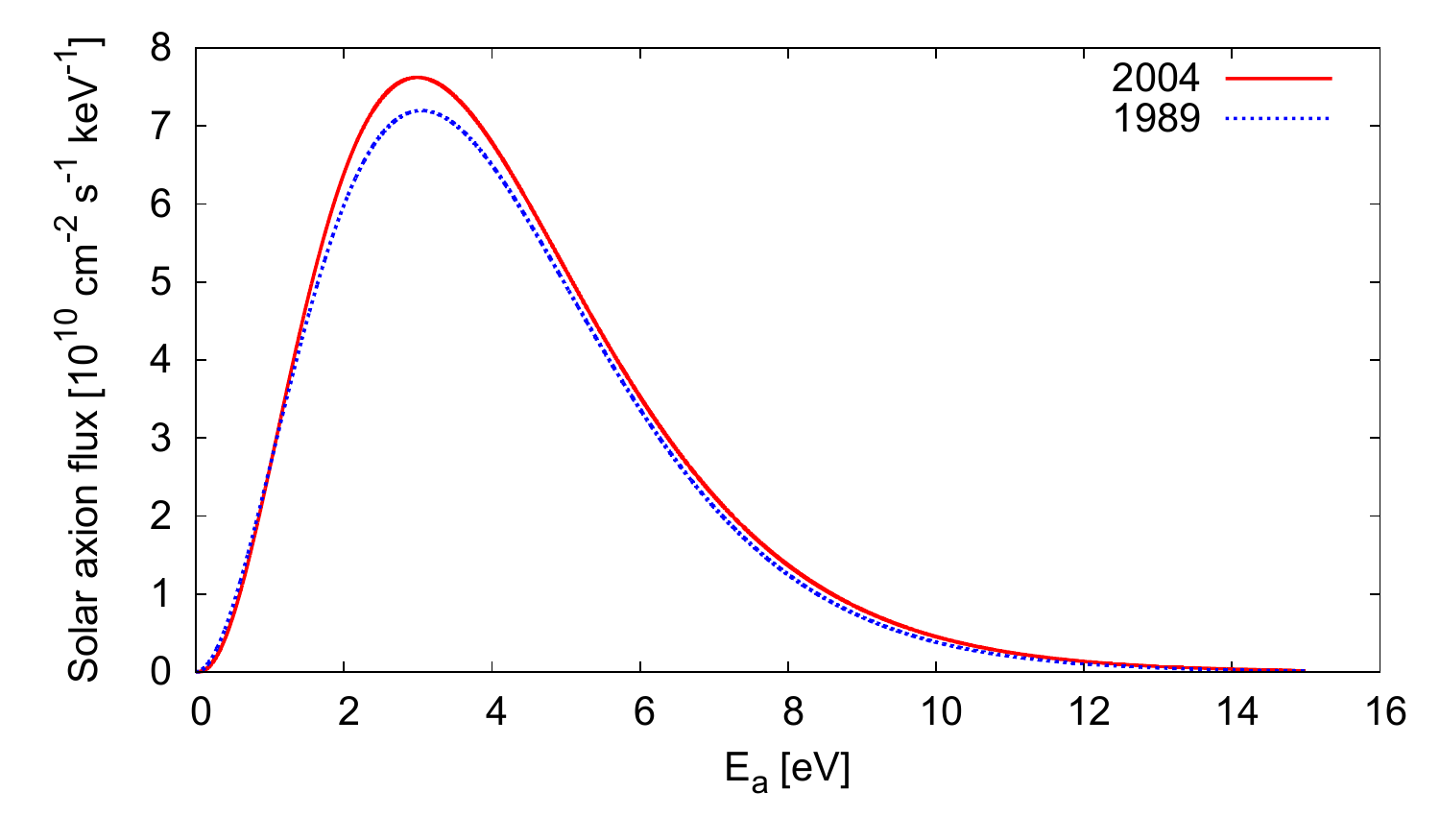}} \par} &
{\centering \resizebox{0.51\textwidth}{!} {\includegraphics{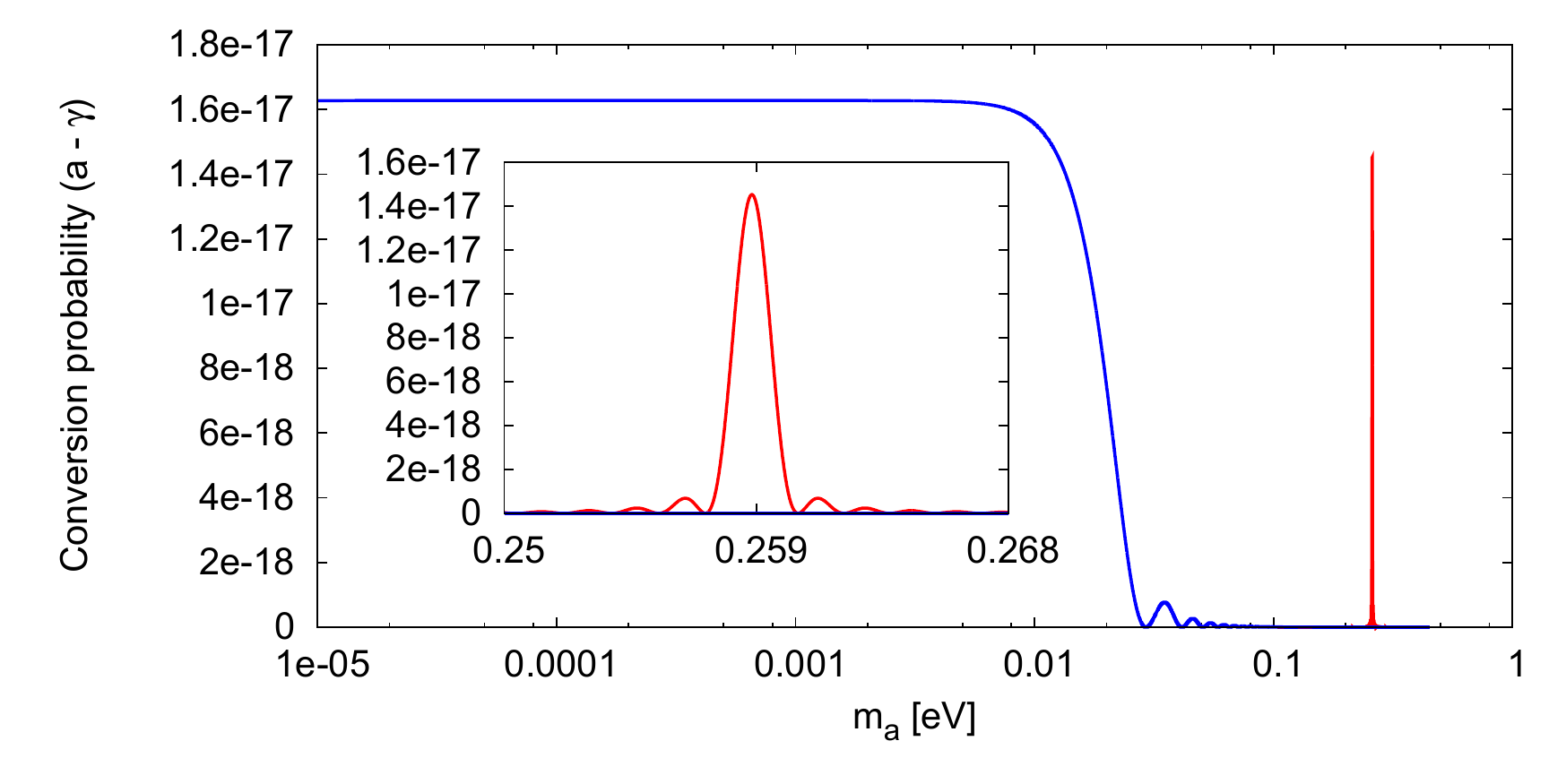}} \par} \\
\end{tabular}
\end{center}
\caption{\fontfamily{ptm}\selectfont{\normalsize{ Expected axion solar flux on Earth (left) and probability conversion of axion to photon in the presence of a transversal magnetic field in CAST conditions (CAST vacuum phase and a resonance given by a fixed density in the magnet bore).  }}}
\label{fi:resumenFluxProb}
\end{figure}

The helioscope search, in which the work presented in this thesis is focused, is based on the well established standard solar model which allows to calculate the expected axion flux generated in the core of the Sun as a function of the axion-photon coupling. The axion solar flux at Earth was calculated by \emph{G. Raffelt} and the probability conversion of axion to photons in the presence of an external magnetic field was given by \emph{K. Bibber} (see Fig.~\ref{fi:resumenFluxProb}). Thus allowing to determine the expected number of converted axions into photons, when the magnet is aligned with the Sun, as a function of the axion-photon coupling, which in absence of signal provides an upper-limit to the axion-photon coupling.

\vspace{0.1cm}

\vspace{0.1cm}

This thesis presents the latest results from CAST (CERN Axion Solar Telescope), an experiment hosted at the European Organization for Nuclear Research (CERN) which is searching for solar axions. CAST consists of a superconducting prototype dipole magnet from the Large Hadron Collider (LHC) which is able to provide an intense magnetic field of almost 9\,T, over a lenght of $9.26$\,m. The magnet is mounted over a rotating platform which allows to continuously follow the Sun during the sunrise and the sunset, each for about $1.5\,$hours. The expected axion signal mean energy is within the X-ray range thus CAST implements X-ray detectors to cover both magnet sides, corresponding to the sunrise side and the sunset side.

\vspace{0.1cm}

\begin{figure}[!ht]
{\centering \resizebox{0.8\textwidth}{!} {\includegraphics{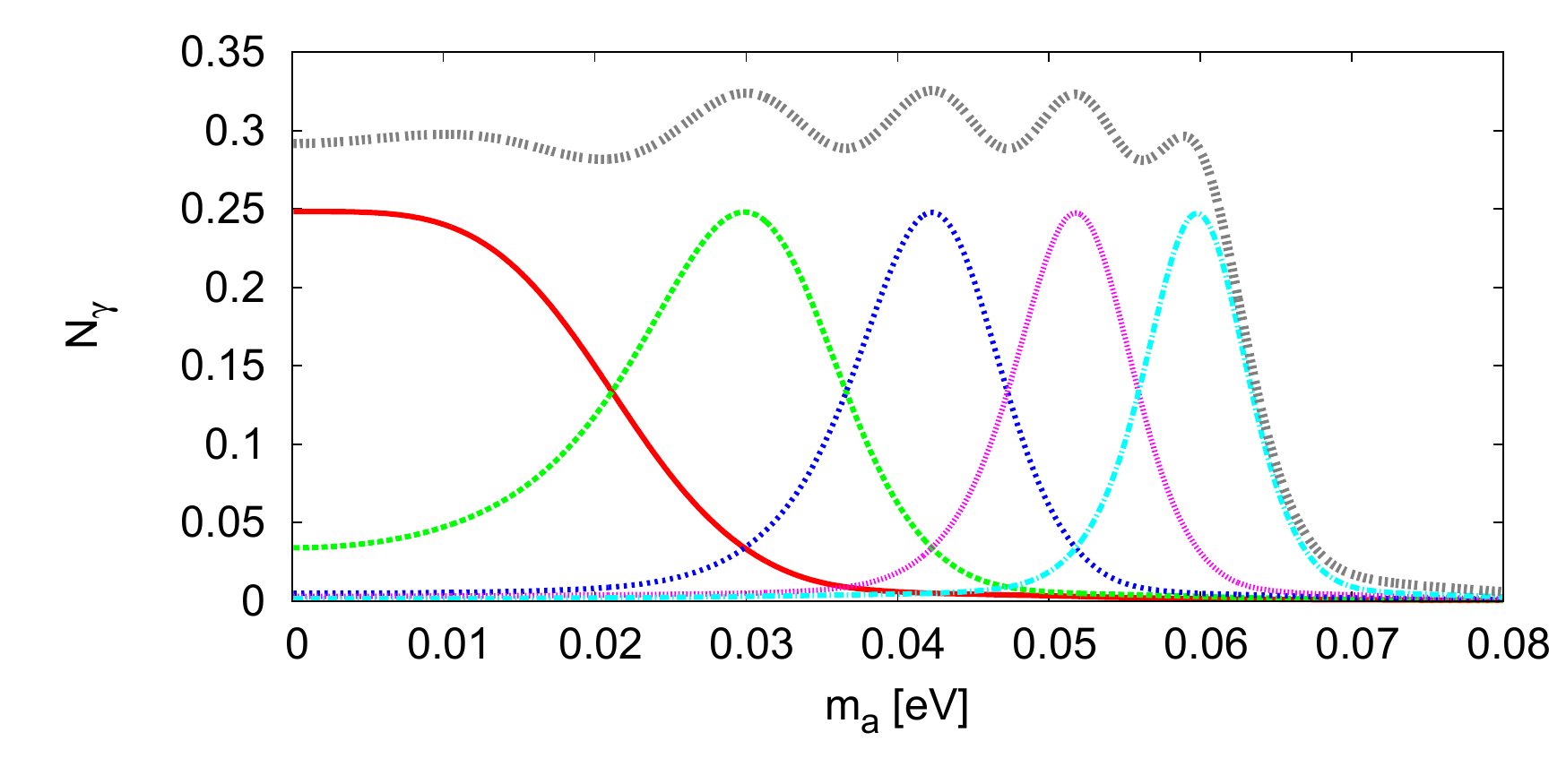}} \par}
\caption{\fontfamily{ptm}\selectfont{\normalsize{ Density steps scanning, in terms of the expected photons for a tracking with one detector at $g_{a\gamma} = 1\cdot10^{-10}$\,GeV$^{-1}$, leading to coverage in a wide axion mass range, first scanning steps are presented in this plot. }}}
\label{fi:resumenNgamma}
\end{figure}

CAST has published the most restrictive axion-photon coupling limit in a wide axion mass range. The CAST experiment started to take data already by 2003. During its first phase (CAST Phase I) the magnet was operating with vacuum inside the magnet bores, conditions in which the sensitivity of the experiment in the axion-photon coupling $g_{a\gamma}$ was enhanced for axion masses up to $m_a \lesssim 0.02$\,eV. The experiment completed its first phase in 2004 improving the current limits on the coupling constant, $g_{a\gamma} < 8.8\cdot10^{-11}\,\mbox{GeV}^{-1} \mbox{at 95\%~CL for}\,m_a\leq 0.02\,\mbox{eV}$. In the second phase (CAST Phase II), the experiment extends its sensitivity to higher axion masses by filling the magnet with a refractive buffer gas that enhances the conversion of axions to photons for a narrow axion mass range which depends on the buffer gas density. The continuous scanning in axion mass is achieved by smoothly increasing the amount of gas inside the magnet bore allowing to be sensitive to axions in a wider axion mass range (see Fig.~\ref{fi:resumenNgamma}).

\begin{figure}[!hb]
{\centering \resizebox{0.8\textwidth}{!} {\includegraphics{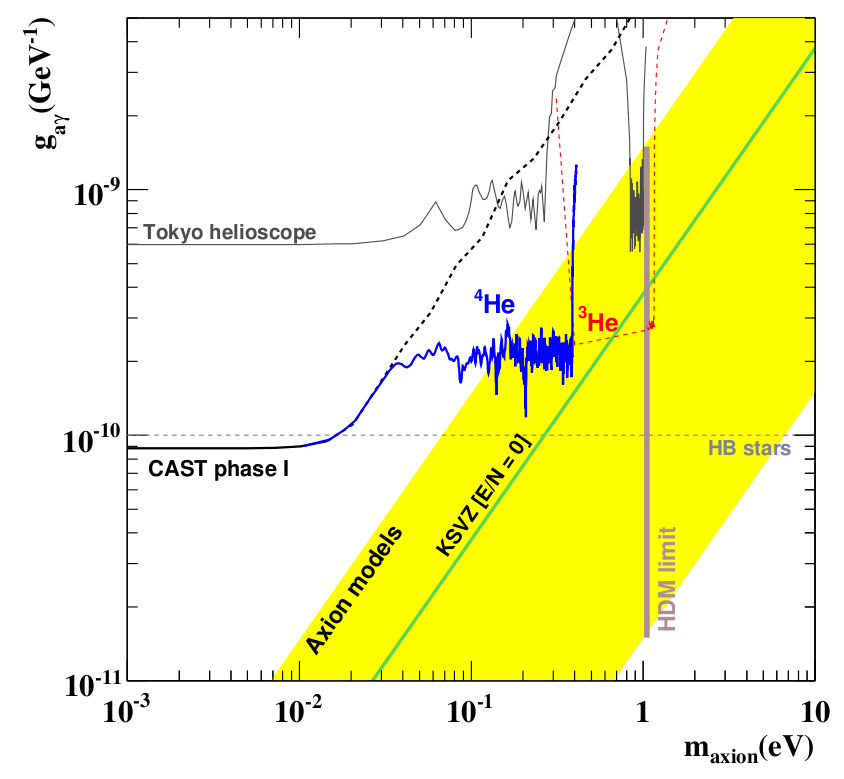}} \par}
\caption{\fontfamily{ptm}\selectfont{\normalsize{Axion-photon coupling limit as a function of the axion mass provided by CAST experiment in the vacuum phase (Phase I) and $^4$He axion mass range coverage. Expected coverage for the $^3$He operating period is also shown. }}}
\label{fi:resumenExclusionHe4}
\end{figure}

The second phase of CAST used $^4$He as buffer gas during the first data taking period which allowed to cover axion masses up to 0.4\,eV. This data taking period finished in 2006 allowing to set a new limit on the axion-photon coupling for the new explored axion mass range of $g_{a\gamma} < 2.17\cdot10^{-10}\,\mbox{GeV}^{-1} \mbox{ at 95\%~CL for }0.02\,\mbox{eV} < m_a < 0.4\,\mbox{eV}$ being the first experiment entering in the axion models favored region for eV-mass scale axions. These results are summarized by representing the excluded $g_{a\gamma}-m_a$ parameter space (see Fig.~\ref{fi:resumenExclusionHe4}) together with other helioscope searches, astrophysical constrains, and favored axion models region.

\vspace{0.1cm}

In order to sweep higher axion masses CAST requires to be operated with $^3$He gas that allows to reach higher gas densities, the saturating pressure of $^3$He above 130\,mbar allows to scan masses up to about 1\,eV. The gas system had to be adapted to work with $^3$He and it required a number of technically challenging upgrades to equip the experiment for the new run.

\vspace{0.1cm}

\begin{figure}[!ht]
\begin{center}
\begin{tabular}{cc}
{\centering \resizebox{0.35\textwidth}{!} {\includegraphics{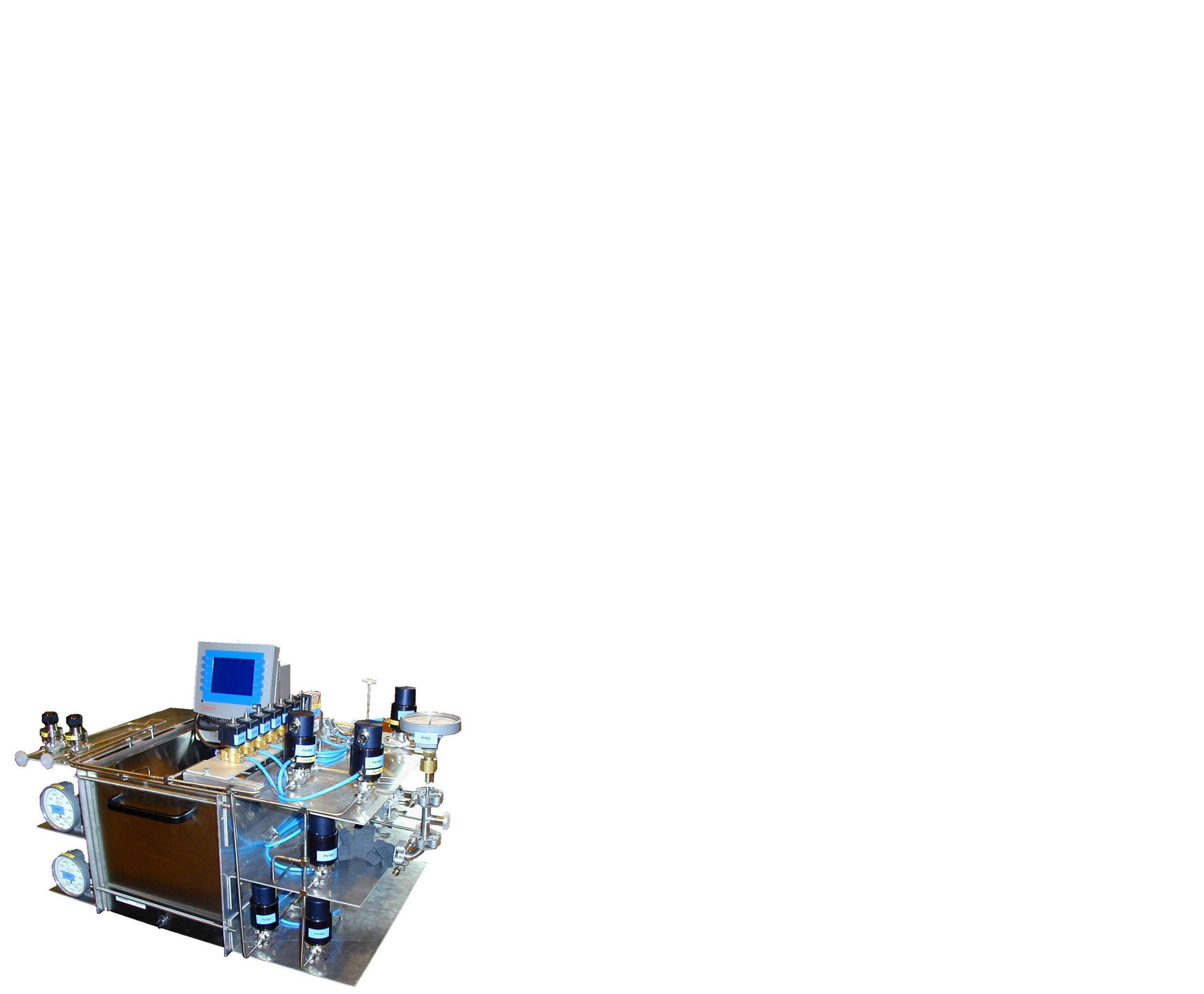}} \par} &
{\centering \resizebox{0.57\textwidth}{!} {\includegraphics{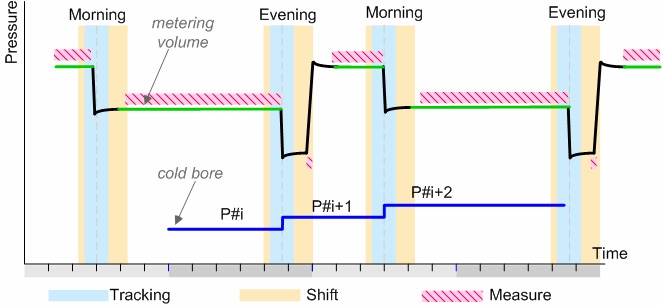}} \par}
\end{tabular}
\end{center}
\caption{\fontfamily{ptm}\selectfont{\normalsize{ A picture of the new metering system developed for the new data taking phase (left), and a plot representing the cold bore filling scheme (right) during the data taking period. }}}
\label{fi:resumenHe3Filling}
\end{figure}

The new $^3$He equipment required to implement a recovery system to avoid the loss of the valuable $^3$He gas in case of a magnet quench, a superconducting magnet protection system which raises the temperature of the magnet thus increasing the pressure of the gas inside the bores. Then, the gas needs to be quickly evacuated, via an expansion volume, in order to protect the thin high X-ray transmission windows containing the gas. Moreover, the required upgrade was used to increase the functionality and reliability of the system. A new metering volumes design was constructed for fastening the magnet bores gas filling process from empty bores state, which requires longer periods of time as the pressure in the bores is higher. Furthermore, the $^3$He pipework connections are controlled by automated electro-pneumatic valves allowing to program the functionality of the system through a Power Line Communications (PLC) system, which in addition allows to monitor an extense distribution of pressure and flow sensors all along the system. The new functionalities of the system include automatic recovery and gas purging system, between others. Furthermore, the new more versatile system allows to define different magnet bore filling schemas, during the data taking periods, which included a slow density ramping during tracking thanks to the accurate mass flow controllers implemented at the metering system connection to the bores. The final bores filling schema carried out during the $^3$He data taking period was the one covering \emph{two} density steps per tracking by increasing the density in the bores to the next step in the middle of the tracking in a short period of time, of about $3$\,minutes (see Fig.~\ref{fi:resumenHe3Filling}).

\vspace{0.1cm}

CAST exploits the availability of the \emph{four} magnet bore apertures by implementing different type of X-ray detector systems reinforcing the detection of a hypothetical signal which should be detected independently at each detection system during sunrise and sunset trackings. The vacuum phase and $^4$He data taking periods were covered by a Time Projection Chamber (TPC) covering both magnet bores in the sunset side, and each bore in the sunrise side being covered by a X-ray telescope focusing in a Charge-Coupled Device (CCD) adapted to operate in the X-ray energy range, and a Micromegas detector.

\vspace{0.1cm}

\begin{figure}[h!]
\begin{tabular}{cc}
\includegraphics[width=0.52\textwidth]{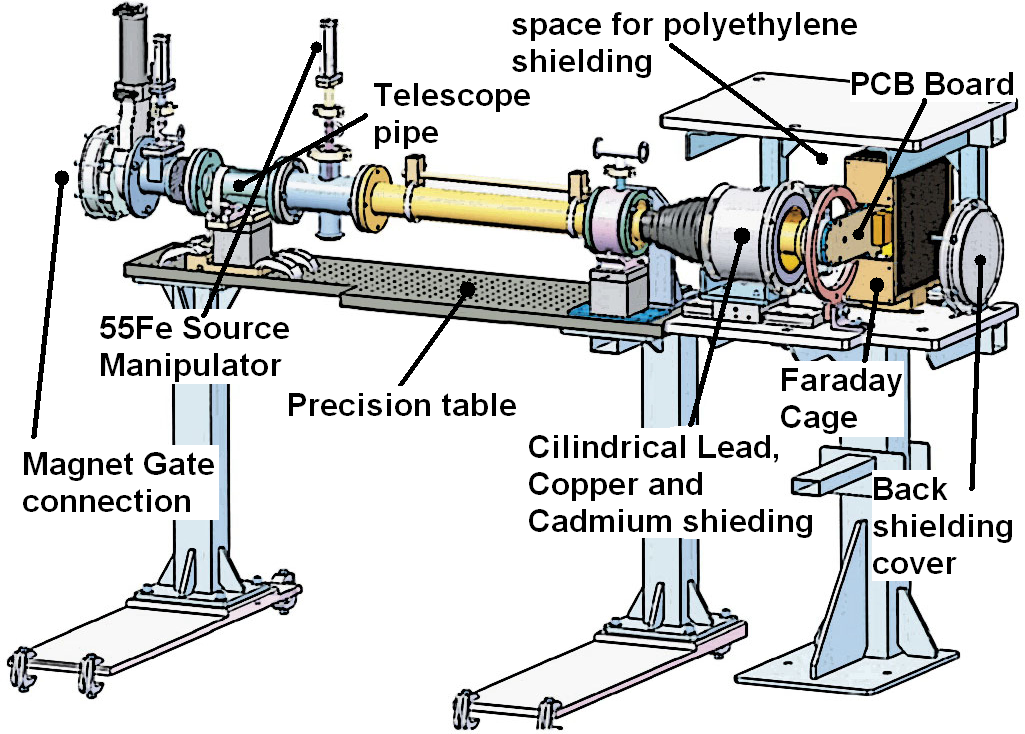} &
\includegraphics[width=0.42\textwidth]{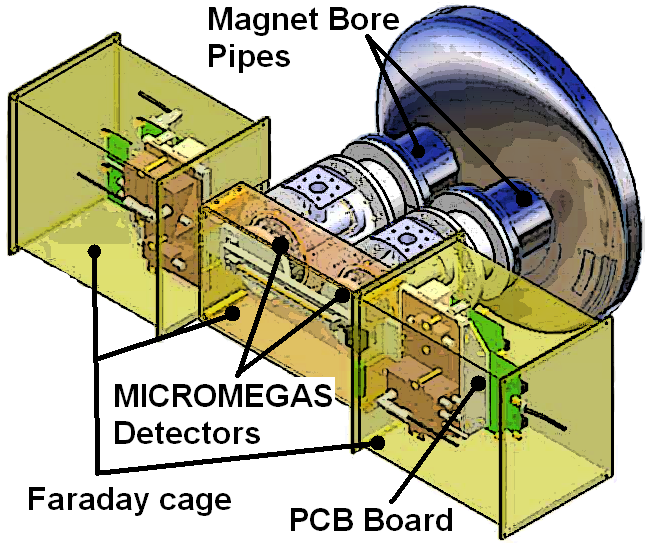} \\
\end{tabular}
\caption{\fontfamily{ptm}\selectfont{\normalsize{ Drawing from the new Micromegas detection line installed at the sunrise side (left) and the new Micromegas detectors system installed at the sunset side (right).  }}}
\label{fi:resumenMicromegasSystems}
\end{figure}

The CAST shutdown in 2007 required to implement the new $^3$He system was used by the Micromegas detector groups, hosted by CEA Saclay and University of Zaragoza, to upgrade the detection systems of CAST. The good performance of Micromegas detectors in terms of stability and lower background pushed the replacement of the TPC taking data at the sunset side by \emph{two} Micromegas detectors, increasing considerably the discovery potential at the sunset side of the experiment. In addition, the Micromegas detection system at the sunrise side was upgraded by installing a new vacuum line, with the capability of holding a future X-ray focusing system, that further improved the monitoring of the main detector parameters. The new sunrise line was particularly designed to allocate a complementary shielding made of lead, copper, cadmium, plexiglass and polyethylene, that allowed to reduce the final background level of the detector by more than a factor~3. Furthermore, the last Micromegas technologies, bulk and microbulk types, have been developed to operate in CAST. These new technologies have improved the robustness of the previous conventional technology by implementing the 2-dimensional strip readout and mesh in one single entity. The new technologies, bulk and microbulk, are able to reach energy resolutions of about 18\% and 12\% at 6\,keV, respectively. Chapter~\ref{chap:micromegas} fully describes these new technologies and the new systems installed, shown in~Figure~\ref{fi:resumenMicromegasSystems}.

\vspace{0.1cm}

\vspace{0.1cm}

\begin{figure}[!ht]\begin{center}
\begin{tabular}{cc}
\includegraphics[width=0.455\textwidth]{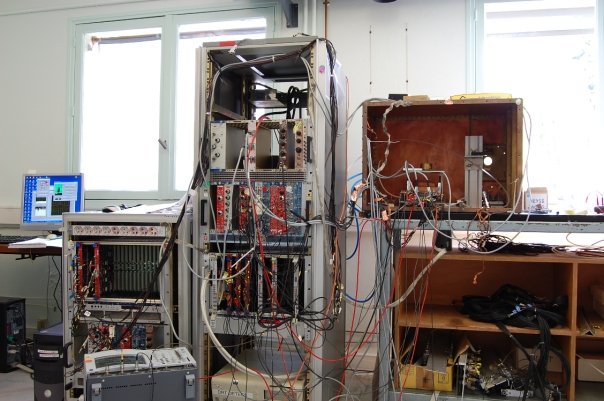} &
\includegraphics[width=0.455\textwidth]{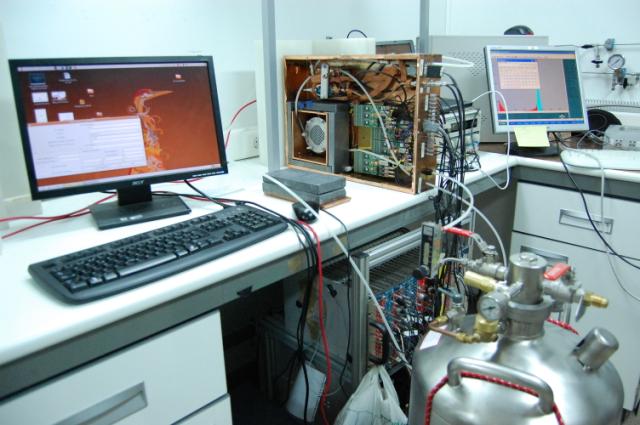} \\
\end{tabular}
\caption{\fontfamily{ptm}\selectfont{\normalsize{On the left, Micromegas acquisition set-up at CEA Saclay. On the right, Micromegas acquisition set-up at Zaragoza.}}}
\label{fi:resumenZaragozaSetup}
\end{center}\end{figure}

The Micromegas detector readout allows to extract useful information that describes the ionizing processes having place at the detector chamber. The time and spatial resolution allow to easily distinguish X-ray events from cosmic muons and electronic noise, which are the main source of triggering events in Micromegas detectors. An extense characterization of these detectors is carried out in CEA Saclay before being finally installed at the CAST magnet bore ends. The chapter~\ref{chap:rawdata} of this thesis fully describes the detector readout and the typical pattern of X-ray event obtained from the parameterization of the temporal signal and spatial information. Moreover, a characterization of a bulk detector has been studied at different conditions; related with pressure, high voltage settings, and isobutane concentration for a Ar+iC$_4$H$_10$ gas mixture. Moreover, a recently built Micromegas acquisition system in Zaragoza (see Fig.~\ref{fi:resumenZaragozaSetup}) was used to perform a basic characterization of a microbulk detector. As well, a model to describe the shape of the temporal mesh pulse was developed and used to characterize different configurations of the electronic set-up concerning the shaping of the signal.

\vspace{0.1cm}

The selection of X-ray events between all the processes recorded by the Micromegas data acquisition is done by applying statistical methods. The readout data of the detector is synthetized to obtain potential discriminant parameters, or observables, which are introduced in the analysis in order to recognize X-ray events. The statistical method allows to reduce significatively the amount of X-ray representative background events by introducing a high acceptance of real X-rays coming from an $^{55}$Fe source. In chapter~\ref{chap:discrimination} are described the potential observables definition and the ones leading to a higher background rejection are presented. \emph{Two} statistical methods are described in this chapter; an modified multivariate analysis and a rejection based on the theory Self-Organized Feature Maps (SOFM) showing the capabilities of background rejection of Micromegas detectors by using different approaches.

\vspace{0.1cm}

\vspace{0.1cm}

The data taking period in 2008 had to deal with an unexpected problem, the presence of a leak on the $^3$He magnet bores circuit that was actually shifting the desired axion mass coverage, thus lossing the reference of density inside the bores due to the unknown nature of the leak. Since the effect of the leak was not recognized till the end of the run it produced some uncovered axion mass regions, due to the required emptying and re-filling of the cold bores, which lead to start back at the expected density in the absence of a leak. An extense analysis of the behaviour of the pressure inside the $^3$He is presented in chapter~\ref{chap:leak}, which allowed to characterize the leak and to establish the relation between the cold bore density and the measured pressure in the bores, which is continously monitored during the data taking periods. Moreover, the influence of the delimiting windows temperature on the gas distribution, being a key parameter, was analysed for different amounts of gas corresponding to the density range covered in 2008. The better understanding of the behaviour of the gas inside the bores allowed to realice of necesary real gas corrections, that start to be non negligible at the starting of the new $^3$He density coverage, which showed better agreement with Computational Fluid Dynamics (CFD) calculations than the ideal gas approach.

\vspace{0.1cm}

\begin{figure}[!ht]
{\centering \resizebox{1.0\textwidth}{!} {\includegraphics{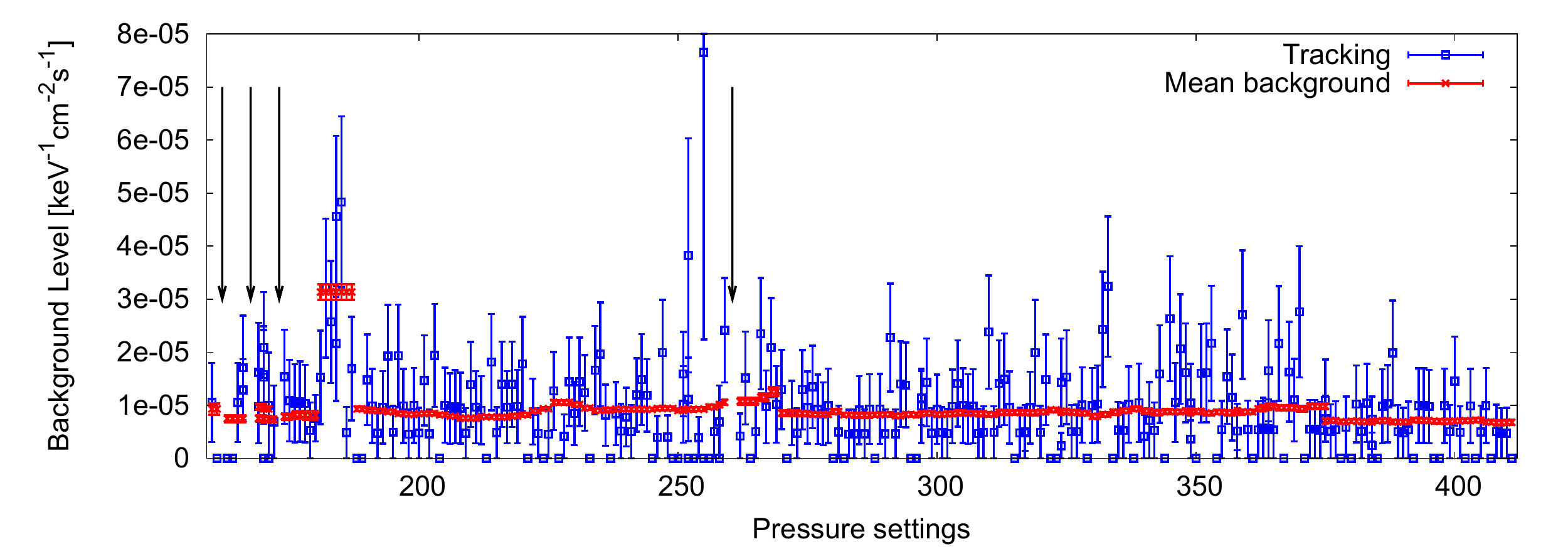}} \par}
\caption{\fontfamily{ptm}\selectfont{\normalsize{Tracking rates and mean background rate defined as a function of the step numbers measured in 2008. }}}
\label{fi:resumenTracking}
\end{figure}

This thesis focuses on the analysis of the background and tracking data of 2008 for the sunrise Micromegas detection line. During 2008 \emph{three} different detectors took data at the sunrise line, since \emph{two} of those detectors had to be replaced due to technical problems. Chapter~\ref{chap:gLimitHe3} describes the coverage of each of those detectors, and the technical issues affecting the data taking period. In addition, this chapter describes the background definition used for each tracking and studies possible systematic effects in background and tracking data. The background, to be compared with the tracking counts, is defined for each tracking day by integrating background data measured in surrounding days. The final tracking counts detected at each cold bore density allow to obtain a limit on the axion-photon coupling in the absence of signal (see Fig.~\ref{fi:resumenTracking}).

\begin{figure}[!ht]
{\centering \resizebox{1.0\textwidth}{!} {\includegraphics{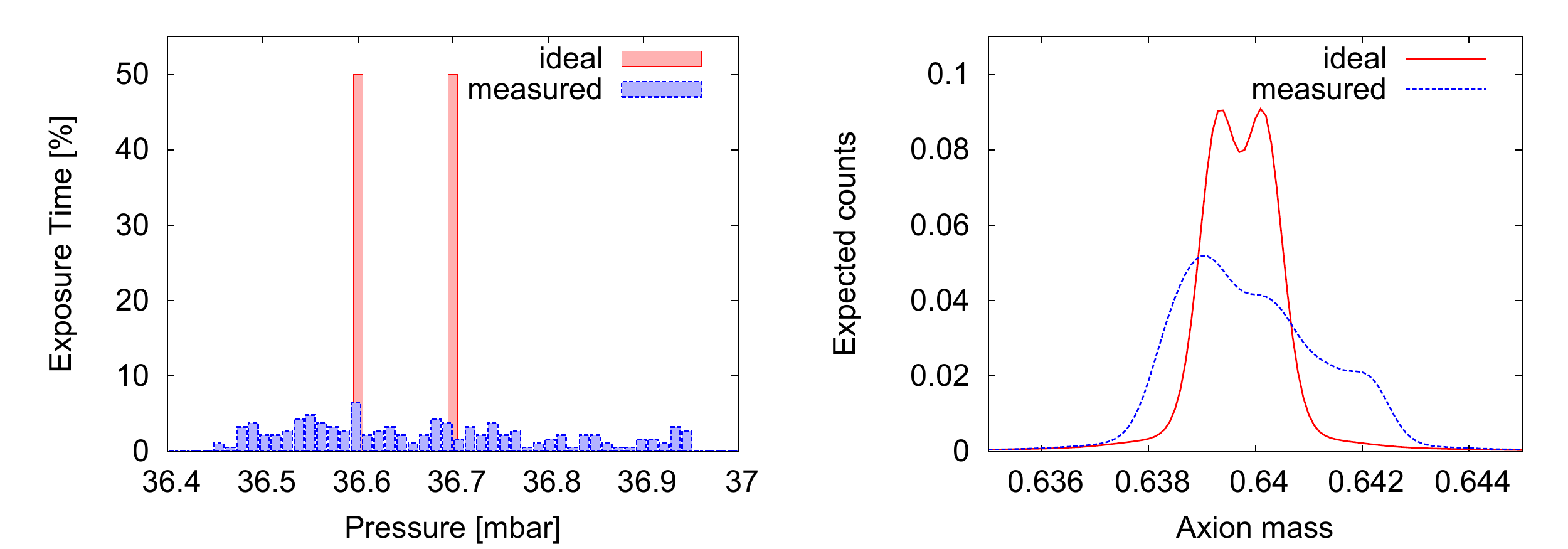}} \par}
\caption{\fontfamily{ptm}\selectfont{\normalsize{ On the left, the ideal scanning step, 2 settings per tracking, together with the measured pressure scanning in percentage for the last tracking in 2008. On the right, the equivalent axion mass resonance at $g_{a\gamma} = 10^{-10}$\,GeV$^{-1}$ for the ideal case and for the generated with the measured pressure.   }}}
\label{fi:resumenTrackingScan}
\end{figure}

The tracking and background data is then introduced in a likelihood method that allows to compute the compatibility of the tracking measured data with the expected background level, the signal is introduced in the likelihood for different values of the axion-photon coupling $g_{a\gamma}$ that allows to obtain $\chi^2(g_{a\gamma})$. In the absence of signal the minimum of $\chi^2$ should be compatible with zero, and an upper limit on $g_{a\gamma}$ can be obtained integrating to a certain confidence level, usually fixed in 95\% C.L. Chapter~\ref{chap:gLimitHe3Final} describes the binned likelihood method used in the previous phases of the experiment to obtain a limit, together with the Micromegas data that lead to a limit in the $^4$He data taking period. The likelihood method in the new $^3$He phase had to be re-adapted in order to take into account the changes in density during tracking movement, due to window temperature changes and gravitational effects on the gas distribution along the magnet bore during tracking. The ideal \emph{two} step coverage per tracking was re-calculated in order to obtain a more realistic axion mass coverage (see Fig.~\ref{fi:resumenTrackingScan}), and the new unbinned likelihood presented allows to take into account the conditions of the bore density at the time of detection of each tracking count.

\begin{figure}[!ht]
{\centering \resizebox{1.0\textwidth}{!} {\includegraphics{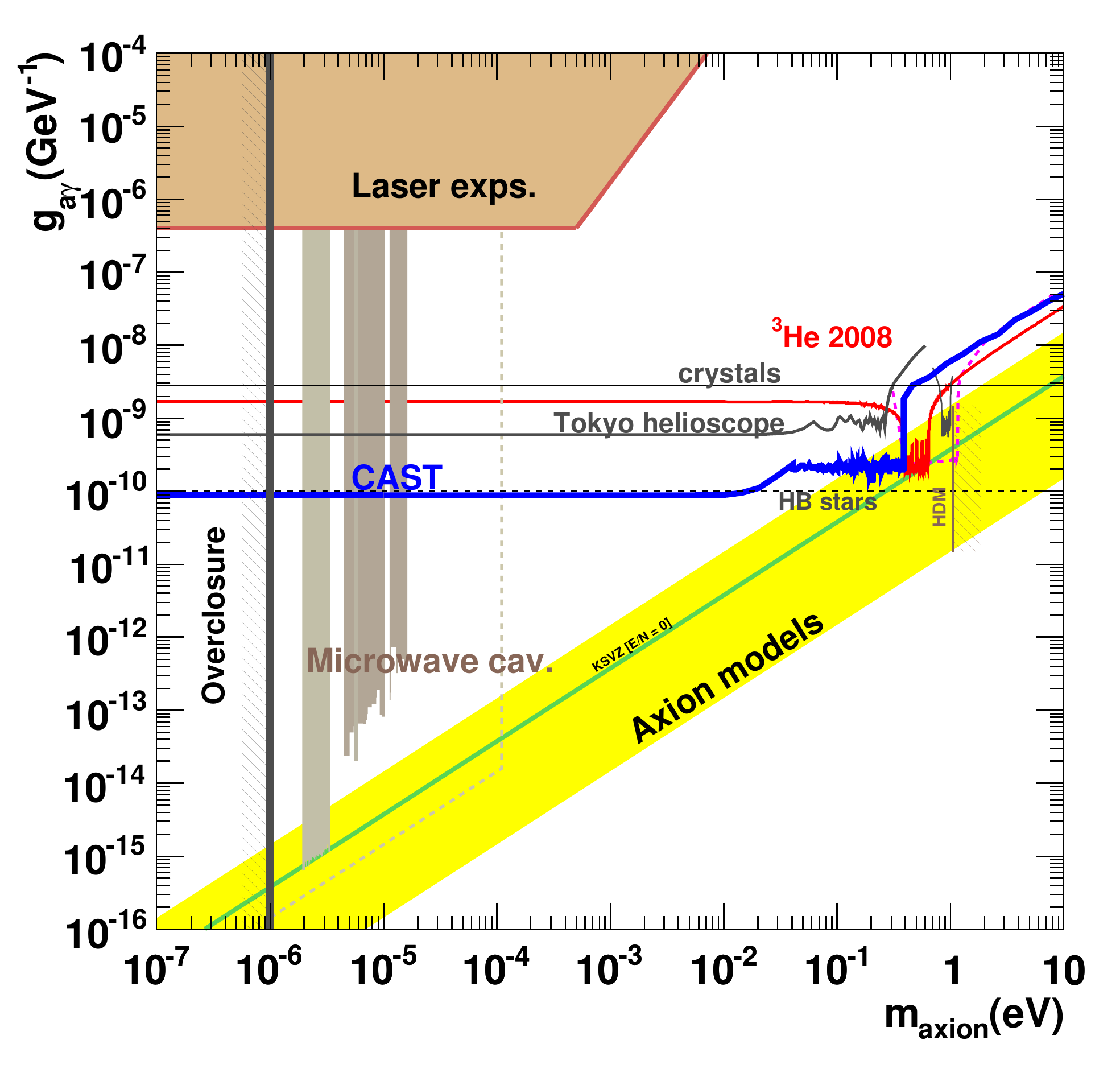}} \par}
\caption{\fontfamily{ptm}\selectfont{\normalsize{ CAST axion-photon coupling limit obtained in 2008 by combinning \emph{three} Micromegas detection lines (red), together with the previous CAST published results (blue). The sensitivity reached in this new axion mass coverage allows to exclude a significant region favoured by different axion models, represented with a yellow band. In addition, the CAST experiment crossed the standard KSVZ axions model line by the end of 2008 data run. The most relevant astrophysical limits and cosmological limits are also shown, HB stars from globular clusters ($g_{a\gamma} < 10^{-10}$\,GeV$^{-1}$), and cold dark matter and hot dark matter overclossures. Laser based searches, microwave cavity searches, and cristal searches are also represented, together with the exclusion of the Tokyo helioscope recently published results. }}}
\label{fi:resumenExclusionFinalBig}
\end{figure}

\vspace{0.1cm}
The tracking and background data from the sunrise Micromegas line presented in this thesis has been combined with the data from the \emph{two} Micromegas detectors running at the sunset side of the experiment, leading to a more restrictive limit on the axion-photon coupling in the new axion mass coverage, given by the first $^3$He data taking period carried out in 2008. The gas density distribution along the magnet bore causes a negative effect on the effective coherence lenght, which in the worst scenario would be reduced to $7.26$\,m. Thus, resulting in a conservative limit of $g_{a\gamma} < 2.7\cdot10^{-10}$\,GeV$^{-1}$ given by the sunrise Micromegas detector, and a combined limit of $g_{a\gamma} < 2.2\cdot10^{-10}$\,GeV$^{-1}$, at the mass regions covered during 2008 data taking, 0.39\,eV $< m_a <$ 0.64\,eV, excluding the uncovered regions due to the effect of the leak. The final result on the coupling for the axion-photon interaction in terms of the axion mass is presented in figure~\ref{fi:resumenExclusionFinalBig}, together with the main limits provided by other axion searches.

\vspace{0.1cm}

The results presented in this thesis conclude that the negative effect of the reduction of the magnet coherence lenght, and the reduced tracking exposure in the new $^3$He phase coverage, by covering \emph{two} settings per tracking, has been counter-balanced by the higher performance introduced by the new Micromegas detectors and detection systems, in terms of lower background level and higher efficiency. Thus placing the detector discovery potential in a level competitive with the X-ray focusing telescope and CCD detection potential. The lower background levels achieved, which placed the detected tracking counts per tracking in few counts per density steps, with a mean rate of about 2 counts per hour, has lead to a combined Micromegas coupling limit at the same level as the combined limit reached in the $^4$He period coverage with \emph{four} detection lines. The future inclusion of the data collected by the CCD detector, and the coverage of the unexplored mass regions which appeared due to the effect of the leak, will push the limit even lower.

\vspace{0.1cm}

In addition, the extended CAST search to higher masses has entered in a region of the $g_{a\gamma}-m_a$ parameter space that has never been explored before, excluding by first time standard KSVZ axions (E/N = 0). Moreover, providing the first experimental limit based in direct axion search excluding standard KSVZ axions for the eV-mass scale. The results presented in this thesis allow to conclude that CAST will exclude standard KSVZ axions for $m_a \gtrsim 0.6$\,eV, when the current CAST physics program reaches its end measuring axion masses close to $1$\,eV, in case no signal is found.

\vspace{0.1cm}

Figure~\ref{fi:resumenExclusionFinalBig} presents the coupling limit obtained for the axion-photon interaction as a final result of the work of this thesis, taking into account \emph{three} Micromegas detection lines with a coherence lenght of $7.26$\,m. The result is presented with the previous published work of the CAST collaboration during the vacuum and $^4$He data taking phases of the experiment, together with the astrophysical and cosmplogical bounds and the main axion searches limits.

\appendix

\chapter{Data summary}
\label{chap:appendix1}

\section{Leak measurements data.}

\begin{table}[!h]
\begin{center}
\begin{tabular}{c|cc|c|cc|c}
        &  \multicolumn{3}{|c|}{Initial} &  \multicolumn{3}{|c}{Final} \\
\hline
Step    &	MV10 [mbar]	&	$\sigma_{MV10}$	&  Bath [C] &	MV10 [mbar]	&   $\sigma_{MV10}$	&  Bath [C]	\\
\hline
\hline
17	&       928.3166	&       0.0003	&       36.12	&       122.3924	&       0.0018	&       36.12	\\
34	&	915.3648	&	0.0006	&	36.13	&	128.3709	&	0.0051	&	36.13	\\
51	&	927.5090	&	0.0010	&	36.12	&	132.2017	&	0.0036	&	36.13	\\
68	&	927.6465	&	0.0023	&	36.14	&	133.7188	&	0.0005	&	36.14	\\
85	&	927.5943	&	0.0131	&	36.14	&	135.5431	&	0.0072	&	36.14	\\
102	&	927.6301	&	0.0045	&	36.15	&	138.0738	&	7.0472	&	36.15	\\
119	&	927.6184	&	0.0045	&	36.16	&	135.5784	&	0.0017	&	36.18	\\
136	&	927.9372	&	0.0016	&	36.12	&	123.4737	&	0.0033	&	36.11	\\
153	&	927.7652	&	0.0055	&	36.12	&	116.4110	&	0.0007	&	36.12	\\
170	&	927.7025	&	0.0067	&	36.09	&	110.3718	&	0.0019	&	36.08	\\
187	&	927.7338	&	0.0110	&	36.09	&	104.2979	&	0.0034	&	36.09	\\
203	&	927.6578	&	0.0056	&	36.11	&	149.5578	&	0.0039	&	36.11	\\
219	&	927.7232	&	0.0040	&	36.12	&	145.2694	&	0.0026	&	36.11	\\
235	&	927.7958	&	0.0068	&	36.12	&	139.9662	&	0.0010	&	36.12	\\
251	&	927.7656	&	0.0023	&	36.11	&	136.1251	&	0.0010	&	36.11	\\
267	&	927.7280	&	0.0088	&	36.09	&	131.6583	&	0.0020	&	36.09	\\
282	&	927.7928	&	0.0030	&	36.09	&	107.0945	&	0.0012	&	36.10	\\
295	&	927.7795	&	0.0036	&	36.11	&	145.3049	&	0.0026	&	36.11	\\
308	&	927.8850	&	0.0011	&	36.11	&	142.2472	&	0.0062	&	36.11	\\
321	&	927.8342	&	0.0070	&	36.10	&	139.8158	&	0.0069	&	36.08	\\
334	&	927.7716	&	0.0018	&	36.10	&	137.2961	&	0.0049	&	36.09	\\
347	&	927.8755	&	0.0056	&	36.12	&	134.2811	&	0.0056	&	36.12	\\
360	&	928.0057	&	0.0012	&	36.12	&	131.9229	&	0.0011	&	36.12	\\
373	&	928.4803	&	0.0053	&	36.26	&	129.3472	&	0.0081	&	36.25	\\
386	&	927.9851	&	0.0002	&	36.26	&	126.8072	&	0.0061	&	36.26	\\

\end{tabular}
\caption{Table for the pondered values of the metering volume for each pressure filling during the $^3He$ tests for a constant heating power of 7.2~W. The metering pressure is pondered with at least 2-3 minutes of data when the pressure was already stabilized.}
\label{ta:MV10_2}
\end{center}
\end{table}

\begin{table}[!h]
\begin{center}
\begin{tabular}{c|cc|c|cc|c}
        &  \multicolumn{3}{|c|}{Initial} &  \multicolumn{3}{|c}{Final} \\
\hline
Step    &	MV10 [mbar]	&	$\sigma_{MV10}$	&  Bath [C] &	MV10 [mbar]	&   $\sigma_{MV10}$	&  Bath [C]	\\
\hline
\hline
17	&       915.3651	&       0.0112	&       36.00	&       122.6870 	&       0.0085	&       36.04	\\
34	&	923.8489	&	0.0029	&	36.02	&	127.8354	&	0.0054	&	36.01	\\
51 	&	927.2777	&	0.0039	&	36.02	&	131.5976	&	0.0020	&	36.02	\\
68 	&	927.3664	&	0.0069	&	36.03	&	133.8112	&	0.0023	&	36.03	\\
85 	&	927.5127   	&	0.0109	&	36.09	&	136.2579	&	0.0044	&	36.07	\\
102 	&	927.3702	&	0.0006	&	36.10	&	137.9163	&	0.0002	&	36.10	\\
119 	&	927.5923	&	0.0011	&	36.10	&	135.7231	&	0.0066	&	36.10	\\
136 	&	927.6290	&	0.0059	&	36.11	&	123.1772	&	0.0008	&	36.10	\\
153 	&	927.6573	&	0.0191	&	36.09	&	116.4418   	&	0.0008	&	36.09	\\
170 	&	927.6390	&	0.0079	&	36.08	&	110.2135	&	0.0021	&	36.08	\\
187 	&	927.6842	&	0.0004	&	36.07	&	104.1225	&	0.0039	&	36.06	\\
203 	&	927.7356  	&	0.0006	&	36.10	&	149.0531	&	0.0023	&	36.08	\\
219 	&	927.8814  	&	0.0094	&	36.11	&	100.2998	&	0.0006	&	36.10	\\
235 	&	927.8283	&	0.0015	&	36.10	&	184.4810	&	0.0027	&	36.10	\\
251 	&	927.6283	&	0.0125	&	36.00	&	135.8333	&	0.0120	&	36.01	\\
267 	&	927.7201	&	0.0112	&	36.02	&	131.3777   	&	0.0022	&	36.01	\\
282 	&	927.8989	&	0.0059	&	36.03	&	106.9358   	&	0.0011	&	36.04	\\
295 	&	927.8325	&	0.0006	&	36.05	&	145.7100	&	0.0075	&	36.05	\\
308 	&	928.0121	&	0.0040	&	36.10	&	142.0492	&	0.0018	&	36.09	\\
321 	&	927.8351	&	0.0074	&	36.08	&	139.3555	&	7.2002	&	36.07	\\
334 	&	927.7627	&	0.0124	&	36.01	&	136.6809	&	0.0071	&	36.01	\\
347 	&	928.1564	&	0.0043	&	36.03	&	133.9579	&	0.0047	&	36.02	\\
360 	&	927.9154	&	0.0022	&	36.02	&	131.4598   	&	0.0053	&	36.01	\\
373 	&	928.2564 	&	0.0003	&	36.10	&	129.2943	&	0.0034	&	36.10	\\
386 	&	928.0397	&	0.0078	&	36.10	&	126.5761	&	0.0029	&	36.10	\\
399 	&	927.7954	&	0.0075	&	36.02	&	124.3798	&	0.0006	&	36.02	\\
412 	&	927.9887	&	0.0078	&	36.02	&	121.4993	&	0.0065	&	36.01	\\
\end{tabular}
\caption{Table for the pondered values of the metering volume for each pressure filling during the $^3He$ tests for a constant heating power of 3.6~W. The metering pressure is pondered with at least 2-3 minutes of data when the pressure was already stabilized.}
\label{ta:MV10_1}
\end{center}
\end{table}

\begin{table}
\begin{center}
\begin{tabular}{c|cc|c|cc|c}
        &  \multicolumn{3}{|c|}{3.6~W Filling series} &  \multicolumn{3}{|c}{7.2~W Filling Series} \\
\hline
Step    &	Moles (1)	&	Moles (2)	&  Pcalc [mbar] &	Moles (1)	&	Moles (2)	&  Pcalc [mbar]		\\ 
\hline
\hline
17	&	0.271283	&	0.271278	&  1.3635321	&	0.275704	&	0.275811	&  1.385751	\\	
34	&	0.543684	&	0.543697	&  2.7326785	&	0.544924	&	0.545144	&  2.738910	\\
51 	&	0.815972	&	0.816003	&  4.1012586	&	0.816997	&	0.817322	&  4.106413	\\
68 	&	1.087524	&	1.087581	&  5.4661396	&	1.088580	&	1.089028	&  5.471450	\\
85 	&	1.358233	&	1.358372	&  6.8267849	&	1.359521	&	1.360091	&  6.833259	\\
102 	&       1.628320	&	1.628547	&  8.1843046	&	1.629600	&	1.630301	&  8.190736	\\
119 	&       1.899234	&	1.899549	&  9.5459775	&	1.900523	&	1.901361	&  9.552453	\\
136 	&       2.174442	&	2.174856	&  10.929235	&	2.175726	&	2.176672	&  10.93568	\\
153 	&	2.451984	&	2.452478	&  12.324220	&	2.453288	&	2.454342	&  12.33077	\\
170 	&	2.731659	&	2.732226	&  13.729930	&	2.732920	&	2.734057	&  13.73627	\\
187 	&	3.013441	&	3.014073	&  15.146232	&	3.014643	&   	3.015862	&  15.15227	\\
203 	&	3.279840	&	3.280562	&  16.485213	&	3.280837	&   	3.282151	&  16.49022	\\
219 	&	3.562962	&	3.563785	&  17.908244	&	3.548511	&	3.549930	&  17.83561	\\
235 	&	3.817275	&	3.818181	&  19.186480	&	3.818025	&	3.819549	&  19.19024	\\
251 	&	4.088253	&	4.089157	&  20.548474	&	4.088852	&	4.090472	&  20.55148	\\
267 	&	4.360766	&	4.361689	&  21.918186	&	4.361211	&	4.362911	&  21.92042	\\
282 	&	4.641698	&	4.642647	&  23.330213	&	4.641998	&	4.643778	&  23.33172	\\
295 	&	4.909320	&	4.910313	&  24.675343	&	4.909689	&	4.911565	&  24.67719	\\
308 	&	5.178212	&	5.179293	&  26.026852	&	5.178462	&	5.180433	&  26.02811	\\
321 	&	5.447982	&	5.449134	&  27.382776	&	5.448056	&	5.450117	&  27.38314	\\
334 	&	5.718705	&	5.719866	&  28.743491	&	5.718491	&	5.720641	&  28.74241	\\
347 	&	5.990476	&	5.991665	&  30.109471	&	5.989977	&	5.992233	&  30.10696	\\
360 	&	6.263028	&	6.264236	&  31.479378	&	6.262315	&	6.264676	&  31.47579	\\
373 	&	6.536368	&	6.537664	&  32.853248	&	6.535571	&	6.538164	&  32.84924	\\
386 	&	6.810564	&	6.811949	&  34.231419	&	6.809528	&	6.812351	&  34.22621	\\
399 	&	7.085499	&	7.086902	&  35.613304		\\
412 	&	7.361485	&	7.362907	&  37.000469		\\

\end{tabular}
\caption{Table for the calculated values of number of moles introduced in the system for each pressure step during the $^3He$ filling tests with heating powers at 3.6~W and 7.2~W. The calculation (1) does reference to the calculated number of moles correcting with the temperature of the metering bath. The calculation (2) it is obtained using a constant thermal bath temperature of $36\,^{\circ}\mathrm{C}$.  }
\label{ta:moles}
\end{center}
\end{table}

\chapter{Detailed calculations.}
\label{chap:appendix2}

\section{Cold bore density as a function of windows temperatures.}\label{app:Pcheck}

The gas density inside the inner part of the magnet cold bore it is affected by the temperature of the windows that contain the buffer gas. The heat transfer coming from the external parts of the magnet disturbs the temperature distribution of the gas.

\vspace{0.2cm}

This appendix describes an analytical approach of the behavior of the gas in stationary conditions. The goal is to obtain a relation that describes the density along the magnet as a function of the windows temperature and the total number of moles introduced in the system.

\vspace{0.2cm}

By definition each point inside the magnet must satisfy the equation of state

\begin{equation}\label{eq:B.GasEquation}
P_m = k \rho(z) T(z)
\end{equation}

\noindent that describes the system pressure without considering the effect of gravity neither \emph{Van der Wals} corrections.

\vspace{0.2cm}

In stationary conditions the temperature distribution will be given by 

\begin{equation}
\nabla ^2 T = 0
\end{equation}

\noindent choosing as boundary conditions a wall at the windows temperature, $T_w$, and supposing the magnet temperature, $T_m$, to be at an infinite distance, one obtains the following approximation to the magnet temperature distribution,

\begin{equation}\label{eq:B.Temperature}
T(z) = T_m + (T_w - T_m) e^{-z / z_o(n) }
\end{equation}

\noindent where $z$ is defined along the axis of the magnet with reference in the cold window wall $T(z=0) = T_w$. The parameter $z_o(n)$ is to be obtained experimentally. It is a parameter that quantifies the effect of windows temperature in terms of penetration length.

\vspace{0.2cm}

In order to relate the total number of moles in the system with the magnet measured parameters we need to integrate to the density distribution along the magnet. Considering the system to be symmetric from the center of the magnet we have the relation

\begin{equation}
n_T/2 = \oint_{V_{cb}} \rho dV_{cb} = \int_0^{L_m/2} \rho(z) A_{cb} dz = \frac{A_{cb}}{k} \int \frac{P_m}{T(z)}
\end{equation}

\vspace{0.2cm}

\noindent and introducing the temperature distribution proposed in relation \ref{eq:B.Temperature} allows us to find an equation that relates the variables measured in the CAST magnet,

\vspace{0.2cm}

\begin{equation}
n_T/2 = \frac{z_o(n) A_{cb} P_m}{k T_m} \left[ \frac{L_m}{2 z_o(n)} - log\left(\frac{T_w}{T_m}\right) \right] 
\end{equation}

\vspace{0.2cm}

\noindent that can be re-formulated to describe the pressure $P_m$ as a function of $n_T$, $T_w$ and $z_o(n)$.

\begin{equation}\label{eq:B.PmVsTw}
P_m = \frac{n_T R T_m} {V_{cb}} \frac{1}{ 1 - \frac{z_o(n)}{L_m/2} log(T_w/T_m) }
\end{equation}

\newpage

\section{Unbinned likelihood equivalence.}\label{app:likelihood}

In this section it is argumented the mathematical equivalence between the binned likelihood method used in $^4He$ data and the new unbinned likelihood method used to analyze the data corresponding to 2008 for the obtention of an upper-limit in $g_{a\gamma}$.

\vspace{0.2cm}

If the density in the cold-bore $\rho_{cb}$ varies too much during a tracking run the likelihood function needs to be modified to adjust to the fact that the axion resonance is not the same during such tracking but it is shifting in mass coverage. A possible solution is to define a new likelihood function that describes a short period of time where the density in the cold bore could be considered constant.

\vspace{0.2cm}

We must choose $\Delta t_k$, from equation \ref{eq:expected}, to be this short period of time, instead of referring to the full tracking length as it was done in $^4He$ analysis. Moreover, by choosing $\Delta t_k$ small enough, we will be able to split the likelihood terms in equation~\ref{eq:Lglobal} that contribute to determine $g_{a\gamma}$ for a specific mass $m_a$ into \emph{two} groups, the ones where no tracking count was measured, and the ones where a single tracking count was detected,

\begin{equation}\label{eq:newL}
L_{m_a}(g_{a\gamma}) = \prod_{k} L_k(n_i = 0) \prod_{k} L_k(n_i = 1)
\end{equation}

\noindent where now $k$ does not make reference to a single pressure step but a short period of time. The product of the likelihoods defined by these periods integrates all the periods that have a significant contribution to the axion mass we want to calculate. Building the likelihood function assuming a poissonian distribution of background and signal, as defined in expression \ref{eq:He4Likelihood}, we can write the $\chi^2$ from the new likelihood \ref{eq:newL} obtained by the relation \ref{eq:Chi2}, and introducing $n_i = 0$ and $n_i = 1$ where corresponds,

\begin{equation}\label{eq:newX}
-\frac{1}{2}\chi^2_{m_a} = log\left( L_{m_a}(g_{a\gamma}) \right) = \sum_{k_{n_i=1}} \left[ - (\mu_{ik} + 1) + log (\mu_{ik}) \right] - \sum_{k_{n_i=0}}  \mu_{ik}
\end{equation}

\vspace{0.2cm}

\noindent where $i$ makes reference to the corresponding energy bin. This expression can be expanded by introducing expression \ref{eq:expected_counts} and \ref{eq:expected}, and written as a function of $\Delta t_k$ and $g_{a\gamma}$,

\begin{multline}
-\frac{1}{2} \chi^2_{m_a} = n_{c} - \sum_{\begin{subarray}{1}k_{n_i}=0\\k_{n_i}=1\end{subarray}} \mu_{ik} + \sum_{k_{n_i=1}} log(\mu_{ik}) = \\
= \cancel{n_{c}} - \sum_{\begin{subarray}{1}k_{n_i}=0\\k_{n_i}=1\end{subarray}} \cancel{b_{ik}} + g_{a\gamma}^4 \Delta t_k \int_{E_o}^{E_f} \frac{dn_{\gamma}}{\Delta t_k \cdot dE} dE\\
 + \sum_{k_{n_i=1}} \left\{ log\left[ \left(\frac{\Delta b_{ik}}{\Delta t_k} + g_{a\gamma}^4 \int_{E_i}^{E_i + \Delta E} \frac{dn_{\gamma}} {\Delta t_k \cdot dE} dE \right)  \right] + \cancel{log(\Delta t_k )} \right\}
\end{multline}

\vspace{0.2cm}

\noindent some terms can be simplified applying the fact that terms that do not depend on $g_{a\gamma}$ since they do not contribute to the obtention of an upper limit on $g_{a\gamma}$\footnote{ $0.95 = \frac{\int_0^{g_{a\gamma}}e^{-\frac{1}{2} \chi^2}}{\int_0^{\infty}  e^{-\frac{1}{2}\chi^2} } =  \frac{\int_0^{g_{a\gamma}}e^{-\frac{1}{2} \chi^2 + C} }{\int_0^{\infty}  e^{-\frac{1}{2}\chi^2} + C} = \frac{\cancel{e^C} \int_0^{g_{a\gamma}}e^{-\frac{1}{2} \chi^2} }{\cancel{e^C}\int_0^{\infty}  e^{-\frac{1}{2}\chi^2}} $ }. To obtain a generalized expression we can take the limit $\Delta t_k \rightarrow 0$, obtaining

\vspace{0.2cm}

\begin{equation}\label{eq:final}
-\frac{1}{2} \chi^2_{m_a} = \underbrace{- g_{a\gamma}^4 \int_E \int_{t_k}  \frac{d^2n_{\gamma}}{dE \cdot dt_k} dE \cdot dt_k}_{Zero\,\,counts\,\,detected\,\,contribution} + \sum_{k_{n_i} = 1} \underbrace{ log \int_{E_i}^{E_i + \Delta E}\left(\frac{ d b_{ik}}{d t_k} + g_{a\gamma}^4 \frac{dn_{\gamma}} { dt_k \cdot dE} \right) dE}_{One\,\,count\,\,detected\,\,contribution}
\end{equation}

\vspace{0.2cm}

\noindent where we find \emph{two} main contribution terms. The first term gives account of the number of counts that should have been detected for a given axion mass $m_a$, integrating to all the cold-bore densities that might have a considerable contribution. In the second term, we find the contribution for each tracking count detected, where background level and the expected signal at the time of detection for the energy bin $i$ is taken into account. 

\vspace{0.2cm}

The expression~\ref{eq:final} can be expressed in the short form given by relation~\ref{eq:newLikelihood}.

\chapter{Ultra low background periods}
\label{chap:appendix5}

During the data taking phase in 2008, an ultra-low background reduction was achieved for relative short periods of time. It must be noticed that these low background periods were observed in both, bulk and microbulk detectors. During these periods the background level reached values below $5\cdot10^{-7}$\,cm$^{-2}$s$^{-1}$keV$^{-1}$.

\vspace{0.2cm}

In order to study the nature of the background reduction achieved, the discrimination analysis was performed with different combinations of observables. It was concluded that all the combinations which were leading to major background reduction included the \emph{pulse width} observable. When this parameter is not introduced in the analysis the lowest background achieved is the nominal background level of about $\sim 1\cdot10^{-5}$\,cm$^{-2}$s$^{-1}$keV$^{-1}$. Therefore, for the data sets analyzed the reduction can be only attributed to the \emph{pulse width} parameter.

\vspace{0.2cm}

\begin{figure}[!ht]
{\centering \resizebox{0.98\textwidth}{!} {\includegraphics[angle=270]{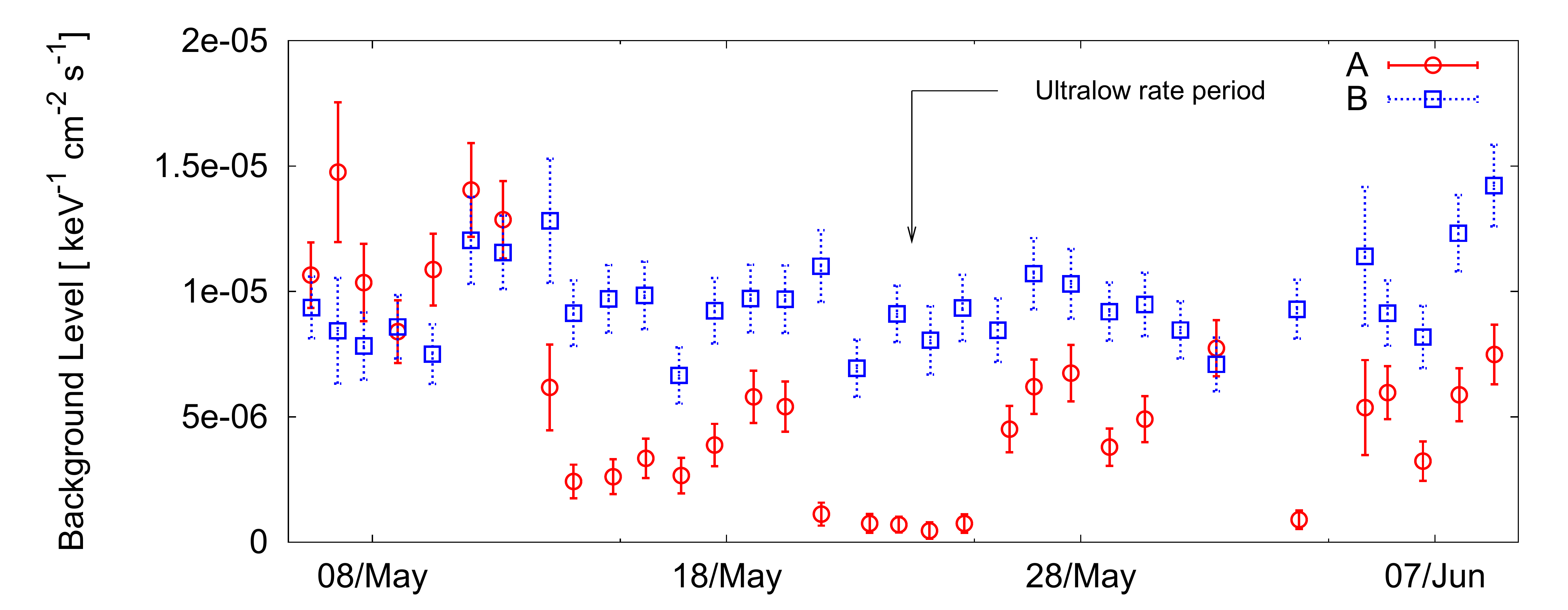}} \par}
\caption{\fontfamily{ptm}\selectfont{\normalsize{ Background rate evolution for a period when an ultra-low background level was achieved. The background level is presented for \emph{two} different discrimination analysis; the combination A is analyzed taking into account the \emph{pulse width} parameter (red circles), and the combination B is obtained by using the same observables combination but excluding the \emph{pulse risetime} parameter.   }}}
\label{fi:lowRateEvolution}
\end{figure}

In this appendix is presented the analysis of one of those periods, which has been performed for two different observables combinations; a combination (A) leading to low background rejection, and combination (B) leading to the nominal background rejection. The period analyzed includes background days when the background level achieved is the same for both combinations (A and B) and periods when ultra-low background levels are observed (see Fig.~\ref{fi:lowRateEvolution}).

\begin{figure}[!ht]
{\centering \resizebox{0.68\textwidth}{!} {\includegraphics[angle=270]{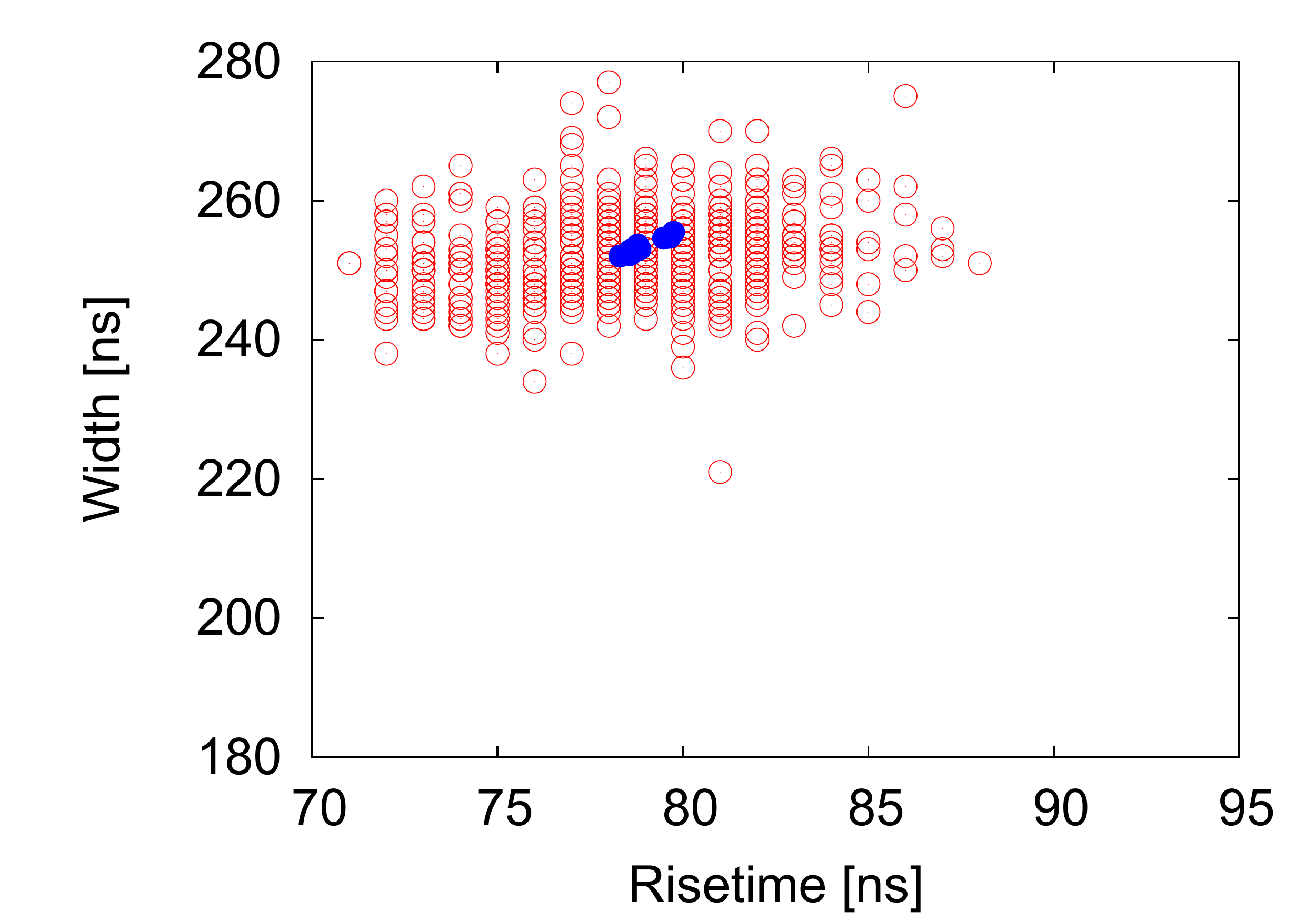}} \par} 
\caption{\fontfamily{ptm}\selectfont{\normalsize{ Pulse risetime versus pulse width scatter plot for days when combination A and combination B background levels were comparable. Background events (empty red circles) and calibration mean values (filled blue circles) are shown.   }}}
\label{fi:rtVsWdNormalBackground}
\end{figure}

\begin{figure}[!ht]
{\centering \resizebox{0.68\textwidth}{!} {\includegraphics[angle=270]{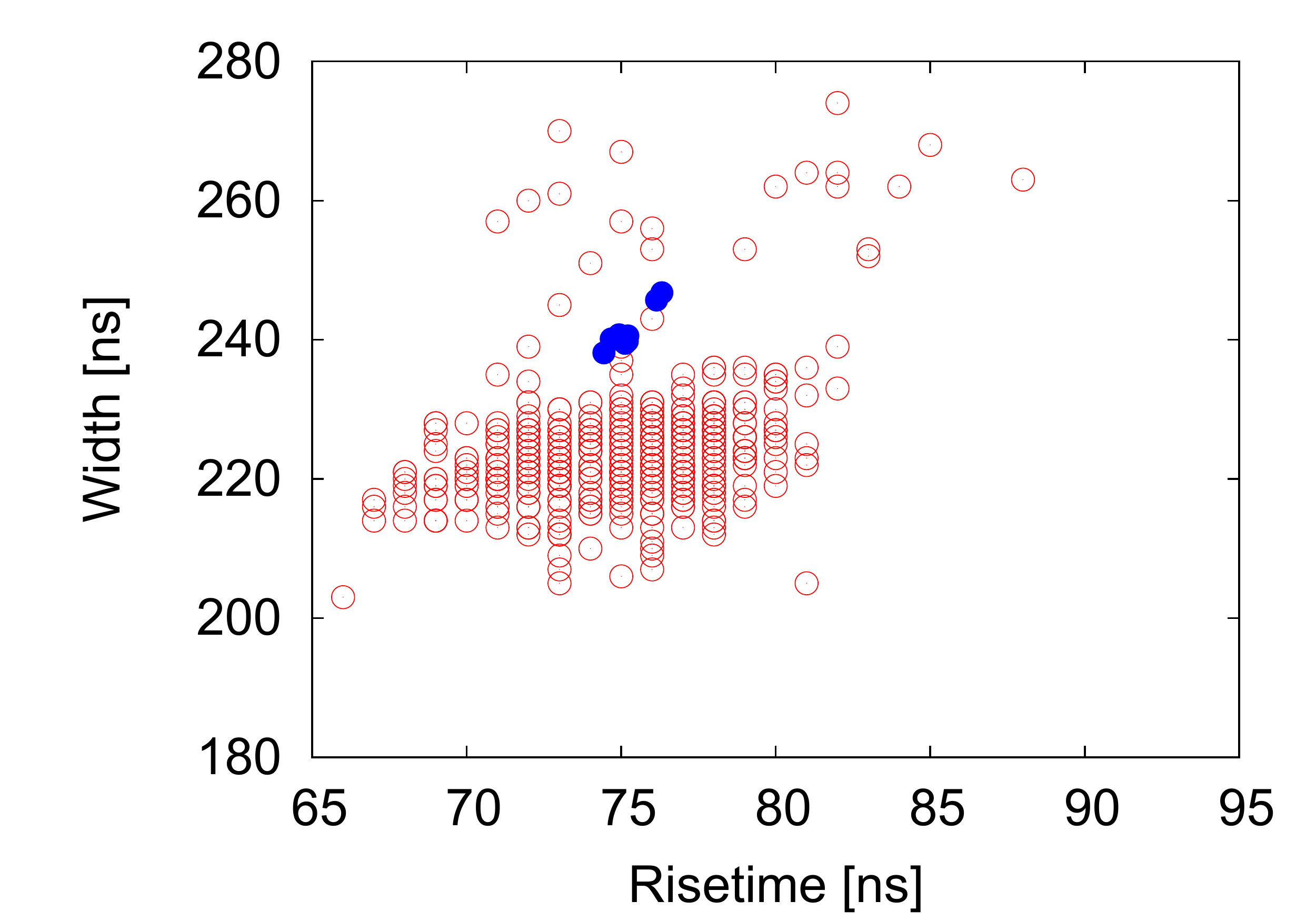}} \par} 
\caption{\fontfamily{ptm}\selectfont{\normalsize{ Pulse risetime versus pulse width scatter plot for days when combination A presented ultra-low background level. Background events (empty red circles) and calibration mean values (filled blue circles) are shown. }}}
\label{fi:rtVsWdLowBackground}
\end{figure}

The data set presented in figure~\ref{fi:lowRateEvolution} has been used to compare the pulse parameters distribution for periods when combination A and combination B had a comparable background level (see Fig.~\ref{fi:rtVsWdNormalBackground}), and when combination A was leading to ultra-low background (see Fig.~\ref{fi:rtVsWdLowBackground}). From these plots is inferred the origin from background reduction, which is clearly due to an offset in the pulse width parameter respect to the mean values obtained during a calibration run. Wether this offset is related with a real physical process or due to systematics in the pulse acquisition is uncertain. 

It must be noticed that the mean pulse width during calibrations changed slightly (about $\sim 10$\,ns) during the days when ultra-low background was achieved, which leads to suspect of changes in the running conditions of the detector during that days, in spite of, the main parameters of the detector remained stable, detector pressure, gas flow, and no bottle change was done during that days. Therefore, if a systematic effect was affecting the data it had to be related with a source which was not monitored, i.e. detector gas quality or humidity. However, it is also difficult to understand the nature of a systematic effect that it is affecting to background events with a different weight than to calibration events, in such way that the background events cloud is biased from the calibration mean values, moreover considering that daily calibration runs during such days were taken at different times during the day.

\vspace{0.2cm}

From a personal point of view, I am inclined to believe that if the background reduction achieved during that days is related with a physical process it must be due to a population of events that produces a pattern similar X-ray events, and that under given circumstances (or conditions) these events are shifted away from the mean X-ray typical values. In spite of this reasoning throws away the idea of background reduction due to natural background reduction, that could be probably coming from changes in the radon abundance around the detector. A possible candidate for this shifting could be related with sparking processes. Based on this, a possible explanation could be given by assuming that a dominant background population it is coming from sparks generated at the anode (drift window or even mesh plane). The sparking phenomena was deeply studied in ref.~\cite{Ectons01}, where ectons, a physical phenomena which involves electron explosion which magnitude is related with adsorption of oxygen or other molecules at the surface. The electron shower produced by ectons could easily generate a similar pattern to X-rays, given its localized production origin, and the shower properties could be related with the abundance of oxygen inside the chamber related with the humidity surrounding the detector.

\vspace{0.2cm}

In spite of any hypothetical explanation, the answer to the observed low background decrease is still a mystery. Further tests and measurements are mandatory to proof the final background level that Micromegas detectors are able to reach. Measurements are undergoing at the Underground Canfranc Laboratory (LSC) which include a set-up where the main parameters of the detector are controlled, such as pressure, flow, temperature, humidity and radon abundance surrounding the detector. Measurements at different conditions are going on, including a shielding of $20$\,cm lead surrounding the detector. Up to now, the lowest background level reached in these measurements place the detector background level at the order of $\sim 10^{-6}$\,cm$^{-1}$keV$^{-1}$s$^{-1}$~\cite{PacoCanfranc}. In addition to the experimental background measurements, detailed Geant4 simulations allow us to understand the nature of the final background of the detector.

\chapter{$^4He$ Phase data.}
\label{chap:appendix3}

\section{$4He$ Tracking and Background Rates}
\begin{table}[h]
\begin{center}
\caption{Counts, times and rates for background and tracking in each pressure setting. }

\begin{tabular}{c|ccc|ccc}
\multicolumn{1}{c} {} & \multicolumn{3}{c} { Tracking } & \multicolumn{3}{c} { Background } \\
\hline
Pressure & Time & Counts & Rate & Time & Counts & Rate  \\
\small{$[mbar]$} & \small{$[hours]$} &  & \small{$[keV^{-1} s^{-1} cm^{-2}]$} & \small{$[hours]$} &  & \small{$[keV^{-1} s^{-1} cm^{-2}]$} \\
\hline \hline
0.0799 &      1.71 &   29 & 4.62e-05 &      10.56 &  168 & 4.35e-5 \\ 
0.1616 &      3.44 &   52 & 4.14e-05 &      21.10 &  390 & 5.05e-5 \\ 
0.2444 &      1.71 &   27 & 4.30e-05 &      31.67 &  553 & 4.77e-5 \\ 
0.3313 &      1.73 &   18 & 2.84e-05 &      31.64 &  561 & 4.85e-5 \\ 
0.4151 &      1.73 &   29 & 4.57e-05 &      31.75 &  532 & 4.58e-5 \\ 
0.4969 &      1.74 &   33 & 5.19e-05 &      31.86 &  625 & 5.36e-5 \\ 
0.5798 &      1.75 &   37 & 5.77e-05 &      21.32 &  449 & 5.75e-5 \\ 
0.6627 &      1.75 &   28 & 4.37e-05 &      21.42 &  441 & 5.63e-5 \\ 
0.7455 &      1.75 &   25 & 3.90e-05 &      21.82 &  384 & 4.81e-5 \\ 
0.8284 &      1.42 &   19 & 3.67e-05 &      43.34 &  861 & 5.43e-5 \\ 
0.9117 &      3.53 &   80 & 6.19e-05 &      43.41 &  905 & 5.70e-5 \\ 
0.9963 &      1.77 &   25 & 3.87e-05 &      43.24 &  889 & 5.62e-5 \\ 
1.0809 &      1.58 &   25 & 4.31e-05 &      32.43 &  589 & 4.96e-5 \\ 
1.1655 &      1.62 &   28 & 4.73e-05 &      32.38 &  508 & 4.29e-5 \\ 
1.2488 &      1.61 &   27 & 4.57e-05 &      32.29 &  488 & 4.13e-5 \\ 
1.3320 &      1.62 &   20 & 3.38e-05 &      21.59 &  311 & 3.94e-5 \\ 
1.4153 &      1.50 &   22 & 4.02e-05 &      21.76 &  350 & 4.40e-5 \\ 
1.4987 &      1.57 &   20 & 3.49e-05 &      21.82 &  374 & 4.68e-5 \\ 
1.5818 &      1.50 &   23 & 4.19e-05 &      32.08 &  530 & 4.51e-5 \\ 
1.6651 &      1.69 &   24 & 3.89e-05 &      31.45 &  551 & 4.79e-5 \\ 
1.7484 &      1.67 &   32 & 5.24e-05 &      20.57 &  364 & 4.84e-5 \\ 
1.8316 &      1.65 &   31 & 5.13e-05 &      20.54 &  466 & 6.20e-5 \\ 
1.9151 &      1.65 &   21 & 3.48e-05 &      20.39 &  445 & 5.96e-5 \\ 
1.9981 &      1.63 &   36 & 6.02e-05 &      30.68 &  639 & 5.69e-5 \\ 
2.0813 &      1.65 &   29 & 4.80e-05 &      30.56 &  547 & 4.89e-5 \\ 
\end{tabular}

\end{center}
\end{table}

\begin{center}
\begin{tabular}{c|ccc|ccc}
\multicolumn{1}{c} {} & \multicolumn{3}{c} { Tracking } & \multicolumn{3}{c} { Background } \\
\hline
Pressure & Time & Counts & Rate & Time & Counts & Rate  \\
\small{$[mbar]$} & \small{$[hours]$} &  & \small{$[keV^{-1} s^{-1}  cm^{-2}]$} & \small{$[hours]$} &  & \small{$[keV^{-1} s^{-1} cm^{-2}]$} \\
\hline \hline
2.1646 &      1.63 &   31 & 5.19e-05 &      30.45 &  558 & 5.01e-5 \\ 
2.2479 &      1.62 &   28 & 4.73e-05 &      40.25 &  718 & 4.87e-5 \\ 
2.3311 &      3.22 &   64 & 5.44e-05 &      35.45 &  629 & 4.85e-5 \\ 
2.4167 &      1.56 &   31 & 5.42e-05 &      25.40 &  431 & 4.64e-5 \\ 
2.4994 &      1.55 &   17 & 2.99e-05 &      10.53 &  159 & 4.13e-5 \\ 
2.5822 &      1.57 &   22 & 3.82e-05 &       5.22 &   82 & 4.29e-5 \\ 
2.6648 &      1.57 &   26 & 4.53e-05 &      34.39 &  636 & 5.05e-5 \\ 
2.7556 &      1.57 &   24 & 4.19e-05 &      39.10 &  735 & 5.14e-5 \\ 
2.8353 &      1.58 &   33 & 5.70e-05 &      43.72 &  804 & 5.02e-5 \\ 
2.9177 &      1.58 &   30 & 5.19e-05 &      48.22 &  886 & 5.02e-5 \\ 
3.0000 &      1.60 &   31 & 5.29e-05 &      56.34 & 1035 & 5.02e-5 \\ 
3.0824 &      1.60 &   30 & 5.12e-05 &      60.26 & 1101 & 4.99e-5 \\ 
3.1647 &      1.60 &   28 & 4.78e-05 &      75.64 & 1387 & 5.01e-5 \\ 
3.2476 &      1.62 &   26 & 4.39e-05 &     119.20 & 2193 & 5.03e-5 \\ 
3.3306 &      1.62 &   32 & 5.41e-05 &     114.20 & 2099 & 5.02e-5 \\ 
3.4138 &      4.73 &   74 & 4.27e-05 &     132.26 & 2423 & 5.01e-5 \\ 
3.4968 &      3.37 &   62 & 5.03e-05 &     135.09 & 2459 & 4.97e-5 \\ 
3.5799 &      3.48 &   76 & 5.97e-05 &     141.43 & 2604 & 5.03e-5 \\ 
3.6630 &      8.81 &  154 & 4.78e-05 &     143.81 & 2615 & 4.97e-5 \\ 
3.7463 &     27.96 &  509 & 4.97e-05 &     146.72 & 2700 & 5.03e-5 \\ 
3.7883 &      1.75 &   25 & 3.90e-05 &     142.01 & 2601 & 5.00e-5 \\ 
3.8290 &     10.40 &  162 & 4.26e-05 &     141.01 & 2598 & 5.03e-5 \\ 
3.9147 &      3.48 &   58 & 4.55e-05 &     136.52 & 2516 & 5.04e-5 \\ 
3.9978 &      5.22 &   96 & 5.02e-05 &     132.58 & 2445 & 5.04e-5 \\ 
4.0810 &      3.47 &   71 & 5.59e-05 &     132.83 & 2464 & 5.07e-5 \\ 
4.1641 &      3.46 &   64 & 5.05e-05 &     121.68 & 2244 & 5.04e-5 \\ 
4.2448 &      1.70 &   31 & 4.99e-05 &      77.27 & 1442 & 5.10e-5 \\ 
4.3280 &      1.70 &   30 & 4.82e-05 &      81.73 & 1532 & 5.12e-5 \\ 
4.4111 &      1.49 &   23 & 4.21e-05 &      58.63 & 1127 & 5.25e-5 \\ 
4.4943 &      1.68 &   25 & 4.06e-05 &      60.45 & 1130 & 5.11e-5 \\ 
4.5775 &      1.68 &   25 & 4.05e-05 &      58.62 & 1062 & 4.95e-5 \\ 
4.6606 &      1.67 &   37 & 6.07e-05 &      56.32 & 1025 & 4.97e-5 \\ 
4.7438 &      1.65 &   36 & 5.98e-05 &      53.68 &  945 & 4.81e-5 \\ 
4.8269 &      1.66 &   25 & 4.13e-05 &      58.67 & 1021 & 4.76e-5 \\ 
4.9101 &      1.43 &   29 & 5.53e-05 &      60.05 & 1053 & 4.79e-5 \\ 
4.9932 &      3.23 &   52 & 4.40e-05 &      65.10 & 1138 & 4.78e-5 \\ 
5.0763 &      3.20 &   56 & 4.78e-05 &      66.06 & 1156 & 4.78e-5 \\ 
5.1594 &      1.60 &   31 & 5.28e-05 &      66.97 & 1151 & 4.70e-5 \\ 
5.2426 &      1.58 &   25 & 4.31e-05 &      68.11 & 1155 & 4.63e-5 \\ 
\end{tabular}
\newpage
\begin{tabular}{c|ccc|ccc}
\multicolumn{1}{c} {} & \multicolumn{3}{c} { Tracking } & \multicolumn{3}{c} { Background } \\
\hline
Pressure & Time & Counts & Rate & Time & Counts & Rate  \\
\small{$[mbar]$} & \small{$[hours]$} &  & \small{$[keV^{-1} s^{-1} cm^{-2}]$} & \small{$[hours]$} &  & \small{$[keV^{-1} s^{-1}  cm^{-2}]$} \\
\hline \hline
5.3257 &      3.17 &   62 & 5.35e-05 &      69.21 & 1147 & 4.53e-5 \\ 
5.4088 &      1.57 &   27 & 4.70e-05 &      70.10 & 1153 & 4.49e-5 \\ 
5.4919 &      1.57 &   23 & 4.01e-05 &      75.60 & 1253 & 4.53e-5 \\ 
5.5752 &      1.57 &   32 & 5.58e-05 &      71.73 & 1203 & 4.58e-5 \\ 
5.6586 &      1.57 &   27 & 4.71e-05 &      67.97 & 1124 & 4.52e-5 \\ 
5.7419 &      1.55 &   21 & 3.70e-05 &      68.85 & 1155 & 4.58e-5 \\ 
5.8252 &      1.55 &   27 & 4.76e-05 &      69.55 & 1176 & 4.62e-5 \\ 
5.9085 &      1.55 &   31 & 5.47e-05 &      70.30 & 1209 & 4.70e-5 \\ 
5.9918 &      1.53 &   24 & 4.27e-05 &      71.21 & 1206 & 4.63e-5 \\ 
6.0751 &      1.53 &   21 & 3.74e-05 &      71.95 & 1211 & 4.60e-5 \\ 
6.1585 &      1.53 &   23 & 4.10e-05 &      72.90 & 1236 & 4.63e-5 \\ 
6.2418 &      1.52 &   22 & 3.96e-05 &      73.98 & 1283 & 4.74e-5 \\ 
6.3253 &      1.52 &   15 & 2.70e-05 &      74.50 & 1298 & 4.76e-5 \\ 
6.4087 &      1.52 &   24 & 4.32e-05 &      75.34 & 1309 & 4.75e-5 \\ 
6.5752 &      1.50 &   29 & 5.30e-05 &      76.28 & 1299 & 4.65e-5 \\ 
6.6585 &      1.48 &   22 & 4.06e-05 &      77.20 & 1293 & 4.58e-5 \\ 
6.7418 &      1.48 &   30 & 5.54e-05 &      77.99 & 1320 & 4.63e-5 \\ 
6.8251 &      1.50 &   31 & 5.65e-05 &      78.64 & 1355 & 4.71e-5 \\ 
6.9084 &      1.48 &   20 & 3.69e-05 &      79.43 & 1386 & 4.77e-5 \\ 
6.9916 &      1.48 &   22 & 4.06e-05 &      80.18 & 1388 & 4.73e-5 \\ 
7.0750 &      1.48 &   33 & 6.08e-05 &      81.06 & 1386 & 4.67e-5 \\ 
7.1583 &      1.49 &   19 & 3.50e-05 &      81.65 & 1395 & 4.67e-5 \\ 
7.2415 &      1.46 &   29 & 5.42e-05 &      82.89 & 1417 & 4.67e-5 \\ 
7.3248 &      1.47 &   24 & 4.46e-05 &      91.03 & 1560 & 4.68e-5 \\ 
7.4081 &      1.47 &   28 & 5.22e-05 &      91.98 & 1525 & 4.53e-5 \\ 
7.4913 &      1.47 &   24 & 4.47e-05 &      93.41 & 1554 & 4.55e-5 \\ 
7.5746 &      1.45 &   25 & 4.71e-05 &      94.61 & 1560 & 4.51e-5 \\ 
7.6579 &      2.90 &   48 & 4.52e-05 &      95.66 & 1586 & 4.53e-5 \\ 
7.7411 &      1.45 &   29 & 5.46e-05 &      96.64 & 1632 & 4.61e-5 \\ 
7.8244 &      2.90 &   49 & 4.62e-05 &      97.60 & 1626 & 4.55e-5 \\ 
7.9077 &      1.45 &   23 & 4.34e-05 &      98.75 & 1660 & 4.59e-5 \\ 
7.9910 &      1.45 &   21 & 3.96e-05 &      99.74 & 1657 & 4.54e-5 \\ 
8.0743 &      1.45 &   24 & 4.52e-05 &     100.91 & 1675 & 4.54e-5 \\ 
8.1575 &      1.45 &   23 & 4.33e-05 &     101.95 & 1715 & 4.60e-5 \\ 
8.2408 &      1.45 &   28 & 5.27e-05 &     103.17 & 1750 & 4.64e-5 \\ 
8.3361 &      1.43 &   26 & 4.96e-05 &     112.43 & 1868 & 4.54e-5 \\ 
8.4092 &      1.45 &   21 & 3.96e-05 &     106.08 & 1729 & 4.45e-5 \\ 
8.4923 &      1.45 &   27 & 5.09e-05 &     107.07 & 1779 & 4.54e-5 \\ 
8.5757 &      1.45 &   24 & 4.52e-05 &     107.90 & 1781 & 4.51e-5 \\ 
\end{tabular}
\newpage

\begin{tabular}[h]{c|ccc|ccc}
\multicolumn{1}{c} {} & \multicolumn{3}{c} { Tracking } & \multicolumn{3}{c} { Background } \\
\hline
Pressure & Time & Counts & Rate & Time & Counts & Rate  \\
\small{$[mbar]$} & \small{$[hours]$} &  & \small{$[keV^{-1} s^{-1} cm^{-2}]$} & \small{$[hours]$} &  & \small{$[keV^{-1} s^{-1} cm^{-2}]$} \\
\hline \hline
8.6590 &      1.45 &   28 & 5.28e-05 &     118.37 & 2009 & 4.64e-5 \\ 
8.7421 &      1.45 &   36 & 6.78e-05 &     119.00 & 2024 & 4.65e-5 \\ 
8.8255 &      3.53 &   62 & 4.80e-05 &     119.89 & 2020 & 4.60e-5 \\ 
8.9087 &      1.46 &   23 & 4.29e-05 &     120.86 & 2027 & 4.58e-5 \\ 
8.9920 &      1.45 &   14 & 2.63e-05 &     121.74 & 1998 & 4.48e-5 \\ 
9.0753 &      1.45 &   26 & 4.90e-05 &     122.77 & 2015 & 4.48e-5 \\ 
9.1585 &      2.20 &   35 & 4.35e-05 &     123.75 & 2030 & 4.48e-5 \\ 
9.2418 &      1.47 &   31 & 5.78e-05 &     124.68 & 2030 & 4.45e-5 \\ 
9.3251 &      1.45 &   27 & 5.09e-05 &     134.34 & 2202 & 4.48e-5 \\ 
9.4084 &      1.46 &   28 & 5.23e-05 &     126.79 & 2107 & 4.54e-5 \\ 
9.4916 &      1.47 &   16 & 2.98e-05 &     118.91 & 1982 & 4.55e-5 \\ 
9.5760 &      1.48 &   20 & 3.68e-05 &     119.66 & 1975 & 4.51e-5 \\ 
9.6604 &      1.47 &   21 & 3.92e-05 &     138.49 & 2322 & 4.58e-5 \\ 
9.7449 &      1.48 &   24 & 4.42e-05 &     129.74 & 2131 & 4.49e-5 \\ 
9.8295 &      2.97 &   37 & 3.41e-05 &     131.16 & 2152 & 4.48e-5 \\ 
9.9145 &      1.48 &   20 & 3.69e-05 &     132.38 & 2202 & 4.55e-5 \\ 
9.9987 &      1.48 &   22 & 4.05e-05 &     133.47 & 2227 & 4.56e-5 \\ 
10.0835 &      1.50 &   27 & 4.92e-05 &     134.78 & 2261 & 4.58e-5 \\ 
10.1682 &      4.52 &   87 & 5.26e-05 &     135.84 & 2263 & 4.55e-5 \\ 
10.2530 &      3.07 &   49 & 4.37e-05 &     137.04 & 2286 & 4.56e-5 \\ 
10.3379 &      1.53 &   30 & 5.35e-05 &     138.10 & 2231 & 4.41e-5 \\ 
10.4228 &      1.55 &   16 & 2.82e-05 &     130.42 & 2068 & 4.33e-5 \\ 
10.5077 &      1.55 &   25 & 4.41e-05 &     131.46 & 2102 & 4.37e-5 \\ 
10.5927 &      1.57 &   23 & 4.01e-05 &     141.31 & 2283 & 4.41e-5 \\ 
10.6778 &      1.56 &   27 & 4.72e-05 &     142.29 & 2318 & 4.45e-5 \\ 
10.7630 &      1.57 &   21 & 3.66e-05 &     125.26 & 2019 & 4.40e-5 \\ 
10.8482 &      1.59 &   29 & 4.99e-05 &     127.21 & 2115 & 4.54e-5 \\ 
10.9335 &      1.59 &   24 & 4.14e-05 &     117.82 & 1967 & 4.56e-5 \\ 
11.0187 &      1.59 &   35 & 6.03e-05 &     108.30 & 1781 & 4.49e-5 \\ 
11.1041 &      1.60 &   34 & 5.80e-05 &     109.03 & 1817 & 4.55e-5 \\ 
11.1894 &      1.60 &   22 & 3.76e-05 &     109.12 & 1786 & 4.47e-5 \\ 
11.2748 &      1.62 &   27 & 4.56e-05 &     109.42 & 1781 & 4.45e-5 \\ 
11.3603 &      1.62 &   38 & 6.42e-05 &     109.95 & 1769 & 4.40e-5 \\ 
11.4458 &      1.61 &   23 & 3.89e-05 &     110.62 & 1851 & 4.57e-5 \\ 
11.5314 &      1.62 &   29 & 4.90e-05 &     121.34 & 2064 & 4.65e-5 \\ 
11.6169 &      1.64 &   23 & 3.82e-05 &     122.03 & 2057 & 4.61e-5 \\ 
11.7026 &      1.65 &   18 & 2.99e-05 &     122.62 & 2027 & 4.52e-5 \\ 
11.7883 &      1.65 &   25 & 4.14e-05 &     122.91 & 2010 & 4.47e-5 \\ 
11.8741 &      1.65 &   19 & 3.14e-05 &     123.43 & 1982 & 4.39e-5 \\ 
\end{tabular}

\end{center}

\begin{center}
\begin{tabular}[h]{c|ccc|ccc}
\multicolumn{1}{c} {} & \multicolumn{3}{c} { Tracking } & \multicolumn{3}{c} { Background } \\
\hline
Pressure & Time & Counts & Rate & Time & Counts & Rate  \\
\small{$[mbar]$} & \small{$[hours]$} &  & \small{$[keV^{-1} s^{-1} cm^{-2}]$} & \small{$[hours]$} &  & \small{$[keV^{-1} s^{-1} cm^{-2}]$} \\
\hline \hline
11.9599 &      1.65 &   36 & 5.97e-05 &     122.94 & 1930 & 4.29e-5 \\ 
12.0457 &      3.34 &   57 & 4.67e-05 &     133.51 & 2115 & 4.33e-5 \\ 
12.1316 &      1.68 &   21 & 3.41e-05 &     144.06 & 2275 & 4.32e-5 \\ 
12.2176 &      1.67 &   24 & 3.93e-05 &     144.44 & 2301 & 4.35e-5 \\ 
12.3036 &      1.68 &   23 & 3.74e-05 &     145.28 & 2369 & 4.46e-5 \\ 
12.3896 &      1.70 &   24 & 3.86e-05 &     146.01 & 2375 & 4.44e-5 \\ 
12.4757 &      1.70 &   33 & 5.31e-05 &     146.31 & 2410 & 4.50e-5 \\ 
12.5618 &      1.72 &   29 & 4.62e-05 &     146.71 & 2403 & 4.48e-5 \\ 
12.6480 &      1.72 &   29 & 4.62e-05 &     137.35 & 2220 & 4.42e-5 \\ 
12.7342 &      1.74 &   29 & 4.54e-05 &     126.89 & 2058 & 4.43e-5 \\ 
12.8204 &      1.73 &   32 & 5.04e-05 &     127.20 & 2056 & 4.42e-5 \\ 
12.9067 &      1.75 &   25 & 3.90e-05 &     138.58 & 2246 & 4.43e-5 \\ 
12.9931 &      1.75 &   25 & 3.90e-05 &     128.10 & 2110 & 4.50e-5 \\ 
13.0795 &      1.75 &   20 & 3.13e-05 &     117.57 & 1927 & 4.48e-5 \\ 
13.1659 &      1.75 &   24 & 3.75e-05 &     107.00 & 1742 & 4.45e-5 \\ 
13.2524 &      1.77 &   29 & 4.49e-05 &      96.45 & 1582 & 4.48e-5 \\ 
13.3390 &      1.76 &   26 & 4.03e-05 &      85.90 & 1372 & 4.36e-5 \\ 
13.4256 &      3.52 &   48 & 3.72e-05 &      75.33 & 1178 & 4.27e-5 \\ 
\hline
Total 	  &	336.63& 5682 & 4.61e-5&    14231.01&	241798&	4.64e-5\\
\end{tabular}

\end{center}

\begin{landscape}
\begin{table}[h!]
\caption[h]{Total background and tracking rates as a function of the energy for set A and set B.  }
\begin{center}

\begin{tabular}{c|cc|cc|cc|cc}

\multicolumn{1}{c|} {} & \multicolumn{4}{c|} { Set A} & \multicolumn{4}{c} { Set B } \\
\hline

\multicolumn{1}{c|} {} & \multicolumn{2}{c} { Tracking } & \multicolumn{2}{c} { Background } & \multicolumn{2}{|c} { Tracking } & \multicolumn{2}{c} { Background } \\
\hline

Energy & Counts & Rate & Counts & Rate & Counts & Rate & Counts & Rate \\
\small{$[keV]$} &   & \small{$[keV^{-1} s^{-1} cm^{-2}]$} &  & \small{$[keV^{-1} s^{-1} cm^{-2}]$} &  & \small{$[keV^{-1} s^{-1} cm^{-2}]$} &  & \small{$[keV^{-1} s^{-1} cm^{-2}]$}  \\
\hline \hline
2.25 &	 79 &	5.38e-5 &	506 &	6.80e-5 &	471 &	6.42e-5 &	1898 &	6.87e-5 \\
2.75 &   88 &	6.00e-5 &	500 &	6.72e-5 &	473 &	6.45e-5 &	1799 &	6.51e-5 \\
3.25 &   82 &	5.59e-5 &	462 &	6.21e-5 &	453 &	6.18e-5 &	1677 &	6.07e-5 \\
3.75 &   65 &	4.43e-5 &	308 &	4.14e-5 &	353 &	4.81e-5 &	1249 &	4.52e-5 \\
4.25 &   48 &	3.27e-5 &	274 &	3.68e-5 &	273 &	3.72e-5 &	1087 &	3.94e-5 \\
4.75 &   46 &	3.14e-5 &	221 &	2.97e-5 &	287 &	3.91e-5 &	1001 &	3.62e-5 \\
5.25 &   42 &	2.86e-5 &	269 &	3.61e-5 &	263 &	3.59e-5 &	1073 &	3.89e-5 \\
5.75 &   46 &	3.14e-5 &	251 &	3.37e-5 &	286 &	3.90e-5 &	1103 &	3.99e-5 \\
6.25 &   46 &	3.14e-5 &	217 &	2.92e-5 &	263 &	3.59e-5 &	1089 &	3.94e-5 \\
6.75 &   50 &	3.41e-5 &	195 &	2.62e-5 &	294 &	4.01e-5 &	978  &  3.54e-5 \\
7.25 &   50 &	3.41e-5 &	366 &	4.92e-5 &	255 &	3.48e-5 &	1044 &	3.78e-5 \\
7.75 &   97 &	6.61e-5 &	586 &	7.87e-5 &	355 &	4.84e-5 &	1250 &	4.53e-5 \\
8.25 & 	115 &	7.84e-5 &	623 &	8.37e-5 &	418 &	5.70e-5 &	1337 &	4.84e-5 \\
8.75 & 	 79 &	5.38e-5 &	415 &	5.58e-5 &	305 &	4.16e-5 &	1267 &	4.59e-5 \\
\end{tabular}

\end{center}
\end{table}
\end{landscape}

\chapter{$^3He$ Phase data.}
\label{chap:appendix4}

\section{$3He$ Tracking and Background Rates}
\begin{table}[h]
\begin{center}
\caption{Counts, times and rates for background and tracking in each pressure setting. }

\begin{tabular}{c|ccc|ccc}
\multicolumn{1}{c} {} & \multicolumn{3}{c} { Tracking } & \multicolumn{3}{c} { Background } \\
\hline
Step & Time & Counts & Rate & Time & Counts & Rate  \\
\small{\#} & \small{$[hours]$} &  & \small{$[keV^{-1} s^{-1} cm^{-2}]$} & \small{$[hours]$} &  & \small{$[keV^{-1} s^{-1} cm^{-2}]$} \\
\hline \hline
160 &	0.73 &	2 &	1.05e-5 &	 78.01 &	168 &	8.24e-6 \\
161 &	0.74 &	0 &	0.00e+0 &	 78.01 &	168 &	8.24e-6 \\
163 &	0.53 &	0 &	0.00e+0 &	143.65 &	278 &	7.40e-6 \\
164 &	0.69 &	0 &	0.00e+0 &	143.65 &	278 &	7.40e-6 \\
165 &	0.73 &	2 &	1.05e-5 &	121.16 &	231 &	7.29e-6 \\
166 &	0.67 &	3 &	1.71e-5 &	121.16 &	231 &	7.29e-6 \\
166 &	1.48 &	5 &	1.29e-5 &	102.67 &	198 &	7.38e-6 \\
169 &	0.71 &	3 &	1.62e-5 &	132.11 &	267 &	7.73e-6 \\
169 &	0.79 &	2 &	9.74e-6 &	135.01 &	343 &	9.72e-6 \\
170 &	0.73 &	3 &	1.58e-5 &	132.11 &	267 &	7.73e-6 \\
170 &	0.73 &	4 &	2.09e-5 &	132.11 &	267 &	7.73e-6 \\
170 &	0.75 &	3 &	1.53e-5 &	135.01 &	343 &	9.72e-6 \\
170 &	0.79 &	0 &	0.00e+0 &	156.27 &	381 &	9.33e-6 \\
171 &	0.23 &	0 &	0.00e+0 &	132.11 &	267 &	7.73e-6 \\
171 &	0.38 &	1 &	9.98e-6 &	132.11 &	267 &	7.73e-6 \\
171 &	0.76 &	0 &	0.00e+0 &	156.27 &	381 &	9.33e-6 \\
172 &	0.56 &	1 &	6.86e-6 &	137.27 &	276 &	7.69e-6 \\
174 &	0.75 &	3 &	1.54e-5 &	 74.96 &	154 &	7.86e-6 \\
175 &	0.70 &	2 &	1.09e-5 &	 74.96 &	154 &	7.86e-6 \\
176 &	0.74 &	2 &	1.04e-5 &	 74.96 &	154 &	7.86e-6 \\
177 &	0.72 &	2 &	1.07e-5 &	 74.96 &	154 &	7.86e-6 \\
178 &	0.74 &	2 &	1.03e-5 &	 74.96 &	154 &	7.86e-6 \\
179 &	0.73 &	1 &	5.26e-6 &	 74.96 &	154 &	7.86e-6 \\
180 &	1.52 &	3 &	7.57e-6 &	 62.21 &	134 &	8.24e-6 \\
181 &	0.75 &	3 &	1.53e-5 &	 53.20 &	436 &	3.14e-5 \\
\end{tabular}

\end{center}
\end{table}

\begin{center}
\begin{tabular}{c|ccc|ccc}
\multicolumn{1}{c} {} & \multicolumn{3}{c} { Tracking } & \multicolumn{3}{c} { Background } \\
\hline
Pressure & Time & Counts & Rate & Time & Counts & Rate  \\
\small{$[mbar]$} & \small{$[hours]$} &  & \small{$[keV^{-1} s^{-1}  cm^{-2}]$} & \small{$[hours]$} &  & \small{$[keV^{-1} s^{-1} cm^{-2}]$} \\
\hline \hline
182 &	0.72 &	6 &	3.21e-5 &	 53.20 &	436 &	3.14e-5 \\
183 &	0.74 &	5 &	2.57e-5 &	 53.20 &	436 &	3.14e-5 \\
184 &	0.71 &	4 &	2.17e-5 &	 53.20 &	436 &	3.14e-5 \\
184 &	0.75 &	9 &	4.56e-5 &	 53.20 &	436 &	3.14e-5 \\
185 &	0.71 &	9 &	4.83e-5 &	 53.20 &	436 &	3.14e-5 \\
186 &	0.79 &	1 &	4.86e-6 &	 53.20 &	436 &	3.14e-5 \\
187 &	0.68 &	3 &	1.69e-5 &	 53.20 &	436 &	3.14e-5 \\
188 &	0.82 &	0 &	0.00e+0 &	177.52 &	432 &	9.31e-6 \\
189 &	0.73 &	0 &	0.00e+0 &	177.52 &	432 &	9.31e-6 \\
190 &	0.78 &	3 &	1.48e-5 &	198.79 &	473 &	9.10e-6 \\
191 &	0.77 &	2 &	9.88e-6 &	198.79 &	473 &	9.10e-6 \\
192 &	0.81 &	1 &	4.75e-6 &	219.95 &	516 &	8.98e-6 \\
193 &	0.79 &	2 &	9.64e-6 &	219.95 &	516 &	8.98e-6 \\
194 &	0.79 &	4 &	1.93e-5 &	241.42 &	555 &	8.80e-6 \\
195 &	0.77 &	1 &	4.94e-6 &	241.42 &	555 &	8.80e-6 \\
196 &	0.79 &	4 &	1.93e-5 &	237.89 &	525 &	8.44e-6 \\
197 &	0.78 &	2 &	9.86e-6 &	237.89 &	525 &	8.44e-6 \\
198 &	0.83 &	1 &	4.59e-6 &	252.74 &	546 &	8.27e-6 \\
199 &	0.77 &	2 &	9.99e-6 &	252.74 &	546 &	8.27e-6 \\
200 &	0.80 &	1 &	4.77e-6 &	256.59 &	566 &	8.44e-6 \\
201 &	0.78 &	3 &	1.47e-5 &	256.59 &	566 &	8.44e-6 \\
202 &	0.83 &	1 &	4.62e-6 &	256.57 &	568 &	8.47e-6 \\
203 &	0.79 &	4 &	1.94e-5 &	256.57 &	568 &	8.47e-6 \\
204 &	0.82 &	0 &	0.00e+0 &	235.52 &	504 &	8.19e-6 \\
205 &	0.76 &	2 &	1.00e-5 &	235.52 &	504 &	8.19e-6 \\
206 &	0.81 &	2 &	9.39e-6 &	246.34 &	513 &	7.97e-6 \\
207 &	0.79 &	2 &	9.74e-6 &	246.34 &	513 &	7.97e-6 \\
208 &	0.81 &	2 &	9.46e-6 &	242.58 &	481 &	7.59e-6 \\
209 &	0.81 &	1 &	4.73e-6 &	242.58 &	481 &	7.59e-6 \\
210 &	0.82 &	3 &	1.39e-5 &	277.48 &	550 &	7.58e-6 \\
211 &	0.79 &	2 &	9.65e-6 &	277.48 &	550 &	7.58e-6 \\
212 &	0.84 &	2 &	9.07e-6 &	277.34 &	568 &	7.84e-6 \\
213 &	0.77 &	0 &	0.00e+0 &	277.34 &	568 &	7.84e-6 \\
214 &	0.84 &	4 &	1.81e-5 &	277.23 &	564 &	7.78e-6 \\
215 &	0.79 &	1 &	4.84e-6 &	277.23 &	564 &	7.78e-6 \\
216 &	0.82 &	3 &	1.39e-5 &	239.01 &	483 &	7.73e-6 \\
217 &	0.79 &	2 &	9.64e-6 &	239.01 &	483 &	7.73e-6 \\
218 &	0.82 &	3 &	1.39e-5 &	239.85 &	500 &	7.98e-6 \\
219 &	0.78 &	2 &	9.79e-6 &	239.85 &	500 &	7.98e-6 \\
220 &	0.86 &	4 &	1.78e-5 &	264.18 &	567 &	8.21e-6 \\
\end{tabular}
\end{center}

\begin{center}
\begin{tabular}{c|ccc|ccc}
\multicolumn{1}{c} {} & \multicolumn{3}{c} { Tracking } & \multicolumn{3}{c} { Background } \\
\hline
Pressure & Time & Counts & Rate & Time & Counts & Rate  \\
\small{$[mbar]$} & \small{$[hours]$} &  & \small{$[keV^{-1} s^{-1}  cm^{-2}]$} & \small{$[hours]$} &  & \small{$[keV^{-1} s^{-1} cm^{-2}]$} \\
\hline \hline
221 &	0.81 &	0 &	0.00e+0 &	264.18 &	567 &	8.21e-6 \\
222 &	0.87 &	2 &	8.82e-6 &	266.16 &	620 &	8.91e-6 \\
223 &	0.82 &	1 &	4.69e-6 &	266.16 &	620 &	8.91e-6 \\
224 &	0.83 &	0 &	0.00e+0 &	268.29 &	656 &	9.36e-6 \\
225 &	0.83 &	1 &	4.58e-6 &	268.29 &	656 &	9.36e-6 \\
226 &	0.90 &	3 &	1.27e-5 &	227.47 &	627 &	1.05e-5 \\
227 &	0.73 &	0 &	0.00e+0 &	227.47 &	627 &	1.05e-5 \\
228 &	0.92 &	1 &	4.15e-6 &	227.47 &	627 &	1.05e-5 \\
229 &	0.80 &	3 &	1.44e-5 &	227.47 &	627 &	1.05e-5 \\
230 &	0.90 &	2 &	8.46e-6 &	241.92 &	641 &	1.01e-5 \\
231 &	0.80 &	3 &	1.44e-5 &	241.92 &	641 &	1.01e-5 \\
232 &	0.93 &	3 &	1.23e-5 &	239.83 &	594 &	9.48e-6 \\
233 &	0.77 &	0 &	0.00e+0 &	239.83 &	594 &	9.48e-6 \\
234 &	0.92 &	4 &	1.66e-5 &	237.62 &	559 &	9.00e-6 \\
235 &	0.78 &	4 &	1.96e-5 &	237.62 &	559 &	9.00e-6 \\
236 &	0.94 &	2 &	8.15e-6 &	217.76 &	519 &	9.12e-6 \\
237 &	0.78 &	0 &	0.00e+0 &	217.76 &	519 &	9.12e-6 \\
238 &	0.91 &	2 &	8.42e-6 &	254.09 &	618 &	9.31e-6 \\
239 &	0.74 &	1 &	5.16e-6 &	254.09 &	618 &	9.31e-6 \\
240 &	0.98 &	2 &	7.78e-6 &	254.09 &	618 &	9.31e-6 \\
241 &	0.76 &	1 &	5.01e-6 &	254.09 &	618 &	9.31e-6 \\
242 &	0.96 &	3 &	1.19e-5 &	266.91 &	649 &	9.30e-6 \\
243 &	0.77 &	3 &	1.48e-5 &	266.91 &	649 &	9.30e-6 \\
244 &	0.97 &	3 &	1.19e-5 &	256.47 &	618 &	9.22e-6 \\
245 &	0.75 &	0 &	0.00e+0 &	256.47 &	618 &	9.22e-6 \\
246 &	0.96 &	1 &	3.98e-6 &	256.44 &	630 &	9.40e-6 \\
247 &	0.77 &	4 &	1.99e-5 &	256.44 &	630 &	9.40e-6 \\
248 &	0.95 &	1 &	4.05e-6 &	256.44 &	630 &	9.40e-6 \\
249 &	0.76 &	0 &	0.00e+0 &	256.44 &	630 &	9.40e-6 \\
250 &	0.72 &	0 &	0.00e+0 &	250.57 &	590 &	9.01e-6 \\
251 &	0.75 &	2 &	1.02e-5 &	250.57 &	590 &	9.01e-6 \\
251 &	0.96 &	4 &	1.59e-5 &	208.20 &	504 &	9.26e-6 \\
252 &	0.07 &	0 &	0.00e+0 &	208.20 &	504 &	9.26e-6 \\
252 &	0.08 &	0 &	0.00e+0 &	208.20 &	504 &	9.26e-6 \\
252 &	0.30 &	3 &	3.83e-5 &	208.20 &	504 &	9.26e-6 \\
252 &	0.69 &	2 &	1.11e-5 &	187.16 &	452 &	9.24e-6 \\
253 &	0.08 &	0 &	0.00e+0 &	187.16 &	452 &	9.24e-6 \\
254 &	0.98 &	1 &	3.91e-6 &	182.11 &	442 &	9.29e-6 \\
255 &	0.11 &	0 &	0.00e+0 &	182.11 &	442 &	9.29e-6 \\
\end{tabular}
\end{center}

\begin{center}
\begin{tabular}{c|ccc|ccc}
\multicolumn{1}{c} {} & \multicolumn{3}{c} { Tracking } & \multicolumn{3}{c} { Background } \\
\hline
Pressure & Time & Counts & Rate & Time & Counts & Rate  \\
\small{$[mbar]$} & \small{$[hours]$} &  & \small{$[keV^{-1} s^{-1}  cm^{-2}]$} & \small{$[hours]$} &  & \small{$[keV^{-1} s^{-1} cm^{-2}]$} \\
\hline \hline
255 &	0.10 &	2 &	7.65e-5 &	182.11 &	442 &	9.29e-6 \\
256 &	0.93 &	0 &	0.00e+0 &	166.75 &	422 &	9.68e-6 \\
257 &	0.76 &	1 &	5.05e-6 &	166.75 &	422 &	9.68e-6 \\
258 &	0.17 &	0 &	0.00e+0 &	166.80 &	442 &	1.01e-5 \\
258 &	0.30 &	0 &	0.00e+0 &	166.80 &	442 &	1.01e-5 \\
258 &	0.56 &	1 &	6.87e-6 &	166.80 &	442 &	1.01e-5 \\
259 &	0.95 &	6 &	2.42e-5 &	154.21 &	425 &	1.05e-5 \\
262 &	0.90 &	1 &	4.24e-6 &	101.68 &	286 &	1.08e-5 \\
263 &	0.76 &	3 &	1.51e-5 &	101.68 &	286 &	1.08e-5 \\
264 &	0.94 &	0 &	0.00e+0 &	101.68 &	286 &	1.08e-5 \\
265 &	0.76 &	1 &	5.02e-6 &	101.68 &	286 &	1.08e-5 \\
266 &	0.81 &	5 &	2.35e-5 &	 80.76 &	243 &	1.15e-5 \\
267 &	0.79 &	2 &	9.72e-6 &	 80.76 &	243 &	1.15e-5 \\
268 &	0.92 &	5 &	2.09e-5 &	 63.25 &	208 &	1.26e-5 \\
269 &	0.75 &	2 &	1.02e-5 &	 63.25 &	208 &	1.26e-5 \\
270 &	0.88 &	3 &	1.30e-5 &	115.34 &	255 &	8.46e-6 \\
271 &	0.82 &	0 &	0.00e+0 &	115.34 &	255 &	8.46e-6 \\
272 &	0.89 &	2 &	8.56e-6 &	160.50 &	348 &	8.30e-6 \\
273 &	0.81 &	1 &	4.74e-6 &	160.50 &	348 &	8.30e-6 \\
274 &	0.89 &	3 &	1.29e-5 &	160.50 &	348 &	8.30e-6 \\
275 &	0.80 &	2 &	9.61e-6 &	160.50 &	348 &	8.30e-6 \\
276 &	0.85 &	3 &	1.35e-5 &	181.65 &	405 &	8.53e-6 \\
277 &	0.83 &	2 &	9.25e-6 &	181.65 &	405 &	8.53e-6 \\
278 &	0.91 &	2 &	8.39e-6 &	202.79 &	440 &	8.30e-6 \\
279 &	0.77 &	2 &	9.92e-6 &	202.79 &	440 &	8.30e-6 \\
280 &	0.88 &	0 &	0.00e+0 &	223.88 &	482 &	8.24e-6 \\
281 &	0.77 &	1 &	4.97e-6 &	223.88 &	482 &	8.24e-6 \\
282 &	0.86 &	0 &	0.00e+0 &	256.36 &	561 &	8.37e-6 \\
283 &	0.84 &	1 &	4.58e-6 &	256.36 &	561 &	8.37e-6 \\
284 &	0.83 &	1 &	4.63e-6 &	256.67 &	557 &	8.30e-6 \\
285 &	0.84 &	2 &	9.10e-6 &	256.67 &	557 &	8.30e-6 \\
286 &	0.83 &	1 &	4.63e-6 &	235.53 &	505 &	8.20e-6 \\
287 &	0.82 &	2 &	9.29e-6 &	235.53 &	505 &	8.20e-6 \\
288 &	0.82 &	0 &	0.00e+0 &	235.66 &	506 &	8.22e-6 \\
289 &	0.82 &	2 &	9.32e-6 &	235.66 &	506 &	8.22e-6 \\
290 &	0.83 &	1 &	4.63e-6 &	226.95 &	487 &	8.21e-6 \\
291 &	0.84 &	5 &	2.28e-5 &	226.95 &	487 &	8.21e-6 \\
292 &	0.83 &	1 &	4.61e-6 &	248.03 &	529 &	8.16e-6 \\
293 &	0.82 &	3 &	1.40e-5 &	248.03 &	529 &	8.16e-6 \\
\end{tabular}
\end{center}

\begin{center}
\begin{tabular}{c|ccc|ccc}
\multicolumn{1}{c} {} & \multicolumn{3}{c} { Tracking } & \multicolumn{3}{c} { Background } \\
\hline
Pressure & Time & Counts & Rate & Time & Counts & Rate  \\
\small{$[mbar]$} & \small{$[hours]$} &  & \small{$[keV^{-1} s^{-1}  cm^{-2}]$} & \small{$[hours]$} &  & \small{$[keV^{-1} s^{-1} cm^{-2}]$} \\
\hline \hline
294 &	0.83 &	3 &	1.38e-5 &	223.97 &	488 &	8.34e-6 \\
295 &	0.80 &	0 &	0.00e+0 &	223.97 &	488 &	8.34e-6 \\
296 &	0.73 &	0 &	0.00e+0 &	245.16 &	536 &	8.37e-6 \\
297 &	0.78 &	2 &	9.87e-6 &	245.25 &	528 &	8.24e-6 \\
297 &	1.68 &	5 &	1.14e-5 &	224.10 &	493 &	8.42e-6 \\
298 &	0.80 &	3 &	1.43e-5 &	245.35 &	547 &	8.53e-6 \\
299 &	0.81 &	1 &	4.73e-6 &	245.35 &	547 &	8.53e-6 \\
300 &	0.83 &	2 &	9.26e-6 &	224.27 &	499 &	8.51e-6 \\
301 &	0.79 &	1 &	4.84e-6 &	224.27 &	499 &	8.51e-6 \\
302 &	0.82 &	1 &	4.67e-6 &	224.33 &	498 &	8.49e-6 \\
303 &	0.78 &	2 &	9.80e-6 &	224.33 &	498 &	8.49e-6 \\
304 &	0.81 &	3 &	1.41e-5 &	224.33 &	489 &	8.34e-6 \\
305 &	0.77 &	2 &	9.93e-6 &	224.33 &	489 &	8.34e-6 \\
306 &	0.82 &	2 &	9.30e-6 &	224.26 &	488 &	8.33e-6 \\
307 &	0.78 &	2 &	9.84e-6 &	224.26 &	488 &	8.33e-6 \\
308 &	0.82 &	1 &	4.67e-6 &	224.30 &	505 &	8.61e-6 \\
309 &	0.78 &	1 &	4.90e-6 &	224.30 &	505 &	8.61e-6 \\
310 &	0.80 &	5 &	2.38e-5 &	245.54 &	551 &	8.59e-6 \\
311 &	0.78 &	1 &	4.90e-6 &	245.54 &	551 &	8.59e-6 \\
312 &	0.81 &	3 &	1.41e-5 &	233.32 &	534 &	8.76e-6 \\
313 &	0.77 &	3 &	1.49e-5 &	233.32 &	534 &	8.76e-6 \\
314 &	0.80 &	2 &	9.59e-6 &	233.53 &	527 &	8.63e-6 \\
315 &	0.77 &	0 &	0.00e+0 &	233.53 &	527 &	8.63e-6 \\
316 &	0.80 &	1 &	4.78e-6 &	233.62 &	526 &	8.61e-6 \\
317 &	0.76 &	1 &	5.03e-6 &	233.62 &	526 &	8.61e-6 \\
317 &	1.60 &	2 &	4.78e-6 &	265.97 &	615 &	8.85e-6 \\
318 &	0.80 &	2 &	9.58e-6 &	233.64 &	530 &	8.68e-6 \\
319 &	0.77 &	4 &	1.99e-5 &	233.64 &	530 &	8.68e-6 \\
320 &	0.78 &	1 &	4.93e-6 &	244.69 &	562 &	8.79e-6 \\
321 &	0.77 &	3 &	1.48e-5 &	244.69 &	562 &	8.79e-6 \\
322 &	0.78 &	0 &	0.00e+0 &	259.58 &	591 &	8.71e-6 \\
323 &	0.77 &	0 &	0.00e+0 &	259.58 &	591 &	8.71e-6 \\
324 &	0.80 &	3 &	1.43e-5 &	238.38 &	545 &	8.75e-6 \\
324 &	1.60 &	1 &	2.39e-6 &	238.47 &	549 &	8.81e-6 \\
325 &	0.75 &	3 &	1.53e-5 &	238.38 &	545 &	8.75e-6 \\
326 &	0.76 &	1 &	5.03e-6 &	259.78 &	568 &	8.37e-6 \\
327 &	0.76 &	1 &	5.06e-6 &	259.78 &	568 &	8.37e-6 \\
328 &	0.76 &	2 &	1.01e-5 &	238.51 &	535 &	8.58e-6 \\
329 &	0.76 &	0 &	0.00e+0 &	238.51 &	535 &	8.58e-6 \\
\end{tabular}
\end{center}

\begin{center}
\begin{tabular}{c|ccc|ccc}
\multicolumn{1}{c} {} & \multicolumn{3}{c} { Tracking } & \multicolumn{3}{c} { Background } \\
\hline
Pressure & Time & Counts & Rate & Time & Counts & Rate  \\
\small{$[mbar]$} & \small{$[hours]$} &  & \small{$[keV^{-1} s^{-1}  cm^{-2}]$} & \small{$[hours]$} &  & \small{$[keV^{-1} s^{-1} cm^{-2}]$} \\
\hline \hline
330 &	0.77 &	2 &	9.97e-6 &	240.44 &	545 &	8.67e-6 \\
331 &	0.75 &	2 &	1.03e-5 &	240.44 &	545 &	8.67e-6 \\
332 &	0.79 &	5 &	2.43e-5 &	296.58 &	681 &	8.79e-6 \\
333 &	0.71 &	6 &	3.24e-5 &	296.58 &	681 &	8.79e-6 \\
334 &	0.76 &	0 &	0.00e+0 &	243.19 &	553 &	8.70e-6 \\
335 &	0.72 &	1 &	5.32e-6 &	243.19 &	553 &	8.70e-6 \\
336 &	0.73 &	1 &	5.23e-6 &	243.19 &	553 &	8.70e-6 \\
337 &	0.75 &	2 &	1.02e-5 &	243.19 &	553 &	8.70e-6 \\
338 &	0.75 &	0 &	0.00e+0 &	229.51 &	523 &	8.72e-6 \\
339 &	0.73 &	2 &	1.05e-5 &	229.51 &	523 &	8.72e-6 \\
340 &	0.93 &	1 &	4.12e-6 &	229.88 &	555 &	9.24e-6 \\
341 &	0.53 &	1 &	7.19e-6 &	229.88 &	555 &	9.24e-6 \\
342 &	0.73 &	1 &	5.26e-6 &	207.19 &	482 &	8.90e-6 \\
343 &	0.72 &	3 &	1.59e-5 &	207.19 &	482 &	8.90e-6 \\
344 &	0.72 &	0 &	0.00e+0 &	228.19 &	529 &	8.87e-6 \\
345 &	0.73 &	5 &	2.63e-5 &	228.19 &	529 &	8.87e-6 \\
346 &	0.73 &	2 &	1.06e-5 &	181.59 &	418 &	8.81e-6 \\
347 &	0.74 &	4 &	2.06e-5 &	181.59 &	418 &	8.81e-6 \\
348 &	0.71 &	3 &	1.61e-5 &	202.97 &	462 &	8.71e-6 \\
349 &	0.73 &	2 &	1.04e-5 &	202.97 &	462 &	8.71e-6 \\
349 &	1.07 &	1 &	3.59e-6 &	223.18 &	509 &	8.73e-6 \\
350 &	0.72 &	0 &	0.00e+0 &	224.24 &	503 &	8.58e-6 \\
351 &	0.72 &	3 &	1.60e-5 &	224.24 &	503 &	8.58e-6 \\
352 &	0.71 &	3 &	1.61e-5 &	244.42 &	559 &	8.75e-6 \\
353 &	0.71 &	4 &	2.17e-5 &	244.42 &	559 &	8.75e-6 \\
354 &	0.73 &	0 &	0.00e+0 &	244.44 &	567 &	8.88e-6 \\
355 &	0.71 &	1 &	5.42e-6 &	244.44 &	567 &	8.88e-6 \\
356 &	0.75 &	3 &	1.54e-5 &	223.10 &	522 &	8.95e-6 \\
357 &	0.67 &	2 &	1.15e-5 &	223.10 &	522 &	8.95e-6 \\
358 &	0.74 &	1 &	5.15e-6 &	223.19 &	515 &	8.83e-6 \\
358 &	1.48 &	2 &	5.16e-6 &	244.70 &	583 &	9.12e-6 \\
359 &	0.71 &	5 &	2.71e-5 &	223.19 &	515 &	8.83e-6 \\
360 &	0.70 &	1 &	5.44e-6 &	222.88 &	534 &	9.17e-6 \\
361 &	0.71 &	0 &	0.00e+0 &	222.88 &	534 &	9.17e-6 \\
362 &	0.71 &	1 &	5.38e-6 &	244.31 &	591 &	9.26e-6 \\
363 &	0.71 &	1 &	5.42e-6 &	244.31 &	591 &	9.26e-6 \\
363 &	0.67 &	1 &	5.70e-6 &	266.11 &	640 &	9.20e-6 \\
364 &	0.70 &	3 &	1.65e-5 &	266.11 &	640 &	9.20e-6 \\
364 &	0.69 &	1 &	5.57e-6 &	235.20 &	571 &	9.29e-6 \\
\end{tabular}
\end{center}

\begin{center}
\begin{tabular}{c|ccc|ccc}
\multicolumn{1}{c} {} & \multicolumn{3}{c} { Tracking } & \multicolumn{3}{c} { Background } \\
\hline
Pressure & Time & Counts & Rate & Time & Counts & Rate  \\
\small{$[mbar]$} & \small{$[hours]$} &  & \small{$[keV^{-1} s^{-1}  cm^{-2}]$} & \small{$[hours]$} &  & \small{$[keV^{-1} s^{-1} cm^{-2}]$} \\
\hline \hline
365 &	0.71 &	1 &	5.37e-6 &	235.20 &	571 &	9.29e-6 \\
366 &	0.71 &	4 &	2.16e-5 &	213.93 &	530 &	9.48e-6 \\
367 &	0.71 &	0 &	0.00e+0 &	213.93 &	530 &	9.48e-6 \\
368 &	0.70 &	3 &	1.63e-5 &	213.93 &	530 &	9.48e-6 \\
369 &	0.70 &	2 &	1.10e-5 &	213.93 &	530 &	9.48e-6 \\
370 &	0.69 &	5 &	2.76e-5 &	171.32 &	427 &	9.54e-6 \\
371 &	0.71 &	0 &	0.00e+0 &	171.32 &	427 &	9.54e-6 \\
372 &	0.70 &	1 &	5.50e-6 &	149.92 &	375 &	9.57e-6 \\
373 &	0.69 &	1 &	5.51e-6 &	149.92 &	375 &	9.57e-6 \\
374 &	0.72 &	1 &	5.34e-6 &	149.92 &	375 &	9.57e-6 \\
375 &	0.70 &	2 &	1.09e-5 &	149.92 &	375 &	9.57e-6 \\
375 &	1.53 &	3 &	7.48e-6 &	160.37 &	290 &	6.92e-6 \\
376 &	0.75 &	1 &	5.10e-6 &	181.53 &	332 &	7.00e-6 \\
377 &	0.71 &	1 &	5.40e-6 &	181.53 &	332 &	7.00e-6 \\
378 &	0.60 &	0 &	0.00e+0 &	202.75 &	375 &	7.08e-6 \\
379 &	0.66 &	1 &	5.83e-6 &	202.75 &	375 &	7.08e-6 \\
380 &	0.76 &	0 &	0.00e+0 &	223.99 &	415 &	7.09e-6 \\
381 &	0.75 &	2 &	1.03e-5 &	223.99 &	415 &	7.09e-6 \\
382 &	0.75 &	1 &	5.11e-6 &	223.99 &	415 &	7.09e-6 \\
383 &	0.73 &	2 &	1.04e-5 &	223.99 &	415 &	7.09e-6 \\
384 &	0.76 &	0 &	0.00e+0 &	245.22 &	451 &	7.04e-6 \\
384 &	1.57 &	1 &	2.44e-6 &	246.84 &	455 &	7.05e-6 \\
384 &	1.55 &	3 &	7.40e-6 &	253.76 &	462 &	6.97e-6 \\
385 &	0.74 &	0 &	0.00e+0 &	245.22 &	451 &	7.04e-6 \\
386 &	0.78 &	2 &	9.83e-6 &	253.97 &	450 &	6.78e-6 \\
387 &	0.74 &	2 &	1.03e-5 &	253.97 &	450 &	6.78e-6 \\
388 &	0.77 &	4 &	1.98e-5 &	253.89 &	462 &	6.96e-6 \\
389 &	0.76 &	1 &	5.02e-6 &	253.89 &	462 &	6.96e-6 \\
390 &	0.80 &	1 &	4.76e-6 &	233.08 &	434 &	7.12e-6 \\
391 &	0.71 &	1 &	5.36e-6 &	233.08 &	434 &	7.12e-6 \\
392 &	0.76 &	0 &	0.00e+0 &	233.14 &	427 &	7.01e-6 \\
393 &	0.77 &	2 &	9.93e-6 &	233.14 &	427 &	7.01e-6 \\
394 &	0.77 &	2 &	9.95e-6 &	233.25 &	431 &	7.07e-6 \\
395 &	0.78 &	2 &	9.79e-6 &	233.25 &	431 &	7.07e-6 \\
396 &	0.78 &	0 &	0.00e+0 &	233.21 &	431 &	7.07e-6 \\
397 &	0.76 &	0 &	0.00e+0 &	233.21 &	431 &	7.07e-6 \\
398 &	0.77 &	2 &	9.91e-6 &	232.99 &	427 &	7.01e-6 \\
399 &	0.76 &	1 &	5.03e-6 &	232.99 &	427 &	7.01e-6 \\
400 &	0.79 &	3 &	1.46e-5 &	241.88 &	438 &	6.93e-6 \\
\end{tabular}
\end{center}

\begin{center}
\begin{tabular}{c|ccc|ccc}
\multicolumn{1}{c} {} & \multicolumn{3}{c} { Tracking } & \multicolumn{3}{c} { Background } \\
\hline
Pressure & Time & Counts & Rate & Time & Counts & Rate  \\
\small{$[mbar]$} & \small{$[hours]$} &  & \small{$[keV^{-1} s^{-1}  cm^{-2}]$} & \small{$[hours]$} &  & \small{$[keV^{-1} s^{-1} cm^{-2}]$} \\
\hline \hline
401 &	0.78 &	1 &	4.91e-6 &	241.88 &	438 &	6.93e-6 \\
402 &	0.78 &	0 &	0.00e+0 &	220.65 &	402 &	6.97e-6 \\
403 &	0.78 &	2 &	9.87e-6 &	220.65 &	402 &	6.97e-6 \\
404 &	0.77 &	0 &	0.00e+0 &	199.61 &	359 &	6.88e-6 \\
405 &	0.77 &	1 &	4.96e-6 &	199.61 &	359 &	6.88e-6 \\
406 &	0.77 &	2 &	9.97e-6 &	178.32 &	327 &	7.02e-6 \\
407 &	0.79 &	0 &	0.00e+0 &	178.32 &	327 &	7.02e-6 \\
408 &	0.76 &	1 &	5.05e-6 &	157.19 &	296 &	7.20e-6 \\
409 &	0.81 &	1 &	4.73e-6 &	157.19 &	296 &	7.20e-6 \\
410 &	0.81 &	1 &	4.75e-6 &	136.00 &	243 &	6.84e-6 \\
411 &	0.78 &	0 &	0.00e+0 &	136.00 &	243 &	6.84e-6 \\
\end{tabular}
\end{center}

\begin{landscape}
\begin{table}[h!]
\caption[h]{Total background and tracking rates as a function of the energy for each of the Micromegas detectors taking data in 2008.  }
\begin{center}

\begin{tabular}{c|cc|c|cc|c|cc|c}

\multicolumn{1}{c|} {} & \multicolumn{3}{c|} { B2 } & \multicolumn{3}{c|} { B4 } & \multicolumn{3}{c} { M10 }  \\
\hline

\multicolumn{1}{c|} {} & \multicolumn{2}{c|} { Tracking } & \multicolumn{1}{c} { Background } & \multicolumn{2}{|c} { Tracking } & \multicolumn{1}{|c|} { Background } & \multicolumn{2}{c} { Tracking } & \multicolumn{1}{|c} { Background } \\
\hline

Energy & Counts & \multicolumn{2}{c|} {Rate} & Counts & \multicolumn{2}{c|} {Rate} & Counts & \multicolumn{2}{c} {Rate} \\
\small{$[keV]$} &   & \multicolumn{2}{c|} {\small{$[keV^{-1} s^{-1} cm^{-2}]$}} &  &  \multicolumn{2}{c|} {\small{$[keV^{-1} s^{-1} cm^{-2}]$}} &  & \multicolumn{2}{c} {\small{$[keV^{-1} s^{-1} cm^{-2}]$}}  \\
\hline \hline
2.25 &	21 &	1.16e-05 	&	1.09e-5		&	17	&	7.20e-6			&	9.21e-6		&	4	&	4.78e-6	&		6.22e-6 \\
2.75 &	22 &	1.21e-05 	&	1.16e-5		&	24	&	1.01e-5			&	1.13e-5		&	1	&	1.19e-6	&		6.97e-6 \\
3.25 &	18 &	9.94e-06 	&	1.07e-5		&	23	&	9.74e-6			&	8.85e-6		&	5	&	5.98e-6	&		6.29e-6 \\
3.75 &	17 &	9.39e-06 	&	9.08e-6		&	19	&	8.04e-6			&	6.33e-6		&	2	&	2.39e-6	&		6.09e-6 \\
4.25 &	10 &	5.52e-06 	&	6.66e-6		&	23	&	9.74e-6			&	6.00e-6		&	2	&	2.39e-6	&		5.14e-6 \\
4.75 &	17 &	9.39e-06 	&	6.56e-6		&	16	&	6.77e-6			&	6.41e-6		&	7	&	8.38e-6	&		5.82e-6 \\
5.25 &	16 &	8.84e-06 	&	7.75e-6		&	15	&	6.35e-6			&	8.69e-6		&	4	&	4.78e-6	&		8.25e-6 \\
5.75 &	9  &    4.97e-06 	&	9.01e-6		&	26	&	1.10e-5			&	1.04e-5		&	5	&	5.98e-6	&		9.33e-6 \\
6.25 &	17 &	9.39e-06 	&	9.87e-6		&	24	&	1.01e-5			&	9.35e-6		&	11	&	1.31e-5	&		8.86e-6 \\
6.75 &	15 &	8.29e-06 	&	9.44e-6		&	19	&	8.04e-6			&	1.04e-5		&	6	&	7.18e-6	&		7.91e-6 \\
7.25 &	16 &	8.84e-06 	&	8.11e-6		&	14	&	5.93e-6			&	8.66e-6		&	9	&	1.07e-5	&		6.90e-6 \\
7.75 &	12 &	6.63e-06 	&	7.31e-6		&	11	&	4.65e-6			&	8.00e-6		&	10	&	1.19e-5	&		6.56e-6 \\
8.25 &	5  &    2.76e-06 	&	5.40e-6		&	14	&	5.93e-6			&	6.85e-6		&	8	&	9.57e-6	&		6.56e-6 \\
8.75 &	8  &    4.42e-06 	&	4.21e-6		&	11	&	4.65e-6			&	6.03e-6		&	7	&	8.38e-6	&		6.36e-6 \\
\end{tabular}

\end{center}
\end{table}
\end{landscape}

\bibliographystyle{unsrt}
\bibliography{Biblio}

\end{document}